\documentclass[12pt,oneside]{article}
\usepackage{amsfonts}
\newtheorem{bphzthm}{Theorem}
\newtheorem{bphzlemma}{Lemma}
\newtheorem{bphzfirstseriesobs}{(\hspace{-0.7ex}}
\newtheorem{bphzobservation}{\hspace{-5.0pt}}
\newtheorem{bphzrule}{Rule}

\begin{document}


\begin{center}
{\large\bf A BPHZ convergence proof in Euclidean position
space\\}
\vspace{0.14cm}
\vspace*{.05in}
{Chris Austin\footnote{Email:
chrisaustin@ukonline.co.uk}\\
\small 33 Collins Terrace, Maryport, Cumbria CA15 8DL,
England\\
}
\end{center}
\begin{center}
{\bf Abstract}
\end{center}
\noindent Two BPHZ convergence theorems are proved directly
in Euclidean position space, without exponentiating the
propagators, making use of the Cluster Convergence Theorem
presented previously.  The first theorem proves the
absolute convergence of arbitrary BPHZ-renormalized Feynman
diagrams, when counterterms are allowed for
one-line-reducible subdiagrams, as well as for
one-line-irreducible subdiagrams.  The second theorem
proves the conditional convergence of arbitrary
BPHZ-renormalized Feynman diagrams, when counterterms are
allowed only for one-line-irreducible subdiagrams.
Although the convergence in this case is only conditional,
there is only one natural way to approach the limit, namely
from propagators smoothly regularized at short distances,
so that the integrations by parts needed to reach an
absolutely convergent integrand can be carried out, without
picking up short-distance surface terms.  Neither theorem
requires translation invariance, but the second theorem
assumes a much weaker property, called ``translation
smoothness''.  Both theorems allow the propagators in the
counterterms to differ, at long distances, from the
propagators in the direct terms.  For massless theories,
this makes it possible to eliminate all the long-distance
divergences from the counterterms, without altering the
propagators in the direct terms.  Massless theories can
thus be studied directly, without introducing a regulator
mass and taking the limit as it tends to zero, and without
infra-red subtractions.

\vspace{0.8cm}

\noindent {\Large\bf Preface}

\vspace{0.3cm}

\noindent BPHZ renormalization was introduced by Bogoliubov
and Parasiuk \cite{Bogoliubov Parasiuk 1,
Bogoliubov Parasiuk 2, Bogoliubov Shirkov}, following
earlier work by Dyson \cite{Dyson 1, Dyson 2}, and Salam
\cite{Salam 1, Salam 2}.  A BPHZ
convergence proof, in the parameter space of the
exponentiated propagators, was given by Hepp \cite{Hepp}.
A power-counting convergence theorem, in Euclidean momentum
space, was proved by Weinberg \cite{Weinberg}, and a
simplified version of the theorem was proved by Hahn and
Zimmermann \cite{Hahn Zimmermann}. The theorem was extended
to Minkowski signature by Zimmermann \cite{Zimmermann
Minkowski}.  These theorems were used to give a BPHZ
convergence proof, in momentum space, by
Zimmermann~\cite{Zimmermann convergence}.

The aim of the present paper, and the preceding one,
\cite{Cluster Convergence Theorem}, is to find out how to
calculate BPHZ-renormalized Feynman diagrams, directly in
Euclidean position space, without exponentiating the
propagators. The original motivation was to study the
renormalization of the Group-Variation Equations
\cite{Group-Variation Equations,
A brief summary of the group-variation equations}.
A brief
discussion of the problems involved, and the way they are
overcome, is given in Section 6.3 of
\cite{Group-Variation Equations}.

The Cluster Convergence Theorem \cite{Cluster Convergence
Theorem} is intended to serve as a position-space
substitute for Weinberg's momentum-space power counting
theorem, giving a guarantee that a position-space integral
will converge, if the integrand satisfies certain bounds.
The present paper gives
a way, for an arbitrary BPHZ-renormalized position-space
integrand, of cutting up the Cartesian product, of the
configuration space of the vertices, and the list of all
the forests, into a finite number of sectors, each of which
is, itself, the Cartesian product, of a subset of the
configuration space of the vertices, and a suitable set of
forests, such that the bounds on the integrand, required to
apply the Cluster Convergence Theorem, can be proved to
hold, in each sector.  The idea of $ \sigma $-clusters,
used in the Cluster
Convergence Theorem, arose from seeking an analogue, in
position space, of the momentum space ``clusters'', used by
't Hooft \cite{'t Hooft}.

What follows this preface is the \LaTeX2e transcription of
my 1993 paper, with references added.  Conventions, some
definitions, and some basic combinatoric results, are given
in Section \ref{Section 2}, on
page \pageref{Section 2}.
The Cluster Convergence Theorem is reviewed in Section
\ref{Section 3}, on
page \pageref{Section 3}.  The sectors
described above are constructed in Section \ref{Section 4},
on
page \pageref{Section 4}.  The BPHZ integrands, for the
forests in such a sector, are combined, by means of the
Taylor remainder formula, into a form suitable for
bounding, in Section \ref{Section 5}, on
page
\pageref{Section 5}.  The main result of Section
\ref{Section 5} is Lemma \ref{Lemma 22}, on
page
\pageref{Lemma 22}.

The First Convergence Theorem, in Section \ref{Section 6},
on
page \pageref{Section 6}, establishes the absolute
convergence, in Euclidean position space, of an arbitrary
BPHZ-renormalized Feynman diagram, when counterterms are
permitted for one-line reducible subdiagrams, as well as
for one-line irreducible subdiagrams.  Translation
invariance is not required.

The Second Convergence Theorem, in Section \ref{Section 7},
on
page \pageref{Section 7}, establishes the conditional
convergence, in Euclidean position space, of an arbitrary
BPHZ-renormalized Feynman diagram, when counterterms are
permitted only for one-line irreducible subdiagrams.
Although the convergence is only conditional, there is only
one natural way to approach the limit, namely by smooth
short-distance regularizations of the propagators, so that
the integrations by parts, needed to reach absolutely
convergent integrands, can be performed without picking up
short-distance surface terms. The result is demonstrated to
be independent of the details, of the smooth
short-distance regularizations of the propagators, provided
they satisfy the broad requirements given. Once the
integrations by parts are done, the short-distance
regularizations of the propagators can be dropped.
Translation invariance is not required, but a much weaker
property, called translation smoothness, is assumed.

The proof of the Second Convergence Theorem, Theorem
\ref{Theorem 2}, on
page \pageref{Theorem 2}, involves the
addition and subtraction of additional counterterms for
one-line reducible subdiagrams, which are then shown to be
conditionally convergent. This method enables the
integrations by parts to be done, after a suitable change
of variables, without crossing the boundaries of the
sectors, that the Cartesian
product of the configuration space of the vertices, and the
list of all the forests, is cut up into, for the
application of the Cluster Convergence Theorem.  The
necessary algebraic identities are established in Lemma
\ref{Lemma 34}, on
page \pageref{Lemma 34}, and Lemma
\ref{Lemma 35}, on
page \pageref{Lemma 35}.
\enlargethispage{\baselineskip}

The results in Sections \ref{Section 2} to \ref{Section 7}
are presented in a general form, without specific reference
to Feynman diagrams, and the application of the results of
these sections, to BPHZ-renormalized Feynman diagrams, is
explained in Section \ref{Section 8}, on
page
\pageref{Section 8}.

Both the convergence theorems allow the use of propagators,
in the counterterms, which differ, at long distances, from
the propagators in the direct terms, so that, for theories
with massless particles, the long-distance divergences, in
the counterterms, can be cut off, without altering the
propagators in the direct terms. The form of the long
distance cutoff, allowed for the propagators, in the
counterterms, is that the propagator is multiplied by a
factor, that is equal to 1, when the distance between the
propagator ends is less than a fixed, nonzero, distance,
then decreases smoothly, from 1, to 0, as the distance
between the propagator ends increases, to a larger fixed
distance, then is equal to 0, when the distance, between
the propagator ends, is greater than the larger fixed
distance. This factor is assumed to be continuously
differentiable, as many times as needed.  For theories with
dimensional transmutation, such as QCD, the fixed, nonzero,
distance, at which the smooth long-distance cutoffs of the
propagators in the counterterms commence, can serve as the
origin of the renormalization mass scale, while for
theories with a dimensional coupling constant, such as
gravity, the long-distance cutoffs would commence at a
numerical multiple of the distance defined by the
dimensional coupling constant.  The idea of
seeking a BPHZ convergence proof, in which the
propagators, in the counterterms, are allowed to differ, at
large distances, from the propagators in the direct terms,
so that, for theories with massless particles, the
non-physical infra-red divergences, which the standard BPHZ
method produces in the counterterms, would not occur, was
suggested by a method used to calculate certain UV
counterterms, by Chetyrkin, Kataev, and Tkachov
\cite{Chetyrkin Kataev Tkachov}.

If long-distance cutoffs on massless propagators in the
counterterms, as allowed by the convergence proofs in the
present paper, are not used, then for theories with
massless particles, it is necessary to use the BPHZL method
\cite{Lowenstein 1, Lowenstein 2, Lowenstein Zimmermann,
Lowenstein 3}, which involves introducing a regulator mass
for massless particles, performing the usual BPHZ
subtractions in the presence of the regulator mass, to
remove the UV divergences, then performing further
subtractions, around the limit of zero regulator mass, to
remove the IR divergences, before the regulator mass is
allowed to tend to zero.  This method was recently used in
\cite{Quadri 2}.

The existence of overlapping divergences means that the way
the BPHZ forests are grouped into good sets of forests,
such that the short-distance divergences cancel between the
members of a good set of forests, has to be different, in
different regions of the configuration space of the
vertices of the diagram, just as it has to be different, in
different regions of the parameter space of the
exponentiated propagators, in Hepp's method, and in
different regions of momentum space, in Zimmerman's method.
In the present paper, the basis of the contruction of the
good sets of forests, for each position-space
configuration, $ x $, of the vertices of the Feynman
diagram, is the definition, on page
\pageref{Start of original page 20}, of a set, $ \Omega
\left( x \right) $, whose members are pairs, $ \left( F, G
\right) $, of forests, such that $ F \subseteq G $, and
each member, $ A $, of $ G $, that is not a member of $ F
$, is, in an appropriate sense, a ``cluster'', in the
position-space configuration, $ x $, of the vertices of the
diagram, after contracting all the members of $ F $ that
are strict subsets of $ A $, and ignoring the positions of
the outer ends of any legs of $ A $, that lie outside a
member of $ F $ that contains $ A $ as a strict subset.
The idea is that a ``cluster'' is a subdiagram whose
vertices are all close to one another, in the
position-space configuration, $ x $, of the vertices of the
diagram, and thus may need to have a counterterm
subtracted, in order to cancel an ultraviolet divergence,
that would otherwise come from somewhere near $ x $, in
the configuration space of the vertices of the diagram.  An
example of what can go wrong, if one does not contract the
members of $ F $, that are strict subsets of $ A $, before
applying the ``cluster'' condition, is given on pages 159
to 160 of \cite{Group-Variation Equations}.  The precise
definition, of a ``cluster'', is chosen so that one obtains
Lemma \ref{Lemma 7}, on page \pageref{Lemma 7}, Lemma
\ref{Lemma 10}, on page \pageref{Lemma 10}, and Lemma
\ref{Lemma 11}, on page \pageref{Lemma 11}.  Lemma
\ref{Lemma 7} states that if $ \left( F, G \right) $ and $
\left( F', G' \right) $ are members of $ \Omega \left( x
\right) $, such that $ \left( F \cup F' \right) \subseteq
\left( G \cap G' \right) $, then $ G \cup G' $ is a forest,
and $ \left( \left( F \cap F' \right), \left( G \cup G'
\right) \right) $ is a member of $ \Omega \left( x \right)
$.  Once one has Lemma \ref{Lemma 7}, it follows by
combinatorics alone, in Lemma \ref{Lemma 8}, on page
\pageref{Lemma 8}, and Lemma \ref{Lemma 9}, on page
\pageref{Lemma 9}, independent of the details of the
definition of a ``cluster'', that for any forest, $ F $,
there is a unique member, $ \left( P, Q \right) $, of $
\Omega \left( x \right) $, such that $ P \subseteq F
\subseteq Q $ holds, and there is no member, $ \left( P',
Q' \right) $, of $ \Omega \left( x \right) $, different
from $ \left( P, Q \right) $, such that $ P' \subseteq P $
and $ Q \subseteq Q' $ both hold.  The good set of forests,
to which the forest, $ F $, belongs, at the position-space
configuration, $ x $, of the vertices of the diagram, is
then the set of all the forests, $ G $, such that $ P
\subseteq G \subseteq Q $ holds.  And if $ \left( P, Q
\right) $ is a member of $ \Omega \left( x \right) $, that
defines a good set of forests, in this way, then Lemma
\ref{Lemma 10} gives lower bounds, used on page
\pageref{Start of original page 112}, on the diameters, in
the position space configuration, $ x $, of the vertices of
the diagram, of the members, $ A $, of $ P $, that have two
or more vertices, after contracting the members of $ P $
that are strict subdiagrams of $ A $, and Lemma
\ref{Lemma 11}, used on page
\pageref{Start of original page 121}, ensures that if $ A $
is either the whole diagram, or a member of $ P $, that has
two or more vertices, then, after contracting the members
of $ P $, that are strict subdiagrams of $ A $, the
connected components of any $ \sigma $-cluster, of the
restriction of the diagram to $ A $, are, under appropriate
conditions, members of $ \left( Q \vdash P \right) $.

It seems possible that analogues of the set $ \Omega \left(
x \right) $, described above, might also be used as the
basis for constructing good sets of forests, in the
parameter space of the exponentiated propagators, and in
momentum space.  The situation is simplest in the space of
the integration parameters of the exponentiated
propagators, usually called the $ \alpha $-parameters,
because the $ R $-operation does not change the value of
the $ \alpha $-parameter of a line, whereas it does change
the lengths of lines, and the diameters of subdiagrams, in
position space, because it changes the positions of the
inner ends of the legs of a subdiagram, when that subdiagram
is contracted, and it also changes the magnitudes of the
line momenta, in momentum space, because it sets the
external momenta of a subdiagram to zero, in the internal
lines of that subdiagram, when that subdiagram is
contracted.

An example of a definition of $ \Omega \left( \alpha
\right) $, in the parameter space of the exponentiated
propagators, broadly analogous to the definition of
$ \Omega \left( x \right) $, and for which an analogue of
Lemma \ref{Lemma 7} can be derived, is as follows.  For
each forest, $ F $, subdiagram, $ A $, and assignment, $
\alpha $, of the values of the $ \alpha $-parameters of the
lines of the diagram, we define $ \lambda \left( F, A,
\alpha \right) $ to be the maximum value of $
\alpha_{\Delta} $, among the internal lines, $ \Delta $, of
$ A $, that are not internal lines of any member, of $ F $,
that is a strict subset of $ A $.  This implies that for
forests, $ F $, and $ G $, such that $ F \subseteq G $,
$ \lambda \left( G, A, \alpha \right) \leq \lambda \left(
F, A, \alpha \right) $ holds.  We then define $ \Omega
\left( \alpha \right) $ to be the set of all ordered pairs,
$ \left( P, Q \right) $ of forests of one-line-irreducible
subdiagrams, such that $ P \subseteq Q $, and such that,
for all $ A \in \left( Q \vdash P \right) $, and for every
line, $ \Delta $, with one end in $ A $, and its other end
not in $ A $, but not outside any member of $Q$, if any,
that contains $ A $ as a strict subset, $ \lambda \left( Q,
A, \alpha \right) < \alpha_{\Delta} $.  One can prove that
this definition implies that, if $ \left( P, Q \right) \in
\Omega \left( \alpha \right) $, and $ A $ is a connected
subdiagram, with at least two vertices, then $ \lambda
\left( Q, A, \alpha \right) = \lambda \left( P, A, \alpha
\right) $.  Hence for all $ F $, such that $ P \subseteq F
\subseteq Q $, $ \lambda \left( F, A, \alpha \right) =
\lambda \left( Q, A, \alpha \right) = \lambda \left( P, A,
\alpha \right) $.  To prove the analogue of Lemma
\ref{Lemma 7}, suppose that $ \left( P, Q \right) \in
\Omega \left( \alpha \right) $ and $ \left( P', Q' \right)
\in \Omega \left( \alpha \right) $, and $ \left( P \cup P'
\right) \subseteq \left( Q \cap Q' \right) $.  Suppose some
member, $ A $, of $ Q \vdash P $, overlaps some member, $
A' $, of $ Q' \vdash P' $.  For any ordered pair, $ \left(
F, C \right) $, of a forest, $ F $, and a member, $ C $, of
$ F $, let $ \mathcal{T} \left( F, C \right) $ denote the
set of all the members, $ D $, of $ F $, such that $ C
\subseteq D $.  Then $ \mathcal{T} \left( F, C \right) $ is
a nonempty finite set, totally ordered by the subset
relation.  Now let $ B $ be the largest member of $
\mathcal{T} \left( Q, A \right) $, that overlaps any member
of $ \mathcal{T} \left( Q', A' \right) $, and let $ B' $ be
the largest member of $ \mathcal{T} \left( Q', A' \right) $
that overlaps any member of $ \mathcal{T} \left( Q, A
\right) $.  Then $ B $ overlaps $ B' $, $ B' $ does not
overlap any member of $ Q $, that contains $ B $ as a
strict subset, and $ B $ does not overlap any member of
$ Q' $, that contains $ B' $ as a strict subset.  Hence
there is a line, $ \Delta $, with one end in $ \left( B
\cap B' \right) $, and its other end in $ B' $, but not
in $ B $, and not outside any member of $ Q $, that
contains $ B $ as a strict subset, and there is a line, $
\Delta' $, with one end in $ \left( B \cap B' \right) $,
and its other end in $ B $, but not in $ B' $, and not
outside any member of $ Q' $, that contains $ B' $ as a
strict subset.  Then $ \left( P, Q \right) \in \Omega
\left( \alpha \right) $ implies that $ \lambda \left( Q,
B, \alpha \right) < \alpha_{\Delta} \leq \lambda \left( P',
B', \alpha \right) $, and $ \left( P', Q' \right) \in
\Omega \left( \alpha \right) $ implies that $ \lambda
\left( Q', B', \alpha \right) < \alpha_{\Delta'} \leq
\lambda \left( P, B, \alpha \right) $.  But $ \left( P \cup
P' \right) \subseteq \left( Q \cap Q' \right) $ implies
that $ P \subseteq \left( P \cup P' \right) \subseteq Q $,
and $ P' \subseteq \left( P \cup P' \right) \subseteq Q' $,
hence $\lambda \left( \left( P \cup P' \right), B, \alpha
\right) < \lambda \left( \left( P \cup P' \right), B',
\alpha \right) $, and $ \lambda \left( \left( P \cup P'
\right), B', \alpha \right) < \lambda \left( \left( P \cup
P' \right), B, \alpha \right) $, which is a contradiction.
Hence no member of $ Q $ overlaps any member of $ Q' $, so
$ \left( Q \cup Q' \right) $ is a forest.  Now let $ A $ be
any member of $ \left( Q \cup Q' \right) \vdash \left( P
\cap P' \right) $, and $ \Delta $ be any line, with one end
in $ A $, and its other end not in $ A $, but not outside
any member of $ \left( Q \cup Q' \right) $, if any, that
contains $ A $ as a strict subset.  Then $ \left( P \cup P'
\right) \subseteq \left( Q \cap Q' \right) $ implies that
either $ A \in Q \vdash P $ or $ A \in Q' \vdash
P' $.  Suppose $ A \in Q \vdash P $.  The end of $ \Delta
$, that is not in $ A $, is not outside any member of $ Q
$, that contains $ A $ as a strict subset, hence
$ \left( P, Q \right) \in \Omega \left( \alpha \right) $
implies $ \lambda \left( Q \cup Q', A, \alpha \right) \leq
\lambda \left( Q, A, \alpha \right) < \alpha_{\Delta} $.
The result $ \lambda \left( Q \cup Q', A, \alpha \right) <
\alpha_{\Delta} $ also follows from $ A \in Q' \vdash P' $.
Hence $\left( \left( P \cap P' \right), \left( Q \cup Q'
\right) \right) \in \Omega \left( \alpha \right) $.

The analogue, in Hepp's convergence proof \cite{Hepp}, of
the good set of forests defined by the pair $ \left( P, Q
\right) $ of forests, as above, is the collection of the
Hepp trees, $ T = \left( \mathfrak{U},\mathcal{M}, \sigma
\right) $, defined on page 308 of \cite{Hepp}, such that $
\mathfrak{U} = Q $, and the set of all the members, $ A $,
of $ \mathfrak{U} $, such that $ \sigma \left( A \right)
\leq 0 $, is $ P $.  It would be interesting to find out
whether or not the good sets of forests defined by $ \Omega
\left( \alpha \right) $, as above, in each Hepp sector,
bear some relation to the partial sums of
counterterms, $ \mathcal{F}_T \left( U \right) $, which
Hepp decomposes the BPHZ-renormalized integrand into, in
Lemma 2.4(a) of \cite{Hepp}, and whose contributions Hepp
bounds, in Lemma 3.2 of \cite{Hepp}, although consideration
of the first good
set of forests, in Fig. 5 of \cite{Hepp}, shows that the
correspondence cannot be exact.  The correspondence with
Hepp's terminology, here, is that the members of $ P $
are the vertices of the graph, together with Hepp's
``twigs'', and the members of $ Q $, that are not members
of $ P $, are Hepp's ``boughs''.  It would also be
interesting to see if a suitable definition of $ \Omega
\left( p \right) $, for momentum space, could be extracted
from Zimmermann's proof of convergence in momentum space,
\cite{Zimmermann convergence}.  The criterion for a
subdiagram to be a
``cluster'', for given momenta in the lines of the diagram,
in Euclidean signature momentum space, would broadly be
that the magnitude of the momentum, in each of its internal
lines, be larger than the magnitude of the momentum, in any
of its legs, but, due to the fact that the $ R $-operation
changes the magnitudes of the momenta, in the internal
lines of subdiagrams, the precise criterion might need to
be chosen carefully, in order to obtain analogues of Lemmas
\ref{Lemma 7}, \ref{Lemma 10}, and \ref{Lemma 11}.
Zimmermann is able to avoid dividing the momentum space of
the diagram into tessellating sectors, because the momentum
space power-counting theorem, whose hypotheses he has to
satisfy, is expressed in terms of the power-counting
behaviour of the integrand, as the loop momenta tend to
infinity, within hyperplanes of various dimensions.  Thus,
on each hyperplane, he can divide up the set of all
forests, into good sets of forests, in a manner appropriate
for bounding the integrand on that hyperplane, without
being concerned about the fact that the hyperplanes
intersect one another, and the way he divides up the set of
all forests, into good sets of forests, at a given point in
the momentum space of the diagram, will in general be
different, according to which hyperplane through that point
he is considering.  The good sets of forests, on a given
hyperplane, in Section 4 of \cite{Zimmermann convergence},
are defined by the pairs $ \left( B,C \right) $, where, in
Zimmermann's terminology, $ C $ is a complete forest, with
respect to that hyperplane, and $ B $ is the base of $ C $.

To make practical use of BPHZ renormalization, for
physically interesting gauge theories, it is necessary to
ensure that the Slavnov-Taylor identities \cite{Slavnov 1,
Slavnov 2, Taylor}, which are the quantum form of the BRST
invariance \cite{Becchi Rouet Stora 1, Tyutin} of the
properly gauge-fixed classical action, are preserved.  The
BRST invariance of the gauge-fixed classical action
implies, if the effects of renormalization are ignored,
that the vacuum expectation values of BRST-invariant
quantities, hence of gauge-invariant quantities, are
independent of the choice of the gauge-fixing term in the
action.  This follows from the fact that, in the standard
framework of Fadeev and Popov \cite{Feynman, DeWitt,
Faddeev Popov} and 't Hooft \cite{'t Hooft unitarity}, the
sum of the gauge-fixing and Faddeev-Popov terms in the
gauge-fixed action is a BRST variation, and, if the effects
of renormalization are ignored, vacuum expectation values
of BRST-invariant quantities are unaltered by the addition
of a BRST variation to the action.  Indeed, in the example
of Yang-Mills theory, if we implement the gauge fixing
by Nakanishi-Lautrup auxiliary fields $ B_{ a } $, (which
simply become Lagrange multipliers, in gauges such as
Landau gauge and radiation gauge, where the gauge-fixing
parameter $ \alpha $ vanishes), we can write the BRST
variation as:
\[
\delta_{\mathrm{BRST}} = \int d^d x \left( \left( D_{\mu}
\phi \right)_{ax} \frac{ \delta }{ \delta A_{ \mu ax } } -
\frac{ 1 }{ 2 } f_{ abc } \phi_{ bx } \phi_{ cx } \frac{
\delta }{ \delta \phi_{ ax } } + iB_{ ax } \frac{ \delta }{
\delta \psi_{ ax } } \right)
\]
which is nilpotent identically, $ \left( \delta_{
\mathrm{BRST} } \right)^2 = 0 $, and independent of the
form of the gauge-fixing term in the action.  Here
$ \left( D_{ \mu } \phi \right)_{ ax } = \partial_{ \mu }
\phi_{ax} + A_{ \mu bx } f_{ abc } \phi_{ cx } $.  Then the
sum of the gauge-fixing and Fadeev-Popov terms in the
action can be written as
\[
\frac{ 1 }{ g^2 } \delta_{ \mathrm{BRST} } \left( \psi_{ a }
\mathcal{F}_{ a } - i \frac{ \alpha }{ 2 } \psi_{ a } B_{ a
} \right)
\]
where $ \mathcal{F}_{ a } $ is the gauge-fixing function,
for example $ \mathcal{F}_{ a } = \partial_{ \mu } A_{ \mu
a } $ for covariant gauges, and $ \mathcal{F}_{ a } =
\partial_{ i } A_{ ia } $ for radiation gauge.  So in a
covariant gauge, we have:
\[
\frac{ 1 }{ g^2 } \left( iB_{ a } \left( \partial_{ \mu }
A_{ \mu a } \right) + \frac{ \alpha }{ 2 } B_{ a } B_{ a }
+ \psi_{ a } \left( \partial_{ \mu } \left( D_{ \mu } \phi
\right)_{ a } \right) \right) = \frac{ 1 }{ g^2 } \delta_{
\mathrm{BRST} } \left( \psi_{ a } \left( \partial_{ \mu }
A_{ \mu a } \right) - i \frac{ \alpha }{ 2 } \psi_{ a } B_{
a } \right)
\]
Thus we can interpolate between different choices of
gauge-fixing function, and different gauge parameters, by
adding different BRST variations to the action.  So to
demonstrate that vacuum expectation values of BRST-invariant
quantities are independent of the choice of the
gauge-fixing function and the gauge parameter, it is
sufficient to demonstrate that vacuum expectation values,
of BRST-invariant quantities, are unaltered by the
addition, to the action, of an infinitesimal multiple of a
BRST variation.  The change of a vacuum expectation value,
resulting from the addition of an infinitesimal term
to the action, is, in Euclidean signature, equal to the
negative of that same vacuum expectation value, with the
infinitesimal addition to the action, now included in the
pre-exponential factor.  Thus the change of the vacuum
expectation value of a product of BRST-invariant
quantities, say $ X, \ldots, Y $, resulting from the
addition, to the action, of an infinitesimal multiple,
say $ \eta $, of a BRST variation, say $ \delta_{
\mathrm{BRST} } Z $, is equal to the vacuum expectation
value of a BRST variation, namely
\[
- \eta \delta_{ \mathrm{BRST} } \left( X \ldots YZ \right)
= - \eta X \ldots Y \left( \delta_{ \mathrm{BRST} } Z
\right)
\]
On the other hand, by making an infinitesimal change of
variables in the functional integral, $ \tilde{A}_{ \mu ax
} = A_{ \mu ax } + \varepsilon \delta_{ \mathrm{BRST} } A_{
\mu ax } $, $ \tilde{\phi}_{ ax } = \phi_{ ax } +
\varepsilon \delta_{ \mathrm{BRST} } \phi_{ ax } $, $
\tilde{\psi}_{ ax } = \psi_{ ax } + \varepsilon \delta_{
\mathrm{BRST} } \psi_{ ax } $, and $ \tilde{B}_{ ax } =
B_{ ax } + \varepsilon \delta_{ \mathrm{BRST} } B_{ ax } $,
where $ \varepsilon $ is an infinitesimal anticommuting
constant,
and using the facts that the value of a functional integral
is unaltered by a change of integration variables, the
action is BRST invariant, and the Jacobian of this
particular change of variables is exactly equal to 1, one
finds that the vacuum expectation value, of a BRST
variation, is equal to 0.  The way in which the effects, of
an infinitesimal change in the choice of the gauge-fixing
term in the action, cancel among groups of Feynman diagrams
contributing to the vacuum expectation value of a physical
quantity, if the effects of renormalization are ignored,
was demonstrated, for Yang-Mills theory, by Mills
\cite{Mills}.  Conversely, if the effects of
renormalization violate BRST invariance, or, in other
words, violate the Slavnov-Taylor identities, the vacuum
expectation values of physical quantities will usually not
be independent of the choice of the gauge-fixing term in
the action, which means that the predictions of the theory
will fail to be Lorentz-covariant, if the calculations are
done in a physical gauge, such as radiation gauge, or the
theory will have negative-norm states, whose effects do not
cancel out, if the calculations are done in a
Lorentz-covariant gauge.  The proof of the cancellation of
the effects of negative-norm states, in a
Lorentz-covariant gauge, is usually called a unitarity
proof, because it implies that the $ S $-matrix, restricted
to the physical states, is unitary.  The unitarity proofs
of 't Hooft for renormalized Yang-Mills theory
\cite{'t Hooft unitarity}, and of Sterman, Townsend, and
van Nieuwenhuizen for renormalized supergravity
\cite{Sterman Townsend van Nieuwenhuizen}, explicitly use
only linear identities, called Ward-Takahashi identities
\cite{Ward, Takahashi}, that depend only on the linear
terms in the BRST variation, but the argument assumes the
validity of the full nonlinear Slavnov-Taylor identities,
that correspond to the full non-linear BRST variation.

The BPHZ-renormalized Feynman diagrams generally violate
the Slavnov-Taylor identities, but, in consequence of the
convergence proof, the violation is by a finite amount.  It
is therefore necessary to seek, at each successive order in
the loop expansion, (i.e. the expansion in powers of $
\hbar $), additional finite local counterterms, which, when
added to the action, restore the validity of the
Slavnov-Taylor identities.  If no such finite local
counterterm exists, the theory is said to have a gauge
``anomaly''~\cite{Steinberger, Fukuda Miyamoto, Adler,
Bell Jackiw, Adler Bardeen, Gross Jackiw, Fujikawa,
Alvarez-Gaume Witten, Atiyah Singer, Zumino, Stora,
Green Schwarz 1, Green Schwarz 2, Alvarez-Gaume Ginsparg 1,
Alvarez-Gaume Ginsparg 2, Callan Harvey,
Itoyama Nair Ren 1, Itoyama Nair Ren 2, Grignani, Fisher,
Harvey}, which results in the shortcomings mentioned above.
The possible anomalies are constrained by the Wess-Zumino
consistency condition \cite{Wess Zumino, Baulieu,
Brandt Dragon Kreuzer, Brandt 1}, which has led to the
complete classification of the possible anomalies, in some
types of gauge theories
\cite{Dubois-Violette Henneaux Talon Viallet,
Barnich Henneaux, Barnich Brandt Henneaux 2,
Barnich Brandt Henneaux 3, Barnich Brandt Henneaux 4,
Brandt 2, Henneaux Knaepen Schomblond,
Barnich Brandt Henneaux 5}.  An explicit example of a
finite counterterm, that restores a Ward-Takahashi
identity for a gauge-field propagator in an Abelian gauge
theory, when the position-space integral, in the
counterterm, is cut off at large distances, is given in
Section 5.4.1 of \cite{Group-Variation Equations}, although
the detailed form of the long-distance cutoff, in that
example, is different from that considered in the present
paper.

The Slavnov-Taylor identities relate vacuum
expectation values (VEVs) or correlation functions that
involve not only the fields themselves, but also, certain
products of the fields at coincident points in position
space, namely the products that occur in the non-linear
terms in the BRST variations of the fields.  This is
necessary in order to have a set of VEVs or correlation
functions that is closed under the BRST variations.
Fortunately, the BRST transformation is nilpotent, when
Nakanishi-Lautrup auxiliary fields are used for the
gauge-fixing, or, otherwise, when the antighost field
equation is used, so the required set of field products
is limited to the BRST variations of the fields.  The
products of fields, at coincident points in position space,
are defined to be Zimmermann normal products
\cite{Zimmermann Brandeis, Zimmermann products 1,
Zimmermann products 2}.  VEVs and
correlation functions involving normal products are defined
by BPHZ renormalization, just as if the normal product
vertices in a diagram were additional vertices in the
action, even though their positions are not integrated
over.  It is therefore natural to introduce
position-dependent ``sources'' for the normal products, and
add the normal products, multiplied by their ``sources'',
to the action.  The Slavnov-Taylor identities take a very
compact form, when such ``sources'' are introduced for the
BRST variations of the fields, and it is this compact form
of the Slavnov-Taylor identities, for the action that has
had the BRST variations of the fields, multiplied by the
``sources'' for the BRST variations, added to it, that is
required to be preserved by renormalization, since this
allows for the fact that the BRST variations also get
infinite correction terms, due to the BPHZ definition of
VEVs and correlation functions involving normal products,
and allows finite counterterms dependent on the ``sources''
for the BRST variations, as well as the finite counterterms
independent of these ``sources'', to be added to the
action, order by order in the loop expansion, in order to
restore the validity of the Slavnov-Taylor identities.

The study of the Slavnov-Taylor identities makes use of a
number of relations, sometimes called quantum action
principles \cite{Lowenstein 4, Lowenstein 5, Lam 1, Lam 2,
Gomes Lowenstein, Lam 3}, that involve not only normal
products, but also, slight generalizations, called
oversubtracted normal products, and anisotropic normal
products.  An oversubtracted normal product is a normal
product, $ X $, that is assigned an integer, $ n $, greater
than the dimension of $ X $, that is to be used, instead of
the dimension of $ X $, in determining the maximum degree
of the Taylor expansion, of the BPHZ counterterm, for any
subdiagram that includes $ X $.  The dimension of a
monomial in the fields and derivatives, here, means the
number of derivatives, plus, for each field, half the
dimension, at short distances, of the corresponding
propagator, which is $ \frac{ \left( d - 2 \right) }{ 2 } $
for a propagating boson, and $ \frac{ \left( d - 1 \right)
}{ 2 } $ for a propagating fermion, in $ d $ dimensions.
An anisotropic normal product is a normal product that is
assigned two or more different degrees of oversubtraction,
depending on which legs of the normal product are internal
lines, and which are external lines, of a subdiagram that
includes the normal product.  The oversubtracted and
anisotropic normal products are related to the ordinary
normal products by linear relations, called Zimmermann
identities \cite{Zimmermann Brandeis,
Zimmermann products 1, Zimmermann products 2},
which state, for example, that the difference between an
oversubtracted normal product, and the corresponding
ordinary normal product, is equal to a linear combination,
with coefficients that vanish, at zeroth order in
perturbation theory, of normal products, of dimension less
than or equal to the integer, $ n $, that is assigned to
the oversubtracted normal product, as above.  Both the
quantum action principles, and the Zimmermann identities,
might need to be modified, when the propagators, in the
counterterms, are allowed to differ, at large distances,
from the propagators in the direct terms, as allowed by the
convergence proofs, in the present paper.

In general, the
additional finite counterterms, that are allowed to be
added to the Lagrangian, at a given order in the loop
expansion, or, in other words, at a given order in the
expansion in powers of $ \hbar $, are the most general
linear combinations of local monomials in the fields and
their derivatives, of dimension not exceeding the maximum
dimension of the BPHZ counterterms that occur at that order
in the loop expansion, and which preserve the subgroup of
the full Poincar\'{e} group preserved by the tree level
Lagrangian, and the global internal symmetries of the tree
level Lagrangian.  Thus, in Lorentz-covariant gauges, the
counterterms are required to be Lorentz-invariant, while in
gauges such as radiation gauge, where some terms in the
tree-level Lagrangian depend on a special direction,
defined by a unit vector, $ n_{ \mu } $, the counterterms
are also allowed to depend on $ n_{ \mu } $.  The
coefficients of the finite counterterms are fixed by
requiring, firstly, that the Slavnov-Taylor identities are
preserved, then, secondly, any further independent
coefficients, whose values affect the physical predictions
of the theory, are fixed, by fitting the predictions of the
theory, to a finite number of experimental measurements,
then, finally, the remaining independent coefficients,
whose values do not affect the physical predictions of the
theory, are fixed, by choosing a finite number of
normalization conditions on the correlation functions or
proper vertices, which serve to stabilize the gauge in
which the calculations are being done.  If, after the
Slavnov-Taylor identities have been preserved, the number
of independent coefficients, whose values affect the
physical predictions of the theory, does not increase with
increasing order in the loop expansion, the theory is said
to be renormalizable, and makes definite predictions, with
no scope for adjustment, for the rates of physical
processes, once the values of the relevant coefficients
have been determined, by fitting the results of a finite,
fixed set of experimental measurements.  An example of a
renormalizable theory is the Standard Model \cite{Rosner}
in $ 3 + 1 $ space-time dimensions, in which case the
relevant coefficients correspond to the masses, mixing
angles, and coupling constants in the tree-level
Lagrangian, whereas gravity and supergravity
\cite{Freedman van Nieuwenhuizen Ferrara, Deser Zumino} in
$ 3 + 1 $ dimensions, and Yang-Mills theory in more than $
4 $ dimensions, are not renormalizable.

By making use of
the quantum action principles, the Zimmermann identities,
and the Wess-Zumino consistency condition, Becchi, Rouet,
and Stora \cite{Becchi Rouet Stora 2, Becchi Rouet Stora 3}
explicitly demonstrated, for the Abelian Higgs-Kibble model
\cite{Higgs 1, Higgs 2, Kibble}, and the SU(2) Higgs-Kibble
model \cite{'t Hooft massive, 't Hooft Veltman massive},
in which all particles are massive, and the
Adler-Bell-Jackiw anomaly \cite{Adler, Bell Jackiw} is
absent, that the Slavnov-Taylor identities could be
preserved, to all orders in the loop expansion, by suitable
choices of the additional finite counterterms, and that
when the Slavnov-Taylor identities are preserved, the
physical $ S $-matrix is unitary, in a class of
Lorentz-covariant gauges, and is independent of the
normalization conditions imposed on the BPHZ-renormalized
proper vertices, that correspond to a specification of the
choice of gauge, within the class of Lorentz-covariant
gauges considered.  The practical techniques that have been
developed, for the explicit calculation of the additional
finite counterterms, that restore the validity of the
Slavnov-Taylor identities, order by order in the loop
expansion, are sometimes referred to as algebraic
renormalization \cite{Piguet Sorella, Grassi 1, Kraus,
Ferrari Grassi, Martin Sanchez-Ruiz, Ferrari Grassi Quadri,
Grassi Hurth Steinhauser 1, Grassi 2,
Grassi Hurth Steinhauser 2, Grassi Hurth Steinhauser 3,
Quadri 1, Quadri 2}.

For Yang-Mills theories, a theorem of
Joglekar and Lee \cite{Joglekar Lee,
Henneaux renormalization} states that a general
BRST-invariant local polynomial, in the fields and their
derivatives, of Fadeev-Popov ghost number zero, is equal to
the sum of a gauge-invariant polynomial in the field
strengths $ F_{ \mu \nu }^{ a } $ and their gauge-covariant
derivatives, and a BRST variation.  If an analogous result
also held for other gauge theories, then the non-discrete
physically significant adjustable parameters of the BPHZ
renormalized theory, which remain undetermined, at a given
order in the loop expansion, after satisfying the
Slavnov-Taylor identities at that order, and thus have to
be fitted to experiment, reducing the predictiveness of the
theory, would be in one-to-one correspondence with the
non-trivial adjustable parameters of the corresponding
(on-shell) gauge-invariant classical action, allowing terms
with higher derivatives, up to and including the maximum
number of derivatives required, in the BPHZ counterterms,
at that order in the loop expansion.  ``Non-trivial'',
here, means that the coefficients of higher-derivative
terms that vanish when the field equations of the zero-loop
theory are satisfied, are not counted, because, for small
values of their coefficients, those terms can be absorbed
by a redefinition of the fields.  Coefficients of total
derivative terms are also not counted.  Thus, in
particular, pure Einstein gravity, in $ 3 + 1 $ dimensions,
gets no new adjustable physical parameters at one loop
order, because $ \sqrt{ -g } R^2 $ and $ \sqrt{ -g } R^{
\mu \nu } R_{ \mu \nu } $ vanish when the vacuum Einstein
equations are satisfied, and $ \sqrt{ -g } ( R^2 - 4 R^{
\mu \nu } R_{ \mu \nu } + R^{ \mu \nu \alpha \beta } R_{
\mu \nu \alpha \beta } ) $ is a total derivative
\cite{'t Hooft Veltman gravity, 't Hooft Erice 02}.
``(On-shell) gauge-invariant'' means that, for gauge
transformations such as the local supersymmetry variations
of supergravity without auxiliary fields
\cite{Freedman van Nieuwenhuizen nonclosure}, whose
commutator includes a trivial gauge transformation, as
defined in Section 2.4 of \cite{Brandt supergravity}, the
gauge variations of the higher derivative terms are only
required to vanish when the field equations that follow
from the zero-loop action are satisfied, because, for small
values of the coefficients of the higher derivative terms,
their gauge variations can, in that case, be cancelled, by
the addition of appropriate additional terms to the gauge
transformation rules.  It might be interesting to look at
supergravity in eleven dimensions
\cite{Cremmer Julia Scherk,
Nicolai Townsend van Nieuwenhuizen, Sagnotti Tomaras,
Naito Osada Fukui, Miemiec Schnakenburg,
de Wit van Niewenhuizen Van Proeyen,
Bautier Deser Henneaux Seminara,
Bandos Azcarraga Picon Varela}, and Ho\v{r}ava-Witten
theory \cite{Horava Witten 1, Horava Witten 2, Witten,
de Alwis, Conrad, Bilal Metzger, Moss 1, Moss 2, Moss 3},
from this point of view.

The compact form of the Slavnov-Taylor identities, for the
action that has had the BRST variations of the fields,
multiplied by the ``sources'' for the BRST variations,
added to it, is sometimes called the Zinn-Justin equation
\cite{Zinn-Justin}, when it is expressed in terms of the
generating functional of the proper vertices, defined in
the standard way, by the application of a Legendre
transformation, with respect to the fields and their
sources, to the generating functional of the connected
Green functions.  The Legendre transformation does not
affect the ``sources'' for the BRST variations, which are
treated, in effect, like position-dependent ``coupling
constants'', that may also have Lorentz and Yang-Mills
indices.  The zero-loop form of the Zinn-Justin equation is
called the Batalin-Vilkovisky master equation
\cite{Kallosh, de Wit van Holten, Batalin Vilkovisky 1,
Batalin Vilkovisky 2, Batalin Vilkovisky 3,
Batalin Vilkovisky 4, Batalin Vilkovisky 5,
Troost van Nieuwenhuizen Van Proeyen, Van Proeyen,
De Jonghe, Barnich Brandt Henneaux 1, Anselmi,
Gomis Paris Samuel, Gomis Weinberg, Fuster Henneaux Maas},
and is thought to provide a comprehensive framework for
studying the gauge-fixing of gauge theories, at the
zero-loop order.  In this context, the ``sources'', for the
BRST variations of the fields, are usually called
``antifields''.  An alternative framework for the BPHZ
renormalization of gauge theories, that appears to focus on
the quantum-corrected Batalin-Vilkovisky master equation,
written in terms of Zimmermann normal products, rather than
on the Slavnov-Taylor identities, has been proposed by De
Jonghe, Par\'{i}s, and Troost\cite{De Jonghe Paris Troost}.
A formal proof, that the quantum-corrected
Batalin-Vilkovisky master equation implies that physical
quantities are independent of the choice of the
gauge-fixing terms in the action, was given by Fuster,
Henneaux, and Maas \cite{Fuster Henneaux Maas}.

The definition of the BPHZ counterterms, in position space,
depends on the choice of the contraction weights, $
\omega_{AB} $, as defined on page
\pageref{Start of original page 13}, where $ A $ and $ B $
are subdiagrams of a Feynman diagram, such that $ B $ is a
subdiagram of $ A $.  However, when the action is
translation-invariant, the change to a counterterm, that
results from a change of the contraction weights, vanishes
when the degree of divergence is zero, and is a total
divergence, when the degree of divergence is greater than
zero.  Indeed, for a subdiagram, $ A $, with $ n $
vertices, and a set of contraction weights, $ \omega $, let
$ \lambda_i \equiv \omega_{ A \left\{ i \right\} } $, $ 1
\leq i \leq n $, where $ i $ runs over the vertices of $ A
$.  Then $ \sum_{i = 1}^n \lambda_i = 1 $, and a typical
overall counterterm, for $ A $, has the form:
\[
\left. C_A \left( z \right) \right|_{\lambda} =
\hspace{-12.2pt} \hspace{14.0cm}
\]
\[
= \hspace{-1.7pt}
 \int d^{ nd } x \delta^d \left( \sum_{ i = 1 }^n
\lambda_i x_i - z \right) I \left( x \right) \sum_{ r = 0
}^N \frac{ 1 }{ r! } \left( \left( \sum_{ j = 1 }^m \left(
\bar{x}_j - z \right) . \hat{y}_j \right)^r E \left( y_1,
\ldots, y_m \right) \right)_{ \hspace{-1.6pt}
 y_1 = \ldots = y_m = z }
\]
Here $ I \left( x \right) = I \left( x_1, \ldots, x_n
\right) $, the internal function of $ A $, includes the
internal propagators of $ A $, which may differ, at long
distances, in the counterterm, from the corresponding
propagators in the direct term, as allowed by the
convergence proofs in this paper.  $ I \left( x \right) $
may also include BPHZ counterterms for strict subdiagrams of
$ A $.  Translation invariance means that, for arbitrary $
u $, $ I \left( x_1 + u, \ldots, x_n + u \right) = I \left(
x_1, \ldots, x_n \right) $.  $ N $ is the degree of
divergence.  $ m $ is the number of legs of $ A $, which is
the number of fields, in the corresponding formal
counterterm, in the Lagrangian.  $ \bar{x}_j $ is the
position of the inner end of the $ j $th leg of $ A $,
\emph{before} the contraction of $ A $.  If none of the
subdiagrams of $ A $ have been contracted, then each $
\bar{x}_j $ will simply be one of the $ x_i $, while if
some of the subdiagrams of $ A $ have been contracted, some
of the $ \bar{x}_j $ might be linear combinations of some
of the $ x_i $, with total weight $ 1 $.  $ \hat{y}_{ j \mu
} $ means $ \frac{ \partial }{ \partial y_{ j \mu } } $,
where $ \mu $ is the Lorentz index.  $ E \left( y_1,
\ldots, y_m \right) $ is the external function of $ A $.
In the Feynman diagram, $ y_1, \ldots, y_m $ are the
positions of the ends, in $ A $, of the propagators that
form the legs of $ A $, and are thus evaluated at $
\bar{x}_1, \ldots, \bar{x}_m $, before the contraction of $
A $, and at $ z, \ldots, z $, where $ z = \sum_{i = 1}^n
\lambda_i x_i $, after the contraction of $ A $.  In the
Feynman diagram, $ E \left( y_1, \ldots, y_m \right) $ also
depends, implicitly, on the positions of the external
vertices of the diagram, and, also, on the positions of any
vertices of the diagram, not in $ A $, that have not yet
been integrated over.  In the counterterm, on the other
hand, $ E \left( y_1, \ldots, y_m \right) $ is the product
of some fields, $ \varphi_1 \left( y_1 \right) \ldots
\varphi_m \left( y_m \right) $, all of which are evaluated
at $ z $, after the differentiations have been performed.

Let $ \omega' $ be another set of contraction weights, and
let $ \lambda'_i \equiv \omega'_{ A \left\{ i \right\} } $,
$ 1 \leq i \leq n $, so that $ \sum_{ i = 1 }^n \lambda'_i
= 1 $.  Let $ x'_i = x_i + \sum_{ k = 1 }^n \left(
\lambda'_k - \lambda_k \right) x_k $, $ 1 \leq i \leq n $.
Hence $ \frac{ \partial x'_i }{ \partial x_j } = \delta_{
ij } + \lambda'_j - \lambda_j $, which has determinant
equal to $ 1 $.  Furthermore, $ \sum_{ i = 1 }^n \lambda_i
x'_i = \sum_{ i = 1 }^n \lambda'_i x_i $, and $ \sum_{ i =
1 }^n \lambda'_i x_i' = \sum_{ i = 1 }^n \left( 2
\lambda_i' - \lambda_i \right) x_i $, hence $ x_i = x_i' +
\sum_{ k = 1 }^n \left( \lambda_k - \lambda'_k \right) x'_k
$.  Hence the above counterterm, evaluated with the
contraction weights $ \omega' $, is:
\[
\left. C_A \left( z \right) \right|_{\lambda'} =
\hspace{-14.9pt} \hspace{14.0cm}
\]
\[
= \hspace{-2.0pt}
\int \hspace{-2.0pt}
d^{ nd } x \delta^d \hspace{-2.0pt} \left( \sum_{ i = 1 }^n
\lambda'_i x_i - z \right) I \left( x \right) \sum_{ r = 0
}^N \frac{ 1 }{ r! } \left( \hspace{-2.0pt}
\left( \sum_{ j = 1 }^m \left(
\bar{x}_j - z \right) . \hat{y}_j \hspace{-2.0pt}
\right)^{ \hspace{-1.0pt} r } \hspace{-1.0pt} E \left( y_1,
\ldots, y_m \right) \hspace{-2.0pt}
\right)_{ \hspace{-2.0pt} y_1 = \ldots = y_m = z } =
\]
\[
= \hspace{-2.0pt}
\int \hspace{-2.0pt}
d^{ nd } x' \delta^d \hspace{-2.0pt}
\left( \sum_{ i = 1 }^n
\lambda_i x'_i - z \right) I \left( x' \right) \sum_{ r = 0
}^N \frac{ 1 }{ r! } \left( \hspace{-2.0pt}
\left( \sum_{ j = 1}^m \left(
\bar{x}'_j + \left( \sum_{ k = 1 }^n \left( \lambda_k -
\lambda'_k \right) x'_k \right) - z \right) . \hat{y}_j
\right)^{ \hspace{-2.0pt} r } \hspace{-2.0pt}
\times \right.
\]
\[
\hspace{12.0cm} \hspace{-44.9pt} \left.
\times E \left( y_1, \ldots, y_m \right)
\rule[-3.0ex]{0pt}{7.0ex} \right)_{ y_1 =
\ldots = y_m = z }
\]
where I used the translation invariance of $ I \left( x
\right) $, and the fact that, since each $ \bar{x}_j $ is a
linear combination, of total weight $ 1 $, of the $ x_i $,
we have $ \bar{x}_j = \bar{x}'_j + \sum_{k = 1}^n \left(
\lambda_k - \lambda'_k \right) x'_k $, for all $ 1 \leq j
\leq m $.

Hence, renaming $ x'_i $ as $ x_i $, we have:
\[
\left. \left. C_A \left( z \right) \right|_{\lambda'} - C_A
\left( z
   \right) \right|_{\lambda} = \hspace{-15.1pt}
\hspace{12.0cm}
\]
\[
= \hspace{-2.0pt}
\int \hspace{-2.0pt}
d^{ nd } x \delta^d \hspace{-2.0pt} \left( \sum_{ i = 1 }^n
\lambda_i x_i - z \right) I \left( x \right) \sum_{ r = 0
}^N \frac{ 1 }{ r! } \left( \left( \left( \sum_{ j = 1 }^m
\left( \bar{x}_j + \left( \sum_{ k = 1 }^n \left( \lambda_k
- \lambda'_k \right) x_k \right) - z \right) . \hat{y}_j
\right)^r \right. \right.
\]
\[
\hspace{7.0cm} \hspace{-27.0pt} \left. \left.
- \left( \sum_{ j = 1 }^m \left( \bar{x}_j - z
\right) . \hat{y}_j \right)^r \right) E \left( y_1, \ldots,
y_m \right) \right)_{ y_1 = \ldots = y_m = z } =
\]
\[
= \int d^{ nd } x \delta^d \left( \sum_{ i = 1 }^n
\lambda_i x_i - z \right) I \left( x \right) \left( \sum_{
s = 1 }^N \frac{ 1 }{ s! } \left( \left( \sum_{ k = 1 }^n
\left( \lambda_k - \lambda'_k \right) x_k \right) . \left(
\sum_{ j = 1 }^m \hat{y}_j \right) \right)^s \times
\right. \hspace{-18.7pt} \hspace{2.0cm}
\]
\[
\hspace{6.0cm} \hspace{-18.9pt} \left. \times \sum_{ t = 0
}^{ N - s } \frac{ 1 }{ t! } \left( \sum_{ j = 1 }^m \left(
\bar{x}_j - z \right) . \hat{y}_j \right)^t E \left( y_1,
\ldots, y_m \right) \right)_{ y_1 = \ldots = y_m = z } =
\]
\[
= \int d^{ nd } x \delta^d \left( \sum_{ i = 1 }^n
\lambda_i x_i \right) I \left( x \right) \left( \sum_{ s =
1 }^N \frac{ 1 }{ s! } \left( \left( \sum_{ k = 1 }^n
\left( \lambda_k - \lambda'_k \right) x_k \right) . \left(
\sum_{ j = 1 }^m \hat{y}_j \right) \right)^s \times
\right. \hspace{-26.5pt} \hspace{3.0cm}
\]
\[
\hspace{7.0 cm} \hspace{-15.6pt} \left. \times \sum_{ t = 0
}^{ N - s } \frac{ 1 }{ t! } \left( \sum_{ j = 1 }^m
\bar{x}_j . \hat{y}_j \right)^t E \left( y_1, \ldots, y_m
\right) \right)_{y_1 = \ldots = y_m = z} =
\]
\[
= \sum_{ s = 1 }^N \frac{ 1 }{ s! } \frac{ \partial }{
\partial z_{ \mu_1 } } \ldots \frac{ \partial }{ \partial
z_{ \mu_s } } \left( \int d^{ nd } x \delta^d \left( \sum_{
i = 1 }^n \lambda_i x_i \right) I \left( x \right) \times
\right. \hspace{-25.8pt} \hspace{7.0cm}
\]
\[
\times \left(
\sum_{ k_1 = 1}^n \left( \lambda_{ k_1 } - \lambda'_{ k_1 }
\right) x_{ k_1
\mu_1 } \right) \ldots \left( \sum_{ k_s = 1 }^n \left(
\lambda_{ k_s } - \lambda'_{ k_s } \right) x_{ k_s \mu_s }
\right) \times
\]
\[
\hspace{7.0cm} \hspace{-23.1pt} \left. \times \left(
\sum_{ t = 0 }^{ N - s } \frac{ 1 }{ t! } \left( \sum_{ j =
1 }^m \bar{x}_j . \hat{y}_j \right)^t E \left( y_1, \ldots,
y_m \right) \right)_{ y_1 = \ldots = y_m = z } \right)
\]
where the second-last step resulted from shifting all the
$ x_i $ by $ z $, and using the translation invariance of
$ I \left( x \right) $, again.

More generally, a change, to the contraction weights, will
result in a change to the finite counterterms that need to
be added, in order to satisfy the Slavnov-Taylor
identities.  The convergence proofs, in the present paper,
are given for an arbitrary, fixed set of contraction
weights, that satisfies conditions (i), (ii), and (iii), on
page \pageref{Start of original page 13}.  These
conditions allow a substantial amount of flexibility in the
choice of the contraction weights, as discussed on pages
\pageref{Start of original page 13} and
\pageref{Start of original page 14}.

For the application of the results to the Group-Variation
Equations \cite{Group-Variation Equations},
the renormalization of the gauge-invariant
quantities, in terms of which the Group-Variation Equations
are expressed, is also required.  It was originally
intended that these would be the VEVs and correlation
functions of generalized Wilson loops formed from paths
consisting of finite numbers of straight line segments.
However, these have rather complicated properties under
renormalization, with infinities depending on the angles
made by the straight line segments, at the vertices where
they meet, and further singularities appearing whenever the
vertices are positioned such that two of the straight line
segments intersect.  The intention was to divide these VEVs
and correlation functions by ``short distance factors'',
consisting of the same VEVs and correlation functions, but
calculated with propagators smoothly cut off at long
distances, and then to re-write the Group-Variation
Equations as equations for these ratios, the
``long-distance factors'', which would not be sensitive to
the fine details of the paths.  It now seems likely,
however, that this program could be carried out in a more
practical manner, by using, as the basic gauge-invariant
quantities, not the generalized Wilson loops formed from
paths consisting of finite numbers of straight line
segments, but rather, the ``point-pinned loops'', formed
from the same sequences of points in four-dimensional
Euclidean space, as the corners of the paths, but having,
between each successive pair of path vertices, not the
straight Wilson line between the two vertices, but rather,
a factor
\[
\left( e^{s \mathcal{ D }^2 } \right)_{ x i, y \bar{ j } }
= \hspace{-23.91pt} \hspace{14.0cm}
\]
\[
\hspace{4.0cm} \hspace{-72.11pt}
 = \left( e^{s \partial^2 }
   \right)_{ x y } \delta_{ i \bar{ j } } + \int^s_0 du
   \int d^4 z \left( e^{ u
   \partial^2 } \right)_{ x z } \left( A_{ i \bar{ j } }
   \left( z \right) . \partial_{ z }
   + \partial_{ z } .A_{ i \bar{ j } } \left( z \right)
   \right) \left( e^{ \left( s - u
   \right) \partial^2 } \right)_{ z y } + \ldots
\]
where $ \mathcal{ D }_{ \mu } $ is covariant derivative for
a scalar field in the SU($ n $) fundamental, $ x $ and $ y
$ are the positions of the two vertices, and $ i $ and $
\bar{ j } $ are the colour indices, to be traced along the
loop as usual.  The idea is that, for small enough real $ s
$, the paths from $ x $ to $ y $, that appear in the sum
over paths, in the path integral representation of the
above factor, will be strongly suppressed by the standard
Gaussian factors,
\[
\left( e^{\left( u_{ n + 1 } - u_{ n } \right) \partial^2}
 \right)_{ z_{ n } z_{ n +
 1 } } = \frac{ e^{ - \frac{ \left( z_{ n + 1 } - z_{ n }
  \right)^2}{ 4 \left( u_{ n + 1 }
   - u_{ n } \right) } } }{\left( 4 \pi \left( u_{ n + 1 }
   - u_n \right) \right)^2},
\]
for paths that wander significantly away
from the straight line from $ x $ to $ y $.  In that case,
the pointed-pinned loop, with a suitably small $ s_{ i } $
for each segment, will have approximately the same
long-distance factor as the generalized Wilson loop formed
from the corresponding straight line segments.  On the
other hand, it seems reasonable to anticipate that the
point-pinned loops might have much simpler properties,
under renormalization, then the corresponding generalized
Wilson loops, formed from the corresponding straight line
segments.  The simplest guess would be that a
point-pinned loop needs only two
divergent ``wave-function renormalization'' factors, when
all the vertices are at different positions in
four-dimensional Euclidean space: a fixed divergent factor,
$ N $, for each vertex, and a factor $ e^{ - m^2 s_{ i } }
$ for each segment, with a fixed divergent ``mass'', $ m $.

It seems possible, that the technique of allowing the
propagators in the counterterms to differ, at long
distances, from the propagators in the direct terms, might
also be implemented in Hepp's method, by introducing an
appropriate upper limit to the parameter integrals, for the
propagators in the counterterms, and checking to see if
this caused any problems in Hepp's convergence proof.
Similarly, it seems possible that the long-distance cutoffs
on the propagators, introduced to define the short-distance
factors in the Group-Variation equations, could also be
introduced, in Hepp's method, by means of an appropriate
upper limit to the parameter integrals.  Furthermore, to
the extent that position-dependent background fields,
coupling to composite operators, have to be introduced, for
example for the purpose of studying the Slavnov-Taylor
identities, or for calculations in position-dependent
gravitational backgrounds, it seems possible that a smooth
position-dependence of the backgrounds might be dealt with
in Hepp's method, by introducing an independent ``source'',
$ j_{ \mu } $, or $ p_{ \mu } $, in the Gaussian exponent,
for each vertex, at which position-dependence of the
coupling constant, or the background, is required, doing
the Gaussian integrals over the vertex positions in the
usual way, then differentiating with respect to the
``sources'', to bring down the required pre-exponential
factors of the vertex positions, before setting the
``sources'' to zero.  In momentum space, the result of
doing the Gaussian integrals over the positions of the
vertices, in the presence of the ``sources'', would look
like Symanzik's formula \cite{Symanzik} \cite{Nakanishi},
with a momentum coming in at every vertex, rather than just
at the ``external'' vertices, of the diagram.  Moreover, it
appears that a tolerable  ``Gaussian'' representation, with
$ n $ integration parameters, of the Wilson area law, for
the VEV of a loop that approximately follows a loop formed
from $ n $ straight line segments, can be obtained by a
simple improvement of the ansatz discussed in Section 8.1.3
of \cite{Group-Variation Equations}, based on Douglas's
formula for the area of a minimal-area spanning surface
\cite{Douglas} \cite{Courant}.
Finally, the limiting process proposed in
\cite{Group-Variation Equations}, for calculating the path
integrals weighted by the window factors, in which the
smoothly-varying long-distance factors are only ``tied'',
or ``stitched'', to the paths, at a finite number of
points, and the path sums between the ``stitch points'',
weighted by the short-distance factors, are done exactly,
in terms of modified Feynman diagrams, with the limiting
process consisting of increasing the number of points at
which the long-distance factors are ``stitched'' to the
paths, is also formulated in terms of exponentiated
propagators.  Thus, it seems possible, that if the
Group-Variation Equations are expressed in terms of the
point-pinned loops, instead of generalized Wilson loops
formed from straight line segments, and if the area-law
ansatz, based on Douglas's formula, can also be extended to
an ansatz for the loop correlation functions, so that it
gives tolerable results in circumstances where a
minimal-area spanning surface, of higher topology, actually
exists, then all the ingredients required to calculate the
right-hand sides of the Group-Variation Equations, for a
realistic ansatz, might actually be available within the
Hepp framework of exponentiated propagators, which would,
in that case, probably be the method of choice, for
practical calculations.  Perhaps it might, after all, be
possible to study the Group-Variation Equations within the
framework of dimensional regularization \cite{'t Hooft
Veltman, Bollini Giambiagi, Ashmore 1, Cicuta Montaldi,
Ashmore 2, Collins}.

Thus if the Group-Variation Equations can, in fact, be
reformulated in terms of point-pinned loops, in the manner
just described, then the original motivation for the
present paper, which was the unsuitability of exponentiated
propagator and momentum space methods, for the calculation
of VEVs and correlation functions of generalized Wilson
loops, formed from straight line segments, is largely
removed.  Perhaps, however, the proof of the possibility
of using propagators in the counterterms, which differ, at
large distances, from the propagators in the direct terms,
so that for theories with massless particles, the
long-distance divergences can be eliminated from the
counterterms, without altering the propagators in the
direct terms, might have some practical utility, and lead
to the proof of analogous results, for the parameter space
and momentum space methods.

A different method of renormalization, in the absence of
translation invariance, was given by Brunetti and
Fredenhagen \cite{Brunetti Fredenhagen}, who deal with the
case of Minkowski signature, and base their construction
on the method of Epstein and Glaser \cite{Epstein Glaser}.
A proof of the BPHZ renormalizability of $ \lambda
\phi^{ 4 } $ field theory, in a Euclidean signature
four-dimensional curved gravitational background, was given
by Bunch \cite{Bunch}.

Remarkable progress towards the analytic solution of
SU($ N $) Yang-Mills theory, at large $ N $, in $ 2 + 1 $
dimensions, has recently been reported in \cite{Leigh Minic
Yelnikov}.

\begin{center}
{\bf Acknowledgement}
\end{center}
\noindent The following \LaTeX2e transcription was
produced semi-automatically from the original text, and
I would like to thank Malcolm Surl for helping to recover
the original data from an obsolete type of floppy disk.

\section{Introduction.}
\label{Section 1}

Many attempts to make QCD into a quantitative theory for
hadrons involve the use of Wilson loops.   When a Feynman
diagram contributing to a Wilson loop expectation value is
calculated, some of the vertices of the diagram must be
integrated along the loop.   It is desirable to be able to
carry out the calculation of the such a diagram directly in
position space, and in particular it is desirable to
understand in as much detail as possible how the
cancellation of short-distance divergences, when the
diagram is renormalized, works in position space.   In an
earlier paper, ``Cluster Convergence Theorem'', we
presented
a power-counting convergence theorem for use in Euclidean
position space, and in this paper we use the Cluster
Convergence Theorem to present a BPHZ convergence proof
directly in Euclidean position space.

In Section \ref{Section 6} of this paper we prove, as
Theorem \ref{Theorem 1}, the
absolute convergence, in Euclidean position space, of
renormalized diagrams when the BPHZ forests are allowed to
include \emph{all} connected subdiagrams, (both
one-line-reducible and one-line-irreducible), that are
divergent by power-counting.

In Section \ref{Section 7} of this paper we prove, as
Theorem \ref{Theorem 2}, the
convergence, not necessarily absolute, in Euclidean
position space, of renormalized diagrams when the BPHZ
forests are only allowed to include one-line-irreducible
subdiagrams.   In this case we give the result both in a
form that involves taking a limit from regularized
propagators, and in a form that makes no use of regularized
propagators.

Our methods enable any renormalized Feynman diagram to be
calculated directly in Euclidean position space, without
any use of momentum space or Gaussian integral techniques.

In Section \ref{Section 8}, ``Applications'', we explain
the derivation
of
our formula for the renormalized integrand, and we show
that for QCD, our form of the R-operation requires no
counterterms whose total number of gauge fields plus
derivatives plus $\frac{3}{2}$ times quark fields is
\label{Start of original page 2}
 greater than $4$, so that power-counting renormalizability
is achieved.

For Theorem \ref{Theorem 1} we make no assumption of
translation
invariance, nor of any related property, and for
Theorem \ref{Theorem 2}
we assume not translation invariance but a much more
general property of translation smoothness.

\section{Preparations.}
\label{Section 2}

Let $A$ and $B$ be any sets.   We shall use the convention
that the notation $A\subseteq B$ means ``$ A $
is a subset of $ B $'', and includes the
possibility that $A=B$.   We shall
write $A\subset B$, and say ``$ A $ is a strict subset
 of $ B $'', to
indicate that $A$ is a subset of $B$ but not equal to $B$.
 The notation $A\,\vdash B$
 (``$ A $ outside $ B $'') means the set
of all the members of $A$ that are \emph{not} members of
$B$.

The word ``ifif'' is short for ``if and only if''.

For any finite set $A$, the notation $\#\left( A\right) $
indicates the number of members of $A$.   If $n$ is an
integer $\geq 0$, an \emph{n-member set} is a finite set
$A$ such that $\#\left( A\right) =n$.

The symbol $\emptyset $ denotes the empty set.

For any set $X$ such that every member of $X$ is an ordered
pair, we define $\mathcal{D} \left( X\right) $, the
\emph{domain} of $X$, to be the set of all the first
components of members of $X$, and $\mathcal{R} \left(
X\right) $, the \emph{range} of $X$, to be the set of all
the second components of members of $X$.

A \emph{map} is a set $M$ whose members are all ordered
pairs and which satisfies the requirement that if $\left(
a,b\right) \in M$ and $\left( e,f\right) \in M$ then $a=e$
implies $b=f$.

Note that if $M$ is a $map_{  } $, then $\mathcal{D} \left(
M\right) $ is finite ifif $M$ is finite and that if $M$ is
finite then $\#\left( \mathcal{D} \left( M\right) \right)
=\#\left( M\right) $ and $\#\left( \mathcal{R} \left(
M\right) \right) \leq \#\left( M\right) $.

If $M$ is a map and $i$ is a member of $\mathcal{D} \left(
M\right) $ then the notation $M_{ i } $ denotes the second
component of the unique member of $M$ whose first component
is $i$.

When a subscript appears on a subscript, for example $x_{
A_{ \alpha } } $, the interpretation is \emph{always} $x_{
\left( A_{ \alpha }\right) } $, \emph{never} $\left( x_{
A } \right)_\alpha  $.

An expression such as $x_{ A\alpha } $, where two
subscripts appear side by side, can arise either as an
abbreviation of $x_{ \left( A,\alpha \right) } $, (which
arises when the ordered pair $\left( A,\alpha \right) $ is
a member of the domain of the map $x$), or as an
abbreviation
of $\left( x_{ A } \right)_\alpha $, (which arises when the
member
\label{Start of original page 3}
 $x_{ A } $ of the range of the map $x$, is itself a map).

For any sets $A$ and $B$, we shall denote by $B^{ A }$ the
set whose members are all the maps whose domain is $A$ and
whose range is a subset of $B$.

A \emph{bijection} is a map $M$ such that if $\left(
a,b\right) \in M$ and $\left( e,f\right) \in M$ then $b=f$
implies $a=e$.

For any ordered pair $\left(M,A\right) $ of a map $M$
and a set $A$, we
define $\downarrow \left( M,A\right) $, the
\emph{restriction of} $M$ \emph{to the domain}
$\mathcal{D} (M)\cap A$, to
be the subset of $M$ whose members are all the members
$(a,b) $ of $M$ such that $a\in A$ holds.   Thus $\downarrow
\left( M,A\right) $ is a map whose domain is $\mathcal{D}
\left( M\right) \cap A$.

For any set $F$ whose members are all themselves sets we
define $\mathcal{U} \left( F\right) $ to be the union of
all the members of $F$.

For any \emph{nonempty} set $F$ such that every member of
$F$ is a set, we define $\mathcal{I} \left( F\right) $ to
be the intersection of all the members of $F$.

A \emph{partition} is a set $F$ such that every member of
$F$ is a set, no member of $F$ is empty, and for every two
distinct members $A$ and $B$ of $F$, $A\cap B=\emptyset $
holds.
  And if $A$ is any set, then a
  \emph{partition of }$ A $ is a
partition $F$ such that $\mathcal{U} \left( F\right) =A$.

For any partition $V$ we define $\Xi \left( V\right) $ to
be the set whose members are all the \emph{nonempty}
subsets $A$ of $\mathcal{U} \left( V\right) $ such that if
$B$ is any member of $V$, then either $B\subseteq A$ holds
or $B\cap A$ is empty.   We note that if $V$ is any
partition, and we define a map $M$ whose domain is the set
of all the nonempty subsets of $V$ by specifying that for
each nonempty subset $X$ of $V$, $M_{ X } \equiv
\mathcal{U} \left( X\right) $, then $M$ is a bijection
whose domain is the set of all the nonempty subsets of $V$
and whose range is $\Xi \left( V\right) $.   Thus if $V$ is
a finite set, then $\Xi \left( V\right) $ is a finite set
and $\#\left( \Xi \left( V\right) \right)
=2^{ \#\left( V\right) } - 1 $.

If $F$ is a set such that every member of $F$ is a set,
then we define $\mathcal{M} \left( F\right) $ to be the set
whose members are all the members $A$ of $F$ such that
there is \emph{no} member $B$ of $F$ such that $B\subset A$
holds, and we define $\mathbb{B} \left( F\right) \equiv
\left( F\,\vdash \mathcal{M} \left( F\right) \right) $.

For any ordered pair $\left( F,i\right) $ of a set $F$
such that every
member of $F$ is a set, and a member $i$ of $\mathcal{U}
\left( F\right) $, we define $\mathcal{C} \left( F,i\right)
$ to be the intersection of all the members $A$ of $F$ such
that $i\in A$ holds, and we note that $i\in \mathcal{C}
\left( F,i\right) $ always holds, and that if $F$ is a
partition, then $\mathcal{C} \left( F,i\right) $ is equal
to the unique member of $F$ that has $i$ as a member.

For any ordered pair $\left(F,A\right) $ of a set
$F$ such that every
member of $F$ is a set, and a set $A$, we define
$\mathcal{P} \left( F,A\right) $ to be the set whose
\label{Start of original page 4}
 members are all the members $B$ of $F$ such that $B\subset
A$ holds and there is \emph{no} member $C$ of $F$ such that
$B\subset C\subset A$ holds.

A \emph{wood} is a set $F$ such that every member of $F$ is
a set, no member of $F$ is empty, $\mathcal{U} \left(
F\right) $ is $finite_{ } $, every member of $\mathcal{U}
\left( F\right) $ is a member of some member of
$\mathcal{M} \left( F\right) $, $\#\left( \mathcal{M} \left(
F\right) \right) \geq 2$ holds, and for every two distinct
members $A$ and $B$ of $F$, at least one of $A\,\vdash B$
and $A\cap B$ and $B\,\vdash A$ is empty.

We note that if $F$ is any wood and $A$ and $B$ are any
members of $F$, then exactly one of the four possibilities
$A=B$, $A\subset B$, $B\subset A$, and $A\cap B=\emptyset $
holds.   Thus if $F$ is a wood and $A$ and $B$ are any
members of $\mathcal{M} \left( F\right) $, then either
$A=B$ holds or $A\cap B=\emptyset $ holds.   Hence
$\mathcal{M}
\left( F\right) $ is a partition.   And furthermore, by the
definition of a wood, \emph{every} member of $\mathcal{U}
\left( F\right) $ is a member of some member of
$\mathcal{M} \left( F\right) $, hence $\mathcal{M} \left(
F\right) $ is a partition of $\mathcal{U} \left( F\right) $.

And we note furthermore that if $F$ is a wood and $i$ is
any member of $\mathcal{U} \left( F\right) $, then
$\mathcal{C} \left( F,i\right) $, which by definition is
the intersection of all the members of $F$ that have $i$ as
a member, is equal to the unique member of $\mathcal{M}
\left( F\right) $ that has $i$ as a member.

If $V$ is any partition such that $\mathcal{U} \left(
V\right) $ is finite and $\#\left( V\right) \geq 2$ holds,
then a \emph{wood of }$ V $ is a wood $F$
such that $\mathcal{M}
\left( F\right) =V$ holds.

We note that if $V$ is any partition such that $\mathcal{U}
\left( V\right) $ is finite and $\#\left( V\right) \geq 2$
holds, then $V$ is itself a wood of $V$, and furthermore if
$F$ is any wood of $V$, then $V\subseteq F$ holds.

If $A$ and $B$ are any sets, then we shall say that ``A
overlaps $B$'' ifif \emph{none} of $A\,\vdash B$, $A\cap B$
and $B\,\vdash A$ is empty.   In other words, $A$ overlaps
$B$ ifif \emph{none} of $A\subseteq B$, $A\cap B=\emptyset
$,
and $B\subseteq A$ are true.

Let $V$ be any partition such that $\mathcal{U} \left(
V\right) $ is finite and $\#\left( V\right) \geq 2$ holds,
and let $F$ and $G$ be any woods of $V$.   Then we may
verify directly from the definition of a wood of $V$, that
$F\cap G$ is also a wood of $V$.

Furthermore, let $V$ be any partition such that
$\mathcal{U} \left( V\right) $ is finite and $\#\left(
V\right) \geq 2$ holds, and let $F$ and $G$ be any woods of
$V$ such that no member of $F$ overlaps any member of $G$.
 Then we may again verify directly from the definition of a
wood of $V$, that $F\cup G$ is also a wood of $V$.

We note that it is true in general that if a nonempty set
$X$ is \emph{totally-ordered} and $finite_{ } $, then $X$
has both a unique maximal, or \emph{largest,} member, and a
unique minimal, or $smallest_{ } $, member.   In
\label{Start of original page 5}
 particular, if $F$ is a nonempty finite set such that
every member of $F$ is itself a set, and for every two
distinct members $A$ and $B$ of $F$, exactly one of
$A\subset B$ and $B\subset A$ holds, then there is a unique
member $Z$ of $F$ such that $A\subseteq Z$ holds for all
members $A$ of $F$, and there is a unique member $Y$ of $F$
such that $Y\subseteq A$ holds for all members $A$ of $F$.
 We shall call $Z$ the \emph{largest} member of $F$, and
$Y$ the \emph{smallest} member of $F$.

Now if $F$ is a wood, $A$ is a set, and $B$ and $C$ are any
two distinct members of $\mathcal{P} \left( F,A\right) $,
then exactly one of $B\subset C$, $B\cap C=\emptyset $, and
$C\subset B$ holds, since both $B$ and $C$ are members of
the wood $F$.   But $B\in \mathcal{P} \left( F,A\right) $
implies that $B\subset C$ cannot hold, and $C\in
\mathcal{P} \left( F,A\right) $ implies that $C\subset B$
cannot hold, hence $B\cap C=\emptyset $ must hold.
Furthermore, every member of $\mathcal{P} \left( F,A\right)
$ is a member of $F$, hence no member of $\mathcal{P}
\left( F,A\right) $ is empty.   Hence $\mathcal{P} \left(
F,A\right) $ is a partition.

Now let $B$ be any member of $F$ such that $B\subset A$
holds, and let $X$ be the set whose members are all the
members $C$ of $F$ such that $B\subseteq C$ and $C\subset
A$ both hold.   Then $B\in X$ holds,  hence $X$ is
nonempty.   Furthermore, if $C$ and $D$ are any two
distinct members of $X$, then $B\subseteq \left( C\cap
D\right) $ holds, hence $C\cap D$ is nonempty, hence
exactly one of $C\subset D$ and $D\subset C$ holds.   Hence
$X$ has a unique member $E$ such that $C\subseteq E$ holds
for all members $C$ of $X$.   Now $E\subset A$ holds by the
definition of $X$, and furthermore, if $K$ is a member of
$F$ such that $E\subset K$ and $K\subset A$ both hold, then
$K$ would be a member of $X$ such that $E\subset K$ holds,
contradicting the fact that $C\subseteq E$ holds for all
members $C$ of $X$.   Hence there is no member $K$ of $F$
such that $E\subset K$ and $K\subset A$ both hold, hence
$E$ is a member of $\mathcal{P} \left( F,A\right) $.
Hence for every member $B$ of $F$ such that $B\subset A$
holds, there is a unique member $E$ of $\mathcal{P} \left(
F,A\right) $ such that $B\subseteq E$ holds.

Now suppose that $A$ is a member of $\left( \Xi \left(
\mathcal{M} \left( F\right) \right) \,\vdash \mathcal{M}
\left( F\right) \right) $, or in other words, suppose that
$A$ is any subset of $\mathcal{U} \left( F\right) $ such
that $A$ neither overlaps any member of $\mathcal{M} \left(
F\right) $ nor is a subset of any member of $\mathcal{M}
\left( F\right) $.   Let $i$ be any member of $A$.   Then
$i$ is a member of $\mathcal{U} \left( F\right) $ hence by
the definition of a wood, the set $\mathcal{C} \left(
F,i\right) $ is a member of $\mathcal{M} \left( F\right) $
hence, since $A$ neither overlaps any member of
$\mathcal{M} \left( F\right) $ nor is a subset of any
member of $\mathcal{M} \left( F\right) $, $\mathcal{C}
\left( F,i\right) \subset A$ holds.   Hence by the
preceding paragraph there is a unique member $E$ of
$\mathcal{P} \left( F,A\right) $ such that $\mathcal{C}
\left( F,i\right) \subseteq E$ holds, hence there is a
member $E$ of $\mathcal{P} \left( F,A\right) $ such that
$i\in E$ holds.   Hence $\mathcal{P} \left( F,A\right) $ is
a partition such that every member of
\label{Start of original page 6}
 $\mathcal{P} \left( F,A\right) $ is a strict subset of
$A$, and if $i$ is any member of $A$ then there exists a
member $E$ of $\mathcal{P} \left( F,A\right) $ such that
$i\in E$ holds.   Hence $\mathcal{P} \left( F,A\right) $ is
a partition of $A$ such that $\#\left( \mathcal{P} \left(
F,A\right) \right) \geq 2$ holds.

For any wood $F$, we define the wood $\bar{F} $ by $\bar{F}
\equiv F\cup \left\{ \mathcal{U} \left( F\right) \right\} $.

For any ordered pair $\left(F,B\right) $ of a wood $F$
and a
\emph{nonempty} set $B$, we define $\mathcal{Y} \left(
F,B\right) $ to be the \emph{smallest} member $A$ of $F$
such that $B\subseteq A$ holds, if any members $A$ of $F$
exist such that $B\subseteq A$ holds, and to be equal to
the empty set $\emptyset $ if there are \emph{no} members
$A$ of
$F$ such that $B\subseteq A$ holds.

We note that if $F$ is any wood, and $B$ is any member of
$F$ such that $B\neq \mathcal{U} \left( F\right) $, then
$B\in \mathcal{P} \left( F,\mathcal{Y} \left(
\left( \bar{ F }
\,\vdash \left\{ B\right\} \right) ,B\right) \right) $
holds,
and furthermore the member $\mathcal{Y} \left(
\left( \bar{ F
} \,\vdash \left\{ B\right\} \right) ,B\right) $ of $\bar{ F
} $ is the \emph{only} member $A$ of $ \bar{ F } $ such that
$B\in \mathcal{P} \left( F,A\right) $ holds.   For
$\mathcal{Y} \left( \left( \bar{ F } \,\vdash
 \left\{ B\right\}
\right) ,B\right) $ is by definition the \emph{smallest}
member of $ \bar{ F } $ to contain $B$ as a \emph{strict}
subset.   Furthermore, since $ \bar{ F } $ is equal to $F$
apart from the fact that $\mathcal{U} \left( F\right) $
\emph{is} a member of $\bar{ F } $, while
$\mathcal{U} \left(
F\right) $ may or may not be a member of $F$, $\mathcal{P}
\left( \bar{ F } ,A\right) $ is equal to $\mathcal{P} \left(
F,A\right) $ for every member $A$ of $\bar{ F } $.   Now
$B\subset \mathcal{Y} \left( \left( \bar{ F }
\,\vdash \left\{
B\right\} \right) ,B\right) $ certainly holds, and moreover
there is no member $C$ of $F$ such that $B\subset C\subset
\mathcal{Y} \left( \left( \bar{ F } \,\vdash \left\{
B\right\}
\right) ,B\right) $ holds, for any such member $C$ of $F$
would be a member of $\bar{ F } $, since
$F\subseteq \bar{ F } $
 holds, and there is no such member $C$ of $ \bar{ F } $
since $\mathcal{Y} \left( \left( \bar{ F } \,\vdash \left\{
B\right\} \right) ,B\right) $ is the \emph{smallest} member
of $ \bar{ F } $ to contain $B$ as a strict subset.   Hence
$B\in \mathcal{P} \left( F,\mathcal{Y} \left( \left( \bar{
F }
\,\vdash \left\{ B\right\} \right) ,B\right) \right) $
holds.
  Now let $A$ be any member of $ \bar{ F } $ such that $B\in
\mathcal{P} \left( F,A\right) $ holds.   Then $B\subset A$
holds and there is \emph{no} member $C$ of $F$ such that
$B\subset C\subset A$ holds.   Now if there was a member
$C$ of $ \bar{ F } $ such that $B\subset C\subset A$ held
then $C$ would be a strict subset of $\mathcal{U} \left(
F\right) $ hence could not be equal to $\mathcal{U} \left(
F\right) $, hence $C$ would be a member of $F$. Hence there
is \emph{no} member $C$ of $ \bar{ F } $ such that $B\subset
C\subset A$ holds, hence $A$ is the \emph{smallest} member
of $ \bar{ F } $ to contain $B$ as a strict subset, hence
$A=\mathcal{Y} \left( \left( \bar{ F } \,\vdash \left\{
B\right\} \right) ,B\right) $ holds.

We note that it follows directly from this that if $F$ is
any wood then $F$ is equal to the disjoint union of
$\mathcal{P} \left( F,A\right) $ for all the members $A$ of
$\mathbb{B} \left( \bar{ F } \right) $, together with the
set
$\mathcal{U} \left( F\right) $ if $\mathcal{U} \left(
F\right) $ is a member of $F$.

We observe that for any wood $F$, the inequality
\[
\#\left( \mathcal{M} \left( F\right) \right) \leq \#\left(
F\right) \leq \left( 2\#\left( \mathcal{M} \left( F\right)
\right) -1\right)
\]
holds.   For $\mathcal{M} \left( F\right) $ is a subset of
$F$, hence $\#\left( \mathcal{M} \left( F\right) \right)
\leq \#\left( F\right) $ certainly holds, and $F$ is a
subset of $\bar{ F } $, hence $\#\left( F\right) \leq
\#\left( \bar{ F
} \right) $ holds, and moreover $\mathcal{M}
\left( \bar{ F }
 \right) =\mathcal{M} \left( F\right) $ holds.

Now for each integer $i\geq 2$ let $n_{ i } $ denote the
number of members $A$
\label{Start of original page 7}
 of $\mathbb{B} \left( \bar{ F } \right) $ such that
$\#\left(
\mathcal{P} \left( F,A\right) \right) =i$.   Then we count
the number of members of $ \bar{ F } $ in two ways:

\vspace{1.0ex}

\noindent (1) \hspace{4.0cm} \hspace{1.4ex} $\#\left(
\bar{ F } \right) =\#\left( \mathcal{M}
\left(
F\right) \right) +\displaystyle \sum_{i\geq 2 } n_{ i } $

\vspace{1.0ex}

\noindent (2)  We note that $ \bar{ F } $ is equal
to the disjoint
union of $\mathcal{P} \left( F,A\right) $ for all the
members $A$ of $\mathbb{B} \left( \bar{ F } \right) $,
together with the member $\mathcal{U} \left( F\right) $ of
$\bar{ F } $.   Hence
\[
\#\left( \bar{ F } \right) =1+\sum_{i\geq 2 } in_{ i
}
\]

From these two equations we deduce directly that
\[
\#\left( \bar{ F } \right) =2\#\left( \mathcal{M}
\left(
F\right) \right) -1-\sum_{i\geq 3 } \left( i-2\right) n_{ i
}
\]

Hence to maximize $\#\left( \bar{ F } \right) $ we take
$n_{ i
} $ as small as possible for all $i\geq 3$, (thus we take
$n_{ i } =0$ for all $i\geq 3 $), hence the maximum
possible value of $\#\left( \bar{ F } \right) $ is $\left(
2\#\left( \mathcal{M} \left( F\right) \right) -1\right) $.

\vspace{1.0ex}

For any ordered triple $\left(F,A,i\right) $ of
a wood $F$, a member $A$
of $\left( \Xi \left( \mathcal{M} \left( F\right) \right)
\,\vdash \mathcal{M} \left( F\right) \right) $, and a
member $i$ of $A$ we define $\mathcal{K} \left(
F,A,i\right) $ to be the unique member $B$ of $\mathcal{P}
\left( F,A\right) $ such that $i\in B$ holds.

For any finite set $A$ we define $\mathcal{Q} \left(
A\right) $ to be the set whose members are all the
two-member subsets of $A$, and we note that $\#\left(
\mathcal{Q} \left( A\right) \right) =
\frac{ 1 }{ 2 } \#\left( A\right)
\left( \#\left( A\right) -1\right) $.

For every ordered pair $\left(A,H\right) $ of a set
$A$, and a set $H$
such that every member of $H$ is a set, we define
$\mathcal{T} \left( A,H\right) $ to be the subset of $A$
whose members are all the members $i$ of $A$ such that
there is \emph{no} member $B$ of $H$ such that $i\in B$ and
$B\subseteq A$ both hold.

For every ordered pair $\left(F,H\right) $ of a wood
$F$, and a set $H$
such that every member of $H$ is a set, we define
$\mathcal{O} \left( F,H\right) $ to be the set whose
members are all the members $i$ of $\mathcal{U} \left(
F\right) $ such that there \emph{exists} a member $A$ of
$F$ such that $i\in A$ holds and there is \emph{no} member
$B$ of $H$ such that $i\in B$ and $B\subseteq A$ both hold.

For every ordered triple $\left(F,H,i\right) $ of
a wood $F$, a set $H$
such that every member of $H$ is a set, and a member $i$ of
$\mathcal{O} \left( F,H\right) $, we define $\mathcal{Z}
\left( F,H,i\right) $ to be the \emph{largest} member $A$
of $F$ such that $i\in A$ holds and there is \emph{no}
member $B$ of $H$ such that $i\in B$ and $B\subseteq A$
both hold.

For any sets $A$ and $B$ we define $\mathbb{K} \left(
A,B\right) $ to be the set whose members are all the sets
$C$ such that $A\subseteq C$ and $C\subseteq B$ both hold,
and we note that $\mathbb{K} \left( A,B\right) $ is equal
to the empty set $\emptyset $ if $A\subseteq B$ does
\emph{not}
hold.
\label{Start of original page 8}

For any ordered
 triple $\left(F,A,B\right) $ of a wood $F$, a
\emph{nonempty} set $A$, and a set $B$, we define
$\mathbb{Y} \left( F,A,B\right) $ to be the set whose
members are all the members $C$ of $F$ such that $A\subset
C$ and $C\subseteq B$ both hold, and we note that
$\mathbb{Y} \left( F,A,B\right) $ is equal to the empty set
$\emptyset $ if $A\subset B$ does \emph{not} hold.

The symbol $\mathbb{N} $ denotes the set of all the finite
integers greater than or equal to zero, the symbol
$\mathbb{Z} $ denotes the set of \emph{all} the finite
integers, and the symbol $\mathbb{R} $ denotes the set of
all the finite real numbers.

If $d$ is an integer $\geq 1$, then $\mathbb{E} _{ d } $
denotes $ d $-dimensional Euclidean space.

We define $\mathbb{S} \left( s\right) $, for all $s\in
\mathbb{R} $, by
\[
\mathbb{S} \left( s\right) \equiv \left\{\begin{array}{cc}
 0 & \textrm{if }s<0\textrm{ holds} \\
 1 & \textrm{if }s\geq
0\textrm{ holds}
\end{array}
\right.
\]

And we define $\mathbb{T} \left( s\right) $, for all $s\in
\mathbb{R} $, by
\[
\mathbb{T} \left( s\right) \equiv \left\{\begin{array}{cc}
 1 & \textrm{if }s<0\textrm{ holds} \\
 0 & \textrm{if }s\geq
0\textrm{ holds}
\end{array}
 \right.
\]

We note that $\mathbb{S} \left( s\right) +\mathbb{T} \left(
s\right) =1$ holds for all $s\in \mathbb{R} $.

Throughout this paper, a variable with the $ \hat{ } $ sign
above it, indicates the partial derivative with respect to
that variable.   For example $\hat{ y } $ is short for $
\frac{ \partial }{ \partial y } $.

\section{Review of the Cluster Convergence Theorem.}
\label{Section 3}

A \emph{greenwood} is a wood $F$ such that every member of
$\mathcal{M} \left( F\right) $ has exactly one member, and
if $A$ is any finite set such that $\#\left( A\right) \geq
2$ holds, then a \emph{greenwood of }$ A $ is a greenwood
$F$
such that $\mathcal{U} \left( F\right) =A$ holds.   We say
a greenwood $F$ is \emph{high} ifif $\mathcal{U} \left(
F\right) \in F$ holds.

If $d$ is an integer $\geq 1$, $x$ is a map such that
$\mathcal{D} \left( x\right) $ is finite, $\#\left(
\mathcal{D} \left( x\right) \right) \geq 2$ holds, and
$\mathcal{R} \left( x\right) $ is a subset of $\mathbb{E}
_{ d } $, and $\sigma $ is a real number such that
$0<\sigma <1$ holds, then a $\sigma $\emph{-cluster
 of }$x $ is a
nonempty subset $A$ of $\mathcal{D} \left( x\right) $ such
that either $\#\left( A\right) =1$ holds, or else $\left|
x_{ i }
-x_{ j } \right| <\sigma \left| x_{ k } -x_{ m } \right| $
holds for all $i\in
A$, $j\in A$, $k\in A$, and $m\in \left( \mathcal{D} \left(
x\right) \,\vdash A\right) $.

We note that every one-member subset of $\mathcal{D} \left(
x\right) $ is always a $\sigma $-cluster of $x$, and that
$\mathcal{D} \left( x\right) $ is always a $\sigma
$-cluster of $x$.

We denote the set whose members are all the $\sigma
$-clusters of $x$ by $\mathcal{F} \left( x,\sigma \right) $.

We note that no two $\sigma $-clusters of $x$ can overlap.
 For if $A$ and
\label{Start of original page 9}
 $B$ are two overlapping subsets of $\mathcal{D} \left(
x\right) $ then $A$ has a member, say $i$, that is not a
member of $B$, $\left( A\cap B\right) $ has at least one
member, say $j$, and $B$ has at least one member, say $k$,
that is not a member of $A$.   Hence $\#\left( A\right) \geq
2$ and $\#\left( B\right) \geq 2$.   Hence if $A$ is a
$\sigma $-cluster of $x$ then $\left| x_{ i } -x_{ j }
\right| <\sigma
\left| x_{ j } -x_{ k } \right| $, and if $B$ is a $\sigma
$-cluster of
$x$ then $\left| x_{ j } -x_{ k } \right| <\sigma \left|
x_{ i } -x_{ j } \right| $,
hence $A$ and $B$ cannot both be $\sigma $-clusters of $x$.

It follows from this that $\mathcal{F} \left( x,\sigma
\right) $ is a high greenwood of $\mathcal{D} \left(
x\right) $.

For any ordered triple $\left( x,\sigma ,F\right) $ whose
first component is a map $x$ such that $\mathcal{D} \left(
x\right) $ is finite, $\#\left( \mathcal{D} \left( x\right)
\right) \geq 2$, and $\mathcal{R} \left( x\right) \subseteq
\mathbb{E} _{ d } $, and whose second component is a real
number $\sigma $ such that $0<\sigma <1$, and whose third
component is a high greenwood $F$ of $\mathcal{D} \left(
x\right) $, we define the number $\mathcal{A} \left(
x,\sigma ,F\right) $ by
\[
\mathcal{A} \left( x,\sigma ,F\right) \equiv
\left\{ \begin{array}{cc}
1 & \textrm{if }\mathcal{F}
\left( x,\sigma \right) =F \\
0 & \textrm{if }\mathcal{F} \left( x,\sigma
\right) \neq F
\end{array} \right\} .
\]

For any finite set $J$ such that $\#\left( J\right) \geq 2$
holds, we define $\mathbb{H} \left( J\right) $ to be the
set whose members are all the high greenwoods $F$ of $J$ .
Then the following identity holds for all $x\in
\mathbb{E} _{ d }^{ J } $ and for all $0<\sigma <1$.
\[
\sum_{
F\in \mathbb{H} \left( J\right) }
 \mathcal{A} \left( x,\sigma
,F\right) =1.
\]
(We note that $\mathcal{A} \left( x,\sigma ,F\right) $ is
$0$ unless $\mathcal{D} \left( x\right) \in F$ holds, for
$\mathcal{D} \left( x\right) $ is always a $\sigma
$-cluster of $ x $.)

For any ordered pair $\left(x,L\right) $ whose first
component is a map
$x$ such that $\mathcal{D} \left( x\right) $ is finite,
$\#\left( \mathcal{D} \left( x\right) \right) \geq 2$, and
$\mathcal{R} \left( x\right) \subseteq \mathbb{E} _{ d } $,
and whose second component is a real number $L\geq 0$, we
define the number $\mathcal{B} \left( x,L\right) $ by
\[
\mathcal{B} \left( x,L\right) \equiv \left\{
\begin{array}{cc}
1 & \textrm{if }\left| x_{ i } -x_{
j }
\right| \leq L\textrm{ holds for all }
i\in \mathcal{D} \left( x\right)
\textrm{ and } j\in
\mathcal{D} \left( x\right) \\
0 & \textrm{otherwise}
\end{array} \right\} ,
\]
and we note that
\[
\mathcal{B} \left( x,L\right) =\prod_{\Delta \equiv
\left\{ i,j\right\} \in \mathcal{Q} \left( \mathcal{D}
\left(
x\right) \right) }  \mathbb{S} \left( L-\left| x_{ i } -x_{
j }
\right| \right)
\]
holds, where $\mathbb{S} \left( u\right) $ was defined on
page \pageref{Start of original page 8}, for all
$u\in \mathbb{R} $, to be equal to $1$ for
$u\geq 0$, and equal to $0$ otherwise.

If $A$ is a finite set then a
\emph{set of powers for }$ A $ is a
map $\alpha $ such that $\mathcal{Q} \left( A\right)
\subseteq \mathcal{D} \left( \alpha \right) $ holds, and
for each member $\Delta $ of $\mathcal{Q} \left( A\right)
$, $\alpha _{ \Delta } \in \mathbb{R} $ holds.

For any ordered pair $\left( x,\alpha \right) $ whose first
component is a map $x$ such that $\mathcal{D} \left(
x\right) $ is finite, $\#\left( \mathcal{D} \left( x\right)
\right) \geq 2$, and $\mathcal{R} \left( x\right) \subseteq
\mathbb{E} _{ d } $, and whose second component is a set of
powers $\alpha $ for $\mathcal{D} \left( x\right) $, we
define
\label{Start of original page 10}
\[
 \Psi \left( x,\alpha \right) \equiv \prod_{\Delta
\equiv \left\{ i,j\right\} \in \mathcal{Q} \left(
\mathcal{D}
\left( x\right) \right) }   \left| x_{ i } -x_{ j } \right|
^{ -\alpha
_{ \Delta } }.
\]

For any ordered pair $\left( \alpha ,A\right) $ such that
$A$ is a finite set and $\alpha $ is a set of powers for
$A$, we define
\[
\Gamma \left( \alpha ,A\right) \equiv \sum_{ \Delta \in
\mathcal{Q} \left( A\right) } \alpha_{ \Delta }.
\]

If $F$ is a greenwood and $d$ is an integer $\geq 1$, then
a \emph{good set of powers for }$ \left( F,d\right) $
is a set of powers $\alpha
$ for $\mathcal{U} \left( F\right) $ such that $\Gamma
\left( \alpha ,A\right) <d\left( \#\left( A\right) -1\right)
$ holds for every member $A$ of $\mathbb{B} \left( F\right)
$.

The Cluster Convergence Theorem can now be stated as
follows:

Let $d$ be any integer $\geq 1$, $J$ be any finite set such
that $\#\left( J\right) \geq 2$, $F$ be any high greenwood
of $J$, $\alpha $ be any good set of powers for
$ \left( F,d\right) $,
$\sigma $ be any real number such that $0<\sigma <1$, and
$L$ be any real number $\geq 0$.   Let $i$ be any member of
$J$, $z$ be any member of $\mathbb{E} _{ d } $, and
$\mathbb{W}$ be the subset of $\mathbb{E} _{ d }^{ J }$
whose members are all the members $x$ of
$\mathbb{E} _{ d }^{ J } $
such that $x_{ i } =z$.   Then the integral of $\mathcal{A}
\left( x,\sigma ,F\right) \mathcal{B} \left( x,L\right)
\Psi \left( x,\alpha \right) $ over $\mathbb{W}$ is finite.

\section{Good Sets of Woods.}
\label{Section 4}

If $A$ is a set, and $U$ is a set such that every member of
$U$ is a set, we shall say that $A$ is $U$\emph{-connected}
ifif for every partition $\left\{ B,C\right\} $ of $A$ into
two nonempty parts $B$ and $C$, there exists a member $E$
of $U$ such that $B\cap E$ and $C\cap E$ are both nonempty.

\begin{bphzlemma} \label{Lemma 1}
\end{bphzlemma}
\vspace{-6.143ex}

\noindent \hspace{10.657ex}{\bf.  }Let $U$ be any
set such that every member of $U$
is a set, and let $X$ be any \emph{nonempty} set such that
every member of $X$ is a $U$-connected set and $\mathcal{I}
\left( X\right) $ is nonempty.   Then $\mathcal{U} \left(
X\right) $ is a $U$-connected set.

\vspace{2.5ex}

\noindent {\bf Proof.}  Let $\left\{ J,K\right\} $
be any partition of
$\mathcal{U} \left( X\right) $ into two nonempty parts.
Now by assumption $\mathcal{I} \left( X\right) $ is
nonempty.   Let $i$ be a member of $\mathcal{I} \left(
X\right) $.   Then $i$ is a member of exactly one member of
$\left\{ J,K\right\} $.   Let $M$ be the member of $\left\{
J,K\right\} $ that has $i$ as a member, and let $N$ be the
other member of $\left\{ J,K\right\} $.   Then $M\cap A$ is
nonempty for \emph{every} member $A$ of $X$, (since $i$ is
a member of every member of X), and furthermore since $N$
is a nonempty subset of $\mathcal{U} \left( X\right) $,
there exists at least one member $B$ of $X$ such that
$N\cap B$ is nonempty.   Let $B$ be a member of $X$ such
that $N\cap B$ is nonempty.   Then $\left\{ \left( M\cap
B\right) ,\left( N\cap B\right) \right\} =\left\{ \left(
J\cap B\right) ,\left( K\cap B\right) \right\} $ is a
partition of $B$ into two
\label{Start of original page 11}
 nonempty parts hence, since $B$ is $U$-connected, there
exists a member $E$ of $U$ such that $E$ intersects both
$\left( M\cap B\right) $ and $\left( N\cap B\right) $, or
in other words, such that $E$ intersects both $\left( J\cap
B\right) $ and $\left( K\cap B\right) $, hence there exists
a member $E$ of $U$ such that $E$ intersects both $J$ and
$K$.

(We note that the example $U=\left\{ \left\{ 1,3\right\}
,\left\{ 2,3\right\} ,\left\{ 1,4\right\} ,\left\{
2,4\right\}
\right\} $, \\
$X=\left\{ \left\{ 1,2,3\right\} ,\left\{
1,2,4\right\} \right\} $, shows that $\mathcal{I} \left(
X\right) $ being nonempty and every member of $X$ being
$U $-connected does \emph{not} imply that $\mathcal{I}
\left(
X\right) $ is $U $-connected.)

\vspace{2.5ex}

If $A$ is a set and $U$ is a set such that every member of
$U$ is a set, then a $U $\emph{-connected component
 of }$A $ is a
nonempty subset $B$ of $A$ such that $B$ is $U$-connected
and
$B$ is \emph{not} a strict subset of any $U$-connected
subset
of $A$.

\begin{bphzlemma} \label{Lemma 2}
\end{bphzlemma}
\vspace{-6.143ex}

\noindent \hspace{10.857ex}{\bf.  }Let $A$ be a
set, let $U$ be a set such that every
member of $U$ is a set, and let $F$ be the set whose
members are all the $U$-connected components of $A$.   Then
$F$ is a partition of $A$.

\vspace{2.5ex}

\noindent {\bf Proof.}  We shall first prove that
any member $i$ of $A$ is
a member of a \emph{unique} $U$-connected component $B$ of
$A$.

Let $i$ be any member of $A$, and let $X$ be the set whose
members are all the $U$-connected subsets $C$ of $A$ such
that $i\in C$ holds.   Then $i$ is a member of $\mathcal{I}
\left( X\right) $, hence $\mathcal{I} \left( X\right) $ is
nonempty, hence by Lemma \ref{Lemma 1},
$\mathcal{U} \left( X\right) $
is $U$-connected.   And furthermore, $\mathcal{U} \left(
X\right) $ is \emph{not} a strict subset of any subset $C$
of $A$ such that $C$ is $U$-connected, for if $C$ is any
$U$-connected subset of $A$ such that $\mathcal{U} \left(
X\right) \subseteq C$ holds, then $C$ is a $U$-connected
subset of $A$ such that $i\in C$ holds, hence $C$ is a
member of $X$, hence $C\subseteq \mathcal{U} \left(
X\right) $ holds, hence $C$ is equal to $\mathcal{U} \left(
X\right) $.   Hence $\mathcal{U} \left( X\right) $ is a
$U$-connected component of $A$, hence $\mathcal{U} \left(
X\right) \in F$ holds.

Furthermore, $\mathcal{U} \left( X\right) $ is the
\emph{only} member $B$ of $F$ such that $i\in B$ holds.
For let $B$ be any member of $F$ such that $i\in B$ holds.
 Then $B$ is a $U$-connected subset of $A$ such that $i\in
B$
holds, hence $B\in X$ holds, hence $B$ is a subset of
$\mathcal{U} \left( X\right) $.   Furthermore, since $B$ is
a $U$-connected component of $A$ by the definition of $F$,
$B$ cannot be a \emph{strict} subset of the $U$-connected
subset $\mathcal{U} \left( X\right) $ of $A$, hence
$B=\mathcal{U} \left( X\right) $ holds.

Finally we note that, since every member of $F$ is a subset
of $A$, $\mathcal{U} \left( F\right) \subseteq A$ holds,
and furthermore, since every member of $F$ is a
$U$-connected
component of $A$, and by definition the empty set
$\emptyset $
is
\label{Start of original page 12}
 \emph{not} a $U$-connected component of $A$, the empty set
$\emptyset $ is \emph{not} a member of $F$.

Hence $F$ is a partition of $A$.

\vspace{2.5ex}

If $d$ is any integer $\geq 1$ and $x$ is any map such that
$\mathcal{D} \left( x\right) $ is finite and $\mathcal{R}
\left( x\right) \subseteq \mathbb{E} _{ d } $ holds, then
by definition $\mathbb{V} \left( x\right) $, the
\emph{convex hull of }$x $, is the set of all members $y$ of
$\mathbb{E} _{ d } $ such that there exists a member $s$ of
$\mathbb{R}^{ \mathcal{D} \left( x\right) } $ such that $s_{
i } \geq 0$ holds for all $i\in \mathcal{D} \left( x\right)
$, $ \sum_{ i\in \mathcal{D} \left( x\right) } s_{ i } =1$
holds, and $ y = \sum_{ i\in \mathcal{D} \left( x\right) }
s_{ i } x_{ i } $ holds.

\begin{bphzlemma} \label{Lemma 3}
\end{bphzlemma}
\vspace{-6.143ex}

\noindent \hspace{10.857ex}{\bf.  }Let $A$ be
any nonempty finite set, $x$ be any
member of $\mathbb{E} _{ dA_{ } } $, and $u$ and $v$ be
any members of the convex hull of $x$.   Then $\left|
u-v\right| \leq
\hspace{0.5ex} \max_{ \hspace{-4.5ex}
\begin{array}{c} \\[-3.6ex]
\scriptstyle{ i\in A } \\[-1.8ex]
\scriptstyle{ j\in A }
\end{array} } \hspace{-0.2ex}
\left| x_{ i } -x_{ j } \right| $.

\vspace{2.5ex}

\noindent {\bf Proof.}  Let $s\in \mathbb{R}^{ A }$
be such that $s_{ i }
\geq 0$ holds for all $i\in A$, $ \sum_{ i\in A } s_{ i }
= 1 $, and $ u = \sum_{ i\in A } s_{ i } x_{ i } $ all
hold, and let $t\in \mathbb{R}^{ A }$ be such that
$t_{ j } \geq 0$ holds for all $j\in A$, $ \sum_{ j\in A }
t_{ j } = 1 $, and $ v = \sum_{ j\in A } t_{ j } x_{ j }
$ all hold. Then
\[
\left| u-v\right| =\left| \left( \sum_{ k\in A }
s_{ k } x_{ k } \right) - \left( \sum_{ m\in A }
t_{ m }
x_{ m } \right) \right| =\left| \sum_{
\begin{array}{c} \\[-4.75ex]
\scriptstyle{ k\in A } \\[-1.75ex]
\scriptstyle{ m\in A }
\end{array} } \hspace{-0.9ex}
s_{ k } t_{ m } \left( x_{ k } -x_{ m }
 \right)
\right| \leq \hspace{2.0cm}
\]
\[
\hspace{2.0cm}
\leq \sum_{ \begin{array}{c} \\[-4.75ex]
\scriptstyle{ k\in A } \\[-1.75ex]
\scriptstyle{ m\in A }
\end{array} } \hspace{-0.9ex}
s_{ k } t_{ m } \left| x_{ k } -x_{
m }
\right| \leq \sum_{ \begin{array}{c} \\[-4.75ex]
\scriptstyle{ k\in A } \\[-1.75ex]
\scriptstyle{ m\in A }
\end{array} } \hspace{-0.9ex}
s_{ k } t_{ m } \max_{ \begin{array}{c} \\[-4.75ex]
\scriptstyle{ i\in A } \\[-1.75ex]
 \scriptstyle{ j\in A }
\end{array} } \left| x_{ i } -x_{ j } \right|
=\max_{ \begin{array}{c} \\[-4.75ex]
\scriptstyle{ i\in A } \\[-1.75ex]
 \scriptstyle{ j\in A }
\end{array} } \left| x_{ i } -x_{ j }
\right| .
\]

\vspace{2.5ex}

We observe that if $V$ is any partition, $A$ is any member
of $\Xi \left( V\right) $, and $F$ is the set whose members
are all the members $B$ of $V$ such that $B\subseteq A$
holds, then $F$ is a partition of $A$.   For if $i$ is any
member of $A$, then $i$ is a member of $\mathcal{U} \left(
V\right) $, hence since $V$ is a partition, there is a
unique member $B$ of $V$ such that $i\in B$ holds, and if
$B$ is the unique member of $V$ such that $i\in B$ holds,
then $B\cap A$ is nonempty since it has the member $i$,
hence by the definition of $\Xi \left( V\right) $,
$B\subseteq A$ holds, hence $B\in F$ holds, and
furthermore, since $F$ is a subset of $V$, $B$ is the
\emph{only} member of $F$ to have $i$ as a member.   Hence
if $i$ is any member of $A$, then there is a unique member
$B$ of $F$ such that $i\in B$ holds, and furthermore, since
every member of $F$ is a subset of $A$, $\mathcal{U} \left(
F\right) \subseteq A$
\label{Start of original page 13}
 holds, hence $F$ is a partition of $A$.

We note, furthermore, that if $V$ is any partition, $A$ is
any member of $\Xi \left( V\right) $, and $F$ is the subset
of $V$ whose members are all the members $B$ of $V$ such
that $B\subseteq A$ holds, then $F$ is the one-member set
$\left\{ A\right\} $ if $A$ is a member of $V$, and $F$ is
equal to $\mathcal{P} \left( V,A\right) $ if $A$ is a
member of $\left( \Xi \left( V\right) \,\vdash V\right) $.

If $V$ is a partition such that $V$ is a finite set then a
\emph{set of contraction weights for }$ V $ is a map
$\omega $ such
that the following three conditions are all satisfied:

\vspace{1.0ex}

\noindent (i) if $A$ and $B$ are any members of $\Xi
\left( V\right) $
such that $B\subseteq A$ holds, then the ordered pair
$\left(A,B\right) $
is a member of $\mathcal{D} \left( \omega \right) $, and
$\omega _{ AB } $ is a real number such that $\omega _{ AB
} \geq 0$ holds.

\vspace{1.0ex}

\noindent (ii) if $A$ is any member of $\Xi
\left( V\right) $, and $F$
is any partition of $A$ such that $F\subseteq \Xi \left(
V\right) $ holds, then $\sum_{B\in F } \omega _{ AB } =1$
holds.

\vspace{1.0ex}

\noindent (iii) if $A$, $B$ and $C$ are any members of
$\Xi \left(
V\right) $ such that $C\subseteq B\subseteq A$ holds, then
$\omega _{ AC } =\omega _{ AB } \omega _{ BC } $ holds.

\vspace{1.0ex}

We note that these conditions have the immediate
consequence that if $A$ and $B$ are any members of $\Xi
\left( V\right) $ such that $B\subseteq A$ holds, and $F$
is any partition of $B$ such that $F\subseteq \Xi \left(
V\right) $ holds, then $\omega _{ AB } =\sum_{C\in F }
\omega _{ AC } $ holds.  For $\sum_{C\in F } \omega _{ AC }
=\sum_{C\in F } \omega _{ AB } \omega _{ BC } =\omega _{ AB
} \sum_{C\in F } \omega _{ BC } $ holds by condition
(iii), and $\sum_{C\in F } \omega _{ BC } =1$ holds
by condition (ii).

And we note that one class of solutions to these conditions
is obtained by making any choice of the real numbers
$\omega _{ AC } $ for $A=\mathcal{U} \left( V\right) $ and
$C\in V$ such that each of these real numbers is
\emph{strictly} greater than $0$ and such that their sum is
equal to $1$, (so that condition (ii) is satisfied for
$A=\mathcal{U} \left( V\right) $), then using the
result of the above paragraph to calculate $\omega _{ AB }
$ when $A=\mathcal{U} \left( V\right) $ and $B$ is any
member of $\Xi \left( V\right) $ as $\omega _{ AB } =
\sum_{C\in
F } \omega _{ AC } $, where $F$ is the partition of
$B$ whose members are all the members $C$ of $V$ such that
$C\subseteq B$ holds, and noting that this $\omega _{ AB }
$ is \emph{strictly} greater than $0$, then finally using
condition (iii) to calculate $\omega _{ BC } $, for any
members $B$ and $C$ of $\Xi \left( V\right) $ such that
$C\subseteq B$ holds, as $ \omega_{ BC } = \frac{ \omega_{
AC } }{ \omega_{ AB } } $.   We may verify directly that
the map $\omega $ constructed in this way satisfies all
the conditions (i), (ii), and
\label{Start of original page 14}
(iii) above, and we observe that this construction gives
the general solution of conditions (i), (ii), and (iii) if
$\omega _{ AB } $ is required to be \emph{strictly} greater
than $0$ for all members $A$ and $B$ of $\Xi \left(
V\right) $ such that $B\subseteq A$ holds.

An example of a solution to conditions (i), (ii) and (iii)
that does \emph{not} require $\omega _{ AB } $ to be
\emph{strictly} greater than $0$ for all members $A$ and
$B$ of $\Xi \left( V\right) $ such that $B\subseteq A$
holds is obtained by choosing a numbering of a finite
number of members of $\mathcal{U} \left( V\right) $ such
that every member of $V$ has at least one numbered member,
and defining $\omega _{ AB } \equiv 1$ if $B$ contains the
highest-numbered member of $A$, and $\omega _{ AB } \equiv
0$ otherwise.

We observe that it also follows directly from the defining
conditions (i), (ii), and (iii) for a set $\omega $ of
contraction weights for $V$, that if $A$ is any member of
$\Xi \left( V\right) $ then $\omega _{ AA } =1$ holds.

For any integer $d\geq 1$ and any ordered pair $\left(
V,\omega \right) $ of a partition $V$ such that $V$ is a
finite set, and a set $\omega $ of contraction weights for
$V$, we define $\mathbb{U} _{ d } \left( V,\omega \right) $
to be the set whose members are all the members $x$ of
$\mathbb{E} _{ d }^{ \Xi \left( V\right) } $ such that for
every member $A$ of $\left( \Xi \left( V\right) \,\vdash
V\right) $, $x_{ A } = \displaystyle \sum_{ { B\in
\mathcal{P} \left(
V,A\right) } } \omega _{ AB } x_{ B } $ holds.

\begin{bphzlemma} \label{Lemma 4}
\end{bphzlemma}
\vspace{-6.143ex}

\noindent \hspace{10.857ex}{\bf.  }Let $V$ be
any partition such that $V$ is a finite
set, $\omega $ be any set of contraction weights for $V$,
$d$ be any integer $\geq 1$, $x$ be any member of
$\mathbb{U} _{ d } \left( V,\omega \right) $, $A$ be any
member of $\Xi \left( V\right) $, and $F$ be any partition
of $A$ such that every member of $F$ is a member of $\Xi
\left( V\right) $.
\enlargethispage{1.0ex}

Then $x_{ A } = \displaystyle \sum_{B\in F } \omega_{ AB }
x_{ B } $ holds.

\vspace{2.5ex}

\noindent {\bf Proof.}  We note first that if $A$ is a
member of $V$ then
the only possible partition $F$ is the partition $F=\left\{
A\right\} $, hence the stated result follows immediately
from the fact that $\omega _{ AA } =1$ holds.

Now suppose $A$ is a member of $\left( \Xi \left( V\right)
\,\vdash V\right) $.   Then we find:
\[
\sum_{B\in F } \omega _{ AB } x_{ B } = \left(
\sum_{B\in \left( F\cap
V\right) } \omega _{ AB } x_{ B } \right) + \left(
\sum_{B\in \left(
F\,\vdash V\right) } \omega _{ AB } x_{ B } \right)
\]
\[
= \left( \sum_{B\in \left( F\cap V\right) } \omega_{ AB }
x_{ B } \right) + \left( \sum_{B\in \left(
F\,\vdash V\right) } \omega _{ AB } \left(
\sum_{C\in \mathcal{P} \left( V,B\right) } \omega _{ BC }
x_{ C } \right) \right)
\]
(by the definition of $\mathbb{U} _{ d } \left( V,\omega
\right) $)
\label{Start of original page 15}
\[
= \left( \sum_{ B\in \left( F\cap V\right) }
\omega _{ AB } x_{ B } \right) + \left( \sum_{ B\in \left(
F\,\vdash V\right) } \left( \sum_{ C\in \mathcal{P}
\left( V,B\right) } \omega _{ AC } x_{ C } \right) \right)
\]
(by condition (iii)
on a set of contraction weights for V.)

But the set
whose members are the set $\left( F\cap V\right) $,
together with the sets $\mathcal{P} \left( V,B\right) $
corresponding to the members $B$ of $\left( F\,\vdash
V\right) $, is a partition of $\mathcal{P} \left(
V,A\right) $, hence
\[
\left( \sum_{ B\in \left( F\cap V\right) } \omega _{ AB }
x_{ B } \right) + \left( \sum_{ B\in \left( F\,\vdash V
\right) } \left( \sum_{ C\in \mathcal{P}
\left( V,B\right) } \omega _{ AC } x_{ C } \right) \right)
= \sum_{B\in
\mathcal{P} \left( V,A\right) } \omega _{ AB } x_{ B
} =x_{ A }
\]
holds.

\vspace{2.5ex}

If $d$ is an integer $\geq 1$ and $V$ is a partition such
that $V$ is a finite set, then we define $\mathbb{F} _{ d }
\left( V\right) $ to be the set whose members are all the
members $x$ of $\mathbb{E} _{ d }^{ \Xi \left( V\right) } $
such that if $A$ is any member of $\Xi \left( V\right) $
and $F$ is any partition of $A$ such that $F\subseteq \Xi
\left( V\right) $ holds, then $x_{ A } \in \mathbb{V}
\left( \downarrow \left( x,F\right) \right) $ holds, or in
other words, $x_{ A } $ is a member of the convex hull of
the $x_{ B } $, $B\in F$.

We observe that if $d$ is any integer $\geq 1$, $V$ is any
partition such that $V$ is a finite set, and $\omega $ is
any set of contraction weights for $V$, then it immediately
follows from Lemma \ref{Lemma 4} together with
the defining properties
of a set of contraction weights for $V$, that $\mathbb{U}
_{ d } \left( V,\omega \right) $ is a subset of $\mathbb{F}
_{ d } \left( V\right) $.

However $x\in \mathbb{F} _{ d } \left( V\right) $ does
\emph{not} imply that there exists a set $\omega $ of
contraction weights for $V$ such that $x$ is a member of
$\mathbb{U} _{ d } \left( V,\omega \right) $.   For example
if $d=1$, $V$ is the partition $\left\{ A,B,C\right\} $
where
$A$, $B$, and $C$ are three different, nonintersecting,
nonempty sets, and $x$ is the member of $\mathbb{F} _{ d }
\left( V\right) $ defined by $x_{ A } =-3$, $x_{ A\cup C }
=-2$, $x_{ C } =-1$, $x_{ A\cup B\cup C } =0$, $x_{ B\cup C
} =1$, $x_{ A\cup B } =2$, and $x_{ B } =3$, then if
$\omega $ was a set of contraction weights for $V$ such
that $x$ was a member of $\mathbb{U} _{ d } \left( V,\omega
\right) $, then from Lemma \ref{Lemma 4}, condition
 (ii) on $\omega $,
and the partition $\left\{ A,\left( B\cup C\right) \right\}
$
we would find $\omega _{ \left( A\cup B\cup C,A\right) } =
\frac{ 1 }{ 4 } $, and from Lemma \ref{Lemma 4},
condition (ii)
on $\omega $,
and the partition $\left\{ \left( A\cup B\right) ,C\right\}
$
we would find $\omega _{ \left( A\cup B\cup C,A\cup
B\right) } = \frac{ 1 }{ 3 } $, and from the definition of
$\mathbb{U} _{ d } \left( V,\omega \right) $, condition
(ii) on $\omega $, and the partition $\left\{ A,B\right\} $
we would find $\omega _{ \left( A\cup B,A\right) } =
\frac{ 1 }{ 6
} $, which gives $\omega _{ \left( A\cup B\cup C,A\cup
B\right) } \omega _{ \left( A\cup B,A\right) } =
\frac{ 1}{ 18 }
 \neq \frac{ 1 }{ 4} $ in violation of condition
 (iii) on $\omega $.
\label{Start of original page 16}

 We observe that if $V$ is any partition, $F$ is any
partition such that $F\subseteq \Xi \left( V\right) $
holds, and $A$ is any member of $\Xi \left( F\right) $,
then $A$ is a subset of $\mathcal{U} \left( V\right) $ such
that if $B$ is any member of $V$, then either $B\subseteq
A$ holds or $B\cap A=\emptyset $ holds, hence $A$ is a
member of
$\Xi \left( V\right) $.

Now let $d$ be any integer $\geq 1$, $V$ be any partition
such that $V$ is a finite set, $F$ be any partition such
that $F\subseteq \Xi \left( V\right) $ holds, $A$ be any
member of $\Xi \left( F\right) $, and $x$ be any member of
$\mathbb{F} _{ d } \left( V\right) $.   Then $x_{ A } \in
\mathbb{V} \left( \downarrow \left( x,F\right) \right) $
holds, or in other words, $x_{ A } $ is a member of the
convex hull of the $x_{ B } $, $B\in F$.   For as shown on
page \pageref{Start of original page 12}, if $G$ is
the subset of $F$ whose members are all
the members $B$ of $F$ such that $B\subseteq A$ holds, then
$G$ is a partition of $A$, hence by the definition of
$\mathbb{F} _{ d } \left( V\right) $, $x_{ A } \in
\mathbb{V} \left( \downarrow \left( x,G\right) \right) $
holds.   And it follows directly from the fact that $G$ is
a subset of $F$ that $\mathbb{V} \left( \downarrow \left(
x,G\right) \right) \subseteq \mathbb{V} \left( \downarrow
\left( x,F\right) \right) $ holds, hence $x_{ A } \in
\mathbb{V} \left( \downarrow \left( x,F\right) \right) $
holds.

For any ordered triple $\left(F,A,x\right) $ of
a wood $F$, a member $A$
of $\left( \Xi \left( \mathcal{M} \left( F\right) \right)
\,\vdash \mathcal{M} \left( F\right) \right) $, and a
member $x$ of $\mathbb{F} _{ d } \left( \mathcal{M} \left(
F\right) \right) $, where $d$ is an integer $\geq 1$, we
define \\
$\mathbb{L} \left( F,A,x\right) \equiv
\hspace{1.5ex} \max_{ \hspace{-7.0ex}
\begin{array}{c} \\[-3.1ex]
\scriptstyle{ B\in
\mathcal{P} \left( F,A\right) } \\[-1.2ex]
\scriptstyle{ C\in \mathcal{P} \left(
F,A\right) }
\end{array} } \hspace{-1.2ex}
\left| x_{ B } -x_{ C } \right| $.

We recall from
page \pageref{Start of original page 7}
 that for any ordered pair
$\left(F,H\right) $ of a
wood $F$ and a set $H$ such that every member of $H$ is a
set, we define $\mathcal{O} \left( F,H\right) $ to be the
set whose members are all the members $i$ of $\mathcal{U}
\left( F\right) $ such that there \emph{exists} a member
$A$ of $F$ such that $i\in A$ holds and there is \emph{no}
member $B$ of $H$ such that $i\in B$ and $B\subseteq A$
both hold.

And we also recall from
page \pageref{Start of original page 7}
 that for any ordered triple
$\left(F,H,i\right) $ of a wood $F$, a set
 $H$ such that every member of
$H$ is a set, and a member $i$ of $\mathcal{O} \left(
F,H\right) $, we define $\mathcal{Z} \left( F,H,i\right) $
to be the \emph{largest} member $A$ of $F$ such that $i\in
A$ holds and there is \emph{no} member $B$ of $H$ such that
$i\in B$ and $B\subseteq A$ both hold.

We note that if $F$ is any wood, $H$ is any
\emph{partition,} and $A$ is any member of $\Xi \left(
\mathcal{M} \left( F\right) \right) $, and $i$ is any
member of $A$ and $j$ is any member of $\left( \mathcal{U}
\left( F\right) \,\vdash A\right) $ such that $\left\{
i,j\right\} $ is a member of $H$, then the member
$\mathcal{C} \left( F,i\right) $ of $\mathcal{M} \left(
F\right) $ intersects $A$ hence is a subset of $A$, hence
$j$ is \emph{not} a member of $\mathcal{C} \left(
F,i\right) $, hence the \emph{unique} member $\left\{
i,j\right\} $ of $H$ that has $i$ as a member, is \emph{not}
a subset of $\mathcal{C} \left( F,i\right) $, hence $i$ is
a member of $\mathcal{O} \left( F,H\right) $, hence
$\mathcal{Z} \left( F,H,i\right) $ is defined, and moreover
$\mathcal{Z} \left( F,H,i\right) $ is the largest member of
$F$ that has $i$ as a member but does \emph{not} have $j$
as a member, and furthermore since $j$ is \emph{not} a
member of $A$, the member $\mathcal{C} \left( F,j\right) $
of $\mathcal{M} \left( F\right) $ is \emph{not} a subset of
$A$, hence $\mathcal{C} \left( F,j\right) $ does \emph{not}
intersect $A$, hence $i$ is \emph{not} a member of
$\mathcal{C} \left( F,j\right) $, hence the \emph{unique}
member $\left\{ i,j\right\} $ of
\label{Start of original page 17}
 $H$ that has $j$ as a member, is \emph{not} a subset of
$\mathcal{C} \left( F,j\right) $, hence $j$ is a member of
$\mathcal{O} \left( F,H\right) $, hence $\mathcal{Z} \left(
F,H,j\right) $ is defined, and moreover $\mathcal{Z} \left(
F,H,j\right) $ is the largest member of $F$ that has $j$ as
a member but does \emph{not} have $i$ as a member.

\begin{bphzlemma} \label{Lemma 5}
\end{bphzlemma}
\vspace{-6.143ex}

\noindent \hspace{10.857ex}{\bf.  }Let $V$ be any
partition such that $\mathcal{U}
\left( V\right) $ is finite and $\#\left( V\right) \geq 2$
holds, let $P$ be any wood of $V$, let $H$ be any partition
such that if $E$ is any member of $H$ such that $E$
intersects \emph{more} than one member of $V$, then $E$ has
\emph{exactly} two members, let $A$ be any member of
$\left( \Xi \left( V\right) \,\vdash V\right) $ such that
$A$ is $\left( V\cup H\right) $-connected, let $\lambda $
be any real number such that $0<\lambda < \frac{ 1 }{ 2 } $
holds, let
$d$ be any integer $\geq 1$, and let $x$ be any member of
$\mathbb{F} _{ d } \left( V\right) $.   Let $F$ be any wood
of $V$ such that $P\subseteq F$ holds, and such that for
every ordered triple $\left(B,i,j\right) $ of a
member $B$ of $\left(
F\,\vdash P\right) $ such that $B\subset A$ holds and $B$
is \emph{not} a subset of any member of $\mathcal{P} \left(
P,A\right) $, and a member $i$ of $B$ and a member $j$ of
$\left( A\,\vdash B\right) $ such that $\left\{ i,j\right\}
\in H$ holds, there exists a wood $G$ of $V$ such that
$P\subseteq G$ holds, $A$ overlaps no member of $G$, and
$\mathbb{L} \left( P,B,x\right) \leq \lambda \left| x_{
\mathcal{Z} \left( G,H,i\right) } -x_{ \mathcal{Z} \left(
G,H,j\right) } \right| $ holds.

Then $\mathbb{L} \left( P,A,x\right) \leq
\left( \frac{ 1 }{ 1-2\lambda }
\right) \mathbb{L} \left( F,A,x\right) $ holds.

\vspace{2.5ex}

\noindent {\bf Proof.}  We first note that, as
shown on
page \pageref{Start of original page 5}, if $B$ is
any member of $F$ such that $B\subset A$ holds, then there
is a unique member $D$ of $\mathcal{P} \left( F,A\right) $
such that $B\subseteq D$ holds.

Let $B$ and $C$ be any members of $\mathcal{P} \left(
P,A\right) $, and let $D$ and $E$ be the members of
$\mathcal{P} \left( F,A\right) $ such that $B\subseteq D$
and $C\subseteq E$ hold.   Then $\left| x_{ B } -x_{ C }
\right| \leq
\left| x_{ B } -x_{ D } \right| +\left| x_{ D } -x_{ E }
\right| +\left| x_{ E } -x_{ C }
\right| $ holds by the triangle inequality.

We next note that if $D=B$ holds, then $\left| x_{ B } -x_{
D }
\right| =0$ holds.

Suppose now that $B\subset D$ holds.   Then $D$ cannot be a
member of $P$, for $B\in \mathcal{P} \left( P,A\right) $
holds, $D\subset A$ holds, and by the definition of
$\mathcal{P} \left( P,A\right) $, there is no member $D$ of
$P$ such that $B\subset D$ and $D\subset A$ both hold.
Hence $D\in \left( F\,\vdash P\right) $ holds.

Now $\mathcal{P} \left( P,D\right) $ is a partition of $D$,
hence by the definition of $\mathbb{F} _{ d } \left(
V\right) $, \\
$x_{ D } \in \mathbb{V} \left( \downarrow
\left( x,\mathcal{P} \left( P,D\right) \right) \right) $
holds, or in other words, $x_{ D } $ is a member of the
convex hull of the $x_{ M } $, $M\in \mathcal{P} \left(
P,D\right) $.   And $B$ is a member of $\mathcal{P} \left(
P,D\right) $, (for $B\subset D$ holds, and there can be no
member $N$ of $P$ such that $B\subset N$ and $N\subset D$
both hold, for such an $N$ would satisfy both $B\subset N$
and $N\subset A$, contradicting $B\in \mathcal{P} \left(
P,A\right)  $).   Hence $x_{ B } $ is certainly a
\label{Start of original page 18}
 member of the convex hull of the $x_{ M } $, $M\in
\mathcal{P} \left( P,D\right) $.   Hence by Lemma
\ref{Lemma 3},
$\left| x_{
B } -x_{ D } \right| \leq \mathbb{L} \left( P,D,x\right) $
holds.
\enlargethispage{0.1ex}

Now $D\subset A$ holds, and $A$ is $\left( V\cup H\right)
$-connected by assumption, hence there exists a member $S$
of $\left( V\cup H\right) $ such that $S$ intersects both
$D$ and $\left( A\,\vdash D\right) $.   And both $D$ and
$\left( A\,\vdash D\right) $ are members of $\Xi \left(
V\right) $, hence no member of $V$ intersects both $D$ and
$\left( A\,\vdash D\right) $, hence there exists a member
$S$ of $H$ such that $S$ intersects both $D$ and $\left(
A\,\vdash D\right) $.   Let $S$ be a member of $H$ such
that $S$ intersects both $D$ and $\left( A\,\vdash D\right)
$.   Then $S$ intersects more than one member of $V$, hence
by assumption $S$ has exactly two members.   Let $i$ be the
member of $S$ that is a member of $D$, and let $j$ be the
member of $S$ that is a member of $\left( A\,\vdash
D\right) $.   Now $D$ is a member of $\left( F\,\vdash
P\right) $ such that $D\subset A$ holds.   And $D$ is
\emph{not} a subset of any member of $\mathcal{P} \left(
P,A\right) $, for $B\subset D$ holds and $B\in \mathcal{P}
\left( P,A\right) $ holds.   Hence by assumption there
exists a wood $G$ of $V$ such that $P\subseteq G$ holds,
$A$ overlaps no member of $G$, and $\mathbb{L} \left(
P,D,x\right) \leq \lambda \left| x_{ \mathcal{Z} \left(
G,H,i\right) } -x_{ \mathcal{Z} \left( G,H,j\right) }
\right| $
holds.   Let $G$ be such a wood of $V$.   Then $\mathcal{Z}
\left( G,H,i\right) \cap A$ has the member $i$ hence is
nonempty, $A\,\vdash \mathcal{Z} \left( G,H,i\right) $ has
the member $j$ hence is nonempty, and $A$ does not overlap
$\mathcal{Z} \left( G,H,i\right) $, hence $\mathcal{Z}
\left( G,H,i\right) \subset A$ holds, and $\mathcal{Z}
\left( G,H,j\right) \cap A$ has the member $j$ hence is
nonempty, $A\,\vdash \mathcal{Z} \left( G,H,j\right) $ has
the member $i$ hence is nonempty, and $A$ does not overlap
$\mathcal{Z} \left( G,H,j\right) $, hence $\mathcal{Z}
\left( G,H,j\right) \subset A$ holds.   Furthermore $H$ is
a partition by assumption, hence $S=\left\{ i,j\right\} $ is
the \emph{only} member of $H$ that has $i$ as a member,
hence $\mathcal{Z} \left( G,H,i\right) $ is the largest
member of $G$ that has $i$ as a member but does not have
$j$ as a member, hence since $P\subseteq G$ holds and $i$
and $j$ are members of \emph{distinct} members of
$\mathcal{P} \left( P,A\right) $, (for $\mathcal{K} \left(
P,A,i\right) $ is a subset of $D$, while $\mathcal{K}
\left( P,A,j\right) $ does not intersect $ D $),
$\mathcal{Z}
\left( G,H,i\right) $ is \emph{not} a strict subset of any
member of $\mathcal{P} \left( P,A\right) $, hence
$\mathcal{Z} \left( G,H,i\right) $ is a member of $\Xi
\left( \mathcal{P} \left( P,A\right) \right) $, hence, as
shown on
page \pageref{Start of original page 16}, the fact
that $x$ is a member of
$\mathbb{F} _{ d } \left( V\right) $ implies that $x_{
\mathcal{Z} \left( G,H,i\right) } $ is a member of the
convex hull of the $x_{ M } $, $M\in \mathcal{P} \left(
P,A\right) $.   And similarly, $x_{ \mathcal{Z} \left(
G,H,j\right) } $ is a member of the convex hull of the
$x_{ M } $, $M\in \mathcal{P} \left( P,A\right) $.   Hence
by Lemma \ref{Lemma 3}, $\left| x_{ \mathcal{Z}
\left( G,H,i\right) }
-x_{
\mathcal{Z} \left( G,H,j\right) } \right| \leq \mathbb{L}
\left(
P,A,x\right) $ holds.   Hence $\mathbb{L} \left(
P,D,x\right) \leq \lambda \mathbb{L} \left( P,A,x\right) $
holds, hence $\left| x_{ B } -x_{ D } \right| \leq \lambda
\mathbb{L}
\left( P,A,x\right) $ holds.

And similarly, $\left| x_{ E } -x_{ C } \right| \leq
\lambda \mathbb{L}
\left( P,A,x\right) $ holds.

Hence $\left| x_{ B } -x_{ C } \right| \leq 2\lambda
\mathbb{L} \left(
P,A,x\right) +\left| x_{ D } -x_{ E } \right| $ holds.
But $\left| x_{ D }
-x_{ E } \right| \leq \mathbb{L} \left( F,A,x\right) $
holds by
the definition of $\mathbb{L} \left( F,A,x\right) $, hence
$\left| x_{ B } -x_{ C } \right| \leq 2\lambda \mathbb{L}
\left(
P,A,x\right) +\mathbb{L} \left( F,A,x\right) $ holds.
\label{Start of original page 19}

 And \emph{this} is true for all members $B$ and $C$ of
$\mathcal{P} \left( P,A\right) $, hence $\mathbb{L} \left(
P,A,x\right) \leq $ \\
$ 2\lambda \mathbb{L} \left( P,A,x\right)
+\mathbb{L} \left( F,A,x\right) $ holds, hence $\mathbb{L}
\left( P,A,x\right) \leq \left( \frac{ 1 }{ 1-2\lambda }
\right) \mathbb{L}
\left( F,A,x\right) $ holds, since by assumption $0<\lambda
< \frac{ 1 }{ 2 }$ holds.

\vspace{2.5ex}

We recall from
page \pageref{Start of original page 6}
 that if $P$ is a wood, then we
define the wood $ \bar{ P } $ by $\bar{ P } \equiv P\cup
\left\{
\mathcal{U} \left( P\right) \right\} $, and we note that if
$A$ is any member of $\Xi \left( \mathcal{M} \left(
P\right) \right) $, then $ \bar{ P } $ has at least one
member, namely $\mathcal{U} \left( P\right) $, that
contains $A$ as a subset, hence it follows directly from
the definition on
page \pageref{Start of original page 6}
 of the function $\mathcal{Y} $,
that $\mathcal{Y} \left( \bar{ P } ,A\right) $ is the
\emph{smallest} member $B$ of $ \bar{ P } $ such that
$A\subseteq B$ holds.

For any ordered septuple $\left( P,F,H,A,\sigma ,R,x\right)
$ of a wood $P$, a wood $F$ such that $\mathcal{M} \left(
F\right) =\mathcal{M} \left( P\right) $ holds and
$P\subseteq F$ holds, a partition $H$, a member $A$ of \\
$\left( \Xi \left( \mathcal{M} \left( P\right) \right)
\,\vdash \mathcal{M} \left( P\right) \right) $, a real
number $\sigma $ such that $0<\sigma \leq \frac{ 1 }{ 8 }$
 holds, a
finite real number $R>0$, and a member $x$ of $\mathbb{F}
_{ d } \left( \mathcal{M} \left( P\right) \right) $, where
$d$ is an integer $\geq 1$, we define the proposition
$\mathbb{M} \left( P,F,H,A,\sigma ,R,x\right) $ as follows:

\vspace{1.5ex}

\noindent $\mathbb{M} (P,F,H,A,\sigma ,R,x) $:
$\mathbb{L} \left(
F,A,x\right) <R$ holds, and for every member $i$ of $A$ and
every member $j$ of $\left( \mathcal{Y} \left( \bar{ P }
,A\right) \,\vdash A\right) $ such that $\left\{ i,j\right\}
$ is a member of $H$, \\
$\mathbb{L} \left( F,A,x\right)
<\sigma \left| x_{ \mathcal{Z} \left( F,H,i\right) } -x_{
\mathcal{Z} \left( F,H,j\right) } \right| $ holds.

\vspace{1.5ex}

For every ordered pair $\left(V,H\right) $ of a partition
$V$ such that
$\mathcal{U} \left( V\right) $ is finite and $\#\left(
V\right) \geq 2$ holds, and a set $H$ such that every
member of $H$ is a set, we define $\mathcal{G} \left(
V,H\right) $ to be the set whose members are all the woods
$F$ of $V$ such that every member $A$ of $F$ is $\left(
V\cup H\right) $-connected.

We recall from
page \pageref{Start of original page 7}
 that for any sets $A$ and $B$ we
define $\mathbb{K} \left( A,B\right) $ to be the set whose
members are all the sets $C$ such that $A\subseteq C$ and
$C\subseteq B$ both hold, and we note that if $V$ is any
partition such that $\mathcal{U} \left( V\right) $ is
finite and $\#\left( V\right) \geq 2$ holds, and $P$ and $Q$
are any woods of $V$ such that $P\subseteq Q$ holds, then
it follows directly from the definition of a wood of $V$
that every member of $\mathbb{K} \left( P,Q\right) $ is
also a wood of $V$.

We note that if $V$ is any partition, then $V$ may be
calculated from $\Xi \left( V\right) $ by $V=\mathcal{M}
\left( \Xi \left( V\right) \right) $, and we also note that
if $d$ is any integer $\geq 1$, $V$ is any partition such
that $V$ is a finite set, and $x$ is any member of
$\mathbb{F} _{ d } \left( V\right) $, then $\Xi \left(
V\right) $ may be calculated from $x$ by $\Xi \left(
V\right) =\mathcal{D} \left( x\right) $, hence $V$ may be
calculated from $x$ by $V=\mathcal{M} \left( \mathcal{D}
\left( x\right) \right) $.

If $d$ is any integer $\geq 1$ and $V$ is any partition
such that $\mathcal{U} \left( V\right) $ is
\label{Start of original page 20}
 finite and $\#\left( V\right) \geq 2$ holds, then for every
ordered quadruple $\left( H,\sigma ,R,x\right) $ of a
partition $H$, a real number $\sigma $ such that $0<\sigma
\leq \frac{ 1 }{ 8 } $ holds, a finite real number $R>0$,
and a member
$x$ of $\mathbb{F} _{ d } \left( V\right) $, we define
$\Omega \left( H,\sigma ,R,x\right) $ to be the set whose
members are all the ordered pairs $\left(P,Q\right) $
of members $P$
and
$Q$ of $\mathcal{G} \left( V,H\right) =\mathcal{G} \left(
\mathcal{M} \left( \mathcal{D} \left( x\right) \right)
,H\right) $ such that $P\subseteq Q$ holds, and for every
member $A$ of $\left( Q\,\vdash P\right) $, there
\emph{exists} a member $F$ of $\mathbb{K} \left( P,Q\right)
$ such that $\mathbb{M} \left( P,F,H,A,\sigma ,R,x\right) $
holds.   (Note that $F$ does \emph{not} have to be the same
for every member $A$ of $\left( Q\,\vdash P\right)
$.)

We note that it immediately follows from this definition
that for every member $F$ of $\mathcal{G} \left( V,H\right)
$, the ordered pair $\left(F,F\right) $ is a member of
$\Omega \left(
H,\sigma ,R,x\right) $.

\vspace{1.0ex}

\begin{bphzlemma} \label{Lemma 6}
\end{bphzlemma}
\vspace{-6.143ex}

\noindent \hspace{10.857ex}{\bf.  }Let $V$ be any
partition such that $\mathcal{U}
\left( V\right) $ is finite and $\#\left( V\right) \geq 2$
holds, and let $H$ be any partition such that if $E$ is any
member of $H$ such that $E$ intersects \emph{more} than one
member of $V$, then $E$ has \emph{exactly} two members.

Let $\sigma $ be any real number such that $0<\sigma \leq
\frac{ 1 }{ 8 } $ holds, and let $\lambda $ be
the real number defined
by $\lambda \equiv \left( \frac{ 1 }{ 4 } \right)
 \left( 1-\sqrt{ 1-8\sigma } \right)
$, so that $0<\lambda \leq  \frac{ 1 }{ 4 } $ holds.

We note that $\lambda $ and $\sigma $ satisfy the equation
$\lambda = \frac{\sigma }{ 1-2\lambda }  $, and that
$0<\sigma <\lambda $ holds.

Let $R$ be any finite real number $>0$, let $d$ be any
integer $\geq 1$, let $x$ be any member of $\mathbb{F} _{ d
} \left( V\right) $, and let $\left(P,Q\right) $ be any
member of $\Omega
\left( H,\sigma ,R,x\right) $.

Then the following results hold:

\vspace{2.5ex}

\noindent {\bf (a)}  Let $A$ be any member of
$\left( Q\,\vdash P\right) $,
let $i$ be any member of $A$ and $j$ be any member of
$\left( \mathcal{Y} \left( \bar{ P } ,A\right) \,\vdash
A\right) $ such that $\left\{ i,j\right\} $ is a member of
$H$, and let $F$ be any member of $\mathbb{K} \left(
P,Q\right) $ such that $\mathbb{M} \left( P,F,H,A,\sigma
,R,x\right) $ holds.

Then $\mathbb{L} \left( P,A,x\right) <\lambda \left| x_{
\mathcal{Z} \left( F,H,i\right) } -x_{ \mathcal{Z} \left(
F,H,j\right) } \right| $ holds.

\vspace{2.5ex}

\noindent {\bf Proof.}  The proof is by induction
on the number of members
$B$ of $\left( Q\,\vdash P\right) $ such that $B\subset A$
holds and $B$ is \emph{not} a subset of any member of
$\mathcal{P} \left( P,A\right) $.

Suppose first there are \emph{no} members $B$ of $\left(
Q\,\vdash P\right) $ such that $B\subset A$ holds and $B$
is \emph{not} a subset of any member of $\mathcal{P} \left(
P,A\right) $.   Then $\mathcal{P} \left( P,A\right)
=\mathcal{P} \left( F,A\right) $ holds, hence $\mathbb{L}
\left( P,A,x\right) =\mathbb{L} \left( F,A,x\right) $
holds, hence $\mathbb{L} \left( P,A,x\right) <\sigma \left|
x_{
\mathcal{Z} \left( F,H,i\right) } -x_{ \mathcal{Z} \left(
F,H,j\right) } \right| $ holds by $\mathbb{M} \left(
P,F,H,A,\sigma ,R,x\right) $.   But $0<\sigma <\lambda $
holds, hence $\mathbb{L} \left( P,A,x\right) < \! \lambda
\left| x_{
\mathcal{Z} \left( F,H,i\right) } -x_{ \mathcal{Z} \left(
F,H,j\right) } \right| $
\label{Start of original page 21}
 holds.

Now, as the induction assumption, assume that for every
member $B$ of $\left( Q\,\vdash P\right) $ such that
$B\subset A$ holds and $B$ is \emph{not} a subset of any
member of $\mathcal{P} \left( P,A\right) $, that if $k$ is
any member of $B$ and $m$ is any member of $\left(
\mathcal{Y} \left( \bar{ P } ,B\right) \,\vdash B\right) $
such that $\left\{ k,m\right\} $ is a member of $H$, and $G$
is any member of $\mathbb{K} \left( P,Q\right) $ such that
$\mathbb{M} \left( P,G,H,B,\sigma ,R,x\right) $ holds, then
$\mathbb{L} \left( P,B,x\right) <\lambda \left| x_{
\mathcal{Z}
\left( G,H,k\right) } -x_{ \mathcal{Z} \left( G,H,m\right)
 } \right| $ holds.
\enlargethispage{0.5ex}

Now $\left( P,Q\right) \in \Omega \left( H,\sigma
,R,x\right) $ implies that for any member $B$ of $\left(
Q\,\vdash P\right) $, there exists a member $G$ of
$\mathbb{K} \left( P,Q\right) $ such that $\mathbb{M}
\left( P,G,H,B,\sigma ,R,x\right) $ holds.   And
$F\subseteq Q$ implies that every member of $\left(
F\,\vdash P\right) $ is a member of $\left( Q\,\vdash
P\right) $.   Hence the induction assumption implies that
for every ordered triple $\left(B,k,m\right) $ of
a member $B$ of
$\left(
F\,\vdash P\right) $ such that $B\subset A$ holds and $B$
is \emph{not} a subset of any member of $\mathcal{P} \left(
P,A\right) $, and a member $k$ of $B$ and a member $m$ of
$\left( \mathcal{Y} \left( \bar{ P } ,B\right) \,\vdash
B\right) $ such that $\left\{ k,m\right\} $ is a member of
$H$, there exists a wood $G$ of $V$ such that $G\in
\mathbb{K} \left( P,Q\right) $ holds, (hence $P\subseteq G$
holds, and $A$ overlaps no member of $G$, since $A$ is a
member of $Q $), and $\mathbb{L} \left( P,B,x\right) \leq
\lambda \left| x_{ \mathcal{Z} \left( G,H,k\right) } -x_{
\mathcal{Z} \left( G,H,m\right) } \right| $ holds.   Now if
$B$
is any member of $\left( F\,\vdash P\right) $ such that
$B\subset A$ holds and $B$ is \emph{not} a subset of any
member of $\mathcal{P} \left( P,A\right) $, then
$A\subseteq \mathcal{Y} \left( \bar{ P } ,B\right) $ holds,
for both $A$ and $\mathcal{Y} \left( \bar{ P } ,B\right) $
are
members of the wood $\bar{ Q } $, hence $A$ does not overlap
$\mathcal{Y} \left( \bar{ P } ,B\right) $, and $A\cap
\mathcal{Y} \left( \bar{ P } ,B\right) $ has the nonempty
subset $B$ hence is nonempty, and $\mathcal{Y}
 \left( \bar{ P }
 ,B\right) \subset A$ cannot hold, for if $\mathcal{Y}
\left( \bar{ P } ,B\right) \subset A$ held then since
$\mathcal{Y} \left( \bar{ P } ,B\right) $ is a member
 of $ \bar{ P }
 $, $\mathcal{Y} \left( \bar{ P } ,B\right) $ would be a
subset of some member of $\mathcal{P} \left( \bar{ P }
,A\right) =\mathcal{P} \left( P,A\right) $, as shown on
page \pageref{Start of original page 5},
which contradicts the assumption that $B$ is not a
subset of any member of $\mathcal{P} \left( P,A\right) $.
Hence the induction assumption implies that for every
ordered triple $\left(B,k,m\right) $ of a
member $B$ of $\left( F\,\vdash
P\right) $ such that $B\subset A$ holds and $B$ is
\emph{not} a subset of any member of $\mathcal{P} \left(
P,A\right) $, and a member $k$ of $B$ and a member $m$ of
$\left( A\,\vdash B\right) $ such that $\left\{ k,m\right\}
\in H$ holds, there exists a wood $G$ of $V$ such that
$P\subseteq G$ holds, $A$ overlaps no member of $G$, and
$\mathbb{L} \left( P,B,x\right) \leq \lambda \left| x_{
\mathcal{Z} \left( G,H,k\right) } -x_{ \mathcal{Z} \left(
G,H,m\right) } \right| $ holds.

Furthermore, it follows directly from $A\in \left(
Q \vdash \! P\right) $ and the definition of $\Omega \left(
H,\sigma ,R,x\right) $ that $A$ is a member of $\left( \Xi
\left( V\right) \,\vdash V\right) $ such that $A$ is
$\left( V\cup H\right) $-connected, and it follows directly
from the definitions of $\sigma $ and $\lambda $ that
$0<\lambda < \frac{ 1 }{ 2 } $ holds.

Hence by Lemma \ref{Lemma 5}, the induction assumption
implies that
$\mathbb{L} \left( P,A,x\right) \leq $ \\
$ \left(
\frac{ 1 }{ 1-2\lambda } \right)
\mathbb{L} \left( F,A,x\right) $ holds.
\label{Start of original page 22}

 Now $\mathbb{M} \left( P,F,H,A,\sigma ,R,x\right) $
implies that $\mathbb{L} \left( F,A,x\right) <\sigma \left|
x_{
\mathcal{Z} \left( F,H,i\right) } -x_{ \mathcal{Z} \left(
F,H,j\right) } \right| $ holds, hence $\mathbb{L} \left(
P,A,x\right) <\left( \frac{ \sigma }{ 1-2\lambda } \right)
\left| x_{ \mathcal{Z}
\left( F,H,i\right) } -x_{ \mathcal{Z} \left( F,H,j\right)
 } \right| $ holds.

But $\left( \frac{ \sigma }{ 1-2\lambda } \right) =\lambda
$ holds, hence
$\mathbb{L} \left( P,A,x\right) <\lambda \left| x_{
\mathcal{Z}
\left( F,H,i\right) } -x_{ \mathcal{Z} \left( F,H,j\right)
 } \right| $ holds, which completes the proof of the
induction step.

\vspace{2.5ex}

\noindent {\bf (b)}  Let $A$ be any member of
$\left( \Xi \left( V\right)
\,\vdash V\right) $ such that $A$ is $\left( V\cup H\right)
$-connected and $A$ overlaps no member of $Q$, and let $F$
be \emph{any} member of $\mathbb{K} \left( P,Q\right) $.
Then $\mathbb{L} \left( P,A,x\right) \leq
\left( \frac{ 1 }{ 1-2\lambda } \right)
 \mathbb{L} \left( F,A,x\right) $ holds.

\vspace{2.5ex}

\noindent {\bf Proof.}  From $\left( P,Q\right)
\in \Omega \left( H,\sigma
,R,x\right) $ it follows that for every member $B$ of
$\left( Q\,\vdash P\right) $, hence for every member $B$ of
$\left( F\,\vdash P\right) $, there exists a wood $G$ of
$V$ such that $G\in \mathbb{K} \left( P,Q\right) $ holds,
(hence $P\subseteq G$ holds, and $G\subseteq Q$ holds,
hence $A$ overlaps no member of G), and $\mathbb{M} \left(
P,G,H,B,\sigma ,R,x\right) $ holds.   Hence by
Lemma \ref{Lemma 6} (a) above it follows that for every
ordered triple $\left(B,i,j\right) $ of a member
 $B$ of $\left( Q\,\vdash
P\right) $, and a member $i$ of $B$ and a member $j$ of
$\left( \mathcal{Y} \left( \bar{ P } ,B\right) \,\vdash
B\right) $ such that $\left\{ i,j\right\} $ is a member of
$H$, there exists a wood $G$ of $V$ such that $P\subseteq
G$ holds, $A$ overlaps no member of $G$, and $\mathbb{L}
\left( P,B,x\right) <\lambda \left| x_{ \mathcal{Z} \left(
G,H,i\right) } -x_{ \mathcal{Z} \left( G,H,j\right) }
\right| $
holds, and this is true in particular when $B$ is a member
of $\left( F\,\vdash P\right) $ such that $B\subset A$
holds and $B$ is \emph{not} a subset of any member of
$\mathcal{P} \left( P,A\right) $.   And if $B$ is a member
of $\left( F\,\vdash P\right) $ such that $B\subset A$
holds and $B$ is \emph{not} a subset of any member of
$\mathcal{P} \left( P,A\right) $, then it follows directly
from the fact that $A$ overlaps no member of $Q$, (hence
that $A$ does not overlap $\mathcal{Y} \left( \bar{ P }
,B\right)  $), that $A\subseteq \mathcal{Y}
\left( \bar{ P
} ,B\right) $ holds.   And furthermore, $0<\lambda < \frac{
1 }{ 2 } $ holds in consequence of the
 definitions of $\sigma $
and $\lambda $.   Hence by Lemma \ref{Lemma 5}, $\mathbb{L}
\left(
P,A,x\right) \leq \left( \frac{ 1 }{ 1-2\lambda } \right)
\mathbb{L} \left(
F,A,x\right) $ holds.

\vspace{2.5ex}

\noindent {\bf (c)}  Let $A$ be any member of
$\left( \Xi \left( V\right)
\,\vdash V\right) $ such that $A$ is $\left( V\cup H\right)
$-connected and $A$ overlaps no member of $Q$, and let $F$
and $G$ be any members of $\mathbb{K} \left( P,Q\right) $.
 Then $\mathbb{L} \left( G,A,x\right) \leq \left(
 \frac{ 1 }{ 1-2\lambda } \right)
 \mathbb{L} \left( F,A,x\right) $ holds.

\vspace{2.5ex}

\noindent {\bf Proof.}  We first note that if
$B$ is any member of
$\mathcal{P} \left( G,A\right) $, then $B$ is a subset of
$A$ such that $B$ overlaps no member of $\mathcal{P} \left(
P,A\right) $, and furthermore $B$ cannot be a strict subset
of any member of $\mathcal{P} \left( P,A\right) $, for if
$D$ was a member of $\mathcal{P} \left( P,A\right) $ such
that $B\subset D$ held, then $D$ would be a member of $G$
such that $B\subset D$ and $D\subset A$ both held,
contradicting $B\in \mathcal{P} \left( G,A\right) $.
Hence $B$ is a member of $\Xi \left( \mathcal{P} \left(
P,A\right) \right) $, hence, as shown on
page \pageref{Start of original page 16}, it
follows directly from the fact that $x$ is a member of
\label{Start of original page 23}
 $\mathbb{F} _{ d } \left( V\right) $, that $x_{ B } $ is a
member of the convex hull of the $x_{ E } ,E\in \mathcal{P}
\left( P,A\right) $.

Now let $B$ and $C$ be any members of $\mathcal{P} \left(
G,A\right) $.   Then as just shown, $x_{ B } $ and $x_{ C }
$ are members of the convex hull of the $x_{ E } ,E\in
\mathcal{P} \left( P,A\right) $.   Hence $\mathbb{L} \left(
G,A,x\right) \leq \mathbb{L} \left( P,A,x\right) $ holds by
Lemma \ref{Lemma 3}, hence $\mathbb{L} \left( G,A,x\right)
\leq
\left( \frac{ 1 }{
1-2\lambda } \right) \mathbb{L} \left( F,A,x\right) $ holds
by Lemma \ref{Lemma 6} (b) above.

\vspace{2.5ex}

\noindent {\bf (d)}  Let $i$ and $j$ be any two members of
$\mathcal{U}
\left( V\right) $ such that $i$ and $j$ are members of
\emph{distinct} members of $V$, (or in other words, such
that $\mathcal{C} \left( V,i\right) \neq \mathcal{C} \left(
V,j\right) $ holds), and such that $\left\{ i,j\right\} $ is
a member of $H$, let $F$ and $G $ be any members of
$\mathbb{K}
\left( P,Q\right) $, and let $T \equiv
\hspace{1.5ex} \max_{ \hspace{-7.0ex}
\begin{array}{c} \\[-3.1ex]
\scriptstyle{ M\in \mathbb{K} \left(
P,Q\right) }
\end{array} } \hspace{-1.2ex}
\left| x_{ \mathcal{Z} \left( M,H,i\right) }
-x_{ \mathcal{Z} \left( M,H,j\right) } \right| $.

Then $\left| x_{ \mathcal{Z} \left( F,H,i\right) } -x_{
\mathcal{Z} \left( G,H,i\right) } \right| \leq \lambda T$
holds.

\vspace{2.5ex}

\noindent {\bf Proof.}  We note that $F\subseteq Q$
and $G\subseteq Q$ both
hold, hence $\mathcal{Z} \left( F,H,i\right) \in Q$ and
$\mathcal{Z} \left( G,H,i\right) \in Q$ both hold, hence
$\mathcal{Z} \left( F,H,i\right) $ and $\mathcal{Z} \left(
G,H,i\right) $ do not overlap.   Furthermore $\mathcal{Z}
\left( F,H,i\right) \cap \mathcal{Z} \left( G,H,i\right) $
has the member $i$ hence is nonempty, hence exactly one of
the three possibilities $\mathcal{Z} \left( F,H,i\right)
=\mathcal{Z} \left( G,H,i\right) $, $\mathcal{Z} \left(
F,H,i\right) \subset \mathcal{Z} \left( G,H,i\right) $, and
$\mathcal{Z} \left( G,H,i\right) \subset \mathcal{Z} \left(
F,H,i\right) $ holds.

Suppose first that $\mathcal{Z} \left( F,H,i\right)
=\mathcal{Z} \left( G,H,i\right) $ holds.   Then $\left| x_{
\mathcal{Z} \left( F,H,i\right) } -x_{ \mathcal{Z} \left(
G,H,i\right) } \right| =0$ holds.

Now suppose that $\mathcal{Z} \left( F,H,i\right) \subset
\mathcal{Z} \left( G,H,i\right) $ holds.   Then
$\mathcal{Z} \left( G,H,i\right) \in P$ cannot hold.   For
if $\mathcal{Z} \left( G,H,i\right) \in P$ holds, then
$\mathcal{Z} \left( G,H,i\right) \in F$ holds since
$P\subseteq F$ holds.   But $\mathcal{Z} \left(
F,H,i\right) $ is by definition the \emph{largest} member
of $F$ that has $i$ as a member but does not contain as a
subset any member of $H$ that has $i$ as a member, hence
$\mathcal{Z} \left( G,H,i\right) \in P$ implies that
$\mathcal{Z} \left( G,H,i\right) \subseteq \mathcal{Z}
\left( F,H,i\right) $ holds, which contradicts the
assumption that $\mathcal{Z} \left( F,H,i\right) \subset
\mathcal{Z} \left( G,H,i\right) $ holds.   Hence the
assumption that $\mathcal{Z} \left( F,H,i\right) \subset
\mathcal{Z} \left( G,H,i\right) $ holds, implies that
$\mathcal{Z} \left( G,H,i\right) \in \left( Q\,\vdash
P\right) $ holds.

Now that $\mathcal{Z} \left( G,H,i\right) \in \left(
Q\,\vdash P\right) $ holds implies that there exists a
member $M$ of $\mathbb{K} \left( P,Q\right) $ such that
$\mathbb{M} \left( P,M,H,\mathcal{Z} \left( G,H,i\right)
,\sigma ,R,x\right) $ holds.   Let $M$ be a member of
$\mathbb{K} \left( P,Q\right) $ such that $\mathbb{M}
\left( P,M,H,\mathcal{Z} \left( G,H,i\right) ,\sigma
,R,x\right) $ holds.   Now \emph{suppose} there exists a
member $A$ of $P$ such that $\mathcal{Z} \left(
G,H,i\right) \subset A$ holds and $j$ is \emph{not} a
member of $A$.   Then since $H$ is a \emph{partition,}
hence $\left\{ i,j\right\} $ is the \emph{only} member of
$H$
that has $i$ as a member, $A$ is a member of $P$, hence a
member of $G$, such that $i\in A$ holds and there is
\emph{no} member $E$ of $H$ such that $i\in E$ and
$E\subseteq A$ both hold, hence $A\subseteq \mathcal{Z}
\left( G,H,i\right) $ holds, which contradicts the
assumption
\label{Start of original page 24}
 that $\mathcal{Z} \left( G,H,i\right) \subset A$ holds.
Hence there is \emph{no} member $A$ of $P$ such that
$\mathcal{Z} \left( G,H,i\right) \subset A$ holds and $j$
is \emph{not} a member of $A$ hence, since $\mathcal{Z}
\left( G,H,i\right) $ is not a member of $P$, $j\in \left(
\mathcal{Y} \left( \bar{ P } ,\mathcal{Z} \left(
G,H,i\right)
\right) \,\vdash \mathcal{Z} \left( G,H,i\right) \right) $
holds hence, by Lemma \ref{Lemma 6} (a) above, $\mathbb{L}
\left( P,\mathcal{Z} \left( G,H,i\right) ,x\right) <\lambda
\left| x_{ \mathcal{Z} \left( M,H,i\right) } -x_{
\mathcal{Z}
\left( M,H,j\right) } \right| \leq \lambda T$ holds.

Now $\mathcal{Z} \left( F,H,i\right) \subset \mathcal{Z}
\left( G,H,i\right) $ holds by assumption, and $\mathcal{Z}
\left( F,H,i\right) $ overlaps no member of $\mathcal{P}
\left( P,\mathcal{Z} \left( G,H,i\right) \right) $, and
furthermore, $\mathcal{Z} \left( F,H,i\right) $ cannot be a
strict subset of any member of $\mathcal{P} \left(
P,\mathcal{Z} \left( G,H,i\right) \right) $.   For
\emph{suppose} $\mathcal{Z} \left( F,H,i\right) $ is a
strict subset of a member $B$ of $\mathcal{P} \left(
P,\mathcal{Z} \left( G,H,i\right) \right) $.   Then $B$
would be a member of $P$, hence a member of $F$, such that
$\mathcal{Z} \left( F,H,i\right) \subset B\subset
\mathcal{Z} \left( G,H,i\right) $ holds, hence $B$ would be
a member of $F$ such that $i\in B$ holds and there is
\emph{no} member $E$ of $H$ such that $i\in E$ and
$E\subseteq B$ both hold, (for there is certainly no member
$E$ of $H$ such that $i\in E$ and $E\subseteq \mathcal{Z}
\left( G,H,i\right) $ both hold), hence $B\subseteq
\mathcal{Z} \left( F,H,i\right) $ would hold, in
contradiction with the assumption that $\mathcal{Z} \left(
F,H,i\right) \subset B$ holds.   Hence $\mathcal{Z} \left(
F,H,i\right) $ is a member of $\Xi \left( \mathcal{P}
\left( P,\mathcal{Z} \left( G,H,i\right) \right) \right) $
hence, as shown on
page \pageref{Start of original page 16}, it
follows directly from the
fact that $x$ is a member of $\mathbb{F} _{ d } \left(
V\right) $, that $x_{ \mathcal{Z} \left( F,H,i\right) } $
is a member of the convex hull of the $x_{ B } $, $B\in
\mathcal{P} \left( P,\mathcal{Z} \left( G,H,i\right)
\right) $.   And $\mathcal{P} \left( P,\mathcal{Z} \left(
G,H,i\right) \right) $ is a partition of $\mathcal{Z}
\left( G,H,i\right) $, hence by the definition of
$\mathbb{F} _{ d } \left( V\right) $, $x_{ \mathcal{Z}
\left( G,H,i\right) } $ is also a member of the convex
hull of the $x_{ B } $, $B\in \mathcal{P} \left(
P,\mathcal{Z} \left( G,H,i\right) \right) $.   Hence by
Lemma \ref{Lemma 3}, $\left| x_{ \mathcal{Z} \left(
F,H,i\right) } -x_{
\mathcal{Z} \left( G,H,i\right) } \right| \leq \mathbb{L}
\left(
P,\mathcal{Z} \left( G,H,i\right) ,x\right) $ holds, hence
by the paragraph above, $\left| x_{ \mathcal{Z} \left(
F,H,i\right) } -x_{ \mathcal{Z} \left( G,H,i\right) }
\right| <\lambda T$ holds.

And if $\mathcal{Z} \left( G,H,i\right) \subset \mathcal{Z}
\left( F,H,i\right) $ holds, then by an analogous
argument, \\
$\left| x_{ \mathcal{Z} \left( F,H,i\right) } -x_{
\mathcal{Z}
\left( G,H,i\right) } \right| <\lambda T$ holds.

Hence, in every case, $\left| x_{ \mathcal{Z} \left(
F,H,i\right)
 } -x_{ \mathcal{Z} \left( G,H,i\right) } \right| <\lambda
T$
holds.

\vspace{2.5ex}

\noindent {\bf (e)}  Let $i$ and $j$ be any two members of
$\mathcal{U}
\left( V\right) $ such that $i$ and $j$ are members of
\emph{distinct} members of $V$, (or in other words, such
that $\mathcal{C} \left( V,i\right) \neq \mathcal{C} \left(
V,j\right) $ holds), and such that $\left\{ i,j\right\} $ is
a member of $H$, let $F$ be any member of $\mathbb{K}
\left( P,Q\right) $, and let \\
$T\equiv
\hspace{1.5ex} \max_{ \hspace{-7.0ex}
\begin{array}{c} \\[-3.1ex]
\scriptstyle{ M\in \mathbb{K}
\left( P,Q\right) }
\end{array} } \hspace{-1.2ex}
\left| x_{ \mathcal{Z} \left(
M,H,i\right) } -x_{ \mathcal{Z} \left( M,H,j\right) }
\right| $.

Then $T\leq \left( \frac{ 1 }{ 1-2\lambda } \right)  \left|
x_{ \mathcal{Z} \left(
F,H,i\right) } -x_{ \mathcal{Z} \left( F,H,j\right) }
\right| $
holds.

\vspace{2.5ex}

\noindent {\bf Proof.}  Let $G$ be any member of
$\mathbb{K} \left(
P,Q\right) $.   Then by the triangle inequality,
\[
\left| x_{ \mathcal{Z} \left( G,H,i\right) } -x_{
\mathcal{Z}
\left( G,H,j\right) } \right| \leq \left\{ \left| x_{
\mathcal{Z} \left(
G,H,i\right) } -x_{ \mathcal{Z} \left( F,H,i\right) }
\right|
+ \right. \hspace{-6.5pt} \hspace{7.0cm}
\]
\[
\hspace{7.0cm} \hspace{-5.0pt} \left. +
\left| x_{ \mathcal{Z} \left( F,H,i\right) } -x_{
\mathcal{Z}
\left( F,H,j\right) } \right| +\left| x_{ \mathcal{Z} \left(
F,H,j\right) } -x_{ \mathcal{Z} \left( G,H,j\right) }
\right| \right\}
\]
 holds.
\label{Start of original page 25}

 And by Lemma \ref{Lemma 6} (d) above, $\left| x_{
\mathcal{Z}
\left( G,H,i\right) } -x_{ \mathcal{Z} \left( F,H,i\right)
 } \right| \leq \lambda T$ holds and \\
  $\left| x_{
\mathcal{Z} \left(
F,H,j\right) } -x_{ \mathcal{Z} \left( G,H,j\right) }
\right| \leq \lambda T$ holds.

Hence $\left| x_{ \mathcal{Z} \left( G,H,i\right) } -x_{
\mathcal{Z} \left( G,H,j\right) } \right| \leq 2\lambda
T+\left| x_{
\mathcal{Z} \left( F,H,i\right) } -x_{ \mathcal{Z} \left(
F,H,j\right) } \right| $ holds.

And, since $G$ is an arbitrary member of $\mathbb{K} \left(
P,Q\right) $, this inequality holds for \emph{every} member
$G$ of $\mathbb{K} \left( P,Q\right) $.   Hence $T\leq
2\lambda T+\left| x_{ \mathcal{Z} \left( F,H,i\right) } -x_{
\mathcal{Z} \left( F,H,j\right) } \right| $ holds, hence
$T\leq
\left( \frac{ 1 }{ 1-2\lambda } \right)  \left| x_{
\mathcal{Z} \left( F,H,i\right)
 } -x_{ \mathcal{Z} \left( F,H,j\right) } \right| $ holds.

\vspace{2.5ex}

\noindent {\bf (f)}  Let $A$ be any member of
$\left( Q\,\vdash P\right) $,
let $i$ and $j$ be any two members of $A$ such that $i$ and
$j$ are members of \emph{distinct} members of $\mathcal{P}
\left( P,A\right) $, (or in other words, such that
$\mathcal{K} \left( P,A,i\right) \neq \mathcal{K} \left(
P,A,j\right) $ holds), and such that $\left\{ i,j\right\} $
is a member of $H$, let $k$ be any member of $A$ and $m$ be
any member of $\left( \mathcal{Y} \left( \bar{ P } ,A\right)
\,\vdash A\right) $ such that $\left\{ k,m\right\} $ is a
member of $H$, and let $F$ be any member of $\mathbb{K}
\left( P,Q\right) $.

Then $\left| x_{ \mathcal{Z} \left( F,H,i\right) } -x_{
\mathcal{Z} \left( F,H,j\right) } \right| <
\left( \frac{ \lambda }{ 1-2\lambda } \right)
  \left| x_{ \mathcal{Z} \left( F,H,k\right) } -x_{
\mathcal{Z} \left( F,H,m\right) } \right| $ holds and \\
$\left| x_{
\mathcal{Z} \left( F,H,i\right) } -x_{ \mathcal{Z} \left(
F,H,j\right) } \right| <\left( \frac{ 1 }{ 2 } \right)
\left| x_{ \mathcal{Z}
\left( F,H,k\right) } -x_{ \mathcal{Z} \left( F,H,m\right)
 } \right| $ holds.

\vspace{2.5ex}

\noindent {\bf Proof.}  We first note that $A$
is a member of $\left(
Q\,\vdash P\right) $ by assumption, hence the fact that
$ \left(P ,Q\right) $ is a member of $\Omega \left(
H,\sigma ,R,x\right) $
implies that there exists a member $G$ of $\mathbb{K}
\left( P,Q\right) $ such that $\mathbb{M} \left(
P,G,H,A,\sigma ,R,x\right) $ holds.   Let $G$ be a member
of $\mathbb{K} \left( P,Q\right) $ such that $\mathbb{M}
\left( P,G,H,A,\sigma ,R,x\right) $ holds.
Then by Lemma 6 (a) above, $\mathbb{L} \left(
P,A,x\right) < $ \\
$\lambda \left| x_{ \mathcal{Z} \left( G,H,k\right)
} -x_{ \mathcal{Z} \left( G,H,m\right) } \right| $ holds.

Now by assumption, $i$ and $j$ are members of
\emph{distinct} members of $\mathcal{P} \left( P,A\right)
$, hence since $\left\{ i,j\right\} $ is a member of $H$ and
$H$ is a partition, $\left\{ i,j\right\} $ is the
\emph{only}
member of $H$ to have $i$ as a member, hence $\mathcal{Z}
\left( F,H,i\right) $ is \emph{not} a strict subset of any
member of $\mathcal{P} \left( P,A\right) $, and similarly
$\left\{ i,j\right\} $ is the \emph{only} member of $H$ to
have $j$ as a member, hence $\mathcal{Z} \left(
F,H,j\right) $ is \emph{not} a strict subset of any member
of $\mathcal{P} \left( P,A\right) $.   Furthermore, $A$ is
a member of $Q$, hence $A$ does not overlap $\mathcal{Z}
\left( F,H,i\right) $, hence since $A\cap \mathcal{Z}
\left( F,H,i\right) $ has the member $i$ hence is nonempty,
and $A\subseteq \mathcal{Z} \left( F,H,i\right) $ cannot
hold since the member $\left\{ i,j\right\} $ of $H$ is a
subset of $A$, $\mathcal{Z} \left( F,H,i\right) \subset A$
holds.   And similarly, $\mathcal{Z} \left( F,H,j\right)
\subset A$ holds.   And furthermore neither $\mathcal{Z}
\left( F,H,i\right) $ nor $\mathcal{Z} \left( F,H,j\right)
$ overlaps any member of $\mathcal{P} \left( P,A\right) $,
hence $\mathcal{Z} \left( F,H,i\right) \in \Xi \left(
\mathcal{P} \left( P,A\right) \right) $ holds and
$\mathcal{Z} \left( F,H,j\right) \in \Xi \left( \mathcal{P}
\left( P,A\right) \right) $ holds hence, as shown on
page \pageref{Start of original page 16},
it follows directly from the fact
\label{Start of original page 26}
 that $x$ is a member of $\mathbb{F} _{ d } \left( V\right)
$, that both $x_{ \mathcal{Z} \left( F,H,i\right) } $ and
$x_{ \mathcal{Z} \left( F,H,j\right) } $ are members of
the convex hull of the $x_{ B } $, $B\in \mathcal{P} \left(
P,A\right) $, hence by Lemma \ref{Lemma 3}, $\left| x_{
\mathcal{Z}
\left(
F,H,i\right) } -x_{ \mathcal{Z} \left( F,H,j\right) }
\right| \leq \mathbb{L} \left( P,A,x\right) $ holds.

Hence $\left| x_{ \mathcal{Z} \left( F,H,i\right) } -x_{
\mathcal{Z} \left( F,H,j\right) } \right| <\lambda \left|
x_{
\mathcal{Z} \left( G,H,k\right) } -x_{ \mathcal{Z} \left(
G,H,m\right) } \right| $ holds.

And by Lemma \ref{Lemma 6} (e) above,
$\left| x_{ \mathcal{Z} \left( G,H,k\right) }
 -x_{ \mathcal{Z} \left(G,H,m\right) } \right| \leq \left(
\frac{ 1 }{ 1-2\lambda } \right)
 \left| x_{ \mathcal{Z} \left( F,H,k\right) } -x_{
\mathcal{Z}
\left( F,H,m\right) } \right|  $ holds, hence \\
$\left| x_{
\mathcal{Z} \left(
F,H,i\right) } -x_{ \mathcal{Z} \left( F,H,j\right) }
\right| <\left( \frac{ \lambda }{ 1-2\lambda }
\right)  \left| x_{ \mathcal{Z} \left(
F,H,k\right) } -x_{ \mathcal{Z} \left( F,H,m\right) }
\right| $
holds.

And finally, $0<\sigma \leq \frac{ 1 }{ 8 } $ holds by
 assumption,
hence $0<\lambda \leq \frac{ 1 }{ 4 } $ holds, hence $\left(
1-2\lambda \right) \geq \frac{ 1 }{ 2 } $ holds,
hence $\left( \frac{ 1 }{ 1-2\lambda
} \right) \leq 2$ holds, hence
$\left( \frac{ \lambda }{ 1-2\lambda } \right)
\leq \frac{ 1 }{ 2 } $ holds, hence \\
$\left| x_{ \mathcal{Z}
\left(
F,H,i\right) } -x_{ \mathcal{Z} \left( F,H,j\right) }
\right| <\left( \frac{ 1 }{ 2 } \right) \left| x_{
\mathcal{Z} \left(
F,H,k\right) } -x_{ \mathcal{Z} \left( F,H,m\right) }
\right| $
holds.

\begin{bphzlemma} \label{Lemma 7}
\end{bphzlemma}
\vspace{-6.143ex}

\noindent \hspace{10.857ex}{\bf.  }Let $V$ be any
partition such that $\mathcal{U}
\left( V\right) $ is finite and $\#\left( V\right) \geq 2$
holds, let $H$ be any partition such that if $E$ is any
member of $H$ such that $E$ intersects \emph{more} than one
member of $V$, then $E$ has \emph{exactly} two members, let
$\sigma $ be any real number such that $0<\sigma \leq
\frac{ 3 }{ 25
}  $ holds, let $R$ be any finite real number $>0$, let
$d$ be any integer $\geq 1$, and let $x$ be any member of
$\mathbb{F} _{ d } \left( V\right) $.

Let $\left( P_{ 1 } ,Q_{ 1 } \right) $ and $\left( P_{ 2 }
,Q_{ 2 } \right) $ be any members of $\Omega \left(
H,\sigma ,R,x\right) $ such that $\left( P_{ 1 } \cup P_{ 2
} \right) \subseteq \left( Q_{ 1 } \cap Q_{ 2 } \right) $
holds.   Then $\left( Q_{ 1 } \cup Q_{ 2 } \right) $ is a
wood, and $\left( \left( P_{ 1 } \cap P_{ 2 } \right)
,\left( Q_{ 1 } \cup Q_{ 2 } \right) \right) $ is a member
of $\Omega \left( H,\sigma ,R,x\right) $.

\vspace{2.5ex}

\noindent {\bf Proof.}  We note first that,
as observed on
page \pageref{Start of original page 4}, it
follows directly from the definition of a wood of $V$, that
$\left( P_{ 1 } \cap P_{ 2 } \right) $ is a wood of $V$.
\enlargethispage{0.5ex}

And we also note that, as observed on
page \pageref{Start of original page 4}, it follows
directly from the definition of a wood of $V$, that $\left(
Q_{ 1 } \cup Q_{ 2 } \right) $ is a wood of $V$,
\emph{unless} there is a member of $Q_{ 1 } $ which
overlaps a member of $Q_{ 2 } $.

Suppose now that a member $A$ of $Q_{ 1 } $ overlaps a
member $B$ of $Q_{ 2 } $.   Then $\left( A\,\vdash B\right)
$, $\left( A\cap B\right) $, and $\left( B\,\vdash A\right)
$ are all nonempty, hence, since both $A$ and $B$ are
$ H $-connected (by the definition of
$\Omega \left( H,\sigma
,R,x\right)  $), there exists a member $i$ of $\left(
A\,\vdash B\right) $ and a member $j$ of $\left( A\cap
B\right) $ such that $\left\{ i,j\right\} \in H$ holds, and
there exists a member $k$ of $\left( A\cap B\right) $ and a
member $m$ of $\left( B\,\vdash A\right) $ such that
$\left\{ k,m\right\} \in H$ holds.

Now the assumption that $A$ overlaps $B$ implies that
$A\notin Q_{ 2 } $ and
\label{Start of original page 27}
 that $B\notin Q_{ 1 } $, since $Q_{ 1 } $ and $Q_{ 2 } $
are woods.   And the assumption that $\left( P_{ 1 } \cup
P_{ 2 } \right) \subseteq \left( Q_{ 1 } \cap Q_{ 2 }
\right) $ holds, implies that every member of $P_{ 1 } $ is
a member of $Q_{ 2 } $, and every member of $P_{ 2 } $ is a
member of $Q_{ 1 } $.   Hence $A\notin Q_{ 2 } $ implies
that $A\notin P_{ 1 } $, hence that $A\in \left( Q_{ 1 }
\,\vdash P_{ 1 } \right) $ holds, and $B\notin Q_{ 1 } $
implies that $B\notin P_{ 2 } $, hence that $B\in \left(
Q_{ 2 } \,\vdash P_{ 2 } \right) $ holds.

Furthermore $\left\{ i,j\right\} $ cannot be a subset of any
member of $\mathcal{P} \left( P_{ 1 } ,A\right) $.   For
$P_{ 1 } \subseteq Q_{ 2 } $ holds and $B\in Q_{ 2 } $
holds, hence $B$ overlaps no member of $P_{ 1 } $.
Suppose there exists a member $E$ of $\mathcal{P} \left(
P_{ 1 } ,A\right) $, hence of $P_{ 1 } $, such that $\left\{
i,j\right\} \subseteq E$ holds.   Then $\left\{ i,j\right\}
\subseteq E\subset A$ holds, hence $i\in \left( E\,\vdash
B\right) $ holds, $j\in \left( E\cap B\right) $ holds, and
$m\in \left( B\,\vdash E\right) $ holds, hence $E$ overlaps
$B$, which is impossible.   Hence $i$ and $j$ are members
of \emph{distinct} members of $\mathcal{P} \left( P_{ 1 }
,A\right) $.

And by an exactly analogous argument, $\left\{ k,m\right\} $
cannot be a subset of any member of $\mathcal{P} \left( P_{
2 } ,B\right) $, hence $k$ and $m$ are members of
\emph{distinct} members of $\mathcal{P} \left( P_{ 2 }
,B\right) $.

Furthermore, $m\in \mathcal{Y} \left( \overline{ P_{ 1 } }
,A\right) $ holds, hence $m\in \left( \mathcal{Y} \left(
\overline{ P_{ 1 } } ,A\right) \,\vdash A\right) $ holds.
For
$i\in \left( A\,\vdash B\right) $ holds, hence $i\in \left(
\mathcal{Y} \left( \overline{ P_{ 1 } } ,A\right) \,\vdash
B\right)
$ holds, hence $\left( \mathcal{Y} \left( \overline{ P_{ 1
} }
,A\right) \,\vdash B\right) $ is nonempty, and $j\in \left(
A\cap B\right) $ holds, hence $j\in \left( \mathcal{Y}
\left( \overline{ P_{ 1 } } ,A\right) \cap B\right) $
holds, hence
$\left( \mathcal{Y} \left( \overline{ P_{ 1 } } ,A\right)
\cap
B\right) $ is nonempty.   But if $m$ was not a member of
$\mathcal{Y} \left( \overline{ P_{ 1 } } ,A\right) $, then
$m$
would be a member of $\left( B\,\vdash \mathcal{Y} \left(
\overline{ P_{ 1 } } ,A\right) \right) $, hence $\left(
\mathcal{Y}
\left( \overline{ P_{ 1 } } ,A\right) \,\vdash B\right) $,
$\left(
\mathcal{Y} \left( \overline{ P_{ 1 } } ,A\right) \cap
B\right) $,
and $\left( B\,\vdash \mathcal{Y} \left( \overline{ P_{ 1 }
}
,A\right) \right) $ would all be nonempty, hence
$\mathcal{Y} \left( \overline{ P_{ 1 } } ,A\right) $ would
overlap
$B$, contradicting the fact, shown above, that $B$ overlaps
no member of $P_{ 1 } $.

And by an exactly analogous argument, $i\in \mathcal{Y}
\left( \overline{ P_{ 2 } } ,B\right) $ holds, hence \\
$i\in \left(
\mathcal{Y} \left( \overline{ P_{ 2 } } ,B\right) \,\vdash
B\right)
$ holds.

Furthermore the assumption that $\left( P_{ 1 } \cup P_{ 2
} \right) \subseteq \left( Q_{ 1 } \cap Q_{ 2 } \right) $
holds, implies that there exists a wood $F$ such that both
$P_{ 1 } \subseteq F\subseteq Q_{ 1 } $ holds and $P_{ 2 }
\subseteq F\subseteq Q_{ 2 } $ holds, (hence that
$\mathbb{K} \left( P_{ 1 } ,Q_{ 1 } \right) \cap \mathbb{K}
\left( P_{ 2 } ,Q_{ 2 } \right) $ is nonempty).   For
example, $F$ could be $P_{ 1 } \cup P_{ 2 } $ or $Q_{ 1 }
\cap Q_{ 2 } $.   Let $F$ be a wood such that both $P_{ 1 }
\subseteq F\subseteq Q_{ 1 } $ and $P_{ 2 } \subseteq
F\subseteq Q_{ 2 } $ hold.   Then by
Lemma \ref{Lemma 6} (f)
 for $\left( P_{ 1 } ,Q_{ 1 } \right) $ and $A$,
$\left| x_{ \mathcal{Z} \left( F,H,i\right) }
 -x_{ \mathcal{Z} \left( F,H,j\right) } \right| < \left(
\frac{
1 }{ 2 }\right)  \left| x_{ \mathcal{Z} \left( F,H,k\right)
}
 -x_{\mathcal{Z}
\left( F,H,m\right) } \right|  $ holds,
whereas by Lemma \ref{Lemma 6} (f) for
$\left( P_{ 2 } ,Q_{ 2 } \right) $ and $B$,
$\left| x_{ \mathcal{Z} \left( F,H,k\right) }
-x_{ \mathcal{Z} \left( F,H,m\right) } \right| <\left(
\frac{ 1 }
{ 2 }
\right) \left| x_{ \mathcal{Z} \left( F,H,i\right) }
 -x_{ \mathcal{Z} \left( F,H,j\right) } \right|  $
holds, which is impossible.   Hence no member of $Q_{ 1 } $
can overlap any member of $Q_{ 2 } $, hence $\left( Q_{ 1 }
\cup Q_{ 2 } \right) $ is a wood of $V$.
\label{Start of original page 28}

Now $\left( P_{ 1 } \cap P_{ 2 } \right) \subseteq
\left( Q_{ 1 } \cup Q_{ 2 } \right) $ certainly holds.

Furthermore, the assumption that $\left( P_{ 1 } \cup P_{ 2
} \right) \subseteq \left( Q_{ 1 } \cap Q_{ 2 } \right) $
holds, implies that \\
$\left( \left( Q_{ 1 } \cup Q_{ 2 }
\right) \,\vdash \left( P_{ 1 } \cap P_{ 2 } \right)
\right) \subseteq \left( \left( Q_{ 1 } \,\vdash P_{ 1 }
\right) \cup \left( Q_{ 2 } \,\vdash P_{ 2 } \right)
\right) $ holds.   For let $A$ be any member of $\left(
\left( Q_{ 1 } \cup Q_{ 2 } \right) \,\vdash \left( P_{ 1 }
\cap P_{ 2 } \right) \right) $.   Suppose first that
$A\notin \left( Q_{ 1 } \cap Q_{ 2 } \right) $.   Then
$\left( P_{ 1 } \cup P_{ 2 } \right) \subseteq \left( Q_{ 1
} \cap Q_{ 2 } \right) $ implies $A$ is \emph{not} a member
of $\left( P_{ 1 } \cup P_{ 2 } \right) $, hence $A$ is a
member of at least one of $Q_{ 1 } $ and $Q_{ 2 } $, and is
a member of neither $P_{ 1 } $ nor $P_{ 2 } $, hence at
least one of $A\in \left( Q_{ 1 } \,\vdash P_{ 1 } \right)
$ and $A\in \left( Q_{ 2 } \,\vdash P_{ 2 } \right) $
holds.   Now suppose that $A\in \left( Q_{ 1 } \cap Q_{ 2 }
\right) $.   Then $A$ is a member of both $Q_{ 1 } $ and
$Q_{ 2 } $, but is \emph{not} a member of \emph{both} $P_{
1 } $ and $P_{ 2 } $, hence again at least one of $A\in
\left( Q_{ 1 } \,\vdash P_{ 1 } \right) $ and $A\in \left(
Q_{ 2 } \,\vdash P_{ 2 } \right) $ holds.

We shall now show that if $A$ is any member of $\left(
\left( Q_{ 1 } \cup Q_{ 2 } \right) \,\vdash \left( P_{ 1 }
\cap P_{ 2 } \right) \right) $, then there exists a member
$F$ of $\mathbb{K} \left( \left( P_{ 1 } \cap P_{ 2 }
\right) ,\left( Q_{ 1 } \cup Q_{ 2 } \right) \right) $ such
that $\mathbb{M} \left( \left( P_{ 1 } \cap P_{ 2 } \right)
,F,H,A,\sigma ,R,x\right) $ holds.

Let $A$ be any member of $\left( \left( Q_{ 1 } \cup Q_{ 2
} \right) \,\vdash \left( P_{ 1 } \cap P_{ 2 } \right)
\right) $.   Then, as just shown, $A$ is a member of at
least one of $\left( Q_{ 1 } \,\vdash P_{ 1 } \right) $ and
$\left( Q_{ 2 } \,\vdash P_{ 2 } \right) $.

Suppose first that $A$ is equal to $\mathcal{U} \left(
V\right) $.   Then $\mathbb{M} \left( \left( P_{ 1 } \cap
P_{ 2 } \right) ,F,H,A,\sigma ,R,x\right) $ reduces to the
requirement that $\mathbb{L} \left( F,A,x\right) <R$ holds.
  Suppose first that $A$ is a member of $\left( Q_{ 1 }
\vdash \! P_{ 1 } \right) $.   Then there exists a member
$F$ of $\mathbb{K} \left( P_{ 1 } ,Q_{ 1 } \right) $ such
that $\mathbb{M} \left( P_{ 1 } ,F,H,A,\sigma ,R,x\right) $
holds, hence such that $\mathbb{L} \left( F,A,x\right) <R$
holds, hence such that \\
$\mathbb{M} \left( \left( P_{ 1 }
\cap P_{ 2 } \right) ,F,H,A,\sigma ,R,x\right) $ holds, and
furthermore $F$ is a member of \\
$\mathbb{K} \left( \left(
P_{ 1 } \cap P_{ 2 } \right) ,\left( Q_{ 1 } \cup Q_{ 2 }
\right) \right) $, since $\left( P_{ 1 } \cap P_{ 2 }
\right) \subseteq P_{ 1 } \subseteq F\subseteq Q_{ 1 }
\subseteq \left( Q_{ 1 } \cup Q_{ 2 } \right) $ holds.
Now suppose that $A$ is not a member of $\left( Q_{ 1 }
\,\vdash P_{ 1 } \right) $.   Then $A$ \emph{is} a member
of $\left( Q_{ 2 } \,\vdash P_{ 2 } \right) $, and by an
exactly analogous argument there exists a member $F$ of
$\mathbb{K} \left( P_{ 2 } ,Q_{ 2 } \right) $ with the
required properties.

We now assume that $A$ is \emph{not} equal to $\mathcal{U}
\left( V\right) $.

We suppose first that $A\in \left( Q_{ 1 } \,\vdash P_{ 1 }
\right) $ holds, and define $S$ to be the unique map which
satisfies the following three requirements.

\vspace{1.0ex}

\noindent (i)  $\mathcal{D} \left( S\right) \subseteq
\mathbb{N} $

\vspace{1.0ex}

\noindent (ii)  $0$ is a member of $\mathcal{D} \left(
S\right) $, and $S_{ 0 } \equiv A$

\vspace{1.0ex}

\noindent (iii)  If $r$ is a member of $\mathcal{D}
\left( S\right) $
and $S_{ r } $ is a member of $\overline{ \left( P_{ 1 }
 \cap P_{ 2 }
\right) } $, then $r$ is the largest member of
$\mathcal{D} \left( s\right) $, while otherwise $\left(
r+1\right) $ is a member of $\mathcal{D} \left( S\right) $
and \\
$S_{ r+1 } \equiv \left\{\begin{array}{cc}
\mathcal{Y} \left( \overline{ P_{ 1 } }
,S_{ r } \right) & \left( r\textrm{ even}\right) \\
 \mathcal{Y} \left(
\overline{ P_{ 2 } } ,S_{ r } \right) &
\left( r\textrm{ odd}\right)
\end{array}  \right. $

\vspace{1.0ex}

Thus $S_{ 1 } =\mathcal{Y} \left( \overline{ P_{ 1 } }
,A\right) $,
and if $\mathcal{Y} \left( \overline{ P_{ 1 } } ,A\right) $
is not
a member of $\overline{ P_{ 2 } } $, then \\
$S_{ 2 } =\mathcal{Y}
\left( \overline{ P_{ 2 } } ,\mathcal{Y} \left( \overline{
P_{ 1 } }
,A\right) \right) $, and so on.
\label{Start of original page 29}

Let $n$ be the largest member of $\mathcal{D} \left(
S\right) $, so $n\geq 1$ holds.

Then if $r\geq 1$ and $r\leq \left( n-1\right) $ both hold,
we have that for $r$ odd, $S_{ r } \in \left( P_{ 1 }
\,\vdash P_{ 2 } \right) $ holds hence $S_{ r } \in \left(
Q_{ 2 } \,\vdash P_{ 2 } \right) $ holds, (since $P_{ 1 }
\subseteq Q_{ 2 } $ holds, while for $r$ even, $S_{ r } \in
\left( P_{ 2 } \,\vdash P_{ 1 } \right) $ holds hence $S_{
r } \in \left( Q_{ 1 } \,\vdash P_{ 1 } \right) $ holds
(since $P_{ 2 } \subseteq Q_{ 1 } $ holds).   (Hence $S_{ r
} \in \left( Q_{ 1 } \,\vdash P_{ 1 } \right) $ holds for
\emph{all} even $r$, $0\leq r\leq \left( n-1\right)
 $.)

Hence the fact that $\left( P_{ 1 } ,Q_{ 1 } \right) $ is a
member of $\Omega \left( H,\sigma ,R,x\right) $ implies
that for all even $r$, $0\leq r\leq \left( n-1\right) $,
there exists a member $F_{ r } $ of $\mathbb{K} \left( P_{
1 } ,Q_{ 1 } \right) $ such that \\
$\mathbb{M} \left( P_{ 1 }
,F_{ r } ,H,S_{ r } ,\sigma ,R,x\right) $ holds, and the
fact that $\left( P_{ 2 } ,Q_{ 2 } \right) $ is a member of
$\Omega \left( H,\sigma ,R,x\right) $ implies that for all
odd $r$, $0\leq r\leq \left( n-1\right) $, there exists a
member $F_{ r } $ of $\mathbb{K} \left( P_{ 2 } ,Q_{ 2 }
\right) $ such that $\mathbb{M} \left( P_{ 2 } ,F_{ r }
,H,S_{ r } ,\sigma ,R,x\right) $ holds.

We now make a fixed choice of an $F_{ r } $ for each $0\leq
r\leq \left( n-1\right) $ such that for $r$ even, $F_{ r }
\in \mathbb{K} \left( P_{ 1 } ,Q_{ 1 } \right) $ holds and
$\mathbb{M} \left( P_{ 1 } ,F_{ r } ,H,S_{ r } ,\sigma
,R,x\right) $ holds, while for $r$ odd, $F_{ r } \in
\mathbb{K} \left( P_{ 2 } ,Q_{ 2 } \right) $ holds and
$\mathbb{M} \left( P_{ 2 } ,F_{ r } ,H,S_{ r } ,\sigma
,R,x\right) $ holds.

We note in particular that from $\mathbb{M} \left( P_{ 1 }
,F_{ 0 } ,H,A,\sigma ,R,x\right) $, that $\mathbb{L} \left(
F_{ 0 } ,A,x\right) \leq R$ holds.

Then for all $0\leq r\leq \left( n-1\right) $ we choose a
member $k_{ r } $ of $S_{ r } $ and a member $m_{ r } $ of
$\left( S_{ r+1 } \,\vdash S_{ r } \right) $ such that
$\left\{ k_{ r } ,m_{ r } \right\} \in H$ holds.   Such a
$\left\{ k_{ r } ,m_{ r } \right\} $ exists since $S_{ r } $
and $\left( S_{ r+1 } \,\vdash S_{ r } \right) $ are
nonempty and $S_{ r+1 } $ is $\left( V\cup H\right)
$-connected, and each member of $V$ is either a subset of
$S_{ r } $ or else does not intersect $S_{ r } $.

Then for all even $r$, $0\leq r\leq \left( n-1\right) $, we
have from $\mathbb{M} \left( P_{ 1 } ,F_{ r } ,H,S_{ r }
,\sigma ,R,x\right) $ that
\[
\mathbb{L} \left( F_{ r } ,S_{ r } ,x\right) <\sigma \left|
x_{
\mathcal{Z} \left( F_{ r } ,H,k_{ r } \right) } -x_{
\mathcal{Z} \left( F_{ r } ,H,m_{ r } \right) } \right|
\]
holds, and for all odd $r$, $0\leq r\leq \left( n-1\right)
$, we have from $\mathbb{M} \left( P_{ 2 } ,F_{ r } ,H,S_{
r } ,\sigma ,R,x\right) $ that
\[
\mathbb{L} \left( F_{ r } ,S_{ r } ,x\right) <\sigma \left|
x_{
\mathcal{Z} \left( F_{ r } ,H,k_{ r } \right) } -x_{
\mathcal{Z} \left( F_{ r } ,H,m_{ r } \right) } \right|
\]
holds again.

Suppose now that $r$ is even hence $F_{ r } \in \mathbb{K}
\left( P_{ 1 } ,Q_{ 1 } \right) $ holds, and $k_{ r } \in
S_{ r } $ and $m_{ r } \in \left( \mathcal{Y} \left(
\overline{ P_{ 1
} } ,S_{ r } \right) \,\vdash S_{ r } \right) $ and $S_{
r+1 } =\mathcal{Y} \left( \overline{ P_{ 1 } } ,S_{ r }
\right) $.
 We shall demonstrate that neither \\
 $\mathcal{Z} \left( F_{
r } ,H,k_{ r } \right) $ nor $\mathcal{Z} \left( F_{ r }
,H,m_{ r } \right) $ is a strict subset of any member $B$
of $\mathcal{P} \left( P_{ 1 } ,S_{ r+1 } \right) $, or in
other words, that neither $\mathcal{Z} \left( F_{ r }
,H,k_{ r } \right) $ nor $\mathcal{Z} \left( F_{ r } ,H,m_{
r } \right) $ is a strict subset of any member $B$ of
$\mathcal{P} \left( P_{ 1 } ,\mathcal{Y} \left(
 \overline{ P_{ 1 } }
 ,S_{ r } \right) \right) $.

Now $P_{ 1 } \subseteq F_{ r } $ holds.   \emph{Suppose}
$\mathcal{Z} \left( F_{ r } ,H,k_{ r } \right) $ is a
strict subset of a
\label{Start of original page 30}
 member $B$ of $\mathcal{P} \left( P_{ 1 } ,S_{ r+1 }
\right) $.   Now $P_{ 1 } \subseteq F_{ r } $ holds hence
$B\in F_{ r } $ holds.   But $\left\{ k_{ r } ,m_{ r }
\right\} $ is a member of $H$, and $H$ is a
\emph{partition,} hence $\left\{ k_{ r } ,m_{ r } \right\} $
is the \emph{only} member of $H$ to have $k_{ r } $ as a
member, hence $\mathcal{Z} \left( F_{ r } ,H,k_{ r }
\right) $ is the \emph{largest} member of $F_{ r } $ to
contain $k_{ r } $ but not $m_{ r } $,  hence, since
$\mathcal{Z} \left( F_{ r } ,H,k_{ r } \right) \subset B$
implies $k_{ r } \in B$, we must have that $m_{ r } \in B$
holds.   But $S_{ r } \in Q_{ 1 } $ holds and $B\in Q_{ 1 }
$ holds hence $S_{ r } $ does not overlap $B$, hence since
$k_{ r } \in S_{ r } $ holds and $m_{ r } \notin S_{ r } $
holds, we must have that $S_{ r } \subset B$ holds.   But
then $B\subset S_{ r+1 } $ contradicts the fact that by
definition $S_{ r+1 } $ is the smallest member of $
\overline{ P_{ 1 } } $ to contain $ S_{ r } $.

Now \emph{suppose} $\mathcal{Z} \left( F_{ r } ,H,m_{ r }
\right) $ is a strict subset of a member $B$ of
$\mathcal{P} \left( P_{ 1 } ,S_{ r+1 } \right) $.   Then by
repeating the same argument with $k_{ r } $ and $m_{ r } $
swapped we again conclude that both $k_{ r } $ and $m_{ r }
$ are members of $B$, hence that $S_{ r } \subset B$ holds,
which again with $B\subset S_{ r+1 } $ contradicts the fact
that by definition $S_{ r+1 } $ is the smallest member of
$ \overline{ P_{ 1 } } $ to contain $ S_{ r } $.

Hence neither $\mathcal{Z} \left( F_{ r } ,H,k_{ r }
\right) $ nor $\mathcal{Z} \left( F_{ r } ,H,m_{ r }
\right) $ is a strict subset of any member $B$ of
$\mathcal{P} \left( P_{ 1 } ,S_{ r+1 } \right) $.

Furthermore, the member $\left\{ k_{ r } ,m_{ r } \right\} $
of $H$ \emph{is} a subset of $S_{ r+1 } $, hence
$\mathcal{Z} \left( F_{ r } ,H,k_{ r } \right) \subset S_{
r+1 } $ holds and $\mathcal{Z} \left( F_{ r } ,H,m_{ r }
\right) \subset S_{ r+1 } $ holds, and furthermore, neither
$\mathcal{Z} \left( F_{ r } ,H,k_{ r } \right) $ nor
$\mathcal{Z} \left( F_{ r } ,H,m_{ r } \right) $ overlaps
any member of $\mathcal{P} \left( P_{ 1 } ,S_{ r+1 }
\right) $.   Hence $\mathcal{Z} \left( F_{ r } ,H,k_{ r }
\right) \in $ \\
$ \Xi \left( \mathcal{P} \left( P_{ 1 } ,S_{ r+1
} \right) \right) $ holds hence, as shown on
page \pageref{Start of original page 16}, it
follows directly from the fact that $x$ is a member of
$\mathbb{F} _{ d } \left( V\right) $, that $x_{ \mathcal{Z}
\left( F_{ r } ,H,k_{ r } \right) } $ is a member of the
convex hull of the $x_{ E } $, $E\in \mathcal{P} \left( P_{
1 } ,S_{ r+1 } \right) $, and similarly $\mathcal{Z} \left(
F_{ r } ,H,m_{ r } \right) \in \Xi \left( \mathcal{P}
\left( P_{ 1 } ,S_{ r+1 } \right) \right) $ holds, hence
$x_{ \mathcal{Z} \left( F_{ r } ,H,m_{ r } \right) } $ is
also a member of the convex hull of the $x_{ E } $, $E\in
\mathcal{P} \left( P_{ 1 } ,S_{ r+1 } \right) $.

Hence by Lemma \ref{Lemma 3},
\[
\left| x_{ \mathcal{Z} \left( F_{ r } ,H,k_{ r } \right) }
-x_{
\mathcal{Z} \left( F_{ r } ,H,m_{ r } \right) } \right| \leq
\mathbb{L} \left( P_{ 1 } ,S_{ r+1 } ,x\right)
\]
holds.   Hence for all even $r$, $0\leq r\leq \left(
n-1\right) $, we have that
$\mathbb{L} \left( F_{ r } ,S_{ r } ,x\right) \!
< \! \sigma \mathbb{L} \left( P_{ 1 } ,S_{ r+1 }
,x\right) $ holds.

And by an exactly analogous argument we find that for all
odd $r$, $0\leq r\leq \left( n-1\right) $, that
\[
\left| x_{ \mathcal{Z} \left( F_{ r } ,H,k_{ r } \right) }
-x_{
\mathcal{Z} \left( F_{ r } ,H,m_{ r } \right) } \right| \leq
\mathbb{L} \left( P_{ 2 } ,S_{ r+1 } ,x\right)
\]
holds, hence that
$\mathbb{L} \left( F_{ r } ,S_{ r } ,x\right)
<\sigma \mathbb{L} \left(P_{ 2 } ,S_{ r+1 }
,x\right) $ holds.

We now select a member $G$ of $\mathbb{K} \left( P_{ 1 }
,Q_{ 1 } \right) \cap \mathbb{K} \left( P_{ 2 } ,Q_{ 2 }
\right) $, for example
\label{Start of original page 31}
 we could take $G=P_{ 1 } \cup P_{ 2 } $.

Let $\lambda $ be the real number defined by $\lambda
\equiv \left( \frac{ 1 }{ 4 } \right)
 \left( 1-\sqrt{ 1-8\sigma } \right) $, so that
$0<\lambda \leq  \frac{ 1 }{ 5 }$ holds.

We note that $\lambda $ and $\sigma $ satisfy the equation
$\lambda = \frac{ \sigma }{ 1-2\lambda } $,
and that
$0<\sigma <\lambda $ holds.

Then by two uses of Lemma \ref{Lemma 6} (c) for each of
the above two inequalities we find that
\[
\left( 1-2\lambda \right) \mathbb{L} \left( G,S_{ r }
,x\right) \leq \mathbb{L} \left( F_{ r } ,S_{ r } ,x\right)
<\sigma \mathbb{L} \left( P_{ 1 } ,S_{ r+1 } ,x\right) \leq
\left( \frac{ \sigma }{ 1-2\lambda } \right) \mathbb{L}
\left( G,S_{ r+1 } ,x\right)
\]
holds for all $r$ even, $0\leq r\leq \left( n-1\right) $,
and that
\[
\left( 1-2\lambda \right) \mathbb{L} \left( G,S_{ r }
,x\right) \leq \mathbb{L} \left( F_{ r } ,S_{ r } ,x\right)
<\sigma \mathbb{L} \left( P_{ 2 } ,S_{ r+1 } ,x\right) \leq
\left( \frac{ \sigma }{ 1-2\lambda } \right) \mathbb{L}
\left( G,S_{ r+1 } ,x\right)
\]
holds for all odd $r$, $0\leq r\leq \left( n-1\right) $.

Hence, since $ \frac{ \sigma }{ 1-2\lambda }
=\lambda $ holds
by the definition of $\lambda $, we have that
\[
\mathbb{L} \left( G,S_{ r } ,x\right) \leq \left(
\frac{ \lambda }{ 1-2\lambda }
 \right) \mathbb{L} \left( G,S_{ r+1 }
,x\right)
\]
holds for all $0\leq r\leq \left( n-1\right) $.

We further note from the preceding inequalities that
$\mathbb{L} \left( F_{ 0 } ,S_{ 0 } ,x\right)
<\lambda \mathbb{L} \left( G,S_{ 1 } ,x\right)
 $ holds, hence for all $m$ such that $1\leq m\leq n$ holds,
we have that
\[
\mathbb{L} \left( F_{ 0 } ,A,x\right) =\mathbb{L} \left(
F_{ 0 } ,S_{ 0 } ,x\right) <\lambda \mathbb{L} \left( G,S_{
1 } ,x\right) \hspace{7.5cm}
\]
\[
\leq \lambda \left( \frac{ \lambda }{ 1-2\lambda }
 \right)^{
m-1 }\mathbb{L} \left( G,S_{ m } ,x\right)
\]
\[
\leq \left( \frac{ \lambda }{ 1-2\lambda } \right)^{
m }\mathbb{L} \left( F_{ m } ,S_{ m } ,x\right)
\]
holds, where in the last step we made a further use of
Lemma \ref{Lemma 6} (c).

Now let $i$ be any member of $A$ and $j$ be any member of
$\left( S_{ n } \,\vdash A\right) = $ \\
$\left( \mathcal{Y}
\left( \overline{ \left( P_{ 1 } \cap P_{ 2 } \right) }
 ,A\right)
\,\vdash A\right) $ such that $\left\{ i,j\right\} \in H$
holds.

Suppose that $j\in \left( S_{ m+1 } \,\vdash S_{ m }
\right) $ holds, where $0\leq m\leq \left( n-1\right) $
holds.   Then we have, by $\mathbb{M} \left( P_{ 1 } ,F_{ m
} ,H,S_{ m } ,\sigma ,R,x\right) $ if $m$ is even, and by
$\mathbb{M} \left( P_{ 2 } ,F_{ m } ,H,S_{ m } ,\sigma
,R,x\right) $ if $m$ is odd, that
$\mathbb{L} \left( F_{ m } ,S_{ m } ,x\right)
<\sigma \left| x_{ \mathcal{Z} \left( F_{ m } ,H,i\right)
} -x_{ \mathcal{Z} \left( F_{ m } ,H,j\right) } \right|  $
holds.

Now if $m=0$, then this is the desired result, namely that
$\mathbb{L} \left( F_{ 0 } ,A,x\right)
< \! $ \\
$ \sigma \left| x_{ \mathcal{Z} \left( F_{ 0 }
,H,i\right) }
-x_{ \mathcal{Z} \left( F_{ 0 } ,H,j\right) } \right|  $
holds.

And if $1\leq m\leq \left( n-1\right) $ holds, we use the
inequality above to obtain
\label{Start of original page 32}
\[
\mathbb{L} \left( F_{ 0 } ,A,x\right) <\sigma
\left( \frac{ \lambda }{
1-2\lambda } \right)^{ m }\left| x_{ \mathcal{Z}
\left( F_{ m } ,H,i\right) } -x_{ \mathcal{Z} \left( F_{ m
} ,H,j\right) } \right|  ,
\]
then we use Lemma \ref{Lemma 6} (e) to obtain
\[
\mathbb{L} \left( F_{ 0 } ,A,x\right) <\sigma
\left( \frac{ 1 }{ 1-2\lambda
 }  \right) \left( \frac{ \lambda }{ 1-2\lambda }
  \right)^{
m }\left| x_{ \mathcal{Z} \left( G,H,i\right) } -x_{
\mathcal{Z}
\left( G,H,j\right) } \right|
\]
then we use Lemma \ref{Lemma 6} (e) again to obtain
\[
\mathbb{L} \left( F_{ 0 } ,A,x\right) <\sigma
\left( \frac{ 1 }{ 1-2\lambda
 } \right)^{ 2 } \left( \frac{ \lambda }{ 1-2\lambda }
 \right)^{
m }\left| x_{ \mathcal{Z} \left( F_{ 0 } ,H,i\right) } -x_{
\mathcal{Z} \left( F_{ 0 } ,H,j\right) } \right|
\]

Now by assumption $0<\sigma \leq \frac{ 3 }{ 25 } $ holds,
 hence
$0<\lambda \leq \frac{ 1 }{ 5 } $ holds, hence $0<
\left( \frac{ \lambda }{ 1-2\lambda }
  \right) \leq \frac{ 1 }{ 3 }$ holds, hence $0<
  \left( \frac{ 1 }{
1-2\lambda } \right)^{ 2 } \left( \frac{ \lambda }
{ 1-2\lambda }
\right)^{ m }\leq \frac{ 25 }{ 27 } <1$ holds for all
 $m\geq 1$,
hence
\[
\mathbb{L} \left( F_{ 0 } ,A,x\right) <\sigma \left| x_{
\mathcal{Z} \left( F_{ 0 } ,H,i\right) } -x_{ \mathcal{Z}
\left( F_{ 0 } ,H,j\right) } \right|
\]
holds also for all $m$ such that $m\geq 1$ and $m\leq
\left( n-1\right) $ both hold.

And this is true for any member $i$ of $A$ and any member
$j$ of $\left( \mathcal{Y} \left(
\overline{ \left( P_{ 1 } \cap P_{ 2
} \right) } ,A\right) \vdash A\right) $ such that
$\left\{ i,j\right\} \in H$ holds.   Hence $\mathbb{M}
\left(
\left( P_{ 1 } \cap P_{ 2 } \right) ,F_{ 0 } ,H,A,\sigma
,R,x\right) $ holds.

Now $\left( P_{ 1 } \cap P_{ 2 } \right) \subseteq P_{ 1 }
\subseteq F_{ 0 } \subseteq Q_{ 1 } \subseteq \left( Q_{ 1
} \cup Q_{ 2 } \right) $ holds, hence \\
$F_{ 0 } \in
\mathbb{K} \left( \left( P_{ 1 } \cap P_{ 2 } \right)
,\left( Q_{ 1 } \cup Q_{ 2 } \right) \right) $ holds.

And if $A\in \left( Q_{ 2 } \,\vdash P_{ 2 } \right) $
holds, then with $F_{ 0 } $ now defined to be a member of
$\mathbb{K} \left( P_{ 2 } ,Q_{ 2 } \right) $ such that
$\mathbb{M} \left( P_{ 2 } ,F_{ 0 } ,H,A,\sigma ,R,x\right)
$ holds, we again conclude, by an exactly analogous
argument, that $\mathbb{M} \left( \left( P_{ 1 } \cap P_{ 2
} \right) ,F_{ 0 } ,H,A,\sigma ,R,x\right) $ holds.

And $\left( P_{ 1 } \cap P_{ 2 } \right) \subseteq P_{ 2 }
\subseteq F_{ 0 } \subseteq Q_{ 2 } \subseteq \left( Q_{ 1
} \cup Q_{ 2 } \right) $ holds, hence again \\
$F_{ 0 } \in
\mathbb{K} \left( \left( P_{ 1 } \cap P_{ 2 } \right)
,\left( Q_{ 1 } \cup Q_{ 2 } \right) \right) $ holds.

Hence if $A$ is any member of $\left( \left( Q_{ 1 } \cup
Q_{ 2 } \right) \,\vdash \left( P_{ 1 } \cap P_{ 2 }
\right) \right) $, then there exists a member $F$ of
$\mathbb{K} \left( \left( P_{ 1 } \cap P_{ 2 } \right)
,\left( Q_{ 1 } \cup Q_{ 2 } \right) \right) $ such that
$\mathbb{M} \left( \left( P_{ 1 } \cap P_{ 2 } \right)
,F,H,A,\sigma ,R,x\right) $ holds.   Hence $\left( \left(
P_{ 1 } \cap P_{ 2 } \right) ,\left( Q_{ 1 } \cup Q_{ 2 }
\right) \right) $ is a member of $\Omega \left( H,\sigma
,R,x\right) $.

\vspace{2.5ex}

If $\left(A,B\right) $ and $\left(C,D\right) $ are ordered
pairs of sets such that
$A\subseteq B$ holds and $C\subseteq D$ holds, we shall say
that $\left(A,B\right) $ and $\left(C,D\right) $
\emph{link} ifif $\left( A\cup
C\right) \subseteq \left( B\cap D\right) $ holds.

We note that if $\left(A,B\right) $ and $\left(C,D\right) $
are ordered pairs of sets
such that $A\subseteq B$ holds and $C\subseteq D$ holds,
then $\left(A,B\right) $ and $\left(C,D\right) $ link ifif
$\mathbb{K} \left(
A,B\right) \cap \mathbb{K} \left( C,D\right) \neq \emptyset
$.
For if $\left( A\cup C\right) \subseteq \left( B\cap
D\right) $ holds, then $A\subseteq \left( A\cup C\right)
\subseteq B$ holds and $C\subseteq \left( A\cup C\right)
\subseteq D$ holds, hence $\left( A\cup C\right) $ is a
member
\label{Start of original page 33}
 of both $\mathbb{K} \left( A,B\right) $ and
$\mathbb{K} \left( C,D\right) $.   And if $E$ is a member
of both $\mathbb{K} \left( A,B\right) $ and $\mathbb{K}
\left( C,D\right) $, then $A\subseteq E$, $C\subseteq E$,
$E\subseteq B$, and $E\subseteq D$ all hold, hence $\left(
A\cup C\right) \subseteq E\subseteq \left( B\cap D\right) $
holds.

We note that linking is \emph{not} an equivalence relation.
  For if $A$ and $B$ are sets such that $A\subset B$ holds,
then $\left(A,A\right) $ and $\left(A,B\right) $ link, and
$\left(A,B\right) $ and $\left(B,B\right) $ link, but
$\left(A,A\right) $ and $\left(B,B\right) $ do \emph{not}
link,
(for if $A\subset B$
holds then $\left(A,A\right) $ links $\left(B,B\right) $
ifif
$B\subseteq A$ holds,
which is not the case).

If $X$ is a set such that every member of $X$ is an ordered
pair $\left(A,B\right) $ of sets such that $A\subseteq B$
holds,
 we shall
say that $X$ is \emph{link-connected} ifif for every
partition $\left\{ Y,Z\right\} $ of $X$ into two nonempty
parts $Y$ and $Z$, there exists a member $\left(A,B\right)
$ of
$Y$ and a
member $\left(C,D\right) $ of $Z$ such that
$\left(A,B\right) $
and $\left(C,D\right) $ link.

\begin{bphzlemma} \label{Lemma 8}
\end{bphzlemma}
\vspace{-6.143ex}

\noindent \hspace{10.857ex}{\bf.  }Let $V$ be any
partition such that $\mathcal{U}
\left( V\right) $ is finite and $\#\left( V\right) \geq 2$
holds, let $H$ be any partition such that if $E$ is any
member of $H$ such that $E$ intersects \emph{more} than one
member of $V$, then $E$ has \emph{exactly} two members, let
$\sigma $ be any real number such that $0<\sigma \leq
\frac{ 3 }{ 25
} $ holds, let $R$ be any finite real number $>0$, let
$d$ be any integer $\geq 1$, and let $x$ be any member of
$\mathbb{F} _{ d } \left( V\right) $.

Let $X$ be any nonempty link-connected subset of $\Omega
\left( H,\sigma ,R,x\right) $, let $P$ be the map whose
domain is $X$, and such that for each member $\alpha $ of
$X$, $P_{ \alpha } $ is the first component of $\alpha $,
and let $Q$ be the map whose domain is $X$, and such that
for each member $\alpha $ of $X$, $Q_{ \alpha } $ is the
second component of $\alpha $.

Then $\bigcup_{ \alpha \in X } Q_{ \alpha } $ is a wood of
$V$, and $\left( \left( \bigcap_{ \alpha \in X }
P_{ \alpha }
\right) ,\left( \bigcup_{ \alpha \in X } Q_{ \alpha }
\right) \right)  $ is a member of $\Omega \left( H,\sigma
,R,x\right) $.

\vspace{2.5ex}

\noindent {\bf Proof.}  We first show, for any
integer $r$ such that $r\geq
1$ and $r\leq \left( \#\left( X\right) -1\right) $ both
hold, that if there exists an $ r $-member subset $Y$ of $X$
such that $\left( \left( \bigcap_{ \alpha \in Y } P_{
\alpha } \right) ,\left( \bigcup_{ \alpha \in Y } Q_{ \alpha
} \right) \right)  $ is a member of $\Omega \left(
H,\sigma ,R,x\right) $, then there exists an $\left(
r+1\right) $-member subset $Z$ of $X$ such that $\left(
\left( \bigcap_{ \alpha \in Z } P_{ \alpha } \right) ,\left(
\bigcup_{ \alpha \in Z } Q_{ \alpha } \right) \right)  $
is a member of $\Omega \left( H,\sigma ,R,x\right) $.

For if $r$ is an integer such that $r\geq 1$ and $r\leq
\left( \#\left( X\right) -1\right) $ both hold, and $Y$ is
an $ r $-member subset of $X$ such that
\label{Start of original page 34}
 $\left( \left( \bigcap_{ \alpha \in Y }
 P_{ \alpha } \right)
,\left( \bigcup_{ \alpha \in Y } Q_{ \alpha } \right)
\right)  $ is a member of $\Omega \left( H,\sigma
,R,x\right) $, then the fact that $X$ is link-connected,
together with the facts that $Y$ and $\left( X\,\vdash
Y\right) $ are nonempty, imply that there exists a member
$\beta $ of $Y$ and a member $\gamma $ of $\left( X\,\vdash
Y\right) $ such that $\beta $ links $\gamma $, or in other
words such that $\left( P_{ \beta } \cup P_{ \gamma }
\right) \subseteq \left( Q_{ \beta } \cap Q_{ \gamma }
\right) $ holds.

Now $\left( \bigcap_{ \alpha \in Y } P_{ \alpha }
\right) \cup P_{
\gamma } \subseteq P_{ \beta } \cup P_{ \gamma }$ and
$Q_{ \beta } \cap Q_{ \gamma } \subseteq \left( \bigcup_{
\alpha \in Y } Q_{ \alpha } \right) \cap Q_{ \gamma } $
both hold by the fundamental properties of sets, hence
$\left( \bigcap_{ \alpha \in Y } P_{ \alpha } \right)
\cup P_{ \gamma
} \subseteq P_{ \beta } \cup P_{ \gamma }
 \subseteq Q_{ \beta }
\cap Q_{ \gamma } \subseteq \left( \bigcup_{ \alpha \in Y }
Q_{ \alpha } \right) \cap Q_{ \gamma } $  holds, hence
 $\left(
\bigcap_{ \alpha \in Y } P_{ \alpha } \right) \cup P_{
\gamma
 } \subseteq \left( \bigcup_{ \alpha \in Y } Q_{ \alpha }
\right) \cap Q_{ \gamma }  $ holds, hence, defining
$Z\equiv Y\cup \left\{ \gamma \right\} $, we have, from
Lemma \ref{Lemma 7}, that $\left( \bigcup_{ \alpha \in Z }
Q_{ \alpha
} \right) $ is a wood of $V$, and $\left( \left( \bigcap_{
\alpha \in Z } P_{ \alpha } \right) ,\left( \bigcup_{ \alpha
\in Z } Q_{ \alpha } \right) \right)  $ is a member of
$\Omega \left( H,\sigma ,R,x\right) $.
\enlargethispage{0.5ex}

Now the assumption that $X$ is a subset of $\Omega \left(
H,\sigma ,R,x\right) $ implies directly that if $Y$ is any
\emph{one-member} subset of $X$, then $\left( \left(
\bigcap_{ \alpha \in Y } P_{ \alpha } \right) ,\left(
\bigcup_{ \alpha \in Y } Q_{ \alpha } \right) \right)   $
is a member of $\Omega \left( H,\sigma ,R,x\right) $, for
if $Y=\left\{ \beta \right\} $, then $\left( \bigcap_{
\alpha
\in Y } P_{ \alpha } \right) =P_{ \beta }  $ and $\left(
\bigcup_{ \alpha \in Y } Q_{ \alpha } \right) =Q_{ \beta
} $.

And the assumption that $X$ is nonempty implies that $X$
has at least one one-member subset.   Hence it follows
directly, by induction, that if $r$ is any integer such
that $1\leq r\leq \#\left( X\right) $ holds, then there
exists an $ r $-member subset $Y$ of $X$ such that
$\bigcup_{
\alpha \in Y } Q_{ \alpha } $ is a wood of $V$, and
$\left( \left( \bigcap_{ \alpha \in Y } P_{ \alpha } \right)
,\left( \bigcup_{ \alpha \in Y } Q_{ \alpha } \right)
\right)  $ is a member of $\Omega \left( H,\sigma
,R,x\right) $.

But the only $\#\left( X\right) $-member subset of $X$ is
$X$ itself, hence $\bigcup_{ \alpha \in X } Q_{ \alpha } $
is a wood of $V$, and $\left( \left( \bigcap_{ \alpha \in X
} P_{ \alpha } \right) ,\left( \bigcup_{ \alpha \in X } Q_{
\alpha } \right) \right)  $ is a member of $\Omega
\left( H,\sigma ,R,x\right) $.
\label{Start of original page 35}

\vspace{2.5ex}

We recall from
page \pageref{Start of original page 2}
that if $X$ is any set such that
every member of $X$ is an ordered pair, then we define
$\mathcal{D} \left( X\right) $ to be the set whose members
are all the first components of members of $X$, and we
define $\mathcal{R} \left( X\right) $ to be the set whose
members are all the second components of members of $X$.

And we recall from
page \pageref{Start of original page 3}
 that if $F$ is any set such that
every member of $F$ is itself a set, then we define
$\mathcal{U} \left( F\right) $ to be the union of all the
members of $F$, and we also recall from
page \pageref{Start of original page 3} that if $G$
is any \emph{nonempty} set such that every member of $G$ is
itself a set, then  we define $\mathcal{I} \left( G\right)
$ to be the intersection of all the members of $G$.

Thus if $X$ is any nonempty set such that every member of
$X$ is an ordered pair, and $P$ is defined to be the map
whose domain is $X$, and such that for each member $\alpha
$ of $X$, $P_{ \alpha } $ is the first component of
$\alpha $, and $Q$ is defined to be the map whose domain is
$X$, and such that for each member $\alpha $ of $X$, $Q_{
\alpha } $ is the second component of $\alpha $, then
$\left( \bigcap_{ \alpha \in X } P_{ \alpha } \right)
=\mathcal{I} \left( \mathcal{D} \left( X\right) \right)
$ holds and $\left( \bigcup_{ \alpha \in X } Q_{ \alpha }
\right) =\mathcal{U} \left( \mathcal{R} \left( X\right)
\right) $ holds.

Hence Lemma \ref{Lemma 8} can be restated as follows:  Let
$V$, $H$,
$\sigma $, $R$, $d$, and $x$ be as in
Lemmas \ref{Lemma 7} and \ref{Lemma 8},
and let $X$ be any nonempty link-connected subset of
$\Omega \left( H,\sigma ,R,x\right) $.   Then $\mathcal{U}
\left( \mathcal{R} \left( X\right) \right) $ is a wood of
$V$, and $\left( \mathcal{I} \left( \mathcal{D} \left(
X\right) \right) ,\mathcal{U} \left( \mathcal{R} \left(
X\right) \right) \right) $ is a member of $\Omega \left(
H,\sigma ,R,x\right) $.

If $X$ is a set such that every member of $X$ is an ordered
pair $\left(A,B\right) $ of sets such that $A\subseteq B$
holds,
 then a
\emph{link-connected component of }$X $ is a nonempty
 subset $Y$
of $X$ such that $Y$ is link-connected and $Y$ is
\emph{not} a strict subset of any link-connected subset of
$X$.

Now given any set $X$ such that every member of $X$ is an
ordered pair $\left(A,B\right) $ of sets such that
$A\subseteq B$ holds,
we may define $U$ to be the set whose members are all the
two-member subsets $\left\{ \alpha ,\beta \right\} $ of $X$
such that $\alpha $ links $\beta $.   Then if $Y$ is any
subset of $X$, $Y$ is link-connected ifif $Y$ is
$U$-connected, for if $\left\{ J,K\right\} $ is any
partition
of $Y$ into two nonempty parts $J$ and $K$, then there
exists a member $\alpha $ of $J$ and a member $\beta $ of
$K$ such that $\alpha $ links $\beta $ ifif there exists a
member $\left\{ \alpha ,\beta \right\} $ of $U$ such that
$J\cap \left\{ \alpha ,\beta \right\} $ and $K\cap \left\{
\alpha ,\beta \right\} $ are both nonempty.   Hence we may
conclude directly from Lemma \ref{Lemma 2} that if $F$ is
the set
whose members are all the link-connected components of $X$,
then $F$ is a partition of $X$.

And in particular we may conclude that if $V$, $H$, $\sigma
$, $R$, $d$, and $x$ are
\label{Start of original page 36}
 as in Lemmas \ref{Lemma 7} and \ref{Lemma 8},
 and $F$ is the set whose members
are all the link-connected components of $\Omega \left(
H,\sigma ,R,x\right) $, then $F$ is a partition of $\Omega
\left( H,\sigma ,R,x\right) $.

If $d$ is any integer $\geq 1$ and $V$ is any partition
such that $\mathcal{U} \left( V\right) $ is finite and
$\#\left( V\right) \geq 2$ holds, then for every ordered
quadruple $\left( H,\sigma ,R,x\right) $ of a partition
$H$, a real number $\sigma $ such that $0<\sigma \leq
\frac{
1 }{ 8 } $ holds, a finite real number $R>0$, and a
 member $x$ of
$\mathbb{F} _{ d } \left( V\right) $, we define $\Delta
\left( H,\sigma ,R,x\right) $ to be the set whose members
are the ordered pairs $\left( \mathcal{I} \left(
\mathcal{D} \left( X\right) \right) ,\mathcal{U} \left(
\mathcal{R} \left( X\right) \right) \right) $ corresponding
to all the link-connected components $X$ of $\Omega \left(
H,\sigma ,R,x\right) $.   Then it immediately follows from
Lemma \ref{Lemma 8} that if $V$, $H$, $\sigma $, $R$, $d$,
and $x$ are
as in Lemmas \ref{Lemma 7} and \ref{Lemma 8}, then $\Delta
\left( H,\sigma
,R,x\right) $ is a subset of $\Omega \left( H,\sigma
,R,x\right) $.

Now it immediately follows from Lemma \ref{Lemma 1} that if
$U$ is any
set such that every member of $U$ is a set, and $A$ and $B$
are any $U$-connected sets such that there exists a member
$E$ of $U$ such that $E$ intersects both $A$ and $B$, then
$A\cup B$ is a $U$-connected set.   For let $E$ be any
member
of $U$ such that $E$ intersects both $A$ and $B$.   Then
the set $E\cap \left( A\cup B\right) $ is certainly
$U$-connected, for if $\left\{ J,K\right\} $ is any
partition
of this set into two nonempty parts then the member $E$ of
$U$ intersects both parts, and furthermore, this set
intersects both $A$ and $B$.   Hence by one application of
Lemma \ref{Lemma 1} we find that the set $A\cup \left(
E\cap \left(
A\cup B\right) \right) $ is $U$-connected, and then by a
second application of Lemma \ref{Lemma 1} we find that the
set $B\cup
A\cup \left( E\cap \left( A\cup B\right) \right) $ is
$U$-connected.   But $\left( E\cap \left( A\cup B\right)
\right) $ is a subset of $A\cup B$, hence $B\cup A\cup
\left( E\cap \left( A\cup B\right) \right) $ is equal to
$A\cup B$.

And from this it immediately follows that if $A$ is a set,
$U$ is a set such that every member of $U$ is a set, and
$B$ and $C$ are any two \emph{distinct} $U$-connected
components of $A$, then there is \emph{no} member $E$ of
$U$ such that $E$ intersects both $B$ and $C$, for if there
was such a member $E$ of $U$, then $B\cup C$ would be
$U$-connected, and both $B$ and $C$ would be strict subsets
of the $U$-connected subset $B\cup C$ of $A$.

Now let $V$, $H$, $\sigma $, $R$, $d$, and $x$ be as in
Lemmas \ref{Lemma 7} and \ref{Lemma 8},
so that $\Delta \left( H,\sigma
,R,x\right) $ is a subset of $\Omega \left( H,\sigma
,R,x\right) $, and let $\left(P,Q\right) $ and
$\left(S,T\right) $ be any two
\emph{distinct} members of $\Delta \left( H,\sigma
,R,x\right) $.   Then $\left(P,Q\right) $ and
$\left(S,T\right) $ are members of
\emph{distinct} link-connected components of $\Omega \left(
H,\sigma ,R,x\right) $, hence it immediately follows from
the above paragraph that $\left(P,Q\right) $ and
$\left(S,T\right) $ do \emph{not}
link.   And this means, as shown on
page \pageref{Start of original page 32}, that
$\mathbb{K} \left( P,Q\right) \cap \mathbb{K} \left(
S,T\right) $ is empty.
\label{Start of original page 37}

 We recall from
page \pageref{Start of original page 19}
that for any ordered
 pair $\left(V,H\right) $
of a partition $V$ such that $\mathcal{U} \left( V\right) $
is finite and $\#\left( V\right) \geq 2$ holds, and a set
$H$ such that every member of $H$ is a set, we define
$\mathcal{G} \left( V,H\right) $ to be the set whose
members are all the woods $F$ of $V$ such that every member
$A$ of $F$ is $\left( V\cup H\right) $-connected.

Now let $V$, $H$, $\sigma $, $R$, $d$, and $x$ be as in
Lemmas \ref{Lemma 7} and \ref{Lemma 8},
and let $F$ be any member of
$\mathcal{G} \left( V,H\right) $.   Then as observed on
page \pageref{Start of original page 20},
it immediately follows from the definition of
$\Omega \left( H,\sigma ,R,x\right) $, that the ordered
pair $\left(F,F\right) $ is a member of $\Omega \left(
H,\sigma
,R,x\right) $.   And as shown above, the set whose members
are all the link-connected components of $\Omega \left(
H,\sigma ,R,x\right) $, is a \emph{partition} of $\Omega
\left( H,\sigma ,R,x\right) $, hence there exists a
\emph{unique} link-connected component $X$ of $\Omega
\left( H,\sigma ,R,x\right) $ such that $\left( F,F\right)
\in X$ holds.   Let $X$ be the unique link-connected
component of $\Omega \left( H,\sigma ,R,x\right) $ such
that $\left( F,F\right) \in X$ holds.   Then $\left(
\mathcal{I} \left( \mathcal{D} \left( X\right) \right)
,\mathcal{U} \left( \mathcal{R} \left( X\right) \right)
\right) $ is a member of $\Delta \left( H,\sigma
,R,x\right) $ and a member of $\Omega \left( H,\sigma
,R,x\right) $ and, since it immediately follows from
$\left( F,F\right) \in X$ that $\mathcal{I} \left(
\mathcal{D} \left( X\right) \right) \subseteq F\subseteq
\mathcal{U} \left( \mathcal{R} \left( X\right) \right) $
holds, $F$ is a member of $\mathbb{K} \left( \mathcal{I}
\left( \mathcal{D} \left( X\right) \right) ,\mathcal{U}
\left( \mathcal{R} \left( X\right) \right) \right) $.   And
by the foregoing, if $\left(P,Q\right) $ and
$\left(S,T\right) $ are any two
\emph{distinct} members of $\Delta \left( H,\sigma
,R,x\right) $, then $\mathbb{K} \left( P,Q\right) \cap
\mathbb{K} \left( S,T\right) =\emptyset $.   Hence if
$\left(P,Q\right) $ is
any member of $\Delta \left( H,\sigma ,R,x\right) $ such
that $\left(P,Q\right) $ is \emph{not} equal to $\left(
\mathcal{I}
\left( \mathcal{D} \left( X\right) \right) ,\mathcal{U}
\left( \mathcal{R} \left( X\right) \right) \right) $, then
$F$ is \emph{not} a member of $\mathbb{K} \left( P,Q\right)
$.

Hence if $V$, $H$, $\sigma $, $R$, $d$, and $x$ are as in
Lemmas \ref{Lemma 7} and \ref{Lemma 8},
and $F$ is any member of $\mathcal{G}
\left( V,H\right) $, then there exists exactly one member
$ \left(P ,Q\right) $ of $\Delta \left( H,\sigma
,R,x\right) $,
specifically the member $\left( P,Q\right) =\left(
\mathcal{I} \left( \mathcal{D} \left( X\right) \right)
,\mathcal{U} \left( \mathcal{R} \left( X\right) \right)
\right) $, where $X$ is the unique link-connected component
of $\Omega \left( H,\sigma ,R,x\right) $ such that $\left(
F,F\right) \in X$ holds, such that $F\in \mathbb{K} \left(
P,Q\right) $ holds.

Furthermore, if $V$, $H$, $\sigma $, $R$, $d$, and $x$ are
as in Lemmas \ref{Lemma 7} and \ref{Lemma 8},
 and $\left(P,Q\right) $ is any
member of
$\Delta \left( H,\sigma ,R,x\right) $, then
$\left(P,Q\right) $ is a
member of $\Omega \left( H,\sigma ,R,x\right) $, hence
every member $F$ of $\mathbb{K} \left( P,Q\right) $ is a
member of $\mathcal{G} \left( V,H\right) $

Hence if $V$, $H$, $\sigma $, $R$, $d$, and $x$ are as in
Lemmas \ref{Lemma 7} and \ref{Lemma 8},
and $Z$ is defined to be the set whose
members are the sets $\mathbb{K} \left( P,Q\right) $, where
$ \left(P ,Q\right) $ is a member of $\Delta \left(
H,\sigma ,R,x\right) $,
then $Z$ is a \emph{partition} of $\mathcal{G} \left(
V,H\right) $.

For any ordered pair $\left(V,H\right) $ of a partition $V$
such
 that
$\mathcal{U} \left( V\right) $ is finite and $\#\left(
V\right) \geq 2$ holds, and a set $H$ such that every
member of $H$ is a set, we define $\mathcal{N} \left(
V,H\right) $ to be the set whose members are all the
ordered pairs $\left(F,G\right) $ of members $F$ and
$G$ of $\mathcal{G}
\left( V,H\right) $ such that $F\subseteq G$ holds.

If $\left( H,\sigma ,R,x\right) $ is an ordered quadruple
such that $\Omega \left( H,\sigma ,R,x\right) $ is
\label{Start of original page 38}
 defined and such that $V=\mathcal{M} \left( \mathcal{D}
\left( x\right) \right) $, $H$, $\sigma $, $R$, $d$,
(determined from $x$ as the number of components of any
member of $\mathcal{R} \left( x\right)  $), and $x$
satisfy the conditions of Lemmas \ref{Lemma 7}
and \ref{Lemma 8}, we shall say
that a member $\left(P,Q\right) $ of $\mathcal{N} \left(
V,H\right) $
\emph{generates a good set of woods for }$(H,\sigma ,R,x)
 $ ifif
$ \left( P,Q\right) $ is a member of
$\Delta \left( H,\sigma ,R,x\right) $,
and we shall say a subset $X$ of $\mathcal{G} \left(
V,H\right) $ is a
\emph{good set of woods for }$(H,\sigma ,R,x) $
ifif $X$ has the form $\mathbb{K} \left( P,Q\right) $,
where $\left(P,Q\right) $ is a member of $\Delta \left(
H,\sigma
,R,x\right) $.

If $V$, $H$, $\sigma $, $R$, and $d$ are as in
Lemmas \ref{Lemma 7}
and \ref{Lemma 8}, then we shall
partition the set of all ordered
pairs $\left(F,x\right) $ of a member $F$ of $\mathcal{G}
\left(
V,H\right) $ and a member $x$ of $\mathbb{F} _{ d } \left(
V\right) $ into a finite number of sectors, the sectors
being in one-to-one correspondence with some subset of
$\mathcal{N} \left( V,H\right) $, such that the sector
associated with the member $\left(P,Q\right) $ of
$\mathcal{N} \left(
V,H\right) $ is the set of all ordered pairs
$\left(F,x\right) $
 of a
member $F$ of $\mathbb{K} \left( P,Q\right) $ and a member
$x$ of $\mathbb{F} _{ d } \left( V\right) $ such that $
\left(P ,Q\right) $
generates a good set of woods for $\left( H,\sigma
,R,x\right) $.

We note that the number of sectors is less than the square
of the total number of woods of $V$, which is itself less
than $2^{ \left( 2^{ \#\left( V\right) } \right) } $.

Now let $V$, $H$, $\sigma $, $R$, $d$, and $x$ be as in
Lemmas \ref{Lemma 7} and \ref{Lemma 8}, and let $F$ be any
member of
$\mathcal{G} \left( V,H\right) $.   Then to identify the
unique member $\left(P,Q\right) $ of $\Delta \left(
H,\sigma ,R,x\right)
$ such that $F\in \mathbb{K} \left( P,Q\right) $ holds, we
define $Y$ to be the set whose members are all the members
$ \left( S,T\right) $ of $\Omega \left( H,\sigma
,R,x\right) $ such that
$S\subseteq F\subseteq T$ holds, or in other words such
that $F\in \mathbb{K} \left( S,T\right) $ holds.   Then
Lemma \ref{Lemma 8} guarantees that there exists a unique
member
$ \left( J,K\right) $ of $Y$ such that $J\subseteq
S\subseteq T\subseteq K$
holds for every member $\left(S,T\right) $ of $Y$, for
$\mathcal{I}
\left( \mathcal{D} \left( Y\right) \right) \subseteq
S\subseteq T\subseteq \mathcal{U} \left( \mathcal{R} \left(
Y\right) \right) $ certainly holds for every member $
\left( S,T\right) $
of $Y$, and by Lemma \ref{Lemma 8} the ordered pair $\left(
\mathcal{I} \left( \mathcal{D} \left( Y\right) \right)
,\mathcal{U} \left( \mathcal{R} \left( Y\right) \right)
\right) $ is a member of $Y$, and furthermore if $ \left(
J,K\right) $ and
$ \left( M,N\right) $ are members of $Y$ such that
$J\subseteq S\subseteq
T\subseteq K$ holds for every member $\left(S,T\right) $ of
$Y$ and
$M\subseteq S\subseteq T\subseteq N$ holds for every member
$ \left( S,T\right) $ of $Y$, then $J\subseteq M$ and
$M\subseteq J$ both
hold hence $M=J$ holds, and $N\subseteq K$ and $K\subseteq
N$ both hold, hence $N=K$ holds, hence the member $\left(
\mathcal{I} \left( \mathcal{D} \left( Y\right) \right)
,\mathcal{U} \left( \mathcal{R} \left( Y\right) \right)
\right) $ is the \emph{only} member $ \left( J,K\right) $
of $Y$ such that
$J\subseteq S\subseteq T\subseteq K$ holds for every member
$ \left( S,T\right) $ of $Y$.   Now by definition $\Delta
\left( H,\sigma
,R,x\right) $ is the set whose members are the ordered
pairs $\left( \mathcal{I} \left( \mathcal{D} \left(
X\right) \right) ,\mathcal{U} \left( \mathcal{R} \left(
X\right) \right) \right) $ corresponding to all the
link-connected components $X$ of $\Omega \left( H,\sigma
,R,x\right) $, and by Lemma \ref{Lemma 8}, every member of
$\Delta
\left( H,\sigma ,R,x\right) $ is a member of $\Omega \left(
H,\sigma ,R,x\right) $, which as shown above, implies that
no two distinct members of $\Delta \left( H,\sigma
,R,x\right) $
\label{Start of original page 39}
 link, hence that if $ \left( J,K\right) $ and $ \left(
M,N\right) $ are any two distinct
members of $\Delta \left( H,\sigma ,R,x\right) $ then
$\mathbb{K} \left( J,K\right) \cap \mathbb{K} \left(
M,N\right) $ is empty.   Now let $X$ be the unique
link-connected component of $\Omega \left( H,\sigma
,R,x\right) $ that has $ \left( F,F\right) $ as a member.
Then $\left(
\mathcal{I} \left( \mathcal{D} \left( X\right) \right)
,\mathcal{U} \left( \mathcal{R} \left( X\right) \right)
\right) $ is a member of $\Omega \left( H,\sigma
,R,x\right) $ by Lemma \ref{Lemma 8} and $\mathcal{I} \left(
\mathcal{D} \left( X\right) \right) \subseteq F\subseteq
\mathcal{U} \left( \mathcal{R} \left( X\right) \right) $
holds, hence $\left( \mathcal{I} \left( \mathcal{D} \left(
X\right) \right) ,\mathcal{U} \left( \mathcal{R} \left(
X\right) \right) \right) $ is a member of $Y$.   And $Y\cap
X$ has the member $ \left( F,F\right) $ hence is nonempty,
hence
by Lemma \ref{Lemma 1}, $Y\cup X$ is link-connected, hence
$Y$ is a subset of
$X$ since $X$ is a link-connected \emph{component} of
$\Omega \left( H,\sigma ,R,x\right) $.   Hence $\mathcal{I}
\left( \mathcal{D} \left( X\right) \right) \subseteq
S\subseteq T\subseteq \mathcal{U} \left( \mathcal{R} \left(
X\right) \right) $ holds for every member $\left(S,T\right)
$ of $Y$,
hence $\left( \mathcal{I} \left( \mathcal{D} \left(
X\right) \right) ,\mathcal{U} \left( \mathcal{R} \left(
X\right) \right) \right) $ is equal to $\left( \mathcal{I}
\left( \mathcal{D} \left( Y\right) \right) ,\mathcal{U}
\left( \mathcal{R} \left( Y\right) \right) \right) $.

We note that if $V$, $H$, $\sigma $, $R$, $d$, and $x$ are
as in Lemmas \ref{Lemma 7} and \ref{Lemma 8}, and
$\left(P,Q\right) $ and
$\left(S,T\right) $ are any two
distinct members of $\Delta \left( H,\sigma ,R,x\right) $,
then it immediately follows from the fact that
$\left(P,Q\right) $ and $ \left( S,T\right) $
 do \emph{not} link, that $P\neq S$ holds and $Q\neq
T$ holds.

\begin{bphzlemma} \label{Lemma 9}
\end{bphzlemma}
\vspace{-6.143ex}

\noindent \hspace{10.857ex}{\bf.  }Let $V$ be any
partition such that $\mathcal{U}
\left( V\right) $ is finite and $\#\left( V\right) \geq 2$
holds, let $H$ be any partition such that if $E$ is any
member of $H$ such that $E$ intersects \emph{more} than one
member of $V$, then $E$ has \emph{exactly} two members, let
$\sigma $ be any real number such that $0<\sigma \leq
\frac{ 3 }{ 25
} $ holds, let $R$ be any finite real number $>0$, let
$d$ be any integer $\geq 1$, and let $x$ be any member of
$\mathbb{F} _{ d } \left( V\right) $.

Then the member $\left(P,Q\right) $ of $\mathcal{N} \left(
V,H\right) $
generates a good set of woods for $\left( H,\sigma
,R,x\right) $ ifif both the following conditions (i) and
(ii) hold.

\vspace{1.0ex}

\noindent (i)  $\left( P,Q\right) \in \Omega \left( H,\sigma
,R,x\right) $ holds.

\vspace{1.0ex}

\noindent (ii)  For all $\left( F,G\right) \in \mathcal{N}
\left(
V,H\right) $ such that $F\subseteq P$ and $Q\subseteq G$
both hold, and $\left( F,G\right) \neq \left( P,Q\right) $,
$ \left( F,G\right) $ is \emph{not} a member of $\Omega
\left( H,\sigma
,R,x\right) $.

\vspace{2.5ex}

\noindent {\bf Proof.}  We note first that
(i) is certainly necessary for
$ \left(P ,Q\right) $ to generate a good set of woods at
$x$.   And (ii) is
also necessary, for if $F\subseteq P$ and $Q\subseteq G$
both hold, and $\left( F,G\right) \neq \left( P,Q\right) $,
then $\left( F\cup P\right) \subseteq \left( G\cap Q\right)
$ holds, hence $ \left( F,G\right) $ links $ \left(P
,Q\right) $, but at least one of
$P\subseteq F$ and $G\subseteq Q$ is false.   Hence if
$ \left(P ,Q\right) $ generates a good set of woods for
$\left( H,\sigma
,R,x\right) $, then $ \left( F,G\right) $ cannot be a
member of $\Omega
\left( H,\sigma ,R,x\right) $ for if $\left(P,Q\right) $
generates a good
set of woods for $\left( H,\sigma ,R,x\right) $, then
$P\subseteq J\subseteq K\subseteq Q$ holds for every member
$ \left( J,K\right) $ of the link-connected component of
$\Omega \left(
H,\sigma ,R,x\right) $ that contains $ \left(P ,Q\right) $.

Now let $X$ be the subset of $\Omega \left( H,\sigma
,R,x\right) $ whose members are all the members $ \left(
J,K\right) $ of
$\Omega \left( H,\sigma ,R,x\right) $ such that $P\subseteq
J\subseteq K\subseteq Q$ holds.   We
\label{Start of original page 40}
 shall show that (i) and (ii) imply that $X$ is the
link-connected component of $\Omega \left( H,\sigma
,R,x\right) $ to which $\left(P,Q\right) $ belongs.

 We first note that, since $P\subseteq J\subseteq
K\subseteq Q$ implies that $\left( J\cup P\right) \subseteq
\left( K\cap Q\right) $ holds, every member $ \left(
J,K\right) $ of $X$ is
\emph{linked} to $ \left(P ,Q\right) $, hence $X$ is
certainly
link-connected.

Now suppose $X$ is a strict subset of $Z$, where $Z$ is a
link-connected subset of $\Omega \left( H,\sigma
,R,x\right) $.   Consider the partition of $Z$ into $X$ and
the nonempty set $\left( Z\,\vdash X\right) $.   Then the
assumption that $Z$ is link-connected implies that there
exists $\left( J,K\right) \in X$ and $\left( F,G\right) \in
\left( Z\,\vdash X\right) $ such
that $ \left( J,K\right) $ links $ \left( F,G\right) $.
And $\left(P,Q\right) $ links $ \left( J,K\right) $,
hence $\left\{ \left(
P,Q\right)
,\left( J,K\right) ,\left( F,G\right) \right\} $ is a
link-connected subset of $\Omega \left( H,\sigma
,R,x\right) $, hence, by Lemma \ref{Lemma 8}, $\left(
\left( P\cap
J\cap F\right) ,\left( Q\cup K\cup G\right) \right) \in
\Omega \left( H,\sigma ,R,x\right) $ holds.   Now $\left(
P\cap J\cap F\right) \subseteq P$ holds, and $Q\subseteq
\left( Q\cup K\cup G\right) $ holds.   And by assumption,
$P\subseteq F\subseteq G\subseteq Q$ does \emph{not} hold,
since $ \left( F,G\right) $ is not a member of $X$.   Hence
at least one of
$\left( P\,\vdash F\right) $ and $\left( G\,\vdash Q\right)
$ is nonempty, hence at least one of $\left( P\,\vdash
\left( P\cap J\cap F\right) \right) $ and $\left( \left(
Q\cup K\cup G\right) \,\vdash Q\right) $ is nonempty, hence
$\left( \left( P\cap J\cap F\right) ,\left( Q\cup K\cup
G\right) \right) $ is not equal to $ \left(P ,Q\right) $.
Hence (ii)
implies that \\
$\left( \left( P\cap J\cap F\right) ,\left(
Q\cup K\cup G\right) \right) $ cannot be a member of
$\Omega \left( H,\sigma ,R,x\right) $, which contradicts
the conclusion drawn from the assumption that $\left(
F,G\right) \in \Omega \left( H,\sigma ,R,x\right) $ holds.
 Hence (ii) implies that $ \left( F,G\right) $ cannot be a
member of
$\Omega \left( H,\sigma ,R,x\right) $.

Hence (i) and (ii) imply that $X$ is the link-connected
component of $\Omega \left( H,\sigma ,R,x\right) $ that
contains $ \left(P ,Q\right) $.

\begin{bphzlemma} \label{Lemma 10}
\end{bphzlemma}
\vspace{-6.143ex}

\noindent \hspace{11.9ex}{\bf.  }Let $V$ be any
partition such that $\mathcal{U}
\left( V\right) $ is finite and $\#\left( V\right) \geq 2$
holds, let $H$ be any partition such that if $E$ is any
member of $H$ such that $E$ intersects \emph{more} than one
member of $V$, then $E$ has \emph{exactly} two members, let
$\sigma $ be any real number such that $0<\sigma \leq
\frac{ 3 }{ 25
} $ holds, let $R$ be any finite real number $>0$, let
$d$ be any integer $\geq 1$, and let $x$ be any member of
$\mathbb{F} _{ d } \left( V\right) $.

Let $\left(P,Q\right) $ generate a good set of woods for
$\left( H,\sigma
,R,x\right) $, and let $A$ be any member of $\mathbb{B}
\left( P\right) $.   Then $\mathbb{M} \left( \left(
P\,\vdash \left\{ A\right\} \right) ,P,H,A,\sigma
,R,x\right)
$ does \emph{not} hold.

\vspace{2.5ex}

\noindent {\bf Proof.}  Suppose that
$\mathbb{M} \left( \left( P\,\vdash
\left\{ A\right\} \right) ,P,H,A,\sigma ,R,x\right) $ does
hold.   We shall prove that this would imply that for every
member $B$ of $\left( Q\,\vdash \left( P\,\vdash \left\{
A\right\} \right) \right) $, there exists a member $F$ of
$\mathbb{K} \left( \left( P\,\vdash \left\{ A\right\}
\right)
,Q\right) $ such that $\mathbb{M} \left( \left( P\,\vdash
\left\{ A\right\} \right) ,F,H,B,\sigma ,R,x\right) $ holds,
hence that $\left( \left( P\,\vdash \left\{ A\right\}
\right)
,Q\right) $ is a member of $\Omega \left( H,\sigma
,R,x\right) $, which by Lemma \ref{Lemma 9} contradicts the
assumption
that
\label{Start of original page 41}
 $ \left(P ,Q\right) $ generates a good set of woods for
$\left( H,\sigma
,R,x\right) $.

For suppose first that $B$ is a member of $\left( Q\,\vdash
P\right) $.   Then the assumption that $\left(P,Q\right) $
generates a
good set of woods for $\left( H,\sigma ,R,x\right) $
implies that there exists a member $F$ of $\mathbb{K}
\left( P,Q\right) $ such that $\mathbb{M} \left(
P,F,H,B,\sigma ,R,x\right) $ holds.   Let $F$ be a member
of $\mathbb{K} \left( P,Q\right) $ such that $\mathbb{M}
\left( P,F,H,B,\sigma ,R,x\right) $ holds.   Then
$\mathbb{L} \left( F,B,x\right) <R$ holds, and for all
$i\in B$ and $j\in \left( \mathcal{Y} \left( \bar{ P }
,B\right) \,\vdash B\right) $ such that $\left\{ i,j\right\}
\in H$ holds, $\mathbb{L} \left( F,B,x\right) <\sigma
\left| x_{
\mathcal{Z} \left( F,H,i\right) } -x_{ \mathcal{Z} \left(
F,H,j\right) } \right| $ holds.   Now if $\mathcal{Y} \left(
\overline{ \left( P\,\vdash \left\{ A\right\} \right) } ,
B\right) $ is
equal to $\mathcal{Y} \left( \bar{ P } ,B\right) $, then
this
implies directly that $\mathbb{M} \left( \left( P\,\vdash
\left\{ A\right\} \right) ,F,H,B,\sigma ,R,x\right) $ holds.
\enlargethispage{0.8ex}

Suppose now that $\mathcal{Y} \left( \overline{
 \left( P\,\vdash
\left\{ A\right\} \right) } ,B\right) $ is \emph{not}
equal to
$\mathcal{Y} \left( \bar{ P } ,B\right) $.   Now by
definition
$\bar{ P } \equiv \left( P\cup \left\{ \mathcal{U} \left(
V\right) \right\} \right) $, and $\mathcal{Y} \left( \bar{
P }
,B\right) $ is the \emph{smallest} member $C$ of
$\bar{ P } $
 such that $B\subseteq C$ holds, hence $\mathcal{Y} \left(
\overline { \left( P\,\vdash \left\{ A\right\} \right) }
,B\right) $ is
equal to $\mathcal{Y} \left( \bar{ P } ,B\right) $ unless
$\mathcal{Y} \left( \bar{ P } ,B\right) $ is not a member
of $\overline { \left( P\,\vdash \left\{ A\right\} \right)
} $, or in other
words, unless $\mathcal{Y} \left( \bar{ P } ,B\right) =A$.
Hence the assumption that $\mathcal{Y} \left(
\overline{ \left( P\,
\vdash \left\{ A\right\} \right) } ,B\right) $ is not equal
to $\mathcal{Y} \left( \bar{ P } ,B\right) $ implies that
$\mathcal{Y} \left( \bar{ P } ,B\right) =A$.   And the
assumption that $\mathcal{Y} \left( \overline{
\left( P\,\vdash \left\{ A\right\} \right) } ,B\right) $
is not equal to $\mathcal{Y}
\left( \bar{ P } ,B\right) $ also implies that $A$ is not
equal to $\mathcal{U} \left( V\right) $, for if
$A=\mathcal{U} \left( V\right) $ then $\overline{
\left( P\,\vdash
\left\{ A\right\} \right) } $ is equal to $\bar{ P } $.
   Let $i$ be
any
member of $B$ and $j$ be any member of $\left( \mathcal{Y}
\left( \overline { \left( P\,\vdash \left\{ A\right\}
\right) } ,B\right)
\,\vdash \mathcal{Y} \left( \bar{ P } ,B\right) \right) $
such
that $\left\{ i,j\right\} \in H$ holds.   Now $\mathcal{Y}
\left( \bar{ P } ,B\right) $ is equal to $A$, and
furthermore,
$\mathcal{Y} \left(
\overline{ \left( P\,\vdash \left\{ A\right\}
\right) } ,B\right) $ is equal to $\mathcal{Y} \left(
\overline{\left(
P\,\vdash \left\{ A\right\} \right) } ,\mathcal{Y}
\left( \bar{ P }
 ,B\right) \right) $, hence $\mathcal{Y} \left(
 \overline{ \left(
P\,\vdash \left\{ A\right\} \right) } ,B\right) $ is equal
to
$\mathcal{Y} \left( \overline{ \left( P\,\vdash
\left\{ A\right\}
\right) } ,A\right) $.   Hence $j$ is a member of $\left(
\mathcal{Y} \left( \overline{ \left( P\,\vdash
\left\{ A\right\}
\right) }
,A\right) \,\vdash A\right) $, hence the assumption that \\
$\mathbb{M} \left( \left( P\vdash \left\{ A\right\}
\right)
,P,H,A,\sigma ,R,x\right) $ holds, implies that $\mathbb{L}
\left( P,A,x\right) <\sigma \left| x_{ \mathcal{Z} \left(
P,H,i\right) } -x_{ \mathcal{Z} \left( P,H,j\right) }
\right| $
holds.   Let $F$ be a member of $\mathbb{K} \left(
P,Q\right) $ such that $\mathbb{M} \left( P,F,H,B,\sigma
,R,x\right) $ holds.   Now $A$ is a member of $P$, hence
$A$ is $\left( V\cup H\right) $-connected, and $B$ is a
member of $\left( Q\,\vdash P\right) $, hence $B$ is a
\emph{strict} subset of $A=\mathcal{Y} \left( \bar{ P }
,B\right) $, hence there exists a member $k$ of $B$ and a
member $m$ of $\left( A\,\vdash B\right) $ such that
$\left\{ k,m\right\} \in H$ holds.   Let $k$ be a member of
$B$ and $m$ be a member of $\left( A\,\vdash B\right) $
such that $\left\{ k,m\right\} \in H$ holds.   Then by
$\mathbb{M} \left( P,F,H,B,\sigma ,R,x\right) $,
$\mathbb{L} \left( F,B,x\right) <\sigma \left| x_{
\mathcal{Z}
\left( F,H,k\right) } -x_{ \mathcal{Z} \left( F,H,m\right)
 } \right| $ holds.   Now $\mathcal{Z} \left( F,H,k\right)
$ is a
subset of $A$, and is \emph{not} a strict subset of any
member of $\mathcal{P} \left( P,A\right) $, for
$\mathcal{Y} \left( \bar{ P } ,B\right) =A$ implies that $B$
is \emph{not} a subset of $\mathcal{K} \left( P,A,k\right)
$, (which by definition is the unique member of
$\mathcal{P} \left( P,A\right) $ that has $k$ as a member),
hence $\mathcal{K} \left( P,A,k\right) \subset B$ holds,
hence $m$ is \emph{not} a member of $\mathcal{K} \left(
P,A,k\right) $, hence since $H$ is a \emph{partition,}
hence
\label{Start of original page 42}
 $\left\{ k,m\right\} $ is the \emph{only} member of $H$
that
has $k$ as a member, $\mathcal{K} \left( P,A,k\right)
\subseteq \mathcal{Z} \left( F,H,k\right) $ holds since
$\mathcal{K} \left( P,A,k\right) $ is a member of $F$.
Hence if $\mathcal{Z} \left( F,H,k\right) $ was a strict
subset of a member $C$ of $\mathcal{P} \left( P,A\right) $
then $\mathcal{K} \left( P,A,k\right) \subset C$ would
hold, which is impossible since $\mathcal{K} \left(
P,A,k\right) $ is a member of $\mathcal{P} \left(
P,A\right) $.   And furthermore $\mathcal{Z} \left(
F,H,k\right) $ overlaps no member of $\mathcal{P} \left(
P,A\right) $, hence $\mathcal{Z} \left( F,H,k\right) $ is a
member of $\Xi \left( \mathcal{P} \left( P,A\right) \right)
$, hence as shown on
page \pageref{Start of original page 16},
$x_{ \mathcal{Z} \left(
F,H,k\right) } $ is a member of the convex hull of the
$x_{ C } $, $C\in \mathcal{P} \left( P,A\right) $.   And
$\mathcal{Z} \left( F,H,m\right) $ is a subset of $A$, and
is \emph{not} a strict subset of any member of $\mathcal{P}
\left( P,A\right) $, for as just shown, $m$ is \emph{not} a
member of $\mathcal{K} \left( P,A,k\right) $, hence
$\mathcal{K} \left( P,A,k\right) $ is not equal to
$\mathcal{K} \left( P,A,m\right) $, hence $k$ is not a
member of $\mathcal{K} \left( P,A,m\right) $, hence since
$H$ is a \emph{partition,} hence $\left\{ k,m\right\} $ is
the \emph{only} member of $H$ that has $m$ as a member,
$\mathcal{K} \left( P,A,m\right) \subseteq \mathcal{Z}
\left( F,H,m\right) $ holds since $\mathcal{K} \left(
P,A,m\right) $ is a member of $F$, hence the fact that
$\mathcal{K} \left( P,A,m\right) $ is a member of
$\mathcal{P} \left( P,A\right) $ implies that $\mathcal{Z}
\left( F,H,m\right) $ cannot be a strict subset of any
member of $\mathcal{P} \left( P,A\right) $.   And
furthermore, $\mathcal{Z} \left( F,H,m\right) $ overlaps no
member of $\mathcal{P} \left( P,A\right) $, hence
$\mathcal{Z} \left( F,H,m\right) $ is a member of $\Xi
\left( \mathcal{P} \left( P,A\right) \right) $, hence as
shown on
page \pageref{Start of original page 16},
$x_{ \mathcal{Z} \left( F,H,m\right) }
$ is a member of the convex hull of the $x_{ C } $, $C\in
\mathcal{P} \left( P,A\right) $.
Hence by Lemma \ref{Lemma 3},
$\left| x_{ \mathcal{Z} \left( F,H,k\right) } -x_{
\mathcal{Z}
\left( F,H,m\right) } \right| <\mathbb{L} \left(
P,A,x\right) $
holds, hence $\mathbb{L} \left( F,B,x\right) <\sigma
\mathbb{L} \left( P,A,x\right) $ holds, hence $\mathbb{L}
\left( F,B,x\right) <\sigma ^{ 2 } \left| x_{ \mathcal{Z}
\left(
P,H,i\right) } -x_{ \mathcal{Z} \left( P,H,j\right) }
\right| $
holds hence, by Lemma \ref{Lemma 6} (e), $\mathbb{L}
\left( F,B,x\right) < \left( \frac{\sigma ^{ 2 } }
{ 1-2\lambda } \right) \left| x_{
\mathcal{Z} \left( F,H,i\right) } -x_{ \mathcal{Z} \left(
F,H,j\right) } \right| $ holds, where $\lambda $ is the real
number defined by $\lambda \equiv \left( \frac{ 1 }
{ 4 } \right)
\left( 1- \sqrt{ 1-8\sigma } \right) $, so that
 $0<\lambda \leq \frac{ 1 }{ 5 } $ holds
since $0<\sigma \leq \frac{ 3 }{ 25 } $ holds.
Now $\lambda $
and $\sigma $ satisfy the equation $\lambda =
\frac{ \sigma }{ 1-2\lambda
} $, and $0<\lambda <1$ holds, hence
$\mathbb{L} \left( F,B,x\right) <\sigma \left| x_{
\mathcal{Z}
\left( F,H,i\right) } -x_{ \mathcal{Z} \left( F,H,j\right)
 } \right| $ holds.

Hence $\mathbb{L} \left( F,B,x\right) <\sigma \left| x_{
\mathcal{Z} \left( F,H,i\right) } -x_{ \mathcal{Z} \left(
F,H,j\right) } \right| $ holds for all $i\in B$ and
\emph{all} \\
$j\in \left( \mathcal{Y} \left( \overline{ \left( P\,\vdash
 \left\{
A\right\} \right) } ,B\right) \,\vdash B\right) $, hence
$\mathbb{M} \left( \left( P\,\vdash \left\{ A\right\}
\right)
,F,H,B,\sigma ,R,x\right) $ holds in this case also.

And finally, for the case $B=A$, $\mathbb{M} \left( \left(
P\,\vdash \left\{ A\right\} \right) ,P,H,A,\sigma
,R,x\right)
$ holds by assumption.

Hence the assumption that $\mathbb{M} \left( \left(
P\,\vdash \left\{ A\right\} \right) ,P,H,A,\sigma
,R,x\right)
$ holds implies that for every member $B$ of $\left(
Q\,\vdash \left( P\,\vdash \left\{ A\right\} \right) \right)
$, there exists a member $F$ of $\mathbb{K} \left( \left(
P\,\vdash \left\{ A\right\} \right) ,Q\right) $ such that
$\mathbb{M} \left( \left( P\,\vdash \left\{ A\right\}
\right)
,F,H,B,\sigma ,R,x\right) $ holds, hence that
\label{Start of original page 43}
  $\left( \left( P\,\vdash \left\{ A\right\} \right)
,Q\right) $ is a member of $\Omega \left( H,\sigma
,R,x\right) $.   And by Lemma \ref{Lemma 9}, this
contradicts the
assumption that $\left(P,Q\right) $ generates a good set of
woods for
$\left( H,\sigma ,R,x\right) $.

\begin{bphzlemma} \label{Lemma 11}
\end{bphzlemma}
\vspace{-6.143ex}

\noindent \hspace{11.9ex}{\bf.  }Let $V$ be any
partition such that $\mathcal{U}
\left( V\right) $ is finite and $\#\left( V\right) \geq 2$
holds, let $H$ be any partition such that if $E$ is any
member of $H$ such that $E$ intersects \emph{more} than one
member of $V$, then $E$ has \emph{exactly} two members, let
$\sigma $ be any real number such that $0<\sigma \leq
\frac{ 3 }{ 25
} $ holds, let $R$ be any finite real number $>0$, let
$d$ be any integer $\geq 1$, and let $x$ be any member of
$\mathbb{F} _{ d } \left( V\right) $.

Let $ \left(P ,Q\right) $ be any member of $\Omega \left(
H,\sigma
,R,x\right) $ such that there is \emph{no} member $(P,G) $
of
$\Omega \left( H,\sigma ,R,x\right) $ such that $Q\subset
G$ holds, let $A$ be any member of $\mathbb{B}
\left( \bar{ P }
 \right) $, let $X$ be any $\sigma $-cluster of
$\downarrow \left( x,\mathcal{P} \left( P,A\right) \right)
$ such that $\#\left( X\right) \geq 2$ holds, and $X$ is
\emph{not} equal to $\mathcal{D} \left( \downarrow \left(
x,\mathcal{P} \left( P,A\right) \right) \right)
=\mathcal{P} \left( P,A\right) $,
and let $B$ be any $\left( V\cup H\right)
$-connected component of $\mathcal{U} \left( X\right) $
such that $B$ is \emph{not} a member of $\mathcal{P} \left(
P,A\right) $, and such that $ \mathbb{L} \left( P, B, x
\right) < R $ holds.

Then $B$ is a member of $\left( Q\,\vdash P\right) $.
\enlargethispage{0.5ex}

\vspace{2.5ex}

\noindent {\bf Proof.}  We first note that by
the definition of a $\sigma
$-cluster of $\downarrow \left( x,\mathcal{P} \left(
P,A\right) \right) $ and the assumption that $\#\left(
X\right) \geq 2$ holds, $X$ is a subset of $\mathcal{D}
\left( \downarrow \left( x,\mathcal{P} \left( P,A\right)
\right) \right) =\mathcal{P} \left( P,A\right) $ such that
$\#\left( X\right) \geq 2$ holds and such that for every
$I\in X$, $J\in X$, $K\in X$, and $M\in \left( \mathcal{P}
\left( P,A\right) \,\vdash X\right) $, $\left| x_{ I } -x_{
J }
\right| <\sigma \left| x_{ K } -x_{ M } \right| $ holds.

Now if $C$ is any member of $X$, then either $C\subseteq B$
holds or $C\cap B=\emptyset $ holds.   For each member of
$X$ is
a member of $P$ hence is $\left( V\cup H\right)
$-connected, and by assumption $B$ is a $\left( V\cup
H\right) $-connected component of $X$, hence is $\left(
V\cup H\right) $-connected.   Thus if $\left( C\,\vdash
B\right) $ and $\left( C\cap B\right) $ were both nonempty,
then by Lemma \ref{Lemma 1}, $\left( C\cup B\right) $ would
be a
$\left( V\cup H\right) $-connected subset of $\mathcal{U}
\left( X\right) $ such that $B\subset \left( C\cup B\right)
$ holds, which contradicts the assumption that $B$ is a
$\left( V\cup H\right) $-connected \emph{component} of
$\mathcal{U} \left( X\right) $.

Furthermore, $B$ is a nonempty subset of $A$, and if $C$ is
any member of $\left( \mathcal{P} \! \left( P,A\right)
\vdash \! X\right) $, then $C\cap B=\emptyset $ holds.

Hence $B$ is a member of $\Xi \left( \mathcal{P} \left(
P,A\right) \right) $, and since by assumption $B$ is
\emph{not} a member of $\mathcal{P} \left( P,A\right) $,
$B$ is a member of $\left( \Xi \left( \mathcal{P} \left(
P,A\right) \right) \,\vdash \mathcal{P} \left( P,A\right)
\right) $, hence every member $C$ of $\mathcal{P} \left(
P,A\right) $ such that $C\subseteq B$ holds, is a
\emph{strict} subset of $B$.

Now let $C$ be any member of $\mathcal{P} \left( P,B\right)
$.   Then $C$ is a member of $P$
\label{Start of original page 44}
 such that $C\subset A$ holds hence, as shown on
page \pageref{Start of original page 5},
there is a unique member $E$ of $\mathcal{P} \left(
P,A\right) $ such that $C\subseteq E$ holds.   Let $E$ be
the unique member of $\mathcal{P} \left( P,A\right) $ such
that $C\subseteq E$ holds.   Then $E$ is a member of
$\mathcal{P} \left( P,A\right) $ such that $E\cap B$ is
nonempty hence, by the foregoing, $E\subset B$ holds, hence
$C\subset E$ cannot hold, (for if $C\subset E$ held then
$E$ would be a member of $P$ such that $C\subset E\subset
B$ held, contradicting $C\in \mathcal{P} \left( P,B\right)
 $), hence $C=E$ holds, hence $C$ is a member of
$\mathcal{P} \left( P,A\right) $, hence since $C$ is a
subset of $B$ hence a subset of $\mathcal{U} \left(
X\right) $ hence cannot be a member of $\left( \mathcal{P}
\left( P,A\right) \,\vdash X\right) $, $C$ is a member of
$X$.

Hence $\mathcal{P} \left( P,B\right) $ is a subset of $X$,
hence since $\mathbb{L} \left( P,B,x\right) $ is by
definition equal to $ \rule{0pt}{1ex}
\hspace{1.5ex} \max_{ \hspace{-7.0ex}
\begin{array}{c} \\[-3.1ex]
\scriptstyle{ I\in \mathcal{P}
\left( P,B\right) } \\[-1.2ex]
\scriptstyle{ J\in \mathcal{P} \left( P,B\right) }
\end{array} } \hspace{-1.2ex}
\left| x_{ I } -x_{ J
} \right| $,
the inequality $\mathbb{L} \left( P,B,x\right) \leq
\hspace{0.5ex} \max_{ \hspace{-4.7ex}
\begin{array}{c} \\[-3.1ex]
\scriptstyle{ I\in X } \\[-1.2ex]
\scriptstyle{ J\in X }
\end{array} } \hspace{-0.8ex}
\left| x_{ I } -x_{ J } \right| $ holds.  Furthermore,
by assumption, $\mathbb{L} \left(
P,B,x\right) <R$ holds.

Now let $i$ be any member of $B$ and $j$ be any member of
$\left( \mathcal{Y} \left( \bar{ P } ,B\right) \,\vdash
B\right) $ such that $\left\{ i,j\right\} $ is a member of
$H$.   Now $\mathcal{Y} \left( \bar{ P } ,B\right) \subseteq
A$ certainly holds, for $\mathcal{Y} \left( \bar{ P }
,B\right) $ is by definition the \emph{smallest} member $D$
of $ \bar{ P } $ such that $B\subseteq D$ holds, and $A$ is
a
member of $ \bar{ P } $ such that $B\subseteq A$ holds.
Hence $j$ is a member of $A$, hence $j$ is a member of
$\left( A\,\vdash B\right) $.   And furthermore, $j$ is not
a member of $\mathcal{U} \left( X\right) $, for if $j$ was
a member of $\mathcal{U} \left( X\right) $, then $j$ would
be a member of some member $C$ of $X$ such that $C$ is
\emph{not} a subset of $B$, hence as shown on
page \pageref{Start of original page 36},
the facts that $B$ and $C$ are $\left( V\cup H\right)
$-connected, together with the fact that $\left\{
i,j\right\}
$ is a member of $H$, would imply by Lemma \ref{Lemma 1}
that $\left(
C\cup B\right) $ is $\left( V\cup H\right) $-connected,
which together with the fact that $C$ is \emph{not} a
subset of $B$, hence that $B$ is a \emph{strict} subset of
$\left( C\cup B\right) $, contradicts the fact that $B$ is
a $\left( V\cup H\right) $-connected \emph{component} of
$\mathcal{U} \left( X\right) $.

Now $\mathcal{K} \left( P,A,i\right) $, which by definition
is the unique member of $\mathcal{P} \left( P,A\right) $
that has $i$ as a member, is a member of $\mathcal{P}
\left( P,A\right) $ that intersects $\mathcal{U} \left(
X\right) $, (since $i$ is a member of $\mathcal{U} \left(
X\right)  $), hence is a member of $X$, and $j$ is
not a member of $\mathcal{U} \left( X\right) $, hence
$\mathcal{K} \left( P,A,j\right) $, which is the unique
member of $\mathcal{P} \left( P,A\right) $ that has $j$ as
a member, is not a subset of $\mathcal{U} \left( X\right)
$, hence $\mathcal{K} \left( P,A,j\right) $ is not a member
of $X$, hence $\mathcal{K} \left( P,A,j\right) $ is a
member of $\left( \mathcal{P} \left( P,A\right) \,\vdash
X\right) $.

Hence $\mathcal{K} \left( P,A,j\right) $ is not equal to
$\mathcal{K} \left( P,A,i\right) $, hence $j$ is not a
member of $\mathcal{K} \left( P,A,i\right) $, hence $\left\{
i,j\right\} $ is not a subset of $\mathcal{K} \left(
P,A,i\right) $, hence since $H$ is a \emph{partition,}
hence $\left\{ i,j\right\} $ is the \emph{only} member of
$H$
that has
\label{Start of original page 45}
 $i$ as a member, $\mathcal{K} \left( P,A,i\right) $ is a
subset of $\mathcal{Z} \left( P,H,i\right) $, which by
definition is the \emph{largest} member of $P$ that has $i$
as a member but does not contain as a subset any member of
$H$ that has $i$ as a member.   Furthermore $\left\{
i,j\right\} $ \emph{is} a subset of $A$, and since
$\mathcal{K} \left( P,A,i\right) $ is a member of
$\mathcal{P} \left( P,A\right) $, there is \emph{no} member
$C$ of $P$ such that $\mathcal{K} \left( P,A,i\right)
\subset C\subset A$ holds, hence $\mathcal{K} \left(
P,A,i\right) $ is the \emph{largest} member of $P$ that has
$i$ as a member but does not contain as a subset any member
of $H$ that has $i$ as a member, hence $\mathcal{K} \left(
P,A,i\right) =\mathcal{Z} \left( P,H,i\right) $ holds.

And by an exactly analogous argument, $\mathcal{K} \left(
P,A,j\right) =\mathcal{Z} \left( P,H,j\right) $ holds.

Hence $\left| x_{ \mathcal{Z} \left( P,H,i\right) } -x_{
\mathcal{Z} \left( P,H,j\right) } \right| $ is equal to
$\left| x_{
\mathcal{K} \left( P,A,i\right) } -x_{ \mathcal{K} \left(
P,A,j\right) } \right| $ hence since, as shown above,
$\mathcal{K} \left( P,A,i\right) $ is a member of $X$ and
$\mathcal{K} \left( P,A,j\right) $ is a member of $\left(
\mathcal{P} \left( P,A\right) \,\vdash X\right) $, the
inequality $ \hspace{3.5ex} \min_{ \hspace{-9.0ex}
\begin{array}{c} \\[-3.1ex]
\scriptstyle{ K\in X } \\[-1.2ex]
\scriptstyle{ M\in \left( \mathcal{P}
\left( P,A\right) \,\vdash X\right) }
\end{array} } \hspace{-4.6ex}
\left| x_{ K } -x_{ M } \right| \leq \left| x_{
\mathcal{Z} \left( P,H,i\right) } -x_{ \mathcal{Z} \left(
P,H,j\right) } \right| $ holds, hence by the fact that $X$
is a
$\sigma $-cluster of $\downarrow\left( x,\mathcal{P}
 \left( P,A\right) \right) $ such
that $\#\left( X\right) \geq 2$ holds,
\[
\mathbb{L} \left( P,B,x\right) \leq
\hspace{0.5ex} \textstyle \max_{ \hspace{-4.7ex}
\begin{array}{c} \\[-3.1ex]
\scriptstyle{ I\in X } \\[-1.2ex]
\scriptstyle{ J\in X }
\end{array} } \hspace{-0.8ex} \left| x_{ I } -x_{
J } \right|  < \hspace{2.5ex} \sigma
\hspace{1.0ex} \textstyle \min_{ \hspace{-9.0ex}
\begin{array}{c} \\[-3.1ex]
\scriptstyle{ K\in X } \\[-1.2ex]
\scriptstyle{ M\in \left( \mathcal{P}
\left( P,A\right) \,\vdash X\right) }
\end{array} } \hspace{-4.6ex}
\left| x_{ K } -x_{ M } \right| \leq \sigma
\left| x_{ \mathcal{Z} \left( P,H,i\right) } -x_{
\mathcal{Z}
\left( P,H,j\right) } \right|
\]
holds, hence $\mathbb{L}
\left(
P,B,x\right) <\sigma \left| x_{ \mathcal{Z} \left(
P,H,i\right)
} -x_{ \mathcal{Z} \left( P,H,j\right) } \right| $ holds.

Hence $\mathbb{M} \left( P,P,H,B,\sigma ,R,x\right) $ holds.

Now by assumption $\left(P,Q\right) $ is a member of
$\Omega \left(
H,\sigma ,R,x\right) $, hence for every member $C$ of
$\left( Q\,\vdash P\right) $, there exists a member $F$ of
$\mathbb{K} \left( P,Q\right) $ such that $\mathbb{M}
\left( P,F,H,C,\sigma ,R,x\right) $ holds.   Hence if $B$
was \emph{not} a member of $\left( Q\,\vdash P\right) $,
then for every member $C$ of $\left( \left( Q\cup \left\{
B\right\} \right) \,\vdash P\right) $ there would exist a
member $F$ of $\mathbb{K} \left( P,\left( Q\cup \left\{
B\right\} \right) \right) $ such that $\mathbb{M} \left(
P,F,H,C,\sigma ,R,x\right) $ holds, hence $\left( P,\left(
Q\cup \left\{ B\right\} \right) \right) $ would be a member
of $\Omega \left( H,\sigma ,R,x\right) $.

But by assumption there is \emph{no} member $ \left( P,G
\right) $ of
$\Omega \left( H,\sigma ,R,x\right) $ such that $Q\subset
G$ holds.   Hence $B$ must be a member of $\left( Q\,\vdash
P\right) $.

\vspace{1.0ex}

\noindent \emph{Additional Note to Lemma 11.}
$\mathbb{M} \left(
P,P,H,\mathcal{U} \left( V\right) ,\sigma ,R,x\right) $
holds whenever \\
$\mathbb{L} \left( P,\mathcal{U} \left(
V\right) ,x\right) <R$ holds, hence with the same
assumptions on $\left(P,Q\right) $
as in Lemma \ref{Lemma 11},
$\mathcal{U} \left(
V\right) \in \left( Q\,\vdash P\right) $ holds whenever
$\mathbb{L} \left( P,\mathcal{U} \left( V\right) ,x\right)
<R$ holds.

We recall from
page \pageref{Start of original page 8}
 that for all $s\in \mathbb{R} $, we
define $\mathbb{T} \left( s\right) $ by
\[
\mathbb{T} \left( s\right) \equiv \left\{ \begin{array}{cc}
1 & \textrm{if }s<0 \textrm{ holds} \\
 0 & \textrm{if }s \geq
0 \textrm{ holds}
\end{array}  \right.
\]

For every ordered septuple $\left( P,F,H,A,\sigma
,R,x\right) $ of a wood $P$, a wood $F$ such that
$\mathcal{M} \left( F\right) =\mathcal{M} \left( P\right) $
holds and $P\subseteq F$ holds, a partition $H$, a
member $A $ of \\
 $\left( \Xi \left( \mathcal{M} \left(
P\right)
\right) \,\vdash \mathcal{M} \left( P\right) \right)
$, a
real number $
\sigma $ such that $ 0<\sigma \leq  \frac{1}{8} $
\label{Start of original page 46}
 holds, a finite real number $R>0$, and a member $x$ of
$\mathbb{F} _{ d } \left( \mathcal{M} \left( P\right)
\right) $, where $d$ is an integer $\geq 1$, we define the
number $\tilde{ \mathbb{M} } \left( P,F,H,A,\sigma
,R,x\right)
$ to be equal to $1$ if $\mathbb{M} \left( P,F,H,A,\sigma
,R,x\right) $ holds, and to be equal to $0$ otherwise.

And for every ordered triple $\left(A,B,H\right) $ of
 a set $A$, a set
$B$ such that $A\subseteq B$ holds, and a set $H$ such that
every member of $H$ is a set, we define $\mathbb{Q} \left(
A,B,H\right) $ to be the set whose members are all the
ordered pairs $\left(i,j\right) $ of a member $i$ of $A$
and a member $j$
of $\left( B\,\vdash A\right) $ such that $\left\{
i,j\right\} $ is a member of $H$.

Now the proposition $\mathbb{M} \left( P,F,H,A,\sigma
,R,x\right) $ can be expressed as:

For all $S\in \mathcal{P} \left( F,A\right) $ and all $T\in
\mathcal{P} \left( F,A\right) $, $\left| x_{ S } -x_{ T }
\right| <R$
holds, and for all $i\in A$ and all $j\in \left(
\mathcal{Y} \left( \bar{ P } ,A\right) \,\vdash A\right) $
such that $\left\{ i,j\right\} $ is a member of $H$, and for
all $S\in \mathcal{P} \left( F,A\right) $ and all $T\in
\mathcal{P} \left( F,A\right) $, $\left| x_{ S } -x_{ T }
\right| <\sigma \left| x_{ \mathcal{Z} \left( F,H,i\right)
} -x_{
\mathcal{Z} \left( F,H,j\right) } \right| $ holds.

Hence $\tilde{ \mathbb{M} } \left( P,F,H,A,\sigma
,R,x\right) $
may be constructed as:
\[
\tilde{ \mathbb{M} } \left( P,F,H,A,\sigma ,R,x\right) =
\left( \prod_{\Delta
\equiv \left\{ S,T\right\} \in \mathcal{Q} \left(
\mathcal{P}
\left( F,A\right) \right) } \mathbb{T} \left( \left| x_{ S
} -x_{
T
} \right| -R\right) \right) \times
\]
\[
\times \left(
\prod_{
\begin{array}{c} \\[-4.6ex]
\scriptstyle{ \left( i,j\right) \in
\mathbb{Q} \left( A,\mathcal{Y} \left( \bar{ P },
A\right), H\right) } \\[-1.3ex]
\scriptstyle{ \Delta
\equiv \left\{ S,T\right\} \in \mathcal{Q} \left(
\mathcal{P}
\left( F,A\right) \right) }
\end{array} } \hspace{-1.9ex}
\mathbb{T} \left( \left| x_{ S } -x_{ T
} \right| -\sigma \left| x_{ \mathcal{Z} \left(
F,H,i\right) } -x_{
\mathcal{Z} \left( F,H,j\right) } \right| \right) \right)
\]

For any
ordered sextuple $\left( P,Q,H,\sigma ,R,x\right) $ of a
wood $P$, a wood $Q$ such that $\mathcal{M} \left( Q\right)
=\mathcal{M} \left( P\right) $ holds and $P\subseteq Q$
holds, a partition $H$ such that if $E$ is any member of
$H$ such that $E$ intersects \emph{more} than one member of
$\mathcal{M} \left( P\right) $, then $E$ has \emph{exactly}
two members, a real number $\sigma $ such that $0<\sigma
\leq \frac{ 3 }{ 25 } $ holds, a real number $R>0$, and
 a member
$x$ of $\mathbb{F} _{ d } \left( \mathcal{M} \left(
P\right) \right) $, where $d$ is an integer $\geq 1$, we
define the real numbers $\mathcal{L} \left( P,Q,H,\sigma
,R,x\right) $, $\mathcal{H} \left( P,Q,H,\sigma ,R,x\right)
$, and $\mathcal{E} \left( P,Q,H,\sigma ,R,x\right) $ by:

\vspace{1.0ex}

\noindent $\mathcal{L} \left( P,Q,H,\sigma ,R,x\right) $ is
equal
to $1$ if $ \left(P ,Q\right) $ is
a member of $\Omega \left( H,\sigma ,R,x\right) $, and
equal to $0$ otherwise,

\vspace{1.0ex}

\noindent $\mathcal{H} \left( P,Q,H,\sigma ,R,x\right) $ is
equal
 to $1$ if $ \left(P ,Q\right) $ is
a member of $\Omega \left( H,\sigma ,R,x\right) $ such that
there is \emph{no} member $ \left(P,G\right) $ of $\Omega
\left( H,\sigma
,R,x\right) $ such that $G\subset Q$ holds, and equal to
$0$ otherwise, and

\vspace{1.0ex}

\noindent $\mathcal{E} \left( P,Q,H,\sigma ,R,x\right) $
is equal to
 $1$ if $ \left(P ,Q\right) $
generates a good set of woods for $\left( H,\sigma
,R,x\right) $, and equal to $0$ otherwise.

\vspace{1.0ex}

We note that it follows directly from the definition of
$\Omega \left( H,\sigma ,R,x\right) $ that \\
$\mathcal{L}
\left( P,Q,H,\sigma ,R,x\right) $, $\mathcal{H} \left(
P,Q,H,\sigma ,R,x\right) $, and $\mathcal{E} \left(
P,Q,H,\sigma ,R,x\right) $ are all equal to $0$ unless
every member of $Q$ is
\label{Start of original page 47}
 $\left( \mathcal{M} \left( P\right) \cup H\right)
$-connected, (which implies that every member of $P$ is
$\left( \mathcal{M} \left( P\right) \cup H\right)
 $-connected), and we now assume that every member of
$Q$ is $\left( \mathcal{M} \left( P\right) \cup H\right)
$-connected.

To construct $\mathcal{L} \left( P,Q,H,\sigma ,R,x\right)
$, we first note that \\
$\prod_{F\in \mathbb{K} \left(
P,Q\right) }  \left( 1-\tilde{ \mathbb{M} } \left(
P,F,H,A,\sigma
,R,x\right) \right) $ is $1$ if $\mathbb{M} \left(
P,F,H,A,\sigma ,R,x\right) $ is \emph{false} for \emph{all}
$F\in \mathbb{K} \left( P,Q\right) $, and $0$ otherwise, or
in other words, it is $0$ if there \emph{exists} an $F\in
\mathbb{K} \left( P,Q\right) $ such that $\mathbb{M} \left(
P,F,H,A,\sigma ,R,x\right) $ is true, and $1$ otherwise.

Hence
\[
\mathcal{L} \left( P,Q,H,\sigma ,R,x\right) =\prod_{
 A\in \left( Q\,\vdash P\right)} \left( 1-\prod_{  F\in
\mathbb{K} \left( P,Q\right)} \left( 1-\tilde{ \mathbb{M} }
\left( P,F,H,A,\sigma ,R,x\right) \right) \right).
\]

For every ordered pair $\left(Q,H\right) $ of a wood $Q$
and a partition
$H$ such that every member of $Q$ is $\left( \mathcal{M}
\left( Q\right) \cup H\right) $-connected, we define
$\tilde{ \mathcal{G} } \left( Q,H\right) $ to be the set
whose
members are all the members $G$ of $\mathcal{G} \left(
\mathcal{M} \left( Q\right) ,H\right) $ such that $Q\subset
G$ holds.   Then $\mathcal{H} \left( P,Q,H,\sigma
,R,x\right) $ may be constructed as:
\[
\mathcal{H} \left( P,Q,H,\sigma ,R,x\right) =\mathcal{L}
\left( P,Q,H,\sigma ,R,x\right)  \prod_{G\in \tilde{
\mathcal{G} }
\left( Q,H\right) } \left( 1-\mathcal{L} \left(
P,G,H,\sigma ,R,x\right) \right)
\]

For every ordered triple $\left(P,Q,H\right) $
of a wood $P$, a wood $Q$
such that $\mathcal{M} \left( Q\right) =\mathcal{M} \left(
P\right) $ holds and $P\subseteq Q$ holds, and a partition
$H$ such that every member of $Q$ is $\left( \mathcal{M}
\left( P\right) \cup H\right) $-connected, we define
$\tilde{ \mathcal{N} } \left( P,Q,H\right) $ to be the set
whose members are all the members $ \left( F,G\right) $ of
$\mathcal{N}
\left( \mathcal{M} \left( P\right) ,H\right) $ such that
$F\subseteq P\subseteq Q\subseteq G$ holds and $\left(
F,G\right) \neq \left( P,Q\right) $.   Then condition (ii)
of Lemma \ref{Lemma 9} may be expressed by the function
$\prod_{
\left( F,G\right) \in \tilde{ \mathcal{N} }
 \left( P,Q,H\right)
}\left( 1-\mathcal{L} \left( F,G,H,\sigma ,R,x\right)
\right) $, which is $0$ unless \emph{no} member $ \left(
F,G\right) $ of
$\tilde{ \mathcal{N} } \left( P,Q,H\right) $ is a member of
$\Omega \left( H,\sigma ,R,x\right) $.

Hence $\mathcal{E} \left( P,Q,H,\sigma ,R,x\right) $ may be
constructed as:
\[
\mathcal{E} \left( P,Q,H,\sigma ,R,x\right) =\mathcal{L}
\left( P,Q,H,\sigma ,R,x\right) \prod_{  \left( F,G\right)
\in \tilde{ \mathcal{N} } \left( P,Q,H\right)} \left(
1-\mathcal{L} \left( F,G,H,\sigma ,R,x\right) \right) .
\]

We observe that the following inequalities hold:
\[
\mathcal{E} \left( P,Q,H,\sigma ,R,x\right) \leq
\mathcal{H} \left( P,Q,H,\sigma ,R,x\right) \leq
\mathcal{L} \left( P,Q,H,\sigma ,R,x\right) .
\]

For every ordered pair $\left(A,B\right) $ of sets $A$ and
$B$, we define
$\mathcal{S} \left( A,B\right) $ to be $1$ if $A\subseteq
B$ holds, and $0$ otherwise.

Then if $V$ is any partition such that $\mathcal{U} \left(
V\right) $ is finite and $\#\left( V\right) \geq 2$ holds,
$H$ is any partition such that if $E$ is any member of $H$
such
\label{Start of original page 48}
 that $E$ intersects \emph{more} than one member of $V$,
then $E$ has \emph{exactly} two members, $\sigma $ is any
real number such that $0<\sigma \leq \frac{ 3 }{ 25 } $
holds,
$R$ is any finite real number $>0$, and $d$ is any integer
$\geq 1$, it follows directly from the discussion on
pages \pageref{Start of original page 35}
to \pageref{Start of original page 39}
that the following identity holds for all
ordered pairs $\left(F,x\right) $ of a member $F$ of
$\mathcal{G} \left(
V,H\right) $ and a member $x$ of $\mathbb{F} _{ d } \left(
V\right) $:
\[
 \sum_{\left( P,Q\right) \in \mathcal{N} \left( V,H\right)
} \mathcal{E} \left( P,Q,H,\sigma ,R,x\right) \mathcal{S}
\left( P,F\right) \mathcal{S} \left( F,Q\right) =1.
\]

In fact, for any given ordered pair $\left(F,x\right) $
 of a member $F$
of $\mathcal{G} \left( V,H\right) $ and a member $x$ of
$\mathbb{F} _{ d } \left( V\right) $, exactly one term in
the left-hand side of this equation is equal to $1$, and
all the remaining terms are equal to $0$, and the term that
is equal to $1$ is the term corresponding to the unique
member $\left(P,Q\right) $ of $\Delta \left( H,\sigma
,R,x\right) $ such
that $F\in \mathbb{K} \left( P,Q\right) $ holds.
\enlargethispage{5.5ex}

\vspace{2.5ex}

We now make the following observations:

\begin{bphzfirstseriesobs} \label{First Series Obs 1}
\end{bphzfirstseriesobs}
\vspace{-6.143ex}

\noindent \hspace{2.65ex}{\bf )  }If $F$ is any wood,
$B$ is any set, and $G$ is any
subset of $F$ such that $\mathcal{P} \left( F,B\right)
\subseteq G$ holds, then $\mathcal{P} \left( G,B\right)
=\mathcal{P} \left( F,B\right) $ holds.   For if $C$ is any
member of $\mathcal{P} \left( F,B\right) $, then $C$ is a
member of $G$ such that $C\subset B$ holds and there is no
member $D$ of $F$ such that $C\subset D\subset B$ holds,
hence there is no member $D$ of $G$ such that $C\subset
D\subset B$ holds, hence $C$ is a member of $\mathcal{P}
\left( G,B\right) $.   Now let $C$ be any member of
$\mathcal{P} \left( G,B\right) $.   Then $C$ is a member of
$G$, hence a member of $F$, such that $C\subset B$ holds,
hence as shown on
page \pageref{Start of original page 5}
 the fact that $F$ is a wood
implies that there is a unique member $E$ of $\mathcal{P}
\left( F,B\right) $ such that $C\subseteq E$ holds.   Let
$E$ be the unique member of $\mathcal{P} \left( F,B\right)
$ such that $C\subseteq E$ holds.   Then $E$ is a member of
$G$, (since by assumption $\mathcal{P} \left( F,B\right) $
is a subset of G), hence $C\subset E$ cannot hold, (for if
$C\subset E$ held then $E$ would be a member of $G$ such
that $C\subset E\subset B$ held, contradicting the fact
that $C$ is a member of $\mathcal{P} \left( G,B\right)
 $), hence $C=E$ holds, hence $C$ is a member of
$\mathcal{P} \left( F,B\right) $.

\newpage

\begin{bphzfirstseriesobs} \label{First Series Obs 2}
\end{bphzfirstseriesobs}
\vspace{-6.143ex}

\noindent \hspace{2.65ex}{\bf )  }If $P$ is any wood,
$B$ is any nonempty subset of
$\mathcal{U} \left( P\right) $, and $G$ is any subset of
$P$ such that $G$ is a wood, $\mathcal{U} \left( G\right) $
is a member of $\bar{ P } $,  and $\mathcal{Y} \left( \bar{
P }
,B\right) $ is a member of $\bar{ G } $, then $\mathcal{Y}
\left( \bar{ G } ,B\right) =\mathcal{Y} \left( \bar{ P }
,B\right) $ holds.   For $ \bar{ G } $ has at least one
member, namely $\mathcal{Y} \left( \bar{ P } ,B\right) $,
that
contains $B$ as a subset, hence by definition $\mathcal{Y}
\left( \bar{ G } ,B\right) $ is the \emph{smallest} member
of
$ \bar{ G } $ that contains $B$ as a subset.   Now the
assumptions that $G$ is a subset of $P$ and that
$\mathcal{U} \left( G\right) $ is a member of $ \bar{ P } $
imply that $ \bar{ G } $ is a subset of $\bar{ P } $.
Hence
there is no member $C$ of $ \bar{ G } $ such that
$B\subseteq
C\subset \mathcal{Y} \left( \bar{ P } ,B\right) $ holds, for
if $C$ was such a member of $ \bar{ G } $ then $C$ would be
a
member of $ \bar{ P } $ such
\label{Start of original page 49}
 that $B\subseteq C\subset \mathcal{Y} \left( \bar{ P }
,B\right) $ held, which contradicts the fact that
$\mathcal{Y} \left( \bar{ P } ,B\right) $ is the
\emph{smallest} member of $ \bar{ P } $ that contains $B$ as
a subset.   Hence $\mathcal{Y} \left( \bar{ P } ,B\right) $
is
the \emph{smallest} member of $ \bar{ G } $ that contains
$B$
as a subset.

\begin{bphzfirstseriesobs} \label{First Series Obs 3}
\end{bphzfirstseriesobs}
\vspace{-6.143ex}

\noindent \hspace{2.65ex}{\bf )  }If $F$ is any wood,
$H$ is any set such that every
member of $H$ is a set, $i$ is any member of $\mathcal{O}
\left( F,H\right) $, and $G$ is any subset of $F$ such that
$G$ is a wood and $\mathcal{Z} \left( F,H,i\right) $ is a
member of $G$, then $i$ is a member of $\mathcal{O} \left(
G,H\right) $ and $\mathcal{Z} \left( G,H,i\right)
=\mathcal{Z} \left( F,H,i\right) $ holds.   For by
definition $\mathcal{O} \left( F,H\right) $ is the set
whose members are all the members $i$ of $\mathcal{U}
\left( F\right) $ such that there \emph{exists} a member
$A$ of $F$ such that $i\in A$ holds and there is \emph{no}
member $B$ of $H$ such that $i\in B$ and $B\subseteq A$
both hold, and $\mathcal{Z} \left( F,H,i\right) $ is the
\emph{largest} member $A$ of $F$ such that $i\in A$ holds
and there is \emph{no} member $B$ of $H$ such that $i\in B$
and $B\subseteq A$ both hold.   And by assumption
$\mathcal{Z} \left( F,H,i\right) $ is a member of $G$,
hence $G$ has at least one member, namely $\mathcal{Z}
\left( F,H,i\right) $, that has $i$ as a member and does
\emph{not} contain as a subset any member of $H$ that has
$i$ as a member, hence $i$ is a member of $\mathcal{U}
\left( G\right) $ and also a member of $\mathcal{O} \left(
G,H\right) $, hence $\mathcal{Z} \left( G,H,i\right) $ is
defined and moreover $\mathcal{Z} \left( F,H,i\right)
\subseteq \mathcal{Z} \left( G,H,i\right) $ holds.   And
furthermore there is \emph{no} member $A$ of $G$ such that
$\mathcal{Z} \left( F,H,i\right) \subset A$ holds and $A$
does \emph{not} contain as a subset any member of $H$ that
has $i$ as a member, for if $A$ \emph{was} such a member of
$G$ then $A$ would be a member of $F$ such that
$\mathcal{Z} \left( F,H,i\right) \subset A$ held and $A$
does \emph{not} contain as a subset any member of $H$ that
has $i$ as a member, and by the definition of $\mathcal{Z}
\left( F,H,i\right) $ there is \emph{no} such member $A$ of
$F$.   Hence $\mathcal{Z} \left( F,H,i\right) $ is the
\emph{largest} member of $G$ that has $i$ as a member and
does \emph{not} contain as a subset any member of $H$ that
has $i$ as a member, hence $\mathcal{Z} \left( G,H,i\right)
=\mathcal{Z} \left( F,H,i\right) $ holds.

\begin{bphzfirstseriesobs} \label{First Series Obs 4}
\end{bphzfirstseriesobs}
\vspace{-6.143ex}

\noindent \hspace{2.65ex}{\bf )  }Let $P$ be any wood,
let $F$ be any wood such that
$\mathcal{M} \left( F\right) =\mathcal{M} \left( P\right) $
holds and $P\subseteq F$ holds, and let $A$ be any member
of $\mathbb{B} \left( \bar{ P } \right) $.   Then $F\cap \Xi
\left( \mathcal{P} \left( P,A\right) \right) $ is a wood of
$\mathcal{P} \left( P,A\right) $.   We note first that, as
shown on
pages \pageref{Start of original page 5} and
\pageref{Start of original page 6},
 $\mathcal{P} \left( P,A\right)
$ is a partition of $A$ such that $\#\left( \mathcal{P}
\left( P,A\right) \right) \geq 2$ holds, hence $\mathcal{U}
\left( \mathcal{P} \left( P,A\right) \right) $ is equal to
$A$ hence is a finite set.    Now certainly no member of
$F\cap \Xi \left( \mathcal{P} \left( P,A\right) \right) $
is empty and no two members of $F\cap \Xi \left(
\mathcal{P} \left( P,A\right) \right) $ overlap, and
moreover $P$ is a subset of $F$ hence every member of
$\mathcal{P} \left( P,A\right) $ is a member of $F\cap \Xi
\left( \mathcal{P} \left( P,A\right) \right) $, hence
$F\cap \Xi \left( \mathcal{P} \left( P,A\right) \right) $
is a subset of $\Xi \left( \mathcal{P} \left( P,A\right)
\right) $ that has $\mathcal{P} \left( P,A\right) $ as a
subset, hence $\mathcal{M} \left( F\cap \Xi \left(
\mathcal{P} \left( P,A\right) \right) \right) $ is equal to
$\mathcal{P} \left( P,A\right) $.   And furthermore, every
member of $\Xi \left( \mathcal{P} \left( P,A\right) \right)
$ is a subset of $A$, hence every member of $F\cap \Xi
\left( \mathcal{P} \left( P,A\right) \right) $ is a subset
of $A$, hence $\mathcal{U} \left( F\cap \Xi \left(
\mathcal{P} \left( P,A\right) \right) \right) $ is a subset
of $A$, hence every member
\label{Start of original page 50}
 of $\mathcal{U} \left( F\cap \Xi \left( \mathcal{P} \left(
P,A\right) \right) \right) $ is a member of some member of
$\mathcal{M} \left( F\cap \Xi \left( \mathcal{P} \left(
P,A\right) \right) \right) =\mathcal{P} \left( P,A\right) $.

\begin{bphzfirstseriesobs} \label{First Series Obs 5}
\end{bphzfirstseriesobs}
\vspace{-6.143ex}

\noindent \hspace{2.65ex}{\bf )  }Let $P$ be any wood, let
$Q$ be any wood such that
$\mathcal{M} \left( Q\right) =\mathcal{M} \left( P\right) $
holds and $P\subseteq Q$ holds, and let $B$ be any member
of $\left( Q\,\vdash P\right) $.   Then there exists a
\emph{unique} member $A$ of $\mathbb{B} \left( \bar{ P }
\right) $ such that $B\in \Xi \left( \mathcal{P} \left(
P,A\right) \right) $ holds, and the unique member $A$ of
$\mathbb{B} \left( \bar{ P } \right) $ with this property is
given by $A=\mathcal{Y} \left( \bar{ P } ,B\right) $.   For
$B$ is a subset of $\mathcal{Y} \left( \bar{ P } ,B\right)
$,
$B$ overlaps no member of $\mathcal{P} \left( P,\mathcal{Y}
\left( \bar{ P } ,B\right) \right) $, and $B$ is \emph{not}
a
strict subset of any member of $\mathcal{P} \left(
P,\mathcal{Y} \left( \bar{ P } ,B\right) \right) $, (for if
$B$ was a strict subset of a member $E$ of $\mathcal{P}
\left( P,\mathcal{Y} \left( \bar{ P } ,B\right) \right) $
then
$E$ would be a member of $ \bar{ P } $ such that $B\subseteq
E\subset \mathcal{Y} \left( \bar{ P } ,B\right) $ held,
which
contradicts the fact that $\mathcal{Y} \left( \bar{ P }
,B\right) $ is the \emph{smallest} member of $ \bar{ P } $
to
contain $B$ as a subset), hence $B$ is a member of $\Xi
\left( \mathcal{P} \left( P,\mathcal{Y} \left( \bar{ P }
,B\right) \right) \right) $.   Now let $A$ be any member of
$ \bar{ P } $ such that $B$ is a member of $\Xi \left(
\mathcal{P} \left( P,A\right) \right) $.   Then $B\subseteq
A$ holds, and each member of $\mathcal{P} \left( P,A\right)
$ is either a subset of $B$ or else does not intersect $B$,
and furthermore since $B$ is a member of $\left( Q\,\vdash
P\right) $ hence is \emph{not} a member of $\mathcal{P}
\left( P,A\right) $, each member of $\mathcal{P} \left(
P,A\right) $ is either a \emph{strict} subset of $B$ or
else does not intersect $B$.   Furthermore $B$ is
\emph{nonempty.}  Let $i$ be any member of $B$.   Then $i$
is a member of $A$, and $\mathcal{K} \left( P,A,i\right) $,
which by definition is the unique member of $\mathcal{P}
\left( P,A\right) $ that has $i$ as a member, is a member
of $\mathcal{P} \left( P,A\right) $ such that $B\cap
\mathcal{K} \left( P,A,i\right) $ is nonempty, hence, as
just shown, $\mathcal{K} \left( P,A,i\right) $ is a
\emph{strict} subset of $B$, hence the fact that
$\mathcal{K} \left( P,A,i\right) $ is a member of
$\mathcal{P} \left( P,A\right) $ implies that there is
\emph{no} member $C$ of $ \bar{ P } $ such that $B\subseteq
C\subset A$ holds, (for any such member $C$ of $ \bar{ P } $
would be a member of $P$ such that $\mathcal{K} \left(
P,A,i\right) \subset C\subset A$ held), hence $A$ is equal
to $\mathcal{Y} \left( \bar{ P } ,B\right) $.

\begin{bphzlemma} \label{Lemma 12}
\end{bphzlemma}
\vspace{-6.143ex}

\noindent \hspace{11.9ex}{\bf.  }Let $P$ be any
wood, let $F$ be any wood such
that $\mathcal{M} \left( F\right) =\mathcal{M} \left(
P\right) $ holds and $P\subseteq F$ holds, let $H$ be any
partition, let $A$ be any member of $\mathbb{B}
\left( \bar{ P }
 \right) $, let $B$ be any member of $\left( \Xi \left(
\mathcal{P} \left( P,A\right) \right) \,\vdash \mathcal{P}
\left( P,A\right) \right) $, let $\sigma $ be any real
number such that $0<\sigma \leq  \frac{ 1 }{ 8 } $ holds,
let $R$ be
any real number $>0$, and let $x$ be any member of
$\mathbb{F} _{ d } \left( \mathcal{M} \left( P\right)
\right) $, where $d$ is an integer $\geq 1$.
\enlargethispage{0.4ex}

Let $Z\equiv \Xi \left( \mathcal{P} \left( P,A\right)
\right) $.

Then the proposition $\mathbb{M} \left( P,F,H,B,\sigma
,R,x\right) $ is equivalent to the proposition \\
$\mathbb{M}
\left( P\cap Z,F\cap Z,H,B,\sigma ,R,\downarrow \left(
x,Z\right) \right) $.

\vspace{2.5ex}

\noindent {\bf Proof.}  We first note that by
observation (\ref{First Series Obs 4})
 on
page \pageref{Start of original page 49},
both $P\cap Z$ and $F\cap Z$ are woods of
$\mathcal{P} \left( P,A\right) $, and $B$ is a member of
\label{Start of original page 51}
 $\left( \Xi \left( \mathcal{M} \left( P\cap Z\right)
\right) \,\vdash \mathcal{M} \left( P\cap Z\right) \right)
$, and we note furthermore that it follows directly from
the definition of $\mathbb{F} _{ d } \left( \mathcal{M}
\left( P\right) \right) $ that $\downarrow \left(
x,Z\right) =\downarrow \left( x,\Xi \left( \mathcal{P}
\left( P,A\right) \right) \right) $ is a member of
$\mathbb{F} _{ d } \left( \mathcal{P} \left( P,A\right)
\right) =\mathbb{F} _{ d } \left( \mathcal{M} \left( P\cap
Z\right) \right) $, as is required for the proposition
$\mathbb{M} \left( P\cap Z,F\cap Z,H,B,\sigma ,R,\downarrow
\left( x,Z\right) \right) $ to be defined.

We next note that $\mathcal{P} \left( F,B\right) $ is a
subset of $Z=\Xi \left( \mathcal{P} \left( P,A\right)
\right) $, for by assumption $B$ is a member of $\left( \Xi
\left( \mathcal{P} \left( P,A\right) \right) \,\vdash
\mathcal{P} \left( P,A\right) \right) $, hence if $C$ is
any member of $\mathcal{P} \left( P,A\right) $ then either
$C\cap B$ is empty or else $C$ is a \emph{strict} subset of
$B$.   Let $D$ be any member of $\mathcal{P} \left(
F,B\right) $ and let $C$ be any member of $\mathcal{P}
\left( P,A\right) $ such that $C\cap D$ is nonempty.   Then
$C$ is a member of $\mathcal{P} \left( P,A\right) $ such
that $C\cap B$ is nonempty, hence $C\subset B$ holds, hence
$D\subset C$ cannot hold, (for if $D\subset C$ held then
$C$ would be a member of $P$, hence a member of $F$, such
that $D\subset C\subset B$ held, contradicting the fact
that $D$ is a member of $\mathcal{P} \left( F,B\right)
 $), hence since $C$ does not overlap $D$,
$C\subseteq D$ holds.   Hence if $D$ is any member of
$\mathcal{P} \left( F,B\right) $ then $D$ is a nonempty
subset of $B$, hence a nonempty subset of $A$, such that
each member of $\mathcal{P} \left( P,A\right) $ is either a
subset of $D$ or else does not intersect $D$, hence $D$ is
a member of $\Xi \left( \mathcal{P} \left( P,A\right)
\right) $, hence $\mathcal{P} \left( F,B\right) $ is a
subset of $\Xi \left( \mathcal{P} \left( P,A\right) \right)
$

Hence $\mathcal{P} \left( F,B\right) $ is a subset of
$F\cap \Xi \left( \mathcal{P} \left( P,A\right) \right)
=F\cap Z$, hence by observation
\ref{First Series Obs 1}) on
page \pageref{Start of original page 48},
$\mathcal{P} \left( F\cap Z,B\right) =\mathcal{P}
\left( F,B\right) $ holds, and it immediately follows from
this, by the definition of $\mathbb{L} \left( F,B,x\right)
$, that $\mathbb{L} \left( F\cap Z,B,\downarrow \left(
x,Z\right) \right) =\mathbb{L} \left( F,B,x\right) $ holds.

And it immediately follows from this, in particular, that
$\mathbb{L} \left( F\cap Z,B,\downarrow \left( x,Z\right)
\right) <R$ holds ifif $\mathbb{L} \left( F,B,x\right) <R$
holds.

We next note that $\mathcal{Y} \left( \bar{ P } ,B\right) $
is
equal to $A$, for $B$ is a \emph{nonempty} subset of $A$
such that each member of $\mathcal{P} \left( P,A\right) $
is either a \emph{strict} subset of $B$ or else does not
intersect $B$, hence there can be no member $C$ of
$\bar{ P } $
 such that $B\subseteq C\subset A$ holds, (for if $C$
was such a member of $ \bar{ P } $ then $C$ would be a
member
of $P$, and if $i$ was any member of $B$, then $\mathcal{K}
\left( P,A,i\right) \subset C\subset A$ would hold,
contradicting the fact that $\mathcal{K} \left(
P,A,i\right) $ is a member of $\mathcal{P} \left(
P,A\right) $), hence $A$ is the \emph{smallest}
member of $ \bar{ P } $ to contain $B$ as a subset.

Now $\mathcal{U} \left( P\cap Z\right) =\mathcal{U} \left(
P\cap \Xi \left( \mathcal{P} \left( P,A\right) \right)
\right) $ is equal to $A$, hence $A=\mathcal{Y}
\left( \bar{ P }
 ,B\right) $ is certainly a member of
 $\overline{ \left( P\cap
Z\right) } $, hence by
observation (\ref{First Series Obs 2})
on
page \pageref{Start of original page 48},
$\mathcal{Y} \left(
\overline{ \left( P\cap Z\right) }
 ,B\right) =\mathcal{Y} \left( \bar{ P } ,B\right) =A$
holds.

Now $H$ is a \emph{partition,} hence as shown on
pages \pageref{Start of original page 16}
and \pageref{Start of original page 17},
if $i$ is any member of $B$ and $j$ is any member
of $\left( A\,\vdash B\right) $ such that $\left\{
i,j\right\} $ is a member of $H$, then both $i$ and $j$ are
members of $\mathcal{O} \left( F,H\right) $, and
\label{Start of original page 52}
 furthermore $\mathcal{Z} \left( F,H,i\right) $ is the
largest member of $F$ that has $i$ as a member but does
\emph{not} have $j$ as a member, and $\mathcal{Z} \left(
F,H,j\right) $ is the largest member of $F$ that has $j$ as
a member but does \emph{not} have $i$ as a member.

Now let $i$ be any member of $B$ and $j$ be any member of
$\left( A\,\vdash B\right) $ such that $\left\{ i,j\right\}
$
is a member of $H$.   Then $\left\{ i,j\right\} $ is a
subset
of $A$ hence $\mathcal{Z} \left( F,H,i\right) \subset A$
holds and $\mathcal{Z} \left( F,H,j\right) \subset A$
holds.   And $\mathcal{K} \left( P,A,i\right) $ is a member
of $\mathcal{P} \left( P,A\right) $ that intersects $B$
hence is a strict subset of $B$, hence $j$ is not a member
of $\mathcal{K} \left( P,A,i\right) $, hence since
$\mathcal{K} \left( P,A,i\right) $ is a member of $F$,
$\mathcal{K} \left( P,A,i\right) \subseteq \mathcal{Z}
\left( F,H,i\right) $ holds, hence $\mathcal{Z} \left(
F,H,i\right) $ cannot be a strict subset of any member of
$\mathcal{P} \left( P,A\right) $, (for if $\mathcal{Z}
\left( F,H,i\right) $ was a strict subset of a member $C$
of $\mathcal{P} \left( P,A\right) $, then $C$ would be a
member of $P$ such that $\mathcal{K} \left( P,A,i\right)
\subset C\subset A$ held, which contradicts the fact that
$\mathcal{K} \left( P,A,i\right) $ is a member of
$\mathcal{P} \left( P,A\right) $), hence since
$\mathcal{Z} \left( F,H,i\right) $ overlaps no member of
$\mathcal{P} \left( P,A\right) $, $\mathcal{Z} \left(
F,H,i\right) $ is a member of $Z=\Xi \left( \mathcal{P}
\left( P,A\right) \right) $, hence $\mathcal{Z} \left(
F,H,i\right) $ is a member of $F\cap Z$, hence by
observation (\ref{First Series Obs 3}) on
page \pageref{Start of original page 49}, $i$ is a
member of $\mathcal{O} \left( F\cap Z,H\right) $ and
$\mathcal{Z} \left( F\cap Z,H,i\right) =\mathcal{Z} \left(
F,H,i\right) $ holds.   And $j$ is \emph{not} a member of
$B$, hence $\mathcal{K} \left( P,A,j\right) $ is \emph{not}
a subset of $B$, hence $\mathcal{K} \left( P,A,j\right) $
does \emph{not} intersect $B$, hence $i$ is \emph{not} a
member of $\mathcal{K} \left( P,A,j\right) $, hence
$\mathcal{K} \left( P,A,j\right) \subseteq \mathcal{Z}
\left( F,H,j\right) $ holds, hence $\mathcal{Z} \left(
F,H,j\right) $ cannot be a strict subset of any member of
$\mathcal{P} \left( P,A\right) $, hence $\mathcal{Z} \left(
F,H,j\right) $ is a member of $Z=\Xi \left( \mathcal{P}
\left( P,A\right) \right) $, hence $\mathcal{Z} \left(
F,H,j\right) $ is a member of $F\cap Z$, hence by
observation (\ref{First Series Obs 3}) on
page \pageref{Start of original page 49}, $j$ is a
member of $\mathcal{O} \left( F\cap Z,H\right) $ and
$\mathcal{Z} \left( F\cap Z,H,j\right) =\mathcal{Z} \left(
F,H,j\right) $ holds.

Hence $\left| x_{ \mathcal{Z} \left( F\cap Z,H,i\right) }
-x_{
\mathcal{Z} \left( F\cap Z,H,j\right) } \right| =\left| x_{
\mathcal{Z}
\left( F,H,i\right) } -x_{ \mathcal{Z} \left( F,H,j\right)
 } \right| $ holds, hence \\
 $\mathbb{L} \left( F\cap
Z,B,\downarrow
\left( x,Z\right) \right) <\sigma \left| x_{ \mathcal{Z}
\left(
F\cap Z,H,i\right) } -x_{ \mathcal{Z} \left( F\cap
Z,H,j\right) } \right| $ holds ifif \\
$\mathbb{L} \left(
F,B,x\right) < \sigma \left| x_{ \mathcal{Z} \left(
F,H,i\right)
} -x_{ \mathcal{Z} \left( F,H,j\right) } \right| $ holds.

\begin{bphzlemma} \label{Lemma 13}
\end{bphzlemma}
\vspace{-6.143ex}

\noindent \hspace{11.9ex}{\bf.  }Let $P$ be any
wood, let $Q$ be any wood such
that $\mathcal{M} \left( Q\right) =\mathcal{M} \left(
P\right) $ holds and $P\subseteq Q$ holds, let $H$ be any
partition such that every member of $Q$ is $\left(
\mathcal{M} \left( P\right) \cup H\right) $-connected and
such that if $E$ is any member of $H$ such that $E$
intersects \emph{more} than one member of $\mathcal{M}
\left( P\right) $, then $E$ has \emph{exactly} two members,
let $\sigma $ be any real number such that $0<\sigma \leq
\frac{ 3 }{ 25 } $ holds, let $R$ be any finite real number
$>0$,
let $d$ be any integer $\geq 1$, and let $x$ be any member
of $\mathbb{F} _{ d } \left( V\right) $.

For each member $A$ of $\mathbb{B} \left( \bar{ P } \right)
$
we define $Z_{ A } \equiv \Xi \left( \mathcal{P} \left(
P,A\right) \right) $.
\enlargethispage{3.0ex}

Then
\label{Start of original page 53}
\[
\mathcal{L} \left( P,Q,H,\sigma ,R,x\right) =  \prod_{A\in
\mathbb{B} \left( \bar{ P } \right) } \mathcal{L}
\left( P\cap Z_{ A } ,Q\cap Z_{ A } ,H,\sigma ,R,\downarrow
\left( x,Z_{ A } \right) \right)
\]
holds and
\[
\mathcal{H} \left( P,Q,H,\sigma ,R,x\right) =   \prod_{A\in
\mathbb{B} \left( \bar{ P } \right) } \mathcal{H}
\left( P\cap Z_{ A } ,Q\cap Z_{ A } ,H,\sigma ,R,\downarrow
\left( x,Z_{ A } \right) \right)
\]
holds.

\vspace{2.5ex}

\noindent {\bf Proof.}  We first note that,
by definition, $\mathcal{L}
\left( P,Q,H,\sigma ,R,x\right) $ is equal to $1$ if for
every member $B$ of $\left( Q\,\vdash P\right) $, there
exists a member $F$ of $\mathbb{K} \left( P,Q\right) $ such
that \\
$\mathbb{M} \left( P,F,H,B,\sigma ,R,x\right) $ holds,
and equal to $0$ otherwise, hence by
observation (\ref{First Series Obs 5}) on
page \pageref{Start of original page 50},
$\mathcal{L} \left( P,Q,H,\sigma
,R,x\right) $ is equal to $1$ if for every member $A$ of
$\mathbb{B} \left( \bar{ P } \right) $, and for every member
$B$ of $\left( Q\,\vdash P\right) \cap \Xi \left(
\mathcal{P} \left( P,A\right) \right) =\left( Q\,\vdash
P\right) \cap Z_{ A } $, there exists a member $F$ of
$\mathbb{K} \left( P,Q\right) $ such that $\mathbb{M}
\left( P,F,H,B,\sigma ,R,x\right) $ holds, and equal to $0$
otherwise, hence by Lemma \ref{Lemma 12}, $\mathcal{L}
\left(
P,Q,H,\sigma ,R,x\right) $ is equal to $1$ if for every
member $A$ of $\mathbb{B} \left( \bar{ P } \right) $, and
for
every member $B$ of $\left( Q\,\vdash P\right) \cap Z_{ A }
$, there exists a member $F$ of $\mathbb{K} \left(
P,Q\right) $ such that $\mathbb{M} \left( P\cap Z_{ A }
,F\cap Z_{ A } ,H,B,\sigma ,R,\downarrow \left( x,Z_{ A }
\right) \right) $ holds, and equal to $0$ otherwise.

Now let $A$ be any member of $\mathbb{B} \left( \bar{ P }
\right) $.

Then by observation
(\ref{First Series Obs 4}) on
page \pageref{Start of original page 49}, both
$P\cap Z_{ A } $ and $Q\cap Z_{ A } $ are woods of
$\mathcal{P} \left( P,A\right) $, and furthermore, every
member of $Q\cap Z_{ A } $ is $\left( \mathcal{P} \left(
P,A\right) \cup H\right) $-connected, for if $C$ is any
member of $Q\cap Z_{ A } $, then $C$ is a member of $Q$
hence is $\left( \mathcal{M} \left( P\right) \cup H\right)
$-connected, hence if $\left\{ J,K\right\} $ is any
partition
of $C$ into two nonempty parts then there exists a member
$D$ of $\left( \mathcal{M} \left( P\right) \cup H\right) $
such that $D$ intersects both $J$ and $K$, and if $D$ is a
member of $H$ then $D$ is a member of $\left( \mathcal{P}
\left( P,A\right) \cup H\right) $, while if $D$ is a member
of $\mathcal{M} \left( P\right) $ then $D$ is a member of
$P$ such that $D\subset A$ holds, hence as shown on
page \pageref{Start of original page 5}
there is a unique member $E$ of $\mathcal{P} \left(
P,A\right) $ such that $D\subseteq E$ holds, and this
member $E$ of $\mathcal{P} \left( P,A\right) $ is a member
of $\left( \mathcal{P} \left( P,A\right) \cup H\right) $
such that $E$ intersects both $J$ and $K$.

Hence, by definition, $\mathcal{L} \left( P\cap Z_{ A }
,Q\cap Z_{ A } ,H,\sigma ,R,\downarrow \left( x,Z_{ A }
\right) \right) $ is equal to $1$ if for every member $B$
of $\left( Q\cap Z_{ A } \right) \,\vdash \left( P\cap Z_{
A } \right) =\left( Q\,\vdash P\right) \cap Z_{ A } $,
there exists a member $G$ of $\mathbb{K} \left( P\cap Z_{ A
} ,Q\cap Z_{ A } \right) $ such that $\mathbb{M} \left(
P\cap Z_{ A } ,G,H,B,\sigma ,R,\downarrow \left( x,Z_{ A }
\right) \right) $ holds, and equal to $0$ otherwise.

Suppose now that $B$ is a member of $\left( Q\,\vdash
P\right) \cap Z_{ A } $ such that there exists a member $G$
of $\mathbb{K} \left( P\cap Z_{ A } ,Q\cap Z_{ A } \right)
$ such that $\mathbb{M} \left( P\cap Z_{ A } ,G,H,B,\sigma
,R,\downarrow \left( x,Z_{ A } \right) \right) $ holds, and
let $G$ be such a member of $\mathbb{K} \left( P\cap Z_{ A
} ,Q\cap Z_{ A } \right) $.   Then $F\equiv P\cup G$ is a
member of $\mathbb{K} \left( P,Q\right) $ such that $F\cap
Z_{ A } =\left( P\cup G\right) \cap Z_{ A } =\left( P\cap
Z_{ A } \right) \cup \left( G\cap Z_{ A } \right) =\left(
P\cap Z_{ A } \right) \cup G=G $ holds,
\label{Start of original page 54}
 hence $\mathbb{M} \left( P\cap Z_{ A } ,F\cap Z_{ A }
,H,B,\sigma ,R,\downarrow \left( x,Z_{ A } \right) \right)
$ holds.

Now suppose that $B$ is a member of $\left( Q\,\vdash
P\right) \cap Z_{ A } $ such that there exists a member $F$
of $\mathbb{K} \left( P,Q\right) $ such that $\mathbb{M}
\left( P\cap Z_{ A } ,Q\cap Z_{ A } ,H,B,\sigma
,R,\downarrow \left( x,Z_{ A } \right) \right) $ holds.
Then $G\equiv F\cap Z_{ A } $ is a member of $\mathbb{K}
\left( P\cap Z_{ A } ,Q\cap Z_{ A } \right) $ such that
$\mathbb{M} \left( P\cap Z_{ A } ,G,H,B,\sigma
,R,\downarrow \left( x,Z_{ A } \right) \right) $ holds.

Hence $\mathcal{L} \left( P\cap Z_{ A } ,Q\cap Z_{ A }
,H,\sigma ,R,\downarrow \left( x,Z_{ A } \right) \right) $
is equal to $1$ if for every member $B$ of $\left(
Q\,\vdash P\right) \cap Z_{ A } $, there exists a member
$F$ of $\mathbb{K} \left( P,Q\right) $ such that \\
$\mathbb{M} \left( P\cap Z_{ A } ,F\cap Z_{ A } ,H,B,\sigma
,R,\downarrow \left( x,Z_{ A } \right) \right) $ holds, and
equal to $0$ otherwise.

And this is true for every member $A$ of $\mathbb{B} \left(
\bar{ P } \right) $, hence
\[
\mathcal{L} \left( P,Q,H,\sigma ,R,x\right) =  \prod_{A\in
\mathbb{B} \left( \bar{ P } \right) } \mathcal{L}
\left( P\cap Z_{ A } ,Q\cap Z_{ A } ,H,\sigma ,R,\downarrow
\left( x,Z_{ A } \right) \right)
\]
holds.

Now $\mathcal{H} \left( P,Q,H,\sigma ,R,x\right) $ is equal
to
\[
\mathcal{L} \left( P,Q,H,\sigma ,R,x\right)\prod_{  F\in
\tilde{ \mathcal{G} } \left( Q,H\right)} \left(
1-\mathcal{L}
\left( P,F,H,\sigma ,R,x\right) \right) ,
\]
where $\tilde{ \mathcal{G} } \left( Q,H\right) $ was
defined on
page \pageref{Start of original page 47}
 to be the set whose members are all the members
$F$ of $\mathcal{G} \left( \mathcal{M} \left( Q\right)
,H\right) $ such that $Q\subset F$ holds.

Suppose $\mathcal{H} \left( P,Q,H,\sigma ,R,x\right) $ is
equal to $1$.   Then $\mathcal{L} \left( P,Q,H,\sigma
,R,x\right) $ is equal to $1$, hence by the result just
obtained, $\mathcal{L} \left( P\cap Z_{ A } ,Q\cap Z_{ A }
,H,\sigma ,R,\downarrow \left( x,Z_{ A } \right) \right) $
is equal to $1$ for every member $A$ of $\mathbb{B} \left(
\bar{ P } \right) $, and for every member $F$ of $
\tilde{\mathcal{G} } \left( Q,H\right) $, $\mathcal{L}
\left( P,F,H,\sigma
,R,x\right) $ is equal to $0$, hence in particular, for
every member $B$ of $\mathbb{B} \left( \bar{ P } \right) $,
and for every member $G$ of $\tilde{ \mathcal{G} }
\left( Q\cap
Z_{ B } ,H\right) $, $\mathcal{L} \left( P,Q\cup G,H,\sigma
,R,x\right) $ is equal to $0$   Now if $B$ is any member of
$\mathbb{B} \left( \bar{ P } \right) $, $G$ is any member of
$\tilde{ \mathcal{G} } \left( Q\cap Z_{ B } ,H\right) $, and
$A$ is any member of $\mathbb{B} \left( \bar{ P } \right) $,
then $\left( Q\cup G\right) \cap Z_{ A } $ is equal to $G$
if $A$ is equal to $B$, and equal to $Q\cap Z_{ A } $ if
$A$ is \emph{not} equal to $B$, hence by the result just
obtained, $\mathcal{L} \left( P,Q\cup G,H,\sigma
,R,x\right) $ is equal to
\[
\mathcal{L} \left( P\cap Z_{ B }
,G,H,\sigma ,R,\downarrow \left( x,Z_{ B } \right) \right)
\prod
_{ A\in \left( \mathbb{B} \left( \bar{ P } \right) \,\vdash
\left\{ B\right\} \right) } \mathcal{L} \left( P\cap Z_{ A
}
,Q\cap Z_{ A } ,H,\sigma ,R,\downarrow \left( x,Z_{ A }
\right) \right),
\]
hence since the fact that $\mathcal{L} \left( P,Q,H,\sigma
,R,x\right) $ is equal to $1$ implies, as just shown, that
$\mathcal{L} \left( P\cap Z_{ A } ,Q\cap Z_{ A } ,H,\sigma
,R,\downarrow \left( x,Z_{ A } \right) \right) $ is equal
to $1$ for every member $A$ of $\mathbb{B} \left( \bar{ P }
\right) $, the fact that $\mathcal{L} \left( P,Q\cup
G,H,\sigma ,R,x\right) $ is equal to $0$ implies that \\
$\mathcal{L} \left( P\cap Z_{ B } ,G,H,\sigma ,R,\downarrow
\left( x,Z_{ B } \right) \right) $ is equal to $0$.   And
this is true for \emph{every} member $G$ of
 $\tilde { \mathcal{G} }
 \left( Q\cap Z_{ B } ,H\right) $, hence since, as just
shown, the fact that $\mathcal{L} \left( P,Q,H,\sigma
,R,x\right) $ is equal to $1$ implies that
\label{Start of original page 55}
$ \mathcal{L} \left( P\cap Z_{ B } ,Q\cap Z_{ B } ,H,\sigma
,R,\downarrow \left( x,Z_{ B } \right) \right) $ is equal
to $1$, it follows that
\[
\mathcal{H} \left( P\cap Z_{ B }
,Q\cap Z_{ B } ,H,\sigma ,R,\downarrow \left( x,Z_{ B }
\right) \right) = \hspace{5.0cm}
\]
\[
=\mathcal{L} \left( P\cap Z_{ B } ,Q\cap Z_{ B }
,H,\sigma ,R,\downarrow \left( x,Z_{ B } \right) \right)
\hspace{-0.6cm}
\prod_{
G\in \tilde{ \mathcal{G} } \left( Q\cap Z_{ B } ,H\right) }
\hspace{-0.6cm}
\left( 1-\mathcal{L} \left( P\cap Z_{ B } ,G,H,\sigma
,R,\downarrow \left( x,Z_{ B } \right) \right) \right)
\]
 is
equal to $1$.   And this is true for \emph{every} member
$B$ of $\mathbb{B} \left( \bar{ P } \right) $, hence the
assumption that $\mathcal{H} \left( P,Q,H,\sigma
,R,x\right) $ is equal to $1$ implies that \\
$   \prod_{B\in
\mathbb{B} \left( \bar{ P } \right) } \mathcal{H}
\left( P\cap Z_{ B } ,Q\cap Z_{ B } ,H,\sigma ,R,\downarrow
\left( x,Z_{ B } \right) \right) $ is equal to $1$.

Now assume that $
 \prod_{B\in \mathbb{B} \left( \bar{ P } \right) }
\mathcal{H}
 \left( P\cap Z_{ B } ,Q\cap Z_{ B }
,H,\sigma ,R,\downarrow \left( x,Z_{ B } \right) \right) $
is equal to $1$, \\
hence that $\mathcal{L} \left(
P,Q,H,\sigma ,R,x\right) =   \prod_{B\in \mathbb{B}
\left( \bar{ P }
\right) } \mathcal{L} \left( P\cap Z_{ B }
,Q\cap Z_{ B } ,H,\sigma ,R,\downarrow \left( x,Z_{ B }
\right) \right) $ is \\
equal to $1$, and that for every
member $B$ of $\mathbb{B} \left( \bar{ P } \right) $, and
for
every member $G$ of $\tilde{ \mathcal{G} } \left( Q\cap Z_{
B }
,H\right) $, $\mathcal{L} \left( P\cap Z_{ B } ,G,H,\sigma
,R,\downarrow \left( x,Z_{ B } \right) \right) $ is equal
to $0$.   Let $F$ be any member of $\tilde{ \mathcal{G} }
\left( Q,H\right) $.   Then by the first part of this
Lemma, $\mathcal{L} \left( P,F,H,\sigma ,R,x\right) $ is
equal to $
\prod_{B\in \mathbb{B} \left( \bar{ P } \right) }
\mathcal{L}
\left( P\cap Z_{ B } ,F\cap Z_{ B }
,H,\sigma ,R,\downarrow \left( x,Z_{ B } \right) \right) $,
and by the fact that $F$ is a member of $\tilde{
\mathcal{G} }
\left( Q,H\right) $, or in other words, that $F$ is a
member of $\mathcal{G} \left( \mathcal{M} \left( Q\right)
,H\right) $ such that $Q\subset F$ holds, there exists at
least one member $B$ of $\mathbb{B} \left( \bar{ P }
\right) $
such that $Q\cap Z_{ B } \subset F\cap Z_{ B } $ holds,
hence such that $F\cap Z_{ B } $ is a member of
$\tilde{ \mathcal{G} } \left( Q\cap Z_{ B } ,H\right) $,
hence
such that $\mathcal{L} \left( P\cap Z_{ B } ,F\cap Z_{ B }
,H,\sigma ,R,\downarrow \left( x,Z_{ B } \right) \right) $
is equal to $0$.   Hence $\mathcal{L} \left( P,F,H,\sigma
,R,x\right) $ is equal to $0$.   And this is true for
\emph{every} member $F$ of $\tilde{ \mathcal{G} } \left(
Q,H\right) $, hence $\mathcal{H} \left( P,Q,H,\sigma
,R,x\right) $ is equal to $1$.

Hence $\mathcal{H} \left( P,Q,H,\sigma ,R,x\right) $ is
equal to $1$ if and only if \\
$\prod_{ A\in \mathbb{B}
\left( \bar{ P }
 \right) } \mathcal{H} \left( P\cap Z_{ A } ,Q\cap Z_{ A
} ,H,\sigma ,R,\downarrow \left( x,Z_{ A } \right) \right)
$ is equal to $1$, hence, since each of these expressions
can take \emph{only} the values $0$ and $1$, they are equal.

\begin{bphzlemma} \label{Lemma 14}
\end{bphzlemma}
\vspace{-6.143ex}

\noindent \hspace{11.9ex}{\bf.  }Let $V$ be any
partition such that $\mathcal{U}
\left( V\right) $ is finite and $\#\left( V\right) \geq 2$
holds, and let $H$ be any partition such that if $E$ is any
member of $H$ such that $E$ intersects \emph{more} than one
member of $V$, then $E$ has \emph{exactly} two members.

Let $\sigma $ be any real number such that $0<\sigma \leq
\frac{ 1 }{ 8 } $ holds, and let $\lambda $ be the real
number defined
by $\lambda \equiv  \left( \frac{ 1 }{ 4 } \right)
 \left( 1- \sqrt{ 1-8\sigma } \right)
$, so that
\label{Start of original page 56}
 $0<\lambda \leq  \frac{ 1 }{ 4 }$ holds.

We note that $\lambda $ and $\sigma $ satisfy the equation
$\lambda = \frac{ \sigma }{ 1-2\lambda } $,
and that $0 < \sigma < \lambda $ holds.

Let $R$ be any finite real number $>0$, let $d$ be any
integer $\geq 1$, let $x$ be any member of $\mathbb{F} _{ d
} \left( V\right) $, and let $\left(P,Q\right) $ be any
member of $\Omega
\left( H,\sigma ,R,x\right) $.

Let $i$ and $j$ be any two members of $\mathcal{U} \left(
V\right) $ such that $\left\{ i,j\right\} $ is a member of
$H$ and is \emph{not} a subset of any member of $V$, let
$J$ be the set whose members are all the $A\in Q$ such that
$\mathcal{Z} \left( P,H,i\right) \subseteq A\subseteq
\mathcal{Z} \left( Q,H,i\right) $ holds, and let $K$ be the
set whose members are all the $B\in Q$ such that
$\mathcal{Z} \left( P,H,j\right) \subseteq B\subseteq
\mathcal{Z} \left( Q,H,j\right) $ holds.
\enlargethispage{1.15ex}

Let $u$ be any member of $\mathbb{R}^{ J }$ such that $u_{
A } \geq 0$ holds for all $A\in J$, and such that $ \sum_{
A\in J } u_{
A } =1$ holds, and let $v$ be any member of $\mathbb{R}^{
K } $ such that $v_{ B } \geq 0$ holds for all $B\in K$, and
such that $ \sum_{ B\in K } v_{ B } =1$ holds.

Then
\[
\left| x_{ \mathcal{Z} \left( P,H,i\right) } -x_{
\mathcal{Z}
\left( P,H,j\right) } \right| \leq \left( \frac{ 1 }{
1-2\lambda } \right)
\left| \left( \sum_{A\in
J } u_{ A } x_{ A } \right)
 - \left( \sum_{B\in K } v_{ B } x_{ B } \right) \right|
\]
holds and
\[
\left| \left( \sum_{A\in J } u_{ A } x_{ A } \right)
 - \left( \sum_{B\in K } v_{ B
} x_{
B } \right) \right| \leq
\left( \frac{ 1 }{ 1-2\lambda } \right)
\left| x_{ \mathcal{Z} \left(
P,H,i\right) } -x_{ \mathcal{Z} \left( P,H,j\right) }
\right|
\]
holds.

\vspace{2.5ex}

\noindent {\bf Proof.}  We first obtain the
stated lower bound on
$\left| \left( \sum_{A\in
J } u_{ A } x_{ A } \right) - \left( \sum_{B\in K }
 v_{ B } x_{ B } \right)
\right| $.   In fact we will prove, defining $T\equiv
\hspace{1.5ex} \max_{ \hspace{-7.0ex}
\begin{array}{c} \\[-3.1ex]
\scriptstyle{ F\in
\mathbb{K} \left( P,Q\right) }
\end{array} } \hspace{-1.2ex}
\left| x_{ \mathcal{Z} \left(
F,H,i\right) } -x_{ \mathcal{Z} \left( F,H,j\right) }
\right| $, that \\
$T\leq \left( \frac{ 1 }{ 1-2\lambda } \right)  \left|
\left( \sum_{A\in J } u_{ A }
x_{ A } \right)
 -\left( \sum_{B\in K } v_{ B } x_{ B } \right) \right| $
  holds.

For, using the triangle inequality, we have, for arbitrary
$F\in \mathbb{K} \left( P,Q\right) $, that
\[
\left| x_{ \mathcal{Z} \left( F,H,i\right) } -x_{
\mathcal{Z}
\left( F,H,j\right) } \right| \leq \left\{
\left| x_{ \mathcal{Z}
\left(
F,H,i\right) } -
\left( \sum_{A\in J } u_{ A } x_{ A } \right) \right| +
\right. \hspace{-2.3pt} \hspace{6.0cm}
\]
\[
\hspace{4.0cm} \left.
+\left| \left( \sum_{A\in J } u_{ A } x_{ A } \right)
 - \left( \sum_{B\in K } v_{
B }
x_{ B } \right) \right|
+\left| \left( \sum_{B\in K } v_{ B } x_{ B } \right) -x_{
\mathcal{Z}
\left( F,H,j\right) } \right| \right\}
\]
holds.
\label{Start of original page 57}

 Now we note that $J$ is the set of all the possible
$\mathcal{Z} \left( G,H,i\right) $ for $G\in \mathbb{K}
\left( P,Q\right) $.   Hence, from Lemma \ref{Lemma 6} (d),
 we have that
\[
\left| x_{ \mathcal{Z} \left( F,H,i\right) }
- \left( \sum_{A \in J
} u_{
A } x_{ A } \right) \right| \leq
 \sum_{A\in J } u_{ A } \left| x_{
\mathcal{Z}
\left( F,H,i\right) } -x_{ A } \right| \leq
\sum_{A\in J } u_{ A } \lambda T=\lambda T
\]
holds, (where we used that $ \sum_{ A\in J } u_{ A } =1$,
and that all $u_{ A
} $ are $\geq 0 $),  and similarly we have that
$\left| \left( \sum_{B\in K } v_{ B } x_{ B } \right)
- x_{ \mathcal{Z} \left(
F,H,j\right) } \right| \leq \lambda T$ holds.

Hence
\[
\left| x_{ \mathcal{Z} \left( F,H,i\right) } -x_{
\mathcal{Z}
\left( F,H,j\right) } \right| \leq 2\lambda T+\left|
\left( \sum_{A\in J } u_{
A } x_{ A } \right)
- \left( \sum_{B\in K } v_{ B } x_{ B } \right) \right|
\]
holds.   And this is true for all $F\in \mathbb{K} \left(
P,Q\right) $, hence \\
$T\leq 2\lambda T+  \left| \left( \sum_{A\in J } u_{A }
x_{A } \right)
- \left( \sum_{B\in K } v_{B }x_{B } \right) \right|  $
holds, hence \\
$T\leq \left( \frac{ 1 }{ 1-2\lambda } \right) \left|
\left( \sum_{A\in J } u_{A } x_{A } \right)
- \left( \sum_{B\in K } v_{B } x_{B } \right) \right|  $
holds.

But $\left| x_{ \mathcal{Z} \left( P,H,i\right) } -x_{
\mathcal{Z} \left( P,H,j\right) } \right| \leq T$ holds,
hence
\[
\left| x_{ \mathcal{Z} \left( P,H,i\right) } -x_{
\mathcal{Z}
\left( P,H,j\right) } \right| \leq \left( \frac{ 1 }{
1-2\lambda } \right)
\left| \left( \sum_{A\in
J } u_{ A } x_{ A } \right)
 - \left( \sum_{B\in K } v_{ B } x_{ B } \right) \right|
\]
holds.

We now obtain the stated upper bound on $\left| \left(
 \sum_{A\in
J }
u_{ A } x_{ A } \right)
 - \left( \sum_{B\in K } v_{ B } x_{ B } \right) \right| $.
 In
fact
\[
\left| \left( \sum_{A\in J } u_{ A } x_{ A } \right)
 - \left( \sum_{B\in K } v_{ B
} x_{ B } \right) \right| =\left| \sum_{
\begin{array}{c} \\[-4.75ex]
\scriptstyle{ A\in J } \\[-1.75ex]
\scriptstyle{ B\in K }
\end{array} } \hspace{-0.9ex}
u_{ A } v_{ B } \left( x_{ A } -x_{ B }
\right) \right| \leq \hspace{-0.3cm} \sum_{
\begin{array}{c} \\[-4.75ex]
\scriptstyle{ A\in J } \\[-1.75ex]
\scriptstyle{ B\in K }
\end{array} } \hspace{-0.9ex}
u_{ A } v_{ B } \left| x_{ A } -x_{
B } \right|
\]
\[
\leq \sum_{
\begin{array}{c} \\[-4.75ex]
\scriptstyle{ A\in J } \\[-1.75ex]
\scriptstyle{ B\in K }
\end{array} } \hspace{-0.9ex}
u_{ A } v_{ B }
\hspace{0.5ex} \textstyle \max_{ \hspace{-4.7ex}
\begin{array}{c} \\[-3.1ex]
\scriptstyle{ C\in J } \\[-1.2ex]
\scriptstyle{ D\in K }
\end{array} } \hspace{-0.8ex}
\left| x_{ C } -x_{ D } \right| \leq T .
\]

Furthermore, by Lemma \ref{Lemma 6} (e), $T\leq
\left( \frac{ 1 }{
1-2\lambda } \right) \left| x_{ \mathcal{Z} \left(
P,H,i\right) }
-x_{ \mathcal{Z} \left( P,H,j\right) } \right| $ holds,
hence \\
$\left| \left( \sum_{A\in J} u_{A } x_{A } \right)
 - \left( \sum_{B\in K} v_{B }
x_{B } \right) \right| \leq \left( \frac{ 1 }{
1-2\lambda } \right) \left| x_{ \mathcal{Z} (P,H,i) } -x_{
\mathcal{Z}
(P,H,j) } \right|  $
 holds.
\label{Start of original page 58}

\begin{bphzlemma} \label{Lemma 15}
\end{bphzlemma}
\vspace{-6.143ex}

\noindent \hspace{11.9ex}{\bf.  }Let $H$ be a set
such that every member of
$H$ is a set, let $n$ be an integer $\geq 1$, let $V$ be a
partition such that $\mathcal{U} \left( V\right) $ is
finite, $\mathcal{U} \left( V\right) $ is $\left( V\cup
H\right) $-connected, and $\#\left( V\right) =n$, and let
$A$ and $B$ be any members of $V$.   Then there exists an
integer $m$ such that $1\leq m\leq n$ holds, and a map $M$
such that $\mathcal{D} \left( M\right) $ is the set of all
the integers $p$ such that $1\leq p\leq m$ holds,
$\mathcal{R} \left( M\right) \subseteq V$ holds, $M_{ 1 }
=A$ holds, $M_{ m } =B$ holds, and such that if $p$ is any
integer such that $1\leq p$ and $p\leq \left( m-1\right) $
both hold, then there exists a member $C$ of $H$ such that
$C\cap M_{ p } \neq \emptyset $ and $C\cap M_{ p+1 } \neq
\emptyset $ both hold.
\enlargethispage{1.8ex}

\vspace{2.5ex}

\noindent {\bf Proof.}  We use induction on
$n$.   The result is obvious
for $n=1$.   Now let $n\geq 2$ hold and let $V$ be a
partition such that $\mathcal{U} \left( V\right) $ is
finite, $\mathcal{U} \left( V\right) $ is $\left( V\cup
H\right) $-connected, and $\#\left( V\right) =n$, and let
$A$ and $B$ be any members of $V$.   Then if $A=B$ the
result is obvious.   Assume now that $A\neq B$.   We
consider the partition of $\mathcal{U} \left( V\right) $
into the nonempty parts $B$ and $\left( \mathcal{U} \left(
V\right) \,\vdash B\right) $.   Then the fact that
$\mathcal{U} \left( V\right) $ is $\left( V\cup H\right)
$-connected implies that there exists a member $C$ of
$\left( V\cup H\right) $ such that $C\cap B\neq \emptyset $
holds and $C\cap \left( \mathcal{U} \left( V\right)
\,\vdash B\right) \neq \emptyset $ holds.   But $V$ is a
partition hence $C\cap B$ and $C\cap \left( \mathcal{U}
\left( V\right) \,\vdash B\right) $ cannot both be nonempty
for any member $C$ of $V$, hence $C$ must be a member of
$H$.   Let $C$ be a member of $H$ such that $C\cap B\neq
\emptyset $ holds and $C\cap \left( \mathcal{U} \left(
V\right)
\,\vdash B\right) \neq \emptyset $ holds.   Then $C\cap
\left(
\mathcal{U} \left( V\right) \,\vdash B\right) \neq
\emptyset $
implies that there exists a member $D$ of the partition $V$
such that $D\neq B$ holds and $C\cap D\neq \emptyset $
holds.
Let $D$ be a member of the partition $V$ such that $D\neq
B$ holds and $C\cap D\neq \emptyset $ holds.   Then $\left(
V\,\vdash \left\{ B\right\} \right) $ is a partition such
that $\mathcal{U} \left( V\,\vdash \left\{ B\right\} \right)
$ is finite, $\mathcal{U} \left( V\,\vdash \left\{ B\right\}
\right) $ is $\left( \left( V\,\vdash \left\{ B\right\}
\right) \cup H\right) $-connected, and $\#\left( V\,\vdash
\left\{ B\right\} \right) =\left( n-1\right) \geq 1$ holds,
and $A$ and $D$ are members of $\left( V\,\vdash \left\{
B\right\} \right) $, hence by the induction assumption there
exists an integer $p$ such that $1\leq p\leq \left(
n-1\right) $ holds, and a map $P$ such that $\mathcal{D}
\left( P\right) $ is the set of all the integers $q$ such
that $1\leq q\leq p$ holds, $\mathcal{R} \left( P\right)
\subseteq \left( V\,\vdash \left\{ B\right\} \right) $
holds,
$P_{ 1 } =A$ holds, $P_{ p } =D$ holds, and such that if
$q$ is any integer such that $1\leq q$ and $q\leq \left(
p-1\right) $ both hold, then there exists a member $E$ of
$H$ such that $E\cap P_{ q } \neq \emptyset $ holds and
$E\cap
P_{ q+1 } \neq \emptyset $ holds.   Let $p$ be such an
integer
and $P$ be such a map.   Then $M\equiv P\cup \left\{ \left(
p+1,B\right) \right\} $ is a map with all the required
properties.

\vspace{2.5ex}

\noindent {\bf Corollary.}  Let $H$ be a set such that
every member of $H$
is a set, let $n$ be an integer $\geq 1$, let $V$ be a
partition such that $\mathcal{U} \left( V\right) $ is
finite, $\mathcal{U} \left( V\right) $ is $\left( V\cup
H\right) $-connected, and $\#\left( V\right) =n$ holds, let
$A$ and $B$ be any members of $V$, let $T$ be any real
number such that $T>0$ holds,
\label{Start of original page 59}
 let $d$ be any integer $\geq 1$, and let $x$ be any member
of $\mathbb{E} _{ dV } $ such that if $C$ and $D$ are any
members of $V$ such that there exists a member $E$ of $H$
such that $C\cap E\neq \emptyset $ holds and $D\cap E\neq
\emptyset $
holds, then $\left| x_{ C } -x_{ D } \right| \leq T$ holds.
  Then $\left| x_{
A } -x_{ B } \right| \leq \left( n-1\right) T$ holds.

\vspace{2.5ex}

\noindent {\bf Proof.}  By Lemma \ref{Lemma 15}
there exists an integer $m$
such that
$1\leq m\leq n$ holds, and a map $M$ such that $\mathcal{D}
\left( M\right) $ is the set of all the integers $p$ such
that $1\leq p\leq m$ holds, $\mathcal{R} \left( M\right)
\subseteq V$ holds, $M_{ 1 } =A$ holds, $M_{ m } =B$ holds,
and such that if $p$ is any integer such that $1\leq p$ and
$p\leq \left( m-1\right) $ both hold, then there exists a
member $E$ of $H$ such that $E\cap M_{ p } \neq \emptyset $
holds and $E\cap M_{ p+1 } \neq \emptyset $ holds.   Let
$m$ be
such an integer and $M$ be such a map.   Then by the
triangle inequality,
$\left| x_{ A } -x_{ B } \right| \leq
\sum_{ p\in \mathcal{D} (M) }
\left| x_{ M_{ p } } - x_{ M_{ p+1 } } \right| \leq
(m-1)T $ holds.

\section{The BPHZ Integrands.}
\label{Section 5}

For any ordered pair $\left( M,\theta \right) $ of a map
$M$ such that $\mathcal{D} \left( M\right) $ is finite, and
every member of $\mathcal{R} \left( M\right) $ is a finite
set, and a map $\theta $ such that $\mathcal{D} \left(
M\right) \subseteq \mathcal{D} \left( \theta \right) $
holds, and $\mathcal{R} \left( \theta \right) $ is a subset
of the set $\mathbb{Z} $ of all the integers, we define
$\mathbb{X} \left( M,\theta \right) $ to be the set whose
members are all the maps $p$ such that $\mathcal{D} \left(
p\right) =\mathcal{U} \left( \mathcal{R} \left( M\right)
\right) $ holds, $\mathcal{R} \left( p\right) \subseteq
\mathbb{N} $ holds, and for every member $A$ of
$\mathcal{D} \left( M\right) $, $ \sum_{\alpha \in M_{ A }
} p_{ \alpha } \leq \theta _{ A } $ holds.

We note that this definition has the immediate consequence
that if, for any member $A$ of $\mathcal{D} \left( M\right)
$, $\theta _{ A } <0$ holds, then $\mathbb{X} \left(
M,\theta \right) $ is the empty set, since there is no map
$p$ with the required properties.

\begin{bphzlemma} \label{Lemma 16}
\end{bphzlemma}
\vspace{-6.143ex}

\noindent \hspace{11.9ex}{\bf.  }Let $M$ be a map
such that $\mathcal{D} \left(
M\right) $ is finite, and every member of $\mathcal{R}
\left( M\right) $ is a finite set, and let $D$ be a map
such that $\mathcal{D} \left( M\right) \subseteq
\mathcal{D} \left( D\right) $ holds, and $\mathcal{R}
\left( D\right) $ is a subset of $\mathbb{Z} $.   Let
$\lambda $ be a map whose domain is $\mathcal{U} \left(
\mathcal{R} \left( M\right) \right) $, such that for each
member $\alpha $ of $\mathcal{U} \left( \mathcal{R} \left(
M\right) \right) $, $\lambda _{ \alpha } $ is a nonempty
finite set, and such that for any two distinct members
$\alpha $ and $\beta $ of $\mathcal{U} \left( \mathcal{R}
\left( M\right) \right) $, $\lambda _{ \alpha } \cap
\lambda _{ \beta } =\emptyset $.

And let $T$ be the map such that $\mathcal{D} \left(
T\right) =\mathcal{D} \left( M\right) $, and for each
member $A$ of $\mathcal{D} \left( T\right) =\mathcal{D}
\left( M\right) $, $T_{ A } \equiv
\bigcup_{ \alpha \in M_{ A } } \lambda _{ \alpha } $.
(Thus $\mathcal{U}
\left( \mathcal{R} \left( T\right) \right) $ is the
disjoint union of all the $\lambda _{ \alpha } $, $\alpha
\in \mathcal{U} \left( \mathcal{R} \left( M\right) \right)
 $.)
\label{Start of original page 60}

Let $Y$ be a map such that $\mathcal{D} \left( Y\right)
=\mathcal{U} \left( \mathcal{R} \left( T\right) \right) =
\bigcup_{ \alpha \in \mathcal{U} \left( \mathcal{R}
\left( M\right)
\right) } \lambda _{ \alpha } $, and such that
for each member $\beta $ of $\mathcal{D} \left( Y\right) $,
$Y_{ \beta } $ is an operator such that the $Y_{ \beta }
$ obey the same commutative, associative and distributive
laws of addition and multiplication as the real numbers.

Then the following identity holds:
\[
 \sum_{n\in \mathbb{X} \left( M,D\right) } \left(
\prod_{\alpha \in
\mathcal{U} \left( \mathcal{R} \left( M\right)
\right) } \left( \left( \frac{1}{ n_{ \alpha } ! }
\right) \left(  \sum_{\beta \in
\lambda _{ \alpha
 } } Y_{ \beta } \right)^{ n_{ \alpha } } \right) \right)
  = \sum_{s\in \mathbb{X} \left(
T,D\right) } \left( \prod_{\beta \in \mathcal{U} \left(
\mathcal{R} \left( T\right) \right) } \left( \frac{1}
{ s_{ \beta
} ! } \right) Y_{ \beta }^{ s_{ \beta  } } \right)
\]

\vspace{2.5ex}

\noindent {\bf Proof.}  For any ordered pair
$\left(X,p\right) $
of a finite set $X$ and
an integer $p\geq 0$, we define $ F \left(X,p\right) $
to be the set whose
members are all the maps $q$ whose domain is $X$, whose
range is a subset of $\mathbb{N} $, and which satisfy the
requirement that $\sum_{\alpha \in X } q_{ \alpha } =p$
holds.   (We note that $F(X,p) $ is a finite set.)
\enlargethispage{0.8ex}

Then by the multinomial theorem for $p\in \mathbb{N} $, we
have
\[
\left( \frac{ 1}{ p! } \right) \left( \sum_{\beta
\in \lambda _{ \alpha } }
Y_{ \beta
} \right)^{ p } = \sum_{q\in F\left( \lambda _{
\alpha } ,p\right) } \left(
 \prod_{\beta \in \lambda _{ \alpha } } \left( \left(
  \frac{ 1}
 { q_{ \beta } !
} \right) Y_{ \beta }^{ q_{ \beta  }} \right) \right)
\]

Hence for any map $n$ such that $\mathcal{D} \left(
n\right) =\mathcal{U} \left( \mathcal{R} \left( M\right)
\right) $ and $\mathcal{R} \left( n\right) \subseteq
\mathbb{N} $ we have
\[
 \prod_{\alpha \in \mathcal{U} \left( \mathcal{R}
  \left( M\right)
\right) } \left( \left( \frac{ 1}{ n_{ \alpha } ! }
\right) \left( \sum_{\beta
\in \lambda
_{ \alpha } } Y_{ \beta } \right)^{ n_{\alpha} }
\right) =
\prod_{\alpha \in
\mathcal{U} \left( \mathcal{R} \left( M\right)
\right) } \left( \sum_{q\in F\left( \lambda _{ \alpha }
,n_{ \alpha }
\right) } \left( \prod_{\beta \in \lambda _{ \alpha } }
\left( \left(
\frac{ 1}{ q_{ \beta } ! } \right) Y_{ \beta }^{
 q_{ \beta } } \right) \right) \right)
\]
\[
\hspace{5.0cm}
= \prod_{\alpha \in \mathcal{U} \left( \mathcal{R} \left(
M\right) \right) } \left( \sum_{q_{ \alpha } \in F\left(
\lambda _{ \alpha } ,n_{ \alpha } \right) } \left(
 \prod_{\beta
\in \lambda _{ \alpha } } \left( \left(
 \frac{ 1}{ \left(q_{
\alpha }
\right) _{
\beta } ! } \right) Y_{ \beta }^{ \left(q_{ \alpha}
\right) _{ \beta } } \right) \right) \right) ,
\]
where in the second form of the right-hand side we rewrote
the dummy variable $q$ as $q_{ \alpha } $ in the factor
associated with the member $\alpha $ of $\mathcal{U} \left(
\mathcal{R} \left( M\right) \right) $.

Now for any ordered pair $\left( n,\lambda \right) $ of a
map $n$ such that $\mathcal{D} \left( n\right) $ is finite
and $\mathcal{R} \left( n\right) \subseteq \mathbb{N} $,
and a map $\lambda $ such that $\mathcal{D} \left( \lambda
\right) =\mathcal{D} \left( n\right) $, and for each member
$\alpha $ of $\mathcal{D} \left( \lambda \right)
=\mathcal{D} \left( n\right) $, $\lambda _{ \alpha } $ is
a finite set, and such that for any two distinct members
$\alpha $ and $\beta $ of $\mathcal{D} \left( \lambda
\right) $, $\lambda _{ \alpha } \cap \lambda _{ \beta }
=\emptyset $, we define $G\left( \lambda ,n\right) $ to be
the
set whose members are all the maps $s$ whose domain is $
\bigcup_{ \alpha \in \mathcal{D} \left( \lambda \right)
} \lambda _{ \alpha } $, and whose range is a subset of
$\mathbb{N} $, and such
\label{Start of original page 61}
 that for each member $\alpha $ of $\mathcal{D} \left(
n\right) =\mathcal{D} \left( \lambda \right) $,
$ \sum_{\beta \in
\lambda _{ \alpha } } s_{ \beta } =n_{ \alpha } $
holds.

Now we observe that in the final form of the right-hand
side of the above formula, in which the dummy variable $q$
has been re-written as $q_{ \alpha } $ in the factor
associated with the member $\alpha $ of $\mathcal{U} \left(
\mathcal{R} \left( M\right) \right) $, we have a sum over
all maps $q$ such that $\mathcal{D} \left( q\right)
=\mathcal{U} \left( \mathcal{R} \left( M\right) \right) $,
and for each member $\alpha $ of $\mathcal{U} \left(
\mathcal{R} \left( M\right) \right) =\mathcal{D} \left(
\lambda \right) =\mathcal{D} \left( n\right) $, $q_{ \alpha
 } $ is a member of $F\left( \lambda _{ \alpha } ,n_{
\alpha } \right) $.   We define the following one-to-one
correspondence between $G\left( \lambda ,n\right) $ and the
set of all maps $q$ such that $\mathcal{D} \left( q\right)
=\mathcal{U} \left( \mathcal{R} \left( M\right) \right)
=\mathcal{D} \left( \lambda \right) =\mathcal{D} \left(
n\right) $, and for each $\alpha \in \mathcal{D} \left(
\lambda \right) $, $q_{ \alpha } \in F\left( \lambda _{
\alpha } ,n_{ \alpha } \right) $ holds:

Given any member $s$ of $G\left( \lambda ,n\right) $, the
corresponding map $q$ is the map whose domain is
$\mathcal{D} \left( \lambda \right) =\mathcal{D} \left(
n\right) $, and such that for each member $\alpha $ of
$\mathcal{D} \left( \lambda \right) $, $q_{ \alpha }
=\downarrow \left( s,\lambda _{ \alpha } \right) $, (or in
other words, $q_{ \alpha } $ is the restriction of the map
$s$ to the domain $\lambda _{ \alpha }  $), which is
a member of $F\left( \lambda _{ \alpha } ,n_{ \alpha }
\right) $ in consequence of the definition of $G\left(
\lambda ,n\right) $.

And given any map $q$ such that $\mathcal{D} \left(
q\right) =\mathcal{D} \left( \lambda \right) $, and for
each member $\alpha $ of $\mathcal{D} \left( \lambda
\right) $, $q_{ \alpha } \in F\left( \lambda _{ \alpha }
,n_{ \alpha } \right) $ holds, we define $s\equiv
\bigcup_{ \alpha \in \mathcal{D} \left( \lambda \right)
} q_{ \alpha } $, which is a member of $G\left( \lambda
,n\right) $ in consequence of the definition of $F\left(
\lambda _{ \alpha } ,n_{ \alpha } \right) $.

And we verify immediately that these two correspondences
are one another's inverses, noting that for $\alpha \in
\mathcal{D} \left( \lambda \right) $ and $\beta \in
\mathcal{D} \left( \lambda \right) $ such that $\alpha \neq
\beta $, $q_{ \alpha } \cap q_{ \beta } =\emptyset $ holds
since $\mathcal{D} \left( q_{ \alpha } \right) \cap
\mathcal{D} \left( q_{ \beta } \right) =\emptyset $ holds by
assumption.

Now in the correspondence from $q$ to $s$ we have that for
any member $\alpha $ of $\mathcal{D} \left( \lambda \right)
$ and any member $\beta $ of $\lambda _{ \alpha } $, that
$s_{ \beta } = \left( q_{ \alpha } \right)_{ \beta } $
holds, hence we have that
\[
 \prod_{\alpha \in \mathcal{D} \left( \lambda \right) }
\left( \sum_{q_{
\alpha } \in F\left( \lambda _{ \alpha } ,n_{ \alpha }
\right) } \left( \prod_{\beta \in \lambda _{ \alpha } }
\left( \left(
\frac{1}{\left( q_{ \alpha }\right) _{ \beta } ! } \right)
Y_{ \beta }^{\left( q_{ \alpha }\right)
_{ \beta } } \right) \right) \right) =
 \sum_{s\in G\left( \lambda ,n\right) }
\left( \prod_{\beta
\in \mathcal{U} \left( \mathcal{R} \left( T\right)
\right) } \left( \frac{ 1}{ s_{ \beta } ! } \right)
Y_{ \beta }^{
s_{ \beta  } } \right)
\]
holds, since $\mathcal{U} \left( \mathcal{R} \left(
T\right) \right) = \bigcup_{ \alpha \in \mathcal{D}
\left( \lambda
\right) } \lambda _{ \alpha } $.

But $\mathcal{D} \left( \lambda \right) =\mathcal{U} \left(
\mathcal{R} \left( M\right) \right) $, hence we have
\[
 \prod_{\alpha \in \mathcal{U} \left( \mathcal{R}
 \left( M\right)
\right) } \left( \left( \frac{1}{ n_{ \alpha } ! } \right)
\left( \sum_{\beta
\in \lambda
_{ \alpha } } Y_{ \beta } \right)^{ n_{\alpha } }
\right) =
 \sum_{s\in G\left(
\lambda ,n\right) } \left( \prod_{\beta \in \mathcal{U}
\left(
\mathcal{R} \left( T\right) \right) }  \left( \frac{1}
{ s_{ \beta
} ! } \right) Y_{ \beta }^{ s_{ \beta } } \right) .
\]

Now $\mathbb{X} \left( T,D\right) $ is the set whose
members are all the maps $s$ such
\label{Start of original page 62}
 that $\mathcal{D} \left( s\right) =\mathcal{U} \left(
\mathcal{R} \left( T\right) \right) = \bigcup_{ \alpha \in
\mathcal{D} \left( \lambda \right) } \lambda _{
\alpha } $, and whose range is a subset of $\mathbb{N}
$, and such that for each member $A$ of $\mathcal{D} \left(
T\right) =\mathcal{D} \left( M\right) $, $
\sum_{\beta \in T_{ A } } s_{ \beta } \leq D_{ A } $ holds.
But by
definition $T_{ A } $ is the disjoint union of the $\lambda
_{ \alpha } $ for $\alpha \in M_{ A } $, that is, $T_{ A }
= \bigcup_{ \alpha \in M_{ A } } \lambda _{ \alpha } $ and,
for $\alpha \in M_{ A } $ and $\beta \in M_{ A } $ such
that $\alpha \neq \beta $, $\lambda _{ \alpha } \cap
\lambda _{ \beta } =\emptyset $ holds.   Hence the
constraint
on $s$ for $A$ may be written $\sum_{\alpha \in M_{ A } }
\sum_{\beta \in
\lambda _{ \alpha } } s_{ \beta } \leq D_{ A } $,
hence if, for all members $\alpha $ of $\mathcal{U} \left(
\mathcal{R} \left( M\right) \right) $, $ \sum_{ \beta \in
\lambda_{ \alpha } } s_{ \beta } =n_{ \alpha } $ holds,
where $n$
is a member of $\mathbb{X} \left( M,D\right) $, then $s$ is
automatically a member of $\mathbb{X} \left( T,D\right) $.

Furthermore, if $s$ is any member of $\mathbb{X} \left(
T,D\right) $, then if we define a map $n$ such that
$\mathcal{D} \left( n\right) =\mathcal{U} \left(
\mathcal{R} \left( M\right) \right) $ and such that for
each member $\alpha $ of $\mathcal{U} \left( \mathcal{R}
\left( M\right) \right) =\mathcal{D} \left( \lambda \right)
$, $n_{ \alpha } \equiv
\sum_{\beta \in \lambda _{ \alpha } } s_{ \beta } $
holds, then $n$ is automatically a
member of $\mathbb{X} \left( M,D\right) $.

Hence $\mathbb{X} \left( T,D\right) $ is equal to $
\bigcup_{ n\in
\mathbb{X} \left( M,D\right) } G\left( \lambda
,n\right) $, hence we may express $ \sum_{s\in \mathbb{X}
\left(
T,D\right) } $ as $
\sum_{n\in \mathbb{X} \left( M,D\right) }
 \sum_{s\in G\left( \lambda ,n\right) } $.

Hence, summing the above formula over $n\in \mathbb{X}
\left( M,D\right) $, we obtain the stated result.

\vspace{2.5ex}

For any maps $X$ and $Y$ such that $\mathcal{R} \left(
X\right) \subseteq \mathbb{R} $ holds, $\mathcal{R} \left(
Y\right) \subseteq \mathbb{R} $ holds, and $\mathcal{D}
\left( X\right) =\mathcal{D} \left( Y\right) $ holds, we
define $X+Y$ to be the map whose domain is $\mathcal{D}
\left( X\right) $, and such that for each member $A$ of
$\mathcal{D} \left( X\right) $, $\left( X+Y\right) _{ A }
\equiv \left( X_{ A } +Y_{ A } \right) $ holds, and we
define X-Y to be the map whose domain is $\mathcal{D}
\left( X\right) $, and such that for each member $A$ of
$\mathcal{D} \left( X\right) $, $\left( X-Y\right) _{ A }
\equiv \left( X_{ A } -Y_{ A } \right) $ holds.

\begin{bphzlemma} \label{Lemma 17}
\end{bphzlemma}
\vspace{-6.143ex}

\noindent \hspace{11.9ex}{\bf.  }Let $V$ be a map
such that $\mathcal{D} \left(
V\right) $ is finite and such that for every member $A$ of
$\mathcal{D} \left( V\right) $, $V_{ A } $ is a finite set,
let $J$ be any subset of $\mathcal{D} \left( V\right) $,
and let $K\equiv \left( \mathcal{D} \left( V\right)
\,\vdash J\right) $.   Let $W$ be the map such that
$\mathcal{D} \left( W\right) =K$ and such that for each
member $A$ of $K$, $W_{ A } =V_{ A } \,\vdash \mathcal{U}
\left( \mathcal{R} \left( \downarrow \left( V,J\right)
\right) \right) =V_{ A } \,\vdash \bigcup_{ B\in J }
V_{ B } $ holds.
\label{Start of original page 63}

Then $\mathcal{U} \left( \mathcal{R} \left( W\right)
\right) =\mathcal{U} \left( \mathcal{R} \left( V\right)
\right) \,\vdash \mathcal{U} \left( \mathcal{R} \left(
\downarrow \left( V,J\right) \right) \right) $ holds.
\enlargethispage{3.0ex}

Furthermore, if $\theta $ is a map such that $\mathcal{D}
\left( V\right) \subseteq \mathcal{D} \left( \theta \right)
$ holds and $\mathcal{R} \left( \theta \right) \subseteq
\mathbb{Z} $ holds, (where $\mathbb{Z} $ is the set of all
the integers), and if for each member $u$ of $\mathbb{N}^{
\mathcal{U} \left( \mathcal{R} \left( \downarrow \left(
V,J\right) \right) \right) } $, $\zeta \left( u\right) $
is a map such that $K\subseteq \mathcal{D} \left( \zeta
\left( u\right) \right) $ holds and such that for each
member $A$ of $K$, $\zeta _{ A } \left( u\right) \equiv
\left( \zeta \left( u\right) \right) _{ A } \equiv
\sum_{{
\alpha \in V_{ A } \cap \mathcal{U} \left( \mathcal{R}
\left( \downarrow \left( V,J\right) \right) \right) } }
u_{ \alpha } $ holds, and if $Y$ is a map such
that $\mathcal{D} \left( Y\right) =\mathcal{U} \left(
\mathcal{R} \left( V\right) \right) $ holds, and such that
for each member $\alpha $ of $\mathcal{D} \left( Y\right)
$, $Y_{ \alpha } $ is an operator such that the $Y_{
\alpha } $ obey the same commutative, associative and
distributive laws of addition and multiplication as the
real numbers, then the following equation holds:
\[
 \sum_{ m\in \mathbb{X} \left( V,\theta \right) } \left(
  \prod_{\alpha \in \mathcal{U} \left( \mathcal{R}
  \left( V\right)
\right) } \left( \frac{ Y_{ \alpha }^{ m_{ \alpha } } }
{ m_{ \alpha
 } ! } \right) \right) = \hspace{9.0cm}
\]
\[
\hspace{2.0cm} = \sum_{ u\in \mathbb{X}
 \left( \downarrow \left(
V,J\right) ,\theta \right) } \,\,\, \sum_{v\in \mathbb{X}
\left( W,\left( \theta -\zeta \left( u\right) \right)
\right) } \left( \prod_{\beta \in \mathcal{U} \left(
\mathcal{R} \left( W\right) \right) } \left( \frac{ Y_{
 \beta
}^{ v_{ \beta } } } { v_{ \beta } ! } \right) \right)
\left( \prod_{\alpha
\in \mathcal{U}
\left( \mathcal{R} \left( \downarrow \left( V,J\right)
\right) \right) } \left( \frac{ Y_{ \alpha }^{ u_{
 \alpha } } }
{ u_{ \alpha } ! } \right) \right)
\]

\vspace{2.5ex}

\noindent {\bf Proof.}  We first prove that
$\mathcal{U} \left( \mathcal{R}
\left( W\right) \right) =\mathcal{U} \left( \mathcal{R}
\left( V\right) \right) \,\vdash \mathcal{U} \left(
\mathcal{R} \left( \downarrow \left( V,J\right) \right)
\right) $.   Let $\alpha $ be any member of $\mathcal{U}
\left( \mathcal{R} \left( V\right) \right) \,\vdash
\mathcal{U} \left( \mathcal{R} \left( \downarrow \left(
V,J\right) \right) \right) $.   Then $\alpha $ is a member
of $V_{ A } $ for some member $A$ of $\mathcal{D} \left(
V\right) =\left( J\cup K\right) $, but $\alpha $ is
\emph{not} a member of $V_{ B } $ for any member $B$ of
$J$, hence $\alpha $ is a member of $V_{ A } $ for some
member $A$ of $K$, hence $\alpha $ is a member of $W_{ A }
=V_{ A } \,\vdash \bigcup_{ B\in J } V_{ B } $ for that
member $A$ of $K$, hence $\alpha $ is a member of
$\mathcal{U} \left( \mathcal{R} \left( W\right) \right)
= \bigcup_{ A\in K } W_{ A } $.   Now let $\alpha $ be any
member of $\mathcal{U} \left( \mathcal{R} \left( W\right)
\right) $.   Then $\alpha $ is a member of $W_{ A } $ for
some member $A$ of $K$, hence $\alpha $ \emph{is} a member
of $\mathcal{U} \left( \mathcal{R} \left( V\right) \right)
$, but is \emph{not} a member of $ \bigcup_{ B\in J }
V_{ B }
=\mathcal{U} \left( \mathcal{R} \left( \downarrow \left(
V,J\right) \right) \right) $, hence $\alpha $ is a member
of $\mathcal{U} \left( \mathcal{R} \left( V\right) \right)
\,\vdash \mathcal{U} \left( \mathcal{R} \left( \downarrow
\left( V,J\right) \right) \right) $.

From this it follows immediately that the domain of each
member $m$ of $\mathbb{X} \left( V,\theta \right) $, namely
$\mathcal{U} \left( \mathcal{R} \left( V\right) \right) $,
is equal to the disjoint union of the domain of each member
$u$ of $\mathbb{X} \left( \downarrow \left( V,J\right)
,\theta \right) $, namely $\mathcal{U} \left( \mathcal{R}
\left( \downarrow \left( V,J\right) \right) \right) $, and
the domain of each member $v$ of \\
$\mathbb{X} \left(
W,\left( \theta -\zeta \left( u\right) \right) \right) $,
namely $\mathcal{U} \left( \mathcal{R} \left( W\right)
\right) $.

We now define a one-to-one correspondence between the
members $m$ of $\mathbb{N}^{ \mathcal{U} \left( \mathcal{R}
\left( V\right) \right) } $, and the set of all ordered
pairs $\left(u,v\right) $ of a member $u$
of $\mathbb{N}^{ \mathcal{U}
\left( \mathcal{R} \left( \downarrow \left( V,J\right)
\right) \right) } $ and a member $v$ of $\mathbb{N}^{
\mathcal{U} \left( \mathcal{R} \left( W\right) \right) }
$, by specifying that to any member $m$ of $\mathbb{N}^{
\mathcal{U} \left( \mathcal{R} \left( V\right) \right) }
$, the corresponding such ordered pair is $\left(
u,v\right) \equiv \left( \downarrow \left( m,\mathcal{U}
\left( \mathcal{R} \left( \downarrow \left( V,J\right)
\right) \right) \right) ,\downarrow \left( m,\mathcal{U}
\left( \mathcal{R} \left( W\right) \right) \right) \right)
$, and that to any
\label{Start of original page 64}
 such ordered pair $\left(u,v\right) $, the
 corresponding member $m$ of
$\mathbb{N}^{ \mathcal{U} \left( \mathcal{R} \left(
V\right) \right) } $ is $m\equiv u\cup v$.   We verify
directly that these two correspondences are one another's
inverses.

Now let $m$ be any member of $\mathbb{X} \left( V,\theta
\right) $.   Then $ \sum_{\alpha \in V_{ A } } m_{ \alpha
} \leq \theta _{ A } $ holds for every member $A$ of
$\mathcal{D} \left( V\right) =\left( J\cup K\right) $,
hence if $u\equiv \downarrow \left( m,\mathcal{U} \left(
\mathcal{R} \left( \downarrow \left( V,J\right) \right)
\right) \right) $ and $v\equiv $ \\
$ \downarrow \left(
m,\mathcal{U} \left( \mathcal{R} \left( W\right) \right)
\right) $, then $ \sum_{\alpha \in V_{ A } } u_{ \alpha }
\leq \theta _{ A } $ holds for every member $A$ of $J$,
hence $u$ is a member of $\mathbb{X} \left( \downarrow
\left( V,J\right) ,\theta \right) $, and $ \left(
\sum_{\alpha \in
\left( V_{ A } \,\vdash \mathcal{U} \left( \mathcal{R}
\left( \downarrow \left( V,J\right) \right) \right)
\right) } v_{ \alpha } \right) + \left(
 \sum_{\alpha \in \left( V_{
A } \cap
\mathcal{U} \left( \mathcal{R} \left( \downarrow \left(
V,J\right) \right) \right) \right) } u_{ \alpha } \right)
\leq \theta _{ A } $ holds for every member $A$ of $K$,
hence $v$ is a member of $\mathbb{X} \left( W,\left( \theta
-\zeta \left( u\right) \right) \right) $.   And if $u$ is
any member of $\mathbb{X} \left( \downarrow \left(
V,J\right) ,\theta \right) $, and $v$ is any member of
$\mathbb{X} \left( W,\left( \theta -\zeta \left( u\right)
\right) \right) $, and we define $m\equiv u\cup v$, then we
directly find that $ \sum_{\alpha \in V_{ A } } m_{ \alpha
} \leq \theta _{ A } $ holds for every member $A$ of
$\left( J\cup K\right) =\mathcal{D} \left( V\right) $,
hence $m$ is a member of $\mathbb{X} \left( V,\theta
\right) $.   The equation stated follows directly from this.

\vspace{2.5ex}

For any map $V$ such that $\mathcal{D} \left( V\right) $ is
finite and for each member $A$ of $\mathcal{D} \left(
V\right) $, $V_{ A } $ is a finite set, we define $\psi
\left( V\right) $ to be the map whose domain is equal to
$\mathcal{D} \left( V\right) $, and such that for each
member $A$ of $\mathcal{D} \left( V\right) $, $\psi _{ A }
\left( V\right) \equiv \left( \psi \left( V\right) \right)
_{ A } $ is the set whose members are all the ordered pairs
$\left( \alpha ,X\right) $ of a member $\alpha $ of
$\mathcal{U} \left( \mathcal{R} \left( V\right) \right) $,
and a subset $X$ of $\mathcal{D} \left( V\right) $ such
that $A\in X$ holds, and for every member $B$ of $X$,
$\alpha \in V_{ B } $ holds.

We note that it follows immediately from this definition
that $\mathcal{U} \left( \mathcal{R} \left( \psi \left(
V\right) \right) \right) $ is the set whose members are all
the ordered pairs $\left( \alpha ,X\right) $ of a member
$\alpha $ of $\mathcal{U} \left( \mathcal{R} \left(
V\right) \right) $, and a \emph{nonempty} subset $X$ of
$\mathcal{D} \left( V\right) $ such that $\alpha \in V_{ B
} $ holds for every member $B$ of $X$.   For if $\left(
\alpha ,X\right) $ is any such ordered pair, then $\left(
\alpha ,X\right) $ is a member of $\psi _{ A } \left(
V\right) $ for every member $A$ of the nonempty set $X$,
hence $\left( \alpha ,X\right) $ is certainly a member of
$\mathcal{U} \left( \mathcal{R} \left( \psi \left( V\right)
\right) \right) $.   And if $\left( \alpha ,X\right) $ is
any member of $\mathcal{U} \left( \mathcal{R} \left( \psi
\left( V\right) \right) \right) $, then there exists a
member $A$ of $\mathcal{D} \left( V\right) $ such that
$\left( \alpha ,X\right) \in \psi _{ A } \left( V\right) $
holds, hence $\alpha \in \mathcal{U} \left( \mathcal{R}
\left( V\right) \right) $ holds, and $A\in X$ holds hence
$X$ is nonempty, and $\alpha \in V_{ B } $ holds for every
member $B$ of $X$.

And we note furthermore that it follows immediately from
the above that if $A$ is any member of $\mathcal{D} \left(
V\right) $, then $\psi _{ A } \left( V\right) $ is the set
whose members are all the members $\left( \alpha ,X\right)
$ of $\mathcal{U} \left( \mathcal{R} \left( \psi \left(
V\right) \right) \right) $ such that $A\in X$
\label{Start of original page 65}
 holds.

For any ordered pair $\left(V,n\right) $ of a map $V$ such
that
$\mathcal{D} \left( V\right) $ is finite and every member
of $\mathcal{R} \left( V\right) $ is a finite set, and a
map $n$ such that $\mathcal{D} \left( V\right) \subseteq
\mathcal{D} \left( n\right) $ holds and $\mathcal{R} \left(
n\right) \subseteq \mathbb{N} $ holds, we define
$\mathbb{A} \left( V,n\right) $ to be the set whose members
are all the maps $m$ such that $\mathcal{D} \left( m\right)
=\mathcal{U} \left( \mathcal{R} \left( \psi \left( V\right)
\right) \right) $, $\mathcal{R} \left( m\right) \subseteq
\mathbb{N} $ holds, and for each member $A$ of $\mathcal{D}
\left( V\right) $, the equation $  \sum_{\left( \alpha,
X\right)
\in \psi _{ A } \left( V\right) } m_{ \left( \alpha
,X\right)} =n_{ A }   $ holds.

We shall often abbreviate $m_{ \left( \alpha ,X\right) } $
to $m_{ \alpha X } $.

\begin{bphzlemma} \label{Lemma 18}
\end{bphzlemma}
\vspace{-6.143ex}

\noindent \hspace{11.9ex}{\bf.  }Let $V$ be a map
such that $\mathcal{D} \left(
V\right) $ is finite and such that for every member $A$ of
$\mathcal{D} \left( V\right) $, $V_{ A } $ is a finite set,
let $C$ be a map such that $\mathcal{U} \left( \mathcal{R}
\left( V\right) \right) \subseteq \mathcal{D} \left(
C\right) $ holds, and such that for every member $\alpha $
of $\mathcal{U} \left( \mathcal{R} \left( V\right) \right)
$, $C_{ \alpha } $ is the set whose members are all the
members $A$ of $\mathcal{D} \left( V\right) $ such that
$\alpha \in V_{ A } $ holds, let $n$ be a map such that
$\mathcal{D} \left( V\right) \subseteq \mathcal{D} \left(
n\right) $ holds and $\mathcal{R} \left( n\right) \subseteq
\mathbb{N} $ holds, and let $f$ be a map whose domain is a
subset of $\mathbb{R}^{ \mathcal{U} \left( \mathcal{R}
\left( V\right) \right) } $ and whose range is a subset of
$\mathbb{R} $, such that there exists an open subset $S$ of
$\mathbb{R}^{ \mathcal{U} \left( \mathcal{R} \left(
V\right) \right) } $ such that $S\subseteq \mathcal{D}
\left(
f\right) $ holds, and such that all derivatives of $f$ of
degree less than or equal to $  \sum_{A\in \mathcal{D}
\left(
V\right) } n_{ A}$ exist and are continuous
throughout $S$.
\enlargethispage{0.5ex}

For any member $\rho $ of $\mathbb{R}^{ \mathcal{D} \left(
V\right) } $ let $p\left( \rho \right) $ be the member of
$\mathbb{R}^{ \mathcal{U} \left( \mathcal{R} \left(
V\right) \right) } $ such that for each member $\alpha $ of
$\mathcal{U} \left( \mathcal{R} \left( V\right) \right) $,
$p_{ \alpha } \left( \rho \right) \equiv \left( p\left(
\rho \right) \right) _{ \alpha } \equiv
\prod_{A\in C_{ \alpha } } \rho _{ A } $.

Then for any member $\rho $ of $\mathbb{R}^{ \mathcal{D}
\left( V\right) } $ such that $p\left( \rho \right) $ is a
member of $S$, the following equation holds:
\[
\left( \prod_{A\in \mathcal{D} \left( V\right) }
\left( \frac{ \hat{ \rho }_{A }^{
n_{ A } } } { n_{ A } ! } \right) \right)
 f\left( p\left( \rho \right)
\right)
= \hspace{9.0cm}
\]
\[
\hspace{2.0cm}
= \sum_{m\in \mathbb{A} \left( V,n\right) } \left(
\left( \prod_{\left(
\alpha ,X\right) \in \mathcal{U} \left( \mathcal{R} \left(
\psi \left( V\right) \right) \right) } \left( \frac{
 \left( \left(
\prod_{B\in
\left( C_{ \alpha } \,\vdash X\right) }
\rho _{ B } \right) \hat{ r }_{ \alpha }
\right) ^{ m_{ \alpha X } } }
{ m_{
\alpha X } ! } \right) \right) f\left(
r\right) \right)_{ r=p\left( \rho \right) }
\]

\vspace{2.5ex}

\noindent {\bf Proof.}  The proof is by induction
on $q\equiv \sum_{ A\in
\mathcal{D} \left( V\right) } n_{ A } $.   The equation is
certainly true for $q=0$.

We shall show that if $q$ is any member of $\mathbb{N} $,
and the result is true for all members $n$ of $\mathbb{N}^{
\mathcal{D} \left( V\right) } $ such that $  \sum_{ n\in
\mathcal{D} \left( V\right) } n_{ A } =q$ holds,
then it is also true for all members $n$ of $\mathbb{N}^{
\mathcal{D} \left( V\right) } $ such that
\label{Start of original page 66}
 $  \sum_{A\in \mathcal{D} \left( V\right) } n_{ A }
=\left( q+1\right)  $ holds.

For any ordered pair $\left( \rho ,m\right) $ of a member
$\rho $ of $\mathbb{R}^{ \mathcal{D} \left( V\right) } $
such
that $p\left( \rho \right) \in S$ holds, and a member $m$
of $\mathbb{N}^{ \mathcal{U} \left( \mathcal{R} \left( \psi
\left( V\right) \right) \right) } $, we define
\[
T\left( \rho ,m\right) \equiv \left( \left(
 \prod_{\left( \alpha
,X\right)
\in \mathcal{U} \left( \mathcal{R} \left( \psi \left(
V\right) \right) \right) } \left( \frac{ \left( \left(
 \prod_{B\in
\left( C_{ \alpha } \,\vdash X\right) } \rho _{
B } \right) \hat{ r }_{ \alpha }\right)^{ m_{ \alpha X } } }
{ m_{ \alpha X } !} \right) \right)
f\left( r\right) \right)_{ r=p\left( \rho
\right) }
\]

Then to prove the induction step it will be sufficient to
prove that if $A$ is any member of $\mathcal{D} \left(
V\right) $, and $n$ is any member of $\mathbb{N}^{
\mathcal{D} \left( V\right) } $ such that the result is true
for $n$, and if we define $u$ to be the member of
$\mathbb{N}^{ \mathcal{D} \left( V\right) } $ such that $u_{
B } \equiv n_{ B } $ holds for all members $B$ of $\left(
\mathcal{D} \left( V\right) \,\vdash \left\{ A\right\}
\right) $ and $u_{ A } \equiv \left( n_{ A } +1\right) $
holds, then
\[
\hat{ \rho }_{ A } \left( \sum_{m\in \mathbb{A}
\left( V,n\right)
}
T\left( \rho ,m\right) \right) =\left( n_{ A } +1\right)
 \sum_{w\in
\mathbb{A} \left( V,u\right) } T\left( \rho
,w\right)
\]
holds.

For any ordered pair $\left(A,m\right) $ of a
member $A$ of $\mathcal{D}
\left( V\right) $ and a member $m$ of $\mathbb{N}^{
\mathcal{U} \left( \mathcal{R} \left( \psi \left( V\right)
\right) \right) } $, we define $E(A,m) $ to be the set whose
members are all the members $\left( \alpha ,X\right) $ of \\
$\mathcal{U} \left( \mathcal{R} \left( \psi \left( V\right)
\right) \right) $ such that $A\in \left( C_{ \alpha }
\,\vdash X\right) $ holds and $m_{ \alpha X } \neq 0$ holds.

And for any ordered triple $\left( A,m,\left( \alpha
,X\right) \right) $ of a member $A$ of $\mathcal{D} \left(
V\right) $, a member $m$ of $\mathbb{N}^{ \mathcal{U}
\left( \mathcal{R} \left( \psi \left( V\right) \right)
\right) } $, and a member $\left( \alpha ,X\right) $ of
$E(A,m) $, we define $s\left( A,m,\left( \alpha ,X\right)
\right) $ to be the member of $\mathbb{N}^{ \mathcal{U}
\left( \mathcal{R} \left( \psi \left( V\right) \right)
\right) } $ such that for all members $\left( \beta
,Y\right)
$ of $\mathcal{U} \left( \mathcal{R} \left( \psi \left(
V\right) \right) \right) $ such that $\left( \beta
,Y\right) \neq \left( \alpha ,X\right) $ holds and $\left(
\beta ,Y\right) \neq \left( \alpha ,X\cup \left\{ A\right\}
\right) $ holds, $s_{ \beta Y } \left( A,m,\left( \alpha
,X\right) \right) \equiv \left( s\left( A,m,\left( \alpha
,X\right) \right) \right) _{ \beta Y } \equiv m_{ \beta Y }
$ holds, and such that $s_{ \alpha X } \left( A,m,\left(
\alpha ,X\right) \right) \equiv \left( m_{ \alpha X }
-1\right) $ holds and $\left( s\left( A,m,\left( \alpha
,X\right) \right) \right) _{ \left( \alpha ,\left( X\cup
\left\{ A\right\} \right) \right) } \equiv \left( m_{
\left(
\alpha ,\left( X\cup \left\{ A\right\} \right) \right) }
+1\right) $ holds.

From this definition it immediately follows that if $A$ is
any member of $\mathcal{D} \left( V\right) $, $m$ is any
member of $\mathbb{A} \left( V,n\right) $, and $\left(
\alpha ,X\right) $ is any member of $ E \left(A,m\right) $
then $s\left(
A,m,\left( \alpha ,X\right) \right) $ is a member of
$\mathbb{A} \left( V,u\right) $, where $u$ is defined for
$A$ and $n$ as above.   For if $B$ is any member of $\left(
\mathcal{D} \left( V\right) \,\vdash \left\{ A\right\}
\right) $, then $\left( \alpha ,X\right) \in \psi _{ B }
\left( V\right) $ holds ifif $B\in X$ holds and $\left(
\alpha ,\left( X\cup \left\{ A\right\} \right) \right) \in
\psi _{ B } \left( V\right) $ holds ifif $B\in X$ holds,
hence $\left( \alpha ,X\right) \in \psi _{ B } \left(
V\right) $ holds ifif $\left( \alpha ,\left( X\cup \left\{
A\right\} \right) \right) \in \psi _{ B } \left( V\right) $
holds, hence
$\sum_{\left(\beta ,Y\right)\in \psi_{ B } \left(V\right) }
s_{ \beta Y } (A,m,(\alpha
,X))=\sum_{\left(\beta ,Y\right)\in \psi_{ B }
\left(V\right) } m_{ \beta Y } =n_{B } =u_{B} $
holds, and furthermore
\label{Start of original page 67}
 the assumption that $A\in \left( C_{ \alpha } \,\vdash
X\right) $ holds implies that $A$ is \emph{not} a member of
$X$, hence $\left( \alpha ,X\right) $ is \emph{not} a
member of $\psi _{ A } \left( V\right) $, whereas $\left(
\alpha ,\left( X\cup \left\{ A\right\} \right) \right) $
is a member of $\psi _{ A } \left( V\right) $, hence
$\sum_{\left(\beta ,Y\right)\in \psi_{ A } \left(V\right)
 } s_{\beta
Y } \left(A,m,\left(\alpha,X\right)\right)= \left(
\sum_{\left(\beta ,Y\right)\in \psi_{ A } \left(V\right) }
m_{\beta Y } \right) +1=\left(n_{A }
+1\right)=u_{A} $
 holds.

For any ordered triple $\left( A,m,\alpha \right) $ of a
member $A$ of $\mathcal{D} \left( V\right) $, a member $m$
of \\
$\mathbb{N}^{ \mathcal{U} \left( \mathcal{R} \left( \psi
\left( V\right) \right) \right) } $, and a member $\alpha
$ of $V_{ A } $, (or in other words a member $\alpha $ of
$\mathcal{U} \left( \mathcal{R} \left( V\right) \right) $
such that $A\in C_{ \alpha } $ holds), we define $t\left(
A,m,\alpha \right) $ to be the member of $\mathbb{N}^{
\mathcal{U} \left( \mathcal{R} \left( \psi \left( V\right)
\right) \right) } $ such that for every member $\left( \beta
,Y\right) $ of $\mathcal{U} \left( \mathcal{R} \left( \psi
\left( V\right) \right) \right) $ such that $\left( \beta
,Y\right) \neq \left( \alpha ,\left\{ A\right\} \right) $
holds, $t_{ \beta Y } \left( A,m,\alpha \right) \equiv m_{
\beta Y } $ holds, and such that $t_{ \alpha \left\{
A\right\} } \left( A,m,\alpha \right) \equiv \left( m_{
\alpha \left\{ A\right\} } +1\right) $ holds.

It immediately follows from this definition that if $A$ is
any member of $\mathcal{D} \left( V\right) $, $m$ is any
member of $\mathbb{A} \left( V,n\right) $, and $\alpha $ is
any member of $V_{ A } $, then $t\left( A,m,\alpha \right)
$ is a member of $\mathbb{A} \left( V,u\right) $.   For if
$B$ is any member of $\left( \mathcal{D} \left( V\right)
\,\vdash \left\{ A\right\} \right) $, then $\left( \alpha
,\left\{ A\right\} \right) $ is \emph{not} a member of $\psi
_{ B } \left( V\right) $, hence $\sum_{ \left( \beta,
Y\right)
\in \psi _{ B } \left( V\right) } t_{ \beta Y }
\left( A,m,\alpha \right) = \sum_{\left( \beta ,Y\right) \in
\psi _{ B } \left( V\right) } m_{ \beta Y } =n_{ B }
=u_{ B }  $ holds, and furthermore $\left( \alpha
,\left\{ A\right\} \right) $ \emph{is} a member of $\psi _{
A
} \left( V\right) $, hence
$\sum_{\left(\beta ,Y\right)\in \psi_{ A } \left(V\right) }
t_{\beta Y } (A,m,\alpha
)= \left( \sum_{\left(\beta ,Y\right)\in \psi_{ A }
 \left(V\right) }
m_{\beta Y } \right) +1=\left(n_{A }
+1\right)=u_{A} $
 holds.

Furthermore, it follows directly from these definitions that
\[
\hat{ \rho }_{ A } T\left( \rho ,m\right) = \left(
 \sum_{\left(
\alpha
,X\right) \in E\left( A,m\right) } \left( m_{ \left(
\alpha ,\left( X\cup \left\{ A\right\} \right) \right) }
+1\right) T\left( \rho ,s\left( A,m,\left( \alpha ,X\right)
\right) \right) \right) + \hspace{-9.0pt} \hspace{3.0cm}
\]
\[
\hspace{8.0cm}
+ \left( \sum_{\alpha \in V_{ A } } \left( m_{
\alpha \left\{ A\right\} } +1\right) T\left( \rho ,t\left(
A,m,\alpha \right) \right) \right)
\]
holds.   For when $\hat{\rho}_{ A } $ acts on
\[
T\left( \rho ,m\right) = \left( \left(
 \prod_{\left( \alpha ,X\right)
\in
\mathcal{U} \left( \mathcal{R} \left( \psi \left( V\right)
\right) \right) } \left( \frac{ \left( \left(
\prod_{B\in \left(
C_{ \alpha } \,\vdash X\right) } \rho _{ B } \right)
 \hat{ r }_{
\alpha }\right) ^{ m_{ \alpha X } } }
{ m_{ \alpha X } ! } \right) \right) f\left( r\right)
\right)_{
r=p\left( \rho \right) },
\]
it can either act on a factor $\rho _{ A } $ in the factor
associated with some member $\left( \alpha ,X\right) $ of
$\mathcal{U} \left( \mathcal{R} \left( \psi \left( V\right)
\right) \right) $, or it can act on $ \left( f\left(
r\right) \right)_{ r=p\left( \rho \right) } = f\left(
p\left( \rho \right) \right) $.

Now if $\hat{ \rho }_{ A } $ acts on the factor associated
with
the member $\left( \alpha ,X\right) $ of $\mathcal{U}
\left( \mathcal{R} \left( \psi \left( V\right) \right)
\right) $, it produces nothing unless $A$ is a member of
$\left( C_{ \alpha } \,\vdash X\right) $ and $m_{ \alpha X
} \neq 0$ holds, or in other words, it produces nothing
unless $\left( \alpha ,X\right) \in E\left( A,m\right) $
holds.   And if $\left( \alpha ,X\right) \in E\left(
A,m\right) $ holds, it produces
\label{Start of original page 68}
 $T\left( \rho ,s\left( A,m,\left( \alpha ,X\right) \right)
\right) $, up to a $\rho $-independent factor.   And to
determine this $\rho $-independent factor, we note firstly
that there is a numerator factor of $m_{ \alpha X } $ from
the exponent $m_{ \alpha X } $.   Now in $T\left( \rho
,m\right) $, the denominator factor in the factor
associated with $\left( \alpha ,X\right) $ is $m_{ \alpha X
} !$, whereas in $T\left( \rho ,s\left( A,m,\left( \alpha
,X\right) \right) \right) $, the denominator factor in the
factor associated with $\left( \alpha ,X\right) $ is
$\left( m_{ \alpha X } -1\right) !$.   Hence the numerator
factor of $m_{ \alpha X } $ combines with the denominator
factor $m_{ \alpha X } !$ in $T\left( \rho ,m\right) $ to
produce the denominator factor $\left( m_{ \alpha X }
-1\right) !$ in $T\left( \rho ,s\left( A,m,\left( \alpha
,X\right) \right) \right) $.   And in $T\left( \rho
,m\right) $, the denominator factor in the factor
associated with $\left( \alpha ,\left( X\cup \left\{
A\right\} \right) \right) $ is $m_{ \left( \alpha ,\left(
X\cup \left\{ A\right\} \right) \right) } !$, whereas in
$T\left( \rho ,s\left( A,m,\left( \alpha ,X\right) \right)
\right) $ the denominator factor in the factor associated
with $\left( \alpha ,\left( X\cup \left\{ A\right\} \right)
\right) $ is $\left( m_{ \left( \alpha ,\left( X\cup \left\{
A\right\} \right) \right) } +1\right) !$.   Hence the
overall factor is $\left( m_{ \left( \alpha ,\left( X\cup
\left\{ A\right\} \right) \right) } +1\right) $.

And if $\hat{ \rho }_{ A } $ acts on $f\left( p\left( \rho
\right) \right) $, we have, by the chain rule for
differentiation,
\[
\hat { \rho }_{ A } f\left( p\left( \rho \right) \right)
= \sum_{\alpha
\in V_{ A } } \left( \prod_{B\in \left( C_{ \alpha }
\,\vdash
\left\{ A\right\} \right) } \rho _{ B } \right)
 \left( \hat{ r }_{
\alpha
} f\left( r\right) \right)_{ r=p\left( \rho \right) },
\]
and the term associated with $\alpha $ in this expression
produces $T\left( \rho ,t\left( A,m,\alpha \right) \right)
$, apart from a $\rho $-independent factor.   And to
determine this $\rho $-independent factor, we note that in
$T\left( \rho ,m\right) $ the denominator factor in the
factor associated with $\left( \alpha ,\left\{ A\right\}
\right) $ is $m_{ \left( \alpha ,\left\{ A\right\} \right)
}
!$, whereas in $T\left( \rho ,t\left( A,m,\alpha \right)
\right) $ the denominator factor in the factor associated
with $\left( \alpha ,\left\{ A\right\} \right) $ is $\left(
m_{ \left( \alpha ,\left\{ A\right\} \right) } +1\right)
!$,
hence the overall factor is $\left( m_{ \left( \alpha
,\left\{ A\right\} \right) } +1\right) $.

We note, furthermore, that the coefficient $\left( m_{
\left( \alpha ,\left( X\cup \left\{ A\right\} \right)
\right)
 } \! + \! 1\right) $ of
 $T\left( \rho ,s\left( A,m,\left( \alpha
,X\right) \right) \right) $ in the above expression for
$\hat{ \rho }_{ A } T\left( \rho ,m\right) $, is equal to
$\left(
s\left( A,m,\left( \alpha ,X\right) \right) \right) _{
\left( \alpha ,\left( X\cup \left\{ A\right\} \right)
\right)
 } $, and that the coefficient $\left( m_{ \left( \alpha
,\left\{ A\right\} \right) } +1\right) $ of $T\left( \rho
,t\left( A,m,\alpha \right) \right) $ in the above
expression for $\hat{ \rho }_{ A } T\left( \rho ,m\right)
$, is
equal to $\left( t\left( A,m,\alpha \right) \right) _{
\alpha \left\{ A\right\} } $.

Thus $\hat{ \rho }_{ A } \left( \sum_{m\in \mathbb{A}
\left(
V,n\right)
} T\left( \rho ,m\right) \right) $ is equal to the sum,
 over the
members $w$ of $\mathbb{A} \left( V,u\right) $, of $T\left(
\rho ,w\right) $ times a $\rho $-independent coefficient.

Now let $w$ be any member of $\mathbb{A} \left( V,u\right)
$.   To determine the total coefficient of $T\left( \rho
,w\right) $, we shall identify all the members $m$ of
$\mathbb{A} \left( V,n\right) $ such that $w$ is equal to
$s\left( A,m,\left( \alpha ,X\right) \right) $ for some
member $\left( \alpha ,X\right) $ of $ E \left(A,m\right)
$, and all the
members $m$ of $\mathbb{A} \left( V,n\right) $ such that
$w$ is equal to $t\left( A,m,\alpha \right) $ for some
member $\alpha $ of $V_{ A } $.
\label{Start of original page 69}

 Let $\left( \alpha ,X\right) $ be any member of
$\mathcal{U} \left( \mathcal{R} \left( \psi \left( V\right)
\right) \right) $.   Then if $m$ is a member of $\mathbb{A}
\left( V,n\right) $ such that $s\left( A,m,\left( \alpha
,X\right) \right) =w$, then we must have $m_{ \beta Y }
=w_{ \beta Y } $ when $\left( \beta ,Y\right) \neq \left(
\alpha ,X\right) $ and $\left( \beta ,Y\right) \neq \left(
\alpha ,\left( X\cup \left\{ A\right\} \right) \right) $,
and
we must also have $m_{ \alpha X } =\left( w_{ \alpha X }
+1\right) $ and $m_{ \left( \alpha ,\left( X\cup \left\{
A\right\} \right) \right) } =\left( w_{ \left( \alpha
,\left( X\cup \left\{ A\right\} \right) \right) } -1\right)
$.   Furthermore we must have $A\in \left( C_{ \alpha }
\,\vdash X\right) $, and $w_{ \left( \alpha ,\left( X\cup
\left\{ A\right\} \right) \right) } \neq 0$.

If these requirements are satisfied, then $\hat{ \rho }_{ A
}
T\left( \rho ,m\right) $ contains $T\left( \rho ,w\right) $
with coefficient $w_{ \left( \alpha ,\left( X\cup \left\{
A\right\} \right) \right) } $, while if these requirements
are not satisfied, then $\hat{ \rho }_{ A } T\left( \rho
,m\right) $ does not contain $T\left( \rho ,w\right) $.

Now let $\alpha $ be any member of $\mathcal{U} \left(
\mathcal{R} \left( V\right) \right) $.   Then if $m$ is a
member of $\mathbb{A} \left( V,n\right) $ such that
$t\left( A,m,\alpha \right) =w$, then we must have $\alpha
\in V_{ A } $, (or equivalently $A\in C_{ \alpha }
$), and we must have $m_{ \beta Y } =w_{ \beta Y } $ for
$\left( \beta ,Y\right) \neq \left( \alpha ,X\right) $, and
we must also have $m_{ \alpha \left\{ A\right\} } =\left(
w_{
\alpha \left\{ A\right\} } -1\right) $.   Furthermore we
must
also have $w_{ \alpha \left\{ A\right\} } \neq 0$.

If these requirements are satisfied, then $\hat{ \rho }_{ A
}
T\left( \rho ,m\right) $ contains $T\left( \rho ,w\right) $
with coefficient $w_{ \alpha \left\{ A\right\} } $, while if
these requirements are not satisfied, then $\hat{ \rho }_{
A }
T\left( \rho ,m\right) $ does not contain $T\left( \rho
,w\right) $.
\enlargethispage{1.2ex}

Thus the total coefficient of $T\left( \rho ,w\right) $ in
$\hat{ \rho }_{ A } \left( \sum_{m\in \mathbb{A} \left(
V,n\right) }
T\left( \rho ,m\right) \right) $ is equal to the sum over
all
members $\left( \alpha ,X\right) $ of $\mathcal{U} \left(
\mathcal{R} \left( \psi \left( V\right) \right) \right) $
such that $A\in \left( C_{ \alpha } \,\vdash X\right) $
holds, of $w_{ \left( \alpha ,\left( X\cup \left\{ A\right\}
\right) \right) } $, plus the sum over all members $\alpha
$ of $\mathcal{U} \left( \mathcal{R} \left( V\right)
\right) $ such that $A\in C_{ \alpha } $ holds, of $w_{
\alpha \left\{ A\right\} } $.   But this is precisely equal
to the sum over all members $\left( \alpha ,Y\right) $ of
$\psi _{ A } \left( V\right) $, of $w_{ \alpha Y } $, and
by the definition of $\mathbb{A} \left( V,u\right) $, this
is equal to $u_{ A } =\left( n_{ A } +1\right) $.

\vspace{0.6cm}

We observe that, if $J$ and $K$ are any two sets such that
$J\cap K=\emptyset $, then \\
$\mathcal{U} \left( \left\{
0\right\}
^{
J } \cup \left\{ 1\right\}^{ K } \right) $ is the unique map
$M$ whose domain is $\left( J\cup K\right) $, and such that
for each member $A$ of $J$, $M_{ A } =0$ holds, and for
each member $A$ of $K$, $M_{ A } =1$ holds.   For $\left\{
0\right\}^{ J } $ is the set whose members are all the
maps $P$
such that $\mathcal{D} \left( P\right) =J$ holds and
$\mathcal{R} \left( P\right) \subseteq \left\{ 0\right\} $
holds.   But there is only one such map, namely the unique
map $P$ such that $\mathcal{D} \left( P\right) =J$ holds,
and $P_{ A } =0$ holds for every member $A$ of $J$.
Similarly the set $\left\{ 1\right\}^{ K } $ has exactly one
member, namely the unique map $Q$ such that $\mathcal{D}
\left( Q\right) =K$ holds, and $Q_{ A } =1$ holds for every
member $A$ of $K$.   Furthermore $P\neq Q$ holds provided
$J$ and $K$ are not both equal to the empty set $\emptyset
$,
for $J\cap K=\emptyset $ implies that $J\neq K$ holds
unless $J$
and $K$ are both equal to the empty set $\emptyset $.
Hence if
$J$ and $K$ are not
\label{Start of original page 70}
 both equal to the empty set $\emptyset $, then the set
$\left\{
0\right\}^{ J } \cup \left\{ 1\right\}^{ K }$ has exactly
 two
members, namely the two maps $P$ and $Q$, whose union is
equal to the map $M$, hence $\mathcal{U} \left( \left\{
0\right\}^{ J } \cup \left\{ 1\right\}^{ K } \right) =M$.
And if $J$ and $K$ are both equal to the empty set
$\emptyset $,
then $\left\{ 0\right\}^{ J } \cup
\left\{ 1\right\}^{ K } $
has
the one member $\emptyset $, hence $\mathcal{U} \left(
\left\{
0\right\}^{ J } \cup \left\{ 1\right\}^{ K } \right)
=\emptyset $
holds, and furthermore the map $M$ is equal to the empty
set $\emptyset $ in this case.

\begin{bphzlemma} \label{Lemma 19}
\end{bphzlemma}
\vspace{-6.143ex}

\noindent \hspace{11.9ex}{\bf.  }Let $V$ be a map
such that $\mathcal{D} \left(
V\right) $ is finite and such that for every member $A$ of
$\mathcal{D} \left( V\right) $, $V_{ A } $ is a finite set,
let $\theta $ be a map such that $\mathcal{D} \left(
V\right) \subseteq \mathcal{D} \left( \theta \right) $
holds and $\mathcal{R} \left( \theta \right) \subseteq
\mathbb{N} $ holds, and let $f$ be a map whose domain is a
subset of $\mathbb{R}^{ \mathcal{U} \left( \mathcal{R}
\left( V\right) \right) } $ and whose range is a subset of
$\mathbb{R} $, such that there exists an open subset $S$ of
$\mathbb{R}^{ \mathcal{U} \left( \mathcal{R} \left(
V\right) \right) } $ such that $S\subseteq \mathcal{D}
\left(
f\right) $ holds, and such that all derivatives of $f$ of
degree less than or equal to $ \sum_{{ A\in \mathcal{D}
\left(
V\right) } } \theta _{ A } $ exist and are
continuous throughout $S$.

For any member $\rho $ of $\mathbb{R}^{ \mathcal{D} \left(
V\right) } $ let $p\left( \rho \right) $ be the member of
$\mathbb{R}^{ \mathcal{U} \left( \mathcal{R} \left(
V\right) \right) } $ such that for each member $\alpha $ of
$\mathcal{U} \left( \mathcal{R} \left( V\right) \right) $,
$p_{ \alpha } \left( \rho \right) \equiv \left( p\left(
\rho \right) \right) _{ \alpha } $ is equal to the product
of $\rho _{ A } $ over all the members $A$ of $\mathcal{D}
\left( V\right) $ such that $\alpha \in V_{ A } $ holds.

Let $J$ be any subset of $\mathcal{D} \left( V\right) $,
and let $K\equiv \left( \mathcal{D} \left( V\right)
\,\vdash J\right) $.   We assume furthermore that $p\left(
\mathcal{U} \left( \left\{ 0\right\}^{ J } \cup \left\{
1\right\}^{ K } \right) \right) $ is a member of $S$.   Then
the following equation holds:
\[
\left( \left(
\prod_{A\in J } \left( \sum_{n_{ A } =0 }^{ \theta _{ A } }
\frac{ \hat{ \rho
}_{ A
}^{ n_{ A } } }{ n_{ A } ! } \right) \right) f\left(
 p\left( \rho \right) \right) \right)_{
\rho =\mathcal{U} \left( \left\{ 0\right\}^{ J } \cup
\left\{
1\right\}^{ K } \right) } = \hspace{-4.7pt} \hspace{7.0cm}
\]
\[
\hspace{5.0cm} \hspace{-26.3pt} = \sum_{m\in \mathbb{X}
\left( \downarrow
\left( V,J\right) ,\theta \right) } \left( \left(
 \prod_{\alpha \in
\mathcal{U} \left( \mathcal{R} \left( \downarrow \left(
V,J\right) \right) \right) } \left( \frac{ \hat{ r }_{
\alpha }^{
 m_{
\alpha } } }{ m_{ \alpha } ! } \right) \right)
f\left( r\right) \right)_{ r=p\left(
\mathcal{U} \left( \left\{ 0\right\}^{ J } \cup \left\{
1\right\}^{ K } \right) \right) }
\]

\vspace{2.5ex}

\noindent {\bf Proof.}  For any ordered pair
$\left(W,n\right) $
 of a map $W$ such that
$\mathcal{D} \left( W\right) $ is finite, and every member
of $\mathcal{R} \left( W\right) $ is a finite set, and a
map $n$ such that $\mathcal{D} \left( W\right) \subseteq
\mathcal{D} \left( n\right) $ holds, and $\mathcal{R}
\left( n\right) $ is a subset of the set $\mathbb{Z} $ of
all the integers, we define $H(W,n) $ to be the set whose
members are all the maps $p$ such that $\mathcal{D} \left(
p\right) =\mathcal{U} \left( \mathcal{R} \left( W\right)
\right) $ holds, $\mathcal{R} \left( p\right) \subseteq
\mathbb{N} $ holds, and for every member $A$ of
$\mathcal{D} \left( W\right) $, $ \sum_{\alpha \in W_{ A }
} p_{ \alpha } =n_{ A } $ holds.

We note that this definition has the immediate consequence
that if, for any member $A$ of $\mathcal{D} \left( W\right)
$, $n_{ A } <0$ holds, then $H(V,n) $ is the empty set,
since
there is no map $p$ with the required properties.
\label{Start of original page 71}

 And we observe that $\mathbb{X} \left( W,n\right) $ is
the disjoint union of $ H \left(W,m\right) $ over the
set whose members are
all the members $m$ of $\mathbb{N}^{ \mathcal{D} \left(
W\right) } $ such that for every member $A$ of $\mathcal{D}
\left( W\right) $, $m_{ A } \leq n_{ A } $ holds.

We define $C$ to be the map whose domain is $\mathcal{U}
\left( \mathcal{R} \left( V\right) \right) $, and such that
for each member $\alpha $ of $\mathcal{U} \left(
\mathcal{R} \left( V\right) \right) $, $C_{ \alpha } $ is
the set whose members are all the members $A$ of
$\mathcal{D} \left( V\right) $ such that $\alpha \in V_{ A
} $ holds, so that $C$ satisfies the requirements on the
map $C$ in Lemma \ref{Lemma 18}, and we
consider Lemma \ref{Lemma 18} when $n$ is
any member of $\mathbb{N}^{ \mathcal{D} \left( V\right) }
=\mathbb{N}^{ \left( J\cup K\right) } $ such that $n_{ A }
=0$ holds for every member $A$ of $K$, and $\rho $ is any
member of $\mathbb{R}^{ \mathcal{D} \left( V\right) } $ such
that $\rho _{ A } =0$ holds for every member $A$ of $J$.

Now $\psi \left( V\right) $ is the map whose domain is
equal to $\mathcal{D} \left( V\right) $, and for each
member $A$ of $\mathcal{D} \left( V\right) $, $\psi _{ A }
\left( V\right) $ is the set whose members are all the
ordered pairs $\left( \alpha ,X\right) $ of a member
$\alpha $ of $\mathcal{U} \left( \mathcal{R} \left(
V\right) \right) $, and a subset $X$ of $\mathcal{D} \left(
V\right) $ such that $A\in X$ holds, and for every member
$B$ of $X$, $\alpha \in V_{ B } $ holds.   And $\mathbb{A}
\left( V,n\right) $ is the set whose members are all the
maps $m$ such that $\mathcal{D} \left( m\right)
=\mathcal{U} \left( \mathcal{R} \left( \psi \left( V\right)
\right) \right) $, $\mathcal{R} \left( m\right) \subseteq
\mathbb{N} $ holds, and for each member $A$ of $\mathcal{D}
\left( V\right) $, the equation $ \sum_{{ \left( \alpha,
X\right)
\in \psi _{ A } \left( V\right) } } m_{ \alpha X }
=n_{ A } $ holds.

From this it immediately follows that if $m$ is any member
of $\mathbb{A} \left( V,n\right) $ in the present case, and
$\left( \alpha ,X\right) $ is any member of $\mathcal{U}
\left( \mathcal{R} \left( V\right) \right) $, then $m_{
\alpha X } $ vanishes unless $X$ contains \emph{no} member
of $K$, or in other words, $m_{ \alpha X } $ vanishes
unless $X\subseteq J$ holds.   Furthermore, the term
associated with $m$ in the sum over $\mathbb{A} \left(
V,n\right) $ vanishes when $\rho _{ A } =0$ holds for every
member $A$ of $J$, unless $m_{ \alpha X } =0$ holds for
every member $\left( \alpha ,X\right) $ of $\mathcal{U}
\left( \mathcal{R} \left( \psi \left( V\right) \right)
\right) $ such that $\left( C_{ \alpha } \,\vdash X\right)
$ contains any member of $J$, or in other words, the term
associated with $m$ in the sum over $\mathbb{A} \left(
V,n\right) $ vanishes when $\rho _{ A } =0$ holds for every
member $A$ of $J$, unless $m_{ \alpha X } =0$ holds for
every member $\left( \alpha ,X\right) $ of $\mathcal{U}
\left( \mathcal{R} \left( \psi \left( V\right) \right)
\right) $ such that $X$ does not contain \emph{all} the
members $A$ of $J$ such that $\alpha \in V_{ A } $ holds.
Thus in the present case the sum over the members $m$ of
$\mathbb{A} \left( V,n\right) $ becomes a sum over maps $m$
whose domain is the set of all the ordered pairs $\left(
\alpha ,X\right) $ of a member $\alpha $ of $\mathcal{U}
\left( \mathcal{R} \left( V\right) \right) $, and a
nonempty  set $X$ whose members are precisely those members
$A$ of $J$ such that $\alpha \in V_{ A } $ holds.   Thus
there is at most one possible ordered pair $\left( \alpha
,X\right) $ for each member $\alpha $ of $\mathcal{U}
\left( \mathcal{R} \left( V\right) \right) $, namely the
ordered pair $\left( \alpha ,X\right) $ where $X$ is the
set of all the members $A$ of $J$ such that $\alpha \in V_{
A } $ holds, \emph{provided this set is nonempty.}  Thus the
set of these ordered pairs $\left( \alpha ,X\right) $ is in
one-to-one correspondence with the set
\label{Start of original page 72}
 $\mathcal{U} \left( \mathcal{R} \left( \downarrow \left(
V,J\right) \right) \right) $, namely the set of all the
members $\alpha $ of $\mathcal{U} \left( \mathcal{R} \left(
V\right) \right) $ such that $\alpha $ is a member of $V_{
A } $ for at least one member $A$ of $J$.   To each such
member $m$ of $\mathbb{A} \left( V,n\right) $ we define the
corresponding member $n$ of $\mathbb{N}^{ \mathcal{U}
\left( \mathcal{R} \left( \downarrow \left( V,J\right)
\right) \right) } $ by the requirement that for each member
$\alpha $ of $\mathcal{U} \left( \mathcal{R} \left(
\downarrow \left( V,J\right) \right) \right) $, $p_{ \alpha
 } =m_{ \alpha X } $ holds, where $X$ is the set of all the
members $A$ of $J$ such that $\alpha \in V_{ A } $ holds,
(and the fact that $\alpha $ is a member of $\mathcal{U}
\left( \mathcal{R} \left( \downarrow \left( V,J\right)
\right) \right) $ guarantees that $X$ is nonempty).

Now the constraints on $m$ associated with the members of
$K$ have been satisfied by requiring $m_{ \alpha X } $ to
be equal to $0$ if $X$ contains any member of $K$, hence
the maps $p$ are only constrained by the constraints
associated with the members of $J$.   And the constraint on
the map $p$ associated with the member $A$ of $J$ is that
the sum of $p_{ \alpha } $ over all members $\alpha $ of
$\mathcal{U} \left( \mathcal{R} \left( \downarrow \left(
V,J\right) \right) \right) $ such that the corresponding
$\left( \alpha ,X\right) $ is a member of $\psi _{ A }
\left( V\right) $, be equal to $n_{ A } $.   But a member
$\left( \alpha ,X\right) $ of $\mathcal{U} \left(
\mathcal{R} \left( \psi \left( V\right) \right) \right) $
is a member of $\psi _{ A } \left( V\right) $ ifif $A\in X$
holds, and in the present case, $X$ is the set of all the
members $B$ of $J$ such that $\alpha \in V_{ B } $ holds,
hence $\left( \alpha ,X\right) $ is a member of $\psi _{ A
} \left( V\right) $ ifif $\alpha \in V_{ A } $ holds, thus
the constraint on the map $p$ associated with the member
$A$ of $J$ is $ \sum_{\alpha \in V_{ A } } p_{ \alpha }
=n_{ A } $.

Thus the maps $p$ are precisely the members of $H\left(
\downarrow \left( V,J\right) ,n\right) $, and the following
equation holds:
\[
\left( \left(
\prod_{A\in J } \left( \frac{ \hat{ \rho }_{ A }^{ n_{ A } }
}{ n_{ A } ! } \right) \right) f\left(
p\left( \rho \right) \right) \right)_{ \rho =\mathcal{U}
\left(
\left\{ 0\right\}^{ J } \cup \left\{ 1\right\}^{ K }
\right) }
= \hspace{-22.0pt} \hspace{8.5cm}
\]
\[
\hspace{4.5cm} \hspace{-14.9pt}
= \sum_{m\in H\left( \downarrow \left( V,J\right) ,
n \right) } \left( \left(
\prod_{\alpha \in \mathcal{U} \left( \mathcal{R}
\left( \downarrow \left( V,J\right) \right) \right) } \left(
\frac{ \hat{ r }_{ \alpha }^{ m_{ \alpha } } }{ m_{ \alpha }
! } \right) \right) f\left( r\right) \right)_{
r=p\left( \mathcal{U} \left( \left\{
0\right\}^{ J } \cup \left\{ 1\right\}^{ K } \right) \right)
}
\]

And summing this equation over all members $n$ of
$\mathbb{N}^{ J } $ such that $n_{ A } \leq \theta _{ A } $
holds for every member $A$ of $J$, we obtain the stated
result.

\vspace{2.5ex}

We recall from
page \pageref{Start of original page 3}
that for any ordered
pair $\left(F,i\right) $ of
a set $F$ such that every member of $F$ is a set, and a
member $i$ of $\mathcal{U} \left( F\right) $, we define
$\mathcal{C} \left( F,i\right) $ to be the intersection of
all the members $A$ of $F$ such that $i\in A$ holds, and we
note that if $F$ is a \emph{wood,} then $\mathcal{C} \left(
F,i\right) $ is equal to the unique member $A$ of
$\mathcal{M} \left( F\right) $ such that $i\in A$ holds.

And we recall from
page \pageref{Start of original page 7} that for any
ordered pair $\left(A,H\right) $
of a set $A$, and a set $H$ such that every member of $H$
is a set, we define
\label{Start of original page 73}
 $\mathcal{T} \left( A,H\right) $ to be the subset of $A$
whose members are all the members $i$ of $A$ such that
there is \emph{no} member $B$ of $H$ such that $i\in B$ and
$B\subseteq A$ both hold.

And we also recall from
page \pageref{Start of original page 7}
 that for any ordered pair
$\left(F,H\right) $ of a wood $F$,
and a set $H$ such that every member
of $H$ is a set, we define $\mathcal{O} \left( F,H\right) $
to be the set whose members are all the members $i$ of
$\mathcal{U} \left( F\right) $ such that there
\emph{exists} a member $A$ of $F$ such that $i\in A$ holds
and there is \emph{no} member $B$ of $H$ such that $i\in B$
and $B\subseteq A$ both hold, and we note that it
immediately follows from this definition that the equation
$\mathcal{O} \left( F,H\right) =\bigcup_{A\in F }
\mathcal{T} \left( A,H\right) $ holds.

Furthermore, we observe that if $F$ is any wood and $H$ is
any set such that every member of $H$ is a set, then
$\mathcal{O} \left( F,H\right) =\mathcal{O} \left(
\mathcal{M} \left( F\right) ,H\right) $ holds.   For
$\mathcal{M} \left( F\right) $ is a subset of $F$, hence
$\mathcal{O} \left( \mathcal{M} \left( F\right) ,H\right)
\subseteq \mathcal{O} \left( F,H\right) $ certainly holds,
and if $i$ is any member of $\mathcal{U} \left( F\right) $
and $A$ is any member of $F$ such that $i\in A$ holds, then
the member $\mathcal{C} \left( F,i\right) $ of $\mathcal{M}
\left( F\right) $ is a subset of $A$, hence if there is no
member $B$ of $H$ such that $i\in B$ and $B\subseteq A$
both hold, then there is certainly no member $B$ of $H$
such that $i\in B$ and $B\subseteq \mathcal{C} \left(
F,i\right) $ both hold.   Hence if there exists a member of
$F$ that has $i$ as a member and does not contain as a
subset any member of $H$ that has $i$ as a member, then
there exists a member of $\mathcal{M} \left( F\right) $
that has $i$ as a member and does not contain as a subset
any member of $H$ that has $i$ as a member.

And we note furthermore that it immediately follows from
the preceding paragraph that if $F$ and $G$ are any woods
such that $\mathcal{M} \left( F\right) =\mathcal{M} \left(
G\right) $ holds, and $H$ is a set such that every member
of $H$ is a set, then $\mathcal{O} \left( F,H\right)
=\mathcal{O} \left( G,H\right) $ holds.

We also recall from
page \pageref{Start of original page 7}
that for any ordered triple
$\left(F,H,i\right) $ of a wood $F$,
a set $H$ such that every member of
$H$ is a set, and a member $i$ of $\mathcal{O} \left(
F,H\right) $, we define $\mathcal{Z} \left( F,H,i\right) $
to be the \emph{largest} member $A$ of $F$ such that $i\in
A$ holds and there is \emph{no} member $B$ of $H$ such that
$i\in B$ and $B\subseteq A$ both hold, and we note that it
immediately follows from this definition that $\mathcal{Z}
\left( F,H,i\right) $ is the \emph{largest} member $A$ of
$F$ such that $i\in \mathcal{T} \left( A,H\right) $ holds.

We recall from
page \pageref{Start of original page 3}
 that for any set $F$ such that
every member of $F$ is a set, we define $\mathbb{B} \left(
F\right) $ to be equal to $\left( F\,\vdash \mathcal{M}
\left( F\right) \right) $.

And we recall from
page \pageref{Start of original page 8}
 that for any ordered triple
$\left(F,A,B\right) $ of a wood
$F$, a \emph{nonempty} set $A$, and a set
$B$, we define $\mathbb{Y} \left( F,A,B\right) $ to be
\label{Start of original page 74}
 the set whose members are all the members $C$ of $F$ such
that $A\subset C$ and $C\subseteq B$ both hold, and we note
that $\mathbb{Y} \left( F,A,B\right) $ is equal to the
empty set $\emptyset $ if $A\subset B$ does \emph{not} hold.
And we note furthermore that if $C$ and $D$ are any two
distinct members of $\mathbb{Y} \left( F,A,B\right) $, then
$C$ and $D$ are members of $F$ hence do not overlap, and
$C\cap D$ contains as a subset the nonempty set $A$, hence
either $C\subset D$ holds or $D\subset C$ holds.

For any ordered pair $\left(F,H\right) $ of a wood $F$
and a set $H$ such
that every member of $H$ is a set, we define $\mathbb{I}
\left( F,H\right) $ to be the map whose domain is
$\mathbb{B} \left( F\right) =\left( F\,\vdash \mathcal{M}
\left( F\right) \right) $, and such that for each member
$A$ of $\mathbb{B} \left( F\right) $, $\mathbb{I} _{ A }
\left( F,H\right) \equiv \left( \mathbb{I} \left(
F,H\right) \right) _{ A } $ is the set whose members are
all the ordered pairs $\left(i,B\right) $
 of a member $i$ of $\mathcal{T}
\left( A,H\right) $ and a member $B$ of $\mathbb{Y} \left(
F,\mathcal{C} \left( F,i\right) ,A\right) $.

\begin{bphzlemma} \label{Lemma 20}
\end{bphzlemma}
\vspace{-6.143ex}

\noindent \hspace{11.9ex}{\bf.  }Let $Q$ be any
wood, let $F$ be any wood such
that $\mathcal{M} \left( F\right) =\mathcal{M} \left(
Q\right) $ holds and $F\subseteq Q$ holds, and let $H$ be
any set such that every member of $H$ is a set.   Then
$ \bigcup_{A\in \mathbb{B} \left( F\right) } \mathbb{I} _{
A } \left( Q,H\right) $ is the set whose members are all
the ordered pairs $\left(i,B\right) $ of
 a member $i$ of $\mathcal{O}
\left( Q,H\right) $ and a member $B$ of $\mathbb{Y} \left(
Q,\mathcal{C} \left( F,i\right) ,\mathcal{Z} \left(
F,H,i\right) \right) $.

\vspace{2.5ex}

\noindent {\bf Proof.}  Let $\left(i,B\right) $
be any member
of $ \bigcup_{A\in \mathbb{B} \left(
F\right) } \mathbb{I} _{ A } \left( Q,H\right) $.
  Then there exists a member $A$ of $F$ such that $\left(
i,B\right) \in \mathbb{I} _{ A } \left( Q,H\right) $ holds.
  Let $A$ be a member of $F$ such that $\left( i,B\right)
\in \mathbb{I} _{ A } \left( Q,H\right) $ holds.   Then
$i\in \mathcal{T} \left( A,H\right) $ holds and $B\in
\mathbb{Y} \left( Q,\mathcal{C} \left( Q,i\right) ,A\right)
$ holds.   Now $i\in \mathcal{T} \left( A,H\right) $
implies directly that $i\in \mathcal{O} \left( F,H\right) $
holds, hence that $i\in \mathcal{O} \left( Q,H\right) $
holds.   And $i\in \mathcal{O} \left( F,H\right) $ implies
that $\mathcal{Z} \left( F,H,i\right) $ is the largest
member $C$ of $F$ such that $i\in \mathcal{T} \left(
F,C\right) $ holds, hence $A\subseteq \mathcal{Z} \left(
F,H,i\right) $ holds.   But $B\in \mathbb{Y} \left(
Q,\mathcal{C} \left( Q,i\right) ,A\right) $ holds, hence
$\mathcal{C} \left( Q,i\right) \subset B\subseteq A$ holds,
hence $\mathcal{C} \left( Q,i\right) \subset B\subseteq
\mathcal{Z} \left( F,H,i\right) $ holds, hence $B\in
\mathbb{Y} \left( Q,\mathcal{C} \left( Q,i\right)
,\mathcal{Z} \left( F,H,i\right) \right) $ holds.

Now let $\left(i,B\right) $ be any ordered pair such that
$i\in
\mathcal{O} \left( Q,H\right) $ holds and \\
$B\in \mathbb{Y}
\left( Q,\mathcal{C} \left( Q,i\right) ,\mathcal{Z} \left(
F,H,i\right) \right) $ holds.   Then $i\in \mathcal{O}
\left( F,H\right) $ holds hence $\mathcal{Z} \left(
F,H,i\right) $ is the largest member $A$ of $F$ such that
$i\in \mathcal{T} \left( A,H\right) $ holds, hence $i\in
\mathcal{T} \left( \mathcal{Z} \left( F,H,i\right)
,H\right) $ holds.   Hence $\left( i,B\right) \in
\mathbb{I}_{ \mathcal{Z} \left( F,H,i\right) } \left(
Q,H\right) $ holds, hence $\left( i,B\right) \in
\bigcup_{A\in
\mathbb{B} \left( F\right) } \mathbb{I} _{ A }
\left( Q,H\right) $ holds.

\vspace{2.5ex}

We now observe that if $Q$ is any wood, $F$ is any wood
such that $\mathcal{M} \left( F\right) =\mathcal{M} \left(
Q\right) $ holds and $F\subseteq Q$ holds, and $H$ is any
set such that
\label{Start of original page 75}
 every member of $H$ is a set, then $\downarrow \left(
\mathbb{I} \left( Q,H\right) ,F\right) $, the restriction
of $\mathbb{I} \left( Q,H\right) $ to the domain
$\mathcal{D} \left( \mathbb{I} \left( Q,H\right) \right)
\cap F= \mathbb{B} \left( F\right) $, is the map whose
domain is $\mathbb{B} \left( F\right) $, and such that for
each member $A$ of $\mathbb{B} \left( F\right) $, \\
$\left(
\downarrow \left( \mathbb{I} \left( Q,H\right) ,F\right)
\right) _{ A } =\mathbb{I} _{ A } \left( Q,H\right) $ holds.

Now if $M$ is any map such that every member of
$\mathcal{R} \left( M\right) $ is a set, then the equation
$\mathcal{U} \left( \mathcal{R} \left( M\right) \right) =
 \bigcup_{A\in \mathcal{D} \left( M\right) } M_{ A } $
holds.

Hence if $Q$ is any wood, $F$ is any wood such that
$\mathcal{M} \left( F\right) =\mathcal{M} \left( Q\right) $
holds and $F\subseteq Q$ holds, and $H$ is any set such
that every member of $H$ is a set, then the equation
$\mathcal{U} \left( \mathcal{R} \left( \downarrow \left(
\mathbb{I} \left( Q,H\right) ,F\right) \right) \right)
= \bigcup_{A\in \mathbb{B} \left( F\right) } \mathbb{I} _{
A } \left( Q,H\right) $ holds.
Hence by Lemma \ref{Lemma 20},
$\mathcal{U} \left( \mathcal{R} \left( \downarrow \left(
\mathbb{I} \left( Q,H\right) ,F\right) \right) \right) $ is
the set whose members are all the ordered pairs
$\left(i,B\right) $ of a
member $i$ of $\mathcal{O} \left( Q,H\right) $ and a member
$B$ of $\mathbb{Y} \left( Q,\mathcal{C} \left( Q,i\right)
,\mathcal{Z} \left( F,H,i\right) \right) $.

Furthermore, if $Q$ is any wood, and $H$ is any set such
that every member of $H$ is a set, then $\downarrow \left(
\mathbb{I} \left( Q,H\right) ,Q\right) =\mathbb{I} \left(
Q,H\right) $ holds, hence $\mathcal{U} \left( \mathcal{R}
\left( \mathbb{I} \left( Q,H\right) \right) \right) $ is
the set whose members are all the ordered pairs
$\left(i,B\right) $ of a
member $i$ of $\mathcal{O} \left( Q,H\right) $ and a member
$B$ of $\mathbb{Y} \left( Q,\mathcal{C} \left( Q,i\right)
,\mathcal{Z} \left( Q,H,i\right) \right) $.

For any ordered triple $\left(P,Q,H\right) $ of a wood $P$,
a wood $Q$
such that $\mathcal{M} \left( Q\right) =\mathcal{M} \left(
P\right) $ holds and $P\subseteq Q$ holds, and a set $H$
such that every member of $H$ is a set, we define
$\mathbb{J} \left( P,Q,H\right) $ to be the set whose
domain is $\left( Q\,\vdash P\right) $, and such that for
each member $A$ of $\left( Q\,\vdash P\right) $,
$\mathbb{J} _{ A } \left( P,Q,H\right) \equiv \left(
\mathbb{J} \left( P,Q,H\right) \right) _{ A } $ is defined
by the equation:
\[
\mathbb{J} _{ A } \left( P,Q,H\right) \equiv \left(
\mathbb{I} _{ A } \left( Q,H\right) \,\vdash \bigcup_{B\in
\mathbb{B} \left( P\right) } \mathbb{I} _{ B }
\left( Q,H\right) \right)
\]

Thus, by Lemma \ref{Lemma 20}, $\mathbb{J} _{ A } \left(
P,Q,H\right)
$ is the set whose members are all the ordered
pairs $\left(i,B\right) $
of a member $i$ of $\mathcal{T} \left( A,H\right) $, and a
member $B$ of $\mathbb{Y} \left( Q,\mathcal{C} \left(
Q,i\right) ,A\right) $ such that $B$ is \emph{not} a member
of $\mathbb{Y} \left( Q,\mathcal{C} \left( Q,i\right)
,\mathcal{Z} \left( P,H,i\right) \right) $.

Now there is no such $B$ unless $\mathcal{Z} \left(
P,H,i\right) \subset A$ holds, hence $i$ is a member of
$\mathcal{T} \left( A,H\right) $ such that $\mathcal{Z}
\left( P,H,i\right) \subset A$ holds.   Now $i\in
\mathcal{T} \left( A,H\right) $ implies $i\in \mathcal{O}
\left( Q,H\right) $, hence $i\in \mathcal{O} \left(
P,H\right) $, hence $\mathcal{Z} \left( P,H,i\right) $ is
the largest member $C$ of $P$ such that $i\in \mathcal{T}
\left( C,H\right) $ holds.   Hence the requirement that
$\mathcal{Z} \left( P,H,i\right) \subset A$ holds, is
equivalent to the requirement that there is \emph{no}
member $C$ of $P$ such that $A\subseteq C$ holds and $i\in
\mathcal{T} \left( C,H\right) $ holds.

Thus $\mathbb{J} _{ A } \left( P,Q,H\right) $ is the set
whose members are all the ordered
pairs $\left(i,B\right) $ of a member
$i$ of $\mathcal{T} \left( A,H\right) $ such that there is
no member $C$ of $P$ such that $A\subseteq C$ holds and
$i\in \mathcal{T} \left( C,H\right) $ holds, and a member
$B$ of
\label{Start of original page 76}
 $\mathbb{Y} \left( Q,\mathcal{C} \left( Q,i\right)
,A\right) $ such that $B$ is not a member of $\mathbb{Y}
\left( Q,\mathcal{C} \left( Q,i\right) ,\mathcal{Z} \left(
P,H,i\right) \right) $.   Now $A\in \left( Q\,\vdash
P\right) $ holds hence the requirement that there is no
member $C$ of $P$ such that $A\subseteq C$ holds and $i\in
\mathcal{T} \left( C,H\right) $ holds, implies that
$\mathcal{Z} \left( P,H,i\right) \subset A$ holds, hence
$B$ is a member of $\mathbb{Y} \left( Q,\mathcal{Z} \left(
P,H,i\right) ,A\right) $.   Furthermore, there is no member
$D$ of $P$ such that $\mathcal{Z} \left( P,H,i\right)
\subset D\subset A$ holds, for if there was such a member
$D$ of $P$ then $i\in D$ would hold and $i\in \mathcal{T}
\left( D,H\right) $ would hold, (for if there was a member
$E$ of $H$ such that $i\in E$ and $E\subseteq D$ both held,
then $i\in E$ and $E\subseteq A$ would both hold,
contradicting $i\in \mathcal{T} \left( A,H\right)
$), and this contradicts the fact that by definition
$\mathcal{Z} \left( P,H,i\right) $ is the \emph{largest}
member $D$ of $P$ such that $i\in \mathcal{T} \left(
D,H\right) $ holds.   Hence $\mathcal{Z} \left(
P,H,i\right) $ is equal to $\mathcal{K} \left( P,A,i\right)
$, which by the definition on
page \pageref{Start of original page 7}, is the largest
member $D$ of $P$ such that $i\in D$ and $D\subset A$ both
hold.

Hence $\mathbb{J} _{ A } \left( P,Q,H\right) $ is the set
whose members are all the ordered pairs
 $\left(i,B\right) $ of a member
$i$ of $\mathcal{T} \left( A,H\right) $ such that there is
no member $C$ of $P$ such that $A\subseteq C$ holds and
$i\in \mathcal{T} \left( C,H\right) $ holds, and a member
$B$ of $\mathbb{Y} \left( Q,\mathcal{K} \left( P,A,i\right)
,A\right) $.

We recall from
page \pageref{Start of original page 6} that for any
ordered pair $\left(F,B\right) $ of
a wood $F$ and a \emph{nonempty} set $B$, we define
$\mathcal{Y} \left( F,B\right) $ to be the \emph{smallest}
member $A$ of $F$ such that $B\subseteq A$ holds, if any
members $A$ of $F$ exist such that $B\subseteq A$ holds,
and to be equal to the empty set $\emptyset $ if there are
\emph{no} members $A$ of $F$ such that $B\subseteq A$ holds.

And we observe that it immediately follows from this
definition and the preceding paragraph, that $\mathbb{J} _{
A } \left( P,Q,H\right) $ is the set whose members are all
the ordered pairs $\left(i,B\right) $ of a member $i$ of
$\left(
\mathcal{T} \left( A,H\right) \,\vdash \mathcal{T} \left(
\mathcal{Y} \left( P,A\right) ,H\right) \right) $, and a
member $B$ of $\mathbb{Y} \left( Q,\mathcal{K} \left(
P,A,i\right) ,A\right) $, for if $i$ is a member of
$\mathcal{T} \left( A,H\right) $ such that there is no
member $C$ of $P$ such that $A\subseteq C$ holds and $i\in
\mathcal{T} \left( C,H\right) $ holds, then either there is
no member $C$ of $P$ such that $A\subseteq C$ holds, in
which case $\mathcal{Y} \left( P,A\right) =\emptyset $
holds and
$\mathcal{T} \left( \mathcal{Y} \left( P,A\right) ,H\right)
=\mathcal{T} \left( \emptyset ,H\right) =\emptyset $ holds,
hence $i$
is a member of $\left( \mathcal{T} \left( A,H\right)
\,\vdash \mathcal{T} \left( \mathcal{Y} \left( P,A\right)
,H\right) \right) =\mathcal{T} \left( A,H\right) $, or else
there is at least one member $C$ of $P$ such that
$A\subseteq C$ holds, in which case $i$ is \emph{not} a
member of $\mathcal{T} \left( C,H\right) $ for any such
member $C$ of $P$, and in particular $i$ is not a member of
$\mathcal{T} \left( \mathcal{Y} \left( P,A\right) ,H\right)
$, (for in this case $\mathcal{Y} \left( P,A\right) $ is by
definition the smallest such member of $ P $),
hence again $i$
is a member of $\left( \mathcal{T} \left( A,H\right)
\,\vdash \mathcal{T} \left( \mathcal{Y} \left( P,A\right)
,H\right) \right) $, and if $i$ is a member of $\left(
\mathcal{T} \left( A,H\right) \,\vdash \mathcal{T} \left(
\mathcal{Y} \left( P,A\right) ,H\right) \right) $, then $i$
is a member of $\mathcal{T} \left( A,H\right) $ but
\emph{not} a member of $\mathcal{T} \left( \mathcal{Y}
\left( P,A\right) ,H\right) $, hence either $\mathcal{Y}
\left( P,A\right) =\emptyset $ holds, in which
\label{Start of original page 77}
 case there is \emph{no} member $C$ of $P$ such that
$A\subseteq C$ holds, hence in particular there is
\emph{no} member $C$ of $P$ such that $A\subseteq C$ and
$i\in \mathcal{T} \left( C,H\right) $ both hold, or else
$\mathcal{Y} \left( P,A\right) \neq \emptyset $ holds, in
which
case there exists at least one member $C$ of $P$ such that
$A\subseteq C$ holds, and $\mathcal{Y} \left( P,A\right) $
is the \emph{smallest} member $C$ of $P$ such that
$A\subseteq C$ holds, and $i$ is \emph{not} a member of
$\mathcal{T} \left( \mathcal{Y} \left( P,A\right) ,H\right)
$, hence since $i\in \mathcal{Y} \left( P,A\right) $
certainly holds in this case, there exists a member $B$ of
$H$ such that $i\in B$ and $B\subseteq \mathcal{Y} \left(
P,A\right) $ both hold, hence for \emph{every} member $C$
of $P$ such that $A\subseteq C$ holds, there exists a
member $B$ of $H$ such that $i\in B$ and $B\subseteq C$
both hold, hence again there is \emph{no} member $C$ of $P$
such that $A\subseteq C$ and $i\in \mathcal{T} \left(
C,H\right) $ both hold.

\begin{bphzlemma} \label{Lemma 21}
\end{bphzlemma}
\vspace{-6.143ex}

\noindent \hspace{11.9ex}{\bf.  }Let $P$ be any
wood, $Q$ be any wood such that
$\mathcal{M} \left( Q\right) =\mathcal{M} \left( P\right) $
holds and $P\subseteq Q$ holds, $F$ be any member of
$\mathbb{K} \left( P,Q\right) $, and $H$ be any set such
that every member of $H$ is a set.   Then
$ \bigcup_{A\in \left(
F\,\vdash P\right) } \mathbb{J} _{ A } \left(
P,Q,H\right) $ is equal to $\left( \bigcup_{A\in
 \mathbb{B} \left( F\right)
 } \mathbb{I} _{ A } \left( Q,H\right) \right) \,\vdash
 \left(
 \bigcup_{A\in \mathbb{B} \left( P\right) } \mathbb{I} _{ A
} \left( Q,H\right) \right) $, and is the set whose members
 are all
the ordered pairs $\left(i,B\right) $ of
a member $i$ of $\mathcal{O}
\left( Q,H\right) $ and a member $B$ of $\mathbb{Y} \left(
Q,\mathcal{Z} \left( P,H,i\right) ,\mathcal{Z} \left(
F,H,i\right) \right) $.

\vspace{2.5ex}

\noindent {\bf Proof.}  For any member $A$ of
$\left( Q\,\vdash P\right) $,
$\mathbb{J} _{ A } \left( P,Q,H\right) $ is by definition
equal to \\
$\left( \mathbb{I} _{ A } \left( Q,H\right)
\,\vdash  \bigcup_{B\in \mathbb{B} \left( P\right) }
\mathbb{I} _{ B } \left( Q,H\right) \right) $.   Now
suppose $\alpha $ is a member of $\mathbb{I} _{ A } \left(
Q,H\right) $ for some member $A$ of $\mathbb{B} \left(
F\right) $, and that $\alpha $ is \emph{not} a member of
$ \bigcup_{B\in \mathbb{B} \left( P\right) } \mathbb{I} _{
B } \left( Q,H\right) $.   Then $A$ is not a member of
$\mathbb{B} \left( P\right) $ hence $A$ is a member of
$\left( F\,\vdash P\right) $, and $\alpha $ is a member of
$\mathbb{J} _{ A } \left( P,Q,H\right) $, hence is a member
of $ \bigcup_{A\in \left( F\,\vdash P\right) } \mathbb{J}
_{ A } \left( P,Q,H\right) $.   Now let $\alpha $ be any
member of $ \bigcup_{B\in \mathbb{B} \left( P\right) }
\mathbb{I} _{ B } \left( Q,H\right) $.   Then $\alpha $ is
\emph{not} a member of $\mathbb{J} _{ A } \left(
P,Q,H\right) $ for any member $A$ of $\left( F\,\vdash
P\right) $, hence $\alpha $ is \emph{not} a member of
$ \bigcup_{A\in
\left( F\,\vdash P\right) } \mathbb{J} _{ A }
\left( P,Q,H\right) $.
\enlargethispage{0.8ex}

Finally we note that it follows directly
from Lemma \ref{Lemma 20}
that $\left( \bigcup_{A\in \mathbb{B} \left( F\right) }
\mathbb{I} _{ A } \left( Q,H\right) \right) \,\vdash
\left( \bigcup_{A\in
\mathbb{B} \left( P\right) } \mathbb{I} _{ A }
\left( Q,H\right) \right) $ is the set whose members are
all the
ordered pairs $\left(i,B\right) $ of a member $i$ of
$\mathcal{O} \left(
Q,H\right) $ and a member $B$ of $\mathbb{Y} \left(
Q,\mathcal{Z} \left( P,H,i\right) ,\mathcal{Z} \left(
F,H,i\right) \right) $.

\vspace{2.5ex}

For any ordered quintuple $\left(P,Q,H,x,r\right) $ of
 a wood $P$, a wood
$Q$
\label{Start of original page 78}
 such that $\mathcal{M} \left( Q\right) =\mathcal{M} \left(
P\right) $ holds and $P\subseteq Q$ holds, a set $H$ such
that every member of $H$ is a set, a member $x$ of
$\mathbb{F} _{ d } \left( \mathcal{M} \left( P\right)
\right) $, where $d$ is an integer $\geq 1$, and a member
$r$ of $\mathbb{R}^{ \mathcal{U} \left( \mathcal{R} \left(
\mathbb{J} \left( P,Q,H\right) \right) \right) } $, (or in
other words, a map $r$ whose domain is the set of all
ordered pairs $\left(i,A\right) $ of a
member $i$ of $\mathcal{O} \left(
P,H\right) =\mathcal{O} \left( Q,H\right) $ and a member
$A$ of $\mathbb{Y} \left( Q,\mathcal{Z} \left( P,H,i\right)
,\mathcal{Z} \left( Q,H,i\right) \right) $, and whose range
is a subset of $\mathbb{R}  $), we define the member
$\mu \left( P,Q,H,x,r\right) $ of $\mathbb{E} _{
d }^{ \mathcal{U} \left( P\right) } $ by:
\[
\mu _{ i } \left( P,Q,H,x,r\right) \equiv
\left\{ \begin{array}{cr}
x_{ \mathcal{Z}
\left( Q,H,i\right) } + \hspace{-0.6cm}
 \displaystyle\sum_{A\in \mathbb{Y} \left(
Q,\mathcal{Z} \left( P,H,i\right) ,\mathcal{Z} \left(
Q,H,i\right) \right) } \hspace{-0.6cm}
r_{ iA } \left( x_{
\mathcal{K} \left( Q,A,i\right) } -x_{ A } \right) &
\textrm{if }i\in
\mathcal{O} \left( P,H\right) \\
 x_{ \mathcal{C} \left(
P,i\right) } &
\hspace{-1.6cm}
\textrm{if }i\in \left( \mathcal{U} \left( P\right)
\,\vdash \mathcal{O} \left( P,H\right) \right)
\end{array} \right.
\]
where $\mu _{ i } \left( P,Q,H,x,r\right) \equiv \left( \mu
\left( P,Q,H,x,r\right) \right) _{ i } $.

For every ordered triple $\left(F,H,x\right) $ of
a wood $F$, a set $H$
such that every member of $H$ is a set, and a member $x$ of
$\mathbb{F} _{ d } \left( \mathcal{M} \left( F\right)
\right) $, where $d$ is an integer $\geq 1$, we define the
member $\eta \left( F,H,x\right) $ of $\mathbb{E} _{
d }^{ \mathcal{U} \left( F\right) } $ by:
\[
\eta _{ i } \left( F,H,x\right) \equiv \left( \eta \left(
F,H,x\right) \right) _{ i } \equiv
\left\{ \begin{array}{cl}
x_{ \mathcal{Z} \left(
F,H,i\right) } & \hspace{0.5cm}
\textrm{if }i\in \mathcal{O} \left( F,H\right) \\
 x_{
\mathcal{C} \left( F,i\right) } & \hspace{0.5cm}
\textrm{if }i\in \left( \mathcal{U}
\left( F\right) \,\vdash \mathcal{O} \left( F,H\right)
\right)
\end{array} \right.
\]

We observe that if $P$ and $Q$ are any woods such that
$\mathcal{M} \left( Q\right) =\mathcal{M} \left( P\right) $
and $P\subseteq Q$ both hold, $F$ is any member of
$\mathbb{K} \left( P,Q\right) $, $H$ is any set such that
every member of $H$ is a set, $x$ is any member of
$\mathbb{F} _{ d } \left( \mathcal{M} \left( P\right)
\right) $, where $d$ is an integer $\geq 1$, and $r$ is the
particular member of $\mathbb{R}^{ \mathcal{U} \left(
\mathcal{R} \left( \mathbb{J} \left( P,Q,H\right) \right)
\right) } $ such that for each member $i$ of $\mathcal{O}
\left( P,H\right) $, $r_{ iB } =0$ holds for all members
$B$ of $\mathbb{Y} \left( Q,\mathcal{Z} \left( P,H,i\right)
,\mathcal{Z} \left( F,H,i\right) \right) $, and $r_{ iB }
=1$ holds for all members $B$ of $\mathbb{Y} \left(
Q,\mathcal{Z} \left( F,H,i\right) ,\mathcal{Z} \left(
Q,H,i\right) \right) $, or in other words, if
$r=\mathcal{U} \left( \left\{ 0\right\}^{ \mathcal{U} \left(
\mathcal{R} \left( \mathbb{J} \left( P,F,H\right) \right)
\right) } \cup \left\{ 1\right\}^{ \mathcal{U} \left(
\mathcal{R} \left( \mathbb{J} \left( F,Q,H\right) \right)
\right) } \right) $, then the equation $\mu \left(
P,Q,H,x,r\right) =\eta \left( F,H,x\right) $ holds.

For any ordered pair $\left(Q,H\right) $ of a
 wood $Q$ and a set $H$ such
that every member of $H$ is a set, we define $\mathbb{G}
\left( Q,H\right) $ to be the map whose domain is the set
of all ordered pairs $\left(i,B\right) $ of
a member $i$ of $\mathcal{O}
\left( Q,H\right) $ and a member $B$ of $Q$, and such that
for each member $\left(i,B\right) $ of $\mathcal{D}
\left( \mathbb{G}
\left( Q,H\right) \right) $, $\mathbb{G}_{ iB }\left(
Q,H\right) $ is the set whose members are all the members
$A$ of $Q$ such that $B\subseteq A$ and $A\subseteq
\mathcal{Z} \left( Q,H,i\right) $ both hold.   (Thus
$\mathbb{G}_{ iB }\left( Q,H\right) $ is the empty set
$\emptyset $
if $B\subseteq \mathcal{Z} \left( Q,H,i\right) $ does not
hold.)

We observe that if $P$ and $Q$ are any woods such that
\label{Start of original page 79}
 $\mathcal{M} \left( Q\right) =\mathcal{M} \left( P\right)
$ and $P\subseteq Q$ both hold, $H$ is any set such that
every member of $H$ is a set, and $(i,B) $ is any member of
$\mathcal{U} \left( \mathcal{R} \left( \mathbb{J} \left(
P,Q,H\right) \right) \right) $, then $\mathbb{G}_{ iB
}\left(
Q,H\right) $ is the set whose members are all the members
$A$ of $\left( Q\,\vdash P\right) $ such that $\left(
i,B\right) \in \mathbb{J} _{ A } \left( P,Q,H\right) $
holds.   For if $A$ is any member of $\left( Q\,\vdash
P\right) $, then by definition, $\mathbb{J} _{ A } \left(
P,Q,H\right) $ is the set of all the ordered
pairs $\left(i,B\right) $ of
a member $i$ of $\mathcal{T} \left( A,H\right) $ such that
there is no member $C$ of $P$ such that $A\subseteq C$
holds and $i\in \mathcal{T} \left( C,H\right) $ holds, and
a member $B$ of $\mathbb{Y} \left( Q,\mathcal{Z} \left(
P,H,i\right) ,A\right) $.   And $\mathcal{U} \left(
\mathcal{R} \left( \mathbb{J} \left( P,Q,H\right) \right)
\right) $ is the set of all ordered
pairs $\left(i,B\right) $ of a member
$i$ of $\mathcal{O} \left( P,H\right) =\mathcal{O} \left(
Q,H\right) $, and a member $B$ of $\mathbb{Y} \left(
Q,\mathcal{Z} \left( P,H,i\right) ,\mathcal{Z} \left(
Q,H,i\right) \right) $.

Let $(i,B) $ be any member of $\mathcal{U} \left(
\mathcal{R}
\left( \mathbb{J} \left( P,Q,H\right) \right) \right) $,
and suppose first that $A$ is any member of $\mathbb{G}_{
iB }\left( Q,H\right) $.   Then $B\subseteq A\subseteq
\mathcal{Z} \left( Q,H,i\right) $ holds, hence \\
$B\in
\mathbb{Y} \left( Q,\mathcal{Z} \left( P,H,i\right)
,A\right) $ holds, and furthermore $\mathcal{Z} \left(
P,H,i\right) \subset A$ holds.   And furthermore $i\in
\mathcal{T} \left( A,H\right) $ holds, for $i\in
\mathcal{Z} \left( P,H,i\right) $ holds hence $i\in A$
holds, and if there was a member $E$ of $H$ such that $i\in
E$ and $E\subseteq A$ both held, then $E$ would be a member
of $H$ such that $i\in E$ and $E\subseteq \mathcal{Z}
\left( Q,H,i\right) $ both held, which is impossible.   And
furthermore there is no member $C$ of $P$ such that
$A\subseteq C$ and $i\in \mathcal{T} \left( C,H\right) $
both hold, for if $C$ is any member of $P$ such that $i\in
\mathcal{T} \left( C,H\right) $ holds, then $C\subseteq
\mathcal{Z} \left( P,H,i\right) $ holds, hence $\mathcal{Z}
\left( P,H,i\right) \subset A$ implies $C\subset A$.
Hence $\left( i,B\right) \in \mathbb{J} _{ A } \left(
P,Q,H\right) $ holds.

Now let $A$ be any member of $\left( Q\,\vdash P\right) $
such that $\left( i,B\right) \in \mathbb{J} _{ A } \left(
P,Q,H\right) $ holds.   Then $i\in \mathcal{T} \left(
A,H\right) $ holds, hence $A\subseteq \mathcal{Z} \left(
Q,H,i\right) $ holds, and $B$ is a member of \\
$\mathbb{Y}
\left( Q,\mathcal{Z} \left( P,H,i\right) ,A\right) $, hence
$B\subseteq A$ holds, hence $A\in \mathbb{G}_{ iB }\left(
Q,H\right) $ holds.

For any ordered quadruple $\left( P,Q,H,\rho \right) $ of a
wood $P$, a wood $Q$ such that $\mathcal{M} \left( Q\right)
=\mathcal{M} \left( P\right) $ holds and $P\subseteq Q$
holds, a set $H$ such that every member of $H$ is a set,
and a member $\rho $ of $\mathbb{R}^{ \left( Q\,\vdash
P\right) } $, we define $\mathcal{X} \left( P,Q,H,\rho
\right) $ to be the member of $\mathbb{R}^{ \mathcal{U}
\left( \mathcal{R} \left( \mathbb{J} \left( P,Q,H\right)
\right) \right) } $ such that for each member $(i,B) $ of
$\mathcal{U} \left( \mathcal{R} \left( \mathbb{J} \left(
P,Q,H\right) \right) \right) $,
\[
\mathcal{X} _{ iB } \left( P,Q,H,\rho \right) \equiv \left(
\mathcal{X} \left( P,Q,H,\rho \right) \right) _{ iB }
\equiv  \prod_{A\in \mathbb{G}_{ iB }\left( Q,H\right) }
\rho
_{ A } .
\]

We observe that if $P$ and $Q$ are any woods such that
$\mathcal{M} \left( Q\right) =\mathcal{M} \left( P\right) $
and $P\subseteq Q$ both hold, $F$ is any member of
$\mathbb{K} \left( P,Q\right) $, $H$ is any set such that
every member of $H$ is a set, $x$ is any member of
$\mathbb{F} _{ d } \left( \mathcal{M} \left( P\right)
\right) $, where $d$ is an integer $\geq 1$, and $\rho $ is
the particular member of $\mathbb{R}^{ \left( Q\,\vdash
P\right) } $ such that $\rho _{ A } =0$ holds for every
member $A$ of
\label{Start of original page 80}
 $\left( F\,\vdash P\right) $, and $\rho _{ A } =1$ holds
for every member $A$ of $\left( Q\,\vdash F\right) $, or in
other words, if $\rho =\mathcal{U} \left( \left\{ 0\right\}
^{
\left( F\,\vdash P\right) } \cup \left\{ 1\right\}^{ \left(
Q\,\vdash F\right) } \right) $, then $\mathcal{X} _{ iB }
\left( P,Q,H,\rho \right) =0$ holds for every member $
\left(i,B\right) $
of $\mathcal{U} \left( \mathcal{R} \left( \mathbb{J} \left(
P,Q,H\right) \right) \right) $ such that $B\in \mathbb{Y}
\left( Q,\mathcal{Z} \left( P,H,i\right) ,\mathcal{Z}
\left( F,H,i\right) \right) $ holds, and $\mathcal{X} _{ iB
} \left( P,Q,H,\rho \right) =1$ holds for every member
$ \left(i,B\right) $ of $\mathcal{U} \left( \mathcal{R}
\left( \mathbb{J}
\left( P,Q,H\right) \right) \right) $ such that $B\in
\mathbb{Y} \left( Q,\mathcal{Z} \left( F,H,i\right)
,\mathcal{Z} \left( Q,H,i\right) \right) $ holds, hence the
equation \\
$\mu \left( P,Q,H,x,\mathcal{X} \left(
P,Q,H,\mathcal{U} \left( \left\{ 0\right\}^{ \left(
F\,\vdash
P\right) } \cup \left\{ 1\right\}^{ \left( Q\,\vdash
F\right) } \right) \right) \right) =\eta \left(
F,H,x\right) $ holds.

We observe that if $P$ and $Q$ are any woods such that
$\mathcal{M} \left( Q\right) =\mathcal{M} \left( P\right) $
holds and $P\subseteq Q$ holds, $H$ is any set such that
every member of $H$ is a set, $x$ is any member of
$\mathbb{F} _{ d } \left( \mathcal{M} \left( P\right)
\right) $, where $d$ is an integer $\geq 1$, $i$ is any
member of $\mathcal{O} \left( P,H\right) $, and $\rho $ is
any member of $\mathbb{R}^{ \left( Q\,\vdash P\right) } $
such that $0\leq \rho _{ A } \leq 1$ holds for every member
$A$ of $\left( Q\,\vdash P\right) $, then $\mu _{ i }
\left( P,Q,H,x,\mathcal{X} \left( P,Q,H,\rho \right)
\right) $ is a member of the convex hull, (in $\mathbb{E}
_{ d }  $), of all the $x_{ A } $, $A\in \left(
\left\{ \mathcal{Z} \left( P,H,i\right) \right\} \cup
\mathbb{Y} \left( Q,\mathcal{Z} \left( P,H,i\right)
,\mathcal{Z} \left( Q,H,i\right) \right) \right) $.   (We
note that $\left( \left\{ \mathcal{Z} \left( P,H,i\right)
\right\} \cup \mathbb{Y} \left( Q,\mathcal{Z} \left(
P,H,i\right) ,\mathcal{Z} \left( Q,H,i\right) \right)
\right) $ is the set of all the members $A$ of $Q$ such
that $\mathcal{Z} \left( P,H,i\right) \subseteq A\subseteq
\mathcal{Z} \left( Q,H,i\right) $ holds.)   For by
definition,
\[
\mu _{ i } \left( P,Q,H,x,\mathcal{X} \left( P,Q,H,\rho
\right) \right) =x_{ \mathcal{Z} \left( Q,H,i\right) }
+ \hspace{-1.6cm}
 \sum_{A\in \mathbb{Y} \left( Q,\mathcal{Z}
\left( P,H,i\right)
,\mathcal{Z} \left( Q,H,i\right) \right) } \hspace{-0.2cm}
 \left(
 \prod_{B\in \mathbb{G} _{ iA } \left( Q,H\right) }
 \hspace{-0.6cm} \rho _{
B } \right)
 \left( x_{ \mathcal{K} \left( Q,A,i\right) } -x_{ A }
\right)
\]
\[
=  \left( \prod_{B\in \mathbb{Y} \left( Q,\mathcal{Z} \left(
P,H,i\right) ,\mathcal{Z} \left( Q,H,i\right) \right)
 } \hspace{-0.6cm} \rho _{ B } \right) x_{ \mathcal{Z}
\left(
P,H,i\right) } + \hspace{-1.6cm} \sum_{A\in \mathbb{Y}
\left(
Q,\mathcal{Z}
\left( P,H,i\right) ,\mathcal{Z} \left( Q,H,i\right)
\right) }  \hspace{-0.2cm} \left(
\prod_{B\in \mathbb{Y} \left( Q,A,\mathcal{Z}
\left( Q,H,i\right) \right) } \rho _{ B } \right) \left(
1-\rho _{ A } \right) x_{ A } ,
\]
hence if we define the member $u$ of $\mathbb{R}^{ \left(
 \mathbb{G}_{
i\mathcal{Z} \left( P,H,i\right) } \left( Q,H\right)
\right) } =$ \\
$=\mathbb{R}^{ \left( \left\{ \mathcal{Z} \left(
P,H,i\right) \right\} \cup \mathbb{Y} \left( Q,\mathcal{Z}
\left( P,H,i\right) ,\mathcal{Z} \left( Q,H,i\right)
\right) \right) } $ by
\[
u_{ \mathcal{Z} \left( P,H,i\right) } \equiv \left(
 \prod_{B\in
\mathbb{Y} \left( Q,\mathcal{Z} \left( P,H,i\right)
,\mathcal{Z} \left( Q,H,i\right) \right) } \rho _{
B } \right)
\]
and by
\[
u_{ A } \equiv \left(
 \prod_{B\in \mathbb{Y} \left( Q,A,\mathcal{Z}
\left( Q,H,i\right) \right) } \rho _{ B } \right) \left(
1-\rho _{ A } \right)
\]
for all $A\in \mathbb{Y} \left( Q,\mathcal{Z} \left(
P,H,i\right) ,\mathcal{Z} \left( Q,H,i\right) \right) $, we
have that
\[
\mu _{ i } \left( P,Q,H,x,\mathcal{X} \left( P,Q,H,\rho
\right) \right) = \sum_{A\in \left(
 \mathbb{G}_{ i\mathcal{Z} \left(
P,H,i\right) } \left( Q,H\right) \right) } u_{ A } x_{ A }
\]
holds and that
\label{Start of original page 81}
\[
 \sum_{A\in \left( \mathbb{G}_{ i\mathcal{Z}
 \left( P,H,i\right) }
\left( Q,H\right) \right) } u_{ A } =1
\]
holds, and furthermore, since $0\leq \rho _{ B } \leq 1$
holds for all members $B$ of $\left( Q\,\vdash P\right) $,
that $u_{ A } \geq 0$ holds for all members $A$ of
$\mathbb{G} _{ i\mathcal{Z} \left( P,H,i\right) } \left(
Q,H\right) $.

We note that if $V$ is any partition such that $\mathcal{U}
\left( V\right) $ is finite and $\#\left( V\right) \geq 2$
holds, $H$ is any partition such that if $E$ is any member
of $H$ such that $E$ intersects \emph{more} than one member
of $V$, then $E$ has \emph{exactly} two members, $\sigma $
is any real number such that $0<\sigma \leq  \frac{ 1 }{ 8
} $ holds,
$R$ is any finite real number $>0$, $d$ is any integer
$\geq 1$, $x$ is any member of $\mathbb{F} _{ d } \left(
V\right) $ such that for every member $A$ of $\mathbb{B}
\left( \bar{ P } \right) $, and for every two
\emph{distinct}
members $B$ and $C$ of $\mathcal{P} \left( P,A\right) $
such that there exists a member $i$ of $B$ and a member $j$
of $C$ such that $\left\{ i,j\right\} $ is a member of $H$,
$\left| x_{ B } -x_{ C } \right| >0$ holds,
$\left(P,Q\right) $ is any
member of
$\Omega \left( H,\sigma ,R,x\right) $, and $y$ is any
member of $\mathbb{E} _{ d }^{ \mathcal{U}
 \left( V\right) } $
such that for every member $i$ of $\mathcal{O} \left(
V,H\right) $, $y_{ i } $ is a member of the convex hull,
(in $\mathbb{E} _{ d }  $), of all the $x_{ A } $,
$A\in \left( \left\{ \mathcal{Z} \left( P,H,i\right)
\right\}
\cup \mathbb{Y} \left( Q,\mathcal{Z} \left( P,H,i\right)
,\mathcal{Z} \left( Q,H,i\right) \right) \right) $, then it
follows directly from Lemma \ref{Lemma 14} that if $i$ and
$j$ are
any two members of $\mathcal{U} \left( V\right) $ such that
$\left\{ i,j\right\} $ is a member of $H$ and is \emph{not}
a
subset of any member of $V$, then $\left| y_{ i } -y_{ j }
\right| >0$
holds.   For if $i$ and $j$ are any two members of
$\mathcal{U} \left( V\right) $ such that $\left\{
i,j\right\}
$ is a member of $H$ and is \emph{not} a subset of any
member of $V$, then $\mathcal{Y} \left( \bar{ P } ,\left\{
i,j\right\} \right) $ is a member of $\mathbb{B}
\left( \bar{ P }
 \right) =\left( \bar{ P } \,\vdash V\right) $, and the
fact
that $H$ is a \emph{partition} implies that $\mathcal{Z}
\left( P,H,i\right) $ is the largest member of $P$ that has
$i$ as a member but does \emph{not} have $j$ as a member,
hence $\mathcal{Z} \left( P,H,i\right) =\mathcal{K} \left(
P,\mathcal{Y} \left( \bar{ P } ,\left\{ i,j\right\} \right)
,i\right) $ holds, and $\mathcal{Z} \left( P,H,j\right) $
is the largest member of $P$ that has $j$ as a member but
does \emph{not} have $i$ as a member, hence $\mathcal{Z}
\left( P,H,j\right) =\mathcal{K} \left( P,\mathcal{Y}
\left( \bar{ P } ,\left\{ i,j\right\}
\right) ,j\right) $ holds,
hence $\mathcal{Z} \left( P,H,i\right) $ and $\mathcal{Z}
\left( P,H,j\right) $ are distinct members of $\mathcal{P}
\left( P,\mathcal{Y} \left( \bar{ P } ,\left\{ i,j\right\}
\right) \right) $, hence by assumption $\left| x_{
\mathcal{Z}
\left( P,H,i\right) } -x_{ \mathcal{Z} \left( P,H,j\right)
 } \right| >0$ holds, hence by Lemma \ref{Lemma 14},
$\left| y_{ i }
-y_{ j } \right| >0$
holds.

If $d$ is an integer $\geq 1$ and $A$ is a finite set then
an \emph{open subset of }$\mathbb{E}_{ d }^{ A } $ is a
subset $S$ of
$\mathbb{E} _{ d }^{ A } $ such that for every member $x$
of $S$
there exists a real number $\varepsilon >0$ such that every
member $y$ of $\mathbb{E} _{ d }^{ A } $ such that
$\left| y_{ i }
-x_{ i } \right| \leq \varepsilon $ holds for every member
$i$ of
$A$, is a member of $S$.

For any ordered quadruple $\left(P,Q,H,u\right) $ of
a wood $P$, a wood
$Q$ such that $\mathcal{M} \left( Q\right) =\mathcal{M}
\left( P\right) $ holds and $P\subseteq Q$ holds, a set $H$
such that every member of $H$ is a set, and a member $u$ of
$\mathbb{N}^{ \mathcal{U} \left( \mathcal{R} \left(
\downarrow \left( \mathbb{I} \left( Q,H\right) ,P\right)
\right) \right) } $, we
\label{Start of original page 82}
 define $\xi \left( P,Q,H,u\right) $ to be the map whose
domain is $\left( Q\,\vdash P\right) $, and such that for
each member $A$ of $\left( Q\,\vdash P\right) $, the
equation
\[
\xi _{ A } \left( P,Q,H,u\right) \equiv \left( \xi \left(
P,Q,H,u\right) \right) _{ A } \equiv  \sum_{\left(
i,B\right)
\in \left( \mathbb{I} _{ A } \left( Q,H\right) \cap
\mathcal{U} \left( \mathcal{R} \left( \downarrow \left(
\mathbb{I} \left( Q,H\right) ,P\right) \right) \right)
\right) } u_{ iB }
\]
holds.

\begin{bphzlemma} \label{Lemma 22}
\end{bphzlemma}
\vspace{-6.143ex}

\noindent \hspace{11.9ex}{\bf.  }Let $P$ and $Q$ be any
woods such that
$\mathcal{M} \left( Q\right) =\mathcal{M} \left( P\right) $
holds and $P\subseteq Q$ holds, let $H$ be any partition,
let $d$ be any integer $\geq 1$, let $S$ be any finite real
number $>0$, let $D$ be any map such that $\mathbb{B}
\left( Q\right) \subseteq \mathcal{D} \left( D\right) $
holds and $\mathcal{R} \left( D\right) \subseteq \mathbb{Z}
$ holds, and let $1\hspace{-0.7516ex}1
 $ be a map such
that $\left( Q\,\vdash P\right) \subseteq \mathcal{D}
\left( 1\hspace{-0.7516ex}1 \right) $ holds and such
that for each member $A$ of $\left( Q\,\vdash P\right) $,
$1\hspace{-0.7516ex}1 _{ A } =1$ holds.

Let $\mathcal{J} $ be a map such that $\mathbb{K} \left(
P,Q\right) \subseteq \mathcal{D} \left( \mathcal{J} \right)
$ holds, and such that for each member $F$ of $\mathbb{K}
\left( P,Q\right) $, $\mathcal{J} _{ F } $ is a map whose
domain is a subset of $\mathbb{E} _{ d }^{\mathcal{U} \left(
P\right) } $ and whose range is a subset of $\mathbb{R} $.

For each member $F$ of $\mathbb{K} \left( P,Q\right) $, and
for each member $y$ of $\mathcal{D} \left( \mathcal{J} _{ F
} \right) $, we define $\mathcal{J} _{ F } \left( y\right)
\equiv \left( \mathcal{J} _{ F } \right)_{ y } $.
\enlargethispage{4.0ex}

Let\hspace{\stretch{1}} $\mathcal{J} $\hspace{\stretch{1}}
satisfy\hspace{\stretch{1}} the\hspace{\stretch{1}}
requirement\hspace{\stretch{1}} that\hspace{\stretch{1}}
if\hspace{\stretch{1}} $F$\hspace{\stretch{1}}
and\hspace{\stretch{1}} $G$\hspace{\stretch{1}}
are\hspace{\stretch{1}} any\hspace{\stretch{1}}
members\hspace{\stretch{1}} of\hspace{\stretch{1}}
$\mathbb{K} \left( P,Q\right) $,\hspace{\stretch{1}} and
\newpage
\noindent
$y$ is any member of $\mathbb{E} _{ d }^{ \mathcal{U} \left(
P\right) _{ } } $, such that for every two-member member
$\left\{ i,j\right\} $ of $H$ such that $\left\{ i,j\right\}
\subseteq \mathcal{O} \left( P,H\right) $ holds, either
$\left| y_{ i } -y_{ j } \right| \leq S$ holds or
$\mathcal{Y} \left(
F,\left\{ i,j\right\} \right) =\mathcal{Y} \left( G,\left\{
i,j\right\} \right) $ holds, then $y$ is a member of
$\mathcal{D} \left( \mathcal{J} _{ F } \right) $ ifif $y$
is a member of $\mathcal{D} \left( \mathcal{J} _{ G }
\right) $, and if $y$ \emph{is} a member of $\mathcal{D}
\left( \mathcal{J} _{ F } \right) $, then $\mathcal{J} _{ F
} \left( y\right) =\mathcal{J} _{ G } \left( y\right) $
holds.

Let $x$ be any member of $\mathbb{F} _{ d } \left(
\mathcal{M} \left( P\right) \right) $ such that $\mathbb{L}
\left( P,A,x\right) <S$ holds for every member $A$ of
$\left( Q\,\vdash P\right) $, and such that there exists an
open subset $Z$ of $\mathbb{E} _{ d }^{ \mathcal{O} \left(
P,H\right) } $ that satisfies the following two
requirements:

\vspace{1.0ex}

\noindent (i) If $y$ is any member of $\mathbb{E} _{ d }^{
\mathcal{U}
\left( V\right) } $, such that for every member $i$
of $\mathcal{O} \left( P,H\right) $, $y_{ i } $ is a member
of the convex hull, (in $\mathbb{E} _{ d }  $), of
all the $x_{ A } $, $A\in \left( \left\{ \mathcal{Z} \left(
P,H,i\right) \right\} \cup \right. $ \\
$ \left. \mathbb{Y} \left( Q,\mathcal{Z}
\left( P,H,i\right) ,\mathcal{Z} \left( Q,H,i\right)
\right) \right) $, then $\downarrow \left( y,\mathcal{O}
\left( P,H\right) \right) $ is a member of $Z$.

\vspace{1.0ex}

\noindent (ii) $\mathcal{J}_{ Q } \left( y\right) $ and
all its derivatives with
respect to the $y_{ i } $, $i\in \mathcal{O} \left(
P,H\right) $, of degree up to and including $ \sum_{A\in
\mathbb{B} \left( Q\right) } \left( \max\left( D_{ A
} ,0\right) +1\right) $, exist and are continuous for all
members $y$ of $\mathbb{E} _{ d }^{ \mathcal{U} \left(
P\right)
} $ such that $\downarrow \left( y,\mathcal{O} \left(
P,H\right) \right) \in Z$ holds and \\
$\downarrow \left(
y,\left( \mathcal{U} \left( P\right) \,\vdash \mathcal{O}
\left( P,H\right) \right) \right) =\downarrow \left( \eta
\left( F,H,x\right) ,\left( \mathcal{U} \left( P\right)
\,\vdash \mathcal{O} \left( P,H\right) \right) \right) =x_{
\mathcal{C} \left( P,i\right) } $ holds.

\vspace{1.0ex}

Let $\mathbb{D} $ be the set whose members are all the
members $\rho $ of
\label{Start of original page 83}
 $\mathbb{R}^{ \left( Q\,\vdash P\right) } $ such that
$0\leq
\rho _{ A } \leq 1$ holds for every member $A$ of $\left(
Q\,\vdash P\right) $.

Then the following equation holds:
\[
 \sum_{F\in \mathbb{K} \left( P,Q\right) } \left( -1\right)
^{ \#\left( F\,\vdash P\right) } \sum_{n\in \mathbb{X}
\left(
\mathbb{I} \left( F,H\right) ,D\right) } \times
\hspace{-0.8pt} \hspace{9.5cm}
\]
\[
\times \left( \left(
  \prod_{\left(
i,A\right) \in \mathcal{U} \left( \mathcal{R} \left(
\mathbb{I} \left( F,H\right) \right) \right) }
\left( \frac{\left( \left( x_{ \mathcal{K} \left(
F,A,i\right) } -x_{ A
} \right) .\hat{ y }_{ i } \right)^{ n_{ iA } } }{n_{ iA }
! } \right) \right)
 \mathcal{J} _{ F } \left( y\right) \right) _{ y=\eta \left(
F,H,x\right) } =
\]
\[
=  \sum_{u\in \mathbb{X} \left( \downarrow \left( \mathbb{I}
\left( Q,H\right) ,P\right) ,D\right) }
\hspace{1.5ex}   \sum_{m\in
\mathbb{A} \left( \mathbb{J} \left( P,Q,H\right) ,\left(
D-\xi \left( P,Q,H,u\right) + 1\hspace{-0.7516ex}1
\right) \right) } \int_{ \mathbb{D} } \left( d^{ \#\left(
Q\,\vdash P\right) } \rho \right) \times \hspace{2.5cm}
\]
\[
\times \left( \left(
\prod_{A\in \left( Q\,\vdash
P\right) } \left( 1-\rho_{ A }\right)^{ \left( D_{ A }
-\xi _{ A } \left( P,Q,H,u\right) \right) } \left( D_{ A }
-\xi _{ A } \left( P,Q,H,u\right) +1\right) \right) \right.
\times
\]
\[
\times \left(
 \prod_{\left(
\left(
i,B\right) ,X\right) \in \mathcal{U} \left( \mathcal{R}
\left( \psi \left( \mathbb{J} \left( P,Q,H\right) \right)
\right) \right) } \left( \prod_{E\in \left( \mathbb{G} _{
iB }
\left( Q,H\right) \,\vdash X\right) } \rho _{ E } \right)^{
m_{
iBX } } \right) \times
\]
\[
\times \left( \left(
\prod_{\left( \left( i,B\right) ,X\right) \in
\mathcal{U} \left( \mathcal{R} \left( \psi \left(
\mathbb{J} \left( P,Q,H\right) \right) \right) \right) }
\left( \frac{\left( \left( x_{ \mathcal{K} \left(
 Q,B,i\right)
 } -x_{ B } \right) .\hat{ y }_{ i } \right)^{
  m_{ iBX } } }{m_{
iBX
} ! } \right) \right) \times \right.
\]
\[
\left. \left. \times \left(
    \prod_{\left( i,A\right) \in \mathcal{U} \left(
\mathcal{R}
\left( \downarrow \left( \mathbb{I} \left( Q,H\right)
,P\right) \right) \right) }
\left( \frac{\left( \left( x_{
\mathcal{K} \left( Q,A,i\right) } -x_{ A } \right) .\hat{ y
}_{ i
} \right)^{ u_{ iA } } }{u_{ iA } ! } \right) \right)
 \mathcal{J} _{ Q } \left(
y\right) \right)_{ y=\mu \left( P,Q,H,x,\mathcal{X} \left(
P,Q,H,\rho \right) \right) } \right)
\]

\vspace{2.5ex}

\noindent {\bf Proof.}  We first note that our
assumptions on $\mathcal{J}
$ imply that if $y$ is any member of $\mathbb{E} _{
d }^{ \mathcal{U} \left( P\right) } $ such that
$\left| y_{ i }
-y_{
j } \right| <S$ holds for every two-member member $\left\{
i,j\right\} $ of $H$ such that $\left\{ i,j\right\}
\subseteq
\mathcal{O} \left( P,H\right) $ and $\mathcal{Y} \left(
Q,\left\{ i,j\right\} \right) \in \left( Q\,\vdash P\right)
$
both hold, then there exists a real number $\varepsilon >0$
such that for every member $F$ of $\mathbb{K} \left(
P,Q\right) $, and for every member $z$ of $\mathbb{E} _{
d }^{ \mathcal{U} \left( P\right) } $ such that
$\left| y_{ i }
-z_{
i } \right| \leq \varepsilon $ holds for every member $i$ of
$\mathcal{O} \left( P,H\right) $, $z$ is a member of
$\mathcal{D} \left( \mathcal{J} _{ F } \right) $ ifif $z$
is a member of $\mathcal{D} \left( \mathcal{J} _{ Q }
\right) $, and if $z$ \emph{is} a member of $\mathcal{D}
\left( \mathcal{J} _{ Q } \right) $, then $\mathcal{J} _{ F
} \left( z\right) =\mathcal{J} _{ Q } \left( z\right) $
holds.

For let $\varepsilon $ equal half the minimum value taken
by $\left( S-\left| y_{ i } -y_{ j } \right| \right) $ when
$\left\{
i,j\right\} $ is allowed to be any two-member member of $H$
such that $\left\{ i,j\right\} \subseteq \mathcal{O} \left(
P,H\right) $ and $\mathcal{Y} \left( Q,\left\{ i,j\right\}
\right) \in \left( Q\,\vdash P\right) $ both hold.   Then
if $z$ is any member of $\mathbb{E} _{ d }^{ \mathcal{U}
\left(
P\right) } $ such that $\left| y_{ i } -z_{ i } \right| \leq
\varepsilon $ holds for every member $i$ of $\mathcal{O}
\left( P,H\right) $, and $\left\{ i,j\right\} $ is any
two-member member of $H$ such that $\left\{ i,j\right\}
\subseteq \mathcal{O} \left( P,H\right) $ and $\mathcal{Y}
\left( Q,\left\{ i,j\right\} \right) \in \left( Q\,\vdash
P\right) $ both hold, we find by the triangle inequality
that
\[
\left| z_{ i } -z_{ j } \right| \leq \left| z_{ i } -y_{ i
} \right| +\left| y_{ i } -y_{ j
} \right| +\left| y_{ j } -z_{ j } \right| \leq \varepsilon
+\left(
S-2\varepsilon \right) +\varepsilon =S
\]
holds.

Now let $z$ be any member of $\mathbb{E} _{ d }^{
\mathcal{U}
\left( P\right) } $ such that $\left| z_{ k } -z_{ m }
\right| \leq S$
holds for every two-member member $\left\{ k,m\right\} $ of
$H$ such that $\left\{ k,m\right\} \subseteq \mathcal{O}
\left( P,H\right) $ and $\mathcal{Y} \left( Q,\left\{
k,m\right\} \right) \in \left( Q\,\vdash P\right) $ both
hold, and let $\left\{ i,j\right\} $ be any two-member
member
of $H$ such that $\left\{ i,j\right\} \subseteq \mathcal{O}
\left( P,H\right) $ holds.   Then either
\label{Start of original page 84}
 $\mathcal{Y} \left( Q,\left\{ i,j\right\} \right) \in
\left(
Q\,\vdash P\right) $ holds or $\mathcal{Y} \left( Q,\left\{
i,j\right\} \right) \in P$ holds or $\mathcal{Y} \left(
Q,\left\{ i,j\right\} \right) \notin Q$ holds.

Suppose first that $\mathcal{Y} \left( Q,\left\{ i,j\right\}
\right) \in \left( Q\,\vdash P\right) $ holds.   Then by
the stated condition on $z$, $\left| z_{ i } -z_{ j }
\right| \leq S$
holds.

Now suppose that $\mathcal{Y} \left( Q,\left\{ i,j\right\}
\right) \in P$ holds.   Then for every member $F$ of
$\mathbb{K} \left( P,Q\right) $, $\mathcal{Y} \left(
F,\left\{ i,j\right\} \right) =\mathcal{Y} \left( Q,\left\{
i,j\right\} \right) $ holds.   For $\mathcal{Y} \left(
Q,\left\{ i,j\right\} \right) \in P$ implies $\mathcal{Y}
\left( Q,\left\{ i,j\right\} \right) \in F$ hence, since
$\mathcal{Y} \left( Q,\left\{ i,j\right\} \right) $ is the
smallest member \emph{of }$ Q $ to contain $i$ and $j$,
it is
certainly the smallest member of $F$ to contain $i$ and
$j$.   Hence $\mathcal{Y} \left( F,\left\{ i,j\right\}
\right) =\mathcal{Y} \left( Q,\left\{ i,j\right\} \right) $
holds.

And finally suppose that $\mathcal{Y} \left( Q,\left\{
i,j\right\} \right) \notin Q$ holds, or in other words,
that \\
$\mathcal{Y} \left( Q,\left\{ i,j\right\} \right)
=\emptyset $
holds.   This implies that \emph{no} member of $Q$ contains
both $i$ and $j$, hence for every member $F$ of $\mathbb{K}
\left( P,Q\right) $, \emph{no} member of $F$ contains both
$i$ and $j$, hence $\mathcal{Y} \left( F,\left\{ i,j\right\}
\right) =\mathcal{Y} \left( Q,\left\{ i,j\right\} \right)
=\emptyset $ holds.

Hence for every member $F$ of $\mathbb{K} \left( P,Q\right)
$, and for every two-member member $\left\{ i,j\right\} $ of
$H$ such that $\left\{ i,j\right\} \subseteq \mathcal{O}
\left( P,H\right) $ holds, either $\left| z_{ i } -z_{ j }
\right| \leq
S$ holds or $\mathcal{Y} \left( F,\left\{ i,j\right\}
\right)
=\mathcal{Y} \left( Q,\left\{ i,j\right\} \right) $ holds,
hence by the assumed properties of $\mathcal{J} $,
$\mathcal{J} _{ F } \left( z\right) $ is defined ifif
$\mathcal{J} _{ Q } \left( z\right) $ is defined, and if
$\mathcal{J} _{ Q } \left( z\right) $ \emph{is} defined,
then $\mathcal{J} _{ F } \left( z\right) =\mathcal{J} _{ Q
} \left( z\right) $ holds .

From this it follows immediately that if $F$ is any member
of $\mathbb{K} \left( P,Q\right) $, and $y$ is any member
of $\mathbb{E} _{ d }^{ \mathcal{U} \left( P\right) } $
such
that $\left| y_{ i } -y_{ j } \right| <S$ holds for all
two-member
members $\left\{ i,j\right\} $ of $H$ such that $\left\{
i,j\right\} \subseteq \mathcal{O} \left( P,H\right) $ and
$\mathcal{Y} \left( Q,\left\{ i,j\right\} \right) \in \left(
Q\,\vdash P\right) $ both hold, then each derivative of
$\mathcal{J} _{ F } $ with respect to the $y_{ i } $, $i\in
\mathcal{O} \left( P,H\right) $, exists at $y$ ifif the
corresponding derivative of $\mathcal{J} _{ Q } $ exists at
$y$, and if the corresponding derivative of $\mathcal{J} _{
Q } $ does exist at $y$, then the corresponding derivative
of $\mathcal{J} _{ F } $ is equal to it.

We now observe that our assumption that $\mathbb{L} \left(
P,A,x\right) <S$ holds for every member $A$ of $\left(
Q\,\vdash P\right) $, implies that if $\left\{ i,j\right\} $
is any two-member member of $H$ such that $\left\{
i,j\right\} \subseteq \mathcal{O} \left( P,H\right) $ and
$\mathcal{Y} \left( Q,\left\{ i,j\right\} \right) \in \left(
Q\,\vdash P\right) $ both hold, then $\left| \eta _{ i }
\left(
F,H,x\right) -\eta _{ j } \left( F,H,x\right) \right| <S$
holds.

For $\left\{ i,j\right\} \in \mathcal{O} \left( P,H\right) $
implies that, by definition, $\eta _{ i } \left(
F,H,x\right) =x_{ \mathcal{Z} \left( F,H,i\right) } $
holds.   Furthermore $\mathcal{Z} \left( F,H,i\right)
\subseteq \mathcal{Y} \left( Q,\left\{ i,j\right\} \right) $
holds, since $\mathcal{Y} \left( Q,\left\{ i,j\right\}
\right) $ is a member of $F$ that contains the member
$\left\{ i,j\right\} $ of $H$ as a subset, and $\mathcal{Z}
\left( F,H,i\right) $ is by definition the largest member
$B$ of $F$ such that $i\in B$ holds and there is \emph{no}
member $C$ of $H$ such that $i\in C$ and $C\subseteq B$
both hold.

Furthermore the assumption that $\left\{ i,j\right\}
\subseteq \mathcal{O} \left( P,H\right) $ holds implies
\label{Start of original page 85}
 directly that $\mathcal{Y} \left( Q,\left\{ i,j\right\}
\right) $ is \emph{not} a member of $\mathcal{M} \left(
P\right) $, hence that $\mathcal{P} \left( P,\mathcal{Y}
\left( Q,\left\{ i,j\right\} \right) \right) $ is a
partition
of $\mathcal{Y} \left( Q,\left\{ i,j\right\} \right) $ such
that $\#\left( \mathcal{P} \left( P,\mathcal{Y} \left(
Q,\left\{ i,j\right\} \right) \right) \right) \geq 2$ holds.

Furthermore the assumption that $H$ is a \emph{partition}
implies that $\mathcal{Z} \left( F,H,i\right) $ is
\emph{not} a strict subset of any member of $\mathcal{P}
\left( P,\mathcal{Y} \left( Q,\left\{ i,j\right\} \right)
\right) $.   For $\mathcal{K} \left( P,\mathcal{Y} \left(
Q,\left\{ i,j\right\} \right) ,i\right) $ is a member of
$Q$,
hence the fact that $\mathcal{Y} \left( Q,\left\{
i,j\right\}
\right) $ is by definition the \emph{smallest} member of
$Q$ that contains $\left\{ i,j\right\} $ as a subset,
implies
that $j$ is not a member of $\mathcal{K} \left(
P,\mathcal{Y} \left( Q,\left\{ i,j\right\} \right) ,i\right)
$.   And the assumption that $H$ is a partition implies
that $\left\{ i,j\right\} $ is the \emph{only} member $C$ of
$H$ such that $i\in C$ holds, hence there is \emph{no}
member $C$ of $H$ such that $i\in C$ and $C\subseteq
\mathcal{K} \left( P,\mathcal{Y} \left( Q,\left\{
i,j\right\}
\right) ,i\right) $ both hold, hence the fact that
$\mathcal{K} \left( P,\mathcal{Y} \left( Q,\left\{
i,j\right\} \right) ,i\right) $ is a member of $F$ implies
that $\mathcal{K} \left( P,\mathcal{Y} \left( Q,\left\{
i,j\right\} \right) ,i\right) \subseteq \mathcal{Z} \left(
F,H,i\right) $ holds.

Hence $\mathcal{Z} \left( F,H,i\right) $ is a member of
$\Xi \left( \mathcal{P} \left( P,\mathcal{Y} \left(
Q,\left\{ i,j\right\} \right) \right) \right) $, hence, as
shown on
page \pageref{Start of original page 16},
$\eta _{ i } \left( F,H,x\right) =x_{
\mathcal{Z} \left( F,H,i\right) } $ is a member of the
convex hull of the $x_{ C } $, $C\in \mathcal{P} \left(
P,\mathcal{Y} \left( Q,\left\{ i,j\right\} \right) \right)
$.

And similarly, $\eta _{ j } \left( F,H,x\right) =x_{
\mathcal{Z} \left( F,H,j\right) } $ is a member of the
convex hull of the $x_{ C } $, $C\in \mathcal{P} \left(
P,\mathcal{Y} \left( Q,\left\{ i,j\right\} \right) \right)
$.

Hence by Lemma \ref{Lemma 3}, $\left| \eta _{ i } \left(
F,H,x\right)
-\eta
_{ j } \left( F,H,x\right) \right| \leq \mathbb{L} \left(
P,\mathcal{Y} \left( Q,\left\{ i,j\right\} \right) ,x\right)
$ holds.   But $\mathcal{Y} \left( Q,\left\{ i,j\right\}
\right) \in \left( Q\,\vdash P\right) $ holds by
assumption, hence the assumption that $\mathbb{L} \left(
P,A,x\right) <S$ holds for every member $A$ of $\left(
Q\,\vdash P\right) $, implies that $\mathbb{L} \left(
P,\mathcal{Y} \left( Q,\left\{ i,j\right\} \right) ,x\right)
<S$ holds, hence that $\left| \eta _{ i } \left(
F,H,x\right)
-\eta _{ j } \left( F,H,x\right) \right| <S$ holds.

Now for each member $F$ of $\mathbb{K} \left( P,Q\right) $
let $U_{ F } $ be defined by:
\[
U_{ F } \equiv  \sum_{n\in \mathbb{X} \left( \mathbb{I}
\left(
F,H\right) ,D\right) } \left( \left(
    \prod_{\left( i,A\right) \in
\mathcal{U} \left( \mathcal{R} \left( \mathbb{I} \left(
F,H\right) \right) \right) }
\left( \frac{\left( \left( x_{
\mathcal{K} \left( F,A,i\right) } -x_{ A } \right) .\hat{ y
}_{ i
} \right)^{ n_{ iA } } }{ n_{ iA } ! } \right) \right)
 \mathcal{J} _{ F } \left(
y\right) \right) _{ y=\eta \left( F,H,x\right) }
\]

Then it follows immediately from the foregoing, together
with our assumptions on $\mathcal{J} $ and our assumptions
on $x$, that for each member $F$ of $\mathbb{K} \left(
P,Q\right) $, the following equation holds:
\[
U_{ F } = \sum_{n\in \mathbb{X} \left( \mathbb{I} \left(
F,H\right) ,D\right) } \left( \left(
   \prod_{\left( i,A\right) \in
\mathcal{U} \left( \mathcal{R} \left( \mathbb{I} \left(
F,H\right) \right) \right) }
\left( \frac{\left( \left( x_{
\mathcal{K} \left( F,A,i\right) } -x_{ A } \right) .\hat{ y
}
_{ i } \right)^{ n_{ iA } } }{ n_{ iA } ! }
\right) \right) \mathcal{J} _{ Q }
\left( y\right) \right) _{ y=\eta \left( F,H,x\right) }
\]
\label{Start of original page 86}

Now let $F$ be any member of $\mathbb{K} \left(
P,Q\right) $.   We use Lemma \ref{Lemma 16}, taking
$M$ of Lemma \ref{Lemma 16}
to be the map $\mathbb{I} \left( F,H\right) $, and $\lambda
$ of Lemma \ref{Lemma 16} to be the map whose domain is
$\mathcal{U}
\left( \mathcal{R} \left( \mathbb{I} \left( F,H\right)
\right) \right) $, (namely the set of all ordered pairs
$\left(i,B\right) $ of a member $i$ of
$\mathcal{O} \left( F,H\right)
=\mathcal{O} \left( P,H\right) $ and a member $B$ of
$\mathbb{Y} \left( F,\mathcal{C} \left( F,i\right)
,\mathcal{Z} \left( F,H,i\right) \right)  $), and
such that for each member $\left(i,B\right) $ of
$\mathcal{D} \left(
\lambda \right) =\mathcal{U} \left( \mathcal{R} \left(
\mathbb{I} \left( F,H\right) \right) \right) $, $\lambda _{
\left( i,B\right) } $ is the set whose members are all the
ordered pairs $\left(i,C\right) $ such that
 $C\in \mathbb{Y} \left(
Q,\mathcal{K} \left( F,B,i\right) ,B\right) $ holds.   (We
note that this satisfies the requirement that
if $\left(i,B\right) $ is a
member of $\mathcal{D} \left( \lambda \right) $
and $\left(j,E\right) $
is a member of $\mathcal{D} \left( \lambda \right) $ such
that $\left( i,B\right) \neq \left( j,E\right) $ holds, or
in other words such that at least one of $i\neq j$ and
$B\neq E$ holds, then $\lambda _{ \left( i,B\right) } \cap
\lambda _{ \left( j,E\right) } =\emptyset $ holds.)

Then the corresponding map $T$, defined in terms of $M$ and
$\lambda $ as specified in Lemma \ref{Lemma 16}, is the map
$\downarrow \left( \mathbb{I} \left( Q,H\right) ,F\right)
$, that is, the restriction of the map $\mathbb{I} \left(
Q,H\right) $ to the domain $\mathcal{D} \left( \mathbb{I}
\left( Q,H\right) \right) \cap F=\mathbb{B} \left( F\right)
$.   For if $A$ is any member of $\mathcal{D} \left(
M\right) $, or in other words, if $A$ is any member of
$\mathcal{D} \left( \mathbb{I} \left( F,H\right) \right)
=\mathbb{B} \left( F\right) $, then by definition $T_{ A }
\equiv  \bigcup_{\alpha \in M_{ A } } \lambda _{ \alpha }
$, which in the present case becomes $T_{ A } \equiv
\bigcup_{\left(
i,B\right) \in \mathbb{I} _{ A } \left( F,H\right) }
 \lambda _{ \left( i,B\right) } $.   But
$\mathbb{I} _{ A } \left( F,H\right) $ is the set of all
ordered pairs $\left(i,B\right) $ such
that $i\in \mathcal{T} \left(
A,H\right) $ holds and $B\in \mathbb{Y} \left(
F,\mathcal{C} \left( F,i\right) ,A\right) $ holds hence,
with $\lambda _{ \left( i,B\right) } $ defined as above,
$T_{ A } $ is the set of all ordered
pairs $\left(i,B\right) $ such that
$i\in \mathcal{T} \left( A,H\right) $ holds and $B\in
\mathbb{Y} \left( Q,\mathcal{C} \left( F,i\right) ,A\right)
$ holds, or in other words such that $B\in \mathbb{Y}
\left( Q,\mathcal{C} \left( Q,i\right) ,A\right) $ holds,
hence $T_{ A } =\mathbb{I} _{ A } \left( Q,H\right) $.

And, as specified in Lemma \ref{Lemma 16}, $\mathcal{D}
\left(
T\right) $ is equal to $\mathcal{D} \left( M\right) $,
hence $T=\downarrow \left( \mathbb{I} \left( Q,H\right)
,F\right) $.

Hence Lemma \ref{Lemma 16} implies directly that the
following
equation holds:
\[
U_{ F } = \hspace{-0.65cm}
 \sum_{m\in \mathbb{X} \left( \downarrow \left(
\mathbb{I} \left( Q,H\right) ,F\right) ,D\right) }
\hspace{-0.2cm} \left( \hspace{-0.15cm} \left(
   \prod_{\left( i,A\right) \in \mathcal{U} \left(
\mathcal{R}
\left( \downarrow \left( \mathbb{I} \left( Q,H\right)
,F\right) \right) \right) } \hspace{-0.2cm}
\left( \frac{\left( \left( x_{
\mathcal{K} \left( Q,A,i\right) } -x_{ A } \right) .\hat{ y
}_{ i
} \right)^{ m_{ iA } } }{ m_{ iA } ! }
\right) \hspace{-0.2cm} \right) \mathcal{J} _{ Q } \left(
y\right) \right)_{ \hspace{-0.2cm}
 y=\eta \left( F,H,x\right) }
\]

We next use Lemma \ref{Lemma 17}, with
the map $V$ of Lemma \ref{Lemma 17} taken as the map \\
$\downarrow \left( \mathbb{I} \left( Q,H\right)
,F\right) $, the set $J$ of Lemma \ref{Lemma 17} taken as
the set
$\mathbb{B} \left( P\right) $, and the set $K$
 of Lemma \ref{Lemma 17}
taken as the set $\left( F\,\vdash P\right) $.

Then the corresponding map $W$, defined as specified in
Lemma \ref{Lemma 17}, is the map $\downarrow \left(
\mathbb{J} \left(
P,Q,H\right) ,\left( F\,\vdash P\right) \right) =\downarrow
\left( \mathbb{J} \left( P,Q,H\right) ,F\right) $.   (This
follows directly from the fact that $\downarrow \left(
\downarrow \left( \mathbb{I} \left( Q,H\right) ,F\right)
,P\right) $ is equal to
\label{Start of original page 87}
 $\downarrow \left( \mathbb{I} \left( Q,H\right) ,P\right)
$.)

Furthermore with $V$, $J$, and $K$ taken as above, and $u$
being any member of $\mathbb{N}^{ \mathcal{U} \left(
\mathcal{R} \left( \downarrow \left( V,J\right) \right)
\right) } $, which in the present case means that $u$ is
any member of $\mathbb{N}^{ \mathcal{U} \left( \mathcal{R}
\left( \downarrow \left( \mathbb{I} \left( Q,H\right)
,P\right) \right) \right) } $, the map $\xi \left(
P,Q,H,u\right) $, defined on
pages \pageref{Start of original page 81} and
\pageref{Start of original page 82}, satisfies
the assumptions made on the map $\zeta \left( u\right) $ in
Lemma \ref{Lemma 17}.

Hence Lemma \ref{Lemma 17} implies directly that the
following
equation holds:
\[
U_{ F } = \sum_{u\in \mathbb{X} \left( \downarrow \left(
\mathbb{I} \left( Q,H\right) ,P\right) ,D\right) }
\hspace{0.2cm}
 \sum_{v\in \mathbb{X} \left( \downarrow \left( \mathbb{J}
\left(
P,Q,H\right) ,F\right) ,\left( D-\xi \left( P,Q,H,u\right)
\right) \right) } \times \hspace{-0.7pt} \hspace{6.0cm}
\]
\[
\times \left( \left(
\prod_{\left( i,A\right) \in
\mathcal{U} \left( \mathcal{R} \left( \downarrow \left(
\mathbb{J} \left( P,Q,H\right) ,F\right) \right)
\right) }
\left( \frac{\left( \left( x_{ \mathcal{K}
\left( Q,A,i\right)
 } -x_{ A } \right) .\hat{ y }_{ i } \right)
 ^{ v_{ iA } } }{ v_{
iA } ! } \right) \right) \right. \times
\]
\[
\hspace{3.0cm} \hspace{-7.8pt} \left. \times \left(
 \prod_{\left( i,A\right) \in \mathcal{U}
 \left( \mathcal{R} \left(
\downarrow \left( \mathbb{I} \left( Q,H\right) ,P\right)
\right) \right) }
\left( \frac{\left( \left( x_{ \mathcal{K}
\left( Q,A,i\right) } -x_{ A } \right) .\hat{ y }_{ i }
\right)^{
u_{ iA } } }{ u_{ iA } ! } \right) \right)
 \mathcal{J} _{ Q } \left( y\right) \right) _{
y=\eta \left( F,H,x\right) }
\]

Now, as shown on
page \pageref{Start of original page 80},
$\eta \left( F,H,x\right) $ is
equal to \\
$\mu \left( P,Q,H,x,\mathcal{X} \left(
P,Q,H,\mathcal{U} \left( \left\{ 0\right\}^{ \left(
F\,\vdash
P\right) } \cup \left\{ 1\right\}^{ \left( Q\,\vdash
F\right) } \right) \right) \right) $, hence the above
equation may be rewritten as
\[
U_{ F } =  \sum_{u\in \mathbb{X} \left( \downarrow \left(
\mathbb{I} \left( Q,H\right) ,P\right) ,D\right) }
\hspace{0.2cm} \sum_{v\in \mathbb{X} \left( \downarrow
\left(
\mathbb{J}
\left( P,Q,H\right) ,F\right) ,\left( D-\xi \left(
P,Q,H,u\right) \right) \right) } \times \hspace{-0.7pt}
\hspace{6.0cm}
\]
\[
\times \left( \left( \left( \prod_{\left(
i,A\right) \in \mathcal{U} \left( \mathcal{R} \left(
\downarrow \left( \mathbb{J} \left( P,Q,H\right) ,F\right)
\right) \right) }
\left( \frac{\left( \left( x_{ \mathcal{K}
\left( Q,A,i\right) } -x_{ A } \right) .\hat{ y }_{ i }
\right)^{
v_{ iA } } }{ v_{ iA } ! } \right) \right) \times
\right. \right.
\]
\[ \left. \left. \times \hspace{-0.15cm}
 \left( \prod_{\left( i,A\right) \in
\mathcal{U}
\left( \mathcal{R} \left( \downarrow \left( \mathbb{I}
\left( Q,H\right) ,P\right) \right) \right) }
\hspace{-0.2cm}
\left( \frac{\left( \left( x_{ \mathcal{K} \left(
 Q,A,i\right) } -x_{ A
} \right) .\hat{ y }_{ i } \right)^{ u_{ iA } } }{ u_{ iA }
! } \right) \hspace{-0.2cm} \right)
\mathcal{J}
_{ Q } \left( y\right) \! \right)_{ \hspace{-0.2cm}
 y=\mu \left( P,Q,H,x,r\right) } \right)_{
 \begin{array}{c} \\
 \hspace{-3.0cm}
\scriptstyle{r=\mathcal{X} \left( P,Q,H,\mathcal{U}
 \left( \left\{
0\right\}^{ \left( F\,\vdash P\right) } \cup \left\{
1\right\}^{ \left( Q\,\vdash F\right) } \right) \right) }
\end{array} }
\]

Hence by the chain rule for differentiating a function of a
function, and noting that
\[
\hat{ r }_{ iA } \mathcal{J} _{ Q } \left( \mu \left(
P,Q,H,x,r\right) \right) = \left(
\left( \left( x_{ \mathcal{K}
\left( Q,A,i\right) } -x_{ A } \right) .\hat{ y }_{ i }
\right)
\mathcal{J} _{ Q } \left( y\right) \right)_{ y=\mu \left(
P,Q,H,x,r\right) }
\]
holds, we have
\[
U_{ F } = \hspace{-0.1cm}
 \sum_{u\in \mathbb{X} \left( \downarrow \left(
\mathbb{I} \left( Q,H\right) ,P\right) ,D\right) }
\hspace{0.2cm}
  \sum_{v\in \mathbb{X} \left( \downarrow \left( \mathbb{J}
\left( P,Q,H\right) ,F\right) ,\left( D-\xi \left(
P,Q,H,u\right) \right) \right) } \hspace{-0.05cm}
\left( \hspace{-0.1cm}
 \left(  \prod_{\left(
i,A\right) \in \mathcal{U} \left( \mathcal{R} \left(
\downarrow \left( \mathbb{J} \left( P,Q,H\right) ,F\right)
\right) \right) } \hspace{-0.05cm}
\left( \frac{\hat{ r }_{ iA }^{ v_{ iA } } }{ v_{ iA }
! } \right) \hspace{-0.05cm}
 \right) \right. \hspace{-0.1cm} \times
\]
\[
\left. \times \hspace{-0.15cm} \left( \hspace{-0.2cm}
\left(
 \prod_{\left( i,A\right) \in \mathcal{U} \left(
\mathcal{R}
\left( \downarrow \left( \mathbb{I} \left( Q,H\right)
,P\right) \right) \right) } \hspace{-0.2cm}
\left( \! \frac{\left( \left( x_{
\mathcal{K} \left( Q,A,i\right) } -x_{ A } \right) .\hat{ y
}_{ i
} \right)^{ u_{ iA } } }{ u_{ iA } ! } \right)
\hspace{-0.2cm} \right)
 \mathcal{J} _{ Q } \left(
y\right) \! \right)_{ \hspace{-0.2cm}
y=\mu \left( P,Q,H,x,r\right) } \right)_{
\begin{array}{c} \\
\hspace{-3.0cm} \scriptstyle{ r=\mathcal{X}
\left( P,Q,H,\mathcal{U} \left( \! \left\{ 0\right\}^{
 \left(
F\,\vdash P\right) } \cup \left\{ 1\right\}^{ \left(
Q\,\vdash F\right) } \right) \right) }
\end{array} }
\]

Hence by Lemma \ref{Lemma 19}, with $V$ of
Lemma \ref{Lemma 19} taken as
$\mathbb{J} \left( P,Q,H\right) $, $J$
of Lemma \ref{Lemma 19} taken
as $\left( F\,\vdash P\right) $, and the function $p\left(
\rho \right) $ of Lemma \ref{Lemma 19} taken as
$\mathcal{X} \left(
P,Q,H,\rho \right) $, and noting that $\downarrow \left(
\mathbb{J} \left( P,Q,H\right) ,\left( F\,\vdash P\right)
\right) =\downarrow \left( \mathbb{J} \left( P,Q,H\right)
,F\right) $, we have
\label{Start of original page 88}
\[
U_{ F } =  \sum_{u\in \mathbb{X} \left( \downarrow \left(
\mathbb{I} \left( Q,H\right) ,P\right) ,D\right)
} \left( \left(
  \prod_{A\in \left( F\,\vdash P\right) } \left(
   \sum_{n_{ A } =0
}^{ \left( D_{ A } -\xi _{ A } \left( P,Q,H,u\right)
\right)}
\frac{\hat{ \rho }_{ A }^{ n_{ A } } }{n_{ A } !}
\right) \right) \times \right. \hspace{-14.5pt}
\hspace{5.0cm}
\]
\[
\times \hspace{-0.15cm} \left. \left(
\hspace{-0.15cm} \left(
\prod_{\left( i,A\right)
\in \mathcal{U} \left( \mathcal{R} \left( \downarrow \left(
\mathbb{I} \left( Q,H\right) ,P\right) \right) \right)
 } \hspace{-0.1cm} \left( \frac{\left(
 \left( x_{ \mathcal{K}
 \left( Q,A,i\right)
 } -x_{ A } \right) .\hat{ y }_{ i } \right)^{
 u_{ iA } } }{u_{
iA }
! } \right) \hspace{-0.15cm} \right)
   \mathcal{J} _{ Q } \left( y\right) \right)_{
    \hspace{-0.1cm} y=\mu \left(
P,Q,H,x,\mathcal{X} \left( P,Q,H,\rho \right) \right) }
\right)_{ \begin{array}{c} \\
\hspace{-3.5cm} \scriptstyle{\rho
=\mathcal{U} \left( \left\{ 0\right\}^{ \left( F\,\vdash
P\right) } \cup \left\{ 1\right\}^{ \left( Q\,\vdash
F\right) } \right) }
\end{array}  }
\]

We now define, for each member $A$ of $\left( Q\,\vdash
P\right) $, the operators $\nu _{ A0 } $ and $\nu _{ A1 } $
by specifying that for any $\rho $-dependent function
$f\left( \rho \right) $, the identities
\[
\nu _{ A0 } f\left( \rho \right) \equiv \left(
f\left( \rho
\right) \right)_{ \rho _{ A } =0 }
\]
and
\[
\nu _{ A1 } f\left( \rho \right) \equiv \left(
 f\left( \rho
\right) \right)_{ \rho _{ A } =1 }
\]
hold.

The above equation for $U_{ F } $ may then be re-written as
\[
U_{ F } = \sum_{u\in \mathbb{X} \left( \downarrow \left(
\mathbb{I} \left( Q,H\right) ,P\right) ,D\right)
} \left( \left( \prod_{A\in \left( Q\,\vdash F\right) }
 \nu _{ A1 } \right) \left(
 \prod_{A\in
\left( F\,\vdash P\right) } \left( \nu _{ A0 }
\sum_{n_{ A }
=0 }^{ \left( D_{ A } -\xi _{ A } \left( P,Q,H,u\right)
\right) } \frac{\hat{ \rho }_{ A }^{ n_{ A } } }
{n_{ A } !} \right) \right) \times \right. \hspace{-14.3pt}
\hspace{1.5cm}
\]
\[
\left. \times \hspace{-0.15cm} \left( \hspace{-0.15cm}
 \left( \hspace{-0.15cm} \left(
\prod_{\left(
i,A\right) \in \mathcal{U} \left( \mathcal{R} \left(
\downarrow \left( \mathbb{I} \left( Q,H\right) ,P\right)
\right) \right) } \hspace{-0.1cm}
\left( \frac{\left( \left( x_{ \mathcal{K}
\left( Q,A,i\right) } -x_{ A } \right) .\hat{ y }_{ i }
\right)^{
u_{ iA } } }{ u_{ iA } ! } \right) \hspace{-0.15cm}
 \right)
 \mathcal{J} _{ Q }
\left( y\right) \right)_{ \hspace{-0.1cm}
y=\mu \left( P,Q,H,x,\mathcal{X} \left( P,Q,H,\rho \right)
\right) } \right) \hspace{-0.15cm} \right)
\]

Hence
\[
 \sum_{F\in \mathbb{K} \left( P,Q\right) }
 \hspace{-0.2cm} \left( -1\right)^{
\#\left( F\,\vdash P\right) } U_{ F } =
\hspace{-0.2cm} \sum_{u\in
\mathbb{X}
\left( \downarrow \left( \mathbb{I} \left( Q,H\right)
,P\right) ,D\right) } \hspace{-0.1cm} \left(
\hspace{-0.15cm} \left(
  \prod_{A\in \left( Q\,\vdash
P\right) } \hspace{-0.1cm}
 \left( \nu _{ A1 } -\nu _{ A0 } \hspace{-0.3cm}
  \sum_{n_{ A }
=0 }^{ \left( D_{ A } -\xi _{ A } \left( P,Q,H,u\right)
\right) } \frac{\hat{ \rho }_{ A }^{ n_{ A } } }
{n_{ A } !} \right) \hspace{-0.15cm} \right)
\hspace{-0.1cm} \times \right.
\]
\[
\left. \times \hspace{-0.1cm} \left( \hspace{-0.15cm}
 \left( \hspace{-0.15cm} \left(
 \prod_{\left(
i,A\right) \in \mathcal{U} \left( \mathcal{R} \left(
\downarrow \left( \mathbb{I} \left( Q,H\right) ,P\right)
\right) \right) } \hspace{-0.1cm}
\left( \frac{\left( \left( x_{ \mathcal{K}
\left( Q,A,i\right) } -x_{ A } \right) .\hat{ y }_{ i }
\right)^{
u_{ iA } } }{ u_{ iA } ! } \right) \hspace{-0.15cm} \right)
 \mathcal{J} _{ Q }
\left( y\right) \right)_{ \hspace{-0.1cm}
y=\mu \left( P,Q,H,x,\mathcal{X} \left( P,Q,H,\rho \right)
\right) } \right) \hspace{-0.15cm} \right)
\]

Now if $g$ is a map whose arguments include the real
variable $s$, then the Taylor remainder identity for the
single variable $s$, derived by repeated integration by
parts, is
\[
\left( g\left( s\right) \right) _{ s=1 } - \left( \sum_{
m=0 }^{ n } \frac{\hat{ s }^{ m }
g\left( s\right) }{m! }  \right)_{ s=0 } = \frac{1 }{ n! }
\int_{ 0 }^{ 1 }
\left( ds\right) \left( 1-s\right)^{ n } \left( \hat{ s }^{
n+1 }
g\left( s\right) \right) .
\]

A sufficient condition for the validity of this identity is
for all derivatives of $g$ of degree up to and including
$\left( n+1\right) $ to exist and be continuous throughout
an open subset of $\mathbb{R} $ that contains the interval
$0\leq s\leq 1$.

We use this identity in turn for each of the $\rho _{ A }
$, $A\in \left( Q\,\vdash P\right) $, to
\label{Start of original page 89}
 obtain:
\[
 \sum_{F\in \mathbb{K} \left( P,Q\right) }
 \left( -1\right)^{
\#\left( F\,\vdash P\right) } U_{ F } =
\hspace{10.9cm}
\]
\[
= \hspace{-0.35cm} \sum_{u\in
\mathbb{X}
\left( \downarrow \left( \mathbb{I} \left( Q,H\right)
,P\right) ,D\right) } \int_{\mathbb{D} }
\hspace{-0.15cm} \left( d^{
\#\left(
Q\,\vdash P\right) } \rho \right) \hspace{-0.1cm}
 \left( \hspace{-0.2cm} \left(
  \prod_{A\in \left(
Q\,\vdash P\right) } \hspace{-0.15cm} \!
\left( \frac{\left( 1 \! - \! \rho _{ A } \right)^{
\left( D_{ A } -\xi _{ A } \left( P,Q,H,u\right) \right) }
\hat{ \rho }_{ A }^{ \left( D_{ A } -\xi _{ A } \left(
P,Q,H,u\right) +1\right) } }{\left( D_{ A } -\xi _{ A }
\left(
P,Q,H,u\right) \right) ! } \right) \hspace{-0.2cm}
 \right) \hspace{-0.15cm} \times \right.
\]
\[
\left. \times \hspace{-0.1cm} \left( \hspace{-0.15cm}
 \left( \hspace{-0.15cm} \left(
   \prod_{\left( i,A\right) \in \mathcal{U} \left(
\mathcal{R} \left( \downarrow \left( \mathbb{I} \left(
Q,H\right) ,P\right) \right) \right) } \hspace{-0.1cm}
\left( \frac{\left(
\left( x_{ \mathcal{K} \left( Q,A,i\right) } -x_{ A }
\right) .\hat{ y }_{ i } \right)^{ u_{ iA } } }{ u_{ iA }
! } \right) \hspace{-0.15cm} \right) \mathcal{J} _{
Q } \left( y\right) \right)_{ \hspace{-0.1cm}
y=\mu \left( P,Q,H,x,\mathcal{X}
\left( P,Q,H,\rho \right) \right) } \right)
\hspace{-0.15cm} \right)
\]

Hence by Lemma \ref{Lemma 18}, with the
map $V$ of Lemma \ref{Lemma 18} taken as
the map $\mathbb{J} \left( P,Q,H\right) $, and the map $C$
of Lemma \ref{Lemma 18} taken as the map $\mathbb{G} \left(
Q,H\right)
$, we have
\[
 \sum_{F\in \mathbb{K} \left( P,Q\right) }
 \left( -1\right)^{
\#\left( F\,\vdash P\right) } U_{ F } =
\hspace{10.9cm}
\]
\[
= \hspace{-0.25cm} \sum_{u\in
\mathbb{X}
\left( \downarrow \left( \mathbb{I} \left( Q,H\right)
,P\right) ,D\right) } \int_{\mathbb{D} }
\hspace{-0.1cm} \left( d^{
\#\left(
Q\,\vdash P\right) } \rho \right) \hspace{-0.1cm}
 \left( \hspace{-0.15cm} \left(
   \prod_{A\in \left(
Q\,\vdash P\right) } \hspace{-0.15cm}
 \left( \left( 1-\rho _{ A } \right) \rule{0pt}{3ex}^{
\hspace{-0.15cm} \left( D_{ A } -\xi _{ A }
\left( P,Q,H,u\right) \right)
 } \hspace{-2.2cm}
\left( D_{ A } -\xi _{ A } \left( P,Q,H,u\right) +1\right)
\right) \hspace{-0.15cm} \right) \hspace{-0.1cm}
 \times \right.
\]
\[
\times \hspace{-0.25cm}
 \sum_{m\in \mathbb{A} \left( \mathbb{J} \left(
P,Q,H\right) ,\left( D-\xi \left( P,Q,H,u\right)
+1\hspace{-0.7516ex}1 \right) \right) }
\hspace{-0.15cm} \left( \hspace{-0.2cm} \left(
\hspace{-0.15cm}  \prod_{\left(
\left( i,B\right) ,X\right) \in \mathcal{U} \left(
\mathcal{R} \left( \psi \left( \mathbb{J} \left(
P,Q,H\right) \right) \right) \right) }
\hspace{-0.15cm} \left( \hspace{-0.1cm}
\frac{ \hspace{-0.1cm} \left( \hspace{-0.2cm} \left(
\displaystyle\prod_{E\in \left(
\mathbb{G} _{ iB } \left( Q,H\right) \,\vdash X\right) }
\hspace{-0.55cm}
\rho _{ E } \right) \hspace{-0.1cm}
 \hat{ r }_{ iB } \hspace{-0.1cm} \right)^{
 \hspace{-0.2cm}
 m_{ iBX } } }{ m_{ iBX } ! } \hspace{-0.05cm} \right)
 \hspace{-0.2cm} \right) \hspace{-0.15cm} \times
 \right.
\]
\[
\left. \left. \times \hspace{-0.15cm}
 \left( \hspace{-0.2cm} \left( \hspace{-0.2cm} \left(
  \prod_{\left( i,A\right) \in \mathcal{U} \left(
\mathcal{R}
\left( \downarrow \left( \mathbb{I} \left( Q,H\right)
,P\right) \right) \right) } \hspace{-0.18cm}
\left( \hspace{-0.05cm} \frac{\left( \left( x_{
\mathcal{K} \left( Q,A,i\right) } \! - \! x_{ A }
\right) .\hat{ y
}_{ i
} \right)^{ u_{ iA } } }{ u_{ iA } ! } \hspace{-0.05cm}
 \right) \hspace{-0.2cm} \right) \hspace{-0.05cm}
 \mathcal{J} _{ Q } \left(
y\right) \hspace{-0.05cm} \right)_{
\hspace{-0.15cm} y=\mu \left( P,Q,H,x,r\right) }
\right) \hspace{-0.2cm} \right)_{ \hspace{-0.15cm}
 r=\mathcal{X}
\left( P,Q,H,\rho \right) } \right)
\]

Finally we again use the chain rule for differentiating a
function of a function to express the $\hat{ r } $
derivatives in
terms of $\hat{ y } $ derivatives, and observe that in
 consequence
of our assumptions on $\mathcal{J} $ and on $x$, each term
in the sum over the members $m$ of the finite set
$\mathbb{A} \left( \mathbb{J} \left( P,Q,H\right) ,\left(
D-\xi \left( P,Q,H,u\right) +1\hspace{-0.7516ex}1
\right) \right) $ is continuous and bounded for all $\rho
\in \mathbb{D} $, so that the sum over the members $m$ of
$\mathbb{A} \left( \mathbb{J} \left( P,Q,H\right) ,\left(
D-\xi \left( P,Q,H,u\right) +1\hspace{-0.7516ex}1
\right) \right) $ may be taken outside the $\rho
$-integral, to obtain the stated result.

\section{First Convergence Theorem.}
\label{Section 6}

We recall from
page \pageref{Start of original page 14}
 that for any integer $d\geq 1$ and
any ordered pair $\left( V,\omega \right) $ of a partition
$V$ such that $V$ is a finite set, and a set $\omega $ of
contraction weights for $V$, we define $\mathbb{U} _{ d }
\left( V,\omega \right) $ to be the set
\label{Start of original page 90}
 whose members are all the members $x$ of $\mathbb{E} _{
d }^{ \Xi \left( V\right) } $ such that for every member
 $A$ of
$\left( \Xi \left( V\right) \,\vdash V\right) $, $x_{ A }
= \sum_{B\in \mathcal{P} \left( V,A\right) } \omega
_{ AB } x_{ B } $ holds.

We observe that if $d$ is any integer $\geq 1$, $V$ is any
partition such that $\mathcal{U} \left( V\right) $ is
finite and $\#\left( V\right) \geq 2$ holds, $\omega $ is
any set of contraction weights for $V$, and $A$ and $B$ are
any two distinct members of $\Xi \left( V\right) $ such
that $A\cap B$ is empty, then the subset of $\mathbb{U} _{
d } \left( V,\omega \right) $ whose members are all the
members $x$ of $\mathbb{U} _{ d } \left( V,\omega \right) $
such that $\left| x_{ A } -x_{ B } \right| =0$ holds is the
same as the
subset of $\mathbb{U} _{ d } \left( V,\omega \right) $
whose members are all the members $x$ of $\mathbb{U} _{ d }
\left( V,\omega \right) $ such that the $d$ independent
linear relations $x_{ \mu A } -x_{ \mu B } =0$, (where $\mu
$ is the $d $-vector index), hold among the $d\#\left(
V\right)
$ independent components of the $x_{ C } $, $C\in V$, hence
the equation $\left| x_{ A } -x_{ B } \right| =0$ defines a
$d\left(
\#\left( V\right) -1\right) $-dimensional hyperplane in the
$d\#\left( V\right) $-dimensional space $\mathbb{U} _{ d }
\left( V,\omega \right) $.   And for any given real number
$\varepsilon >0$ there exists an open subset $S_{
\varepsilon } $ of $\mathbb{U} _{ d } \left( V,\omega
\right) $ such that this $d\left( \#\left( V\right)
-1\right) $-dimensional hyperplane is a subset of $S_{
\varepsilon } $ and the $d\#\left( V\right) $-volume of
$S_{ \varepsilon } $ is $\leq \varepsilon $, for we cut up
$\mathbb{U} _{ d } \left( V,\omega \right) $ into a
cardinality-$\mathbb{N} $ collection of finite-volume
sectors labelled by a bijection from $\mathbb{N} $, then
for each $n\in \mathbb{N} $ we choose an open subset $T_{ n
} $ of $\mathbb{U} _{ d } \left( V,\omega \right) $ such
that the intersection of this $d\left( \#\left( V\right)
-1\right) $-dimensional hyperplane with the $n^{ th }$
sector
is a subset of $T_{ n } $ and the $d\#\left( V\right)
$-volume of $T_{ n } $ is $\leq 2^{ -\left( n+1\right) }
\varepsilon $, then define $S_{ \varepsilon } $ to be the
union of all the $T_{ n } $, $n\in \mathbb{N} $.

And furthermore, if $d$, $V$, and $\omega $ are as in the
preceding paragraph, then the set whose members are
\emph{all} the two-member subsets $\left\{ A,B\right\} $ of
$\Xi \left( V\right) $ such that $A\cap B=\emptyset $
holds, is
a finite set, hence for any given real number $\varepsilon
>0$ there exists an open subset $S_{ \varepsilon } $ of
$\mathbb{U} _{ d } \left( V,\omega \right) $ such that the
$d\#\left( V\right) $-volume of $S_{ \varepsilon } $ is
$\leq \varepsilon $ and $S_{ \varepsilon } $ contains as a
subset the set of all members $x$ of $\mathbb{U} _{ d }
\left( V,\omega \right) $ such that there \emph{exist} two
distinct members $A$ and $B$ of $\Xi \left( V\right) $ such
that $A\cap B$ is empty and the equation $\left| x_{ A }
-x_{ B }
\right| =0$ holds.

And we observe furthermore that if $d$, $V$, and $\omega $
are as in the two preceding paragraphs, $b$ is any member
of $\mathbb{E} _{ d } $, $h$ is any member of $\mathcal{U}
\left( V\right) $, and $\mathbb{W}$ is the subset of
$\mathbb{U} _{ d } \left( V,\omega \right) $ whose members
are all the members $x$ of $\mathbb{U} _{ d } \left(
V,\omega \right) $ such that $x_{ \mathcal{C} \left(
V,h\right) } =b$ holds, then for any given real number
$\varepsilon >0$ there exists an open subset $S_{
\varepsilon } $ of the $d\left( \#\left( V\right) -1\right)
$-dimensional space $\mathbb{W}$ such that the $d\left(
\#\left( V\right) -1\right) $-volume of $S_{ \varepsilon }
$ is $\leq \varepsilon $, and $S_{ \varepsilon } $
contains as a subset the set of all members $x$ of
\label{Start of original page 91}
 $\mathbb{W}$ such that there \emph{exist} two distinct
members $A$ and $B$ of $\Xi \left( V\right) $ such that
$A\cap B$ is empty and the equation $\left| x_{ A } -x_{ B
} \right| =0$
holds.

\begin{bphzlemma} \label{Lemma 23}
\end{bphzlemma}
\vspace{-6.143ex}

\noindent \hspace{11.9ex}{\bf.  }Let $n$ be an
integer $\geq 1$ and let $S$ be a
subset of $\mathbb{R}^{ n } $.   Let $f$ be a map such that
$\left( \mathbb{R}^{ n } \,\vdash S\right) $ is a subset of
$\mathcal{D} \left( f\right) $, $\mathcal{R} \left(
f\right) $ is a subset of $\mathbb{R} $, and such that if
$T$ is any subset of $\mathbb{R}^{ n }$ such that the volume
of $T$ is well-defined and finite, and such that for every
member $x$ of $S$ and every member $y$ of $T$, $\left|
x-y\right| >0$
holds, then $\int_{ T } d^{ n } xf\left( x\right) $ is
well-defined
and finite and $\int_{ T } d^{ n } x\left| f\left( x\right)
\right| $ is
well-defined and finite.

Suppose further that there exists a map $F$ such that
$\left( \mathbb{R}^{ n } \,\vdash S\right) $ is a subset of
$\mathcal{D} \left( F\right) $, $\mathcal{R} \left(
F\right) $ is a subset of $\mathbb{R} $, $F\left( x\right)
\geq \left| f\left( x\right) \right| $ holds for all $x\in
\left(
\mathbb{R}^{ n } \,\vdash S\right) $, and $\int_{ \left(
\mathbb{R}^{ n } \,\vdash S\right) } d^{ n } xF\left(
x\right) $ is well-defined and finite as an improper
Riemann integral.
\enlargethispage{0.15ex}

Then $\int_{ \left( \mathbb{R}^{ n } \,\vdash S\right) }
d^{
n } xf\left( x\right) $ is well-defined and finite as an
improper Riemann integral, and is moreover absolutely
convergent.

\vspace{2.5ex}

\noindent {\bf Proof.}  Let $F$ be a map such
that $\left( \mathbb{R}^{ n }
\,\vdash S\right) \subseteq \mathcal{D} \left( F\right) $,
$\mathcal{R} \left( F\right) \subseteq \mathbb{R} $,
$F\left( x\right) \geq \left| f\left( x\right) \right| $
holds for all
$x\in \left( \mathbb{R}^{ n } \,\vdash S\right) $, and
$\int_{
\left( \mathbb{R}^{ n } \,\vdash S\right) } d^{ n }
xF\left( x\right) $ is well-defined and finite as an
improper Riemann integral.   This last assumption implies
that there exists a map $u$ such that $\mathcal{D} \left(
u\right) =\mathbb{N} $, and for each member $m$ of
$\mathbb{N} $, $u_{ m } $ is a subset of $\left( \mathbb{R}
^{ n } \,\vdash S\right) $ such that $\int_{ u_{ m } } d^{
n } xF\left( x\right) $ is well-defined and finite,
and for any members $m$ and $p$ of $\mathbb{N} $ such that
$m\leq p$ holds, $u_{ m } \subseteq u_{ p } $ holds, and
such that if $T$ is any subset of $\left( \mathbb{R}^{ n }
\,\vdash S\right) $ such that $\left| x-y\right| >0$ holds
for every
member $x$ of $S$ and every member $y$ of $T$, then there
exists a member $k$ of $\mathbb{N} $ such that for all
members $m$ of $\mathbb{N} $ such that $m\geq k$ holds,
$T\subseteq u_{ m } $ holds, and such that for any given
$\varepsilon >0$, there exists a member $k$ of $\mathbb{N}
$ such that for all members $l$ of $\mathbb{N} $ such that
$l\geq k$ holds, and all members $m$ of $\mathbb{N} $ such
that $m\geq l$ holds, $\left| \left( \int_{ u_{ m } } d^{
 n } xF\left(
x\right) \right) - \left( \int_{ u_{ l } } d^{ n } xF
\left( x\right) \right) \right| \leq
\varepsilon $ holds, or in other words, $\int_{ u_{ m }
\,\vdash u_{ l } } d^{ n } xF\left( x\right) \leq
\varepsilon $ holds.   Let $u$ be such a map.   Then
$\left| \int_{
 u_{ m } \,\vdash u_{ l } } d^{ n } xf\left( x\right)
\right| \leq \varepsilon $ holds and $\int_{ u_{ m }
\,\vdash u_{ l } } d^{ n } x\left| f\left( x\right) \right|
\leq
\varepsilon $ holds, hence the sequences
\label{Start of original page 92}
 $\int_{ u_{ m } } d^{ n } xf\left( x\right)  $
and $\int_{ u_{ m } } d^{ n } x\left| f\left( x\right)
\right|  $
are Cauchy sequences hence convergent.

\vspace{2.5ex}

We note that, as is well known, if an improper Riemann
integral $\int_{ \left( \mathbb{R}^{ n } \,\vdash S\right)
 } d
^{ n } xf\left( x\right) $ is absolutely convergent then its
value is independent of the particular choice of the map
$u$ used to define it, provided that $u$ has the properties
listed in the proof of Lemma \ref{Lemma 23}.   For suppose
 that $\int_{
\left( \mathbb{R}^{ n } \,\vdash S\right) } d^{ n }
xf\left( x\right) $ is absolutely convergent and that $u$
and $v$ are two possibly different such maps.   Then for
given $\varepsilon >0$ let $k$ be a member of $\mathbb{N} $
such that for all members $l$ and $m$ of $\mathbb{N} $ such
that $m\geq l\geq k$ holds, both $\int_{ u_{ m }
\,\vdash u_{ l
} } d^{ n } x\left| f\left( x\right) \right| \leq \frac{
\varepsilon }{ 2
} $ and $\int_{ v_{ m } \,\vdash v_{ l } } d^{ n } x\left|
f\left(
x\right) \right| \leq \frac{ \varepsilon }{ 2 } $ hold, and
let
$m\geq k$ be such that $u_{ k } \subseteq v_{ m } $ and
$v_{ k } \subseteq u_{ m } $ both hold.   Then for any
$p\geq m$ and any $q\geq m$, $\left( u_{ p } \,\vdash v_{ q
} \right) \subseteq \left( u_{ p } \,\vdash u_{ k } \right)
$ holds, (for any member of $u_{ p } $ that is $\notin v_{
q } $, is not a member of $u_{ k }  $), hence $\int
_{ u_{ p } \,\vdash v_{ q } } d^{ n } x\left| f\left(
x\right)
\right| \leq \int_{ u_{ p } \,\vdash u_{ k } } d^{
 n } x\left|
f\left(
x\right) \right| \leq \frac{ \varepsilon }{ 2 } $ holds,
and
similarly
$\left( v_{ q } \,\vdash u_{ p } \right) \subseteq \left(
v_{ q } \,\vdash v_{ k } \right) $ holds hence
$\int_{ v_{ q }
\,\vdash u_{ p } } d^{ n } x\left| f\left( x\right) \right|
\leq \int_{ v_{
q } \,\vdash v_{ k } } d^{ n } x\left| f\left( x\right)
\right| \leq \frac{
\varepsilon }{ 2 } $ holds, hence both sequences of
integrals
converge to the same limit.

\begin{bphzlemma} \label{Lemma 24}
\end{bphzlemma}
\vspace{-6.143ex}

\noindent \hspace{11.9ex}{\bf.  }Let $V$ be a
finite set such that $\#\left(
V\right) \geq 2$, let $x$ be a member of $\mathbb{E} _{
d }^{ V } $, and let $\lambda $ be any member of
$\mathbb{R}^{ V }$ such that $\sum_{i\in V } \lambda _{ i }
=1$.   Let $i$ be any member of $V$, and let the member $z$
of $\mathbb{E} _{ d }^{ V } $ be defined by $z_{ i } \equiv
\sum_{j\in
V } \lambda _{ j } x_{ j } $, and $z_{ j } \equiv
\left( x_{ j } -x_{ i } \right) $ for $j\neq i$.
\enlargethispage{1.9ex}

Then the Jacobian $\det \frac{\partial z_{ j } }{\partial
x_{ k } }
$ is equal to $1$, and $x_{ i } =z_{ i } -\sum_{j\neq
i } \lambda _{ j } z_{ j } $.

\vspace{2.5ex}

\noindent {\bf Proof.}  We first note that
\[
\sum_{j\neq i } \lambda _{ j } z_{ j } =\sum_{j\neq i }
\lambda _{ j } \left( x_{ j } -x_{ i } \right) =\sum_{j\in
V } \lambda _{ j } \left( x_{ j } -x_{ i } \right)
=z_{ i } -x_{ i } .
\]

And to calculate the Jacobian we note that the
transformation from $x$ to $z$ is equal to the compound of
a transformation from $x$ to $w$, where $w_{ i } \equiv x_{
i } $, and $w_{ j } \equiv \left( x_{ j } -x_{ i } \right)
$ for $j\neq i$, followed by the transformation from $w$ to
$z$ defined by $z_{ i } \equiv w_{ i } +\sum_{j\neq i }
\lambda _{ j } w_{ j } $, and $z_{ j } \equiv w_{ j } $ for
$j\neq i$.   And if we write these transformations as
matrices with
\label{Start of original page 93}
 $i$ in the first rows and columns, then the transformation
from $x$ to $w$ is lower triangular with $1$ in every
leading diagonal position, and the transformation from $w$
to $z$ is upper triangular with $1$ in every leading
diagonal position.

\begin{bphzthm} \label{Theorem 1}
\end{bphzthm}
\vspace{-6.143ex}

\noindent \hspace{12.4ex}{\bf.  }Let $V$ be a partition
such that $\mathcal{U}
\left( V\right) $ is finite and $\#\left( V\right) \geq 2$
holds, and let $H$ be a partition such that $\mathcal{U}
\left( V\right) $ is $\left( V\cup H\right) $-connected and
such that if $E$ is any member of $H$ such that $E$
intersects \emph{more} than one member of $V$, then $E$ has
\emph{exactly} two members.   (Hence \emph{no} member of
$H$ intersects \emph{more} than two members of $V $.)

Let $W$ be the subset of $H$ whose members are all the
members $E$ of $H$ such that $E$ intersects exactly two
members of $V$.   (Thus $W$ is a partition such that every
member $E$ of $W$ has exactly two members.)

Let $d$ be an integer $\geq 1$.

Let $Z$ be the subset of $\mathbb{E} _{ d }^{
\mathcal{U} \left(
V\right) } $ whose members are all the members $y$ of
$\mathbb{E} _{ d }^{ \mathcal{U} \left( V\right) } $
such that
$\left| y_{ i } -y_{ j } \right| =0$ holds for at least one
member
$\left\{ i,j\right\} $ of $W$.

Let $\theta $ be a member of $\mathbb{Z}^{ W } $.

For each member $A$ of $\left( \Xi \left( V\right) \,\vdash
V\right) $, we define
\[
D_{ A } \equiv \left( \sum_{\Delta \in
\left( W\cap \mathcal{Q}
\left(
A\right) \right) } \theta _{ \Delta } \right) -d\left(
\#\left( \mathcal{P} \left( V,A\right) \right) -1\right) .
\]

Let $N\equiv  \sum_{A\in \left( \Xi \left( V\right) \,\vdash
V\right) } \left( 1+\max\left( D_{ A } ,0\right)
\right) $.

Let $M$ be a finite real number $\geq 0$, and let $S$ and
$T$ be finite real numbers such that $0<S<T$ holds.

Let $\mathcal{J} $ be a map whose domain is $\mathcal{G}
\left( V,H\right) $, and such that for each member $F$ of
$\mathcal{G} \left( V,H\right) $, $\mathcal{J} _{ F } $ is
a map whose domain is $\left( \mathbb{E}_{ d }^{
\mathcal{U}
\left( V\right) } \,\vdash Z\right) $ and whose range is a
subset of $\mathbb{R} $.

For each member $F$ of $\mathcal{G} \left( V,H\right) $,
and for each member $y$ of $\left( \mathbb{E}_{ d }^{
 \mathcal{U}
\left( V\right) } \,\vdash Z\right) $, we define
$\mathcal{J} _{ F } \left( y\right) \equiv \left(
\mathcal{J} _{ F } \right) _{ y } $.

Let $\mathcal{J} $ satisfy the requirement that if $F$ and
$G$ are members of $\mathcal{G} \left( V,H\right) $, and
$y$ is a member of $\left( \mathbb{E} _{ d }^{ \mathcal{U}
\left( V\right) } \,\vdash Z\right) $, such that for
every member $\left\{ i,j\right\} $ of $W$, either $\left|
y_{ i }
-y_{ j } \right| \leq S$ holds or $\mathcal{Y} \left(
F,\left\{
i,j\right\} \right) =\mathcal{Y} \left( G,\left\{
i,j\right\}
\right) $ holds, then $\mathcal{J} _{ F } \left( y\right)
=\mathcal{J} _{ G } \left( y\right) $ holds.

Let $\mathcal{J} $ also satisfy the requirement that for
each member $F$ of $\mathcal{G} \left( V,H\right) $,
$\mathcal{J} _{ F } \left( y\right) $ and all its
derivatives with respect to the $y_{ i } $, $i\in
\mathcal{O} \left( V,H\right) $, of degree up to and
including $N$, exist and are continuous
\label{Start of original page 94}
 for all $y\in \left( \mathbb{E} _{ d }^{ \mathcal{U}
  \left(
V\right) } \,\vdash Z\right) $.

For any ordered pair $\left(i,j\right) $ of a map $i$ such
that
$\mathcal{D} \left( i\right) $ is finite and $\mathcal{R}
\left( i\right) \subseteq \mathcal{O} \left( V,H\right) $
holds, and a member $j$ of $\mathcal{O} \left( V,H\right)
$, let $\nu _{ ij } $ denote the number of members $\alpha
$ of $\mathcal{D} \left( i\right) $ such that $i_{ \alpha
} =j$ holds.

And let $\mathcal{J} $ also satisfy the requirement that if
$i$ is any map such that $\mathcal{D} \left( i\right) $ is
finite, $\#\left( \mathcal{D} \left( i\right) \right) \leq
N$, and $\mathcal{R} \left( i\right) \subseteq \mathcal{O}
\left( V,H\right) $, and $u$ is any map such that
$\mathcal{D} \left( i\right) \subseteq \mathcal{D} \left(
u\right) $, and for each member $\alpha $ of $\mathcal{D}
\left( i\right) $, $u_{ \alpha } $ is a unit $d $-vector,
then the following inequality holds for all $F\in
\mathcal{G} \left( V,H\right) $ and for all $y\in \left(
\mathbb{E} _{ d }^{ \mathcal{U} \left( V\right) } \,\vdash
Z\right) $:
\[
\left| \left( \prod_{\alpha \in \mathcal{D}
\left( i\right) }
\left( u_{ \alpha } .\hat{ y }_{ i_{ \alpha } } \right)
\right) \mathcal{J}
_{ F } \left( y\right) \right| \leq M \prod_{\Delta \equiv
\left\{
j,k\right\} \in W } \left| y_{ j } -y_{ k } \right|^{
-\left(
\theta
_{ \Delta } +\nu _{ ij } +\nu _{ ik } \right) } .
\]

And let $\mathcal{J} $ also satisfy the requirement that if
$y$ is any member of $\left( \mathbb{E} _{ d }^{
\mathcal{U}
\left( V\right) } \,\vdash Z\right) $ such that
$\left| y_{
i } -y_{ j } \right| \geq T$ holds for any member $\left\{
i,j\right\} $ of $W$, then $\mathcal{J} _{ F } \left(
y\right) =0$ holds for all members $F$ of $\mathcal{G}
\left( V,H\right) $.

Let $\omega $ be any set of contraction weights for $V$,
let $h$ be any member of $\mathcal{U} \left( V\right) $,
let $O\equiv \mathcal{C} \left( V,h\right) $, let $b$ be
any member of $\mathbb{E} _{ d } $, and let $\mathbb{W}$ be
the subset of $\mathbb{U} _{ d } \left( V,\omega \right) $
whose members are all the members $x$ of $\mathbb{U} _{ d }
\left( V,\omega \right) $ such that $x_{ O } =b$ holds.
Then the following integral is finite and absolutely
convergent:
\[
\int_{ \mathbb{W} } \left( \prod_{A\in \left(
V\,\vdash \left\{
O\right\}
\right) } d^{ d } x_{ A } \right) \left(
 \sum_{F\in \mathcal{G}
\left(
V,H\right) } \left( -1\right)^{ \#\left( \mathbb{B}
\left( F\right) \right) } \sum_{n\in \mathbb{X} \left(
\mathbb{I} \left( F,H\right) ,D\right) }
\times \right. \hspace{4cm}
\]
\[
\hspace{2cm} \left. \times \left( \left(
   \prod_{\left(
i,A\right) \in \mathcal{U} \left( \mathcal{R} \left(
\mathbb{I} \left( F,H\right) \right) \right) }
\left( \frac{\left( \left( x_{ \mathcal{K}
\left( F,A,i\right) } -x_{ A
} \right) .\hat{ y }_{ i } \right)^{ n_{ iA } } }{ n_{ iA }
! } \right) \right)
\mathcal{J}
_{ F } \left( y\right) \right)_{ y=\eta
\left( F,H,x\right) } \right)
\]

\vspace{2.5ex}

\noindent {\bf Proof.}  We choose a real number
$\sigma $ such that
$0<\sigma \leq \frac{3}{25}  $ holds, and we define the
real number $\lambda $ by $\lambda \equiv \left(\frac{ 1}{4}
\right) \left( 1-\sqrt{1-8\sigma}\right) $, so that
$0<\lambda \leq \frac{1}{5}$ holds.

We note that $\lambda $ and $\sigma $ satisfy the equation
$\lambda =\frac{ \sigma}{ 1-2\lambda } $, and that
$0<\sigma <\lambda $ holds.

And we choose a real number $R$ such that $0<R\leq \left(
1-2\lambda \right) S$ holds.

We recall from
page \pageref{Start of original page 48}
 that for any member $F$ of
$\mathcal{G} \left( V,H\right) $, and any member $x$ of
$\mathbb{U} _{ d } \left( V,\omega \right) $, the following
identity holds:
\[
 \sum_{\left( P,Q\right) \in \mathcal{N} \left( V,H\right)
} \mathcal{E} \left( P,Q,H,\sigma ,R,x\right) \mathcal{S}
\left( P,F\right) \mathcal{S} \left( F,Q\right) =1
\]

Inserting this identity into the integrand of the above
integral, and noting that $\mathcal{S} \left( P,F\right)
\mathcal{S} \left( F,Q\right) $ is $0$ unless $F\in
\mathbb{K} \left( P,Q\right) $ holds we find that the above
integral is equal to
\label{Start of original page 95}
\[
 \int_{ \mathbb{W} } \left(  \prod_{A\in \left(
 V\,\vdash \left\{ O\right\}
\right) } d^{ d } x_{ A } \right) \hspace{-0.1cm} \left(
  \sum_{\left( P,Q\right)
\in
\mathcal{N} \left( V,H\right) } \hspace{-0.16cm}
 \mathcal{E} \left(
P,Q,H,\sigma ,R,x\right)
\sum_{F\in \mathbb{K} \left( P,Q\right) }
\left( -1\right)^{ \#\left( \mathbb{B} \left(
F\right) \right) } \hspace{-0.1cm}
 \sum_{n\in \mathbb{X} \left( \mathbb{I}
\left( F,H\right) ,D\right) }
\times \right.
\]
\[
\hspace{2.7cm} \left. \times \left( \left(
  \prod_{\left( i,A\right)
\in \mathcal{U} \left( \mathcal{R} \left( \mathbb{I} \left(
F,H\right) \right) \right) }
\left( \frac{\left( \left( x_{
\mathcal{K} \left( F,A,i\right) } -x_{ A } \right) .\hat{ y
}_{ i
} \right)^{ n_{ iA } } }{ n_{ iA } ! } \right) \right)
 \mathcal{J} _{ F } \left(
y\right) \right)_{ y=\eta \left( F,H,x\right) } \right)
\]

We shall prove that for each member $\left(P,Q\right) $ of
$\mathcal{N}
\left( V,H\right) $, the integral over $\mathbb{W}$ of the
term associated with $\left(P,Q\right) $ in the integrand
of this
integral, is finite and absolutely convergent.

We first use Lemma \ref{Lemma 22} to conclude that the
term associated
with $\left(P,Q\right) $ in the integrand of the above
integral is equal
to
\[
\left( -1\right)^{ \#\left( \mathbb{B} \left( P\right)
\right) } \mathcal{E} \left( P,Q,H,\sigma ,R,x\right)
\times \hspace{9.0cm}
\]
\[
\times   \sum_{u\in \mathbb{X} \left( \downarrow \left(
\mathbb{I} \left( Q,H\right) ,P\right) ,D\right) }
\hspace{0.2cm}
  \sum_{m\in \mathbb{A} \left( \mathbb{J} \left(
P,Q,H\right)
,\left( D-\xi \left( P,Q,H,u\right)
+1\hspace{-0.7516ex}1 \right) \right) } \int_{
\mathbb{D} } \left( d^{ \#\left( Q\,\vdash P\right) }
 \rho \right) \times
\]
\[
\times \left( \left(
 \prod_{A\in \left( Q\,\vdash P\right) } \left( 1-\rho _{ A
} \right)^{ \left( D_{ A } -\xi _{ A } \left( P,Q,H,u\right)
\right) } \left( D_{ A } -\xi _{ A } \left( P,Q,H,u\right)
+1\right) \right) \times \right.
\]
\[
\times \left(
\prod_{\left( \left( i,B\right) ,X\right) \in
\mathcal{U} \left( \mathcal{R} \left( \psi \left(
\mathbb{J} \left( P,Q,H\right) \right) \right) \right) }
\left(
  \prod_{E\in \left( \mathbb{G} _{ iB } \left( Q,H\right)
\,\vdash X\right) } \rho _{ E } \right)^{m_{ iBX } }
\right) \times
\]
\[
\times \left( \left(
  \prod_{\left( \left( i,B\right) ,X\right) \in
\mathcal{U} \left( \mathcal{R} \left( \psi \left(
\mathbb{J} \left( P,Q,H\right) \right) \right) \right)
 } \left( \frac{\left( \left( x_{ \mathcal{K} \left(
 Q,B,i\right)
 } -x_{ B } \right) .\hat{ y }_{ i }
 \right)^{ m_{ iBX } } }{ m_{
iBX }
! } \right) \right) \times \right.
\]
\[
\left. \left. \times \left(
 \prod_{\left( i,A\right) \in \mathcal{U} \left(
\mathcal{R}
\left( \downarrow \left( \mathbb{I} \left( Q,H\right)
,P\right) \right) \right) }
\left( \frac{\left( \left( x_{
\mathcal{K} \left( Q,A,i\right) } -x_{ A }
\right) .\hat{ y
}_{ i
} \right)^{ u_{ iA } } }{ u_{ iA } ! } \right) \right)
 \mathcal{J} _{ Q } \left(
y\right) \right)_{ y=\mu \left( P,Q,H,x,\mathcal{X} \left(
P,Q,H,\rho \right) \right) } \right)
\]
where $1\hspace{-0.7516ex}1 $ is a map such that $\left(
Q\,\vdash P\right) \subseteq \mathcal{D} \left(
1\hspace{-0.7516ex}1 \right) $ holds and such that for
each member $A$ of $\left( Q\,\vdash P\right) $,
$1\hspace{-0.7516ex}1 _{ A } =1$ holds, and $\mathbb{D}
$ is the set of all members $\rho $ of $\mathbb{R}^{ \left(
Q\,\vdash P\right) } $ such that $0\leq \rho _{ A } \leq 1$
holds for every member $A$ of $\left( Q\,\vdash P\right) $.

We shall prove, for each ordered
pair $\left(u,m\right) $ of a member $u$
of $\mathbb{X} \left( \downarrow \left( \mathbb{I} \left(
Q,H\right) ,P\right) ,D\right) $ and a member $m$ of
$\mathbb{A} \left( \mathbb{J} \left( P,Q,H\right) ,\left(
D-\xi \left( P,Q,H,u\right) +1\hspace{-0.7516ex}1
\right) \right) $, that the integral over $\mathbb{W}$ of
the term in the above formula associated with $u$ and $m$
is finite and absolutely convergent.

Let $\tilde{Z}$ denote the subset of $\mathbb{W}$ whose
members
are all the members $x$ of $\mathbb{W}$ such that $\left|
x_{ A }
-x_{ B } \right| =0$ holds for at least one member $\left\{
A,B\right\} $ of $\mathcal{Q} \left( \Xi \left( V\right)
\right) $ such that $A\cap B=\emptyset $.   Now if
$\varepsilon
$ is any given real number $>0$, then as observed on
page \pageref{Start of original page 90},
there exists an open subset $K$ of $\mathbb{W}$ such
that $\tilde{ Z } \subseteq K$ holds, and the $d\left(
\#\left(
V\right) -1\right) $-volume of $K$ is $\leq \varepsilon $.
 And our assumptions on $\mathcal{J} $ guarantee that the
integral over $\left( \mathbb{W} \,\vdash K\right) $ is
finite, hence by Lemma \ref{Lemma 23}, it is sufficient
to prove, for
some $F(x) $ such that for all $x\in \left( \mathbb{W}
\,\vdash \tilde{ Z } \right) $, $F(x) $ is greater than or
equal
to the absolute value of the term under consideration, that
\label{Start of original page 96}
 the integral over $\mathbb{W}$ of $F(x) $ converges.

 For each ordered pair $\left(y,i\right) $
 of a member $y$ of
$\mathbb{E} _{ d }^{ \mathcal{U} \left( V\right) } $ and a
member $i$ of $\mathcal{O} \left( V,H\right) $, we define
$a(y,i) $ to be equal to $\left| y_{ i } -y_{ j } \right| $
if
the
partition $W$ contains the member $\left\{ i,j\right\} $,
and
to be equal to $1$ if $i$ is not a member of $\mathcal{U}
\left( W\right) $.

Let $u$ be any member of $\mathbb{X} \left( \downarrow
\left( \mathbb{I} \left( Q,H\right) ,P\right) ,D\right) $
and $m$ be any member of \\
$\mathbb{A} \left( \mathbb{J}
\left( P,Q,H\right) ,\left( D-\xi \left( P,Q,H,u\right)
+1\hspace{-0.7516ex}1 \right) \right) $.   Then by the
assumed properties of $\mathcal{J} $, the following
inequality holds:
\[
\left| \left( \left( \left(
  \prod_{\left( \left( i,B\right) ,X\right) \in
\mathcal{U}
\left( \mathcal{R} \left( \psi \left( \mathbb{J} \left(
P,Q,H\right) \right) \right) \right) }
\left( \frac{\left(
\left( x_{ \mathcal{K} \left( Q,B,i\right) } -x_{ B }
\right) .\hat{ y }_{ i } \right)^{ m_{ iBX } } }
{ m_{ iBX } ! } \right) \right)
\times \right. \right. \right. \hspace{-2.3pt}
\hspace{4.0cm}
\]
\[
\left. \left. \left. \times \hspace{-0.05cm} \left(
 \prod_{\left(
i,A\right) \in \mathcal{U} \left( \mathcal{R} \left(
\downarrow \left( \mathbb{I} \left( Q,H\right) ,P\right)
\right) \right) } \hspace{-0.08cm}
\left( \frac{\left( \left( x_{ \mathcal{K}
\left( Q,A,i\right) } -x_{ A } \right) .\hat{ y }_{ i }
\right)^{
u_{ iA } } }{ u_{ iA } ! } \right) \hspace{-0.08cm} \right)
 \mathcal{J} _{ Q }
\left( y\right) \right)_{ \hspace{-0.1cm}
y=\mu \left( P,Q,H,x,\mathcal{X} \left( P,Q,H,\rho \right)
\right) } \right) \right| \leq
\]
\[
\leq M \left( \left(
  \prod_{\left( \left( i,B\right) ,X\right) \in
\mathcal{U}
\left( \mathcal{R} \left( \psi \left( \mathbb{J} \left(
P,Q,H\right) \right) \right) \right) }
\left( \frac{
\left| x_{ \mathcal{K} \left( Q,B,i\right) }
 -x_{ B } \right|
 }{a\left(
y,i\right) } \right)^{ m_{ iBX } } \right)
\times \right.
\]
\[
\times \left(
 \prod_{\left( i,A\right) \in
\mathcal{U}
\left( \mathcal{R} \left( \downarrow \left( \mathbb{I}
\left( Q,H\right) ,P\right) \right) \right) }
\left( \frac{
\left| x_{ \mathcal{K} \left( Q,A,i\right) }
 -x_{ A } \right|
 }{a\left(
y,i\right) } \right)^{ u_{ iA } } \right) \times
\]
\[
\hspace{1.0cm} \left.
\times \left( \prod_{\Delta \equiv \left\{ j,k\right\}
 \in W
} \left| y_{ j } -y_{ k } \right| ^{-\theta _{ \Delta } }
\right) \left(
\prod_{\Delta \equiv
\left\{ j,k\right\} \in W } \mathbb{S} \left(
T-\left| y_{ j } -y_{ k } \right| \right) \right)
\right)_{ y=\mu \left(
P,Q,H,x,\mathcal{X}
\left( P,Q,H,\rho \right) \right) }
\]

For any ordered pair $\left(x,i\right) $ of a
member $x$ of $\mathbb{U}
_{ d } \left( V,\omega \right) $ and a member $i$ of
$\mathcal{O} \left( V,H\right) $, we define
$b\left(x,i\right) $ to be
equal to $\left| x_{ \mathcal{Z} \left( P,H,i\right) }
-x_{ \mathcal{Z}
\left( P,H,j\right) } \right| $ if $i$ is a member of the
member
$\left\{ i,j\right\} $ of the partition $W$, and to be equal
to $1$ if $i$ is not a member of any member of $W$.

Now if $\left\{ i,j\right\} \in W$ then by
page \pageref{Start of original page 80} and
Lemma \ref{Lemma 14} we have, for any finite real number
$\alpha $,
that
\[
\left| \mu _{ i } \left( P,Q,H,x,\mathcal{X} \left(
P,Q,H,\rho
\right) \right) -\mu _{ j } \left( P,Q,H,x,\mathcal{X}
\left( P,Q,H,\rho \right) \right) \right|^{ -\alpha }
 \leq \hspace{2.0cm}
\]
\[
\hspace{7.0cm} \leq \left( \frac{ 1 }{
1-2\lambda } \right)^{ \left| \alpha \right| } \left|
x_{ \mathcal{Z} \left(
P,H,i\right) } -x_{ \mathcal{Z} \left( P,H,j\right) }
 \right| ^{
-\alpha }
\]
holds for all $\rho \in \mathbb{D} $ and for all $x$ such
that $\mathcal{E} \left( P,Q,H,\sigma ,R,x\right) \neq 0$,
hence for all $\rho \in \mathbb{D} $ and for all $x$ such
that $\mathcal{E} \left( P,Q,H,\sigma ,R,x\right) \neq 0$,
the above expression is bounded above by
\[
M \left( \prod_{\left( \left( i,B\right) ,X\right)
\in \mathcal{U}
\left( \mathcal{R} \left( \psi \left( \mathbb{J} \left(
P,Q,H\right) \right) \right) \right) }
\left( \frac{
\left| x_{ \mathcal{K} \left( Q,B,i\right) } -x_{ B }
 \right|
 }{\left(
1-2\lambda \right) b\left( x,i\right) } \right)^{
 m_{ iBX } } \right) \times \hspace{5.0cm}
\]
\[
\times \left(
\prod_{\left(
i,A\right) \in \mathcal{U} \left( \mathcal{R} \left(
\downarrow \left( \mathbb{I} \left( Q,H\right) ,P\right)
\right) \right) }
\left( \frac{\left| x_{ \mathcal{K} \left(
Q,A,i\right) } -x_{ A } \right| }{\left( 1-2\lambda
\right)
b\left(
x,i\right) } \right)^{ u_{ iA } } \right) \left(
 \prod_{\Delta \equiv \left\{ j,k\right\}
\in
W } \left( \frac{ 1 }{ 1-2\lambda } \right)^{ \left| \theta
_{ \Delta }
\right| } \right) \times
\]
\label{Start of original page 97}
\[
 \times \hspace{-0.15cm} \left( \prod_{\Delta
\equiv \left\{ j,k\right\} \in W }
\hspace{-0.3cm} \left| x_{ \mathcal{Z}
\left( P,H,j\right) } -x_{ \mathcal{Z}
\left( P,H,k\right) }
\right| ^{-\theta _{ \Delta } } \hspace{-0.1cm} \right)
\hspace{-0.2cm}
\left(  \prod_{\Delta \equiv
\left\{ j,k\right\}
\in W } \hspace{-0.3cm} \mathbb{S}
\left( T-\left( 1 \! - \! 2\lambda
\right) \! \left| x_{ \mathcal{Z} \left( P,H,j\right) }
-x_{ \mathcal{Z}
\left( P,H,k\right) } \right| \right) \hspace{-0.1cm}
 \right)
\]

The absolute value of the integrand has now been bounded by
the product of a factor that depends on $\rho $ but not on
$x$, and a factor that depends on $x$ but not on $\rho $,
and furthermore the $\rho $-dependent factor is the product
of finite powers, all $\geq 0$, of $\left( 1-\rho _{ A }
\right) $ and $\rho _{ A } $, where $A$ is a member of
$\left( Q\,\vdash P\right) $, and the $\rho $ integration
domain is $0\leq \rho _{ A } \leq 1$ for all $A\in \left(
Q\,\vdash P\right) $, hence the $\rho $-integral is finite
and absolutely convergent.

Now for any member $A$ of $\left( \Xi \left( V\right)
\,\vdash V\right) $, the inequality $\left| x_{ \mathcal{K}
\left(
Q,A,i\right) } -x_{ A } \right| \leq \mathbb{L} \left(
P,A,x\right)
$ holds.   Hence it is now sufficient to prove that if $u$
is any member of $\mathbb{X} \hspace{-0.08cm}
 \left( \downarrow \hspace{-0.16cm} \left(
\mathbb{I} \left( Q,H\right) ,P\right) ,D\right) $ and $m$
is any member of $\mathbb{A} \left( \mathbb{J} \left(
P,Q,H\right) ,\left( D-\xi \left( P,Q,H,u\right)
+1\hspace{-0.7516ex}1 \right) \right) $, then the
integral over $\mathbb{W}$ of the quantity $I_{ um } $,
defined as follows, is finite:
\[
I_{ um } \equiv \mathcal{E} \left( P,Q,H,\sigma ,R,x\right)
\left(
   \prod_{\left( \left( i,A\right) ,X\right) \in \mathcal{U}
\left( \mathcal{R} \left( \psi \left( \mathbb{J} \left(
P,Q,H\right) \right) \right) \right) }
\left(\frac{\mathbb{L}
\left( P,A,x\right)}{b\left( x,i\right)}\right)^{
  m_{ iAX } } \right) \times \hspace{2.0cm}
\]
\[
\hspace{6.0cm} \times \left(
   \prod_{\left( i,A\right) \in \mathcal{U} \left(
\mathcal{R}
\left( \downarrow \left( \mathbb{I} \left( Q,H\right)
,P\right) \right) \right) } \left( \frac{\mathbb{L} \left(
P,A,x\right) }{b\left( x,i\right) } \right) ^{  u_{ iA } }
\right) \Theta \hspace{0.2cm}
\Phi
\]
where we also define
\[
\Theta \equiv \left(
\prod_{\Delta \equiv \left\{ j,k\right\} \in W }
\left| x_{ \mathcal{Z} \left( P,H,j\right) }
 -x_{ \mathcal{Z}
\left( P,H,k\right) } \right| ^{-\theta _{ \Delta  } }
\right)
\]
and
\[
\Phi \equiv \left(
\prod_{\Delta \equiv \left\{ j,k\right\} \in W }
\mathbb{S} \left( T-\left( 1-2\lambda \right)
\left| x_{ \mathcal{Z} \left( P,H,j\right) }
 -x_{ \mathcal{Z} \left(
P,H,k\right) } \right| \right) \right) .
\]

Let $u$ be any member of $\mathbb{X} \left( \downarrow
\left( \mathbb{I} \left( Q,H\right) ,P\right) ,D\right) $,
$m$ be any member of \\
$\mathbb{A} \left( \mathbb{J} \left(
P,Q,H\right) ,\left( D-\xi \left( P,Q,H,u\right)
+1\hspace{-0.7516ex}1 \right) \right) $, and $G$ be the
set whose members are all the maps $g$ such that
$\mathcal{D} \left( g\right) =\mathbb{B} \left( Q\right) $,
and for each member $A$ of $\mathbb{B} \left( Q\right) $,
$g_{ A } $ is a member of $\mathcal{Q} \left( \mathcal{P}
\left( P,A\right) \right) $.

For any ordered triple $\left(x,g,A\right) $ of a member
$x$ of
$\mathbb{E} _{ d }^{ \Xi \left( V\right) } $, a member $g$
of $G$, and a member $A$ of $\mathbb{B} \left( Q\right) $,
we define $c\left( x,g,A\right) \equiv \left| x_{ J } -x_{
K }
\right| $, where $J$ and $K$ are the two members of $g_{ A
} $, or
in other words, $g_{ A } =\left\{ J,K\right\} $.

And for any member $g$ of $G$, we define
\[
U_{ g } \equiv \mathcal{E} \left( P,Q,H,\sigma ,R,x\right)
\left(
   \prod_{\left( \left( i,B\right) ,X\right) \in \mathcal{U}
\left( \mathcal{R} \left( \psi \left( \mathbb{J} \left(
P,Q,H\right) \right) \right) \right) } \left( \frac{c\left(
x,g,B\right) }{b\left( x,i\right) }\right) ^{   m_{ iBX } }
\right) \times \hspace{2.0cm}
\]
\[
\hspace{5.0cm} \times \left( \prod_{\left(
i,B\right) \in \mathcal{U} \left( \mathcal{R} \left(
\downarrow \left( \mathbb{I} \left( Q,H\right) ,P\right)
\right) \right) } \left( \frac{c\left( x,g,B\right)
}{b\left(
x,i\right) }\right)^{ u_{ iB } } \right) \Theta
\hspace{0.2cm} \Phi
\]
\label{Start of original page 98} \label{Original 98}

 Now for any member $x$ of $\mathbb{E} _{ d }^{ \Xi \left(
V\right) } $ and for any member $A$ of $\mathbb{B} \left(
Q\right) $, $\mathbb{L} \left( P,A,x\right) $ is by
definition equal to $ \hspace{0.7cm} \max
\rule[-2ex]{0pt}{2ex}_{
\hspace{-1.6cm}\Delta \equiv \left\{ J,K\right\}
\in
\mathcal{Q} \left( P,A\right) } \hspace{-0.8cm}
 \left| x_{ J }
-x_{ K
} \right| $, hence if $x$ is any member of $\mathbb{E} _{
d }^{ \Xi
\left( V\right) } $ then the following inequality holds:
\[
I_{ um } \leq \sum_{g\in G } U_{ g }
\]

(For every term in the right-hand side is $\geq 0$, and
there is always at least one term in the right-hand side
that is equal to the left-hand side.)

We shall prove that if $g$ is any member of the finite set
$G$, then the integral of $U_{ g } $ over $\mathbb{W}$ is
finite.

Let $g$ be any member of $G$.   We first observe that,
apart from the factors \\
$\mathcal{E} \left( P,Q,H,\sigma
,R,x\right) $ and $\Phi $, $U_{ g } $ is a product of
factors each of which has the form $\left| x_{ J } -x_{ K }
\right|^{ \alpha } $, where $\alpha $ is a finite real
number, and $J$
and $K$ are members of $P$ such that the following two
conditions hold:

\vspace{1.0ex}

\noindent (i)  $ J\neq K $

\vspace{1.0ex}

\noindent (ii)  If $A$ is any member of $P$,
then $J\subset A$ holds
ifif $K\subset A$ holds.

\vspace{1.0ex}

For $\left| x_{ J } -x_{ K } \right| $ is either $ c
\left(x,g,B\right) $ for some
member $B$ of $\mathbb{B} \left( Q\right) $, which means
that $\left\{ J,K\right\} \in \mathcal{Q} \left(
\mathcal{P}
\left( P,B\right) \right) $ holds, or else it is
$\left| x_{ \mathcal{Z} \left( P,H,i\right) } -x_{
\mathcal{Z}
\left(
P,H,j\right) } \right| $ for some member $\left\{
i,j\right\}
$ of
$W$.

In the first case, $J\neq K$ follows directly from $\left\{
J,K\right\} \in \mathcal{Q} \left( \mathcal{P} \left(
P,B\right) \right) $.   Let $A$ be any member of $P$.
Then if $B\subseteq A$ holds, $J\subset A$ and $K\subset A$
both hold.   Now suppose that $B\subseteq A$ does
\emph{not} hold.   Then since both $A$ and $B$ are members
of the wood $Q$, either $A\subset B$ holds or $A\cap
B=\emptyset
$ holds.   And if $A\subset B$ holds, the fact that $J$ is
a member of $\mathcal{P} \left( P,B\right) $ implies that
$J\subset A$ cannot hold, and the fact that $K$ is a member
of $\mathcal{P} \left( P,B\right) $ implies that $K\subset
A$ cannot hold.   And if $A\cap B=\emptyset $ holds, then
$J\cap
A=\emptyset $ holds and $K\cap A=\emptyset $ holds, hence
since
neither $J$ nor $K$ is empty, neither $J\subset A$ nor
$K\subset A$ can hold.

And in the second case, $i$ is a member of $\mathcal{Z}
\left( P,H,i\right) $ and $j$ is a member of $\mathcal{Z}
\left( P,H,j\right) $, but $\left\{ i,j\right\} $ is not a
subset of $\mathcal{Z} \left( P,H,i\right) $, hence
$\mathcal{Z} \left( P,H,i\right) $ is not equal to
$\mathcal{Z} \left( P,H,j\right) $.   Now let $A$ be any
member of $P$.   Then if $\left\{ i,j\right\} \subseteq A$
holds, $\left( A\,\vdash \mathcal{Z} \left( P,H,i\right)
\right) $ has the member $j$ hence is nonempty, and $\left(
A\cap \mathcal{Z} \left( P,H,i\right) \right) $ has the
member $i$ hence is nonempty, hence $\mathcal{Z} \left(
P,H,i\right) \subset A$ holds since $\mathcal{Z} \left(
P,H,i\right) $ and $A$ do not overlap, and $\left(
A\,\vdash \mathcal{Z} \left( P,H,j\right) \right) $ has the
member $i$ hence is nonempty, and $\left( A\cap \mathcal{Z}
\left( P,H,j\right) \right) $ has the member $j$ hence is
nonempty, hence $\mathcal{Z} \left( P,H,j\right) \subset A$
holds since
\label{Start of original page 99} \label{Original 99}
 $\mathcal{Z} \left( P,H,j\right) $ and $A$ do not overlap.
  Now suppose that $\left\{ i,j\right\} $ is \emph{not} a
subset of $A$.   Then at least one of $i$ and $j$ is not a
member of $A$.   Suppose first that $i$ is \emph{not} a
member of $A$, and $j$ \emph{is} a member of $A$.   Then
$\mathcal{Z} \left( P,H,i\right) \subset A$ does not hold.
 And by definition, $\mathcal{Z} \left( P,H,j\right) $ is
the largest member of $P$ to have $j$ as a member, but not
to have as a subset any member of $H$ that has $j$ as a
member.   But $H$ is a partition, hence $\left\{ i,j\right\}
$ is the only member of $H$ to have $j$ as a member, hence
$\mathcal{Z} \left( P,H,j\right) $ is the largest member of
$P$ to have $j$ as a member but not have $i$ as a member,
hence $A\subseteq \mathcal{Z} \left( P,H,j\right) $ holds,
hence $\mathcal{Z} \left( P,H,j\right) \subset A$ does not
hold.   And if $i$ \emph{is} a member of $A$, and $j$ is
\emph{not} a member of $A$, an analogous argument shows
again that neither $\mathcal{Z} \left( P,H,i\right) \subset
A$ holds nor $\mathcal{Z} \left( P,H,j\right) \subset A$
holds.   And finally, if neither $i$ nor $j$ is a member of
$A$, then neither $\mathcal{Z} \left( P,H,i\right) \subset
A$ nor $\mathcal{Z} \left( P,H,j\right) \subset A$ can hold.

We next note that (i) and (ii)
imply directly that $J\cap K=\emptyset $ holds.   For
taking
$A$ as $J$ in (ii) shows that $K\subset J$
does not hold, and taking $A$ as $K$ in (ii)
shows that $J\subset K$ does not hold, hence since $J$ and
$K$ do not overlap, and $J$ is not equal to $K$, $J\cap
K=\emptyset $ must hold.

We now observe that (i) and (ii)
imply that $J$ and $K$ are distinct members of
$\mathcal{P} \left( P,\mathcal{Y} \left( \bar{ P } ,J\cup
K\right) \right) $.   For $J\neq K$ and $J$ and $K$ are
nonempty, hence $J\subset \mathcal{Y} \left( \bar{ P }
,J\cup
K\right) $ and $K\subset \mathcal{Y} \left( \bar{ P } ,J\cup
K\right) $ both hold.   Furthermore, there is no member $C$
of $P$ such that $J\subset C\subset \mathcal{Y} \left(
\bar{ P }
,J\cup K\right) $ holds, for if there was such a member
$C$ of $P$, then (ii) and $J\subset C$ would
imply that $K\subset C$, hence $\left( J\cup K\right)
\subseteq C$, which contradicts $C\subset \mathcal{Y}
\left( \bar{ P } ,J\cup K\right) $ since by definition
$\mathcal{Y} \left( \bar{ P } ,J\cup K\right) $ is the
\emph{smallest} member of $ \bar{ P } $ to have $\left(
J\cup
K\right) $ as a subset.   Hence $J\in \mathcal{P} \left(
P,\mathcal{Y} \left( \bar{ P } ,J\cup K\right) \right) $
holds.   And by an analogous argument, $K\in \mathcal{P}
\left( P,\mathcal{Y} \left( \bar{ P } ,J\cup K\right)
\right)
$ holds.

Hence $\left\{ J,K\right\} $ is a member of $\mathcal{Q}
\left( \mathcal{P} \left( P,\mathcal{Y} \left( \bar{ P }
,J\cup K\right) \right) \right) $.

We now define $\alpha $ to be the map whose domain is $
 \bigcup_{A\in \mathbb{B} \left( \bar{ P } \right) }
\mathcal{Q} \left( \mathcal{P} \left( P,A\right) \right)
$, and such that for each member $\left\{ J,K\right\} $ of
$\mathcal{D} \left( \alpha \right) $, $\alpha_{ \left\{ J,K
\right\} } $ is equal to the negative of the total power of
$\left| x_{ J } -x_{ K } \right| $ in the expression
\[
\left(
 \prod_{\left( \left( i,B\right) ,X\right) \in \mathcal{U}
\left( \mathcal{R} \left( \psi \left( \mathbb{J} \left(
P,Q,H\right) \right) \right) \right) }
\hspace{-0.15cm} \left( \frac{c\left(
x,g,B\right) }{b\left( x,i\right) } \right)^{ \! m_{ iBX } }
\! \right) \hspace{-0.2cm} \left(
\prod_{\left(
i,B\right) \in \mathcal{U} \left( \mathcal{R} \left(
\downarrow \left( \mathbb{I} \left( Q,H\right) ,P\right)
\right) \right) }
\hspace{-0.15cm} \left( \frac{c\left( x,g,B\right)
}{b\left(
x,i\right) } \right)^{ \! u_{ iB } } \! \right) \Theta .
\]

Then for each member $A$ of $\mathbb{B} \left( \bar{ P }
\right) $, $\alpha $ is a set of powers for $A$, and for
each member $x$ of $\mathbb{E} _{ d }^{ \Xi \left( V\right)
} $, the following equation holds:
\label{Start of original page 100}
\[
U_{ g } =\mathcal{E} \left( P,Q,H,\sigma ,R,x\right)
\left(
  \prod_{A\in \mathbb{B} \left( \bar{ P } \right) } \Psi
\left( \downarrow \left( x,\mathcal{P} \left( P,A\right)
\right) ,\alpha \right) \right) \Phi
\]

Now let $A$ be any member of $\mathbb{B} \left( Q\right) $,
and let $\left\{ j,k\right\} $ be any member of $W$.   Then
$\left\{ \mathcal{Z} \left( P,H,j\right) ,\mathcal{Z} \left(
P,H,k\right) \right\} $ is a member of $\mathcal{Q} \left(
\mathcal{P} \left( P,A\right) \right) $ ifif $\left\{
j,k\right\} \subseteq A$ holds and $\left\{ j,k\right\} $ is
\emph{not} a subset of any member of $\mathcal{P} \! \left(
P,A\right) $.   For suppose first that $\left\{ \mathcal{Z}
\left( P,H,j\right) ,\mathcal{Z} \left( P,H,k\right)
\right\} $ is a member of $\mathcal{Q} \left( \mathcal{P}
\left( P,A\right) \right) $.   Then $\left\{ j,k\right\}
\subseteq A$ certainly holds, and furthermore $k$ is not a
member of $\mathcal{Z} \left( P,H,j\right) $ and $j$ is not
a member of $\mathcal{Z} \left( P,H,k\right) $, hence
$\left\{ j,k\right\} $ is not a subset of any member of
$\mathcal{P} \left( P,A\right) $.   Now suppose that
$\left\{ j,k\right\} \subseteq A$ holds and $\left\{
j,k\right\} $ is \emph{not} a subset of any member of
$\mathcal{P} \left( P,A\right) $.   Then $\left( A\cap
\mathcal{Z} \left( P,H,j\right) \right) $ has the member
$j$ hence is nonempty, and $\left( A\,\vdash \mathcal{Z}
\left( P,H,j\right) \right) $ has the member $k$ hence is
nonempty, hence $\mathcal{Z} \left( P,H,j\right) \subset A$
holds since $\mathcal{Z} \left( P,H,j\right) $ and $A$ do
not overlap.   And $H$ is a partition hence $\left\{
j,k\right\} $ is the only member of $H$ to have $j$ as a
member, hence the fact that $\left\{ j,k\right\} $ is not a
subset of $\mathcal{K} \left( P,A,j\right) $ implies that
$\mathcal{K} \left( P,A,j\right) \subseteq \mathcal{Z}
\left( P,H,j\right) $ holds.   But $\mathcal{K} \left(
P,A,j\right) \subset \mathcal{Z} \left( P,H,j\right) $
cannot hold since $\mathcal{K} \left( P,A,j\right) $ is a
member of $\mathcal{P} \left( P,A\right) $, hence
$\mathcal{K} \left( P,A,j\right) =\mathcal{Z} \left(
P,H,j\right) $ holds hence $\mathcal{Z} \left( P,H,j\right)
$ is a member of $\mathcal{P} \left( P,A\right) $.   And by
an analogous argument, $\mathcal{Z} \left( P,H,k\right) $
is a member of $\mathcal{P} \left( P,A\right) $.

It follows immediately from this, together with the
definition of $\alpha $, that if $A$ is any member of
$\mathbb{B} \left( Q\right) $, then the contribution to
$\Gamma \left( \alpha ,\mathcal{P} \left( P,A\right)
\right) $ from $\Theta $ is equal to
\[
\left(
 \sum_{\Delta \in W\cap \mathcal{Q} \left( A\right) }
\theta _{ \Delta } \right)
 -  \sum_{C\in \left( \mathcal{P} \left(
P,A\right) \,\vdash V\right) } \left(
  \sum_{\Delta \in W\cap
\mathcal{Q} \left( C\right) } \theta _{ \Delta }
\right),
\]
where we noted that there is no need to include members $C$
of $\left( \mathcal{P} \left( P,A\right) \cap V\right) $ in
the outer summation in the second term, since by assumption
no member of $W$ is a subset of any member of $V$, hence
$W\cap \mathcal{Q} \left( C\right) $ is empty for every
member $C$ of $V$.
\enlargethispage{0.5ex}

Now for any member $A$ of $\left( \Xi \left( V\right)
\,\vdash V\right) $, $D_{ A } $ was defined on
page \pageref{Start of original page 93} by
\[
D_{ A } \equiv \left(
\sum_{\Delta \in W\cap \mathcal{Q} \left( A\right) }
\theta _{ \Delta } \right) -d\left( \#\left( \mathcal{P}
\left( V,A\right) \right) -1\right) ,
\]
hence if $A$ is any member of $\mathbb{B} \left( Q\right)
$, then the contribution to $\Gamma \left( \alpha
,\mathcal{P} \left( P,A\right) \right) $ from $\Theta $ is
equal to
\[
\left( D_{ A } +d\left( \#\left( \mathcal{P} \left(
V,A\right) \right) -1\right) \right) -   \sum_{C\in \left(
\mathcal{P} \left( P,A\right) \,\vdash V\right) }
\left( D_{ C } +d\left( \#\left( \mathcal{P} \left(
V,C\right) \right) -1\right) \right) .
\]

Now
\label{Start of original page 101}
\[
\#\left( \mathcal{P} \left( V,A\right) \right) =
\left( \sum_{C\in
\left( \mathcal{P} \left( P,A\right) \,\vdash V\right) }
\#\left( \mathcal{P} \left( V,C\right) \right) \right)
+\#\left( \mathcal{P} \left( P,A\right) \cap V\right)
\]
hence
\[
d\left( \#\left( \mathcal{P} \left( V,A\right) \right)
-1\right) - \left(
 \sum_{C\in \left( \mathcal{P} \left( P,A\right)
\,\vdash V\right) } d\left( \#\left( \mathcal{P}
\left( V,C\right) \right) -1\right) \right) =
\]
\[
\hspace{4.0cm}
=d\left( \#\left( \mathcal{P} \left( P,A\right) \cap
V\right) -1+\#\left( \mathcal{P} \left( P,A\right) \,\vdash
V\right) \right)
\]
\[
\hspace{4.0cm}
=d\left( \#\left( \mathcal{P} \left( P,A\right) \right)
-1\right) ,
\]
hence the contribution to $\Gamma \left( \alpha
,\mathcal{P} \left( P,A\right) \right) $ from $\Theta $ is
equal to
\[
d\left( \#\left( \mathcal{P} \left( P,A\right) \right)
-1\right) +D_{ A } - \left(
 \sum_{C\in \left( \mathcal{P} \left(
P,A\right) \,\vdash V\right) } D_{ C } \right) .
\]

\vspace{2.5ex}

We now make the following observations:

\begin{bphzobservation} \label{Observation 1}
\end{bphzobservation}
\vspace{-6.143ex}

\noindent \hspace{1.3ex}{\bf ) }Let $A$ be any member of
$\mathbb{B} \left( Q\right) $
and $B$ be any member of $\mathbb{B} \left( Q\right) $ such
that $B\subseteq A$ holds and $B$ is \emph{not} a subset of
any member of $\mathcal{P} \left( P,A\right) $.   Then
$\mathcal{P} \left( P,B\right) \subseteq \mathcal{P} \left(
P,A\right) $ holds.   For let $J$ be any member of
$\mathcal{P} \left( P,B\right) $.   Then $J\subset B$
holds, and there is no member $C$ of $P$ such that
$J\subset C\subset B$ holds.   Hence $J\subset A$ holds.
Now suppose that $E$ is a member of $P$ such that $J\subset
E\subset A$ holds.   Then $E\cap B$ has the nonempty subset
$J$ hence is nonempty, hence either $E\subset B$ holds or
$B\subseteq E$ holds, since $E$ does not overlap $B$.   But
$E\subset B$ cannot hold, since $J\subset E$ holds and $J$
is a member of $\mathcal{P} \left( P,B\right) $.   Now let
$X$ be the set whose members are all the members $Y$ of $P$
such that $E\subseteq Y$ and $Y\subset A$ both hold.   Then
$X$ is a finite set such that if $Y$ and $Z$ are any
members of $X$, then exactly one of $Y\subset Z$, $Y=Z$,
and $Z\subset Y$ holds, and furthermore $E$ is a member of
$X$ hence $X$ is nonempty, hence there is a unique member
$F$ of $X$ such that $Y\subseteq F$ holds for every member
$Y$ of $X$.   Then $F$ is a member of $\mathcal{P} \left(
P,A\right) $ and $E\subseteq F$ holds, hence the assumption
that $B$ is \emph{not} a subset of any member of
$\mathcal{P} \left( P,A\right) $ implies that $B\subseteq
E$ cannot hold.   Hence there is \emph{no} member $E$ of
$P$ such that $J\subset E\subset A$ holds, hence $J$ is a
member of $\mathcal{P} \left( P,A\right) $.

\begin{bphzobservation} \label{Observation 2}
\end{bphzobservation}
\vspace{-6.143ex}

\noindent \hspace{1.3ex}{\bf ) }Let $A$ be any member
of $\mathbb{B} \left( Q\right) $
and let $B$ be any member of $\mathbb{B} \left( Q\right) $
such that $B\subseteq A$ holds and $B$ is \emph{not} a
subset of any member of $\mathcal{P} \left( P,A\right) $.
Then it follows directly from
observation \ref{Observation 1}) above
that $\mathcal{Q} \left( \mathcal{P} \left( P,B\right)
\right) \subseteq \mathcal{Q} \left( \mathcal{P} \left(
P,A\right) \right) $ holds.

\begin{bphzobservation} \label{Observation 3}
\end{bphzobservation}
\vspace{-6.143ex}

\noindent \hspace{1.3ex}{\bf ) }Let $A$ be any member of
$\mathbb{B} \left( Q\right) $
and $\left\{ i,j\right\} $ be any member
of $W$ such that \\
$\left\{ \mathcal{Z} \left( P,H,i\right) ,\mathcal{Z} \left(
P,H,j\right) \right\} $ is a member of $\mathcal{Q} \left(
\mathcal{P} \left( P,A\right) \right) $.   Then $\left\{
i,j\right\} \subseteq A$ holds, and $\left\{ i,j\right\} $
is
\emph{not} a subset of $\mathcal{K} \left( P,A,i\right) $.
 For
\label{Start of original page 102}
 $i\in \mathcal{Z} \left( P,H,i\right) $ holds and
$\mathcal{Z} \left( P,H,i\right) \subset A$ holds hence
$i\in A$ holds, and $j\in \mathcal{Z} \left( P,H,j\right) $
holds and $\mathcal{Z} \left( P,H,j\right) \subset A$ holds
hence $j\in A$ holds.   And by assumption $\mathcal{Z}
\left( P,H,i\right) $ is a member of $\mathcal{P} \left(
P,A\right) $, hence $\mathcal{Z} \left( P,H,i\right) $ is
the unique member of $\mathcal{P} \left( P,A\right) $ to
have $i$ as a member, hence $\mathcal{Z} \left(
P,H,i\right) =\mathcal{K} \left( P,A,i\right) $.   But
$\left\{ i,j\right\} $ is a member of $W$ hence a member of
$H$, hence $\left\{ i,j\right\} $ is \emph{not} a subset of
$\mathcal{Z} \left( P,H,i\right) $, hence $\left\{
i,j\right\} $ is not a subset of $\mathcal{K} \left(
P,A,i\right) $.

\begin{bphzobservation} \label{Observation 4}
\end{bphzobservation}
\vspace{-6.143ex}

\noindent \hspace{1.3ex}{\bf ) }By
page \pageref{Start of original page 64} and
page \pageref{Start of original page 79},
$\mathcal{U} \left( \mathcal{R}
\left( \psi \left( \mathbb{J} \left( P,Q,H\right) \right)
\right) \right) $ is the set whose members are all the
ordered pairs $\left( \left( i,B\right) ,X\right) $ of a
member $ \left(i,B\right) $ of $\mathcal{U} \left(
\mathcal{R} \left(
\mathbb{J} \left( P,Q,H\right) \right) \right) $ and a
nonempty subset $X$ of $\mathbb{G} _{ iB } \left(
Q,H\right) $.   And by Lemma \ref{Lemma 21}, $\mathcal{U}
\left(
\mathcal{R} \left( \mathbb{J} \left( P,Q,H\right) \right)
\right) $ is the set whose members are all the ordered
pairs $\left(i,B\right) $ of a member $i$ of $\mathcal{O}
\left(
Q,H\right) =\mathcal{O} \left( V,H\right) $ and a member
$B$ of $\mathbb{Y} \left( Q,\mathcal{Z} \left( P,H,i\right)
,\mathcal{Z} \left( Q,H,i\right) \right) $.
\enlargethispage{1.5ex}

\begin{bphzobservation} \label{Observation 5}
\end{bphzobservation}
\vspace{-6.143ex}

\noindent \hspace{1.3ex}{\bf ) }Let $A$ be any member of
$\mathbb{B} \left( Q\right) $
and $\left( \left( i,B\right) ,X\right) $ be
any member of \\
$\mathcal{U} \left( \mathcal{R} \left( \psi \left(
\mathbb{J} \left( P,Q,H\right) \right) \right) \right) $
such that $i\in \mathcal{U} \left( W\right) $ holds and
$\mathcal{C} \left( W,i\right) \subseteq A$ holds and
$\mathcal{C} \left( W,i\right) $ is \emph{not} a subset of
$\mathcal{K} \left( P,A,i\right) $.   Then $B\subseteq A$
holds and $B$ is \emph{not} a subset of any member of
$\mathcal{P} \left( P,A\right) $.   For by
observation \ref{Observation 4}) above $B$ is a member of
$\mathbb{Y} \left(
Q,\mathcal{Z} \left( P,H,i\right) ,\mathcal{Z} \left(
Q,H,i\right) \right) $.   Now $\mathcal{C} \left(
W,i\right) \subseteq A$ implies $i\in A$ and $\mathcal{Z}
\left( P,H,i\right) \subset B$ implies $i\in B$, hence
$A\cap B$ is nonempty hence either $B\subseteq A$ holds or
$A\subset B$ holds since both $B$ and $A$ are members of
$Q$ hence do not overlap.   But $B\subseteq \mathcal{Z}
\left( Q,H,i\right) $ holds hence $\mathcal{C} \left(
W,i\right) $ is \emph{not} a subset of $B$ hence $A\subset
B$ does not hold hence $B\subseteq A$ holds.   Now $i$ is a
member of $\mathcal{K} \left( P,A,i\right) $, $W$ is a
subset of $H$, $H$ is a partition, and by assumption
$\mathcal{C} \left( W,i\right) $ is \emph{not} a subset of
$\mathcal{K} \left( P,A,i\right) $, hence $i$ is a member
of $\mathcal{T} \left( \mathcal{K} \left( P,A,i\right)
,H\right) $, hence $\mathcal{K} \left( P,A,i\right) $ is a
subset of $\mathcal{Z} \left( P,H,i\right) $ since by
definition $\mathcal{Z} \left( P,H,i\right) $ is the
\emph{largest} member $C$ of $P$ such that $i\in
\mathcal{T} \left( C,H\right) $ holds.   But $\mathcal{Z}
\left( P,H,i\right) \subset B$ holds hence $\mathcal{K}
\left( P,A,i\right) \subset B$ holds, hence $B$ is
\emph{not} a subset of any member of $\mathcal{P} \left(
P,A\right) $, for if $B$ was a subset of a member $E$ of
$\mathcal{P} \left( P,A\right) $, then $\mathcal{K} \left(
P,A,i\right) \subset E$ would hold, which is impossible
since $\mathcal{K} \left( P,A,i\right) $ is a member of
$\mathcal{P} \left( P,A\right) $ and $\mathcal{P} \left(
P,A\right) $ is a partition.

\begin{bphzobservation} \label{Observation 6}
\end{bphzobservation}
\vspace{-6.143ex}

\noindent \hspace{1.3ex}{\bf ) }Let $\left(
\left( i,B\right) ,X\right) $ be any member
of $\mathcal{U} \left( \mathcal{R} \left( \psi \left(
\mathbb{J} \left( P,Q,H\right) \right) \right) \right) $.
Then the contributions of the factor $\left(
\frac{ c\left( x,g,B\right) }{
b\left( x,i\right) } \right)^{ m_{ iBX } } $ to the
map $\alpha $ are as
follows:

The numerator contributes $\left( -m_{ iBX } \right) $ to
$\alpha_{ \left\{ J,K\right\} } $, where $J$ and $K$
are the two
members of the member $g_{ B } $ of $\mathcal{P} \left(
P,B\right) $, and makes no contribution to $\alpha _{
\Delta } $ for any member $\Delta $ of $\mathcal{D} \left(
\alpha \right) $ other than the
\label{Start of original page 103}
 member $g_{ B } =\left\{ J,K\right\} $.

And if $i\in \mathcal{U} \left( W\right) $ holds, then
$b(x,i) $ is equal to $\left| x_{ \mathcal{Z}
\left( P,H,i\right) }
-x_{ \mathcal{Z} \left( P,H,j\right) } \right| $,
where $j$ is
the
other member of $\mathcal{C} \left( W,i\right) $, (or in
other words, where $\mathcal{C} \left( W,i\right) =\left\{
i,j\right\} $), and the denominator makes the
contribution $\left( +m_{ iBX } \right) $ to
$\alpha_{ \left\{
\mathcal{Z} \left( P,H,i\right) ,\mathcal{Z} \left(
P,H,j\right) \right\} } $, and no contribution to $\alpha _{
\Delta } $ for any member $\Delta $ of $\mathcal{D} \left(
\alpha \right) $ other than the member $\left\{ \mathcal{Z}
\left( P,H,i\right) ,\mathcal{Z} \left( P,H,j\right)
\right\} $.

And if $i$ is \emph{not} a member of $\mathcal{U} \left(
W\right) $, then $b(x,i) $ is equal to $1$, and the
denominator makes no contribution to $\alpha _{ \Delta } $
for any member $\Delta $ of $\mathcal{D} \left( \alpha
\right) $.

\begin{bphzobservation} \label{Observation 7}
\end{bphzobservation}
\vspace{-6.143ex}

\noindent \hspace{1.3ex}{\bf ) }Let $A$ be any member
of $\mathbb{B} \left( Q\right) $
and $\left( \left( i,B\right) ,X\right) $ be any
member of \\
$\mathcal{U} \left( \mathcal{R} \left( \psi \left(
\mathbb{J} \left( P,Q,H\right) \right) \right) \right) $.
Then the contribution of the factor $\left(
\frac{ c\left( x,g,B\right) }{
b\left( x,i\right) } \right)^{ m_{ iBX } } $ to \\
$\Gamma \left( \alpha
,\mathcal{P} \left( P,A\right) \right) $ is $\leq 0$.   For
$m_{ iBX } $ is $\geq 0$ hence by
observation \ref{Observation 6})
above the contribution of the numerator is always $\leq 0$.
  And also by observation \ref{Observation 6}), the
denominator
makes no contribution to the map $\alpha $ unless $i\in
\mathcal{U} \left( W\right) $ holds, and if $i\in
\mathcal{U} \left( W\right) $ does hold then it contributes
$+m_{ iBX } $ to $\alpha_{ \left\{ \mathcal{Z} \left(
P,H,i\right) ,\mathcal{Z} \left( P,H,j\right) \right\} } $,
where $\mathcal{C} \left( W,i\right) =\left\{ i,j\right\} $,
and nothing else.   Now suppose the denominator does
contribute.   Then $\left\{ \mathcal{Z} \left( P,H,i\right)
,\mathcal{Z} \left( P,H,j\right) \right\} $ is a member of
$\mathcal{Q} \left( \mathcal{P} \left( P,A\right) \right)
$, hence by observation \ref{Observation 3}) above, $\left\{
i,j\right\} \subseteq A$ holds and $\left\{ i,j\right\}
\subseteq \mathcal{K} \left( P,A,i\right) $ does \emph{not}
hold, hence by observation \ref{Observation 5}) above,
$B\subseteq
A$ holds and $B$ is \emph{not} a subset of any member of
$\mathcal{P} \left( P,A\right) $, hence by
observation \ref{Observation 2}) above, $\mathcal{Q} \left(
\mathcal{P} \left(
P,B\right) \right) \subseteq \mathcal{Q} \left( \mathcal{P}
\left( P,A\right) \right) $ holds hence the member $g_{ B }
$ of $\mathcal{Q} \left( \mathcal{P} \left( P,B\right)
\right) $ is a member of $\mathcal{Q} \left( \mathcal{P}
\left( P,A\right) \right) $, hence by
observation \ref{Observation 6})
 above the numerator contributes $-m_{ iBX } $ to $\Gamma
\left( \alpha ,\mathcal{P} \left( P,A\right) \right) $,
hence the total contribution of the factor $\left(
\frac{ c\left(
x,g,B\right) }{ b\left( x,i\right) }
\right)^{ m_{ iBX } } $ to $\Gamma
\left( \alpha ,\mathcal{P} \left( P,A\right) \right) $ is
$\leq 0$.
\enlargethispage{4.0ex}

\begin{bphzobservation} \label{Observation 8}
\end{bphzobservation}
\vspace{-6.143ex}

\noindent \hspace{1.3ex}{\bf ) }Let $A$ be any member
of $\left( Q\,\vdash P\right) $ and
$\left( \left( i,B\right) ,X\right) $ be any member of
$\psi _{ A } \left( \mathbb{J} \left( P,Q,H\right) \right)
$.
\newpage
\noindent Then the factor $\left( \frac{ c\left(
x,g,B\right) }{ b\left(
x,i\right) } \right)^{ m_{ iBX } } $
contributes $-m_{ iBX } $ to $\Gamma
\left( \alpha ,\mathcal{P} \left( P,A\right) \right) $.
For by
page \pageref{Start of original page 64}
$\psi _{ A } \left( \mathbb{J} \left(
P,Q,H\right) \right) $ is the set of all ordered pairs
$\left( \left( i,B\right) ,X\right) $ such that $\left(
i,B\right) \in \mathbb{J} _{ A } \left( P,Q,H\right) $
holds and $X$ is a subset of $\mathbb{G} _{ iB } \left(
Q,H\right) $ such that $A\in X$ holds, and by
page \pageref{Start of original page 76},
$\mathbb{J} _{ A } \left( P,Q,H\right) $ is the set of all
ordered pairs $\left(i,B\right) $ of a member $i$
of $\left( \mathcal{T}
\left( A,H\right) \,\vdash \right. $ \\
$\left. \mathcal{T} \left( \mathcal{Y}
\left( P,A\right) ,H\right) \right) $ and a member $B$ of
$\mathbb{Y} \left( Q,\mathcal{K} \left( P,A,i\right)
,A\right) $, and by
page \pageref{Start of original page 78},
$\mathbb{G} _{ iB } \left(
Q,H\right) $ is the set of all members $C$ of $Q$ such that
\label{Start of original page 104}
 $B\subseteq C$ and $C\subseteq \mathcal{Z} \left(
Q,H,i\right) $ both hold.

Now $B\in \mathbb{Y} \left( Q,\mathcal{K} \left(
P,A,i\right) ,A\right) $ implies that $B\subseteq A$ holds
and $B$ is \emph{not} a subset of any member of
$\mathcal{P} \left( P,A\right) $, hence the member $g_{ B }
$ of $\mathcal{Q} \left( \mathcal{P} \left( P,B\right)
\right) $ is a member of $\mathcal{Q} \left( \mathcal{P}
\left( P,A\right) \right) $ hence by
observations \ref{Observation 2})
 and \ref{Observation 6})
above the numerator contributes $-m_{ iBX
} $ to $\Gamma \left( \alpha ,\mathcal{P} \left( P,A\right)
\right) $.   And $i\in \mathcal{T} \left( A,H\right) $
implies that \emph{no} member of $W$ that contains $i$ as a
member is a subset of $A$, hence if $i$ \emph{is} a member
of $\mathcal{U} \left( W\right) $, and $\mathcal{C} \left(
W,i\right) =\left\{ i,j\right\} $, then by
observation \ref{Observation 3}) above $\left\{ \mathcal{Z}
\left( P,H,i\right)
,\mathcal{Z} \left( P,H,j\right) \right\} $ is \emph{not} a
subset of $\mathcal{Q} \left( \mathcal{P} \left( P,A\right)
\right) $, hence the denominator does \emph{not} contribute
to $\Gamma \left( \alpha ,\mathcal{P} \left( P,A\right)
\right) $, and if $i$ is \emph{not} a member of
$\mathcal{U} \left( W\right) $, then $b\left( x,i\right)
=1$ hence the denominator certainly does not contribute
to $\Gamma
\left( \alpha ,\mathcal{P} \left( P,A\right) \right) $.

\begin{bphzobservation} \label{Observation 9}
\end{bphzobservation}
\vspace{-6.143ex}

\noindent \hspace{1.3ex}{\bf ) }It follows directly from
observation \ref{Observation 7}) above
that if $A$ is any member of $\mathbb{B} \left( P\right) $,
then the total contribution to $\Gamma \left( \alpha
,\mathcal{P} \left( P,A\right) \right) $ from the factor \\
$ \displaystyle \left( \prod_{\left( \left( i,B\right),
X\right) \in
\mathcal{U}
\left( \mathcal{R} \left( \psi \left( \mathbb{J} \left(
P,Q,H\right) \right) \right) \right) } \left( \frac{c\left(
x,g,B\right) }{b\left( x,i\right) }
\right)^{ m_{ iBX } } \right) $ is $\leq 0$.

\begin{bphzobservation} \label{Observation 10}
\end{bphzobservation}
\vspace{-6.143ex}

\noindent \hspace{2.6ex}{\bf ) }If $A$ is any member
of $\left( Q\,\vdash P\right) $,
then the total contribution to $\Gamma \left( \alpha
,\mathcal{P} \left( P,A\right) \right) $ from the factor
$\displaystyle \left( \prod_{\left( \left( i,B\right),
X\right) \in
\mathcal{U}
\left( \mathcal{R} \left( \psi \left( \mathbb{J} \left(
P,Q,H\right) \right) \right) \right) } \left( \frac{c\left(
x,g,B\right) }{b\left( x,i\right) }
\right)^{ m_{ iBX } } \right) $ is \\
$\leq
-\left( D_{ A } -\xi _{ A } \left( P,Q,H,u\right) +1\right)
$.   For by observation \ref{Observation 7}) above the
contribution
of every individual factor in this expression is $\leq 0$,
and by observation \ref{Observation 8}), for each member
$\left(
\left( i,B\right) ,X\right) $ of $\psi _{ A } \left(
\mathbb{J} \left( P,Q,H\right) \right) $, the factor
$\left( \frac{ c\left( x,g,B\right) }
{ b\left( x,i\right) } \right)^{ m_{ iBX } } $
contributes $-m_{ iBX } $, hence the total contribution is
$\leq - \displaystyle \sum_{\left( \left( i,B\right),
X\right) \in
\psi _{ A }
\left( \mathbb{J} \left( P,Q,H\right) \right) } m_{
iBX } $.   But $m$ is a member of \\
$\mathbb{A} \left(
\mathbb{J} \left( P,Q,H\right) ,\left( D-\xi \left(
P,Q,H,u\right) +1\hspace{-0.7516ex}1 \right) \right) $,
hence by the definition of $\mathbb{A} $ on
page \pageref{Start of original page 65} and
the definition of the map $1\hspace{-0.7516ex}1 $ on
page \pageref{Start of original page 95}, \\
$\displaystyle \sum_{\left( \left( i,B\right),
X\right)
\in \psi _{ A
} \left( \mathbb{J} \left( P,Q,H\right) \right) }
m_{ iBX } =\left( D_{ A } -\xi _{ A } \left( P,Q,H,u\right)
+1\right) $ holds.
\enlargethispage{1.0ex}

\begin{bphzobservation} \label{Observation 11}
\end{bphzobservation}
\vspace{-6.143ex}

\noindent \hspace{2.6ex}{\bf ) }By Lemma \ref{Lemma 20},
$\mathcal{U} \left( \mathcal{R}
\left(
\downarrow \left( \mathbb{I} \left( Q,H\right) ,P\right)
\right) \right) $ is the set of all the
ordered pairs $\left(i,B\right) $
of a member $i$ of $\mathcal{O} \left( V,H\right)
=\mathcal{O} \left( Q,H\right) $ and a member $B$ of
$\mathbb{Y} \left( Q,\mathcal{C} \left( V,i\right)
,\mathcal{Z} \left( P,H,i\right) \right) $.   (We
\newpage
\noindent note that
for any member $i$ of $\mathcal{U} \left( V\right) $,
$\mathcal{C} \left( Q,i\right) $ is by definition the
unique member $A$ of $\mathcal{M} \left( Q\right) =V$ such
that $i\in A$ holds, hence $\mathcal{C} \left( Q,i\right)
=\mathcal{C} \left( V,i\right)  $.)
\label{Start of original page 105}

\begin{bphzobservation} \label{Observation 12}
\end{bphzobservation}
\vspace{-6.143ex}

\noindent \hspace{2.6ex}{\bf ) }Let $A$ be any member
of $\mathbb{B} \left(
Q\right) $ and $ \left(i,B\right) $ be any member of
$\mathcal{U} \left(
\mathcal{R} \left( \downarrow \left( \mathbb{I} \left(
Q,H\right) ,P\right) \right) \right) $ such that $i\in
\mathcal{U} \left( W\right) $ holds and $\mathcal{C} \left(
W,i\right) \subseteq A$ holds and $\mathcal{C} \left(
W,i\right) $ is \emph{not} a subset of $\mathcal{K} \left(
P,A,i\right) $.   Then either $B\subseteq A$ holds and $B$
is \emph{not} a subset of any member of $\mathcal{P} \left(
P,A\right) $, or else $i\in \left( \mathcal{T} \left(
\mathcal{K} \left( P,A,i\right) ,H\right) \,\vdash
\mathcal{T} \left( A,H\right) \right) $ holds and $B\in
\mathbb{Y} \left( Q,\mathcal{C} \left( V,i\right)
,\mathcal{K} \left( P,A,i\right) \right) $ holds.   For
$\mathcal{C} \left( W,i\right) \subseteq A$ implies
$\mathcal{Z} \left( P,H,i\right) \subset A$ hence $B\subset
A$ holds by observation \ref{Observation 11}).   Now
$\mathcal{K}
\left( P,A,i\right) \cap B$ has the member $i$ hence is
nonempty hence either $\mathcal{K} \left( P,A,i\right)
\subset B$ holds or $B\subseteq \mathcal{K} \left(
P,A,i\right) $ holds, since $B$ is a member of $Q$ hence
does not overlap $\mathcal{K} \left( P,A,i\right) $.   And
if $\mathcal{K} \left( P,A,i\right) \subset B$ holds then
$B\subseteq A$ holds and $B$ is \emph{not} a subset of any
member of $\mathcal{P} \left( P,A\right) $.   Now suppose
that $B\subseteq \mathcal{K} \left( P,A,i\right) $ holds.
Then $B$ is a member of $Q$ such that $\mathcal{C} \left(
V,i\right) \subset B$ and $B\subseteq \mathcal{K} \left(
P,A,i\right) $ both hold, hence $B\in \mathbb{Y} \left(
Q,\mathcal{C} \left( V,i\right) ,\mathcal{K} \left(
P,A,i\right) \right) $ holds.   Now by assumption
$\mathcal{C} \left( W,i\right) $ is \emph{not} a subset of
$\mathcal{K} \left( P,A,i\right) $, hence since $W\subseteq
H$ holds and $H$ is a partition, there is \emph{no} member
$C$ of $H$ such that $i\in C$ and $C\subseteq \mathcal{K}
\left( P,A,i\right) $ both hold, hence $i\in \mathcal{T}
\left( \mathcal{K} \left( P,A,i\right) ,H\right) $ holds,
and furthermore $\mathcal{C} \left( W,i\right) \subseteq A$
implies $i\notin \mathcal{T} \left( A,H\right) $, hence
$i\in \left( \mathcal{T} \left( \mathcal{K} \left(
P,A,i\right) ,H\right) \,\vdash \mathcal{T} \left(
A,H\right) \right) $ holds.

\begin{bphzobservation} \label{Observation 13}
\end{bphzobservation}
\vspace{-6.143ex}

\noindent \hspace{2.6ex}{\bf ) }Let $ \left(i,B\right) $
be any member of $\mathcal{U} \left(
\mathcal{R} \left( \downarrow \left( \mathbb{I} \left(
Q,H\right) ,P\right) \right) \right) $.   Then the
contributions of the factor $\left( \frac{
c\left( x,g,B\right) }{ b\left(
x,i\right) } \right)^{ u_{ iB } } $ to the map
$\alpha $ are as follows:
\enlargethispage{2.5ex}

The numerator contributes $\left( -u_{ iB } \right) $ to
$\alpha_{ \left\{ J,K\right\} } $, where $J$ and $K$
are the two
members of the member $g_{ B } $ of $\mathcal{Q} \left(
\mathcal{P} \left( P,B\right) \right) $, (or in other
words, the member $g_{ B } $ of $\mathcal{Q} \left(
\mathcal{P} \left( P,B\right) \right) $ is equal to $\left\{
J,K\right\}  $), and makes no contribution to $\alpha
_{ \Delta } $ for any member $\Delta $ of $\mathcal{D}
\left( \alpha \right) $ other than the member $g_{ B }
=\left\{ J,K\right\} $.

And if $i\in \mathcal{U} \left( W\right) $ holds and
$\mathcal{C} \left( W,i\right) =\left\{ i,j\right\} $, then
$b(x,i) $ is equal to \\
$\left| x_{ \mathcal{Z}
\left( P,H,i\right) }
-x_{ \mathcal{Z} \left( P,H,j\right) } \right| $, and the
denominator
makes the contribution $\left( +u_{ iB } \right) $ to \\
$\alpha_{ \left\{ \mathcal{Z} \left( P,H,i\right)
,\mathcal{Z}
\left( P,H,j\right) \right\} } $, and makes no contribution
to
$\alpha _{ \Delta } $ for any member $\Delta $ of
$\mathcal{D} \left( \alpha \right) $ other than the member
$\left\{ \mathcal{Z} \left( P,H,i\right) ,\mathcal{Z} \left(
P,H,j\right) \right\} $.

And if $i$ is \emph{not} a member of $\mathcal{U} \left(
W\right) $, then $b\left( x,i\right) =1$ holds and the
denominator makes no contribution to $\alpha _{ \Delta } $
for any member $\Delta $ of $\mathcal{D} \left( \alpha
\right) $.

\begin{bphzobservation} \label{Observation 14}
\end{bphzobservation}
\vspace{-6.143ex}

\noindent \hspace{2.6ex}{\bf ) }Let $A$ be any member of
$\mathbb{B} \left( Q\right) $
and $ \left(i,B\right) $ be any member of $\mathcal{U}
\left( \mathcal{R}
\left( \downarrow \left( \mathbb{I} \left( Q,H\right)
,P\right) \right) \right) $.   Then the contribution of the
factor
\label{Start of original page 106}
 $\left( \frac{ c\left( x,g,B\right) }{ b\left(
 x,i\right) } \right)^{ u_{ iB } } $ to
$\Gamma \left( \alpha ,\mathcal{P} \left( P,A\right)
\right) $ is $\leq 0$ unless $i\in \left( \mathcal{T}
\left( \mathcal{K} \left( P,A,i\right) ,H\right) \,\vdash
\mathcal{T} \left( A,H\right) \right) $ and $B\in
\mathbb{Y} \left( Q,\mathcal{C} \left( V,i\right)
,\mathcal{K} \left( P,A,i\right) \right) $ both hold, in
which case the contribution is $\leq u_{ iB } $.

For $u_{ iB } \geq 0$ holds hence by
observation \ref{Observation 6})
 above the contribution of the numerator is always $\leq
0$.   And also by observation \ref{Observation 6}), the
denominator
makes no contribution to the map $\alpha $ unless $i\in
\mathcal{U} \left( W\right) $ holds, and if $i\in
\mathcal{U} \left( W\right) $ does hold it contributes
$+u_{ iB } $ to $\alpha_{ \left\{ \mathcal{Z} \left(
P,H,i\right) ,\mathcal{Z} \left( P,H,j\right) \right\} } $,
where $\mathcal{C} \left( W,i\right) =\left\{ i,j\right\}
 $,
and nothing else.

Now suppose the denominator does contribute to $\Gamma
\left( \alpha ,\mathcal{P} \left( P,A\right) \right) $.
Then \\
$\left\{ \mathcal{Z} \left( P,H,i\right) ,\mathcal{Z}
\left( P,H,j\right) \right\} \in \mathcal{Q} \left(
\mathcal{P} \left( P,A\right) \right) $ holds, hence by
observation \ref{Observation 3}) above, $\left\{ i,j\right\}
\subseteq
A$ holds and $\left\{ i,j\right\} \subseteq \mathcal{K}
\left( P,A,i\right) $ does \emph{not} hold.   Hence by
observation \ref{Observation 12}) above, either $B\subseteq
A$ holds
and $B$ is \emph{not} a subset of any member of
$\mathcal{P} \left( P,A\right) $, or else $i\in \left(
\mathcal{T} \left( \mathcal{K} \left( P,A,i\right)
,H\right) \,\vdash \mathcal{T} \left( A,H\right) \right) $
holds and $B\in \mathbb{Y} \left( Q,\mathcal{C} \left(
V,i\right) ,\mathcal{K} \left( P,A,i\right) \right) $
holds.   And if $B\subseteq A$ holds and $B$ is \emph{not}
a subset of any member of $\mathcal{P} \left( P,A\right) $,
then by observation \ref{Observation 2}) above,
$\mathcal{Q} \left(
\mathcal{P} \left( P,B\right) \right) \subseteq \mathcal{Q}
\left( \mathcal{P} \left( P,A\right) \right) $ holds, hence
the member $g_{ B } $ of $\mathcal{Q} \left( \mathcal{P}
\left( P,B\right) \right) $ is a member of $\mathcal{Q}
\left( \mathcal{P} \left( P,A\right) \right) $, hence the
numerator contributes $-u_{ iB } $ to $\Gamma \left( \alpha
,\mathcal{P} \left( P,A\right) \right) $, hence the total
contribution of the factor $\left( \frac{ c\left(
x,g,B\right) }{ b\left(
x,i\right) } \right)^{ u_{ iB } } $ is $\leq 0$.

And finally, by observation \ref{Observation 13}) above,
the total
contribution to $\Gamma \left( \alpha ,\mathcal{P} \left(
P,A\right) \right) $ from the factor $\left( \frac{ c\left(
x,g,B\right) }{
b\left( x,i\right) } \right)^{ u_{ iB } } $ is
\emph{always} $\leq u_{ iB
} $.

\begin{bphzobservation} \label{Observation 15}
\end{bphzobservation}
\vspace{-6.143ex}

\noindent \hspace{2.6ex}{\bf ) }Let $A$ be any member of
$\mathbb{B} \left( Q\right) $.
 Then $A\cap \mathcal{T} \left( \mathcal{Y} \left(
P,A\right) ,H\right) $ is equal to $\mathcal{T} \left(
A,H\right) \cap \mathcal{T} \left( \mathcal{Y} \left(
P,A\right) ,H\right) $.   For $\mathcal{T} \left(
A,H\right) \subseteq A$ holds, hence $\mathcal{T} \left(
A,H\right) \cap \mathcal{T} \left( \mathcal{Y} \left(
P,A\right) ,H\right) $ is certainly a subset of $A\cap
\mathcal{T} \left( \mathcal{Y} \left( P,A\right) ,H\right)
$.   Now let $i$ be any member of $A\cap \mathcal{T} \left(
\mathcal{Y} \left( P,A\right) ,H\right) $.   Then $A\cap
\mathcal{T} \left( \mathcal{Y} \left( P,A\right) ,H\right)
$ is nonempty, hence $\mathcal{Y} \left( P,A\right) \neq
\emptyset $, hence $A\subseteq \mathcal{Y} \left(
P,A\right) $
holds.   And if $C$ was a member of $H$ such that $i\in C$
and $C\subseteq A$ both held, then $C$ would be a member of
$H$ such that $i\in C$ and $C\subseteq \mathcal{Y} \left(
P,A\right) $ both held, hence $i\in \mathcal{T} \left(
\mathcal{Y} \left( P,A\right) ,H\right) $ implies that
there is no member $C$ of $H$ such that $i\in C$ and
$C\subseteq A$ both hold, hence $i\in \mathcal{T} \left(
A,H\right) $ holds, hence $i\in \mathcal{T} \left(
A,H\right) \cap \mathcal{T} \left( \mathcal{Y} \left(
P,A\right) ,H\right) $ holds.

\begin{bphzobservation} \label{Observation 16}
\end{bphzobservation}
\vspace{-6.143ex}

\noindent \hspace{2.6ex}{\bf ) }Let $A$ be any member of
$\mathbb{B} \left( Q\right) $,
$i$ be any member of $A\cap \mathcal{T} \left( \mathcal{Y}
\left( P,A\right) ,H\right) $, and $B$ be any member of
$\mathbb{Y} \left( Q,\mathcal{K} \left( P,A,i\right)
,A\right) $.   Then
\label{Start of original page 107}
 $\left( i,B\right) \in \mathcal{U} \left( \mathcal{R}
\left( \downarrow \left( \mathbb{I} \left( Q,H\right)
,P\right) \right) \right) $ holds, and the factor $\left(
\frac{ c\left(
x,g,B\right) }{ b\left( x,i\right) } \right)^{ u_{ iB } } $
contributes
$\left( -u_{ iB } \right) $ to $\Gamma \left( \alpha
,\mathcal{P} \left( P,A\right) \right) $.

For $i\in A$ holds and $i\in \mathcal{T} \left( \mathcal{Y}
\left( P,A\right) ,H\right) $ holds, hence $\mathcal{Y}
\left( P,A\right) \neq \emptyset $ and $i\in \mathcal{O}
\left(
V,H\right) $ holds.   And $\mathcal{K} \left( P,A,i\right)
\subset B$ implies $B$ is \emph{not} a member of
$\mathcal{M} \left( P\right) =V$, hence $\mathcal{C} \left(
V,i\right) \subset B$ holds.   And $i\in \mathcal{T} \left(
\mathcal{Y} \left( P,A\right) ,H\right) $ implies
furthermore that $\mathcal{Y} \left( P,A\right) \subseteq
\mathcal{Z} \left( P,H,i\right) $ holds, hence $A\subseteq
\mathcal{Z} \left( P,H,i\right) $ holds, hence $B\subseteq
\mathcal{Z} \left( P,H,i\right) $ holds, hence $B\in
\mathbb{Y} \left( Q,\mathcal{C} \left( V,i\right)
,\mathcal{Z} \left( P,H,i\right) \right) $ holds, hence
$\left( i,B\right) \in \mathcal{U} \left( \mathcal{R}
\left( \downarrow \left( \mathbb{I} \left( Q,H\right)
,P\right) \right) \right) $ holds.

Now $B\in \mathbb{Y} \left( Q,\mathcal{K} \left(
P,A,i\right) ,A\right) $ implies that $B\subseteq A$ holds
and that $B$ is \emph{not} a subset of any member of
$\mathcal{P} \left( P,A\right) $, hence by
observation \ref{Observation 2}) above the member $g_{ B }
$ of $\mathcal{Q}
\left( \mathcal{P} \left( P,B\right) \right) $ is a member
of $\mathcal{Q} \left( \mathcal{P} \left( P,A\right)
\right) $, hence by observation \ref{Observation 13}) above
the
numerator contributes $\left( -u_{ iB } \right) $ to
$\Gamma \left( \alpha ,\mathcal{P} \left( P,A\right)
\right) $.   And $i\in \mathcal{T} \left( A,H\right) $
implies that \emph{no} member of $W$ that contains $i$ as a
member is a subset of $A$, hence if $i\in \mathcal{U}
\left( W\right) $ and $\mathcal{C} \left( W,i\right)
=\left\{ i,j\right\} $ then by
observation \ref{Observation 3}) above
$\left\{ \mathcal{Z} \left( P,H,i\right) ,\mathcal{Z} \left(
P,H,j\right) \right\} $ is \emph{not} a member of
$\mathcal{Q} \left( \mathcal{P} \left( P,A\right) \right)
$, hence the denominator does not contribute to $\Gamma
\left( \alpha ,\mathcal{P} \left( P,A\right) \right) $.

\begin{bphzobservation} \label{Observation 17}
\end{bphzobservation}
\vspace{-6.143ex}

\noindent \hspace{2.6ex}{\bf ) }Let $A$ be any member of
$\mathbb{B} \left( Q\right) $.
 Then the total contribution to $\Gamma \left( \alpha
,\mathcal{P} \left( P,A\right) \right) $ from the factor
$ \left( \displaystyle \prod_{\left( i,B\right) \in
\mathcal{U} \left(
\mathcal{R}
\left( \downarrow \left( \mathbb{I} \left( Q,H\right)
,P\right) \right) \right) } \left( \frac{c\left(
x,g,B\right) }{b\left( x,i\right) } \right)^{ u_{ iB } }
\right) $
is
\[
\leq \left( \sum_{C\in \left( \mathcal{P} \left( P,A\right)
\,\vdash
V\right) } \hspace{0.2cm} \sum_{\left( i,B\right) \in
\left(
\mathbb{I} _{ C } \left( Q,H\right) \,\vdash \mathbb{I} _{
A } \left( Q,H\right) \right) } u_{ iB } \right) -
\left( \sum_{i\in
\left( A\cap \mathcal{T} \left( \mathcal{Y} \left(
P,A\right) ,H\right) \right) } \hspace{0.2cm}
  \sum_{B\in \mathbb{Y}
\left( Q,\mathcal{K} \left( P,A,i\right) ,A\right) }
u_{ iB } \right) .
\]

For by observation \ref{Observation 14}) above the factor
$\left( \frac{ c\left(
x,g,B\right) }{ b\left( x,i\right) } \right)^{ u_{ iB } } $
does \emph{not}
give a contribution $\geq 0$ unless $i\in \left(
\mathcal{T} \left( \mathcal{K} \left( P,A,i\right)
,H\right) \,\vdash \mathcal{T} \left( A,H\right) \right) $
and $B\in \mathbb{Y} \left( Q,\mathcal{C} \left( V,i\right)
,\mathcal{K} \left( P,A,i\right) \right) $ both hold, and
in that case the contribution is $\leq \! u_{ iB } $.   But
$i\in \left( \mathcal{T} \left( \mathcal{K} \left(
P,A,i\right) ,H\right) \vdash \mathcal{T} \left(
A,H\right) \right) $ and $B\in \mathbb{Y} \left(
Q,\mathcal{C} \left( V,i\right) ,\mathcal{K} \left(
P,A,i\right) \right) $ imply that there exists a member $C$
of $\mathcal{P} \left( P,A\right) $, namely $C=\mathcal{K}
\left( P,A,i\right) $, such that $i\in \left( \mathcal{T}
\left( C,H\right) \,\vdash \mathcal{T} \left( A,H\right)
\right) $ holds and \\
$B\in \mathbb{Y} \left( Q,\mathcal{C}
\left( V,i\right) ,C\right) $ holds, and furthermore $C$ is
not a member of $V$, since if $C$ was a member of $V$ then
$\mathcal{C} \left( V,i\right) $ would be equal to $C$,
hence $\mathcal{C} \left( V,i\right) \subset B\subseteq C$
could not hold, which contradicts $B\in \mathbb{Y} \left(
Q,\mathcal{C} \left( V,i\right) ,\mathcal{K} \left(
P,A,i\right) \right) $.   And conversely, if $C$ is a
member of $\left( \mathcal{P} \left( P,A\right) \,\vdash
V\right) $ and $i$ is a member of $\left( \mathcal{T}
\left( C,H\right) \,\vdash \mathcal{T} \left( A,H\right)
\right) $ and $B$ is a
\label{Start of original page 108}
 member of $\mathbb{Y} \left( Q,\mathcal{C} \left(
V,i\right) ,C\right) $, then $i\in \left( \mathcal{T}
\left( \mathcal{K} \left( P,A,i\right) ,H\right) \,\vdash
\mathcal{T} \left( A,H\right) \right) $ holds and $B\in
\mathbb{Y} \left( Q,\mathcal{C} \left( V,i\right)
,\mathcal{K} \left( P,A,i\right) \right) $ holds.   (We
note that $i\in \mathcal{T} \left( \mathcal{K} \left(
P,A,i\right) ,H\right) $ implies $i\in \mathcal{O} \left(
V,H\right) $ and $\mathcal{K} \left( P,A,i\right) \subseteq
\mathcal{Z} \left( P,H,i\right) $, hence $B\in \mathbb{Y}
\left( Q,\mathcal{C} \left( V,i\right) ,\mathcal{K} \left(
P,A,i\right) \right) $ implies $B\in \mathbb{Y} \left(
Q,\mathcal{C} \left( V,i\right) ,\mathcal{Z} \left(
P,H,i\right) \right) $, hence by
observation \ref{Observation 11})
above, $\left( i,B\right) $ is a member of $\mathcal{U}
\left( \mathcal{R}
\left( \downarrow \left( \mathbb{I} \left( Q,H\right)
,P\right) \right) \right) $.)
\enlargethispage{0.85ex}

And furthermore, if $C$ is any member of $\left(
\mathcal{P} \left( P,A\right) \,\vdash V\right) $, then the
set whose members are all the ordered
pairs $\left(i,B\right) $ of a
member $i$ of $\left( \mathcal{T} \left( C,H\right)
\,\vdash \mathcal{T} \left( A,H\right) \right) $ and a
member $B$ of $\mathbb{Y} \left( Q,\mathcal{C} \left(
V,i\right) ,C\right) $, is equal to $\left( \mathbb{I} _{ C
} \left( Q,H\right) \,\vdash \mathbb{I} _{ A } \left(
Q,H\right) \right) $.   For by definition $\mathbb{I} _{ C
} \left( Q,H\right) $ is the set of all
ordered pairs $\left(i,B\right) $
of a member $i$ of $\mathcal{T} \left( C,H\right) $ and a
member $B$ of $\mathbb{Y} \left( Q,\mathcal{C} \left(
V,i\right) ,C\right) $, hence if $i$ is any member of
$\left( \mathcal{T} \left( C,H\right) \,\vdash \mathcal{T}
\left( A,H\right) \right) $ and $B$ is any member of
$\mathbb{Y} \left( Q,\mathcal{C} \left( V,i\right)
,C\right) $, then $\left( i,B\right) $ is certainly a
member of
$\mathbb{I} _{ C } \left( Q,H\right) $, and furthermore $i$
is \emph{not} a member of $\mathcal{T} \left( A,H\right) $,
hence $\left( i,B\right) $ is \emph{not} a member of
$\mathbb{I} _{ A }
\left( Q,H\right) $, hence $\left( i,B\right) $ is a member
of $\left(
\mathbb{I} _{ C } \left( Q,H\right) \,\vdash \mathbb{I} _{
A } \left( Q,H\right) \right) $.   And if $\left(
i,B\right) $ is any
member of $\left( \mathbb{I} _{ C } \left( Q,H\right)
\,\vdash \mathbb{I} _{ A } \left( Q,H\right) \right) $,
then $i\in \mathcal{T} \left( C,H\right) $ holds and $B\in
\mathbb{Y} \left( Q,\mathcal{C} \left( V,i\right) ,C\right)
$ holds.   But $B\in \mathbb{Y} \left( Q,\mathcal{C} \left(
V,i\right) ,C\right) $ implies $B\in \mathbb{Y} \left(
Q,\mathcal{C} \left( V,i\right) ,A\right) $, hence $\left(
i,B\right) \notin \mathbb{I} _{ A } \left( Q,H\right) $
implies $i\notin \mathcal{T} \left( A,H\right) $, hence
$i\in \left( \mathcal{T} \left( C,H\right) \,\vdash
\mathcal{T} \left( A,H\right) \right) $ holds.

Now let $i$ be any member of $A\cap \mathcal{T} \left(
\mathcal{Y} \left( P,A\right) ,H\right) $ and $B$ be any
member of \\
$\mathbb{Y} \left( Q,\mathcal{K} \left(
P,A,i\right) ,A\right) $.   Then by
observation \ref{Observation 15})
 above, $i$ is a member of $\mathcal{T} \left( A,H\right)
\cap \mathcal{T} \left( \mathcal{Y} \left( P,A\right)
,H\right) $, and by observation \ref{Observation 16})
above, $\left(
i,B\right) \in \mathcal{U} \left( \mathcal{R} \left(
\downarrow \left( \mathbb{I} \left( Q,H\right) ,P\right)
\right) \right) $ holds and the factor $\left( \frac{
c\left(
x,g,B\right) }{ b\left( x,i\right) } \right)^{ u_{ iB } } $
contributes
$\left( -u_{ iB } \right) $ to $\Gamma \left( \alpha
,\mathcal{P} \left( P,A\right) \right) $.   (We note that
$i\in \mathcal{T} \left( A,H\right) \cap \mathcal{T} \left(
\mathcal{Y} \left( P,A\right) ,H\right) $ implies directly
that $i$ is \emph{not} a member of \\
$\left( \mathcal{T}
\left( \mathcal{K} \left( P,A,i\right) ,H\right) \,\vdash
\mathcal{T} \left( A,H\right) \right) $.)

\begin{bphzobservation} \label{Observation 18}
\end{bphzobservation}
\vspace{-6.143ex}

\noindent \hspace{2.6ex}{\bf ) }Let $A$ be any member of
$\left( Q\,\vdash P\right) $.
Then the following identity holds:
\[
\xi _{ A } \left( P,Q,H,u\right) = \hspace{12.5cm}
\]
\[
= \left( \sum_{C\in \left(
\mathcal{P} \left( P,A\right) \,\vdash V\right) }
\hspace{0.2cm}
\sum_{\left( i,B\right) \in \left( \mathbb{I} _{ C } \left(
Q,H\right) \cap \mathbb{I} _{ A } \left( Q,H\right)
\right) } u_{ iB } \right) + \left( \sum_{i\in
\left( A\cap \mathcal{T}
\left(
\mathcal{Y} \left( P,A\right) ,H\right) \right) }
\hspace{0.2cm}
 \sum_{B\in \mathbb{Y} \left( Q,\mathcal{K} \left(
P,A,i\right)
,A\right) } u_{ iB } \right)
\]

For $\xi _{ A } \left( P,Q,H,u\right) $ was defined on
page \pageref{Start of original page 82} by
\[
\xi _{ A } \left( P,Q,H,u\right) \equiv
\sum_{\left( i,B\right)
\in \left( \mathbb{I} _{ A } \left( Q,H\right) \cap
\mathcal{U} \left( \mathcal{R} \left( \downarrow \left(
\mathbb{I} \left( Q,H\right) ,P\right) \right) \right)
\right)  } u_{ iB } .
\]

Now $\mathbb{I} _{ A } \left( Q,H\right) \cap \mathcal{U}
\left( \mathcal{R} \left( \downarrow \left( \mathbb{I}
\left( Q,H\right) ,P\right) \right) \right) $ is
the set of all ordered pairs $\left(i,B\right) $ such that
$i\in
\mathcal{T} \left( A,H\right) $ and $B\in \mathbb{Y} \left(
Q,\mathcal{C} \left( V,i\right) ,\left( A\cap \mathcal{Z}
\left( P,H,i\right) \right) \right) $ both hold.   And
$\mathcal{K} \left( P,A,i\right) \subseteq A$ certainly
holds.   And furthermore
\label{Start of original page 109}
 $i\in \mathcal{T} \left( A,H\right) $ implies that there
is no member $E$ of $H$ such that $i\in E$ and $E\subseteq
A$ both hold, hence there is no member $E$ of $H$ such that
$i\in E$ and $E\subseteq \mathcal{K} \left( P,A,i\right) $
both hold, hence $i\in \mathcal{T} \left( \mathcal{K}
\left( P,A,i\right) ,H\right) $ holds, hence $\mathcal{K}
\left( P,A,i\right) \subseteq \mathcal{Z} \left(
P,H,i\right) $ holds since $\mathcal{Z} \left( P,H,i\right)
$ is by definition the \emph{largest} member $C$ of $P$
such that $i\in \mathcal{T} \left( C,H\right) $ holds.

Hence $\mathcal{K} \left( P,A,i\right) \subseteq \left(
A\cap \mathcal{Z} \left( P,H,i\right) \right) $ holds,
hence $\mathbb{Y} \left( Q,\mathcal{C} \left( V,i\right)
,\left( A\cap \mathcal{Z} \left( P,H,i\right) \right)
\right) $ is equal to the disjoint union of $\mathbb{Y}
\left( Q,\mathcal{C} \left( V,i\right) ,\mathcal{K} \left(
P,A,i\right) \right) $ and \\
$\mathbb{Y} \left( Q,\mathcal{K}
\left( P,A,i\right) ,\left( A\cap \mathcal{Z} \left(
P,H,i\right) \right) \right) $.

Now if $\mathcal{Z} \left( P,H,i\right) \subset A$ holds
then $\mathcal{Z} \left( P,H,i\right) =\mathcal{K} \left(
P,A,i\right) $ holds, since $\mathcal{K} \left(
P,A,i\right) \subseteq \mathcal{Z} \left( P,H,i\right) $
holds as just shown above, and $\mathcal{K} \left(
P,A,i\right) \subset \mathcal{Z} \left( P,H,i\right)
\subset A$ cannot hold since $\mathcal{K} \left(
P,A,i\right) $ is a member of $\mathcal{P} \left(
P,A\right) $, hence $\mathbb{Y} \left( Q,\mathcal{K} \left(
P,A,i\right) ,\left( A\cap \mathcal{Z} \left( P,H,i\right)
\right) \right) $ is empty unless $A\subseteq \mathcal{Z}
\left( P,H,i\right) $ holds.   And $A\subseteq \mathcal{Z}
\left( P,H,i\right) $ implies $\mathcal{Y} \left(
P,A\right) \neq \emptyset $ and $\mathcal{Y} \left(
P,A\right)
\subseteq \mathcal{Z} \left( P,H,i\right) $ hence $i\in
\mathcal{T} \left( \mathcal{Y} \left( P,A\right) ,H\right)
$ holds hence $i\in A\cap \mathcal{T} \left( \mathcal{Y}
\left( P,A\right) ,H\right) $ holds.   And conversely,
$i\in A\cap \mathcal{T} \left( \mathcal{Y} \left(
P,A\right) ,H\right) $ implies $\mathcal{Y} \left(
P,A\right) \neq \emptyset $ (for $\mathcal{T} \left(
\emptyset
,H\right) =\emptyset  $) and $\mathcal{Y} \left(
P,A\right) \subseteq \mathcal{Z} \left( P,H,i\right) $,
hence $A\subseteq \mathcal{Z} \left( P,H,i\right) $,
hence \\
$\mathbb{Y} \left( Q,\mathcal{K} \left( P,A,i\right)
,\left( A\cap \mathcal{Z} \left( P,H,i\right) \right)
\right) =\mathbb{Y} \left( Q,\mathcal{K} \left(
P,A,i\right) ,A\right) $.   Hence the set of all ordered
pairs $\left(i,B\right) $ of a member $i$ of $\mathcal{T}
\left(
A,H\right) $ and a member $B$ of \\
$\mathbb{Y} \left(
Q,\mathcal{K} \left( P,A,i\right) ,\left( A\cap \mathcal{Z}
\left( P,H,i\right) \right) \right) $ is equal to the set
of all ordered pairs $\left(i,B\right) $ of a member $i$ of
$A\cap
\mathcal{T} \left( \mathcal{Y} \left( P,A\right) ,H\right)
$ and a member $B$ of $\mathbb{Y} \left( Q,\mathcal{K}
\left( P,A,i\right) ,A\right) $.   Hence the following
identity holds:
\[
\xi _{ A } \left( P,Q,H,u\right) = \hspace{12.5cm}
\]
\[
\hspace{1.0cm} = \left(
 \sum_{i\in \mathcal{T}
\left( A,H\right) } \hspace{0.2cm}
  \sum_{B\in \mathbb{Y} \left(
Q,\mathcal{C} \left( V,i\right) ,\mathcal{K} \left(
P,A,i\right) \right) } u_{ iB } \right) + \left(
 \sum_{i\in \left( A\cap
\mathcal{T} \left( \mathcal{Y} \left( P,A\right) ,H\right)
\right) } \hspace{0.2cm} \sum_{B\in \mathbb{Y}
\left( Q,\mathcal{K}
\left( P,A,i\right) ,A\right) } u_{ iB } \right).
\]

But the first term in the right-hand side of this identity
is equal to \\
\enlargethispage{4.0ex}
$ \left( \displaystyle
 \sum_{C\in \left( \mathcal{P} \left(
P,A\right)
\,\vdash V\right) } \hspace{0.2cm} \displaystyle
\sum_{\left( i,B\right) \in \left(
\mathbb{I} _{ C } \left( Q,H\right) \cap \mathbb{I} _{ A }
\left( Q,H\right) \right) } u_{ iB } \right) $.   For if
$i\in \mathcal{T} \left( A,H\right) $ holds and \\
$B\in
\mathbb{Y} \left( Q,\mathcal{C} \left( V,i\right)
,\mathcal{K} \left( P,A,i\right) \right) $ holds then
$\left( i,B\right) \in \mathbb{I} _{ A } \left( Q,H\right)
$ holds and $i\in \mathcal{K} \left( P,A,i\right) $ holds,
and furthermore $\mathcal{K} \left( P,A,i\right) $ is not
equal to $\mathcal{C} \left( V,i\right) $ hence
$\mathcal{K} \left( P,A,i\right) $ is not a member of $V$,
and furthermore there is no member $E$ of $H$ such that
$i\in E$ and $E\subseteq \mathcal{K} \left( P,A,i\right) $
both hold, (for any such $E$ would satisfy both $i\in E$
and $E\subseteq A$, contradicting $i\in \mathcal{T} \left(
A,H\right)  $), hence $i\in \mathcal{T} \left(
\mathcal{K} \left( P,A,i\right) ,H\right) $ and $\left(
i,B\right) \in \mathbb{I} _{ \mathcal{K} \left(
P,A,i\right) } \left( Q,H\right) $ both hold, and
conversely if $C$ is any member of $\left( \mathcal{P}
\left( P,A\right) \,\vdash V\right) $ and
$\left( i,B\right) $ is any
member of $\left( \mathbb{I} _{ C } \left( Q,H\right) \cap
\mathbb{I} _{ A } \left( Q,H\right) \right) $, then $i\in
\mathcal{T} \left( C,H\right) \cap \mathcal{T} \left(
A,H\right) $ holds hence $i\in C\cap \mathcal{T} \left(
A,H\right) $ holds, and $B\in \mathbb{Y} \left(
Q,\mathcal{C} \left( V,i\right) ,\mathcal{K} \left(
P,A,i\right) \right) $ holds.

\begin{bphzobservation} \label{Observation 19}
\end{bphzobservation}
\vspace{-6.143ex}

\noindent \hspace{2.6ex}{\bf ) }Let $A$ be any member of
$\left( Q\,\vdash P\right) $.
Then the following inequality
\label{Start of original page 110}
 holds:
\[
\Gamma \left( \alpha ,\mathcal{P} \left( P,A\right) \right)
\leq d\left( \#\left( \mathcal{P} \left( P,A\right) \right)
-1\right) -1-  \sum_{C\in \left( \mathcal{P} \left(
P,A\right)
\,\vdash V\right) } \left( D_{ C } -
\sum_{\left( i,B\right) \in
\mathbb{I} _{ C } \left( Q,H\right) } u_{ iB } \right).
\]

For by pages \pageref{Start of original page 100} and
\pageref{Start of original page 101}
the contribution from $\Theta $ is equal to
\[
d\left( \#\left( \mathcal{P} \left( P,A\right) \right)
-1\right) +D_{ A } - \left( \sum_{C\in \left(
\mathcal{P} \left(
P,A\right) \,\vdash V\right) } D_{ C } \right) ,
\]
and by observation \ref{Observation 10}) above the total
contribution of the factor \\
$\left( \displaystyle
 \prod_{\left(
\left( i,B\right)
,X\right) \in \mathcal{U} \left( \mathcal{R} \left( \psi
\left( \mathbb{J} \left( P,Q,H\right) \right) \right)
\right) } \left( \frac{c\left( x,g,B\right) }{b\left(
x,i\right) } \right)^{ m_{ iBX } } \right) $ is
$\leq -\left(
D_{ A } -\xi _{ A }
\left( P,Q,H,u\right) +1\right) $, which
\newpage
\noindent by observation \ref{Observation 18}) above is
\[
\leq -\left( D_{ A } +1\right) + \hspace{-4.5pt}
\hspace{12.8cm}
\]
\[
+ \left( \sum_{C\in \left(
\mathcal{P}
\left( P,A\right) \,\vdash V\right) } \hspace{0.2cm}
  \sum_{\left(
i,B\right) \in \left( \mathbb{I} _{ C } \left( Q,H\right)
\cap \mathbb{I} _{ A } \left( Q,H\right) \right) }
u_{ iB } \right) + \left( \sum_{i\in \left(
A\cap \mathcal{T} \left(
\mathcal{Y}
\left( P,A\right) ,H\right) \right) } \hspace{0.2cm}
  \sum_{B\in
\mathbb{Y} \left( Q,\mathcal{K} \left( P,A,i\right)
,A\right) } u_{ iB } \right) ,
\]
and by observation \ref{Observation 17}) above, the
total contribution from
the factor \\
$ \left( \displaystyle
 \prod_{\left( i,B\right) \in \mathcal{U}
\left(
\mathcal{R} \left( \downarrow \left( \mathbb{I} \left(
Q,H\right) ,P\right) \right) \right) } \left( \frac{c\left(
x,g,B\right) }{b\left( x,i\right) }
\right)^{ u_{ iB } } \right) $ is
\[
\leq \left( \sum_{C\in \left( \mathcal{P} \left( P,A\right)
\,\vdash
V\right) } \hspace{0.2cm} \sum_{\left( i,B\right)
\in \left(
\mathbb{I} _{ C } \left( Q,H\right) \,\vdash \mathbb{I} _{
A } \left( Q,H\right) \right) } u_{ iB } \right)
 - \left( \sum_{i\in
\left( A\cap \mathcal{T} \left( \mathcal{Y} \left(
P,A\right) ,H\right) \right) } \hspace{0.2cm}
 \sum_{B\in \mathbb{Y}
\left( Q,\mathcal{K} \left( P,A,i\right) ,A\right) }
u_{ iB } \right) ,
\]
and adding these three bounds and noting that for each
member $C$ of $\left( \mathcal{P} \left( P,A\right)
\,\vdash V\right) $, we have the identity
\[
\left(
\sum_{\left( i,B\right) \in \left( \mathbb{I} _{ C } \left(
Q,H\right) \cap \mathbb{I} _{ A } \left( Q,H\right)
\right) } u_{ iB } \right) + \left(
 \sum_{\left( i,B\right) \in
\left( \mathbb{I}
_{ C } \left( Q,H\right) \,\vdash \mathbb{I} _{ A } \left(
Q,H\right) \right) } u_{ iB } \right) = \sum_{
\left( i,B\right) \in
\mathbb{I} _{ C } \left( Q,H\right) } u_{ iB } ,
\]
we obtain the stated result.

\begin{bphzobservation} \label{Observation 20}
\end{bphzobservation}
\vspace{-6.143ex}

\noindent \hspace{2.6ex}{\bf ) }Let $A$ be any member of
$\mathbb{B} \left( P\right) $.
 Then the following inequality holds:
\[
\Gamma \left( \alpha ,\mathcal{P} \left( P,A\right) \right)
\leq \left(
 d\left( \#\left( \mathcal{P} \left( P,A\right) \right)
-1\right) +D_{ A } - \left(
 \sum_{C\in \left( \mathcal{P} \left(
P,A\right) \,\vdash V\right) } D_{ C } \right) + \right.
\hspace{-3.65pt} \hspace{3.0cm}
\]
\[
\hspace{1.0cm} \left. + \left(
 \sum_{C\in \left( \mathcal{P} \left( P,A\right) \,\vdash
V\right) } \hspace{0.2cm}
  \sum_{\left( i,B\right) \in \left(
\mathbb{I} _{ C } \left( Q,H\right) \,\vdash \mathbb{I} _{
A } \left( Q,H\right) \right) } u_{ iB } \right)
 - \left( \sum_{i\in
\mathcal{T} \left( A,H\right) } \hspace{0.2cm}
  \sum_{B\in \mathbb{Y}
\left( Q,\mathcal{K} \left( P,A,i\right) ,A\right) }
u_{ iB } \right) \right)
\]

This follows directly from
pages \pageref{Start of original page 100} and
\pageref{Start of original page 101} and
observations \ref{Observation 9})
and \ref{Observation 17}) above, after
noting that when $A\in \mathbb{B} \left( P\right) $ holds,
the sum over
\label{Start of original page 111}
 $i\in \left( A\cap \mathcal{T} \left( \mathcal{Y} \left(
P,A\right) ,H\right) \right) $ in the second term in the
result of observation \ref{Observation 17}) may be
re-written as a
sum over $i\in \mathcal{T} \left( A,H\right) $, since $A\in
\mathbb{B} \left( P\right) $ implies $\mathcal{Y} \left(
P,A\right) =A$ and $\left( A\cap \mathcal{T} \left(
\mathcal{Y} \left( P,A\right) ,H\right) \right)
 =\mathcal{T} \left(
A,H\right) $.

\begin{bphzobservation} \label{Observation 21}
\end{bphzobservation}
\vspace{-6.143ex}

\noindent \hspace{2.6ex}{\bf ) }Let $\left\{ J,K\right\} $
be any member of $\mathcal{D}
\left( \alpha \right) $ such that there is \emph{no} member
$\left\{ j,k\right\} $ of $W$ such that $j\in J$ and $k\in
K$
both hold.   Then $\alpha _{ \left\{ J,K\right\} } \leq 0$
holds.   For by the definition on\hspace{\stretch{1}}
page\hspace{\stretch{1}}
\pageref{Start of original page 99}\hspace{\stretch{1}}
of\hspace{\stretch{1}} the\hspace{\stretch{1}}
map\hspace{\stretch{1}} $\alpha $,\hspace{\stretch{1}}
contributions\hspace{\stretch{1}} $>0$\hspace{\stretch{1}}
to\hspace{\stretch{1}} $\alpha _{ \left\{
J,K\right\} } $\hspace{\stretch{1}}
can\hspace{\stretch{1}} only\hspace{\stretch{1}}
come\hspace{\stretch{1}} from\hspace{\stretch{1}}
powers\hspace{\stretch{1}} of
\newpage
\noindent $\left| x_{ J
} -x_{
K } \right| $ in the factor $\Theta $, and from any
$b\left( x,i\right) $
such
that $b\left( x,i\right) $ is equal to $\left| x_{ J } -x_{
K } \right| $.
  But it
directly follows from the definition of the factor $\Theta
$ on
page \pageref{Start of original page 97}
 that $\Theta $ does not include any nonzero
power of any $\left| x_{ J } -x_{ K } \right| $ such that
there is
\emph{no} member $\left\{ j,k\right\} $ of $W$ such that
$j\in J$ and $k\in K$ both hold, and it directly follows
from the definition of $b\left( x,i\right) $ on
page \pageref{Start of original page 96}
that $b\left( x,i\right) $ is
not equal to any $\left| x_{ J } -x_{ K } \right| $ such
that there is
\emph{no} member $\left\{ j,k\right\} $ of $W$ such that
$j\in J$ and $k\in K$ both hold.

\vspace{2.5ex}

Now let $A$ be any member of $\mathbb{B} \left( P\right) $
and $x$ be any member of $\mathbb{U} _{ d } \left( V,\omega
\right) $ such that $\mathcal{E} \left( P,Q,H,\sigma
,R,x\right) $ is nonzero.   Then by Lemma \ref{Lemma 10},
$\mathbb{M}
\left( \left( P\,\vdash \left\{ A\right\} \right)
,P,H,A,\sigma ,R,x\right) $ does \emph{not} hold.   Hence
either $\mathbb{L} \left( P,A,x\right) \geq R$ holds or
else there exists a member $i$ of $A$ and a member $j$ of
$\left( \mathcal{Y} \left( \overline{ \left( P\,\vdash
\left\{
A\right\}
\right) } ,A\right) \,\vdash A\right) $ such that
$\left\{ i,j\right\} \in W$ holds and $\mathbb{L} \left(
P,A,x\right) \geq \sigma \left| x_{ \mathcal{Z} \left(
P,H,i\right) }
-x_{ \mathcal{Z} \left( P,H,j\right) }
 \right| $ holds.   (We
note that
if $A=\mathcal{U} \left( V\right) $ holds then the only
possibility is $\mathbb{L} \left( P,A,x\right) \geq
R $.)

We now define $I$ to be the set whose members are all the
maps $i$ such that $\mathcal{D} \left( i\right) =\mathbb{B}
\left( P\right) $, and such that for each member $A$ of
$\mathbb{B} \left( P\right) $, $i_{ A } $ is a member of
$\mathcal{Q} \left( \mathcal{P} \left( P,A\right) \right)
$, and we define $F$ to be the set whose members are all
the maps $f$ such that $\mathcal{D} \left( f\right)
=\mathbb{B} \left( P\right) $, and for each member $A$ of
$\mathbb{B} \left( P\right) $, $f_{ A } $ is either $R$ or
else is an ordered pair $\left(j,k\right) $
of a member $j$ of $A$
 and a
member $k$ of $\left( \mathcal{Y} \left(
\overline{ \left( P\,\vdash
\left\{ A\right\} \right) } ,A\right) \,\vdash A\right)
$
such that $\left\{ j,k\right\} \in W$ holds.   And for each
member $i$ of $I$ and each member $A$ of $\mathbb{B} \left(
P\right) $ we define $e(x,i,A) $ to be $\left| x_{ J } -x_{
K
} \right| $
where $i_{ A } =\left\{ J,K\right\} $, and for each member
$f$ of $F$ and each member $A$ of $\mathbb{B} \left(
P\right) $ we define $h(x,f,A) $ to be $R$ if $f_{ A } =R$,
and otherwise to be $\sigma \left| x_{ \mathcal{Z} \left(
P,H,j\right) } -x_{ \mathcal{Z} \left( P,H,k\right) }
 \right| $
where
$f_{ A } =\left( j,k\right) $.

Then if $r$ is any member of $\mathbb{R}^{ \mathbb{B}
\left( P\right) } $ such that $r_{ A } \geq 0$ holds for all
$A\in \mathbb{B} \left( P\right) $, the following
inequality holds for all $x\in \mathbb{U} _{ d } \left(
V,\omega \right) $ such that $\mathcal{E} \left(
P,Q,H,\sigma ,R,x\right) \neq 0$:
\label{Start of original page 112}
\[
1\leq \sum_{\begin{array}{c} \\[-4.5ex]
\scriptstyle{ f\in F } \\[-1.5ex]
\scriptstyle{ i\in I }
\end{array} }  \prod_{A\in \mathbb{B} \left(
P\right) } \left( \frac{e\left( x,i,A\right) }{h\left(
x,f,A\right) } \right)^{ r_{ A } }
\]
(for there is always at least one term in the right-hand
side that is $\geq 1$, and every term in the right-hand
side is $\geq 0 $.)

We now define $r$ to be the member of $\mathbb{R}^{
\mathbb{B} \left( P\right) } $ such that for each member $A$
of $\mathbb{B} \left( P\right) $,
\[
r_{ A } \equiv \left( D_{ A } - \sum_{\left( i,B\right) \in
\mathbb{I} _{
A } \left( Q,H\right) } u_{ iB } \right) + \frac{ \#\left(
\mathbb{B} \left( P\right) \cap \Xi \left( \mathcal{P}
\left( V,A\right) \right) \right) }{ \#\left( \mathbb{B}
\left( \bar{ P } \right) \right) } ,
\]
where we note that $\mathbb{B} \left( P\right) \cap \Xi
\left( \mathcal{P} \left( V,A\right) \right) $ is the set
of all the members $B$ of $\mathbb{B} \left( P\right) $
such that $B\subseteq A$ holds.

Now $u$ is a member of $\mathbb{X} \left( \downarrow \left(
\mathbb{I} \left( Q,H\right) ,P\right) ,D\right) $, hence
$ \sum_{\left( i,B\right) \in \mathbb{I} _{ A }
\left( Q,H\right) } u_{ iB } \leq D_{ A } $ holds for
every member $A$
of $\mathbb{B} \left( P\right) $, hence $r_{ A } \geq 0$
holds for every member $A$ of $\mathbb{B} \left( P\right) $.

We now define, for each ordered pair
$\left(i,f\right) $ of a member $i$
of $I$ and a member $f$ of $F$,
\[
E_{ if } \equiv U_{ g } \prod_{A\in \mathbb{B} \left(
P\right)
 } \left( \frac{e\left( x,i,A\right) }{h\left( x,f,A\right)
}
 \right)^ {r_{ A } } \hspace{-4.8pt} \hspace{10.0cm}
\]
\[
=\mathcal{E} \left( P,Q,H,\sigma ,R,x\right) \left(
 \prod_{A\in
\mathbb{B} \left( \bar{ P } \right) } \Psi \left(
\downarrow \left( x,\mathcal{P} \left( P,A\right) \right)
,\alpha \right) \right) \left(
  \prod_{A\in \mathbb{B} \left( P\right) }
\left( \frac{e\left( x,i,A\right) }{h\left( x,f,A\right) }
\right)^{ r_{ A } } \right) \Phi
\]

Then it follows immediately from the above observations
that the following inequality holds for all $x\in
\mathbb{U} _{ d } \left( V,\omega \right) $:
\[
U_{ g } \leq \sum_{\begin{array}{c} \\[-4.5ex]
\scriptstyle{ f\in F } \\[-1.5ex]
\scriptstyle{ i\in I }
\end{array} } E_{ if } .
\]

We shall prove that for every ordered pair
$\left(i,f\right) $ of a
member $i$ of the finite set $I$ and a member $f$ of the
finite set $F$, the integral of $E_{ if } $ over
$\mathbb{W}$ is finite.

Let $i$ be any member of $I$ and $f$ be any member of $F$.

We first note that it follows directly from the definition
of $I$ that for each member $A$ of $\mathbb{B} \left(
P\right) $, $i_{ A } $ is a member of $\mathcal{Q} \left(
\mathcal{P} \left( P,A\right) \right) $, and hence $i_{ A }
$ is a member of $\mathcal{D} \left( \alpha \right) =
\bigcup_{B\in
\mathbb{B} \left( \bar{ P } \right) } \mathcal{Q}
\left( \mathcal{P} \left( P,B\right) \right) $.

Now let $A$ be any member of $\mathbb{B} \left( P\right) $
such that $f_{ A } $ is not equal to $R$.   Then $f_{ A } $
is an ordered pair $\left(j,k\right) $ of a
member $j$ of $A$ and a
member $k$ of $\left( \mathcal{Y} \left( \overline{
 \left( P\,\vdash
\left\{ A\right\} \right) } ,A\right) \,\vdash A\right)
$
such that $\left\{ j,k\right\} \in W$ holds.   We note first
that it follows immediately from this that $\mathcal{Z}
\left( P,H,j\right) $ is equal to $A$.
\label{Start of original page 113}
 For $W\subseteq H$ holds and $H$ is a partition, hence
$\mathcal{Z} \left( P,H,j\right) $ is the \emph{largest}
member $B$ of $P$ such that $j\in B$ holds and $\left\{
j,k\right\} $ is \emph{not} a subset of $B$.   Hence
$A\subseteq \mathcal{Z} \left( P,H,j\right) $ holds.   And
$k\in \left( \mathcal{Y} \left( \overline{
 \left( P\,\vdash \left\{
A\right\} \right) } ,A\right) \,\vdash A\right) $
implies that for every member $B$ of $P$ such that
$A\subset B$ holds, $k\in B$ holds, hence $\left\{
j,k\right\} \subseteq B$ holds, hence $A$ is the largest
member of $P$ to have $j$ as a member but not to have
$\left\{ j,k\right\} $ as a subset, hence $A=\mathcal{Z}
\left( P,H,j\right) $.

Now $f_{ A } \neq R$ implies
that $A\neq \mathcal{U} \left( V\right) $, hence there
exists at least one member $B$ of $\mathbb{B} \left( \bar{
P }
\right) $, namely $B=\mathcal{U} \left( V\right) $, such
that $A\subset B$ holds.   Let $B$ be the \emph{smallest}
member of $\mathbb{B} \left( \bar{ P } \right) $ such that
$A\subset B$ holds.   Then $A\in \mathcal{P} \left(
P,B\right) $ holds, hence $\mathcal{Z} \left( P,H,j\right)
\in \mathcal{P} \left( P,B\right) $ holds.   And
furthermore, $\mathcal{Z} \left( P,H,k\right) \in
\mathcal{P} \left( P,B\right) $ holds.   For $W\subseteq H$
holds and $H$ is a partition, hence $\mathcal{Z} \left(
P,H,k\right) $ is the largest member $C$ of $P$ such that
$k\in C$ holds and $\left\{ j,k\right\} $ is \emph{not} a
subset of $C$, hence $j$ is \emph{not} a member of $C$.
(We note that this implies immediately that $\mathcal{Z}
\left( P,H,k\right) $ is \emph{not} equal to $\mathcal{Z}
\left( P,H,j\right) $.)   Now $k\in \left(
\mathcal{Y} \left( \overline{
 \left( P\,\vdash \left\{ A\right\}
\right)
 } ,A\right) \,\vdash A\right) $ implies that $k$ is a
member of \emph{every} member $E$ of $ \bar{ P } $ such that
$A\subset E$ holds, hence $k\in B$ holds, hence $\left\{
j,k\right\} \subseteq B$ holds, hence $\mathcal{Z} \left(
P,H,k\right) \subset B$ holds.   Now suppose there was a
member $E$ of $P$ such that $\mathcal{Z} \left(
P,H,k\right) \subset E\subset B$ held.   Then the fact that
$\mathcal{Z} \left( P,H,k\right) $ is the \emph{largest}
member of $P$ to have $k$ as a member but not have $j$ as a
member, implies that $j\in E$ holds, hence $\left\{
j,k\right\} \subseteq E$ holds, hence $A\subset E$ holds,
and this contradicts the fact that by definition, $B$ is
the \emph{smallest} member of $ \bar{ P } $ to contain $A$
as
a strict subset.   Hence there is \emph{no} member $E$ of
$P$ such that $\mathcal{Z} \left( P,H,k\right) \subset
E\subset B$ holds, hence $\mathcal{Z} \left( P,H,k\right)
\in \mathcal{P} \left( P,B\right) $ holds, hence $\left\{
\mathcal{Z} \left( P,H,j\right) ,\mathcal{Z} \left(
P,H,k\right) \right\} \in \mathcal{Q} \left( \mathcal{P}
\left( P,B\right) \right) $ holds, hence $\left\{
\mathcal{Z} \left( P,H,j\right) ,\mathcal{Z} \left(
P,H,k\right) \right\} $ is a member of $\mathcal{D} \left(
\alpha \right) =\bigcup_{B\in \mathbb{B} \left( \bar{ P }
 \right)
 } \mathcal{Q} \left( \mathcal{P} \left(
P,B\right) \right) $.

We now define $\beta $ to be the map whose domain is equal
to $\mathcal{D} \left( \alpha \right) = $ \\
$\bigcup_{B\in
 \mathbb{B}
\left( \bar{ P } \right) } \mathcal{Q} \left(
\mathcal{P} \left( P,B\right) \right) $, and such that for
each member $\left\{ J,K\right\} $ of $\mathcal{D} \left(
\beta \right) =\mathcal{D} \left( \alpha \right) $,
$\beta_{
\left\{ J,K\right\} } $ is equal to $\alpha_{ \left\{
J,K\right\} } $
plus the negative of the total power of $\left| x_{ J }
-x_{ K }
\right| $ in the expression $\prod_{A\in \mathbb{B} \left(
P\right)
 } \left( \frac{e\left( x,i,A\right) }{h\left( x,f,A\right)
 } \right)^{ r_{ A } } $.

And we define $\kappa $ to be the product, over the members
$A$ of $\mathbb{B} \left( P\right) $, of $\frac{ 1 }{ R } $
if $f_{ A } $ is equal to $R$, and of
$\frac{ 1 }{ \sigma } $ if
$f_{ A } $ is not equal to $R$, and we note that since
$\sigma $ is a finite real number $>0$ and $R$ is a finite
real number $>0$, $\kappa $ is a finite real number $>0$.
\enlargethispage{1.1ex}

Then it follows immediately from these definitions that the
\label{Start of original page 114}
 following equation holds:
\[
E_{ if } =\kappa \mathcal{E} \left( P,Q,H,\sigma
,R,x\right) \left(
\prod_{A\in \mathbb{B} \left( \bar{ P } \right) }
\Psi \left( \downarrow \left( x,\mathcal{P}
\left( P,A\right) \right) ,\beta \right) \right) \Phi
\]

\vspace{2.5ex}

We now make the following observations:

\begin{bphzobservation} \label{Observation 22}
\end{bphzobservation}
\vspace{-6.143ex}

\noindent \hspace{2.6ex}{\bf ) }Let $A$ be any member of
$\mathbb{B} \left( P\right) $.
 Then it follows directly from the definition of the set
$I$ that $i_{ A } =\left\{ J,K\right\} $ is a member of
$\mathcal{Q} \left( \mathcal{P} \left( P,A\right) \right)
$, hence that the factor $e\left( x,i,A\right)^{ r_{ A } }
=\left| x_{ J } -x_{ K } \right|^{ r_{ A } } $ makes the
contribution
$-r_{ A } $ to $\Gamma \left( \beta ,\mathcal{P} \left(
P,A\right) \right) $.

\begin{bphzobservation} \label{Observation 23}
\end{bphzobservation}
\vspace{-6.143ex}

\noindent \hspace{2.6ex}{\bf ) }Let $A$ be any member of
$\mathbb{B} \left( Q\right) $
and $C$ be any member of $\mathbb{B} \left( P\right) $ such
that $f_{ C } $ is not equal to $R$, so that $f_{ C } $ is
an ordered pair $\left(j,k\right) $ of a member
$j$ of $C$ and a member
$k$ of $\left( \mathcal{Y} \left( \overline{
 \left( P\,\vdash \left\{
C\right\} \right) } ,C\right) \,\vdash C\right) $ such
that $\left\{ j,k\right\} \in W$ holds.   Then $\left\{
\mathcal{Z} \left( P,H,j\right) ,\mathcal{Z} \left(
P,H,k\right) \right\} $ is \emph{not} a member of
$\mathcal{Q} \left( \mathcal{P} \left( P,A\right) \right) $
unless $C\in \left( \mathcal{P} \left( P,A\right) \,\vdash
V\right) $ holds.   For if \\
$\left\{ \mathcal{Z} \left(
P,H,j\right) ,\mathcal{Z} \left( P,H,k\right) \right\} \in
\mathcal{Q} \left( \mathcal{P} \left( P,A\right) \right) $
holds, then by observation \ref{Observation 3}) above,
$\left\{
j,k\right\} \subseteq A$ holds and $\left\{ j,k\right\} $ is
\emph{not} a subset of $\mathcal{K} \left( P,A,j\right) $.
 And $W\subseteq H$ holds hence $\left\{ j,k\right\}
\subseteq A$ implies $\mathcal{Z} \left( P,H,j\right)
\subset A$, and $H$ is a partition hence $\left\{
j,k\right\}
$ is the \emph{only} member of $H$ to have $j$ as a member,
hence the fact that $\left\{ j,k\right\} $ is \emph{not} a
subset of $\mathcal{K} \left( P,A,j\right) $ implies that
$\mathcal{K} \left( P,A,j\right) \subseteq \mathcal{Z}
\left( P,H,j\right) $ holds.   But $\mathcal{K} \left(
P,A,j\right) \subset \mathcal{Z} \left( P,H,j\right)
\subset A$ cannot hold since $\mathcal{K} \left(
P,A,j\right) $ is a member of $\mathcal{P} \left(
P,A\right) $, hence $\mathcal{K} \left( P,A,j\right)
=\mathcal{Z} \left( P,H,j\right) $ holds, hence
$\mathcal{Z} \left( P,H,j\right) $ is a member of
$\mathcal{P} \left( P,A\right) $.

And furthermore, as shown on
pages \pageref{Start of original page 112} and
\pageref{Start of original page 113}, the
facts that $C\in \mathbb{B} \left( P\right) $ holds, $j\in
C$ holds, and $k\in \left( \mathcal{Y} \left( \overline{
 \left(
P\,\vdash \left\{ C\right\} \right) } ,C\right) \,\vdash
C\right) $ holds, imply directly that $C\subseteq
\mathcal{Z} \left( P,H,j\right) $ holds, and that for every
member $B$ of $P$ such that $C\subset B$ holds, $\left\{
j,k\right\} \subseteq B$ holds, hence that $\mathcal{Z}
\left( P,H,j\right) =C$ holds.   Hence $C$ is a member of
$\mathcal{P} \left( P,A\right) $.

And furthermore, $C$ is a member of $\mathbb{B} \left(
P\right) $, hence $C$ is not a member of $V$, hence $C$ is
a member of $\left( \mathcal{P} \left( P,A\right) \,\vdash
V\right) $.
\enlargethispage{4.5ex}

\begin{bphzobservation} \label{Observation 24}
\end{bphzobservation}
\vspace{-6.143ex}

\noindent \hspace{2.6ex}{\bf ) }Let $A$ be any member of
$\mathbb{B} \left( P\right) $.
 Then it follows directly from
 observations \ref{Observation 22})
and \ref{Observation 23})
above that the contribution to $\Gamma
\left( \beta ,\mathcal{P} \! \left( P,A\right) \right) $
from
the factor $\left( \prod_{A\in \mathbb{B} \left( P\right) }
\! \left(
\frac{e\left( x,i,A\right) }{h\left( x,f,A\right) }
\right)^{ r_{ A } } \right)
$ is
$\leq \left( \displaystyle
  \sum_{C\in \left( \mathcal{P} \left( P,A\right)
\,\vdash
V\right) } r_{ C } \right) -r_{ A } $.

And by the definition of the map $r$,
\label{Start of original page 115}
\[
\left(
\sum_{C\in \left( \mathcal{P} \left( P,A\right) \,\vdash
V\right) } r_{ C } \right) -r_{ A } = \hspace{-1.3pt}
\hspace{11.0cm}
\]
\[
= \left\{ \left(
 \sum_{C\in \left(
\mathcal{P} \left( P,A\right) \,\vdash V\right) }
\left( \left(
D_{ C } - \sum_{\left( i,B\right) \in \mathbb{I} _{ C }
\left(
Q,H\right) } u_{ iB } \right) + \frac{
 \#\left( \mathbb{B} \left(
P\right) \cap \Xi \left( \mathcal{P} \left( V,C\right)
\right) \right) }{ \#\left( \mathbb{B} \left( \bar{ P }
\right)
\right) } \right) \right) \right.
\]
\[
\left.
- \left( \left(
D_{ A } - \sum_{\left( i,B\right) \in \mathbb{I} _{ A }
\left(
Q,H\right) } u_{ iB } \right) + \frac{
 \#\left( \mathbb{B} \left(
P\right) \cap \Xi \left( \mathcal{P} \left( V,A\right)
\right) \right) }{ \#\left( \mathbb{B} \left( \bar{ P }
\right)
\right) } \right) \right\}
\]
\[
= \left(
 \sum_{C\in \left( \mathcal{P} \left( P,A\right) \,\vdash
V\right) } \left( D_{ C } - \sum_{\left( i,B\right) \in
\mathbb{I} _{ C } \left( Q,H\right) } u_{ iB } \right)
\right) - \left( D_{
A } - \sum_{\left( i,B\right) \in \mathbb{I} _{ A } \left(
Q,H\right) } u_{ iB } \right) - \frac{ 1 }{
\#\left( \mathbb{B}
\left( \bar{ P } \right) \right) }
\]

\begin{bphzobservation} \label{Observation 25}
\end{bphzobservation}
\vspace{-6.143ex}

\noindent \hspace{2.6ex}{\bf ) }Let $A$ be any member of
$\mathbb{B} \left( P\right) $.
 Then the following inequality holds:
\[
\Gamma \left( \beta ,\mathcal{P} \left( P,A\right) \right)
\leq d\left( \#\left( \mathcal{P} \left( P,A\right) \right)
-1\right) - \frac{ 1 }{ \#\left( \mathbb{B} \left(
\bar{ P }
\right)
\right) } .
\]
\newpage

For by observations \ref{Observation 20})
and \ref{Observation 24}) above we
have
\[
\Gamma \left( \beta ,\mathcal{P} \left( P,A\right) \right)
\leq \left\{
d\left( \#\left( \mathcal{P} \left( P,A\right) \right)
-1\right) + \left(
 \sum_{\left( i,B\right) \in \mathbb{I} _{ A }
\left(
Q,H\right) } u_{ iB } \right) \right. \hspace{4.0cm}
\]
\[
 - \left(
 \sum_{C\in \left( \mathcal{P} \left( P,A\right) \,\vdash
V\right) } \hspace{0.2cm}
 \sum_{\left( i,B\right) \in \left(
\mathbb{I} _{ C } \left( Q,H\right) \cap \mathbb{I} _{ A }
\left( Q,H\right) \right) } u_{ iB } \right)
 - \left( \sum_{i\in
\mathcal{T} \left( A,H\right) } \hspace{0.2cm}
 \sum_{B\in \mathbb{Y}
\left( Q,\mathcal{K} \left( P,A,i\right) ,A\right) }
u_{ iB } \right)
\]
\[
\left. \hspace{12.0cm} - \frac{ 1 }{
\#\left( \mathbb{B} \left( \bar{ P }
\right) \right) } \right\}
\]

And by using the identity
\[
\left(
 \sum_{\left( i,B\right) \in \mathbb{I} _{ A } \left(
 Q,H\right)
 } u_{ iB } \right) = \hspace{-9.6pt} \hspace{12.0cm}
\]
\[
\hspace{2.0cm}
  = \left( \sum_{i\in \mathcal{T} \left( A,H\right)
 } \hspace{0.2cm}
  \sum_{B\in \mathbb{Y} \left( Q,\mathcal{C} \left(
V,i\right) ,\mathcal{K} \left( P,A,i\right) \right)
} u_{ iB } \right) + \left(
 \sum_{i\in \mathcal{T} \left( A,H\right) }
 \hspace{0.2cm}
 \sum_{B\in \mathbb{Y} \left( Q,\mathcal{K} \left(
P,A,i\right)
,A\right) } u_{ iB } \right)
\]
and noting that the first term in the right-hand side of
this identity is equal to $\left( \displaystyle
 \sum_{C\in \left( \mathcal{P}
\left( P,A\right) \,\vdash V\right) } \hspace{0.2cm}
 \displaystyle \sum_{\left(
i,B\right) \in \left( \mathbb{I} _{ C } \left( Q,H\right)
\cap \mathbb{I} _{ A } \left( Q,H\right) \right) }
u_{ iB } \right) $, we obtain the stated result.
\enlargethispage{4.5ex}

\begin{bphzobservation} \label{Observation 26}
\end{bphzobservation}
\vspace{-6.143ex}

\noindent \hspace{2.6ex}{\bf ) }Let $A$ be any member of
$\left( Q\,\vdash P\right) $.
Then it follows directly from
observation \ref{Observation 23})
above that the contribution to $\Gamma \left( \beta
,\mathcal{P} \left( P,A\right) \right) $ from the factor
$\left( \prod_{A\in \mathbb{B} \left( P\right) }
\left( \frac{e\left(
x,i,A\right) }{h\left( x,f,A\right) }
\right)^{ r_{ A } } \right) $
is
$\leq \left( \displaystyle
 \sum_{C\in \left( \mathcal{P} \left( P,A\right)
\,\vdash
V\right) } r_{ C } \right) $.

And by the definition of the map $r$,
\[
\left(
\sum_{C\in \left( \mathcal{P} \left( P,A\right) \,\vdash
V\right) } r_{ C } \right) = \hspace{-8.3pt}
\hspace{12.0cm}
\]
\[
\hspace{2.0cm} \hspace{-6.9pt} = \left(
 \sum_{C\in \left( \mathcal{P}
\left( P,A\right) \,\vdash V\right) } \left( \left(
 D_{ C }
-\sum_{\left( i,B\right) \in \mathbb{I} _{ C }
\left( Q,H\right)
 } u_{ iB } \right)
  + \frac{\#\left( \mathbb{B} \left( P\right)
\cap \Xi \left( \mathcal{P} \left( V,C\right) \right)
\right)}{\#\left( \mathbb{B} \left( \bar{ P } \right)
\right)} \right) \right)
.
\]

\begin{bphzobservation} \label{Observation 27}
\end{bphzobservation}
\vspace{-6.143ex}

\noindent \hspace{2.6ex}{\bf ) }Let $A$ be any member of
$\left( Q\,\vdash P\right) $.
Then the following inequality holds:
\label{Start of original page 116}
\[
\Gamma \left( \beta ,\mathcal{P} \left( P,A\right)
\right) \leq d\left( \#\left( \mathcal{P} \left( P,A\right)
\right) -1\right) -
\frac{ 1 }{ \#\left( \mathbb{B} \left( \bar{
P }
\right) \right) } .
\]
\newpage

For by observations \ref{Observation 19})
and \ref{Observation 26}) above we
have
\[
\Gamma \left( \beta ,\mathcal{P} \left( P,A\right) \right)
\leq \left( d\left( \#\left( \mathcal{P} \left( P,A\right)
\right) -1\right) -1\right) + \left(
 \sum_{C\in \left( \mathcal{P}
\left( P,A\right) \,\vdash V\right) }\frac{\#\left(
\mathbb{B} \left( P\right) \cap \Xi \left( \mathcal{P}
\left( V,C\right) \right) \right)}{\#\left( \mathbb{B}
\left( \bar{ P } \right) \right)} \right)
\]
\[
\hspace{3.0cm}
=\left( d\left( \#\left( \mathcal{P} \left( P,A\right)
\right) -1\right) -1\right) + \frac{\#\left( \mathbb{B}
\left(
P\right) \cap \Xi \left( \mathcal{P} \left( V,A\right)
\right) \right)}{\#\left( \mathbb{B} \left( \bar{ P }
\right)
\right)}
\]

Now $A$ is a member of $\left( Q\,\vdash P\right) $, hence
$\#\left( \mathbb{B} \left( P\right) \cap \Xi \left(
\mathcal{P} \left( V,A\right) \right) \right) $ is the
number of members $B$ of $\mathbb{B} \left( P\right) $ such
that $B\subset A$ holds.   And this number is $\leq \left(
\#\left( \mathbb{B} \left( \bar{ P } \right) \right)
-1\right)
$, since the member $\mathcal{U} \left( V\right) $ of
$\mathbb{B} \left( \bar{ P } \right) $ is not a
\emph{strict}
subset of \emph{any} member of $\Xi \left( V\right) $, and
in particular is not a \emph{strict} subset of $A$.   And
the stated result follows directly from this.

\begin{bphzobservation} \label{Observation 28}
\end{bphzobservation}
\vspace{-6.143ex}

\noindent \hspace{2.6ex}{\bf ) }Let $A$ be any member of
$\mathbb{B} \left( Q\right) $.
 Then directly from
 observations \ref{Observation 25}) and
\ref{Observation 27})
above, the following inequality holds:
\[
\Gamma \left( \beta ,\mathcal{P} \left( P,A\right) \right)
\leq d\left( \#\left( \mathcal{P} \left( P,A\right) \right)
-1\right) - \frac{ 1 }{ \#\left(
\mathbb{B} \left( \bar{ P }
\right)
\right) } .
\]

\begin{bphzobservation} \label{Observation 29}
\end{bphzobservation}
\vspace{-6.143ex}

\noindent \hspace{2.6ex}{\bf ) }Let $\left\{ J,K\right\} $
be any member of $\mathcal{D}
\left( \beta \right) =\mathcal{D} \left( \alpha \right) $
such that there is \emph{no} member $\left\{ j,k\right\} $
of
$W$ such that $j\in J$ and $k\in K$ both hold.   Then
$\beta _{ \left\{ J,K\right\} } \leq 0$ holds.   For $\alpha
_{ \left\{ J,K\right\} } \leq 0$ holds by
observation \ref{Observation 21}) above, and by the
definition on
page \pageref{Start of original page 111} of
$h\left( x,f,A\right) $, $h\left( x,f,A\right) $
is either equal to $R$ or else is equal
to $\left| x_{ B } -x_{ C } \right| $, where $B$ and $C$
are members of
$P$ such that there \emph{does} exist a member $\left\{
j,k\right\} $ of $W$ such that $j\in B$ and $k\in C$ both
hold, hence it directly follows from the definition on
page \pageref{Start of original page 113}
of the map $\beta $ that $\beta _{ \left\{ J,K\right\}
} \leq 0$ holds.

\vspace{2.5ex}

Now the factor $\Phi $ was defined on
page \pageref{Start of original page 97} by \\
\label{Phi on original page 116}
\hspace{\stretch{1}}$
\Phi \equiv \left( \displaystyle
\prod_{\Delta \equiv \left\{ j,k\right\} \in W
 } \mathbb{S} \left( T-\left( 1-2\lambda \right)
\left| x_{ \mathcal{Z} \left( P,H,j\right) }
 -x_{ \mathcal{Z} \left(
P,H,k\right) } \right| \right) \right) .
$\hspace{\stretch{1}}

Now every member $A$ of $\mathbb{B} \left( \bar{ P }
\right) $
is $\left( V\cup W\right) $-connected, hence every member
$A$ of $\mathbb{B} \left( \bar{ P } \right) $ is $\left(
\mathcal{P} \left( P,A\right) \cup W\right) $-connected.

Let $L\equiv \left( \frac{ T }{ 1-2\lambda } \right)
\hspace{0.9ex}
\max \rule[-2.4ex]{0pt}{2.4ex}_{ \hspace{-5.5ex}
A\in \mathbb{B} \left( \bar{ P } \right) } \hspace{-0.5ex}
\left( \#\left( \mathcal{P}
\left( P,A\right) \right) -1\right) $.
\enlargethispage{5.0ex}

Now as shown on pages
\pageref{Original 98}-\pageref{Original 99},
$\left\{ j,k\right\} \in W$
implies that $\left\{ \mathcal{Z} \hspace{-1.0pt}
\left( P,H,j\right), \mathcal{Z} \hspace{-1.0pt}
\left( P,H,k\right) \right\} \hspace{-1.1pt} \in
\bigcup_{A\in
\mathbb{B} \left( \bar{ P } \right) } \mathcal{Q}
\left( \mathcal{P} \left( P,A\right) \right) $ holds.
Hence it follows directly from Lemma \ref{Lemma 15} that \\
\label{Start of original page 117}
\hspace{\stretch{1}}$
\Phi \leq \displaystyle
 \prod_{A\in \mathbb{B} \left( \bar{ P } \right) }
 \left( \displaystyle \prod_{\Delta \equiv \left\{
 J,K\right\} \in  \mathcal{Q} \left(
\mathcal{P} \left( P,A\right) \right) }
\mathbb{S} \left( L-\left| x_{ J } -x_{ K } \right| \right)
\right)
$\hspace{\stretch{1}}
\newpage
holds.   Hence, using the definition of $\mathcal{B} $ on
page \pageref{Start of original page 9}, we have
\[
\Phi \leq  \prod_{A\in \mathbb{B} \left( \bar{ P } \right) }
\mathcal{B} \left( \downarrow \left( x,\mathcal{P} \left(
P,A\right) \right) ,L\right) .
\]

And by
page \pageref{Start of original page 47} we have
\[
\mathcal{E} \left( P,Q,H,\sigma ,R,x\right) \leq
\mathcal{H} \left( P,Q,H,\sigma ,R,x\right) ,
\]
hence by Lemma \ref{Lemma 13} we have
\[
\mathcal{E} \left( P,Q,H,\sigma ,R,x\right) \leq
\hspace{11.5cm}
\]
\[
\hspace{2.0cm} \hspace{-6.0pt} \leq \prod_{ A\in
\mathbb{B} \left( \bar{ P } \right) }
 \mathcal{H} \left( P\cap
\Xi \left( \mathcal{P} \left( P,A\right) \right) ,Q\cap \Xi
\left( \mathcal{P} \left( P,A\right) \right) ,H,\sigma
,R,\downarrow \left( x,\Xi \left( \mathcal{P} \left(
P,A\right) \right) \right) \right) .
\]

For each member $A$ of $\mathbb{B} \left( \bar{ P } \right)
$
we define $Z_{ A } \equiv \Xi \left( \mathcal{P} \left(
P,A\right) \right) $.

Then it immediately follows from the foregoing that the
following inequality holds:
\[
E_{ if } \leq \hspace{14.0cm}
\]
\[
\leq \kappa \hspace{-0.25cm} \prod_{A\in \mathbb{B}
\left( \bar{ P } \right)
 } \hspace{-0.25cm}
 \left( \mathcal{H} \! \left( P\! \cap \! Z_{ A } ,
 Q\! \cap \!
Z_{ A } ,\! H,\! \sigma ,\! R,\downarrow \! \! \left(
x,\! Z_{ A } \right)
\right) \Psi \! \left( \downarrow \! \!
\left( x,\! \mathcal{P} \! \left( \!
P,\! A\right) \right) \! ,\! \beta \right) \mathcal{B}
\! \left(
\downarrow \! \! \left( x,\! \mathcal{P} \!
\left( P,\! A\right) \right) \!
,\! L\right) \right)
\]

Now by definition, $\mathbb{U} _{ d } \left( V,\omega
\right) $ is the set of all members $x$ of $\mathbb{E} _{
d }^{ \Xi \left( V\right) } $ such that for
each member $A$ of
$\left( \Xi \left( V\right) \,\vdash V\right) $, $x_{ A }
= \displaystyle
 \sum_{B\in \mathcal{P} \left( V,A\right) } \omega
_{ AB } x_{ B } $ holds.

We now choose a map $S$ such that $\mathcal{D} \left(
S\right) =\mathbb{B} \left( \bar{ P } \right) $, and such
that
for each member $A$ of $\mathbb{B} \left( \bar{ P } \right)
$,
$S_{ A } $ is a member of $\mathcal{P} \left( P,A\right) $.
  And for every member $B$ of $\left( P\,\vdash \left\{
\mathcal{U} \left( V\right) \right\} \right) $, we define
$z_{ B } \equiv \left( x_{ B } -x_{ S_{ A } } \right) $,
where $A$ is the \emph{smallest} member $C$ of $ \bar{ P } $
such that $B\subset C$ holds, or in other words, where
$A=\mathcal{Y} \left( \overline{
 \left( P\,\vdash \left\{ B\right\}
\right) } ,B\right) $.   (We note that $x_{ S_{ A } } $
means $x_{ \left( S_{ A } \right) } $, in
accordance with our
general rule for interpreting a subscript on a subscript,
stated on
page \pageref{Start of original page 2}.)

This means that if $A$ is any member of $\mathbb{B} \left(
\bar{ P } \right) $, and $B$ and $C$ are any members of
$\mathcal{P} \left( P,A\right) $, then the equation $\left(
x_{ B } -x_{ C } \right) =\left( z_{ B } -z_{ C } \right) $
holds.

We note that if $B$ is a member of $\mathcal{R} \left(
S\right) $, then $S_{ \mathcal{Y} \left( \overline{
 \left( P\,\vdash
\left\{ B\right\} \right) } ,B\right) } =B$ holds, hence
$z_{ B } =0$ holds.

Now $ \bar{ P } $ is equal to the disjoint union of $\left\{
\mathcal{U} \left( V\right) \right\} $ and all the
$\mathcal{P} \left( P,A\right) $, $A\in \mathbb{B}
\left( \bar{ P
} \right) $, hence
\[
\#\left( \bar{ P } \right) =1+ \sum_{A\in \mathbb{B}
\left( \bar{ P }
\right) } \#\left( \mathcal{P} \left( P,A\right)
\right)
\]
holds hence
\label{Start of original page 118}
\[
\#\left( V\right) =\#\left( \bar{ P } \right) -\#\left(
\mathbb{B} \left( \bar{ P } \right) \right) =1+
\sum_{A\in \mathbb{B}
\left( \bar{ P } \right) } \left( \#\left( \mathcal{P}
\left( P,A\right) \right) -1\right)
\]
holds hence
\[
\#\left( V\right) =1+\#\left( P\,\vdash \left( \mathcal{R}
\left( S\right) \cup \left\{ \mathcal{U} \left( V\right)
\right\} \right) \right)
\]
holds.

And furthermore, the $x_{ A } $, $A\in V$, may be expressed
in terms of the $z_{ B } $, $B\in \left( P\,\vdash \left(
\mathcal{R} \left( S\right) \cup \left\{ \mathcal{U} \left(
V\right) \right\} \right) \right) $, together with
$x_{ \mathcal{U} \left( V\right) } $.

For if $A$ is any member of $\mathbb{B} \left( \bar{ P }
\right) $, then $x_{ A } = \displaystyle
 \sum_{B\in \mathcal{P} \left(
P,A\right) } \omega _{ AB } x_{ B } $ holds by
Lemma \ref{Lemma 4}.   And as noted above, for each member
$B$ of
$\mathcal{P} \left( P,A\right) $, $z_{ B } =\left( x_{ B }
-x_{ S_{ A } } \right) $ holds.   Hence
\[
 \sum_{B\in \mathcal{P} \left( P,A\right) } \omega _{
AB } z_{ B } = \sum_{B\in \mathcal{P} \left( P,A\right)
 } \omega _{ AB } \left( x_{ B } -x_{ S_{ A } } \right)
=\left( x_{ A } -x_{ S_{ A } } \right)
\]
holds, hence $x_{ S_{ A } } =x_{ A } - \displaystyle
\sum_{B\in \mathcal{P}
\left(
P,A\right) } \omega _{ AB } z_{ B } $ holds,
hence for all members $B$ of $\mathcal{P} \left( P,A\right)
$ we have
\[
x_{ B } =z_{ B } +x_{ S_{ A } } = x_{ A } + \left( z_{ B }
 - \sum_{C\in
\mathcal{P} \left( P,A\right) } \omega _{ AC }
z_{ C } \right) .
\]

From this we find directly that, for any member $i$ of
$\mathcal{U} \left( V\right) $,
\[
x_{ \mathcal{C} \left( V,i\right) }
 =x_{ \mathcal{U} \left(
V\right) } + \sum_{A\in \mathbb{Y} \left( \bar{ P } ,
\mathcal{C} \left(
V,i\right) ,\mathcal{U} \left( V\right) \right) }
\left(
z_{ \mathcal{K} \left( P,A,i\right) }
 - \sum_{C\in \mathcal{P}
\left(
P,A\right) } \omega _{ AC } z_{ C } \right) .
\]

Furthermore, the linear transformation from the variables
$x_{ A } $, $A\in V$, to the variables $z_{ B } $, $B\in
\left( P\,\vdash \left( \mathcal{R} \left( S\right) \cup
\left\{ \mathcal{U} \left( V\right) \right\} \right) \right)
$, and $x_{ \mathcal{U} \left( V\right) } $,
may be realized as
a sequence of linear transformations, each associated with
one member of $\mathbb{B} \left( \bar{ P } \right) $, and
each
of the type considered in Lemma \ref{Lemma 24}.   Hence the
linear
transformation from the variables $x_{ A } $, $A\in V$, to
the variables $z_{ B } $, $B\in \left( P\,\vdash \left(
\mathcal{R} \left( S\right) \cup \left\{ \mathcal{U} \left(
V\right) \right\} \right) \right) $, and $x_{ \mathcal{U}
\left( V\right) } $, has determinant equal to $1$.

And furthermore, the particular equation of the above form
when the member $i$ of $\mathcal{U} \left( V\right) $ is
taken to be $h$, where $h$ is the particular member of
$\mathcal{U} \left( V\right) $ such that $x_{ \mathcal{C}
\left( V,h\right) } $ has the fixed value $b$ in the
definition of the integration domain $\mathbb{W} $, may be
used to express $x_{ \mathcal{U} \left( V\right) } $
in terms of
$x_{ \mathcal{C} \left( V,h\right) } $
and the $z_{ B } $, $B\in
\left( P\,\vdash \left( \mathcal{R} \left( S\right) \cup
\left\{ \mathcal{U} \left( V\right) \right\} \right) \right)
$, and thus transform from the variables $x_{ \mathcal{U}
\left( V\right) } $ and the $z_{ B } $, $B\in \left(
P\,\vdash \left( \mathcal{R} \left( S\right) \cup \left\{
\mathcal{U} \left( V\right) \right\} \right) \right) $, to
the variables $x_{ \mathcal{C} \left( V,h\right) } $ and
the
$z_{ B } $, $B\in \left( P\,\vdash \left( \mathcal{R}
\left( S\right) \cup \left\{ \mathcal{U} \left( V\right)
\right\} \right) \right) $.   And this final linear
transformation has
\label{Start of original page 119}
 the form of a triangular linear transformation with every
diagonal element equal to $1$, and thus has determinant
equal to $1$.

We now make this linear transformation, with determinant
equal to $1$, from the integration variables $x_{ A } $,
$A\in \left( V\,\vdash \left\{ \mathcal{C} \left( V,h\right)
\right\} \right) $, to the variables $z_{ B } $, \\
$B\in
\left( P\,\vdash \left( \mathcal{R} \left( S\right) \cup
\left\{ \mathcal{U} \left( V\right) \right\} \right) \right)
$.   Then the above bound on $E_{ if } $ shows that the
integral of $E_{ if } $ over $\mathbb{W}$ is bounded by the
finite real number $\kappa $ multiplied by the product,
over the members $A$ of $\mathbb{B} \left( \bar{ P } \right)
$, of the integral
\[
\int \left(
  \prod_{B\in \left( \mathcal{P} \left( P,A\right) \,\vdash
\left\{ S_{ A } \right\} \right) } \hspace{-15.2pt}
 d^{ d } z_{ B } \right)
\left\{
\mathcal{H} \left( P\cap Z_{ A } ,Q\cap Z_{ A } ,H,\sigma
,R,\downarrow \left( z,Z_{ A } \right) \right) \Psi \left(
\downarrow \left( z,\mathcal{P} \left( P,A\right) \right)
,\beta \right) \times \right.
\]
\[
\hspace{10.0cm} \left. \times \mathcal{B} \left( \downarrow
\left( z,\mathcal{P} \left( P,A\right) \right) ,L\right)
\right\}.
\]
(We note that with our new choice of independent
integration variables, $z_{ S_{ A } } \equiv 0 $ holds, and
the members of $\downarrow \left( z,Z_{ A } \right) $
depend \emph{only} on the $z_{ B } $, $B\in \mathcal{P}
\left( P,A\right) $, or in other words, \emph{only} on the
$z_{ B } $, $B\in \left( \mathcal{P} \left( P,A\right)
\,\vdash \left\{ S_{ A } \right\} \right) $.)

Let $A$ be any member of $\mathbb{B} \left( \bar{ P }
\right)
$.   We insert into the above integral for $A$ the identity
\[
1= \sum_{F\in \mathbb{H} \left( \mathcal{P} \left(
P,A\right)
\right) } \mathcal{A} \left( \downarrow \left(
z,\mathcal{P} \left( P,A\right) \right) ,\sigma ,F\right) ,
\]
where by the definition of the function $\mathcal{A} $ on
page \pageref{Start of original page 9},
 $\mathcal{A} \left( \downarrow \left(
z,\mathcal{P} \left( P,A\right) \right) ,\sigma ,F\right) $
is equal to $1$ if the set $\mathcal{F} \left( \downarrow
\left( z,\mathcal{P} \left( P,A\right) \right) ,\sigma
\right) $ of all the $\sigma $-clusters of $\downarrow
\left( z,\mathcal{P} \left( P,A\right) \right) $ is equal
to $F$, and equal to $0$ otherwise, and $\mathbb{H} \left(
\mathcal{P} \left( P,A\right) \right) $ is the set whose
members are all the greenwoods $F$ of $\mathcal{P} \left(
P,A\right) $ such that $\mathcal{P} \left( P,A\right) \in
F$ holds.

(We note that the greenwoods of $\mathcal{P} \left(
P,A\right) $ are \emph{not} the same as the woods of
$\mathcal{P} \left( P,A\right) $.   Rather if $F$ is any
greenwood of $\mathcal{P} \left( P,A\right) $, then the set
whose members are all the $\mathcal{U} \left( X\right) $,
$X\in F$, is a wood of $\mathcal{P} \left( P,A\right) $,
and if $G$ is any wood of $\mathcal{P} \left( P,A\right) $,
then the set whose members are all the one-member subsets
of $\mathcal{P} \left( P,A\right) $, together with all the
$\mathcal{P} \left( P,B\right) $, $B\in \mathbb{B} \left(
G\right) $, is a greenwood of $\mathcal{P} \left(
P,A\right) $.   This situation arises because we worked
with the greenwoods of a set, rather than the woods of a
partition, in ``Cluster Convergence Theorem Theorem''.)

Let $F$ be any member of the finite set $\mathbb{H} \left(
\mathcal{P} \left( P,A\right) \right) $.   We shall show
that the following integral is finite:
\label{Start of original page 120}
\[
\int \left( \prod_{B\in \left( \mathcal{P}
\left( P,A\right) \,\vdash
\left\{ S_{ A } \right\} \right) } \hspace{-27.0pt}
 d^{ d } z_{ B } \right)
\left\{
\mathcal{A} \left( \downarrow \! \left( z,\mathcal{P} \left(
P,A\right) \right) ,\sigma ,F\right) \mathcal{H} \left(
P\cap Z_{ A } ,Q\cap Z_{ A } ,H,\sigma ,R,\downarrow \!
 \left(
z,Z_{ A } \right) \right) \times \right.
\]
\[
\hspace{7.0cm} \left. \times \Psi \left(
\downarrow \left( z,\mathcal{P} \left( P,A\right) \right)
,\beta \right) \mathcal{B} \left( \downarrow \left(
z,\mathcal{P} \left( P,A\right) \right) ,L\right) \right\}
\]

We first note that by Lemma \ref{Lemma 11} and the
Additional Note on
page \pageref{Start of original page 45},
if $Y $ is a nonempty subset of a member $X $ of $F $ such
that
$\mathcal{U}
\left( Y\right) $ is \emph{not} a member of $P $, (hence
$\#\left( Y\right) \geq
2$ holds, and if $A=\mathcal{U} \left( V\right)$ and
$\mathcal{U}
\left( V\right) \notin P$ do not \emph{both} hold, then
$Y$ is \emph{not}
equal to $\mathcal{P} \left( P,A\right) $), $
\mathcal{U} \left( Y\right) $ is $\left( V\cup H\right)
$-connected, and $Y$ is \emph{not}
a strict subset of any subset $Z$ of $X$ such that
$\mathcal{U} \left(
Z\right) $ is $\left( V\cup H\right) $-connected, then
either $\mathcal{U} \left( Y\right) \in \left( Q\,\vdash
P\right) $ holds or else the product $\mathcal{A} \left(
\downarrow \left( z,\mathcal{P} \left( P,A\right) \right)
,\sigma ,F\right) \mathcal{H} \left( P\cap Z_{ A } ,Q\cap
Z_{ A } ,H,\sigma ,R,\downarrow \left( z,Z_{ A } \right)
\right) $ vanishes whenever $\mathbb{L} \left( P\cap Z_{ A
} ,\mathcal{U} \left( Y\right) ,\downarrow \left( z,Z_{ A }
\right) \right) <R$ holds.   For the factor $\mathcal{A}
\left( \downarrow \left( z,\mathcal{P} \left( P,A\right)
\right) ,\sigma ,F\right) $ vanishes unless $X$ is a $\sigma
$-cluster of $\downarrow \left( z,\mathcal{P}
\left( P,A\right)
\right) $.   And the factor \\
$\mathcal{H} \left( P\cap Z_{ A
} ,Q\cap Z_{ A } ,H,\sigma ,R,\downarrow \left( z,Z_{ A }
\right) \right) $ vanishes unless $\left( \left( P\cap Z_{
A } \right) ,\left( Q\cap Z_{ A } \right) \right) $ is a
member of $\Omega \left( H,\sigma ,R,\downarrow \left(
z,Z_{ A } \right) \right) $, and there is \emph{no} wood
$G$ of $\mathcal{P} \left( P,A\right) $ such that $\left(
Q\cap Z_{ A } \right) \subset G$ holds and $\left( \left(
P\cap Z_{ A } \right) ,G\right) \in \Omega \left( H,\sigma
,R,\downarrow \left( z,Z_{ A } \right) \right) $ holds.
But by Lemma \ref{Lemma 11} this means that if
$\mathcal{A} \left(
\downarrow \left( z,\mathcal{P} \left( P,A\right) \right)
,\sigma ,F\right) \neq 0$, so $X$ is a $\sigma $-cluster of
$\downarrow \left( z,\mathcal{P} \left( P,A\right) \right)
$, and $\mathcal{H} \left( P\cap Z_{ A } ,Q\cap Z_{ A }
,H,\sigma ,R,\downarrow \left( z,Z_{ A } \right) \right)
\neq 0$, then either $\mathbb{L} \left( P\cap Z_{ A }
,\mathcal{U} \left( Y\right) ,\downarrow \left( z,Z_{ A }
\right) \right) \geq R$ holds or $\mathcal{U} \left(
Y\right) \in \left( Q\,\vdash P\right) $ holds.

(We note in passing that by Lemma \ref{Lemma 6} (f) the
factor \\
$\mathcal{A} \left( \downarrow \left( z,\mathcal{P}
\left( P,A\right) \right) ,\sigma ,F\right) \mathcal{H}
\left( P\cap Z_{ A } ,Q\cap Z_{ A } ,H,\sigma ,R,\downarrow
\left( z,Z_{ A } \right) \right) $ vanishes if any such $Y$
overlaps any member of $\left( Q\,\vdash P\right) $, but we
do not use this.)

Let $M $ be the set whose members are all the nonempty
subsets $Y $ of members
$X$ of $F$ such that $\mathcal{U} \left( Y\right) $ is
\emph{not}
a member of $Q$, (hence $\#\left( Y\right) \geq 2$
holds), $\mathcal{U} \left( Y\right) $ is $\left( V\cup
H\right) $-connected, and $Y$ is \emph{not} a strict subset
of any subset $Z$ of $X$ such that $\mathcal{U}
\left( Z\right)
$ is $\left( V\cup H\right)$-connected.  Let $T$ be the
set whose members are all the maps $s $ such that
$\mathcal{D} \left(
s\right) =M$, and for each member $Y$ of $M$, $s_{ Y } $ is
a
member of $\mathcal{Q} \left( \mathcal{P} \left(
P,\mathcal{U} \left( Y\right) \right) \right) $.   And for
each member $s$ of $T$ and each member $Y$ of $M$ we define
$t\left( z,s,Y\right) \equiv \left| z_{ J } -z_{ K }
\right| $, where
$J$ and $K$ are the two members of $s_{ Y } $, or in
other words, $s_{ Y
} =\left\{ J,K\right\} $.

Then if $v$ is any member of
$\mathbb{R}^{ M }$
such that $v_{ Y } \geq 0$ holds for every
member $Y$ of $M$, it follows directly from the above that
the following inequality holds for all $z$ such that
$\mathcal{A} \left( \downarrow \left( z,\mathcal{P} \left(
P,A\right) \right) ,\sigma ,F\right) \mathcal{H} \left(
P\cap Z_{ A } ,Q\cap Z_{ A } ,H,\sigma ,R,\downarrow \left(
z,Z_{ A } \right) \right) \neq 0$:
\[
1\leq \sum_{s\in S } \prod_{Y\in M }
\left( \frac{t\left( z,s,Y\right) }{ R } \right)^{
v_{ Y } }
\]

We now choose a real number $\eta $ such that $\eta \geq
\frac{1}{\#(\mathbb{B} ( \bar{ P } )) } $holds,
\label{Start of original page 121}
 and define $v$ to be the member of $\mathbb{R}^{ M } $
 such
that for each member $Y$ of $M$, \\
$v_{ Y } \equiv \max\left(
\left( \eta +\Gamma \left( \beta ,Y\right) -d\left( \#\left(
Y\right) -1\right) \right) ,\eta \right) $ holds, and
multiply the integrand of the above integral for the member
$A$ of $\mathbb{B} \left( \bar{ P } \right) $ by the above
inequality with this choice of the map $v$.

Let $s$ be any member of $T$.   We shall show that the
following integral is finite:
\[
\int \hspace{-0.67pt} \left( \prod_{B\in \left(
\mathcal{P} \left( P,A\right) \,\vdash
\left\{ S_{ A } \right\} \right) } \hspace{-27.0pt}
 d^{ d } z_{ B }
\right) \hspace{-1.21pt} \left\{
\mathcal{A} \left( \downarrow \!
\left( z,\mathcal{P} \left(
P,A\right) \right) ,\sigma ,F\right) \mathcal{H} \left(
P\cap Z_{ A } ,Q\cap Z_{ A } ,H,\sigma ,R,\downarrow \!
 \left(
z,Z_{ A } \right) \right) \rule{0pt}{4.0ex} \times \right.
\]
\[
\hspace{3.0cm} \hspace{-8.2pt} \left. \times \Psi \left(
\downarrow \left( z,\mathcal{P} \left( P,A\right) \right)
,\beta \right) \left( \prod_{Y\in M }
\left( \frac{t\left( z,s,Y\right) }{R } \right)^
{ v_{ Y } } \right) \mathcal{B} \left( \downarrow \left(
z,\mathcal{P}
\left( P,A\right) \right) ,L\right) \right\}.
\]

We first note that the factor $\mathcal{H} \left( P\cap Z_{
A } ,Q\cap Z_{ A } ,H,\sigma ,R,\downarrow \left( z,Z_{ A }
\right) \right) $ satisfies $0\leq \mathcal{H} \left( P\cap
Z_{ A } ,Q\cap Z_{ A } ,H,\sigma ,R,\downarrow \left( z,Z_{
A } \right) \right) \leq 1$ for all $z$, and the other
factors in the integrand are all $\geq 0$ for all $z$,
hence we may drop the factor \\
$\mathcal{H} \left( P\cap Z_{
A } ,Q\cap Z_{ A } ,H,\sigma ,R,\downarrow \left( z,Z_{ A }
\right) \right) $, and we now do that.

We now define $\gamma $ to be the map whose domain is
$\mathcal{Q} \left( \mathcal{P} \left( P,A\right) \right)
$, and such that for each member $\left\{ J,K\right\} $ of
$\mathcal{Q} \left( \mathcal{P} \left( P,A\right) \right)
$, $\gamma_{ \left\{ J,K\right\} } $
is equal to $\beta_{ \left\{
J,K\right\} } $ plus the negative of the total power of
$\left| z_{
J } -z_{ K } \right| $ in the factor $\prod_{Y\in M }
\left( \frac{t\left(
z,s,Y\right)}{R} \right)^{ v_{ Y } } $.

Now the contribution of the factor $\prod_{Y\in M } \left(
\frac{t\left( z,s,Y\right) }{R } \right)^
{ v_{ Y } } $ to $\gamma_{ \left\{
J,K\right\} } $ is $\leq 0$ for all $\left\{ J,K\right\} \in
\mathcal{Q} \left( \mathcal{P} \left( P,A\right) \right) $,
hence $\gamma_{ \left\{ J,K\right\} } \leq
\beta_{ \left\{
J,K\right\} } $ holds for all $\left\{ J,K\right\} \in
\mathcal{Q} \left( \mathcal{P} \left( P,A\right) \right) $.
  Hence by the above observation \ref{Observation 28}), on
page \pageref{Start of original page 116},
 if $B$ is any member of $Q\cap \left( \Xi \left(
\mathcal{P} \left( P,A\right) \right) \,\vdash \mathcal{P}
\left( P,A\right) \right) $, then $\Gamma \left(\gamma ,
\mathcal{P} \left( P,B\right) \right) \leq d \left(
\# \left( \mathcal{P}
\left( P,B\right) \right) - 1 \right) - \frac{ 1 }{
\# \left(\mathbb{B} \left( \bar{ P } \right) \right) } $
 holds.

And by construction, if $Y$ is any member of $M$, then
\[
\Gamma \left( \gamma ,Y\right) \leq \Gamma \left( \beta
,Y\right) -\max\left( \left( \eta +\Gamma \left( \beta
,Y\right) -d\left( \#\left( Y\right) -1\right) \right) ,\eta
\right)
\]
\[
\leq d\left( \#\left( Y\right) -1\right) -\eta
\hspace{4.0cm}
\]
\[
\leq d\left( \#\left( Y\right) -1\right) - \frac{ 1 }{
 \#\left(
\mathbb{B} \left( \bar{ P } \right) \right) }
\hspace{2.0cm}
\]
holds.

Hence if $Y$ is \emph{any} subset of a member $X$ of $F$
such that $\#\left( Y\right) \geq 2$ holds, $\mathcal{U}
\left( Y\right) $ is $\left( V\cup H\right) $-connected,
and $Y$ is \emph{not} a strict subset of any subset $Z$ of
$X$ such that $\mathcal{U} \left( Z\right) $ is $\left(
V\cup H\right) $-connected, then
\[
\Gamma \left( \gamma ,Y\right) \leq d\left( \#\left(
Y\right) -1\right) - \frac{ 1 }{ \#\left( \mathbb{B}
\left( \bar{
P }
\right) \right) }
\]
\label{Start of original page 122}
 holds.

Now if $\left\{ J,K\right\} $ is any member of $\mathcal{Q}
\left( \mathcal{P} \left( P,A\right) \right) $ such that
there is no member $\left\{ j,k\right\} $ of $W$ such that
$j\in J$ holds and $k\in K$ holds, then by
observation \ref{Observation 29}) on
page \pageref{Start of original page 116}, together
with the fact that
$\gamma_{ \left\{ B,C\right\} } \leq \beta_{
 \left\{ B,C\right\} } $
holds for every member $\left\{ B,C\right\} $ of
$\mathcal{Q}
\left( \mathcal{P} \left( P,A\right) \right) $, we find
that $\gamma_{ \left\{ J,K\right\} } \leq 0$ holds.

We define the set $U$ to be the subset of $\mathcal{Q}
\left( \mathcal{P} \left( P,A\right) \right) $ such that if
$\left\{ B,C\right\} $ is a member of $\mathcal{Q} \left(
\mathcal{P} \left( P,A\right) \right) $, then $\left\{
B,C\right\} $ is a member of $U$ ifif there exists a member
$\left\{ i,j\right\} $ of $W$ such that $i\in B$ holds and
$j\in C$ holds.   It then follows that if $X$ is any
nonempty subset of $\mathcal{P} \left( P,A\right) $, then
$\mathcal{U} \left( X\right) $ is $\left( V\cup H\right)
$-connected ifif $X$ is $U$-connected.   For suppose first
that $\mathcal{U} \left( X\right) $ is $\left( V\cup
H\right) $-connected.   Then if $\left\{ J,K\right\} $ is
any
partition of $\mathcal{U} \left( X\right) $, there exists a
member $C$ of $\left( V\cup H\right) $ such that $C\cap J$
and $C\cap K$ are both nonempty.   Let $\left\{ Y,Z\right\}
$
be any partition of $X$.   Then $\left\{ \mathcal{U} \left(
Y\right) ,\mathcal{U} \left( Z\right) \right\} $ is a
partition of $\mathcal{U} \left( X\right) $, hence there
exists a member $C$ of $\left( V\cup H\right) $ such that
$C\cap \mathcal{U} \left( Y\right) $ and $C\cap \mathcal{U}
\left( Z\right) $ are both nonempty.   Now if $E$ is any
member of $V$, then either $E\subseteq \mathcal{U} \left(
Y\right) $ holds or $E\cap \mathcal{U} \left( Y\right)
=\emptyset $ holds, and either $E\subseteq \mathcal{U}
\left(
Z\right) $ holds or $E\cap \mathcal{U} \left( Z\right) =0$
holds, hence $C$ cannot be a member of $V$, hence $C$ is a
member of $H$, and furthermore, since $C$ has at least two
members and is not a subset of any member of $V$, $C$ is a
member of $W$.   And $C\cap \mathcal{U} \left( Y\right)
\neq \emptyset $ implies that there exists a member $J$ of
$Y$
such that $C\cap J\neq \emptyset $, and $C\cap \mathcal{U}
\left( Z\right) \neq \emptyset $ implies that there exists a
member $K$ of $Z$ such that $C\cap K\neq \emptyset $, and
by the
definition of $U$, $\left\{ J,K\right\} $ is a member of
$U$.
  Hence there exists a member $\left\{ J,K\right\} $ of $U$
such that $\left\{ J,K\right\} \cap Y\neq \emptyset $ and
$\left\{
J,K\right\} \cap Z\neq \emptyset $.   Hence $X$ is
$U$-connected.

Now assume that $X$ is $U$-connected.   Then if $\left\{
Y,Z\right\} $ is any partition of $X$, there exists a member
$\left\{ J,K\right\} $ of $U$ such that $\left\{ J,K\right\}
\cap Y\neq \emptyset $ and $\left\{ J,K\right\} \cap Z\neq
\emptyset
$,
hence there exists a member $\left\{ J,K\right\} $ of $U$
such that $J\in Y$ holds and $K\in Z$ holds.   Let $\left\{
E,F\right\} $ be any partition of $\mathcal{U} \left(
X\right) $ into two nonempty parts.   Suppose first that
some member $C$ of $X$ intersects both $E$ and $F$.   Then
$C\subseteq \mathcal{U} \left( X\right) $ holds, hence
$\left\{ \left( C\cap E\right) ,\left( C\cap F\right)
\right\} $ is a partition of $C$ into two nonempty parts
$\left( C\cap E\right) $ and  $\left( C\cap F\right) $,
hence the fact that $C$ is a member of $\mathcal{P} \left(
P,A\right) $, hence that $C$ is $\left( V\cup H\right)
$-connected, implies that there exists a member $G$ of
$\left( V\cup H\right) $ such that $G$ intersects both
$\left( C\cap E\right) $ and $\left( C\cap F\right) $,
hence $G$ intersects both $E$ and $F$.   Now suppose that
no member $C$ of $X$ intersects both $E$ and $F$.   Then if
$B$ is any member of $C$, exactly
\label{Start of original page 123}
 one of $B\subseteq E$ and $B\subseteq F$ holds, and the
fact that $E$ is nonempty implies that there is at least
one member $B$ of $X$ such that $B\subseteq E$ holds, and
the fact that $F$ is nonempty implies that there is at
least one member $B$ of $X$ such that $B\subseteq F$ holds,
hence if we define $Y$ to be the set whose members are all
the members $B$ of $X$ such that $B\subseteq E$ holds and
we define $Z$ to be the set whose members are all the
members $B$ of $X$ such that $B\subseteq F$ holds, then
$\left\{ Y,Z\right\} $ is a partition of $X$ into two
nonempty parts, hence there exists a member $\left\{
J,K\right\} $ of $U$ such that $J\in Y$ holds and $K\in Z$
holds.   Now $\left\{ J,K\right\} \in U$ implies that there
exists a member $C$ of $H$ such that $C\cap J$ and $C\cap
K$ are both nonempty, hence, since $E=\mathcal{U} \left(
Y\right) $ holds and $F=\mathcal{U} \left( Z\right) $
holds, $C\cap E$ and $C\cap F$ are both nonempty.   Hence
$\mathcal{U} \left( X\right) $ is $\left( V\cup H\right)
$-connected.

It follows directly from this that if $X$ is any nonempty
subset of $\mathcal{P} \left( P,A\right) $, and $Y$ is any
nonempty subset of $X$ such that $\mathcal{U} \left(
Y\right) $ is $\left( V\cup H\right) $-connected and $Y$ is
\emph{not} a strict subset of any subset $Z$ of $X$ such
that $\mathcal{U} \left( Z\right) $ is $\left( V\cup
H\right) $-connected, then $Y$ is a $U$-connected component
of $X$, hence by Lemma \ref{Lemma 2}, if $F$ is the set
whose members
are all the nonempty subsets $Y$ of $X$ such that
$\mathcal{U} \left( Y\right) $ is $\left( V\cup H\right)
$-connected, and $Y$ is \emph{not} a strict subset of any
subset $Z$ of $X$ such that $\mathcal{U} \left( Z\right) $
is $\left( V\cup H\right) $-connected, then $F$ is a
partition of $X$.

Now let $X$ be any member of $F$ such that $\#\left(
F\right) \geq 2$ holds.   Then it follows immediately from
the above that $\Gamma \left( \gamma ,X\right) $ is equal
to the sum of $\Gamma \left( \gamma ,Y\right) $ over all
the nonempty subsets $Y$ of $X$ such that $\mathcal{U}
\left( Y\right) $ is $\left( V\cup H\right) $-connected and
$Y$ is \emph{not} a strict subset of any subset $Z$ of $X$
such that $\mathcal{U} \left( Z\right) $ is $\left( V\cup
H\right) $-connected, plus the sum of $\gamma_{ \left\{
J,K\right\} } $ over members $\left\{ J,K\right\} $ of
$\mathcal{Q} \left( \mathcal{P} \left( P,A\right) \right) $
such that $J$ and $K$ are members of two \emph{distinct}
such subsets $Y$ of $X$.   Now if $J$ is a member of such a
subset $Y_{ 1 } $ of $X$ and $K$ is a member of a different
such subset $Y_{ 2 } $ of $X$, then there can be no member
$\left\{ j,k\right\} $ of $W$ such that $j\in J$ holds and
$k\in K$ holds, for $W\subseteq H$ holds hence if there was
such a member of $W$ then, by Lemma \ref{Lemma 1} and
page \pageref{Start of original page 36},
$\left( Y_{ 1 } \cup Y_{ 2 } \right) $ would be a $\left(
V\cup H\right) $-connected subset of $X$ such that $Y_{ 1 }
\subset \left( Y_{ 1 } \cup Y_{ 2 } \right) $ holds and
$Y_{ 2 } \subset \left( Y_{ 1 } \cup Y_{ 2 } \right) $
holds, which contradicts the assumed properties of $Y_{ 1 }
$ and $Y_{ 2 } $.   Hence $\gamma_{ \left\{ J,K\right\} }
\leq 0$ holds.

Thus if $X$ is any member of $F$ such that $\#\left(
X\right) \geq 2$ holds, then
\[
\Gamma \left( \gamma ,X\right) \leq d\left( \#\left(
X\right) -n\right) - \frac{ n }{ \#\left(
\mathbb{B} \left( \bar{
P }
\right) \right) }
\]
holds, where $n$ is the number of $U$-connected components
of
$X$.   But
\label{Start of original page 124}
 $n\geq 1$ holds, hence
\[
\Gamma \left( \gamma ,X\right) \leq d\left( \#\left(
X\right) -1\right) - \frac{ 1 }{ \#\left(
\mathbb{B} \left( \bar{
P }
\right) \right) }
\]
holds.

Hence by the Cluster Convergence Theorem, the integral
\[
\int \!
 \left( \prod_{B\in \left( \mathcal{P} \left( P,A\right)
\,\vdash
\left\{ S_{ A } \right\} \right) } \hspace{-25.0pt}
 d^{ d } z_{ B } \right)
\mathcal{A} \left( \downarrow \! \! \left( z,\mathcal{P}
\left(
P,A\right) \right) ,\sigma ,F\right) \Psi \left( \downarrow
\! \! \left( z,\mathcal{P} \left( P,A\right) \right) ,\gamma
\right) \mathcal{B} \left( \downarrow \! \! \left(
z,\mathcal{P}
\left( P,A\right) \right) ,L\right) ,
\]
or in other words, the integral
\[
\int \left( \prod_{B\in \left( \mathcal{P} \left( P,A\right)
\,\vdash
\left\{ S_{ A } \right\} \right) } d^{ d } t_{ B } \right)
\mathcal{A} \left( t,\sigma ,F\right) \Psi \left( t,\gamma
\right) \mathcal{B} \left( t,L\right) ,
\]
is finite.

And this concludes the proof of Theorem \ref{Theorem 1}.

\section{Second Convergence Theorem.}
\label{Section 7}

If $A$ is a set, and $U$ is a set such that every member of
$U$ is a set, then a $U $\emph{-key of }$A $ is a member
$E$ of $U$
such that there exists a partition $\left\{ B,C\right\} $ of
$A$ into two nonempty parts $B$ and $C$ such that $B\cap
E\neq \emptyset $, $C\cap E\neq \emptyset $, and $E$ is the
\emph{only} member of $U$ to have nonempty intersection
with both $B$ and $C$.

If $A$ is a set, $U$ is a set such that every member of $U$
is a set, and $V$ is a subset of $U$, then we shall say
that $A$ is $U $\emph{-firm over }$V $ ifif $A$ is
$U$-connected and
every $U$-key of $A$ is a member of $V$.

If $A$ is a set, $U$ is a set such that every member of $U$
is a set, and $V$ is a subset of $U$, then a
$U $\emph{-firm over }$V $\emph{ component of }$A $
is a nonempty subset $B$ of
$A$ such that $B$ is $U$-firm over $V$ and $B$ is \emph{not}
a strict subset of any subset of $A$ that is $U$-firm over
$V$.

If $H$ is a set such that every member of $H$ is a set, and
$F$ is a wood, then we shall say that $F$ is
$H $\emph{-principal} ifif every member $A$ of $\mathbb{B}
\left( F\right) $ is $\left( \mathcal{M} \left( F\right)
\cup H\right) $-connected, every member $A$ of $\mathbb{B}
\left( F\right) $ has at least one $\left( \mathcal{M}
\left( F\right) \cup H\right) $-key $E$ such that $E\notin
\mathcal{M} \left( F\right) $ holds, and if $A$ is any
member of $\mathbb{B} \left( F\right) $ and $B$ is any
member of $F$ such that $B\subset A$ holds, then $B$ is a
subset of an $\left( \mathcal{M} \left( F\right) \cup
H\right) $-firm over $\mathcal{M} \left( F\right)
$ component of $A $.

For every ordered pair $\left(F,H\right) $ of a
 wood $F$ and a set $H$
such that every member of $H$ is a set, we define
$\mathbb{P} \left( F,H\right) $ to be the set whose members
are the members of $\mathcal{M} \left( F\right) $, together
with any members $A$ of
\label{Start of original page 125}
 $\mathbb{B} \left( F\right) $ that satisfy both the
following requirements:

\vspace{1.0ex}

\noindent (i)  $ A $ has at least one $\left(
\mathcal{M} \left( F\right)
\cup H\right) $-key $E$ such that $E\notin \mathcal{M}
\left( F\right) $ holds

\vspace{1.0ex}

\noindent (ii)  if $B$ is any member of $F$ such
that $B\subset A$
holds, then $B$ is a subset of an $\left( \mathcal{M}
\left( F\right) \cup H\right) $-firm over $\mathcal{M}
\left( F\right) $ component of $A$.

\vspace{1.0ex}

We note that it follows immediately from this definition
that if $V$ is any partition such that $\mathcal{U} \left(
V\right) $ is finite and $\#\left( V\right) \geq 2$ holds,
$H$ is any set such that every member of $H$ is a set, and
$F$ is any member of $\mathcal{G} \left( V,H\right) $, (or
in other words, $F$ is any wood of $V$ such that every
member of $F$ is $\left( V\cup H\right)
$-connected), then $\mathbb{P} \left( F,H\right) $
is an $H$-principal wood
of $V$.

For any ordered pair $\left(V,H\right) $ of
 a partition $V$ such that
$\mathcal{U} \left( V\right) $ is finite and $\#\left(
V\right) \geq 2$ holds, and a set $H$ such that every
member of $H$ is a set, we define $\mathcal{V} \left(
V,H\right) $ to be the set whose members are all the woods
$F$ of $V$ such that every member $A$ of $\mathbb{B} \left(
F\right) $ is $\left( V\cup H\right) $-firm over $V$, and
we define $\mathcal{W} \left( V,H\right) $ to be the set
whose members are all the $H $-principal woods of $V$.

For any ordered pair $\left(G,H\right) $ such that $H$ is a
set such that
every member of $H$ is a set, and $G$ is an $H $-principal
wood, we define $\mathbb{O} \left( G,H\right) $ to be the
set whose members are all the members $F$ of $\mathcal{G}
\left( \mathcal{M} \left( G\right) ,H\right) $ such that
$G\subseteq \mathbb{P} \left( F,H\right) $ holds.   Thus
$\mathbb{O} \left( G,H\right) $ is the set whose members
are all the members $F$ of $\mathcal{G} \left( \mathcal{M}
\left( G\right) ,H\right) $ such that $G\subseteq F$ holds,
and if $A$ is any member of $\mathbb{B} \left( G\right) $
and $B$ is any member of $F$ such that $B\subset A$ holds,
then $B$ is a subset of an $\left( \mathcal{M} \left(
F\right) \cup H\right) $-firm over $\mathcal{M} \left(
F\right) $ component of $A$.

\begin{bphzlemma} \label{Lemma 25}
\end{bphzlemma}
\vspace{-6.143ex}

\noindent \hspace{11.9ex}{\bf.  }If $U$ is a set
such that every member of $U$ is
a set, $A$ is a nonempty $U$-connected set, $E$ is a
$U$-key of
$A$, and $\left\{ B,C\right\} $ is a partition of $A$ into
two nonempty parts $B$ and $C$ such that $B\cap E\neq
\emptyset
$, $C\cap E\neq \emptyset $, and $E$ is the \emph{only}
member
of $U$ to have nonempty intersection with both $B$ and $C$,
then $B$ is $U$-connected.

\vspace{2.5ex}

\noindent {\bf Proof.}  Suppose $B$ is not
$U$-connected.   Then there
exists
a partition $\left\{ J,K\right\} $ of $B$ into two nonempty
parts $J$ and $K$ such that \emph{no} member $S$ of $U$
intersects both $J$ and $K$.   Let $\left\{ J,K\right\} $ be
such a partition of $B$.   Now $E$ is a member of $U$ hence
$E$ cannot intersect both $J$ and $K$.   Suppose for
definiteness that $J\cap E=\emptyset $.   Then $\left\{
J,\left(
C\cup K\right) \right\} $ is a partition of $A$ into two
nonempty parts $J$ and $\left( C\cup K\right) $ such that
no member $S$ of $U$ intersects both $J$ and $\left( C\cup
K\right) $.   For if $S$ is a member of $U$ such that $S$
intersects both $J$ and $\left( C\cup K\right) $,
\label{Start of original page 126}
 then $S\neq E$ since $J\cap E=\emptyset $.   Hence $S$ is a
member of $U$, different from $E$, such that $J\cap S\neq
\emptyset $, hence $B\cap S\neq \emptyset $, and $\left(
C\cup
K\right) \cap S\neq \emptyset $, hence $S$ intersects at
least
one of $C$ and $K$.   But $S\neq E$ implies $C\cap
S=\emptyset
$, since $B\cap S\neq \emptyset $, and $J\cap S\neq
\emptyset $
implies $K\cap S=\emptyset $, since by assumption \emph{no}
member of $U$ intersects both $J$ and $K$.

Hence, as stated, $\left\{ J,\left( C\cup K\right) \right\}
$
is a partition of $A$ into two nonempty parts such that
\emph{no} member of $U$ intersects both parts.   But this
contradicts the assumption that $A$ is $U$-connected.

\begin{bphzlemma} \label{Lemma 26}
\end{bphzlemma}
\vspace{-6.143ex}

\noindent \hspace{11.9ex}{\bf.  }If $U$ is a set
such that every member of $U$ is
a set, $A$ is a $U$-connected set, $S$ and $T$ are
\emph{distinct} $U$-keys of $A$, $\left\{ B,C\right\} $ is a
partition of $A$ such that $B\cap S\neq \emptyset $, $C\cap
S\neq \emptyset $, and $S$ is the \emph{only} member of $U$
to
have nonempty intersection with both $B$ and $C$, and
$B\cap T=\emptyset $ holds, and $\left\{ D,E\right\} $ is a
partition of $A$ such that $D\cap T\neq \emptyset $, $E\cap
T\neq \emptyset $, and $T$ is the \emph{only} member of $U$
to
have nonempty intersection with both $D$ and $E$, and
$E\cap S=\emptyset $ holds, then $B\cap E=\emptyset $ holds.

\vspace{2.5ex}

\noindent {\bf Proof.}  Suppose that $B\cap
E\neq \emptyset $ holds.   Now
$B\cap D$ is nonempty since $B=\left( B\cap D\right) \cup
\left( B\cap E\right) $, $B\cap S\neq \emptyset $, and
$E\cap
S=\emptyset $ all hold, hence $\left\{ \left( B\cap D\right)
,\left( B\cap E\right) \right\} $ is a partition of $B$,
hence since $B$ is $U$-connected by Lemma \ref{Lemma 25},
there is a
member $W$ of $U$ such that $\left( B\cap D\right) \cap
W\neq \emptyset $ and $\left( B\cap E\right) \cap W\neq
\emptyset $
both hold, and $W\neq T$ holds since $B\cap T=\emptyset $
holds
by assumption.   But this implies that $W$ is a member of
$U$, different from $T$, such that $D\cap W\neq \emptyset $
and
$E\cap W\neq \emptyset $ both hold, and by assumption there
is
no such member of $U$.

\begin{bphzlemma} \label{Lemma 27}
\end{bphzlemma}
\vspace{-6.143ex}

\noindent \hspace{11.9ex}{\bf.  }Let $U$ be a set
such that every member of $U$ is
a set, let $A$ be a nonempty $U$-connected set, and let $i$
and $j$ be any members of $A$ such that $i\neq j$.   Then
there exists a finite integer $n\geq 0$ and a map $M$ such
that $\mathcal{D} \left( M\right) $ is the set of all the
integers $r$ such that $0\leq r\leq n$ holds, and for each
member $r$ of $\mathcal{D} \left( M\right) $, $M_{ r } $ is
a member of $U$, and for all $r$ such that $0\leq r$ and
$r\leq \left( n-1\right) $ both hold, $A\cap \left( M_{ r }
\cap M_{ r+1 } \right) $ is nonempty, and $i\in M_{ 0 } $
holds and $j\in M_{ n } $ holds.

\vspace{2.5ex}

\noindent {\bf Proof.}  Let $B$ be the subset
of $A$ whose members are all
the members $k$ of $A$ such that there \emph{exists} a
finite integer $n\geq 0$ and a map $M$ such that
$\mathcal{D} \left( M\right) $ is the set of all the
integers $r$ such that $0\leq r\leq n$ holds, and for each
member $r$ of $\mathcal{D} \left( M\right) $, $M_{ r } $ is
a member of $U$, and for all $r$ such that $0\leq r$ and
$r\leq \left( n-1\right) $ both hold, $A\cap \left( M_{ r }
\cap M_{ r+1 } \right) $ is
\label{Start of original page 127}
 nonempty, and $i\in M_{ 0 } $ holds and $k\in M_{ n } $
holds.   Then $i$ is a member of $B$, since $A$ has at
least two members, hence there exists a member $E$ of $U$
such that $i\in E$ holds.   Hence $B$ is nonempty.   Now
\emph{suppose} $A\,\vdash B$ was nonempty.   Then $\left\{
B,\left( A\,\vdash B\right) \right\} $ would be a partition
of $A$ into two nonempty parts, hence since $A$ is
$U$-connected, there would exist a member $E$ of $U$ such
that $E\cap B$ and $E\cap \left( A\,\vdash B\right) $ were
both nonempty.   Let $E$ be a member of $U$ such that
$E\cap B$ and $E\cap \left( A\,\vdash B\right) $ are both
nonempty, and let $k$ be any member of $E\cap B$ and let
$l$ be any member of $E\cap \left( A\,\vdash B\right) $.
Now by the definition of $B$, there exists a finite integer
$n\geq 0$ and a map $M$ such that $\mathcal{D} \left(
M\right) $ is the set of all the integers $r$ such that
$0\leq r\leq n$ holds, and for each member $r$ of
$\mathcal{D} \left( M\right) $, $M_{ r } $ is a member of
$U$, and for all $r$ such that $0\leq r$ and $r\leq \left(
n-1\right) $ both hold, $A\left( M_{ r } \cap M_{ r+1 }
\right) $ is nonempty, and $i\in M_{ 0 } $ holds and $k\in
M_{ n } $ holds.   Let $n$ be such a finite integer, let
$M$ be such a map, and let $N\equiv M\cup \left\{ \left(
n+1,E\right) \right\} $.   Then $N$ is a map such that
$\mathcal{D} \left( N\right) $ is the set of all the
integers $r$ such that $0\leq r\leq \left( n+1\right) $
holds, and for every member $r$ of $\mathcal{D} \left(
N\right) $, $N_{ r } $ is a member of $U$, and for all $r$
such that $0\leq r$ and $r\leq n$ both hold, $A\cap \left(
N_{ r } \cap N_{ r+1 } \right) $ is nonempty, (for $N_{ n }
\cap N_{ n+1 } $ has $k$ as a member), and $i\in N_{ 0 } $
holds and $l\in N_{ n+1 } $ holds.   Hence $l\in B$ holds
by the definition of $B$, and this contradicts the
assumption that $l\in \left( A\,\vdash B\right) $ holds.
Hence \emph{no} member of $U$ intersects both $B$ and
$\left( A\,\vdash B\right) $, hence the assumption that
$\left( A\,\vdash B\right) $ is nonempty contradicts the
assumption that $A$ is $U$-connected.

\begin{bphzlemma} \label{Lemma 28}
\end{bphzlemma}
\vspace{-6.143ex}

\noindent \hspace{11.9ex}{\bf.  }Let $U$ be a set
such that every member of $U$ is
a set, and let $A$ be a nonempty $U$-connected set.

For any ordered pair $\left(i,X\right) $ of a member $i$ of
$A$ and a
subset $X$ of $U$, let $\uparrow \left( i,X\right) $ denote
the set whose members are all the members $j$ of $A$ such
that if $E$ is any member of $X$, and $\left\{ B,C\right\} $
is any partition of $A$ into two nonempty parts such that
$B\cap E\neq \emptyset $ holds, $C\cap E\neq \emptyset $
holds, and
$E$ is the \emph{only} member of $U$ to intersect both $B$
and $C$, then $j$ is a member of the \emph{same} member of
$\left\{ B,C\right\} $ as $i$, or in other words,
$\mathcal{C} \left( \left\{ B,C\right\} ,j\right)
=\mathcal{C} \left( \left\{ B,C\right\} ,i\right) $ holds.

Let $i$ be any member of $A$.

We define a binary relation, written $\to $, among the
$U$-keys of $A$ as follows:

If $S$ and $T$ are $U$-keys of $A$, then $S\to T$ holds ifif
there \emph{exists} a partition $\left\{ B,C\right\} $ of
$A$
into two nonempty parts $B$ and $C$ such that $T$
\label{Start of original page 128}
 intersects both $B$ and $C$, $T$ is the \emph{only} member
of $U$ to intersect both $B$ and $C$, and $S$ does
\emph{not} intersect the member of $\left\{ B,C\right\} $ of
which $i$ is a member.

We note that this definition has the immediate consequence
that $S\to S$ does \emph{not} hold for any $U$-key $S$ of
$A$, and that the definition of the relation $\to $ depends
implicitly on the member $i$ of $A$.
\enlargethispage{0.5ex}

Then the following results hold:

\vspace{2.5ex}

\noindent {\bf (i)}  Let $S$ and $T$ be any $U$-keys of $A$
such that $S\to T$
holds.   Then $T\to S$ does \emph{not} hold.   For $S\to T$
implies that there exists a partition $\left\{ B,C\right\} $
of $A$ into two nonempty parts such that $T$ intersects
both $B$ and $C$, $T$ is the \emph{only} member of $U$ to
intersect both $B$ and $C$, $i\in C$ holds, and $S$ does
\emph{not} intersect $C$.   Let $\left\{ B,C\right\} $ be
such a partition of $A$, and let $\left\{ J,K\right\} $ be
any partition of $A$ into two parts such that $S$
intersects both $J$ and $K$, $S$ is the \emph{only} member
of $U$ to intersect both $J$ and $K$, and $i\in K$ holds.
Now $S$ is not equal to $T$, for $T$ intersects both $B$
and $C$, but $S$ does \emph{not} intersect $C$.   Hence by
the definition of $\left\{ J,K\right\} $, $T$ does not
intersect both $J$ and $K$.   Now if $T\cap K$ were empty,
then by Lemma \ref{Lemma 26}, $C\cap K$ would be empty.
But $C\cap
K$ has the member $i$, hence $T\cap J$ must be empty.   But
$T$ is a $U$-key of $A$, hence $T\cap A$ is nonempty, hence
$T$ certainly intersects at least one of $J$ and $K$, hence
$T$ intersects $K$, hence $T$ intersects the member of
$\left\{ J,K\right\} $ of which $i$ is a member.   And this
is true for \emph{every} partition $\left\{ J,K\right\} $ of
$A$ into two parts such that $S$ intersects both parts and
$S$ is the \emph{only} member of $U$ to intersect both
parts.   Hence $T\to S$ does \emph{not} hold.

\vspace{2.5ex}

\noindent {\bf (ii)}  Let $R$, $S$ and $T$ be any
$U$-keys of $A$ such that
$R\to S$ holds and $S\to T$ holds.   Then $R\to T$ holds.
For $R\to S$ implies there exists a partition $\left\{
B,C\right\} $ of $A$ into two nonempty parts such that $i\in
C$ holds, $S$ intersects both $B$ and $C$, $S$ is the
\emph{only} member of $U$ to intersect both $B$ and $C$,
and $R$ does \emph{not} intersect $C$.   Let $\left\{
B,C\right\} $ be such a partition of $A$.   And $S\to T$
implies there exists a partition $\left\{ J,K\right\} $ of
$A$ into two nonempty parts such that $i\in K$ holds, $T$
intersects both $J$ and $K$, $T$ is the \emph{only} member
of $U$ to intersect both $J$ and $K$, and $S$ does
\emph{not} intersect $K$.   Let $\left\{ J,K\right\} $ be
such a partition of $A$.   Now $T$ intersects $K$ and $S$
does \emph{not} intersect $K$, hence $T$ is not equal to
$S$, hence since $S$ is the \emph{only} member of $U$ to
intersect both $B$ and $C$, $T$ does \emph{not} intersect
both $B$ and $C$.   Now if $T\cap C$ was empty, then by
Lemma \ref{Lemma 26}, $C\cap K$ would be empty, but $C\cap
K$ has the
member $i$, hence
\label{Start of original page 129}
 $T\cap C$ cannot be empty, hence $T\cap B$ is empty, hence
by Lemma \ref{Lemma 26}, $B\cap K$ is empty.   Now every
member of $B$
is either a member of $J$ or a member of $K$, and $B\cap K$
is empty hence no member of $B$ is a member of $K$, hence
every member of $B$ is a member of $J$, hence $B\subseteq
J$ holds.   Now $R$ is a $U$-key of $A$ such that $R$ does
not intersect $C$, hence $R$ \emph{does} intersect $B$, and
$B\subseteq J$ holds, hence $R$ intersects $J$.   And $R\to
S$ and $S\to T$ both hold, hence by (i) above, $R$ is
\emph{not} equal to $T$, hence $R$ does \emph{not}
intersect both $J$ and $K$, hence since $R$ intersects $J$,
$R$ does \emph{not} intersect $K$.   Hence $\left\{
J,K\right\} $ is a partition of $A$ into two nonempty parts
such that $i\in K$ holds, $T$ intersects both $J$ and $K$,
$T$ is the \emph{only} member of $U$ to intersect both $J$
and $K$, and $R$ does \emph{not} intersect $K$.   Hence
$R\to T$ holds.

\vspace{2.5ex}

\noindent {\bf (iii)}  Let $k$ be any member of $A$
such that $k\neq i$
holds, let $S$ be any $U$-key of $A$ such that there exists
a
partition $\left\{ B,C\right\} $ of $A$ into two parts such
that $k\in B$ holds, $i\in C$ holds, $S$ intersects both
$B$ and $C$, and $S$ is the \emph{only} member of $U$ to
intersect both $B$ and $C$, and let $T$ be any $U$-key of
$A$
such that there exists a partition $\left\{ J,K\right\} $ of
$A$ into two parts such that $k\in J$ holds, $i\in K$
holds, $T$ intersects both $J$ and $K$, and $T$ is the
\emph{only} member of $U$ to intersect both $J$ and $K$.
Then either $S=T$ holds or $S\to T$ holds or $T\to S$
holds.   For suppose $S\neq T$ holds, and let $\left\{
B,C\right\} $ and $\left\{ J,K\right\} $ be partitions of
$A$
with the properties just specified.   Then one of $B\cap T$
and $C\cap T$ is empty, and one of $J\cap S$ and $K\cap S$
is empty.   Now $C\cap K$ has the member $i$ so is
nonempty, hence by Lemma \ref{Lemma 26}, $C\cap T$ and
$K\cap S$
cannot both be empty, and $B\cap J$ has the member $k$ so
is nonempty, hence by Lemma \ref{Lemma 26}, $B\cap T$ and
$J\cap S$
cannot both be empty, hence either $B\cap T$ and $K\cap S$
are both empty, or $C\cap T$ and $J\cap S$ are both empty.
 And if $K\cap S$ is empty then $S\to T$ holds, while if
$C\cap T$ is empty, then $T\to S$ holds.

\vspace{2.5ex}

\noindent {\bf (iv)}  Let $k$ be any member of
$A$ such that $k\neq i$ holds,
let $S$ be any $U$-key of $A$ such that there exists a
partition $\left\{ B,C\right\} $ of $A$ into two parts such
that $k\in B$ holds, $i\in C$ holds, $S$ intersects both
$B$ and $C$, and $S$ is the \emph{only} member of $U$ to
intersect both $B$ and $C$, and let $T$ be any $U$-key of
$A$
such that $S\to T$ holds.   Then there exists a partition
$\left\{ J,K\right\} $ of $A$ such that $k\in J$ holds,
$i\in
K$ holds, $T$ intersects both $J$ and $K$, and $T$ is the
\emph{only} member of $U$ to intersect both $J$ and $K$.
For let $\left\{ B,C\right\} $ be a partition of $A$ with
the
properties just specified.   Now $S\to T$ implies that
there exists a partition $\left\{ J,K\right\} $ of $A$ into
two parts such that $i\in K$ holds, $T$ intersects both $J$
and $K$,
\label{Start of original page 130}
 $T$ is the \emph{only} member of $U$ to intersect both $J$
and $K$, and $S$ does \emph{not} intersect $K$.   Let
$\left\{ J,K\right\} $ be such a partition of $A$.   Now $T$
intersects $K$ and $S$ does not intersect $K$, hence $T$ is
not equal to $S$, hence one of $B\cap T$ and $C\cap T$ is
empty.   And $C\cap K$ has the member $i$ hence is
nonempty, hence by Lemma \ref{Lemma 26}, $C\cap T$ cannot
be empty,
hence $B\cap T$ is empty, hence by Lemma \ref{Lemma 26},
$B\cap K$ is
empty, hence since $k$ is a member of $B$, $k$ is
\emph{not} a member of $K$, hence $k$ is a member of $J$.

\vspace{2.5ex}

\noindent {\bf (v)}  Let $k$ be
any member of $A$ such that $k\neq i$ holds,
let $S$ be any $U$-key of $A$ such that there exists a
partition $\left\{ B,C\right\} $ of $A$ into two parts such
that $k\in B$ holds, $i\in C$ holds, $S$ intersects both
$B$ and $C$, and $S$ is the \emph{only} member of $U$ to
intersect both $B$ and $C$, let $n$ be any finite integer
$\geq 0$, and let $M$ be any map such that $\mathcal{D}
\left( M\right) $ is the set of all the integers $r$ such
that $0\leq r\leq n$ holds, and for every member $r$ of
$\mathcal{D} \left( M\right) $, $M_{ r } $ is a member of
$U$, and for all $r$ such that $0\leq r$ and $r\leq \left(
n-1\right) $ both hold, $A\cap \left( M_{ r } \cap M_{ r+1
} \right) $ is nonempty, and $i\in M_{ 0 } $ holds and
$k\in M_{ n } $ holds.   Then $S\in \mathcal{R} \left(
M\right) $ holds.   For $S$ is the \emph{only} member of
$U$ to intersect both $B$ and $C$, hence if $S\in
\mathcal{R} \left( M\right) $ does \emph{not} hold, then
\emph{no} member of $\mathcal{R} \left( M\right) $
intersects both $B$ and $C$.   Suppose now that $S$ is
\emph{not} a member of $\mathcal{R} \left( M\right) $, so
\emph{no} member of $\mathcal{R} \left( M\right) $
intersects both $B$ and $C$.   Then for any $r$ such that
$0\leq r$ and $r\leq \left( n-1\right) $ both hold,
$A\left( M_{ r } \cap M_{ r+1 } \right) $ is nonempty,
hence if $M_{ r } $ intersects $C$ but not $B$, then $M_{
r+1 } $ also intersects $C$ but not $B$.   But $M_{ 0 } $
intersects $C$ but not $B$, hence by induction, $M_{ n } $
intersects $C$ but not $B$, and this contradicts the fact
that $k$ is a member of $B$.

\vspace{2.5ex}

\noindent {\bf (vi)}  Let $X$ be any subset
of $U$ such that every member of
$X$ is a $U$-key of $A$, and let $k$ be any member of $A$
such that there \emph{exists} a member $S$ of $X$ and a
partition $\left\{ B,C\right\} $ of $A$ into two parts such
that $k\in B$ holds, $i\in C$ holds, $S$ intersects both
$B$ and $C$, and $S$ is the \emph{only} member of $U$ to
intersect both $B$ and $C$.   Then there exists a unique
member $T$ of $X$ with the properties that:

\vspace{1.0ex}

\noindent (a)  there exists a partition
$\left\{ J,K\right\} $ of $A$
into two parts such that $k\in J$ holds, $i\in K$ holds,
$T$ intersects both $J$ and $K$, and $T$ is the \emph{only}
member of $U$ to intersect both $J$ and $K$, and

\vspace{1.0ex}

\noindent (b)  there is \emph{no} member $R$ of
$X$ such that $T\to R$
holds.

\vspace{2.5ex}

\noindent {\bf Proof.}  Let $Y$ be the set
whose members are all the
members $S$ of $X$ such that there exists a partition
$\left\{ B,C\right\} $ of $A$ into two parts such that $k\in
B$ holds, $i\in C$ holds, $S$ intersects both $B$ and $C$,
and $S$ is the \emph{only} member of $U$ to intersect both
$B$ and $C$.    Now by assumption
\label{Start of original page 131}
 there \emph{exists} a member $S$ of $X$ and a partition
$\left\{ B,C\right\} $ of $A$ such that $k\in B$ holds,
$i\in
C$ holds, $S$ intersects both $B$ and $C$, and $S$ is the
\emph{only} member of $U$ to intersect both $B$ and $C$.
Hence $Y$ is nonempty.

Now let $S$ be any member of $Y$ and let $R$ be any member
of $X$ such that $S\to R$ holds.   Then by (iv) above,
there exists a partition $\left\{ J,K\right\} $ of $A$ into
two parts such that $k\in J$ holds, $i\in K$ holds, $R$
intersects both $J$ and $K$, and $R$ is the \emph{only}
member of $U$ to intersect both $J$ and $K$.   Hence $R$ is
a member of $Y$.

Now by Lemma \ref{Lemma 27}, there exists a finite integer
$n\geq 0$
and a map $M$ such that $\mathcal{D} \left( M\right) $ is
the set of all the integers $r$ such that $0\leq r\leq n$
holds, and for every member $r$ of $\mathcal{D} \left(
M\right) $, $M_{ r } $ is a member of $U$, and for all $r$
such that $0\leq r$ and $r\leq \left( n-1\right) $ both
hold, $A \cap \left( M_{ r } \cap M_{ r+1 } \right) $ is
nonempty, and $i\in M_{ 0 } $ holds and $k\in M_{ n } $
holds.   Let $n$ be such a finite integer and let $M$ be
such a map.   Then by (v) above, every member of $Y$ is a
member of $\mathcal{R} \left( M\right) $.   Hence $Y$ is a
finite set.

Now by (i), (ii), and (iii) above, the set $Y$ is totally
ordered by the relation $\to $, hence since $Y$ is a finite
set, there exists a unique member $T$ of $Y$ such that
there is no member $R$ of $Y$ such that $T\to R$ holds.
Let $T$ be this unique member of $Y$.

Now by the definition of $Y$, $T$ is a member of $X$ such
that there exists a partition $\left\{ J,K\right\} $ of $A$
into two parts such that $k\in J$ holds, $i\in K$ holds,
$T$ intersects both $J$ and $K$, and $T$ is the \emph{only}
member of $U$ to intersect both $J$ and $K$.   And as noted
above, if $R$ is any member of $X$ such that $T\to R$
holds, then $R$ is a member of $Y$, hence since there is no
member $R$ of $Y$ such that $T\to R$ holds, there is no
member $R$ of $X$ such that $T\to R$ holds.

And finally, if $S$ is any member of $X$ such that there
exists a partition $\left\{ J,K\right\} $ of $A$ into two
parts such that $k\in J$ holds, $i\in K$ holds, $S$
intersects both $J$ and $K$, and $S$ is the \emph{only}
member of $U$ to intersect both $J$ and $K$, and there is
\emph{no} member $R$ of $X$ such that $S\to R$ holds, then
$S$ is a member of $Y$, hence $S$ is equal to $T$.

\vspace{2.5ex}

\noindent {\bf (vii)}  Let $X$ be any subset of
$U$ such that every member of
$X$ is a $U$-key of $A$, and let $Z$ be the subset of $X$
whose members are all the members $T$ of $X$ such that
there is \emph{no} member $R$ of $X$ such that $T\to R$
holds.   Then $\uparrow \left( i,X\right) =\uparrow \left(
i,Z\right) $ holds.   For $\uparrow \left( i,X\right)
\subseteq \uparrow \left( i,Z\right) $ certainly holds.
Now let $k$ be any member of $\left( A\,\vdash \uparrow
\left( i,X\right) \right) $.   Then there exists a member
$S$ of $X$ and a partition $\left\{ B,C\right\} $ of $A$
such
that $k\in B$ holds, $i\in C$ holds, $S$ intersects both
$B$ and $C$, and $S$ is the \emph{only}
\label{Start of original page 132}
 member of $U$ to intersect both $B$ and $C$.   Then by
(vi) above there exists a member $T$ of $X$ such that there
exists a partition $\left\{ J,K\right\} $ of $A$ into two
parts such that $k\in J$ holds, $i\in K$ holds, $T$
intersects both $J$ and $K$, and $T$ is the \emph{only}
member of $U$ to intersect both $J$ and $K$, and such that
there is \emph{no} member $R$ of $X$ such that $T\to R$
holds.   And this $T$ is a member of $Z$, hence $k$ is
\emph{not} a member of $\uparrow \left( i,Z\right) $.

\vspace{2.5ex}

\noindent {\bf (viii)}  Let $E$ be any $U$-key
of $A$.   Then $\left\{
\uparrow
\left( i,\left\{ E\right\} \right) ,\left( A\,\vdash
\uparrow
\left( i,\left\{ E\right\} \right) \right) \right\} $ is a
partition of $A$ into two nonempty parts such that $E$
intersects both parts, and $E$ is the \emph{only} member of
$U$ to intersect both parts.   For by definition, $\uparrow
\left( i,\left\{ E\right\} \right) $ is the set whose
members
are all the members $j$ of $A$ such that for every
partition $\left\{ B,C\right\} $ of $A$ into two parts such
that $E$ intersects both $B$ and $C$, and $E$ is the
\emph{only} member of $U$ to intersect both $B$ and $C$,
$j$ is a member of the \emph{same} member of $\left\{
B,C\right\} $ as $i$ is.   Hence $i$ is a member of
$\uparrow \left( i,\left\{ E\right\} \right) $, hence
$\uparrow \left( i,\left\{ E\right\} \right) $ is nonempty.
And since $E$ is a $U$-key of $A$, there \emph{exists} a
partition $\left\{ B,C\right\} $ of $A$ into two nonempty
parts such that $E$ intersects both $B$ and $C$, and $E$ is
the \emph{only} member of $U$ to intersect both $B$ and
$C$, hence since every member of the member of $\left\{
B,C\right\} $ of which $i$ is \emph{not} a member, is a
member of $\left( A\,\vdash \uparrow \left( i,\left\{
E\right\} \right) \right) $, $\left( A\,\vdash \uparrow
\left( i,\left\{ E\right\} \right) \right) $ is nonempty.
Hence $\left\{ \uparrow \left( i,\left\{ E\right\} \right)
,\left( A\,\vdash \uparrow \left( i,\left\{ E\right\}
\right)
\right) \right\} $ is a partition of $A$ into two nonempty
parts hence, since $A$ is $U$-connected, there \emph{exists}
a member $R$ of $U$ such that $R$ intersects both $\uparrow
\left( i,\left\{ E\right\} \right) $ and $\left( A\,\vdash
\uparrow \left( i,\left\{ E\right\} \right) \right) $.   Let
$R$ be any member of $U$ such that $R$ intersects both
$\uparrow \left( i,\left\{ E\right\} \right) $ and $\left(
A\,\vdash \uparrow \left( i,\left\{ E\right\} \right)
\right)
$.   Now if $\left\{ B,C\right\} $ is any partition of $A$
such that $i\in C$ holds, $E$ intersects both $B$ and $C$,
and $E$ is the \emph{only} member of $U$ to intersect both
$B$ and $C$, then $\uparrow \left( i,\left\{ E\right\}
\right) \subseteq C$ holds, hence $R$ intersects $C$, hence
if $R$ is not equal to $E$, then $\left( A\cap R\right)
\subseteq C$ holds.   And this is true for \emph{every}
partition $\left\{ B,C\right\} $ of $A$ such that $i\in C$
holds, $E$ intersects both $B$ and $C$, and $E$ is the
\emph{only} member of $U$ to intersect both $B$ and $C$,
hence if $R\neq E$ holds, and $j$ is any member of $\left(
A\cap R\right) $, then for \emph{every} partition $\left\{
B,C\right\} $ of $A$ such that $E$ intersects both $B$ and
$C$, and $E$ is the \emph{only} member of $U$ to intersect
both $B$ and $C$, $j$ is a member of the \emph{same} member
of $\left\{ B,C\right\} $ as $i$ is, hence by the definition
of $\uparrow \left( i,\left\{ E\right\} \right) $, $j\in
\uparrow \left( i,\left\{ E\right\} \right) $ holds.   Hence
if $R\neq E$ holds then $\left( A\cap R\right) \subseteq
\uparrow \left( i,\left\{ E\right\} \right) $ holds, which
contradicts the fact that $\left( A\,\vdash \uparrow \left(
i,\left\{ E\right\} \right) \right) \cap R$ is nonempty.
Hence $R=E$ holds and $E$ is the \emph{only} member of $U$
to intersect both $\uparrow \left( i,\left\{ E\right\}
\right) $ and $\left( A\,\vdash \uparrow \left( i,\left\{
E\right\} \right) \right) $.

\vspace{2.5ex}

\noindent {\bf (ix)}  Let $E$ be any $U$-key
of $A$.   Then $E$ is \emph{not}
a
$U$-key of $\uparrow \left( i,\left\{ E\right\} \right) $.
\label{Start of original page 133}
For if there exists a partition $\left\{ B,C\right\} $ of
$\uparrow \left( i,\left\{ E\right\} \right) $
 into two parts such that $i\in C$ holds, $E$ intersects
both $B$ and $C$, and $E$ is the \emph{only} member of $U$
to intersect both $B$ and $C$, let $\left\{ B,C\right\} $ be
such a partition of $\uparrow \left( i,\left\{ E\right\}
\right) $.   Then $\left\{ \left( B\cup \left( A\,\vdash
\uparrow \left( i,\left\{ E\right\} \right) \right) \right)
,C\right\} $ is a partition of $A$ into two nonempty parts
such that $i\in C$ holds, $E$ intersects both parts, and
$E$ is the \emph{only} member of $U$ to intersect both
parts, for by assumption $E$ is the only member of $U$ to
intersect both $B$ and $C$, and by (viii) above, together
with the fact that $C\subseteq \uparrow \left( i,\left\{
E\right\} \right) $ holds, $E$ is the \emph{only} member of
$U$ to intersect both $C$ and $\left( A\,\vdash \uparrow
\left( i,\left\{ E\right\} \right) \right) $.   Hence by the
definition of $\uparrow \left( i,\left\{ E\right\} \right)
$,
no member of $B$ could be a member of $\uparrow \left(
i,\left\{ E\right\} \right) $, in contradiction with the
assumption that $B$ is a nonempty subset of $\uparrow
\left( i,\left\{ E\right\} \right) $.

\vspace{2.5ex}

\noindent {\bf (x)}  Let $X$ be any subset of
$U$ such that every member of
$X$ is a $U$-key of $A$, and let $T$ be any member of $X$
such that there is \emph{no} member $R$ of $X$ such that
$T\to R$ holds.   Then $T\cap \uparrow \left( i,X\right) $
is nonempty, and $\left( T\cap \uparrow \left( i,\left\{
T\right\} \right) \right) \subseteq \uparrow \left(
i,X\right) $ holds.   For by (viii) above, $T\cap \uparrow
\left( i,\left\{ T\right\} \right) $ is nonempty.
Let $j$ be any member of $T\cap \uparrow \left( i,\left\{
T\right\} \right) $.   \emph{Suppose} $j\in \left(
A\,\vdash \uparrow \left( i,X\right) \right) $ holds.
Then there exists a member $E$ of $X$ and a partition
$\left\{ B,C\right\} $ of $A$ into two parts such that $j\in
B$ holds, $i\in C$ holds, $E$ intersects both $B$ and $C$,
and $E$ is the \emph{only} member of $U$ to intersect both
$B$ and $C$.   Let $E$ be such a member of $X$ and $\left\{
B,C\right\} $ be such a partition of $A$.   Suppose first
$E\neq T$.   Then $T$ only intersects \emph{one} of $B$ and
$C$.   And $j\in B$ implies $B\cap T\neq \emptyset $ hence
$C\cap T=\emptyset $, hence $T\to E$ holds, but by
assumption
$T\to R$ does not hold for \emph{any} member $R$ of $X$.
Hence $E=T$ must hold.   But $j\in T\cap \uparrow \left(
i,\left\{ T\right\} \right) $ implies that $j$ is a
member of the \emph{same} member of $\left\{ B,C\right\} $
as
$i$ for \emph{all} partitions $\left\{ B,C\right\} $ of $A$
into two parts such that $T$ intersects both $B$ and $C$,
and $T$ is the \emph{only} member of $U$ to intersect both
$B$ and $C$.   Hence there is \emph{no} member $E$ of $X$
such such that there exists a partition $\left\{ B,C\right\}
$ of $A$ into two parts such that $j\in B$ holds, $i\in C$
holds, $E$ intersects both $B$ and $C$, and $E$ is the
\emph{only} member of $U$ to intersect both $B$ and $C$.
Hence $j\in \uparrow \left( i,X\right) $ holds, and since
this is true for \emph{all} members $j$ of $T\cap \uparrow
\left( i,\left\{ T\right\} \right) $, $\left( T\cap \uparrow
\left( i,\left\{ T\right\} \right) \right) \subseteq
\uparrow
\left( i,X\right) $ holds.   And finally, since every
member of the nonempty set $\left( T\cap \uparrow \left(
i,\left\{ T\right\} \right) \right) $ is a member of $T$,
$T\cap \uparrow \left( i,X\right) $ is nonempty.

\vspace{2.5ex}

\noindent {\bf (xi)}  Let $X$ be any subset
of $U$ such that every member of
$X$ is a $U$-key of $A$, and let $Z$ be the subset of $X$
whose members are all the members $T$ of $X$ such that
there is \emph{no} member $R$ of $X$ such that $T\to R$
\label{Start of original page 134}
 holds.   Then $A$ is equal to the disjoint union of
$\uparrow \left( i,X\right) =\uparrow \left( i,Z\right) $,
and the sets $\left( A\,\vdash \uparrow \left( i,\left\{
T\right\} \right) \right) $ for all the members $T$ of $Z$,
(or in other words, $\uparrow \left( i,X\right) $ and the
sets $\left( A\,\vdash \uparrow \left( i,\left\{ T\right\}
\right) \right) $ for all the members $T$ of $Z$, are all
distinct from one another, and the set whose members are
all these sets, is a partition of A).   For let $S$ and $T$
be any two distinct members of $Z$, and if $\left(
A\,\vdash \uparrow \left( i,\left\{ S\right\} \right)
\right)
\cap \left( A\,\vdash \uparrow \left( i,\left\{ T\right\}
\right) \right) $ is nonempty  let $k$ be any member of
$\left( A\,\vdash \uparrow \left( i,\left\{ S\right\}
\right)
\right) \cap \left( A\,\vdash \uparrow \left( i,\left\{
T\right\} \right) \right) $.   Then by (iii) above either
$S\to T$ holds or $T\to S$ holds, contradicting the
assumption that $S$ and $T$ are distinct members of $Z$.
Hence $\left( A\,\vdash \uparrow \left( i,\left\{ S\right\}
\right) \right) \cap \left( A\,\vdash \uparrow \left(
i,\left\{ T\right\} \right) \right) $ is empty.   Now by
definition, $\uparrow \left( i,Z\right) $ is $ \bigcap_{
T\in Z }
\uparrow \left( i,\left\{ T\right\} \right) $.   Hence if
$k$ is a member of $\left( A\,\vdash \uparrow \left(
i,\left\{ T\right\} \right) \right) $ for some member $T$ of
$Z$ then $k$ is \emph{not} a member of $\uparrow \left(
i,Z\right) $.   Hence all these sets are mutually disjoint.
  Now let $k$ be any member of $A$.   Suppose first there
\emph{exists} a member $T$ of $Z$ such that there exists a
partition $\left\{ B,C\right\} $ of $A$ into two parts such
that $k\in B$ holds, $i\in C$ holds, $T$ intersects both
$B$ and $C$, and $T$ is the \emph{only} member of $U$ to
intersect both $B$ and $C$.   Then $k$ is not a member of
$\uparrow \left( i,\left\{ T\right\} \right) $ hence $k\in
\left( A\,\vdash \uparrow \left( i,\left\{ T\right\} \right)
\right) $ holds.   Now suppose that for \emph{every} member
$T$ of $Z$, and \emph{every} partition $\left\{ B,C\right\}
$
of $A$ into two parts such that $T$ intersects both $B$ and
$C$, and $T$ is the \emph{only} member of $U$ to intersect
both $B$ and $C$, $k$ and $i$ are members of the
\emph{same} member of $\left\{ B,C\right\} $.   Then $k\in
\uparrow \left( i,Z\right) $ holds.

\vspace{2.5ex}

\noindent {\bf (xii)}  Let $X$ be any subset
of $U$ such that every member of
$X$ is a $U$-key of $A$, let $Z$ be the subset of $X$ whose
members are all the members $E$ of $X$ such that there is
\emph{no} member $R$ of $X$ such that $E\to R$ holds, and
let $S$ and $T$ be any two distinct members of $Z$.   Then
\emph{no} member of $U$ intersects both $\left( A\,\vdash
\uparrow \left( i,\left\{ S\right\} \right) \right) $ and
$\left( A\,\vdash \uparrow \left( i,\left\{ T\right\}
\right)
\right) $.   For $\left( A\,\vdash \uparrow \left( i,\left\{
S\right\} \right) \right) \cap \left( A\,\vdash \uparrow
\left( i,\left\{ T\right\} \right) \right) =\emptyset $
holds by
(xi) above, and it immediately follows from this that
$\left( A\,\vdash \uparrow \left( i,\left\{ S\right\}
\right)
\right) \subseteq \uparrow \left( i,\left\{ T\right\}
\right)
$ holds and $\left( A\,\vdash \uparrow \left( i,\left\{
T\right\} \right) \right) \subseteq \uparrow \left(
i,\left\{
S\right\} \right) $ holds.  And by (viii) above, $S$ is the
\emph{only} member of $U$ to intersect both $\left(
A\,\vdash \uparrow \left( i,\left\{ S\right\} \right)
\right)
$ and \mbox{$ \uparrow \left( i,\left\{ S\right\} \right)
$,} and $T$
is the \emph{only} member of $U$ to intersect both $\left(
A\,\vdash \uparrow \left( i,\left\{ T\right\} \right)
\right)
$ and $\uparrow \left( i,\left\{ T\right\} \right) $.
Hence
if a member $R$ of $U$ intersected both $\left( A\,\vdash
\uparrow \left( i,\left\{ S\right\} \right) \right) $ and
$\left( A\,\vdash \uparrow \left( i,\left\{ T\right\}
\right)
\right) $ then $R$ would have to be equal to both $S$ and
$T$, which is impossible since by assumption $S\neq T$.

\vspace{2.5ex}

\noindent {\bf (xiii)}  Let $X$ be any subset
of $U$ such that every member
of $X$ is a $U$-key of $A$.   Then \emph{no} member of $X$
is
a $U$-key of $\uparrow \left( i,X\right) $.   For if
\label{Start of original page 135}
 there exists a member $S$ of $X$ such that $S$ is a $U$-key
of $\uparrow \left( i,X\right) $, let $S$ be such a member
of $X$.   Now if there is any member $E$ of $X$ such that
$S\to E$ holds, let $E$ be such a member of $X$.   Then
there exists a partition $\left\{ B,C\right\} $ of $A$ into
two parts such that $i\in C$ holds, $E$ intersects both $B$
and $C$, $E$ is the \emph{only} member of $U$ to intersect
both $B$ and $C$, and $S$ does \emph{not} intersect $C$.
Hence \emph{no} member of $S$ is a member of $\uparrow
\left( i,X\right) $, hence $S\cap \uparrow \left(
i,X\right) =\emptyset $, hence $S$ is \emph{not} a $U$-key
of
$\uparrow \left( i,X\right) $, contrary to assumption.
Hence there is \emph{no} member $E$ of $X$ such that $S\to
E$ holds, hence by (x) above, $S$ \emph{does} intersect
$\uparrow \left( i,X\right) $.   Now by (ix) above, $S$ is
\emph{not} a $U$-key of $\uparrow \left( i,\left\{ S\right\}
\right) $.   Let $Z$ be the set whose members are all the
members $T$ of $X$ such that there is \emph{no} member $E$
of $X$ such that $T\to E$ holds.   Then as just shown, $S$
is a member of $Z$.   Hence by (xi) above, $\uparrow \left(
i,\left\{ S\right\} \right) $ is equal to the disjoint union
of $\uparrow \left( i,Z\right) =\uparrow \left( i,X\right)
$, and all the sets $\left( A\,\vdash \uparrow \left(
i,\left\{ T\right\} \right) \right) $ for the members $T$ of
$Z$ such that $T\neq S$.   Now \emph{suppose} $\left\{
B,C\right\} $ is a partition of $\uparrow \left( i,X\right)
=\uparrow \left( i,Z\right) $ into two parts such that $S$
intersects both $B$ and $C$, and $S$ is the \emph{only}
member of $U$ to intersect both $B$ and $C$.   Then since
by (x) above, every member of $Z$ does intersect $\uparrow
\left( i,X\right) $, every member of $Z$ other than $S$
intersects exactly \emph{one} of $B$ and $C$.   Let $J$ be
the set equal to the disjoint union of $B$, and the sets
$\left( A\,\vdash \uparrow \left( i,\left\{ T\right\}
\right)
\right) $ for all the members $T$ of $Z$ such that $T\neq
S$ and $T\cap B\neq \emptyset $ both hold, and let $K$ be
the
set equal to the disjoint union of $C$, and the sets
$\left( A\,\vdash \uparrow \left( i,\left\{ R\right\}
\right)
\right) $ for all the members $R$ of $Z$ such that $R\neq
S$ and $R\cap C\neq \emptyset $ both hold.   Then $\left\{
J,K\right\} $ is a partition of $\uparrow \left( i,\left\{
S\right\} \right) $ into two nonempty parts such that $S$
intersects both parts.   Now suppose some member $E$ of
$U$, such that $E\neq S$ holds, intersects both $J$ and
$K$.   Then $E$ does not intersect both $B$ and $C$, since
by assumption $S$ is the \emph{only} member of $U$ to
intersect both $B$ and $C$.   And by (xii) above, $E$
cannot intersect both $\left( A\,\vdash \uparrow \left(
i,\left\{ T\right\} \right) \right) $ and $\left( A\,\vdash
\uparrow \left( i,\left\{ R\right\} \right) \right) $ for
any
two distinct members $T$ and $R$ of $Z$.   Hence either $E$
intersects both $B$ and a set $\left( A\,\vdash \uparrow
\left( i,\left\{ R\right\} \right) \right) $, where $R$ is a
member of $Z$ such that $R\neq S$ and $R\cap C\neq
\emptyset $
both hold, or else $E$ intersects both $C$ and a set
$\left( A\,\vdash \uparrow \left( i,\left\{ T\right\}
\right)
\right) $, where $T$ is a member of $Z$ such that $T\neq S$
and $T\cap B\neq \emptyset $ both hold.   But in the first
case
$E$ must also intersect $\uparrow \left( i,\left\{ R\right\}
\right) $, since $B$ is a subset of $\uparrow \left(
i,\left\{ R\right\} \right) $, hence by (viii) above, $E=R$
holds, hence $E$ is a member of $Z$ different from $S$,
hence by assumption $E$ does \emph{not} intersect both $B$
and $C$, but since $E$ is equal to $R$ this contradicts the
assumptions that $E\cap B$ and $R\cap C$
\label{Start of original page 136}
 are both nonempty.   And in the second case $E$ must also
intersect $\uparrow \left( i,\left\{ T\right\} \right) $,
since $C$ is a subset of $\uparrow \left( i,\left\{
T\right\}
\right) $, hence by (viii) above, $E=T$ holds, hence $E$ is
a member of $Z$ different from $S$, hence by assumption $E$
does not intersect both $B$ and $C$, but since $E$ is equal
to $T$ this contradicts the assumptions that $E\cap C$ and
$T\cap B$ are both nonempty.   Hence there is \emph{no}
member $E$ of $U$ such that $E\neq S$ holds and $E$
intersects both $J$ and $K$, hence the assumption that $S$
is a $U$-key of $\uparrow \left( i,X\right) $ implies that
$S$ is a $U$-key of $\uparrow \left( i,\left\{ S\right\}
\right) $, in contradiction with (ix) above.

\vspace{2.5ex}

\noindent {\bf (xiv)}  Let $X$ be any subset
of $U$ such that every member of
$X$ is a $U$-key of $A$, and let $\left\{ B,C\right\} $ be
any
partition of $\uparrow \left( i,X\right) $ into two parts
such that \emph{no} member of $X$ intersects both parts.
Let $Z$ be the subset of $X$ whose members are all the
members $S$ of $X$ such that there is \emph{no} member $R$
of $X$ such that $S\to R$ holds.   Let $J$ be the set equal
to the disjoint union of $B$, and the sets $\left(
A\,\vdash \uparrow \left( i,\left\{ T\right\} \right)
\right)
$ for all the members $T$ of $Z$ such that $B\cap T\neq
\emptyset $ holds, and let $K$ be the set equal to the
disjoint
union of $C$, and the sets $\left( A\,\vdash \uparrow
\left( i,\left\{ R\right\} \right) \right) $ for all the
members $R$ of $Z$ such that $C\cap R\neq \emptyset $ holds.
Then $\left\{ J,K\right\} $ is a partition of $A$, and if
$E$
is \emph{any} member of $U$ such that $E$ intersects both
$J$ and $K$, then $E$ intersects both $B$ and $C$.   For
(x) and (xi) above, together with the assumption that
\emph{no} member of $X$ intersects both $B$ and $C$, hence
that \emph{no} member of $Z$ intersects both $B$ and $C$,
imply directly that $\left\{ J,K\right\} $ is a partition of
$A$.   Now let $E$ be any member of $U$ such that $E$
intersects both $J$ and $K$.   Then since by (xii) above,
$E$ does not intersect both $\left( A\,\vdash \uparrow
\left( i,\left\{ T\right\} \right) \right) $ and $\left(
A\,\vdash \uparrow \left( i,\left\{ R\right\} \right)
\right)
$ for any two distinct members $T$ and $R$ of $Z$, either
$E$ intersects both $B$ and $C$, or $E$ intersects both $B$
and the set $\left( A\,\vdash \uparrow \left( i,\left\{
R\right\} \right) \right) $, where $R$ is some member of $Z$
such that $R\cap C$ is nonempty, or else $E$ intersects
both $C$ and the set $\left( A\,\vdash \uparrow \left(
i,\left\{ T\right\} \right) \right) $, where $T$ is some
member of $Z$ such that $T\cap B$ is nonempty.   But if $E$
intersects both $B$ and the set $\left( A\,\vdash \uparrow
\left( i,\left\{ R\right\} \right) \right) $, where $R$ is a
member of $Z$, then $E$ also intersects $\uparrow \left(
i,\left\{ R\right\} \right) $ since $B\subseteq \uparrow
\left( i,\left\{ R\right\} \right) $ holds, hence by (viii)
above $E$ is equal to $R$, hence $R$ intersects $B$, hence
since by assumption no member of $Z$ intersects both $B$
and $C$, $R$ does \emph{not} intersect $C$.   And if $E$
intersects both $C$ and the set $\left( A\,\vdash \uparrow
\left( i,\left\{ T\right\} \right) \right) $, where $T$ is a
member of $Z$, then $E$ also intersects $\uparrow \left(
i,\left\{ T\right\} \right) $ since $C\subseteq \uparrow
\left( i,\left\{ T\right\} \right) $ holds, hence by (viii)
above $E$ is equal to $T$, hence $T$ intersects $C$, hence
since by assumption no member of $Z$ intersects both $B$
and $C$, $T$ does \emph{not}
\label{Start of original page 137}
 intersect $B$.   Hence $E$ must intersect both $B$ and $C$.

\vspace{2.5ex}

\noindent {\bf (xv)}  Let $X$ be any subset
of $U$ such that every member of
$X$ is a $U$-key of $A$.   Then $\uparrow \left( i,X\right)
$
is $U$-connected.   For if there exists a partition $\left\{
B,C\right\} $ of $\uparrow \left( i,X\right) $ into two
nonempty parts such that \emph{no} member of $U$ intersects
both parts, let $\left\{ B,C\right\} $ be such a partition
of
$\uparrow \left( i,X\right) $.   Then no member of $X$
intersects both $B$ and $C$.   Let $J$ and $K$ be the sets
constructed from $B$ and $C$ as described in (xiv) above.
Then by (xiv) above, $\left\{ J,K\right\} $ is a partition
of
$A$ into two nonempty parts such that \emph{no} member of
$U$ intersects both parts, which is impossible since $A$ is
$U$-connected.

\vspace{2.5ex}

\noindent {\bf (xvi)}  Let $X$ be any subset
of $U$ such that every member of
$X$ is a $U$-key of $A$, and let $E$ be any member of $U$.
Then $E$ is a $U$-key of $\uparrow \left( i,X\right) $ ifif
$E$ is a $U$-key of $A$ such that $E$ intersects $\uparrow
\left( i,X\right) $ and $E$ is \emph{not} a member of $X$.
 For suppose first that $E$ is a $U$-key of $\uparrow \left(
i,X\right) $.   Then by (xiii) above, $E$ is \emph{not} a
member of $X$.   Let $\left\{ B,C\right\} $ be a partition
of
$\uparrow \left( i,X\right) $ into two parts such that $E$
intersects both $B$ and $C$, and $E$ is the \emph{only}
member of $U$ to intersect both $B$ and $C$.   Then no
member of $X$ intersects both $B$ and $C$.   Let $J$ and
$K$ be the sets constructed from $B$ and $C$ as described
in (xiv) above.   Then by (xiv) above, $\left\{ J,K\right\}
$
is a partition of $A$ into two parts such that $E$
intersects both parts and $E$ is the \emph{only} member of
$U$ to intersect both parts.   Hence $E$ is a $U$-key of
$A$.
  And furthermore, $E$ certainly intersects $\uparrow
\left( i,X\right) $ and, as already noted, $E$ is
\emph{not} a member of $X$.   Now let $E$ be any $U$-key of
$A$ such that $E$ intersects $\uparrow \left( i,X\right) $
and $E$ is \emph{not} a member of $X$.   Let $Z$ be the
subset of $X$ whose members are all the members $T$ of $X$
such that there is \emph{no} member $R$ of $X$ such that
$T\to R$ holds.   Then $E$ does not intersect $\left(
A\,\vdash \uparrow \left( i,\left\{ T\right\} \right)
\right)
$ for \emph{any} member $T$ of $Z$, for if $T$ was a member
of $Z$ such that $E$ intersected $\left( A\,\vdash \uparrow
\left( i,\left\{ T\right\} \right) \right) $, then since
$\uparrow \left( i,X\right) \subseteq \uparrow \left(
i,\left\{ T\right\} \right) $ holds, $E$ would intersect
both
$\left( A\,\vdash \uparrow \left( i,\left\{ T\right\}
\right)
\right) $ and $\uparrow \left( i,\left\{ T\right\} \right)
$,
hence by (viii) above $E$ would be equal to $T$, which
contradicts the assumption that $E$ is \emph{not} a member
of $X$.   Hence by (xi) above, $E\cap A$ is a subset of
$\uparrow \left( i,X\right) =\uparrow \left( i,Z\right) $.
 Now the assumption that $E$ is a $U$-key of $A$ implies
that
there exists a partition $\left\{ J,K\right\} $ of $A$ into
two parts such that $E$ intersects both $J$ and $K$, and
$E$ is the \emph{only} member of $U$ to intersect both $J$
and $K$.   Let $\left\{ J,K\right\} $ be such a partition of
$A$.   Then $J\cap E$ is nonempty.   Let $j$ be any member
of $J\cap E$.   Then $j$ is a member of $E\cap A$ hence
since, as just shown, every member of $E\cap A$ is a member
\label{Start of original page 138}
 of $\uparrow \left( i,X\right) $, $j$ is a member of
$\uparrow \left( i,X\right) $, hence $J\cap \uparrow \left(
i,X\right) $ is nonempty and $E$ intersects $\left( J\cap
\uparrow \left( i,X\right) \right) $.   And $K\cap E$ is
nonempty hence again, since every member of $E\cap A$ is a
member of $\uparrow \left( i,X\right) $, hence every member
of $K\cap E$ is a member of $\uparrow \left( i,X\right) $,
$K\cap \uparrow \left( i,X\right) $ is nonempty and $E$
intersects $\left( K\cap \uparrow \left( i,X\right) \right)
$.   Hence $\left\{ \left( J\cap \uparrow \left( i,X\right)
\right) ,\left( K\cap \uparrow \left( i,X\right) \right)
\right\} $ is a partition of $\uparrow \left( i,X\right) $
into two nonempty parts such that $E$ intersects both parts
and $E$ is the \emph{only} member of $U$ to intersect both
parts.   Hence $E$ is a $U$-key of $\uparrow \left(
i,X\right) $.

\vspace{2.5ex}

We recall from
page \pageref{Start of original page 124}
 that if $A$ is a set, $U$ is a
set such that every member of $U$ is a set, and $V$ is a
subset of $U$, then a $ U $\emph{-firm over }$ V
$\emph{ component of }$ A $ is a
nonempty subset $B$ of $A$ such that $B$ is $U$-firm over
$V$
and $B$ is \emph{not} a strict subset of any subset of $A$
that is $U$-firm over $V$.

\begin{bphzlemma} \label{Lemma 29}
\end{bphzlemma}
\vspace{-6.143ex}

\noindent \hspace{11.9ex}{\bf.  }Let $U$ be a set
such that every member of $U$ is
a set, $V$ be a subset of $U$, $A$ be a nonempty
$U$-connected set, $i$ be any member of $A$, and $B$ be the
subset of $A$ whose members are all the members $j$ of $A$
such that for every $U$-key $E$ of $A$ such that $E\notin V$
holds, and every partition $\left\{ J,K\right\} $ of
$A$ into
two parts such that $E$ intersects both $J$ and $K$, and
$E$ is the \emph{only} member of $U$ to intersect both $J$
and $K$, $j$ is a member of the \emph{same} member of
$\left\{ J,K\right\} $ as $i$ is.   Then $B$ is a
$U$-firm over
$V$ component of $A$.

\vspace{2.5ex}

\noindent {\bf Proof.}  We define $X$ to be
the set whose members are all
the $U$-keys $E$ of $A$ such that $E\notin V$ holds, and
observe that the set $B$ as defined above is then equal to
the set $\uparrow \left( i,X\right) $, where the function
$\uparrow \left( i,X\right) $ is defined
as in Lemma \ref{Lemma 28}.
Hence by (xv) of Lemma \ref{Lemma 28}, $B$ is
$U$-connected, and by
(xvi) of Lemma \ref{Lemma 28}, every $U$-key of $B$ is a
member of $V$.
 Hence $B$ is $U$-firm over $V$.

Now let $C$ be any subset of $A$ such that $B\subset C$
holds.   Then $C\,\vdash B$ is nonempty.   Let $j$ be any
member of $C\,\vdash B$.   Then by the definition of $B$
there exists a $U$-key $E$ of $A$ such that $E\notin V$, and
a partition $\left\{ J,K\right\} $ of $A$ into two parts
such
that $j\in J$ holds, $i\in K$ holds, $E$ intersects both
$J$ and $K$, and $E$ is the \emph{only} member of $ U
\rule[-2.2ex]{0pt}{2.2ex} $ to
intersect both $J$ and $K$.   Let $E$ be such a $U$-key of
$A$ and $\left\{ J,K\right\} $ be such a partition of $A$.
Then $j$ is a member of $J\cap C$ hence $J\cap C$ is
nonempty, and $i$ is a member of $K\cap C$ hence $K\cap C$
is nonempty, hence $\left\{ \left( J\cap C\right) ,\left(
K\cap C\right) \right\} $ is a partition of $C$ into two
nonempty parts.   And
\label{Start of original page 139}
 if $E$ intersects both $J\cap C$ and $K\cap C$, then $E$
is a $U$-key of $C$ such that $C\notin V$, hence $C$ is not
$U$-firm over $V$, while if $E$ does \emph{not} intersect
both $J\cap C$ and $K\cap C$, then \emph{no} member of $U$
intersects both $J\cap C$ and $K\cap C$, hence $C$ is not
$U$-connected hence again $C$ is not $U$-firm over $V$.
Hence
$B$ is a $U$-firm over $V$ subset of $A$ such that $i\in B$
holds, and $B$ is not a strict subset of any $U$-firm over
$V$ subset of $A$, hence $B$ is a $U$-firm over $V$
component
of $A$.

\begin{bphzlemma} \label{Lemma 30}
\end{bphzlemma}
\vspace{-6.143ex}

\noindent \hspace{11.9ex}{\bf.  }Let $U$ be a set
such that every member of $U$ is
a set, $V$ be a subset of $U$, and $B$ and $C$ be any two
distinct nonempty $U$-firm over $V$ sets such that $B\cap C$
is nonempty.   Then $B\cup C$ is $U$-firm over $V$.

\vspace{2.5ex}

\noindent {\bf Proof.}  Let $\left\{ J,K\right\} $
be any partition of
$B\cup
C$ into two parts.  Now by assumption $B\cap C$ is
nonempty.   Let $i$ be a member of $B\cap C$.   Then $i$ is
a member of exactly one member of $\left\{ J,K\right\} $.
Let $M$ be the member of $\left\{ J,K\right\} $ that has $i$
as a member, and let $N$ be the other member of $\left\{
J,K\right\} $.   Then both $B\cap M$ and $C\cap M$ are
nonempty, and at least one of $B\cap N$ and $C\cap N$ is
nonempty.   Let $W$ be a member of $\left\{ B,C\right\} $
such that $W\cap N$ is nonempty.   Then $\left\{ \left(
W\cap M\right) ,\left( W\cap N\right) \right\} $ is a
partition of $W$ into two nonempty parts hence the fact
that $W$ is $U$-firm over $V$ implies that there exists a
member $E$ of $U$ such that $E$ intersects both $\left(
W\cap M\right) $ and $\left( W\cap N\right) $, and
furthermore since every $U$-key of $W$ is a member of $V$,
either $E\in V$ holds or else $E$ is \emph{not} the only
member of $U$ to intersect both $\left( W\cap M\right) $
and $\left( W\cap N\right) $.   Now $\left\{ \left( W\cap
M\right) ,\left( W\cap N\right) \right\} =\left\{ \left(
W\cap J\right) ,\left( W\cap K\right) \right\} $ holds hence
$E$ is a member of $U$ that intersects both $J$ and $K$,
and furthermore either $E\in V$ holds or else $E$ is
\emph{not} the only member of $U$ to intersect both $J$ and
$K$.

\begin{bphzlemma} \label{Lemma 31}
\end{bphzlemma}
\vspace{-6.143ex}

\noindent \hspace{11.9ex}{\bf.  }Let $U$ be a set
such that every member of $U$ is
a set, $V$ be a subset of $U$, and $A$ be a nonempty
$U$-connected set.   Then the set whose members are all the
$U$-firm over $V$ components of $A$, is a partition of $A$.

\vspace{2.5ex}

\noindent {\bf Proof.}  Let $F$ be the set whose
members are all the
$U$-firm
over $V$ components of $A$.   Then it follows directly from
the definition of a $U$-firm over $V$ component of $A$ that
no member of $F$ is empty.   And it follows directly from
Lemma \ref{Lemma 30} that no two distinct members of $F$
intersect one
another, for if $B$ and $C$ are any two distinct nonempty
$U$-firm over $V$ subsets of $A$ such that $B\cap C$ is
nonempty, then by
\label{Start of original page 140}
 Lemma \ref{Lemma 30}, $B\cup C$ is $U$-firm over $V$,
hence both $B$
and $C$ are strict subsets of the $U$-firm over $V$ subset
$\left( B\cup C\right) $ of $A$, hence neither $B$ nor $C$
is a $U$-firm over $V$ component of $A$.   And furthermore,
every member of $F$ is a subset of $A$, hence $\mathcal{U}
\left( F\right) \subseteq A$ holds, and if $i$ is any
member of $A$, then by Lemma \ref{Lemma 29} there
\emph{exists} a
member $B$ of $F$ such that $i\in B$ holds, hence
$A\subseteq \mathcal{U} \left( F\right) $ holds, hence
$A=\mathcal{U} \left( F\right) $ holds.

\begin{bphzlemma} \label{Lemma 32}
\end{bphzlemma}
\vspace{-6.143ex}

\noindent \hspace{11.9ex}{\bf.  }Let $U$ be a set
such that every member of $U$ is
a set, $V$ be a subset of $U$, $A$ be a nonempty
$U$-connected set, and $f$ be any member of $A$.

We define a binary relation, written $\to $, among the
$U$-keys of $A$ as follows:

If $S$ and $T$ are $U$-keys of $A$, then $S\to T$ holds ifif
there \emph{exists} a partition $\left\{ J,K\right\} $ of
$A$
into two nonempty parts $J$ and $K$ such that $T$
intersects both $J$ and $K$, $T$ is the \emph{only} member
of $U$ to intersect both $J$ and $K$, and $S$ does
\emph{not} intersect the member of $\left\{ J,K\right\} $ of
which $f$ is a member.

We note that this definition has the immediate consequence
that $S\to S$ does \emph{not} hold for any $U$-key $S$ of
$A$, and that the definition of the relation $\to $ depends
implicitly on the member $f$ of $A$.

Let $B$ be any $U$-firm over $V$ component of $A$ such that
$f$ is \emph{not} a member of $B$, and let $Z$ be the set
whose members are all the $U$-keys $E$ of $A$ such that
$E\notin V$ and $E\cap B\neq \emptyset $ both hold.

Then there exists a \emph{unique} member $T$ of $Z$ such
that there exists a partition $\left\{ J,K\right\} $ of $A$
into two nonempty parts such that $B\subseteq J$ holds,
$f\in K$ holds, $T$ intersects both $J$ and $K$, and $T$ is
the \emph{only} member of $U$ to intersect both $J$ and
$K$.   And this unique member $T$ of $Z$ has the properties
that $S\to T$ holds for every member $S$ of $Z$ different
from $T$, $T\to S$ does \emph{not} hold for \emph{any}
member $S$ of $Z$, and if $R$ and $S$ are any members of
$Z$ such that $S\to R$ holds, then $R=T$ holds.

\vspace{2.5ex}

\noindent {\bf Proof.}  We first note that
the relation $\to $ just defined
is the same as the relation $\to $
defined in Lemma \ref{Lemma 28} if
we choose the member $i$ of $A$ with respect to which the
relation $\to $ was defined in Lemma \ref{Lemma 28}, to be
equal to
$f$.   Hence by (i) of Lemma \ref{Lemma 28}, if $S$ and $T$
are any
$U$-keys of $A$, \emph{at most one} of $S\to T$ and $T\to S$
can hold.

Now for any ordered pair $\left(i,X\right) $ of
a member $i$ of $A$ and a
subset $X$ of $U$, let $\uparrow \left( i,X\right) $
denote, as in Lemma \ref{Lemma 28}, the set whose members
are all the
members $j$ of $A$ such that if $E$ is any member of $X$,
and $\left\{ J,K\right\} $
\label{Start of original page 141}
 is any partition of $A$ into two parts such that $E$
intersects both $J$ and $K$, and $E$ is the \emph{only}
member of $U$ to intersect both $J$ and $K$, then $j$ is a
member of the \emph{same} member of $\left\{ J,K\right\} $
as
$i$ is.

And let $X$ be the set whose members are all the $U$-keys
$E$
of $A$ such that $E$ is \emph{not} a member of $V$.   Then
if $i$ is any member of the $U$-firm over $V$ component $B$
of $A$, it follows directly from
Lemmas \ref{Lemma 29} and \ref{Lemma 31} that
$B$ is equal to $\uparrow \left( i,X\right) $.

And furthermore, the set $Z$, defined above to be the set
whose members are all the $U$-keys $E$ of $A$ such that
$E\notin V$ and $E\cap B\neq \emptyset $ both hold, is
equal to
the subset of $X$ whose members are all the members $S$ of
$X$ such that there is \emph{no} member $R$ of $X$ such
that there exists a partition $\left\{ J,K\right\} $ of $A$
into two parts such that $R$ intersects both $J$ and $K$,
$B\subseteq K$ holds, $R$ intersects both $J$ and $K$, $R$
is the \emph{only} member of $U$ to intersect both $J$ and
$K$, and $S$ does \emph{not} intersect $K$.   For by
definition $Z$ is equal to the subset of $X$ whose members
are all the members $E$ of $X$ such that $E\cap B\neq
\emptyset
$.   Let $i$ be any member of $B$.   We now
use Lemma \ref{Lemma 28}
with the $i$ of Lemma \ref{Lemma 28} identified with this
member $i$
of $B$, (so the relation $\to $ of Lemma \ref{Lemma 28} is
now
\emph{not} the same as the relation $\to $ defined above).
 Now it follows directly from the definition of $B=\uparrow
\left( i,X\right) $ that if $R$ is any member of $X$, and
$\left\{ J,K\right\} $ is any partition of $A$ into two
parts
such that $R$ intersects both parts, and $R$ is the
\emph{only} member of $U$ to intersect both parts, then
every member of $B=\uparrow \left( i,X\right) $ is a member
of the \emph{same} member of $\left\{ J,K\right\} $,
or in other words, $B=\uparrow \left( i,X\right) $ is a
\emph{subset} of one of the two members of $\left\{
J,K\right\} $.   Hence by (x) of Lemma \ref{Lemma 28}, if
$S$ is any
member of $X$ such that there is \emph{no} member $R$ of
$X$ such that there exists a partition
$\left\{ J,K\right\} $
of $A$ into two parts such that $R$ intersects both parts,
$R$ is the \emph{only} member of $U$ to intersect both
parts, $B\subseteq K$ holds, and $S$ does \emph{not}
intersect $K$, then $S$ \emph{does} intersect $B=\uparrow
\left( i,X\right) $, hence $S$ is a member of $Z$.   And
conversely, if $S$ is a member of $Z$, then $S$ intersects
$B=\uparrow \left( i,X\right) $, hence there is \emph{no}
partition $\left\{ J,K\right\} $ of $A$ into two parts such
that $B\subseteq K$ holds and $S$ does not intersect $K$.

Hence by (vii) of Lemma \ref{Lemma 28}, $B=\uparrow \left(
i,X\right)
=\uparrow \left( i,Z\right) $ holds, and by (xi) of
Lemma \ref{Lemma 28}, $A$ is equal to the disjoint union of
$B=\uparrow \left( i,X\right) =\uparrow \left( i,Z\right)
$, and the sets $\left( A\,\vdash \uparrow \left( i,\left\{
S\right\} \right) \right) $ for all the members $S$ of $Z$.
 Hence since $f$ is not a member of $B$, there is a unique
member $T$ of $Z$ such that $f\in \left( A\,\vdash \uparrow
\left( i,\left\{ T\right\} \right) \right) $ holds.   Let
$T$
be the unique member of $Z$ such that $f\in \left(
A\,\vdash \uparrow \left( i,\left\{ T\right\} \right)
\right)
$ holds.   Now by (viii) of Lemma \ref{Lemma 28}, $T$
intersects
\label{Start of original page 142}
 both $\left( A\,\vdash \uparrow \left( i,\left\{ T\right\}
\right) \right) $ and $\uparrow \left( i,\left\{ T\right\}
\right) $, and $T$ is the \emph{only} member of $U$ to
intersect both $\left( A\,\vdash \uparrow \left( i,\left\{
T\right\} \right) \right) $ and $\uparrow \left( i,\left\{
T\right\} \right) $.   And furthermore, it follows directly
from the definition of $\uparrow \left( i,X\right) =B$ that
$\uparrow \left( i,X\right) \subseteq \uparrow \left(
i,\left\{ T\right\} \right) $ holds.   Hence $\left\{ \left(
A\,\vdash \uparrow \left( i,\left\{ T\right\} \right)
\right)
,\uparrow \left( i,\left\{ T\right\} \right) \right\} $ is a
partition of $A$ into two parts such that $f\in \left(
A\,\vdash \uparrow \left( i,\left\{ T\right\} \right)
\right)
$ holds, $B\subseteq \uparrow \left( i,\left\{ T\right\}
\right) $ holds, $T$ intersects both parts, and $T$ is the
\emph{only} member of $U$ to intersect both parts.   And if
$S$ is any member of $Z$ different from $T$, then $S$
intersects $B=\uparrow \left( i,X\right) $, hence $S$
intersects $\uparrow \left( i,\left\{ T\right\} \right) $,
hence $S$ does \emph{not} intersect $\left( A\,\vdash
\uparrow \left( i,\left\{ T\right\} \right) \right) $, hence
$S\to T$ holds, where the relation $\to $ is defined with
reference to $f$ as in the statement of this Lemma.   And
furthermore, since $S\to T$ and $T\to S$ do not \emph{both}
hold for \emph{any} member $S$ of $Z$, and $S\to S$ does
not hold for \emph{any} $U$-key $S$ of $A$, $T\to S$ does
\emph{not} hold for \emph{any} member $S$ of $Z$.   And
finally, let $R$ and $S$ be any members of $Z$ such that
$S\to R$ holds.   Then there exists a partition $\left\{
J,K\right\} $ of $A$ into two nonempty parts such that $R$
intersects both $J$ and $K$, $R$ is the \emph{only} member
of $U$ to intersect both $J$ and $K$, $f$ is a member of
$K$, and $S$ does \emph{not} intersect $K$.   Let $\left\{
J,K\right\} $ be such a partition of $A$.   Then $B=\uparrow
\left( i,X\right) $ is either a subset of $J$ or a subset
of $K$, hence since $S$ intersects $B$ and $S$ does
\emph{not} intersect $K$, $B$ is a subset of $J$.   Now
\emph{suppose} that $R\neq T$ holds.   Then since $T$
intersects $B$, $T$ intersects $J$, hence $R\neq T$ implies
that $T$ does \emph{not} intersect $K$.   Hence $T\to R$
holds which contradicts the fact that $R\to T$ holds.

\vspace{2.5ex}

We recall from
page \pageref{Start of original page 124} that for every
ordered pair $\left(F,H\right) $
a wood $F$ and a set $H$ such that every member of $H$ is a
set, we define $\mathbb{P} \left( F,H\right) $ to be the
set whose members are the members of $\mathcal{M} \left(
F\right) $, together with any members $A$ of $\mathbb{B}
\left( F\right) $ that satisfy both the following
requirements:

\vspace{1.0ex}

\noindent (i)  $ A $ has at least one $\left(
\mathcal{M} \left( F\right)
\cup H\right) $-key $E$ such that $E\notin \mathcal{M}
\left( F\right) $ holds

\vspace{1.0ex}

\noindent (ii)  if $B$ is any member of $F$
such that $B\subset A$
holds, then $B$ is a subset of an $\left( \mathcal{M}
\left( F\right) \cup H\right) $-firm over $\mathcal{M}
\left( F\right) $ component of $A$.

\vspace{1.0ex}

And we recall from
page \pageref{Start of original page 125}
 that for any ordered pair
$\left(V,H\right) $ of a partition
 $V$ such that $\mathcal{U} \left(
V\right) $ is finite and $\#\left( V\right) \geq 2$ holds,
and a set $H$ such that every member of $H$ is a set, we
define $\mathcal{V} \left( V,H\right) $ to be the set whose
members are all the woods $F$ of $V$ such that every member
$A$ of $\mathbb{B} \left( F\right) $ is $\left( V\cup
H\right) $-firm over $V$.
\label{Start of original page 143}
\enlargethispage{3.0ex}

\begin{bphzlemma} \label{Lemma 33}
\end{bphzlemma}
\vspace{-6.143ex}

\noindent \hspace{11.9ex}{\bf.  }Let $V$ be any
partition such that
$\mathcal{U} \left( V\right) $ is finite and $\#\left(
V\right) \geq 2$ holds, and let $H$ be any set such that
every member of $H$ is a set.   Then $\mathcal{V} \left(
V,H\right) $ is the set whose members are all the members
$F$ of $\mathcal{G} \left( V,H\right) $ such that
$\mathbb{B} \left( \mathbb{P} \left( F,H\right) \right)
=\emptyset $ holds.

\vspace{2.5ex}

\noindent {\bf Proof.}  We note first that if
every member of $\mathbb{B}
\left( F\right) $ is $\left( V\cup H\right) $-firm over
$V$, then $\mathbb{B} \left( \mathbb{P} \left( F,H\right)
\right) =\emptyset $ certainly holds.   Hence to complete
the
proof it is sufficient to prove that if $\mathbb{B} \left(
\mathbb{P} \left( F,H\right) \right) =\emptyset $ holds then
every member of $\mathbb{B} \left( F\right) $ is $\left(
V\cup H\right) $-firm over $V$.

Let $F$ be any member of $\mathcal{G} \left( V,H\right) $
such that $\mathbb{B} \left( \mathbb{P} \left( F,H\right)
\right) =\emptyset $ holds.

We first note that if $A$ is any member of $F$ such that
there is \emph{no} member $B$ of $F$ such that $B\subset A$
holds, then $A$ is a member of $V$, hence $A$ is
\emph{certainly} $\left( V\cup H\right) $-firm over $V$.

We shall next show that if $A$ is any member of $F$ such
that every member $B$ of $F$ such that $B\subset A$ holds,
is $\left( V\cup H\right) $-firm over $V$, then $A$ is
$\left( V\cup H\right) $-firm over $V$.

\emph{Suppose} $A$ is a member of $F$ such that
every member $B$
of $F$ such that $B\subset A$ holds, is $\left( V\cup
H\right) $-firm over $V$, and $A$ is \emph{not} $\left(
V\cup H\right) $-firm over $V$.   Then $A$ is \emph{not} a
member of $V=\mathcal{M} \left( F\right) $, hence $A\in
\mathbb{B} \left( F\right) $ holds, and $A$ has at least
one $\left( V\cup H\right) $-key $E$ such that $E\notin V$
holds.   Let $B$ be any member of $F$ such that $B\subset
A$ holds.   Then by assumption, $B$ is $\left( V\cup
H\right) $-firm over $V$.   Now $B$ is a member of $F$,
hence $B$ is nonempty, and by Lemma \ref{Lemma 31}, the set
whose
members are all the $\left( V\cup H\right) $-firm over $V$
components of $A$, is a partition of $A$.   Let $i$ be a
member of $B$, and let $C$ be the unique $\left( V\cup
H\right) $-firm over $V$ component of $A$ of which $i$ is a
member.   Then $B\cap C$ has the member $i$ hence is
nonempty, hence by Lemma \ref{Lemma 30}, $B\cup C$ is
$\left( V\cup
H\right) $-firm over $V$, hence the fact that $C$ is a
$\left( V\cup H\right) $-firm over $V$ \emph{component} of
$A$ implies that $B\subseteq C$ holds.   Hence $A$ is a
member of $\mathbb{B} \left( F\right) $ such that $A$ has
at least one $\left( V\cup H\right) $-key $E$ such that
$E\notin V$ holds, and such that if $B$ is any member of
$F$ such that $B\subset A$ holds, then $B$ is a subset of a
$\left( V\cup H\right) $-firm over $V$ component of $A$.
Hence $A$ is a member of $\mathbb{B} \left( \mathbb{P}
\left( F,H\right) \right) $, in contradiction with the
assumption that $\mathbb{B} \left( \mathbb{P} \left(
F,H\right) \right) =\emptyset $ holds.

Hence, as stated, if $A$ is any member of $F$ such that
every member $B$ of $F$ such that $B\subset A$ holds, is
$\left( V\cup H\right) $-firm over $V$, then $A$ is $\left(
V\cup H\right) $-firm over $V$.

Now for each member $A$ of $F$, let $n_{ A } $ denote the
number of members $B$ of $F$ such that $B\subset A$ holds.
 Then $n_{ A } \geq 0$ holds, and if $n_{ A } =0$ holds
\label{Start of original page 144}
 then $A$ is $\left( V\cup H\right) $-firm over $V$.

Now let $m$ be any integer $\geq 0$ and suppose that every
member $B$ of $F$ such that $n_{ B } \leq m$ holds is
$\left( V\cup H\right) $-firm over $V$.   Then every member
$A$ of $F$ such that $n_{ A } \leq \left( m+1\right) $
holds is $\left( V\cup H\right) $-firm over $V$.   For if
$n_{ A } \leq \left( m+1\right) $ holds, then every member
$B$ of $F$ such that $B\subset A$ holds satisfies $n_{ B }
\leq m$ hence is $\left( V\cup H\right) $-firm over $V$,
and as just shown this implies that $A$ is $\left( V\cup
H\right) $-firm over $V$.

Hence, by induction, every member $A$ of $F$ is $\left(
V\cup H\right) $-firm over $V$.

\vspace{2.5ex}

We recall from
page \pageref{Start of original page 124}
 that if $H$ is a set such that
every member of $H$ is a set, and $F$ is a wood, then we
say that $F$ is $H $-principal ifif every member $A$ of
$\mathbb{B} \left( F\right) $ is $\left( \mathcal{M} \left(
F\right) \cup H\right) $-connected, every member $A$ of
$\mathbb{B} \left( F\right) $ has at least one $\left(
\mathcal{M} \left( F\right) \cup H\right) $-key $E$ such
that $E\notin \mathcal{M} \left( F\right) $ holds, and if
$A$ is any member of $\mathbb{B} \left( F\right) $ and $B$
is any member of $F$ such that $B\subset A$ holds, then $B$
is a subset of an $\left( \mathcal{M} \left( F\right) \cup
H\right) $-firm over $\mathcal{M} \left( F\right) $
component of $A$.

And we recall from
page \pageref{Start of original page 125}
 that for any ordered pair
$\left(V,H\right) $ of a partition $V$ such that
$\mathcal{U} \left(
V\right) $ is finite and $\#\left( V\right) \geq 2$ holds,
and a set $H$ such that every member of $H$ is a set, we
define $\mathcal{W} \left( V,H\right) $ to be the set whose
members are all the $H $-principal woods of $V$.

And we also recall from
page \pageref{Start of original page 125}
 that for any ordered pair
$\left(G,H\right) $ such that $H$ is a set such that
every member of $H$
is a set, and $G$ is an $H $-principal wood, we define
$\mathbb{O} \left( G,H\right) $ to be the set whose members
are all the members $F$ of $\mathcal{G} \left( \mathcal{M}
\left( G\right) ,H\right) $ such that $G\subseteq
\mathbb{P} \left( F,H\right) $ holds, and we note that
$\mathbb{O} \left( G,H\right) $ is the set whose members
are all the members $F$ of $\mathcal{G} \left( \mathcal{M}
\left( G\right) ,H\right) $ such that $G\subseteq F$ holds,
and if $A$ is any member of $\mathbb{B} \left( G\right) $
and $B$ is any member of $F$ such that $B\subset A$ holds,
then $B$ is a subset of an $\left( \mathcal{M} \left(
F\right) \cup H\right) $-firm over $\mathcal{M} \left(
F\right) $ component of $A$.

\begin{bphzlemma} \label{Lemma 34}
\end{bphzlemma}
\vspace{-6.143ex}

\noindent \hspace{11.9ex}{\bf.  }Let $V$ be any
partition such that $\mathcal{U}
\left( V\right) $ is finite and $\#\left( V\right) \geq 2$
holds, let $H$ be a set such that every member of $H$ is a
set, and let $\mathcal{J} $ be a map such that $\mathcal{G}
\left( V,H\right) \subseteq \mathcal{D} \left( \mathcal{J}
\right) $ holds and $\mathcal{R} \left( \mathcal{J} \right)
\subseteq \mathbb{R} $ holds.   Then the following identity
holds:
\[
  \sum_{F\in \mathcal{V} \left( V,H\right) } \mathcal{J}
_{ F } =  \sum_{G\in \mathcal{W} \left( V,H\right) }
\left( -1\right)^{ \#\left( \mathbb{B} \left( G\right)
\right) }
 \sum_{F\in \mathbb{O} \left( G,H\right) } \mathcal{J} _{ F
} .
\]

\vspace{2.5ex}

\noindent {\bf Proof.}  We first note that
if $G$ is any member of
$\mathcal{W} \left( V,H\right) $ and $F$ is any member of
$\mathbb{K} \left( V,G\right) $, (or in other words, if $F$
is any wood of $V$
\label{Start of original page 145}
 such that $F\subseteq G$ holds), then it follows directly
from the definition of a principal wood of $V$ that $F\in
\mathcal{W} \left( V,H\right) $ holds.

Now let $F$ be any member of $\mathcal{G} \left( V,H\right)
$.   We calculate the coefficient of $F$ in each side of
the above equation.

Now by Lemma \ref{Lemma 33}, the coefficient of $F$ in the
left-hand
side is $1$ if $\mathbb{B} \left( \mathbb{P} \left(
F,H\right) \right) =\emptyset $ holds, and $0$ otherwise.

In the right-hand side we calculate the coefficient of $F$
in terms of $\#\left( \mathbb{B} \left( \mathbb{P} \left(
F,H\right) \right) \right) $.   Now $F$ is a member of
$\mathbb{O} \left( G,H\right) $ for all members $G$ of
$\mathcal{W} \left( V,H\right) $ such that $G\subseteq
\mathbb{P} \left( F,H\right) $ holds, and for no member $G$
of $\mathcal{W} \left( V,H\right) $ such that $G$ is
\emph{not} a subset of $\mathbb{P} \left( F,H\right) $,
hence the coefficient of $F$ in the right-hand side is
equal to the sum of $\left( -1\right)^{ \#\left( \mathbb{B}
\left( G\right) \right) } $ over all members $G$ of
$\mathcal{W} \left( V,H\right) $ such that $G\subseteq
\mathbb{P} \left( F,H\right) $ holds.   But $\mathbb{P}
\left( F,H\right) \rule[-2.0ex]{0pt}{2.0pt} $ is a
principal wood of $V$ hence, as
noted above, \emph{every} wood $G$ of $V$ such that
$G\subseteq \mathbb{P} \left( F,H\right) $ holds is a
member of $\mathcal{W} \left( V,H\right) $, hence the
coefficient of $F$ in the right-hand side is equal to the
sum of $\left( -1\right)^{ \#\left( \mathbb{B}
\left( G\right)
\right) } $ over all members $G$ of $\mathbb{K} \left(
V,\mathbb{P} \left( F,H\right) \right) $, which is equal to
$1$ if $\#\left( \mathbb{B} \left( \mathbb{P} \left(
F,H\right) \right) \right) =0$ holds, and equal to $0$
otherwise.

\begin{bphzlemma} \label{Lemma 35}
\end{bphzlemma}
\vspace{-6.143ex}

\noindent \hspace{11.9ex}{\bf.  }Let $V$ be any
partition such that $\mathcal{U}
\left( V\right) $ is finite and $\#\left( V\right) \geq 2$
holds, let $H$ be any set such that every member of $H$ is
a set and $\mathcal{U} \left( V\right) $ is $\left( V\cup
H\right) $-connected, let $G$ be any member of $\mathcal{W}
\left( V,H\right) $, and let $Y$ be the set whose members
are all the $\left( V\cup H\right) $-firm over $V$
components of the members of $\mathbb{B} \left( G\right) $,
together with the set $\mathcal{U} \left( V\right) $.

Then the following results hold:

\vspace{2.5ex}

\noindent {\bf (i)}  Let $B$ be any member
of $\left( Y\,\vdash \left( V\cup
\left\{ \mathcal{U} \left( V\right) \right\} \right) \right)
$.   Then $\mathcal{Y} \left( G,B\right) $ is a member of
$\mathbb{B} \left( G\right) $, $B$ is a $\left( V\cup
H\right) $-firm over $V$ component of $\mathcal{Y} \left(
G,B\right) $, and moreover $\mathcal{Y} \left( G,B\right) $
is the \emph{only} member $A$ of $G$ such that $B$ is a
$\left( V\cup H\right) $-firm over $V$ component of $A$.
For $B\neq \mathcal{U} \left( V\right) $ holds hence by the
definition of $Y$ there \emph{exists} a member $A$ of $G$
such that $B$ is a $\left( V\cup H\right) $-firm over $V$
component of $A$, hence in particular there \emph{exists} a
member $A$ of $G$ such that $B\subseteq A$ holds, hence by
definition $\mathcal{Y} \left( G,B\right) $ is the
\emph{smallest} member $A$ of $G$ such that $B\subseteq A$
holds, and moreover, since $B$ is not a member of $V$,
$\mathcal{Y} \left( G,B\right) $ is a member of $\mathbb{B}
\left( G\right) =\left( G\,\vdash V\right) $.   Now let $A$
be any member of $G$ such that $B$ is a $\left( V\cup
H\right) $-firm over $V$ component of $A$.   Then $A$ is a
member of $G$ such that $B\subseteq A$ holds, hence
$\mathcal{Y} \left( G,B\right) \subseteq A$ holds, hence
$A$ is a member of $\mathbb{B} \left( G\right) $.   Now if
$\mathcal{Y} \left( G,B\right) \subset A$ held, then by the
definition of an $H $-principal wood of $V$,
\label{Start of original page 146}
 $\mathcal{Y} \left( G,B\right) $ would be a subset of a
$\left( V\cup H\right) $-firm over $V$ component of $A$,
hence since $\mathcal{Y} \left( G,B\right) $ is a member of
$\mathbb{B} \left( G\right) $, hence by the definition of
an $H $-principal wood of $V$, $\mathcal{Y} \left(
G,B\right)
$ is \emph{not} $\left( V\cup H\right) $-firm over $V$,
hence $B$ is a \emph{strict} subset of $\mathcal{Y} \left(
G,B\right) $, $B$ could \emph{not} be a $\left( V\cup
H\right) $-firm over $V$ component of $A$.   Hence
$\mathcal{Y} \left( G,B\right) \subset A$ cannot hold,
hence $\mathcal{Y} \left( G,B\right) =A$ holds.

\vspace{2.5ex}

\noindent {\bf (ii)}  $Y\subseteq \Xi
\left( V\right) $ holds,
or in other words, every
member $B$ of $Y$ is a nonempty subset of $\mathcal{U}
\left( V\right) $ such that $B$ neither overlaps any member
of $V$ nor is a strict subset of any member of $V$.   For
every member of $Y$ is certainly a nonempty subset of
$\mathcal{U} \left( V\right) $, and moreover $V$ is a
subset of $\Xi \left( V\right) $ and $\mathcal{U} \left(
V\right) $ is a member of $\Xi \left( V\right) $.   Now let
$B$ be any member of $\left( Y\,\vdash \left( V\cup \left\{
\mathcal{U} \left( V\right) \right\} \right) \right) $.
Then by (i) above, $\mathcal{Y} \left( G,B\right) $ is a
member of $\mathbb{B} \left( G\right) $, and $B$ is a
$\left( V\cup H\right) $-firm over $V$ component of
$\mathcal{Y} \left( G,B\right) $.   Let $i$ be any member
of $B$.   Then since $\mathcal{Y} \left( G,B\right) $ is
$\left( V\cup H\right) $-connected, it follows directly
from Lemmas \ref{Lemma 29} and \ref{Lemma 31} that $B$ is
equal to the set
whose members are all the members $j$ of $\mathcal{Y}
\left( G,B\right) $ such that for every $\left( V\cup
H\right) $-key $E$ of $\mathcal{Y} \left( G,B\right) $ such
that $E\notin V$ holds, and every partition $\left\{
J,K\right\} $ of $\mathcal{Y} \left( G,B\right) $ into two
parts such that $E$ intersects both $J$ and $K$, and $E$ is
the \emph{only} member of $\left( V\cup H\right) $ to
intersect both $J$ and $K$, $j$ is a member of the
\emph{same} member of $\left\{ J,K\right\} $ as $i$ is.
Let
$C$ be any member of $V$ and let $j$ and $k$ be any members
of $C$.   Then if $\left\{ J,K\right\} $ is any partition of
$\mathcal{Y} \left( G,B\right) $ into two nonempty parts
such that some member $E$ of $\left( V\cup H\right) $ such
that $E\notin V$ holds is the \emph{only} member of $\left(
V\cup H\right) $ to intersect both parts, $j$ and $k$ must
be members of the \emph{same} member of $\left\{ J,K\right\}
$, since otherwise the member $C$ of $V$ would intersect
both parts.   Hence $j$ is a member of $B$ ifif $k$ is a
member of $B$, hence either \emph{every} member of $C$ is a
member of $B$ or \emph{no} member of $C$ is a member of
$B$, hence either $C\subseteq B$ holds or $C\cap
B=\emptyset $
holds.

\vspace{2.5ex}

\noindent {\bf (iii)}  The set $G\cup Y$ is
a wood of $V$.   For $G$ is a
wood of $V$, and by (ii) above, every member of $Y$ is a
member of $\Xi \left( V\right) $, hence $G\cup Y$ will be a
wood of $V$ provided that no member of $Y$ overlaps any
member of $G\cup Y$.   Now certainly no member of $V$
overlaps any member of $G\cup Y$, and $\mathcal{U} \left(
V\right) $ overlaps no member of $G\cup Y$.   Now let $B$
be any member of $\left( Y\,\vdash \left( V\cup \left\{
\mathcal{U} \left( V\right) \right\} \right) \right) $.
Then by (i) above, $\mathcal{Y} \left( G,B\right) $ is a
member of $\mathbb{B} \left( G\right) $, and $B$ is a
$\left( V\cup H\right) $-firm over $V$ component of
$\mathcal{Y} \left( G,B\right) $.   Let $C$ be any member
of $G\cup Y$.   Then $C$ is either equal to $\mathcal{U}
\left( V\right) $ or else is a member of $G$ or else is a
$\left( V\cup H\right) $-firm over $V$
\label{Start of original page 147}
 component of some member of $G$.   Now if $C$ is equal to
$\mathcal{U} \left( V\right) $ then $\mathcal{Y} \left(
G,B\right) \subseteq C$ holds, and if $C$ is a member of
$G$ then either $\mathcal{Y} \left( G,B\right) \cap
C=\emptyset
$ holds or $\mathcal{Y} \left( G,B\right) \subseteq C$
holds or $C\subset \mathcal{Y} \left( G,B\right) $ holds.
Now suppose that $C$ is a $\left( V\cup H\right) $-firm
over $V$ component of a member $A$ of $G$.   Then either
$\mathcal{Y} \left( G,B\right) \cap A=\emptyset $ holds or
$\mathcal{Y} \left( G,B\right) \subseteq A$ holds or
$A\subset \mathcal{Y} \left( G,B\right) $ holds, and if
$\mathcal{Y} \left( G,B\right) \cap A=\emptyset $ holds then
$\mathcal{Y} \left( G,B\right) \cap C=\emptyset $ holds and
if
$A\subset \mathcal{Y} \left( G,B\right) $ holds then
$C\subset \mathcal{Y} \left( G,B\right) $ holds.   Suppose
now that $\mathcal{Y} \left( G,B\right) \subseteq A$ holds.
  Then either $\mathcal{Y} \left( G,B\right) =A$ holds, in
which case $C\subseteq \mathcal{Y} \left( G,B\right) $
holds, (hence either $C=\mathcal{Y} \left( G,B\right) $
holds, hence $\mathcal{Y} \left( G,B\right) \subseteq C$
holds, or else $C\subset \mathcal{Y} \left( G,B\right) $
holds), or else $\mathcal{Y} \left( G,B\right) \subset A$
holds.   Suppose now that $\mathcal{Y} \left( G,B\right)
\subset A$ holds.   Then by the definition of an
$H $-principal wood of $V$, $\mathcal{Y} \left( G,B\right) $
is a subset of a $\left( V\cup H\right) $-firm over $V$
component of $A$, hence since, by Lemma \ref{Lemma 31}, the
distinct
$\left( V\cup H\right) $-firm over $V$ components of $A$ do
not intersect one another, either $\mathcal{Y} \left(
G,B\right) \cap C=\emptyset $ holds or $\mathcal{Y} \left(
G,B\right) \subseteq C$ holds.   Hence in every case,
either $\mathcal{Y} \left( G,B\right) \cap C=\emptyset $
holds
or $\mathcal{Y} \left( G,B\right) \subseteq C$ holds or
$C\subset \mathcal{Y} \left( G,B\right) $ holds.   And if
$\mathcal{Y} \left( G,B\right) \cap C=\emptyset $ holds then
$B\cap C=\emptyset $ holds, and if $\mathcal{Y} \left(
G,B\right) \subseteq C$ holds then $B\subseteq C$ holds,
while if $C\subset \mathcal{Y} \left( G,B\right) $ holds
then $C$ is certainly not equal to $\mathcal{U} \left(
V\right) $, hence $C$ is either a member of $G$ or else is
a $\left( V\cup H\right) $-firm over $V$ component of some
member of $G$, and if $C$ is a member of $G$ then by the
definition of an $H $-principal wood of $V$, $C$ is a subset
of a $\left( V\cup H\right) $-firm over $V$ component of
$\mathcal{Y} \left( G,B\right) $, hence either $B\cap
C=\emptyset $ holds or $C\subseteq B$ holds, since $B$ is a
$\left( V\cup H\right) $-firm over $V$ component of
$\mathcal{Y} \left( G,B\right) $, and if $C$ is a $\left(
V\cup H\right) $-firm over $V$ component of a member $A$ of
$G$, then either $A=\mathcal{Y} \left( G,B\right) $ holds
or $A\subset \mathcal{Y} \left( G,B\right) $ holds, (since
$A$ does not overlap $\mathcal{Y} \left( G,B\right) $,
$A\cap \mathcal{Y} \left( G,B\right) $ has the nonempty
subset $C$, and if $\mathcal{Y} \left( G,B\right) \subset
A$ held then by the definition of an $H $-principal wood of
$V$, $\mathcal{Y} \left( G,B\right) $ would be a subset of
a $\left( V\cup H\right) $-firm over $V$ component of $A$,
hence $C\subset \mathcal{Y} \left( G,B\right) $ implies
that $C$ could \emph{not} be equal to a $\left( V\cup
H\right) $-firm over $V$ component of $ A $), and if
$A=\mathcal{Y} \left( G,B\right) $ holds, then both $B$ and
$C$ are $\left( V\cup H\right) $-firm over $V$ components
of $\mathcal{Y} \left( G,B\right) $, hence either $B=C$
holds or $B\cap C=\emptyset $ holds, and if $A\subset
\mathcal{Y} \left( G,B\right) $ holds, then by the
definition of an $H $-principal wood of $V$, $A$ is a subset
of a $\left( V\cup H\right) $-firm over $V$ component of
$\mathcal{Y} \left( G,B\right) $,  hence either $B\cap
C=\emptyset $ holds or $C\subseteq B$ holds, since $B$ is a
$\left( V\cup H\right) $-firm over $V$ component of
$\mathcal{Y} \left( G,B\right) $.

\vspace{2.5ex}

\noindent {\bf (iv)}  Let $B$ and $C$ be any
two distinct members of $\left(
Y\,\vdash G\right) $.   Then $\Xi \left( \mathcal{P} \left(
G,B\right) \right) \cap \Xi \left( \mathcal{P} \left(
G,C\right) \right) =\emptyset $ holds.   For by the
definition
of the
\label{Start of original page 148}
 function $\Xi $ on
page \pageref{Start of original page 3},
$\Xi \left( \mathcal{P} \left(
G,B\right) \right) $ is the set whose members are all the
subsets $D$ of the set $\mathcal{U} \left( \mathcal{P}
\left( G,B\right) \right) =B$ such that $D$ neither
overlaps any member of $\mathcal{P} \left( G,B\right) $ nor
is a strict subset of any member of $\mathcal{P} \left(
G,B\right) $, (so that if $S$ is any member of $\mathcal{P}
\left( G,B\right) $, then either $S\subseteq D$ holds or
$S\cap D=\emptyset $ holds).   And similarly, $\Xi \left(
\mathcal{P} \left( G,C\right) \right) $ is the set whose
members are all the nonempty subsets $D$ of $C$ such that
if $S$ is any member of $\mathcal{P} \left( G,C\right) $,
then either $S\subseteq D$ holds or $S\cap D=\emptyset $
holds.
 Now since $B$ and $C$ are two \emph{distinct} members of
$\left( Y\,\vdash G\right) $, at most one of $B$ and $C$
can be equal to $\mathcal{U} \left( V\right) $.   Suppose
$B=\mathcal{U} \left( V\right) $ holds.   Then $C$ is a
member of $\left( Y\,\vdash \left( V\cup \left\{ \mathcal{U}
\left( V\right) \right\} \right) \right) $, hence by (i)
above, $\mathcal{Y} \left( G,C\right) $ is a member of
$\mathbb{B} \left( G\right) $, and $C$ is a $\left( V\cup
H\right) $-firm over $V$ component of $\mathcal{Y} \left(
G,C\right) $.   Hence since, by the definition of an
$H $-principal wood of $V$, \emph{no} member of $\mathbb{B}
\left( G\right) $ is $\left( V\cup H\right) $-firm over
$V$, $C$ is a \emph{strict} subset of the member
$\mathcal{Y} \left( G,C\right) $ of $G$.   Now $B$ is a
member of $\left( Y\,\vdash G\right) $, hence
$B=\mathcal{U} \left( V\right) $ implies $\mathcal{U}
\left( V\right) $ is \emph{not} a member of $G$.   Hence
every member of $G$ is a \emph{strict} subset of
$\mathcal{U} \left( V\right) $ hence, by
page \pageref{Start of original page 5}, every
member of $G$ is a subset of some member of $\mathcal{P}
\left( G,\mathcal{U} \left( V\right) \right) =\mathcal{P}
\left( G,B\right) $, hence since $C$ is a \emph{strict}
subset of some member of $G$, every member of $\Xi \left(
\mathcal{P} \left( G,C\right) \right) $ is a \emph{strict}
subset of some member of $\mathcal{P} \left( G,B\right) $,
hence since \emph{no} member of $\Xi \left( \mathcal{P}
\left( G,B\right) \right) $ is a \emph{strict} subset of
any member of $\mathcal{P} \left( G,B\right) $, $\Xi \left(
\mathcal{P} \left( G,B\right) \right) \cap \Xi \left(
\mathcal{P} \left( G,C\right) \right) =\emptyset $ holds.
Now
suppose \emph{neither} $B$ \emph{nor} $C$ is equal to
$\mathcal{U} \left( V\right) $.   Then $B$ is a member of
$\left( Y\,\vdash \left( V\cup \left\{ \mathcal{U} \left(
V\right) \right\} \right) \right) $ hence by (i) above,
$\mathcal{Y} \left( G,B\right) $ is a member of $\mathbb{B}
\left( G\right) $, $B$ is a $\left( V\cup H\right) $-firm
over $V$ component of $\mathcal{Y} \left( G,B\right) $, and
$\mathcal{Y} \left( G,B\right) $ is the \emph{only} member
$A$ of $G$ such that $B$ is a $\left( V\cup H\right) $-firm
over $V$ component of $A$, and $C$ is a member of $\left(
Y\,\vdash \left( V\cup \left\{ \mathcal{U} \left( V\right)
\right\} \right) \right) $ hence by (i) above, $\mathcal{Y}
\left( G,C\right) $ is a member of $\mathbb{B} \left(
G\right) $, $C$ is a $\left( V\cup H\right) $-firm over $V$
component of $\mathcal{Y} \left( G,C\right) $, and
$\mathcal{Y} \left( G,C\right) $ is the \emph{only} member
$A$ of $G$ such that $C$ is a $\left( V\cup H\right) $-firm
over $V$ component of $A$.   Now since $G$ is a wood,
either $\mathcal{Y} \left( G,B\right) \cap \mathcal{Y}
\left( G,C\right) =\emptyset $ holds or $\mathcal{Y} \left(
G,B\right) =\mathcal{Y} \left( G,C\right) $ holds or
$\mathcal{Y} \left( G,B\right) \subset \mathcal{Y} \left(
G,C\right) $ holds or $\mathcal{Y} \left( G,C\right)
\subset \mathcal{Y} \left( G,B\right) $ holds.   Now if
$\mathcal{Y} \left( G,B\right) \cap \mathcal{Y} \left(
G,C\right) =\emptyset $ holds then $\Xi \left( \mathcal{P}
\left( G,B\right) \right) \cap \Xi \left( \mathcal{P}
\left( G,C\right) \right) =\emptyset $ holds.   Suppose now
that
$\mathcal{Y} \left( G,B\right) =\mathcal{Y} \left(
G,C\right) $ holds.   Then $B$ and $C$ are two distinct
$\left( V\cup H\right) $-firm over $V$ components of $A$
hence by Lemma \ref{Lemma 31}, $B\cap C=\emptyset $ holds
hence $\Xi
\left(
\mathcal{P} \left( G,B\right) \right) \cap \Xi \left(
\mathcal{P} \left( G,C\right) \right) =\emptyset $ holds.
Now
suppose $\mathcal{Y} \left( G,B\right) \subset \mathcal{Y}
\left( G,C\right) $ holds.   Then by the definition of an
$H $-principal wood of $V$, $\mathcal{Y} \left( G,B\right) $
is a subset of a $\left( V\cup H\right) $-firm over $V$
component of $\mathcal{Y} \left( G,C\right) $, hence since
$\mathcal{Y} \left( G,B\right) $ is a member of $\mathbb{B}
\left( G\right) $ hence is \emph{not}
\label{Start of original page 149}
 $\left( V\cup H\right) $-firm over $V$, hence cannot be
\emph{equal} to any $\left( V\cup H\right) $-firm over $V$
component of $\mathcal{Y} \left( G,C\right) $, $\mathcal{Y}
\left( G,B\right) $ is a \emph{strict} subset of some
$\left( V\cup H\right) $-firm over $V$ component of
$\mathcal{Y} \left( G,C\right) $, hence since $C$ is a
$\left( V\cup H\right) $-firm over $V$ component of
$\mathcal{Y} \left( G,C\right) $, either $\mathcal{Y}
\left( G,B\right) \cap C=\emptyset $ holds or $\mathcal{Y}
\left( G,B\right) $ is a \emph{strict} subset of $C$.   Now
if $\mathcal{Y} \left( G,B\right) \cap C=\emptyset $ holds
then
$B\cap C=\emptyset $ holds hence $\Xi \left( \mathcal{P}
\left(
G,B\right) \right) \cap \Xi \left( \mathcal{P} \left(
G,C\right) \right) =\emptyset $ certainly holds.   Now
suppose
that $\mathcal{Y} \left( G,B\right) \subset C$ holds.
Then, by
page \pageref{Start of original page 5},
$\mathcal{Y} \left( G,B\right) $ is a
subset of some member of $\mathcal{P} \left( G,C\right) $
hence, since $B$ is a $\left( V\cup H\right) $-firm over
$V$ component of $\mathcal{Y} \left( G,B\right) $ and
$\mathcal{Y} \left( G,B\right) $ is a member of $\mathbb{B}
\left( G\right) $ hence is \emph{not} $\left( V\cup
H\right) $-firm over $V$, hence $B\subset \mathcal{Y}
\left( G,B\right) $ holds, $B$ is a \emph{strict} subset of
some member of $\mathcal{P} \left( G,C\right) $, hence
every member of $\Xi \left( \mathcal{P} \left( G,B\right)
\right) $ is a \emph{strict} subset of some member of
$\mathcal{P} \left( G,C\right) $, hence since \emph{no}
member of $\Xi \left( \mathcal{P} \left( G,C\right) \right)
$ is a \emph{strict} subset of any member of $\mathcal{P}
\left( G,C\right) $, $\Xi \left( \mathcal{P} \left(
G,B\right) \right) \cap \Xi \left( \mathcal{P} \left(
G,C\right) \right) =\emptyset $ holds.   And finally, if
$\mathcal{Y} \left( G,C\right) \subset \mathcal{Y} \left(
G,B\right) $ holds, then $\Xi \left( \mathcal{P} \left(
G,B\right) \right) \cap \Xi \left( \mathcal{P} \left(
G,C\right) \right) =\emptyset $ holds again by an analogous
argument.

\vspace{2.5ex}

\noindent {\bf (v)}  Let $B$ be any member of
$\left( Y\,\vdash G\right) $
and $C$ be any member of $\left( Y\cap G\right) $.   Then
$C$ is \emph{not} a member of $\Xi \left( \mathcal{P}
\left( G,B\right) \right) $.   For by the definition of an
$H $-principal wood of $V$, \emph{no} member of $\mathbb{B}
\left( G\right) $ is $\left( V\cup H\right) $-firm over
$V$, hence the members of $\left( Y\cap G\right) $ are any
members of $V$ that are $\left( V\cup H\right) $-firm over
$V$ components of members of $\mathbb{B} \left( G\right) $,
together with the set $\mathcal{U} \left( V\right) $ if
$\mathcal{U} \left( V\right) $ \emph{is} a member of $G$.
Suppose first that $B$ is equal to $\mathcal{U} \left(
V\right) $.   Then $\mathcal{U} \left( V\right) $ is
\emph{not} a member of $G$, hence every member of $G$ is a
\emph{strict} subset of $\mathcal{U} \left( V\right) $,
hence by
page \pageref{Start of original page 5},
every member of $G$ is a subset of some
member of $\mathcal{P} \left( G,\mathcal{U} \left( V\right)
\right) =\mathcal{P} \left( G,B\right) $, and furthermore,
$C$ is a $\left( V\cup H\right) $-firm over $V$ component
of some member of $\mathbb{B} \left( G\right) $, hence $C$
is a \emph{strict} subset of some member of $\mathbb{B}
\left( G\right) $, hence $C$ is a \emph{strict} subset of
some member of $\mathcal{P} \left( G,B\right) $, hence $C$
is \emph{not} a member of $\Xi \left( \mathcal{P} \left(
G,B\right) \right) $.   Now assume $B$ is \emph{not} equal
to $\mathcal{U} \left( V\right) $, so $B$ is a member of
$\left( Y\,\vdash \left( V\cup \left\{ \mathcal{U} \left(
V\right) \right\} \right) \right) $, hence by (i) above,
$\mathcal{Y} \left( G,B\right) $ is a member of $\mathbb{B}
\left( G\right) $, $B$ is a $\left( V\cup H\right) $-firm
over $V$ component of $\mathcal{Y} \left( G,B\right) $, and
$\mathcal{Y} \left( G,B\right) $ is the \emph{only} member
$A$ of $G$ such that $B$ is a $\left( V\cup H\right) $-firm
over $V$ component of $A$.   Suppose first that $C$ is
equal to $\mathcal{U} \left( V\right) $.   Then since $B$
is \emph{not} equal to $\mathcal{U} \left( V\right) $,
hence $B$ is a \emph{strict} subset of $\mathcal{U} \left(
V\right) $, every member of $\Xi \left( \mathcal{P} \left(
G,B\right) \right) $ is a \emph{strict} subset of
$\mathcal{U} \left( V\right) $, hence $\mathcal{U} \left(
V\right) $ is certainly not a member of $\Xi \left(
\mathcal{P} \left( G,B\right) \right) $.   Now assume $C$
is \emph{not} equal to $\mathcal{U} \left( V\right) $, so
$C$ is a member of $V$ that is a $\left( V\cup H\right)
$-firm over $V$ component of some member of $\mathbb{B}
\left( G\right) $.   Let $A$ be a member of
\label{Start of original page 150}
 $\mathbb{B} \left( G\right) $ such that $C$ is a $\left(
V\cup H\right) $-firm over $V$ component of $A$.   Then
since $G$ is a wood, either $\mathcal{Y} \left( G,B\right)
\cap A=\emptyset $ holds or $\mathcal{Y} \left( G,B\right)
=A$
holds or $\mathcal{Y} \left( G,B\right) \subset A$ holds or
$A\subset \mathcal{Y} \left( G,B\right) $ holds, and if
$\mathcal{Y} \left( G,B\right) \cap A=\emptyset $ holds
then $C$
is certainly not a member of $\Xi \left( \mathcal{P} \left(
G,B\right) \right) $.   Suppose now that $\mathcal{Y}
\left( G,B\right) =A$ holds.   Then $B$ is a $\left( V\cup
H\right) $-firm over $V$ component of $\mathcal{Y} \left(
G,B\right) $ such that $B$ is \emph{not} a member of $V$,
and $C$ is a $\left( V\cup H\right) $-firm over $V$
component of $\mathcal{Y} \left( G,B\right) $ such that $C$
\emph{is} a member of $V$, hence since,
by Lemma \ref{Lemma 31},
distinct $\left( V\cup H\right) $-firm over $V$ components
of $\mathcal{Y} \left( G,B\right) $ do not intersect one
another, $C$ is \emph{not} a member of $\Xi \left(
\mathcal{P} \left( G,B\right) \right) $.   Now suppose that
$\mathcal{Y} \left( G,B\right) \subset A$ holds.   Then by
the definition of an $H $-principal wood of $V$,
$\mathcal{Y}
\left( G,B\right) $ is a subset of some $\left( V\cup
H\right) $-firm over $V$ component of $A$, and since
$\mathcal{Y} \left( G,B\right) $ is a member of $\mathbb{B}
\left( G\right) $ and $C$ is a member of $V$, $\mathcal{Y}
\left( G,B\right) $ is a subset of some $\left( V\cup
H\right) $-firm over $V$ component of $A$ that is
\emph{not} equal to $C$ hence,
again by Lemma \ref{Lemma 31},
$\mathcal{Y} \left( G,B\right) $ does \emph{not} intersect
$C$, hence $C$ is not a member of $\Xi \left( \mathcal{P}
\left( G,B\right) \right) $.   And finally suppose that
$A\subset \mathcal{Y} \left( G,B\right) $ holds.   Then by
the definition of an $H $-principal wood of $V$, $A$ is a
subset of some $\left( V\cup H\right) $-firm over $V$
component of $\mathcal{Y} \left( G,B\right) $, hence either
$A\cap B=\emptyset $ holds or $A\subseteq B$ holds, and if
$A\cap B=\emptyset $ holds then $C$ is certainly not a
member of
$\Xi \left( \mathcal{P} \left( G,B\right) \right) $.
Suppose now that $A\subseteq B$ holds.   Then since $A$ is
a member of $\mathbb{B} \left( G\right) $ hence is
\emph{not} $\left( V\cup H\right) $-firm over $V$, and $B$
\emph{is} $\left( V\cup H\right) $-firm over $V$, $A$ is a
\emph{strict} subset of $B$ hence, by
page \pageref{Start of original page 5}, $A$ is a
subset of some member of $\mathcal{P} \left( G,B\right) $
hence, since $C$ is a member of $V$ hence is a
\emph{strict} subset of $A$, $C$ is a \emph{strict} subset
of some member of $\mathcal{P} \left( G,B\right) $, hence
$C$ is not a member of $\Xi \left( \mathcal{P} \left(
G,B\right) \right) $.

\vspace{2.5ex}

\noindent {\bf (vi)}  Let $F$ be any member
of $\mathbb{O} \left( G,H\right)
$, or in other words, let $F$ be any member of $\mathcal{G}
\left( V,H\right) $ such that $G\subseteq \mathbb{P} \left(
F,H\right) $ holds, and let $B$ be any member of $\left(
Y\,\vdash G\right) $.   Then $F\cap \Xi \left( \mathcal{P}
\left( G,B\right) \right) $ is a member of $\mathcal{G}
\left( \mathcal{P} \left( G,B\right) ,H\right) $.   For no
member of $F\cap \Xi \left( \mathcal{P} \left( G,B\right)
\right) $ is empty, and since $G\subseteq F$ holds, every
member of $\mathcal{P} \left( G,B\right) $ is a member of
$F$, hence every member of $\mathcal{P} \left( G,B\right) $
is a member of $F\cap \Xi \left( \mathcal{P} \left(
G,B\right) \right) $ hence, since no two members of $F$
overlap one another, hence no two members of $F\cap \Xi
\left( \mathcal{P} \left( G,B\right) \right) $ overlap one
another, $F\cap \Xi \left( \mathcal{P} \left( G,B\right)
\right) $ is a wood of $\mathcal{P} \left( G,B\right) $.
And furthermore the fact that every member of $F$ is
$\left( V\cup H\right) $-connected implies that every
member of $F\cap \Xi \left( \mathcal{P} \left( G,B\right)
\right) $ is  $\left( \mathcal{P} \left( G,B\right) \cup
H\right) $-connected, for if $C$ is any member of $F\cap
\Xi \left( \mathcal{P} \left( G,B\right) \right) $ and
$\left\{ J,K\right\} $ is any partition of $C$ into two
nonempty parts, then there exists a member $E$ of $\left(
V\cup H\right) $ such that $E$ intersects both $J$ and $K$.
\label{Start of original page 151}
 Let $E$ be a member of $\left( V\cup H\right) $ such that
$E$ intersects both $J$ and $K$.   Then if $E$ is a member
of $H$, $E$ is a member of $\left( \mathcal{P} \left(
G,B\right) \cup H\right) $, while if $E$ is a member of
$V$, then $E$ is either a subset of or disjoint from each
member of $F$, hence since $C$ is a member of $\Xi \left(
\mathcal{P} \left( G,B\right) \right) $ and $E$ intersects
$C$, $E$ intersects some member of $\mathcal{P} \left(
G,B\right) $ hence is a subset of that member of
$\mathcal{P} \left( G,B\right) $, and if $S$ is the member
of $\mathcal{P} \left( G,B\right) $ that contains $E$ as a
subset, then $S$ is a member of $\left( \mathcal{P} \left(
G,B\right) \cup H\right) $ that intersects both $J$ and $K$.

\vspace{2.5ex}

\noindent {\bf (vii)}  Let $F$ be any member of
$\mathbb{O} \left( G,H\right)
$, or in other words, let $F$ be any member of $\mathcal{G}
\left( V,H\right) $ such that $G\subseteq \mathbb{P} \left(
F,H\right) $ holds.   Then $F$ is equal to the disjoint
union of the sets $F\cap \Xi \left( \mathcal{P} \left(
G,B\right) \right) $ associated with all the members $B$ of
$\left( Y\,\vdash G\right) $, together with the set $\left(
Y\cap G\right) $, or in other words, the set whose members
are the sets $F\cap \Xi \left( \mathcal{P} \left(
G,B\right) \right) $ associated with all the members $B$ of
$\left( Y\,\vdash G\right) $, together with the set $\left(
Y\cap G\right) $, is a partition of $F$.   For if $B$ and
$C$ are any two distinct members of $\left( Y\,\vdash
G\right) $, then $\Xi \left( \mathcal{P} \left( G,B\right)
\right) \cap \Xi \left( \mathcal{P} \left( G,C\right)
\right) =\emptyset $ holds by (iv) above, hence $\left(
F\cap
\Xi \left( \mathcal{P} \left( G,B\right) \right) \right)
\cap \left( F\cap \Xi \left( \mathcal{P} \left( G,C\right)
\right) \right) =\emptyset $ certainly holds.   And if $B$
is
any member of $\left( Y\,\vdash G\right) $ and $C$ is any
member of $\left( Y\cap G\right) $, then by (v) above, $C$
is \emph{not} a member of $\Xi \left( \mathcal{P} \left(
G,B\right) \right) $, hence $C$ is certainly not a member
of $F\cap \Xi \left( \mathcal{P} \left( G,B\right) \right)
$, hence if $B$ is any member of $\left( Y\,\vdash G\right)
$, then $\left( F\cap \Xi \left( \mathcal{P} \left(
G,B\right) \right) \right) \cap \left( Y\cap G\right)
=\emptyset
$ certainly holds.   Now $G\subseteq \mathbb{P} \left(
F,H\right) $ implies that if $A$ is any member of
$\mathbb{B} \left( G\right) $, and $C$ is any member of $F$
such that $C\subset A$ holds, then $C$ is a subset of some
$\left( V\cup H\right) $-firm over $V$ component of $A$.
Now let $C$ be any member of $F$, and suppose first that
there is \emph{no} member $A$ of $G$ such that $C\subset A$
holds.   Then either $C$ is equal to $\mathcal{U} \left(
V\right) $, or else $C\subset \mathcal{U} \left( V\right) $
holds and $\mathcal{U} \left( V\right) $ is \emph{not} a
member of $G$.   Suppose first that $C$ is equal to
$\mathcal{U} \left( V\right) $.   Then if $\mathcal{U}
\left( V\right) $ \emph{is} a member of $G$, $C=\mathcal{U}
\left( V\right) $ is a member of $\left( Y\cap G\right) $,
while if $\mathcal{U} \left( V\right) $ is \emph{not} a
member of $G$, then $\mathcal{U} \left( V\right) $ is a
member of $\left( Y\,\vdash G\right) $, and $C=\mathcal{U}
\left( V\right) $ is a member of $\Xi \left( \mathcal{P}
\left( G,\mathcal{U} \left( V\right) \right) \right) $
Now suppose that $C\subset \mathcal{U} \left( V\right) $
holds, hence that $\mathcal{U} \left( V\right) $ is
\emph{not} a member of $G$, since by assumption there is
\emph{no} member $A$ of $G$ such that $C\subset A$ holds.
Then $\mathcal{U} \left( V\right) $ is a member of $\left(
Y\,\vdash G\right) $.   Now $C\subseteq \mathcal{U} \left(
V\right) $ certainly holds.   Let $S$ be any member of
$\mathcal{P} \left( G,\mathcal{U} \left( V\right) \right)
$.   Then $S$ is a member of $F$ since $G\subseteq F$
holds, hence $S$ does not overlap $C$, hence since
$C\subset S$ does \emph{not} hold by assumption, either
$S\subseteq C$ holds or $S\cap C=\emptyset $ holds.   Hence
$C$
is a member of $\Xi \left( \mathcal{P} \left( G,\mathcal{U}
\left( V\right) \right) \right) $.   Now let $C$ be any
member of $F$ such that there \emph{does} exist a member
$A$ of $G$ such that $C\subset A$ holds.   Let $A$ be
\label{Start of original page 152}
 the \emph{smallest} member of $G$ to contain $C$ as a
\emph{strict} subset.   Now $C\subset A$ holds hence $A$
cannot be a member of $V$, hence $A$ is a member of
$\mathbb{B} \left( G\right) $ hence, as noted above, the
fact that $G$ is a subset of $\mathbb{P} \left( F,H\right)
$ implies that $C$ is a subset of some $\left( V\cup
H\right) $-firm over $V$ component of $A$.   Let $B$ be the
$\left( V\cup H\right) $-firm over $V$ component of $A$
that contains $C$ as a subset.   Then if $B$ is a member of
$V$, the facts that $C\subseteq B$ holds and that $C$ is a
member of $F$, hence that $C$ is a member of $\Xi \left(
V\right) $, imply that $C=B$ holds, and the facts that $A$
is a member of $\mathbb{B} \left( G\right) $ and that $B$
is a member of $V$, hence that $B$ is a member of $G$,
imply that $B$ is a member of $\left( Y\cap G\right) $,
hence that $C$ is a member of $\left( Y\cap G\right) $.
Now suppose that $B$ is \emph{not} a member of $V$, and let
$S$ be any member of $\mathcal{P} \left( G,B\right) $.
Then $S$ is a member of $F$ hence $S$ does not overlap $C$,
and $C\subset S$ cannot hold, for if $C\subset S$
\emph{did} hold then $A$ would \emph{not} be the
\emph{smallest} member of $G$ to contain $C$ as a
\emph{strict} subset, contrary to assumption.   Hence since
$C\subseteq B$ implies that $C$ is a subset of $\mathcal{U}
\left( \mathcal{P} \left( G,B\right) \right) =B$, $C$ is a
member of $\Xi \left( \mathcal{P} \left( G,B\right) \right)
$.

\vspace{2.5ex}

\noindent {\bf (viii)}  Let $J$ be any map
such that $\mathcal{D} \left(
J\right) =\left( Y\,\vdash G\right) $ holds, and such that
for each member $B$ of $\left( Y\,\vdash G\right) $, $J_{ B
} $ is a member of $\mathcal{G} \left( \mathcal{P} \left(
G,B\right) ,H\right) $.   Then the set $F\equiv \left(
 \bigcup_{B\in \left(
Y\,\vdash G\right) } J_{ B } \right) \cup \left( Y\cap
G\right) $ is a member of $\mathbb{O} \left( G,H\right)
$, or in other words, $F$ is a member of $\mathcal{G}
\left( V,H\right) $ such that $G\subseteq F$ holds, and if
$A$ is any member of $\mathbb{B} \left( G\right) $ and $C$
is any member of $F$ such that $C\subset A$ holds, then $C$
is a subset of some $\left( V\cup H\right) $-firm over $V$
component of $A$.

For we first note that $\mathcal{U} \left( F\right)
\subseteq \mathcal{U} \left( V\right) $ holds, and
furthermore, if $B$ is any member of $\left( Y\,\vdash
G\right) $, then $J_{ B } $ is a subset of $\Xi \left(
\mathcal{P} \left( G,B\right) \right) $, and $\mathcal{P}
\left( G,B\right) $ is a partition that is a subset of $\Xi
\left( V\right) $, hence by the observation at the top of
page \pageref{Start of original page 16},
$J_{ B } $ is a subset of $\Xi \left( V\right) $.

Hence, since $\left( Y\cap G\right) $ is certainly a subset
of $\Xi \left( V\right) $, $F$ is a subset of $\Xi \left(
V\right) $.

We next note that $G$ is certainly a member of $\mathbb{O}
\left( G,H\right) $, hence by (vii) above, (with the wood
$F$ of (vii) above taken as $ G $), $G$ is equal to the
disjoint union of the sets $G\cap \Xi \left( \mathcal{P}
\left( G,B\right) \right) $ for the members $B$ of $\left(
Y\,\vdash G\right) $, together with the set $\left( Y\cap
G\right) $.   But for $B\in \left( Y\,\vdash G\right) $ the
set $G\cap \Xi \left( \mathcal{P} \left( G,B\right) \right)
$ is equal to $\mathcal{P} \left( G,B\right) $, and hence
is a subset of $J_{ B } $, hence $G$ \emph{is} a subset of
$F$.

Hence, in particular, $V$ is a subset of $F$.

Now let $C$ and $D$ be any members of $F$.   We first note
that if
\label{Start of original page 153}
 either $C$ or $D$ is a member of $\left( Y\cap G\right) $,
then since $\left( Y\cap G\right) $ is a subset of $V\cup
\left\{ \mathcal{U} \left( V\right) \right\} $, and $F$ is a
subset of $\Xi \left( V\right) $, and no member of $\Xi
\left( V\right) $ overlaps either $\mathcal{U} \left(
V\right) $ or any member of $V$, $C$ and $D$ do \emph{not}
overlap.   Suppose now that neither $C$ nor $D$ is a member
of $\left( Y\cap G\right) $.   Then since, by (iv) above,
the sets $\Xi \left( \mathcal{P} \left( G,B\right) \right)
$ for distinct members $B$ of $\left( Y\,\vdash G\right) $
do not intersect one another, there exists a unique member
$S$ of $\left( Y\,\vdash G\right) $ such that $C$ is a
member of $J_{ S } $, and there exists a unique member $T$
of $\left( Y\,\vdash G\right) $ such that $D$ is a member
of $J_{ T } $.   Let $S$ be the unique member of $\left(
Y\,\vdash G\right) $ such that $C\in J_{ S } $ holds, and
let $T$ be the unique member of $\left( Y\,\vdash G\right)
$ such that $D\in J_{ T } $ holds.   Then if $S$ is equal
to $T$, both $C$ and $D$ are members of the wood $J_{ S }
=J_{ T } $, hence $C$ does not overlap $D$.   Suppose now
that $S$ is \emph{not} equal to $T$.   Then since, by (iii)
above, $G\cup Y$ is a wood of $V$, either $S\cap
T=\emptyset $
holds or $S\subset T$ holds or $T\subset S$ holds.   We
first note that if $S\cap T=\emptyset $ holds, then $C\cap
D=\emptyset $ holds.   Suppose now that $S\subset T$ holds.
Then $S$ is certainly not equal to $\mathcal{U} \left(
V\right) $, and $S$ is a member of $\left( Y\,\vdash
G\right) $ hence $S$ is not a member of $V$, hence by (i)
above, $\mathcal{Y} \left( G,S\right) $ is a member of
$\mathbb{B} \left( G\right) $, and $S$ is a $\left( V\cup
H\right) $-firm over $V$ component of $\mathcal{Y} \left(
G,S\right) $.   Suppose first that $T$ is equal to
$\mathcal{U} \left( V\right) $.   Then since $T$ is a
member of $\left( Y\,\vdash G\right) $, $\mathcal{U} \left(
V\right) $ is \emph{not} a member of $G$, hence
$\mathcal{Y} \left( G,S\right) $ is a \emph{strict} subset
of $\mathcal{U} \left( V\right) $, hence by
page \pageref{Start of original page 5}, there
is a unique member $E$ of $\mathcal{P} \left( G,\mathcal{U}
\left( V\right) \right) =\mathcal{P} \left( G,T\right) $
such that $\mathcal{Y} \left( G,S\right) \subseteq E$
holds.   Let $E$ be the unique member of $\mathcal{P}
\left( G,T\right) $ such that $\mathcal{Y} \left(
G,S\right) \subseteq E$ holds.   Then since $D$ is a member
of $\Xi \left( \mathcal{P} \left( G,T\right) \right) $,
either $E\subseteq D$ holds or $E\cap D=\emptyset $ holds,
and
if $E\subseteq D$ holds then $C\subseteq D$ holds, while if
$E\cap D=\emptyset $ holds then $C\cap D=\emptyset $ holds.
  Now
suppose that $T$ is \emph{not} equal to $\mathcal{U} \left(
V\right) $, hence that $T$ is a member of $\left( Y\,\vdash
\left( V\cup \left\{ \mathcal{U} \left( V\right) \right\}
\right) \right) $.   Then by (i) above, $\mathcal{Y} \left(
G,T\right) $ is a member of $\mathbb{B} \left( G\right) $,
and $T$ is a $\left( V\cup H\right) $-firm over $V$
component of $\mathcal{Y} \left( G,T\right) $.   Now
$S\subset T$ implies that $\mathcal{Y} \left( G,S\right)
\subseteq \mathcal{Y} \left( G,T\right) $ holds, and by
Lemma \ref{Lemma 31} the distinct $\left( V\cup H\right)
$-firm over
$V$ components of $\mathcal{Y} \left( G,T\right) $ do not
intersect one another, hence $\mathcal{Y} \left( G,S\right)
$ cannot equal $\mathcal{Y} \left( G,T\right) $, (for if
$\mathcal{Y} \left( G,S\right) =\mathcal{Y} \left(
G,T\right) $ held then $S$ and $T$ would be two distinct
$\left( V\cup H\right) $-firm over $V$ components of
$\mathcal{Y} \left( G,T\right) $ such that $S\cap T=S$ is
nonempty), hence $\mathcal{Y} \left( G,S\right) \subset
\mathcal{Y} \left( G,T\right) $ holds, hence by the
definition of an $H $-principal wood of $V$, $\mathcal{Y}
\left( G,S\right) $ is a subset of a $\left( V\cup H\right)
$-firm over $V$ component of $\mathcal{Y} \left( G,T\right)
$, hence since $\mathcal{Y} \left( G,S\right) \cap T$ has
the nonempty subset $S$, $\mathcal{Y} \left( G,S\right)
\subseteq T$ holds by Lemma \ref{Lemma 31}.   Now
$\mathcal{Y} \left(
G,S\right) $ is a member of $\mathbb{B} \left( G\right) $
hence is \emph{not} $\left( V\cup H\right) $-firm over $V$,
hence $\mathcal{Y} \left( G,S\right) $ is not equal
\label{Start of original page 154}
 to $T$, hence $\mathcal{Y} \left( G,S\right) \subset T$
holds, hence by
page \pageref{Start of original page 5}
 there exists a unique member $E$
of $\mathcal{P} \left( G,T\right) $ such that $\mathcal{Y}
\left( G,S\right) \subseteq E$ holds, (just as in the case
where $T$ was equal to $\mathcal{U} \left( V\right)
$).   Let $E$ be the unique member of $\mathcal{P} \left(
G,T\right) $ such that $\mathcal{Y} \left( G,S\right)
\subseteq E$ holds.   Then since $D$ is a member of $\Xi
\left( \mathcal{P} \left( G,T\right) \right) $, either
$E\subseteq D$ holds or $E\cap D=\emptyset $ holds, and if
$E\subseteq D$ holds then $C\subseteq D$ holds, while if
$E\cap D=\emptyset $ holds then $C\cap D=\emptyset $ holds.

And if $T\subset S$ holds, we find by an exactly analogous
argument that either $D\subseteq C$ holds or $D\cap
C=\emptyset
$ holds.

Hence $F$ is a subset of $\Xi \left( V\right) $ such that
$V\subseteq F$ holds and no two members of $F$ overlap,
hence $F$ is a wood of $V$, and furthermore $G\subseteq F$
holds, as shown on
page \pageref{Start of original page 152}.

Now every member of $G$ is $\left( V\cup H\right)
$-connected by the definition of an $H $-principal wood of
$V$, hence in particular every member of $\left( Y\cap
G\right) $ is $\left( V\cup H\right) $-connected, and
furthermore $\mathcal{U} \left( V\right) $ is $\left( V\cup
H\right) $-connected by assumption.   Now let $B$ be any
member of $\left( Y\,\vdash G\right) $ and let $C$ be any
member of $J_{ B } $.   Then $C$ is a member of $\Xi \left(
\mathcal{P} \left( G,B\right) \right) $ such that $C$ is
$\left( \mathcal{P} \left( G,B\right) \cup H\right)
$-connected.   Let $\left\{ J,K\right\} $ be any partition
of
$C$ into two nonempty parts.   Then there exists a member
$E$ of $\left( \mathcal{P} \left( G,B\right) \cup H\right)
$ such that $E$ intersects both $J$ and $K$.   Let $E$ be a
member of $\left( \mathcal{P} \left( G,B\right) \cup
H\right) $ such that $E$ intersects both $J$ and $K$.
Then if $E$ is a member of $H$, $E$ is a member of $\left(
V\cup H\right) $ such that $E$ intersects both $J$ and $K$.
  Suppose now that $E$ is a member of $\mathcal{P} \left(
G,B\right) $.   Then since $E$ intersects $C$, and $C$ is a
member of $\Xi \left( \mathcal{P} \left( G,B\right) \right)
$, $E\subseteq C$ holds, hence $\left\{ \left( E\cap
J\right) ,\left( E\cap K\right) \right\} $ is a partition of
$E$ into two nonempty parts, hence since $E$ is a member of
$G$ hence $E$ is $\left( V\cup H\right) $-connected, there
exists a member $M$ of $\left( V\cup H\right) $ such that
$M$ intersects both $\left( E\cap J\right) $ and $\left(
E\cap K\right) $, and any such member $M$ of $\left( V\cup
H\right) $ is a member of $\left( V\cup H\right) $ that
intersects both $J$ and $K$.

Hence every member of $F$ is $\left( V\cup H\right)
$-connected, hence $F$ is a member of $\mathcal{G} \left(
V,H\right) $.

Now let $A$ be any member of $\mathbb{B} \left( G\right) $
and $C$ be any member of $F$ such that $C\subset A$ holds.
 Suppose first that $C$ is a member of $\left( Y\cap
G\right) $.   Then $C$ is a member of $G$, hence by the
definition of an $H $-principal wood of $V$, $C$ is a subset
of a $\left( V\cup H\right) $-firm over $V$ component of
$A$.   Now suppose there exists a member $B$ of $\left(
Y\,\vdash G\right) $ such that $C$ is a member of $J_{ B }
$.   Then since, by (iv) above, the sets $\Xi \left(
\mathcal{P} \left( G,B\right) \right) $ for distinct
members $B$ of $\left( Y\,\vdash G\right) $ do not
intersect one another, there exists a \emph{unique} member
$B$ of $\left( Y\,\vdash G\right) $ such that $C$ is a
member of $J_{ B } $.
\label{Start of original page 155}
 Let $B$ be the unique member of $\left( Y\,\vdash G\right)
$ such that $C$ is a member of $J_{ B } $.   Now by (iii)
above, $B$ does not overlap $A$, and $B\cap A$ has the
nonempty subset $C$, hence $B\cap A$ is nonempty, and $C$
is a member of $\Xi \left( \mathcal{P} \left( G,B\right)
\right) $, hence $A\subset B$ cannot hold, for if $A\subset
B$ held then by
page \pageref{Start of original page 5},
$A$ would be a subset of some
member of $\mathcal{P} \left( G,B\right) $, hence $C\subset
A$ would imply that $C$ is a \emph{strict} subset of some
member of $\mathcal{P} \left( G,B\right) $, which
contradicts $C\in \Xi \left( \mathcal{P} \left( G,B\right)
\right) $.   Hence $B\subseteq A$ holds hence, since $A$
\emph{is} a member of $G$, and $B$ is \emph{not} a member
of $G$, hence $B$ is \emph{not} equal to $A$, $B\subset A$
holds.   Hence $B$ is certainly not equal to $\mathcal{U}
\left( V\right) $, hence since $B$ is a member of $\left(
Y\,\vdash G\right) $ hence $B$ is a member of $\left(
Y\,\vdash V\right) $, it follows from (i) above that
$\mathcal{Y} \left( G,B\right) $ is a member of $\mathbb{B}
\left( G\right) $, and that $B$ is a $\left( V\cup H\right)
$-firm over $V$ component of $\mathcal{Y} \left( G,B\right)
$.   Now $\mathcal{Y} \left( G,B\right) $ does not overlap
$A$, hence since $B\subset A$ holds and $\mathcal{Y} \left(
G,B\right) $ is the \emph{smallest} member of $G$ to
contain $B$ as a subset, $\mathcal{Y} \left( G,B\right)
\subseteq A$ holds.   Suppose first that $\mathcal{Y}
\left( G,B\right) =A$ holds.   Then $B$ is a $\left( V\cup
H\right) $-firm over $V$ component of $A=\mathcal{Y} \left(
G,B\right) $, hence since $C$ is a subset of $B$, $C$ is a
subset of a $\left( V\cup H\right) $-firm over $V$
component of $A$.   Now suppose that $\mathcal{Y} \left(
G,B\right) \subset A$ holds.   Then by the definition of an
$H $-principal wood of $V$, $\mathcal{Y} \left( G,B\right) $
is a subset of a $\left( V\cup H\right) $-firm over $V$
component of $A$, hence since $C$ is a subset of
$\mathcal{Y} \left( G,B\right) $, $C$ is a subset of a
$\left( V\cup H\right) $-firm over $V$ component of $A$.

Hence $F$ is a member of $\mathbb{O} \left( G,H\right) $.

\vspace{2.5ex}

\noindent {\bf (ix)}  Let $X$ be the set
whose members are all the maps $J$
such that $\mathcal{D} \left( J\right) $ is equal to
$\left( Y\,\vdash G\right) $, and for each member $B$ of
$\left( Y\,\vdash G\right) $, $J_{ B } $ is a member of
$\mathcal{G} \left( \mathcal{P} \left( G,B\right) ,H\right)
$, let $E$ be the map such that $\mathcal{D} \left(
E\right) =X$ holds and such that for each member $J$ of
$X$, $E_{ J } \equiv \left( \bigcup_{B\in \left( Y\,\vdash
G\right)
 } J_{ B } \right) \cup \left( Y\cap G\right) $ holds, and
let $M$ be the map whose domain is equal to $\mathbb{O}
\left( G,H\right) $, and such that for each member $F$ of
$\mathbb{O} \left( G,H\right) $, $M_{ F } $ is the map
whose domain is $\left( Y\,\vdash G\right) $, and such that
for each member $B$ of $\left( Y\,\vdash G\right) $, $M_{
FB } \equiv \left( M_{ F } \right)_{ B } $ is equal
to $F\cap \Xi \left(
\mathcal{P} \left( G,B\right) \right) $.   Then $E$ is a
\emph{bijection} whose domain is $X$ and whose range is
equal to $\mathbb{O} \left( G,H\right) $, $M$ is a
\emph{bijection} whose domain is $\mathbb{O} \left(
G,H\right) $ and whose range is equal to $X$, and $M$ is
the inverse of $E$ and $E$ is the inverse of $M$.

For by (viii) above, $\mathcal{R} \left( E\right) \subseteq
\mathbb{O} \left( G,H\right) $ holds.   Now let $F$ be any
member of $\mathbb{O} \left( G,H\right) $.   Then by (vi)
above, for each member $B$ of $\left( Y\,\vdash G\right) $,
$F\cap \Xi \left( \mathcal{P} \left( G,B\right) \right) $
is a member of $\mathcal{G} \left( \mathcal{P} \left(
G,B\right) ,H\right) $.   Hence $\mathcal{R} \left(
M\right) $ is a subset of $X$.   And furthermore, by (vii)
above, for each member $F$
\label{Start of original page 156}
 of $\mathbb{O} \left( G,H\right) $,
\[
E_{ M_{ F } } = \left( \bigcup_{B\in \left( Y\,\vdash
G\right) } M_{
FB } \right) \cup \left( Y\cap G\right) = \left( \bigcup_{
B\in \left(
Y\,\vdash
G\right) } \left( F\cap \Xi \left( \mathcal{P}
\left( G,B\right) \right) \right) \right) \cup \left( Y\cap
G\right)
\]
is equal to $F$.   Thus \emph{every} member $F$ of
$\mathbb{O} \left( G,H\right) $ is a member of $\mathcal{R}
\left( E\right) $, hence $\mathcal{R} \left( E\right) $ is
equal to $\mathbb{O} \left( G,H\right) $.   Now let $J$ and
$K$ be any two \emph{distinct} members of $X$.   Then
since, for every member $B$ of $\left( Y\,\vdash G\right)
$, $J_{ B } \subseteq \Xi \left( \mathcal{P} \left(
G,B\right) \right) $ holds and $K_{ B } \subseteq \Xi
\left( \mathcal{P} \left( G,B\right) \right) $ holds, and
by (iv) above, the sets $\Xi \left( \mathcal{P} \left(
G,B\right) \right) $ for \emph{distinct} members $B$ of
$\left( Y\,\vdash G\right) $ do \emph{not} intersect one
another, and by (v) above, the set $\left( Y\cap G\right) $
does \emph{not} intersect the set $\Xi \left( \mathcal{P}
\left( G,B\right) \right) $ for any member $B$ of $\left(
Y\,\vdash G\right) $, and furthermore, the fact that $J$ is
\emph{not} equal to $K$ implies that there is at least one
member $B$ of $\left( Y\,\vdash G\right) $ such that $J_{ B
} $ is \emph{not} equal to $K_{ B } $, $E_{ J } $ is
\emph{not} equal to $E_{ K } $.   Hence for each member $F$
of $\mathbb{O} \left( G,H\right) $, there is
\emph{exactly one} member $J$ of $X$
such that $E_{ J } $ is
equal to $F$, hence, since  $\mathcal{R} \left( E\right) $
is a subset of $\mathbb{O} \left( G,H\right) $, $E$ is a
\emph{bijection} whose domain is $X$ and whose range is
equal to $\mathbb{O} \left( G,H\right) $.   Finally we note
that the facts that $\mathcal{R} \left( E\right) $ is a
subset of $\mathbb{O} \left( G,H\right) $, $\mathcal{R}
\left( M\right) $ is a subset of $X$, and that for each
member $F$ of $\mathbb{O} \left( G,H\right) $, $E_{ M_{ F }
} $ is equal to $F$, and that for any two distinct members
$J$ and $K$ of $X$, $E_{ J } $ is \emph{not} equal to $E_{
K } $, together imply that $\mathcal{R} \left( M\right) $
is \emph{equal} to $X$, for if $\left( X\,\vdash
\mathcal{R} \left( M\right) \right) $ was nonempty and $J$
was a member of $\left( X\,\vdash \mathcal{R} \left(
M\right) \right) $, then since $M_{ E_{ J } } $ \emph{is} a
member of $\mathcal{R} \left( M\right) $, $J$ and $M_{ E_{
J } } $ would be two distinct members of $X$ such that $E_{
J } =E_{ M_{ E_{ J } } } $ held.   Hence $M$ is a
\emph{bijection} whose domain is $\mathbb{O} \left(
G,H\right) $ and whose range is equal to $X$, and $M$ is
the inverse of $E$ and $E$ is the inverse of $M$.

\begin{bphzlemma} \label{Lemma 36}
\end{bphzlemma}
\vspace{-6.143ex}

\noindent \hspace{11.9ex}{\bf.  }If $U$ is a set
such that every member of $U$ is
a set, $A$ is a $U$-connected set, $E$ is a $U$-key of $A$
such
that $\#\left( E\right) =2$, $\left\{ B,C\right\} $ is a
partition of $A$ such that $B\cap E\neq \emptyset $, $C\cap
E\neq \emptyset $, and $E$ is the \emph{only} member of $U$
to
have nonempty intersection with both $B$ and $C$, and
$\left\{ J,K\right\} $ is a partition of $A$ such that
$J\cap
E\neq \emptyset $, $K\cap E\neq \emptyset $, and $E$ is the
\emph{only} member of $U$ to have nonempty intersection
with both $J$ and $K$, then $\left\{ B,C\right\} =\left\{
J,K\right\} $.

\vspace{2.5ex}

\noindent {\bf Proof.}  Suppose $\left\{ B,C\right\}
\neq \left\{
J,K\right\}
$.   Then either $B\cap J$ and $B\cap K$ are both nonempty,
or $C\cap J$ and $C\cap K$ are both nonempty.   For $B\cap
J$ and $C\cap J$ cannot both be empty since $\left( B\cap
J\right) \cup \left( C\cap J\right) =J$, and $B\cap K$ and
\label{Start of original page 157}
 $C\cap K$ cannot both be empty since $\left( B\cap
K\right) \cup \left( C\cap K\right) =K$, and if $B\cap J$
and $C\cap K$ are both empty then $B=K$ holds and $C=J$
holds hence $\left\{ B,C\right\} =\left\{ J,K\right\} $
holds,
contrary to assumption, and if $B\cap K$ and $C\cap J$ are
both empty then $B=J$ holds and $C=K$ holds hence again
$\left\{ B,C\right\} =\left\{ J,K\right\} $ holds, contrary
to
assumption, hence every case where at least one of $B\cap
J$ and $B\cap K$ is empty \emph{and} at least one of $C\cap
J$ and $C\cap K$ is empty, is excluded.
\enlargethispage{0.1ex}

Suppose now for definiteness that $B\cap J$ and $B\cap K$
are both nonempty.   Then since $B$ is $U$-connected by
Lemma \ref{Lemma 25}, there exists a member $S$ of $U$ such
that
$\left( B\cap J\right) \cap S\neq \emptyset $ and $\left(
B\cap
K\right) \cap S\neq \emptyset $.   Now this implies that
$S$ has
at least two \emph{distinct} members that are members of
$B$, hence since by assumption $\#\left( E\right) =2$ holds
and one of the members of $E$ is a member of $C$ hence not
a member of $B$, $S$ is not equal to $E$.   But this
contradicts the assumption that $E$ is the \emph{only}
member of $U$ to have nonempty intersection with both $J$
and $K$.

\begin{bphzthm} \label{Theorem 2}
\end{bphzthm}
\vspace{-6.143ex}

\noindent \hspace{12.4ex}{\bf.  }Let $V$ be a partition
such that $\mathcal{U}
\left( V\right) $ is finite and $\#\left( V\right) \geq 2$
holds, and let $H$ be a partition such that $\mathcal{U}
\left( V\right) $ is $\left( V\cup H\right) $-connected and
such that if $E$ is any member of $H$ such that $E$
intersects \emph{more} than one member of $V$, then $E$ has
\emph{exactly} two members.   (Hence \emph{no} member of
$H$ intersects \emph{more} than two members of $V $.)

Let $W$ be the subset of $H$ whose members are all the
members $E$ of $H$ such that $E$ intersects exactly two
members of $V$.   (Thus $W$ is a partition such that every
member $E$ of $W$ has exactly two members.)

Let $d$ be an integer $\geq 1$.

Let $Z$ be the subset of $\mathbb{E} _{ d }^{
\mathcal{U} \left(
V\right) } $ whose members are all the members $y$ of
$\mathbb{E} _{ d }^{ \mathcal{U} \left( V\right) } $
such that
$\left| y_{ i } -y_{ j } \right| =0$ holds for at least one
member
$\left\{ i,j\right\} $ of $W$.

Let $\theta $ be a member of $\mathbb{Z}^{ W }$ such that
$\theta _{ \Delta } <d$ holds for every member $\Delta $
of $W$.

Let $g$ be a map such that $\mathcal{D} \left( g\right) $
is finite and $\mathcal{R} \left( g\right) \subseteq
\mathcal{U} \left( V\right) $ holds, and let $c$ be a map
such that $\mathcal{D} \left( g\right) \subseteq
\mathcal{D} \left( c\right) $ holds and for each member
$\alpha $ of $\mathcal{D} \left( g\right) $, $c_{ \alpha }
$ is a unit $d $-vector.

For any ordered pair $\left(i,j\right) $ of a map $i$ such
that
$\mathcal{D} \left( i\right) $ is finite and $\mathcal{R}
\left( i\right) \subseteq \mathcal{U} \left( V\right) $
holds, and a member $j$ of $\mathcal{U} \left( V\right) $,
let $\nu _{ ij } $ denote the number of members $\alpha $
of $\mathcal{D} \left( i\right) $ such that $i_{ \alpha }
=j$ holds.

For each member $A$ of $\left( \Xi \left( V\right) \,\vdash
V\right) $, we define
\label{Start of original page 158}
\[
D_{ A } \equiv \left( \sum_{\Delta \equiv
\left\{ j,k\right\} \in \left(
W\cap \mathcal{Q} \left( A\right) \right) }
\left( \theta _{ \Delta } +\nu _{ gj } +\nu _{ gk }
\right) \right)
 - d \left( \#\left( \mathcal{P} \left( V,A\right)
\right) -1\right).
\]

Let $N\equiv \#\left( \mathcal{D} \left( g\right) \right)
+ \sum_{A\in \left( \Xi \left( V\right) \,\vdash V\right) }
\left( 1+\max\left( D_{ A } ,0\right) \right) $.

Let $M$ be a finite real number $\geq 0$, and let $S$ and
$T$ be finite real numbers such that $0<S<T$ holds.

Let $\mathcal{J} $ be a map whose domain is $\mathcal{G}
\left( V,H\right) $, and such that for each member $F$ of
$\mathcal{G} \left( V,H\right) $, $\mathcal{J} _{ F } $ is
a map whose domain is $\left( \mathbb{E} _{ d }^{
\mathcal{U}
\left( V\right) } \,\vdash Z\right) $ and whose range
is a subset of $\mathbb{R} $.

For each member $F$ of $\mathcal{G} \left( V,H\right) $,
and for each member $y$ of $\left( \mathbb{E} _{
d }^{ \mathcal{U} \left( V\right) } \,\vdash Z\right) $,
we define $\mathcal{J} _{ F } \left( y\right) \equiv \left(
\mathcal{J} _{ F } \right) _{ y } $.

Let $\mathcal{J} $ satisfy the requirement that if $F$ and
$G$ are members of $\mathcal{G} \left( V,H\right) $, and
$y$ is a member of $\left( \mathbb{E} _{ d }^{ \mathcal{U}
\left( V\right) } \,\vdash Z\right) $, such that for
every member $\left\{ i,j\right\} $ of $W$, either $\left|
y_{ i }
-y_{ j } \right| \leq S$ holds or $\mathcal{Y} \left(
F,\left\{
i,j\right\} \right) =\mathcal{Y} \left( G,\left\{
i,j\right\}
\right) $ holds, then $\mathcal{J} _{ F } \left( y\right)
=\mathcal{J} _{ G } \left( y\right)
\rule[-1.7ex]{0pt}{1.7ex} $ holds.

Let $\mathcal{J} $ also satisfy the requirement that for
each member $F$ of $\mathcal{G} \left( V,H\right) $,
$\mathcal{J} _{ F } \left( y\right) $ and all its
derivatives with respect to the $y_{ i } $, $i\in
\mathcal{U} \left( V\right) $, of degree up to and
including $N$, exist and are continuous for all $y\in
\left( \mathbb{E} _{ d }^{ \mathcal{U} \left( V\right) }
\,\vdash Z\right) $.

For each member $S$ of $W$, we define the $ d $-vector
differential operator $t_{ S } $ by $t_{ S } \equiv \left(
\hat{ y }_{ j }
+\hat{ y }_{ k } \right) $, where $j$ and $k$ are the two
members
of $S$, or in other words, $S=\left\{ j,k\right\} $, and
the $ d $-vector index has not been shown.

And let $\mathcal{J} $ also satisfy the requirement that if
$i$ is any map such that $\mathcal{D} \left( i\right) $ is
finite and $\mathcal{R} \left( i\right) \subseteq
\mathcal{U} \left( V\right) $ holds, $u$ is any map such
that $\mathcal{D} \left( i\right) \subseteq \mathcal{D}
\left( u\right) $ holds and for each member $\alpha $ of
$\mathcal{D} \left( i\right) $, $u_{ \alpha } $ is a unit
$d $-vector, $E$ is any map such that $\mathcal{D} \left(
E\right) $ is finite and $\mathcal{R} \left( E\right)
\subseteq W$ holds, and $v$ is any map such that
$\mathcal{D} \left( E\right) \subseteq \mathcal{D} \left(
v\right) $ holds and for each member $\beta $ of
$\mathcal{D} \left( E\right) $, $v_{ \beta } $ is a unit
$d $-vector, and if the maps $i$ and $E$ furthermore satisfy
the requirement that $\#\left( \mathcal{D} \left( i\right)
\right) +\#\left( \mathcal{D} \left( E\right) \right) \leq
N$ holds, then the following inequality holds for all
$F\in
\mathcal{G} \left( V,H\right) $ and for all $y\in \left(
\mathbb{E} _{ d }^{ \mathcal{U} \left( V\right) } \,\vdash
Z\right) $:
\[
\left| \left(
 \prod_{\beta \in \mathcal{D} \left( E\right) }
\left( v_{ \beta } .t_{ E_{ \beta } } \right) \right)
\left( \prod_{\alpha
\in
\mathcal{D} \left( i\right) } \left( u_{ \alpha
} .\hat{ y }_{ i_{ \alpha } } \right) \right)
 \mathcal{J} _{ F } \left(
y\right) \right| \leq M \prod_{\Delta \equiv \left\{
j,k\right\}
\in
W } \left| y_{ j } -y_{ k } \right|^{ -\left( \theta _{
\Delta }
+\nu _{ ij } +\nu _{ ik } \right) } ,
\]
where in accordance with the definition on
page \pageref{Start of original page 158}, and
for any member $m$ of $\mathcal{U} \left( V\right) $, $\nu
_{ im } $ denotes the number of members $\alpha $ of
$\mathcal{D} \left( i\right) $ such that $i_{ \alpha } =m$
holds.

And let $\mathcal{J} $ also satisfy the requirement that if
$y$ is any member
\label{Start of original page 159}
 of $\left( \mathbb{E} _{ d }^{ \mathcal{U}
 \left( V\right) }
\,\vdash Z\right) $ such that $\left| y_{ i } -y_{ j }
\right| \geq
T$ holds for any member $\left\{ i,j\right\} $ of $W$, then
$\mathcal{J} _{ F } \left( y\right) =0$ holds for all
members $F$ of $\mathcal{G} \left( V,H\right) $.

Let $\tilde { \mathcal{J} } $ be any map such that
$\mathcal{D}
\left( \tilde{ \mathcal{J} } \right) =\mathbb{N} $ holds,
and
such that for each member $k $ of $\mathbb{N} $,
$\tilde{ \mathcal{J} }
_{ k } $ is a map such that $\tilde{ \mathcal{J} } _{ k } $
has all the properties
 assumed above for the map $\mathcal{J}
$, and such that furthermore, for every
member $k $ of $\mathbb{N} $, and every member $F $ of
$\mathcal{G} \left(
V,H\right) $ ,and \emph{every} member $y $ of $\mathbb{E}_{
d }^{ \mathcal{U} \left( V\right) } $, $\mathcal{J} _{ kF
} \left( y\right) \equiv \left( \left( \mathcal{J} _{ k }
\right) _{ F } \right) _{ y } $ is defined, continuous and
continuously differentiable with respect to the $y_{ i } $,
$i\in
\mathcal{U} \left( V\right) $, up to total degree $N$.

And let $\tilde{ \mathcal{J} } $
 satisfy the requirement that for any given real number
$\varepsilon >0$ and any given real number $r>0$, there
exists a member $k$ of $\mathbb{N} $ such that for all
integers
$m\geq
k,$ and for all members $y $ of $\mathbb{E} _{
 d }^{ \mathcal{U} \left(
V\right) } $ such that $\left| y_{ i } -y_{ j } \right| \geq
r $ holds for all members $\left\{ i,j\right\}
$ of $W $, and for all maps $i $ and
$u$ such that $\mathcal{D} \left( i\right) $ is finite,
$\#\left( \mathcal{D} \left( i\right) \right) \leq N$ holds,
$\mathcal{R} \left( i\right) \subseteq \mathcal{U} \left(
V\right) $ holds, $\mathcal{D} \left( i\right) \subseteq
\mathcal{D} \left( u\right) $ holds, and for each member
 $\alpha$ of $
\mathcal{D} \left( i\right) $, $u_{ \alpha }
$ is a unit $d $-vector,  $\left| \left(
\prod_{ \alpha \in \mathcal{D}
\left( i\right)} u_{ \alpha } .\hat{y}_{ i_{ \alpha } }
\right)
 \left( \tilde{ \mathcal{J} }
_{ mF } \left( y\right) -\mathcal{J} _{ F } \left( y\right)
\right) \right| \leq \varepsilon  $ holds for all members
$F$
of $\mathcal{G} \left( V,H\right) $.

Let $\omega $ be any set of contraction weights for $V$,
let $h$ be any member of $\mathcal{U} \left( V\right) $,
let $O\equiv \mathcal{C} \left( V,h\right) $, let $b$ be
any member of $\mathbb{E} _{ d } $, and let $\mathbb{W}$ be
the subset of $\mathbb{U} _{ d } \left( V,\omega \right) $
whose members are all the members $x$ of $\mathbb{U} _{ d }
\left( V,\omega \right) $ such that $x_{ O } =b$ holds.

For each member $k$ of $\mathbb{N} $, let $I_{ k } $ be
defined by:
\[
I_{ k }\equiv \int_{ \mathbb{W} } \left( \prod_{A\in
 \left( V\,\vdash \left\{
O\right\} \right) } d^{ d } x_{ A } \right) \left(
 \sum_{F\in
\mathcal{V}
\left( V,H\right) } \left( -1\right)^{ \#\left(
\mathbb{B} \left( F\right) \right) } \sum_{n\in \mathbb{X}
\left( \mathbb{I} \left( F,H\right) ,D\right) }
\times \rule{0pt}{5.5ex} \right. \hspace{-24.0pt}
\hspace{5.0cm}
\]
\[
\left. \times \left( \hspace{-0.05cm}
\left(
  \prod_{\left( i,A\right) \in \mathcal{U} \left(
  \mathcal{R} \left(
\mathbb{I} \left( F,H\right) \right) \right) }
\hspace{-0.05cm}
\left( \frac{
\left( \left( x_{ \mathcal{K} \left( F,A,i\right) } -x_{ A
} \right) .\hat{ y }_{ i } \right)^{ n_{ iA } } }{
n_{ iA } ! } \right) \hspace{-0.05cm} \right)
\hspace{-0.05cm} \left(
\prod_{\alpha \in
\mathcal{D} \left( \! g\right) } c_{ \alpha } .\hat{ y
}_{ g_{
\alpha } } \! \right) \tilde{ \mathcal{J} } _{ kF }
\left( y\right) \right)_{ \hspace{-0.05cm}
y=\eta
\left( F,H,x\right) } \right)
\]

Then for each member $k$ of $\mathbb{N} $, $I_{ k } $ is a
finite real number, and the $I_{ k } $, $k\in \mathbb{N} $,
form a Cauchy sequence, or in other words, for any given
real number $\varepsilon >0$, there exists a member $k$ of
$\mathbb{N} $ such that for all members $l$ and $m$ of
$\mathbb{N} $ such that $l\geq k$ and $m\geq k$ both hold,
$\left| I_{ l } -I_{ m } \right| \leq \varepsilon $
holds.
(We note that by the completeness of $\mathbb{R} $, it
follows directly
from this that there exists a unique finite real number $I$
such that for any given real number $\varepsilon >0$, there
exists a member $k$ of $\mathbb{N} $ such that for all
members $l$ of $\mathbb{N} $ such that $l\geq k$ holds,
$\left| I_{ l } -I\right| \leq \varepsilon $ holds.)

\vspace{2.5ex}

\noindent {\bf Proof.}  We first note that it
follows directly from the
assumed properties of $\tilde{ \mathcal{J} } $
that for each member $k $ of $\mathbb{N} $ , $I_{ k } $
is a finite real number,
\label{Start of original page 160}
hence it remains to prove that the $I_{ k } $,
$k\in \mathbb{N} $, form a Cauchy sequence.

For each member $k$ of $\mathbb{N} $, and for each
$H $-principal wood $G$ of $V$, we define
$\tilde{ I } _{ kG } $
to be given by the same expression as $I_{ k } $, but with
the sum over the members $F$ of $\mathcal{V} \left(
V,H\right) $ replaced by the sum over the members $F$ of
$\mathbb{O} \left( G,H\right) $.
Then by Lemma \ref{Lemma 34}, the
following identity holds for each member $k$ of $\mathbb{N}
$:
\[
I_{ k } =   \sum_{G\in \mathcal{W} \left( V,H\right) }
\left( -1\right)^{ \#\left( \mathbb{B} \left( G\right)
\right) } \tilde{ I }_{ kG } .
\]

Now $\mathcal{W} \left( V,H\right) $ is a finite set, hence
it will be sufficient to prove that for every $H $-principal
wood $G$ of $V$, the $\tilde{ I }_{ kG } $, $k\in
\mathbb{N} $, form
a Cauchy sequence.   We shall prove that if $G$ is any
$H $-principal wood of $V$, then the $\tilde{ I }_{ kG } $,
$k\in
\mathbb{N} $, form a Cauchy sequence.

Let $G$ be any $H $-principal wood of $V$, let $Y$ be the
set
whose members are all the $\left( V\cup H\right) $-firm
over $V$ components of members of $\mathbb{B} \left(
G\right) $, together with the set $\mathcal{U} \left(
V\right) $, let $X$ be the set whose members are all the
maps $J$ such that $\mathcal{D} \left( J\right) $ is equal
to $\left( Y\,\vdash G\right) $, and for each member $B$ of
$\left( Y\,\vdash G\right) $, $J_{ B } $ is a member of
$\mathcal{G} \left( \mathcal{P} \left( G,B\right) ,H\right)
$, and let $E$ be the map such that $\mathcal{D} \left(
E\right) =X$ holds and such that for each member $J$ of
$X$, $E_{ J } \equiv \left( \bigcup_{B\in \left(
Y\,\vdash G\right)
 } J_{ B } \right) \cup \left( Y\cap G\right) $ holds.
Then by Lemma \ref{Lemma 35} (ix), and for each member
$k$ of $\mathbb{N} $, $\tilde{ I }_{ kG }$ is equal to
\[
\int_{ \mathbb{W} } \left(
 \prod_{A\in \left( V\,\vdash
\left\{ O\right\}
\right) } d^{ d } x_{ A } \right) \left(
 \sum_{J\in X } \left(
-1\right)^{ \#\left( \mathbb{B} \left( E_{ J } \right)
\right) }
 \sum_{n\in \mathbb{X} \left( \mathbb{I} \left( E_{J}
,H\right)
,D\right) } \rule{0pt}{5.5ex}
\times \right. \hspace{-16.0pt} \hspace{6.0cm}
\]
\[
\left. \times \hspace{-3.0pt}
 \left( \hspace{-5.0pt} \left(
 \prod_{\left( i,A\right) \in \mathcal{U}
\left( \mathcal{R} \left( \mathbb{I} \left( E_{ J },
H\right)
\right) \right) } \hspace{-5.0pt}
\left( \frac{\left( \left( x_{ \mathcal{K} \left(
E_{ J },A,i\right) } -x_{ A } \right) .\hat{ y }_{ i }
\right)^{
n_{
iA } } }{ n_{ iA } ! } \right) \hspace{-5.0pt}
\right) \hspace{-5.0pt} \left(
\prod_{\alpha \in \mathcal{D} \left( g\right)
 } \hspace{-3.0pt}
 c_{ \alpha } .\hat{ y }_{ g_{ \alpha } } \hspace{-3.0pt}
 \right)
  \tilde{ \mathcal{J} } _{
kE_{ J }
} \left( y\right) \right)_{ \hspace{-5.0pt}
y=\eta \left( E_{ J },H,x\right) }
\right)
\]

We now choose a real number $\sigma $ such that
$0<\sigma \leq \frac{3}{25}  $ holds, and we define the
real number $\lambda $ by $\lambda \equiv  \left(\frac{1}{4}
\right) \left( 1-\sqrt{1-8\sigma} \right) $, so
that $0<\lambda \leq \frac{1}{5}
$ holds.

We note that $\lambda $ and $\sigma $ satisfy the equation
$\lambda =\frac{\sigma}{1-2\lambda } $, and that
$0<\sigma <\lambda $ holds.

And we choose a real number $R$ such that $0<R\leq \left(
1-2\lambda \right) S$ holds.

Then for any member $B$ of $\left( Y\,\vdash G\right) $,
for any member $J_{ B } $ of $\mathcal{G} \left(
\mathcal{P} \left( G,B\right) ,H\right) $, and for any
member $x$ of $\mathbb{U} _{ d } \left( V,\omega \right) $,
the following identity holds by
page \pageref{Start of original page 48}:
\[
\sum_{\left( P_{ B } ,Q_{ B } \right) \in \mathcal{N}
\left(
\mathcal{P} \left( G,B\right) ,H\right) }
\mathcal{E} \left( P_{ B } ,Q_{ B } ,H,\sigma ,R,\downarrow
\left( x,\Xi \left( \mathcal{P} \left( G,B\right) \right)
\right) \right) \mathcal{S} \left( P_{ B } ,J_{ B } \right)
\mathcal{S} \left( J_{ B } ,Q_{ B } \right) =1
\]
\label{Start of original page 161}

We define $ \tilde{ X }$ to be the set whose members are
all
the ordered pairs $\left(P,Q\right) $ of members
of $X$ such that for
every member $B$ of $\left( Y\,\vdash G\right) $, $\left(
P_{ B } ,Q_{ B } \right) $ is a member of $\mathcal{N}
\left( \mathcal{P} \left( G,B\right) ,H\right) $.   (Thus
$\tilde { X } $ is the set whose members are all
 the ordered pairs
$\left(P,Q\right) $ of members of $X$ such that for
every member $B$ of
$\left( Y\,\vdash G\right) $, $P_{ B } \subseteq Q_{ B } $
holds.)   Then it immediately follows from the
preceding identity that for every member $J$ of $X$, and
for every member $x$ of $\mathbb{U} _{ d } \left( V,\omega
\right) $, the following identity holds:
\[
   \sum_{\left( P,Q\right) \in \tilde{ X } }  \prod_{B\in
   \left(
Y\,\vdash G\right) } \left( \mathcal{E} \left( P_{
B } ,Q_{ B } ,H,\sigma ,R,\downarrow \left( x,\Xi \left(
\mathcal{P} \left( G,B\right) \right) \right) \right)
\mathcal{S} \left( P_{ B } ,J_{ B } \right) \mathcal{S}
\left( J_{ B } ,Q_{ B } \right) \right) =1
\]

Now by (iv) and (v) of Lemma \ref{Lemma 35}, $ \prod_{B\in
\left( Y\,\vdash
G\right) } \mathcal{S} \left( P_{ B } ,J_{ B }
\right) =\mathcal{S} \left( E_{ P } ,E_{ J } \right) $
holds and $  \prod_{B\in \left( Y\,\vdash G\right) }
\mathcal{S} \left( J_{ B } ,Q_{ B } \right) =\mathcal{S}
\left( E_{ J } ,E_{ Q } \right) $ holds.

Hence if we insert the above identity into a sum over the
members $J$ of $X$, and take the sum over the
 members $ \left(P ,Q\right) $
of $\tilde{ X } $ outside the sum over the members
 $J$ of $X$,
then in the term associated with the member
 $ \left(P ,Q\right) $ of $ \tilde { X }
 $, the sum over the members $J$ of $X$ reduces to a sum
over those members $J$ of $X$ such that $E_{ J } \in
\mathbb{K} \left( E_{ P } ,E_{ Q } \right) $ holds.   And
furthermore, by Lemma \ref{Lemma 35} (ix), it directly
follows from the fact that $\left(P,Q\right) $ is a member
of $\tilde{ X
} $,
that for \emph{every} member $F$ of $\mathbb{K} \left( E_{
P } ,E_{ Q } \right) $, there is \emph{exactly one} member
$J$ of $X$ such that $E_{ J } =F$ holds, and the member $J$
of $X$ corresponding to the member $F$ of $\mathbb{K}
\left( E_{ P } ,E_{ Q } \right) $ is given by $J=M_{ F } $,
where $M$ is the map defined in Lemma \ref{Lemma 35} (ix).
   Hence in the term associated with the member
$\left(P,Q\right) $ of
$\tilde{ X } $, the sum over the members $J$ of $X$ reduces
to
the sum over the members $F$ of $\mathbb{K} \left( E_{ P }
,E_{ Q } \right) $.

We now define $\tilde{ \mathbb{O} } \left( G,H\right) $ to
be the set whose members are all the ordered
pairs $\left(P,Q\right) $ of
members $P$ and $Q$ of $\mathbb{O} \left( G,H\right) $ such
that $P\subseteq Q$ holds.   Then it immediately follows
from (iv), (v), and (ix) of Lemma \ref{Lemma 35},
that $\tilde { \mathbb{O} }
 \left( G,H\right) $ is equal to the set of all ordered
pairs $\left(P,Q\right) $ of members of
 $\mathbb{O} \left( G,H\right) $
such that for every member $B$ of $\left( Y\,\vdash
G\right) $, $\left( P\cap \Xi \left( \mathcal{P} \left(
G,B\right) \right) \right) \subseteq \left( Q\cap \Xi
\left( \mathcal{P} \left( G,B\right) \right) \right) $
holds, and furthermore that the set $\tilde{ \mathbb{O} }
\left( G,H\right) $ is in one-to-one correspondence with
the set $\tilde{ X } $, with the correspondences being
given by
the map $E$ and the map $M$ defined in
Lemma \ref{Lemma 35} (ix).

Hence for each member $k$ of $\mathbb{N} $,
$\tilde{ I }_{ kG }$ is
equal to
\label{Start of original page 162}
\[
\int_{ \mathbb{W} } \left( \prod_{A\in \left( V\,\vdash
\left\{ O\right\}
\right) } d^{ d } x_{ A } \right) \left(
 \sum_{\left( P,Q\right)
\in
\tilde{ \mathbb{O} } \left( G,H\right) } \left(
\rule{0pt}{5.5ex} \left( \prod_{B\in \left(
Y\,\vdash G\right) } \times \right. \right. \right.
\hspace{6.0cm}
\]
\[
\left. \times \mathcal{E} \left( \left(
P\cap \Xi \left( \mathcal{P} \left( G,B\right) \right)
\right) ,\left( Q\cap \Xi \left( \mathcal{P} \left(
G,B\right) \right) \right) ,H,\sigma ,R,\downarrow \left(
x,\Xi \left( \mathcal{P} \left( G,B\right) \right) \right)
\right) \rule[-3.0ex]{0pt}{8.0ex} \right) \times
\]
\[
\times \sum_{F\in \mathbb{K} \left( P,Q\right) } \left(
-1\right)^{ \#\left( \mathbb{B} \left( F\right) \right) }
 \sum_{n\in \mathbb{X} \left( \mathbb{I} \left( F,H\right)
,D\right) } \left( \left(
   \prod_{\left( i,A\right) \in \mathcal{U}
\left( \mathcal{R} \left( \mathbb{I} \left( F,H\right)
\right) \right) }
\left( \frac{\left( \left( x_{ \mathcal{K}
\left( F,A,i\right) } -x_{ A } \right) .\hat{ y }_{ i }
\right)^{
n_{ iA } } }{ n_{ iA } ! } \right) \right) \times \right.
\]
\[
\hspace{8.0cm} \left. \left. \left. \times
\left( \prod_{\alpha \in \mathcal{D}
\left( g\right)
 } c_{ \alpha } .\hat{ y }_{ g_{ \alpha } } \right)
  \tilde{ \mathcal{J} } _{
kF
} \left( y\right) \right)_{ y=\eta \left( F,H,x\right) }
\right) \right)
\]

We shall prove that for each member $\left(P,Q\right) $ of
the finite set
$\tilde{ \mathbb{O} } \left( G,H\right) $, the integrals
over
$\mathbb{W}$ of the term associated with $\left(P,Q\right)
$ in the
integrands of these integrals for the members $k$ of
$\mathbb{N} $, form a Cauchy sequence.

We first use Lemma \ref{Lemma 22} to conclude that the
term
associated with $\left(P,Q\right) $ in the integrand of the
above
integral for the member $k$ of $\mathbb{N} $ is equal to
\[
\left( -1\right)^{ \#\left( \mathbb{B} \left( P\right)
\right) } \times \hspace{12.0cm}
\]
\[
\times \left( \prod_{B\in \left( Y\,\vdash G\right) } \!
\mathcal{E} \left( \left( P\cap \Xi \left( \mathcal{P}
\left( G,B\right) \right) \right) ,\left( Q\cap \Xi \left(
\mathcal{P} \left( G,B\right) \right) \right) ,H,\sigma
,R,\downarrow \! \left( x,\Xi \left( \mathcal{P} \left(
G,B\right) \right) \right) \right) \right) \times
\]
\[
\times  \sum_{u\in \mathbb{X} \left( \downarrow \left(
\mathbb{I} \left( Q,H\right) ,P\right) ,D\right)
} \hspace{0.2cm}
 \sum_{m\in \mathbb{A} \left( \mathbb{J} \left(
P,Q,H\right)
,\left( D-\xi \left( P,Q,H,u\right)
+1\hspace{-0.7516ex}1 \right) \right) } \int_{
\mathbb{D} } \left( d^{ \#\left( Q\,\vdash P\right) }
 \rho
\right) \times \hspace{4.0cm}
\]
\[
\times \left( \left(
 \prod_{A\in \left( Q\,\vdash P\right) } \left( 1-\rho _{ A
} \right)^{ \left( D_{ A } -\xi _{ A } \left( P,Q,H,u\right)
\right) } \left( D_{ A } -\xi _{ A } \left( P,Q,H,u\right)
+1\right) \right) \rule{0pt}{5.5ex} \times \right.
\]
\[
\times \left(
 \prod_{\left( \left( i,B\right) ,X\right) \in
\mathcal{U} \left( \mathcal{R} \left( \psi \left(
\mathbb{J} \left( P,Q,H\right) \right) \right) \right)
 } \left( \prod_{E\in \left( \mathbb{G} _{ iB }
 \left( Q,H\right)
\,\vdash X\right) } \rho _{ E } \right)^{ m_{ iBX } }
\right) \times
\]
\[
\times \left( \left(
\prod_{\left( \left( i,B\right) ,X\right) \in
\mathcal{U} \left( \mathcal{R} \left( \psi \left(
\mathbb{J} \left( P,Q,H\right) \right) \right) \right)
 } \left( \frac{
  \left( \left( x_{ \mathcal{K} \left( Q,B,i\right)
 } -x_{ B } \right) .\hat{ y }_{ i }
 \right)^{ m_{ iBX } } }{ m_{
iBX }
! } \right) \right) \times \right.
\]
\[
\times \left(
\prod_{\left( i,A\right) \in \mathcal{U} \left(
\mathcal{R}
\left( \downarrow \left( \mathbb{I} \left( Q,H\right)
,P\right) \right) \right) }
\left( \frac{\left( \left( x_{
\mathcal{K} \left( Q,A,i\right) } -x_{ A }
\right) .\hat{ y
}_{ i
} \right)^{ u_{ iA } } }{ u_{ iA } ! } \right) \right)
\times
\]
\[
\hspace{8.0cm} \hspace{-45.0pt} \left. \left. \times
\left( \prod_{\alpha \in \mathcal{D}
\left(
g\right) } c_{ \alpha } .\hat{ y }_{ g_{ \alpha } }
\right)
\tilde{ \mathcal{J} }
_{ kQ } \left( y\right) \right)_{ y=\mu \left(
P,Q,H,x,\mathcal{X} \left( P,Q,H,\rho \right) \right) }
\right)
\]
where $1\hspace{-0.7516ex}1 $ is a map such that $\left(
Q\,\vdash P\right) \subseteq \mathcal{D} \left(
1\hspace{-0.7516ex}1 \right) $ holds and such that for
each member $A$ of $\left( Q\,\vdash P\right) $,
$1\hspace{-0.7516ex}1 _{ A } =1$ holds, and $\mathbb{D}
$ is the set of all members $\rho $ of $\mathbb{R}^{ \left(
Q\,\vdash P\right) } $ such that $0\leq \rho _{ A } \leq 1$
holds for every member $A$ of $\left( Q\,\vdash P\right) $.

We shall prove, for each ordered
pair $\left(u,m\right) $ of a member $u$
of $\mathbb{X} \left( \downarrow \left( \mathbb{I} \left(
Q,H\right) ,P\right) ,D\right) $ and a member $m$ of
$\mathbb{A} \left( \mathbb{J} \left( P,Q,H\right) ,\left(
D-\xi \left( P,Q,H,u\right) +1\hspace{-0.7516ex}1
\right) \right) $, that the integrals over $\mathbb{W} $,
of
the term in the above formula associated with $u$ and $m$,
for the members $k$ of $\mathbb{N} $, form a Cauchy
sequence.
\label{Start of original page 163}

 Let $\left(P,Q\right) $ be any member of $\tilde{
\mathbb{O} } \left(
G,H\right) $, let $u$ be any member of $\mathbb{X} \left(
\downarrow \! \left( \mathbb{I} \left(
Q,\! H\right) ,\! P\right)
,\! D\right) $, and let $m$ be any member of $\mathbb{A}
\left( \mathbb{J} \left( P,Q,H\right) ,\left( D-\xi \left(
P,Q,H,u\right) +1\hspace{-0.7516ex}1 \right) \right) $.

We first note that in consequence of the assumed regularity
of $\tilde{ \mathcal{J} } _{ kQ } \left( y\right) $, we may
swap the order of the $\rho $-integration and the
$x $-integration, and we now do this, so that we now do the
$x $-integrals before the $\rho $-integrals.

For any member $E$ of $W$, we define the differential
operator $t_{ E } $, as on
page \pageref{Start of original page 158},
by $t_{ E } \equiv
\left( \hat{ y }_{ j } +\hat{ y }_{ k } \right) $, where
$j$ and $k$ are
the two members of $E$, or in other words, $E=\left\{
j,k\right\} $.

For any ordered septuple $\left(B,i,n,j,s,E,v\right) $
of a map $B$ such
that $\mathcal{D} \left( B\right) $ is finite and
$\mathcal{R} \left( B\right) \subseteq \mathbb{B} \left(
Q\right) $ holds, a map $i$ such that $\mathcal{D} \left(
B\right) \subseteq \mathcal{D} \left( i\right) $ holds and
for each member $\alpha $ of $\mathcal{D} \left( B\right)
$, $i_{ \alpha } $ is a member of $B_{ \alpha } $, a map
$n$ such that $\mathcal{D} \left( B\right) \subseteq
\mathcal{D} \left( n\right) $ holds and for each member
$\alpha $ of $\mathcal{D} \left( B\right) $, $n_{ \alpha }
$ is a unit $d $-vector, a map $j$ such that $\mathcal{D}
\left( j\right) $ is finite and $\mathcal{R} \left(
j\right) \subseteq \mathcal{U} \left( V\right) $ holds, a
map $s$ such that $\mathcal{D} \left( j\right) \subseteq
\mathcal{D} \left( s\right) $ holds and for each member
$\beta $ of $\mathcal{D} \left( j\right) $, $s_{ \beta } $
is a unit $d $-vector, a map $E$ such that $\mathcal{D}
\left(
E\right) $ is finite and $\mathcal{R} \left( E\right)
\subseteq W$ holds, and a map $v$ such that $\mathcal{D}
\left( E\right) \subseteq \mathcal{D} \left( v\right) $
holds and for each member $\gamma $ of $\mathcal{D} \left(
E\right) $, $v_{ \gamma } $ is a unit $d $-vector, we
define,
for every member $k$ of $\mathbb{N} $, and for every member
$\rho $ of $\mathbb{D} $, (or in other words, for every
member $\rho $ of $\mathbb{R}^{
\left( Q\,\vdash P\right) } $
such that $0\leq \rho _{ A } \leq 1$ holds for every member
$A$ of $\left( Q\,\vdash P\right) $),
\[
I \rule{0pt}{2.55ex}^{ \hspace{-0.75ex} \circ }_{ k }
\left( B,i,n,j,s,E,v,\rho \right) \equiv
\int_{
\mathbb{W} } \left( \prod_{A\in \left( V\,\vdash \left\{
O\right\} \right)
 } d^{ d } x_{ A } \right) \times \hspace{6.0cm}
\]
\[
\times \hspace{-2.0pt} \left( \hspace{-2.0pt} \left(
  \prod_{B\in \left( Y\,\vdash G\right)
 } \hspace{-6.0pt}
 \mathcal{E} \left( \left( P\cap \Xi \left(
\mathcal{P} \left( G,B\right) \right) \right) ,\left( Q\cap
\Xi \left( \mathcal{P} \left( G,B\right) \right) \right)
,H,\sigma ,R,\downarrow \hspace{-3.0pt}
 \left( x,\Xi \left( \mathcal{P}
\left( G,B\right) \right) \right) \right) \right)
\hspace{-2.0pt} \rule{0pt}{5.5ex} \times
\right.
\]
\[
\times \left( \left(
 \prod_{\alpha \in \mathcal{D} \left( B\right) }
\left( \left( x_{ \mathcal{K} \left( Q,B_{ \alpha } ,i_{
\alpha } \right) } -x_{ B_{ \alpha } }
\right) .n_{ \alpha }
\right) \right) \left(
  \prod_{\beta \in \mathcal{D} \left( j\right) }
\left( s_{ \beta } .\hat{ y }_{ j_{ \beta } }
\right) \right)
\left( \prod_{\gamma \in
\mathcal{D} \left( E\right) }
\left( v_{ \gamma } .t_{
E_{ \gamma } } \right) \right) \times \right.
\]
\[
\hspace{10.0cm} \hspace{-15.2pt} \left. \left. \times
\rule{0pt}{5.0ex}
 \tilde{ \mathcal{J} } _{ kQ } \left(
y\right) \right)_{ y=\mu \left( P,Q,H,x,\mathcal{X} \left(
P,Q,H,\rho \right) \right) } \right)
\]

(We note that the definition of
$I \rule{0pt}{2.55ex}^{ \hspace{-0.75ex} \circ } _{ k }
\left(
B,i,n,j,s,E,v,\rho \right) $ depends implicitly on $G$,
$H$, $P$, and $Q $.)

Then by introducing a complete set of $d$ orthonormal unit
$d $-vectors, (for example the $d$ unit $d $-vectors
parallel to
the coordinate axes in the positive directions), we find
that for each member $k$ of $\mathbb{N} $, the term in our
integral associated with the member $u$ of $\mathbb{X}
\left( \downarrow \left( \mathbb{I} \left( Q,H\right)
,P\right) ,D\right) $ and the member $m$ of
\label{Start of original page 164}
 $\mathbb{A} \left( \mathbb{J} \left( P,Q,H\right) ,\left(
D-\xi \left( P,Q,H,u\right) +1\hspace{-0.7516ex}1
\right) \right) $, is equal to the integral with respect to
$\rho $, over $\mathbb{D} $, of the $\rho $-dependent factor
\[
\left(
 \prod_{A\in \left( Q\,\vdash P\right) } \left( 1-\rho _{ A
} \right)^{ \left( D_{ A } -\xi _{ A } \left( P,Q,H,u\right)
\right) } \left( D_{ A } -\xi _{ A } \left( P,Q,H,u\right)
+1\right) \right) \times \hspace{3.0cm}
\]
\[
\hspace{4.0cm} \times \left(
 \prod_{\left( \left( i,A\right) ,X\right) \in
\mathcal{U} \left( \mathcal{R} \left( \psi \left(
\mathbb{J} \left( P,Q,H\right) \right) \right) \right)
 } \left( \prod_{C\in \left( \mathbb{G} _{ iA }
 \left( Q,H\right)
\,\vdash X\right) } \rho _{ C } \right)^{m_{ iAX } }
\right)
\]
multiplied by the sum, over a \emph{finite} number of
ordered septuples $\left(B,i,n,j,s,E,v\right) $ as
on
page \pageref{Start of original page 163}, of a
finite coefficient, independent of $\rho $ and $k$, times
$I \rule{0pt}{2.55ex}^{ \hspace{-0.75ex} \circ } _{ k }
\left( B,i,n,j,s,E,v,\rho \right) $, and
that furthermore, and by analogy with observations
\ref{Observation 1})
to \ref{Observation 21})
on
pages \pageref{Start of original page 101} to
\pageref{Start of original page 111}, every
septuple $\left(B,i,n,j,s,E,v\right) $ that
occurs in this sum satisfies
the following condition:

For each member $A$ of $\mathbb{B} \left( Q\right) $, the
number of members $\beta $ of $\mathcal{D} \left( j\right)
$ such that $j_{ \beta } \in \mathcal{U} \left( W\right) $
and $\left\{ \mathcal{Z} \left( P,H,j_{ \beta } \right)
,\mathcal{Z} \left( P,H,l\right) \right\} \in \mathcal{Q}
\left( \mathcal{P} \left( P,A\right) \right) $ both hold,
where $l$ is the other member of the unique member
$\mathcal{C} \left( W,j_{ \beta } \right) $ of $W$ that
has $j_{ \beta } $ as a member, is less than or equal to
the number of members $\alpha $ of $\mathcal{D} \left(
B\right) $ such that $B_{ \alpha } \in \left( \Xi \left(
\mathcal{P} \left( P,A\right) \right) \,\vdash \mathcal{P}
\left( P,A\right) \right) $ holds, plus the number of
members $\alpha $ of $\mathcal{D} \left( g\right) $ such
that $g_{ \alpha } \in \mathcal{U} \left( W\right) $ and \\
$\left\{ \mathcal{Z} \left( P,H,g_{ \alpha } \right)
,\mathcal{Z} \left( P,H,l\right) \right\} \in \mathcal{Q}
\left( \mathcal{P} \left( P,A\right) \right) $ both hold,
where $l$ is the other member of the unique member
$\mathcal{C} \left( W,g_{ \alpha } \right) $ of $W$ that
has $g_{ \alpha } $ as a member, plus
\[
\left\{ \begin{array}{c}
\left( -\left( D_{ A } +1\right) + \left(
\displaystyle \sum_{C\in \left( \mathcal{P}
\left( P,A\right) \,\vdash V\right) } \hspace{0.2cm}
\displaystyle \sum_{\left(
i,K\right) \in \mathbb{I} _{ C } \left( Q,H\right) }
u_{ iK } \right) \right) \hspace{1.0cm}
 \mathrm{if}\,\,A\in \left( Q\,\vdash P\right)
\mathrm{holds}, \\[5.5ex]
\left( \left( \displaystyle \sum_{C\in
\left( \mathcal{P} \left( P,A\right) \,\vdash V\right)
 } \hspace{0.2cm} \displaystyle
  \sum_{\left( i,K\right) \in \left( \mathbb{I} _{ C }
\left( Q,H\right) \,\vdash \mathbb{I} _{ A } \left(
Q,H\right) \right) } u_{ iK } \right) - \left(
\displaystyle \sum_{i\in \mathcal{T}
\left( A,H\right) } \hspace{0.2cm}
\displaystyle \sum_{K\in \mathbb{Y} \left(
Q,\mathcal{K} \left( P,A,i\right) ,A\right) } u_{ iK
} \right) \right) \rule[-7.0ex]{0pt}{7.0ex}
_{ \hspace{-4.0cm}
\textstyle{ \mathrm{if}\,\,A\in \mathbb{B} \left( P\right)
\mathrm{holds.} } }
\end{array} \right.
\]

(We recall that the map $g$ was introduced on
page \pageref{Start of original page 157},
and occurs in the differential operator $ \left(
 \prod_{\alpha \in \mathcal{D} \left( g\right) }
c_{ \alpha } .\hat{ y }_{ g_{
\alpha } } \right) $ in the integand of
the integral on
page \pageref{Start of original page 159}.)

We note that we may also assume that all septuples
$\left(B,i,n,j,s,E,v\right) $ occurring in the above
sum are such that
$\mathcal{D} \left( E\right) $ is equal to the empty set
$\emptyset $, (so that $E$ is equal to the empty map
$\emptyset
 $), but we do not use this fact, except in so far as
it ensures that the total number of derivatives acting on
$\tilde{ \mathcal{J} } _{ kQ } \left( y\right) $ in the
\label{Start of original page 165}
 integrand of any occurring $I \rule{0pt}{2.55ex}^{
 \hspace{-0.75ex} \circ } _{ k } \left(
B,i,n,j,s,E,v,\rho \right) $ is not greater than the
integer $N$ defined on
page \pageref{Start of original page 158}.

We shall prove that if $\left(B,i,n,j,s,E,v\right) $ is
any ordered
septuple as on
page \pageref{Start of original page 163},
such that the above condition
holds, and such that $\left( \#\left( \mathcal{D}
\left( j\right) \right) +\#\left( \mathcal{D} \left(
E\right) \right) \right) $ is not greater than the integer
$N$ defined on
page \pageref{Start of original page 158},
then the integrals with respect
to $\rho $, over $\mathbb{D} $, of the $\rho $-dependent
factor displayed at the bottom of
page \pageref{Start of original page 164}, times
$I \rule{0pt}{2.55ex}^{ \hspace{-0.75ex} \circ }
_{ k } \left( B,i,n,j,s,E,v,\rho \right) $, for the members
$k$ of $\mathbb{N} $, form a Cauchy sequence.

Let $\rho $ be any member of $\mathbb{D} $, or in other
words, let $\rho $ be any member of $\mathbb{R}^{ \left(
Q\,\vdash P\right) } $ such that $0\leq \rho _{ A } \leq 1$
holds for every member $A$ of $\left( Q\,\vdash P\right) $.
  We define new integration variables for the integration
with respect to $x$ over $\mathbb{W} $, in two steps, as
follows:

We first follow
pages \pageref{Start of original page 117} to
\pageref{Start of original page 119}
exactly, and choose a
map $S$ such that $\mathcal{D} \left( S\right) =\mathbb{B}
\left( \bar{ P } \right) $, and such that for every member
$A$
of $\mathbb{B} \left( \bar{ P } \right) $, $S_{ A } $ is a
member of $\mathcal{P} \left( P,A\right) $, and then
define, for every member $B$ of $\left( P\,\vdash \left\{
\mathcal{U} \left( V\right) \right\} \right) $, $z_{ B }
\equiv \left( x_{ B } -x_{ S_{ A } } \right) $, where $A$
is the \emph{smallest} member $C$ of $ \bar{ P } $ such that
$B\subset C$ holds, or in other words, where $A=\mathcal{Y}
\left( \overline{ \left( P\,\vdash \left\{ B\right\}
\right) } ,B\right) $
holds, and then choose, as our first new set of integration
variables, the $z_{ B } $, $B\in \left( P\,\vdash \left(
\mathcal{R} \left( S\right) \cup \left\{ \mathcal{U} \left(
V\right) \right\} \right) \right) $, exactly as on
pages \pageref{Start of original page 118}
and \pageref{Start of original page 119},
and note furthermore, exactly as on
pages \pageref{Start of original page 118}
and \pageref{Start of original page 119},
that the linear transformation to this
first new set of integration variables has determinant
equal to $1$.

We next define a wood $J$ of $V$ such that $P\subseteq J$
holds and $J\cup Q$ is a wood of $V$, as follows.   We
first choose a map $f$ such that $\mathcal{D} \left(
f\right) $ is equal to $\mathbb{B} \left( G\right) $, and
such that for every member $A$ of $\mathbb{B} \left(
G\right) $, $f_{ A } $ is a member of $S_{ A } $, and we
note that, since $P$ is a member of $\mathbb{O} \left(
G,H\right) $, hence for each member $A$ of $\mathbb{B}
\left( G\right) $, $S_{ A } $ is a subset of a $\left(
V\cup H\right) $-firm over $V$ component of $A$, this
definition has the immediate consequence that for every
member $A$ of $\mathbb{B} \left( G\right) $, $S_{ A } $ is
a subset of the $\left( V\cup H\right) $-firm over $V$
component of $A$ that has $f_{ A } $ as a member.   Then we
define $J$ to be the set whose members are the members of
$P$ plus, for each ordered pair
$\left(A,T\right) $ of a member $A$ of
$\mathbb{B} \left( G\right) $ and a $\left( V\cup H\right)
$-key $T$ of $A$ such that $T$ is \emph{not} a member of
$V$, the set $\left( A\,\vdash \uparrow \left( f_{ A }
,\left\{ T\right\} \right) \right) $, where
$\uparrow \left(
f_{ A } ,\left\{ T\right\} \right) $ is
defined to be the set
whose members are all the members $l$ of $A$ such that for
every partition $\left\{ B,C\right\} $ of $A$ into two
nonempty parts $B$ and $C$ such that $T$ intersects both
$B$ and $C$ and $T$ is the \emph{only} member of $\left(
V\cup H\right) $ to intersect both $B$ and $C$, $l$ is a
member
\label{Start of original page 166}
 of the \emph{same} member of $\left\{ B,C\right\} $ as $f_{
A } $ is.   (We note that if $A$ is any $\left( V\cup
H\right) $-connected member of $\Xi \left( V\right) $,
hence in particular if $A$ is any member of $\mathbb{B}
\left( G\right) $, and $T$ is any $\left( V\cup H\right)
$-key of $A$ such that $T$ is \emph{not} a member of $V$,
then it follows directly from our assumptions on $V$ and
$H$ that $T$ is a member of $W$ hence has exactly two
members, hence by Lemma \ref{Lemma 36} there is exactly one
partition
$\left\{ B,C\right\} $ of $A$ into two parts such that $T$
intersects both $B$ and $C$ and $T$ is the \emph{only}
member of $\left( V\cup H\right) $ to intersect both $B$
and $C $.)

To check that $J$ and $J\cup Q$ are woods of $V$, we first
note that if $A$ is any member of $\mathbb{B} \left(
G\right) $ and $T$ is any $\left( V\cup H\right) $-key of
$A$ such that $T$ is \emph{not} a member of $V$, then each
$\left( V\cup H\right) $-firm over $V$ component of $A$ is
either a subset of $\left( A\,\vdash \uparrow \left( f_{ A
} ,\left\{ T\right\} \right) \right) $ or else does not
intersect $\left( A\,\vdash \uparrow \left( f_{ A } ,\left\{
T\right\} \right) \right) $, hence by
Lemma \ref{Lemma 35} (ii), Lemma \ref{Lemma 31}, and
page \pageref{Start of original page 16}, $\left( A\,\vdash
\uparrow \left( f_{ A } ,\left\{ T\right\} \right) \right) $
is a member of $\Xi \left( V\right) $.   Hence every member
of $J$ is a member of $\Xi \left( V\right) $ hence, since
$V$ is certainly a subset of $J$, it remains to check that
no two members of $J\cup Q$ overlap one another.   Now
certainly no two members of $Q$ overlap one another.
Suppose now that $B$ is any member of $Q$, $A$ is any
member of $\mathbb{B} \left( G\right) $, and $T$ is any
$\left( V\cup H\right) $-key of $A$ such that $T$ is
\emph{not} a member of $V$.   Then either $B\cap
A=\emptyset $
holds or $A\subseteq B$ holds or $B\subset A$ holds, and if
$B\cap A=\emptyset $ holds then $B\cap \left( A\,\vdash
\uparrow
\left( f_{ A } ,\left\{ T\right\} \right) \right)
=\emptyset $
holds and if $A\subseteq B$ holds then $\left( A\,\vdash
\uparrow \left( f_{ A } ,\left\{ T\right\} \right) \right)
\subseteq B$ holds, while if $B\subset A$ holds then since
$Q$ is a member of $\mathbb{O} \left( G,H\right) $, $B$ is
a subset of a $\left( V\cup H\right) $-firm over $V$
component of $A$, hence either $B\subseteq \left( A\,\vdash
\uparrow \left( f_{ A } ,\left\{ T\right\} \right) \right) $
holds or $B\cap \left( A\,\vdash \uparrow \left( f_{ A }
,\left\{ T\right\} \right) \right) =\emptyset $ holds.   Now
suppose that $A$ and $B$ are two \emph{distinct} members of
$\mathbb{B} \left( G\right) $, $T$ is any $\left( V\cup
H\right) $-key of $A$ such that $T$ is \emph{not} a member
of $V$, and $R$ is any $\left( V\cup H\right) $-key of $B$
such that $R$ is \emph{not} a member of $V$.   Then either
$B\cap A=\emptyset $ holds or $A\subset B$ holds or
$B\subset A$
holds, and if $B\cap A=\emptyset $ holds then $\left(
B\,\vdash
\uparrow \left( f_{ B } ,\left\{ R\right\} \right) \right)
\cap \left( A\,\vdash \uparrow \left( f_{ A } ,\left\{
T\right\} \right) \right) =\emptyset $ holds, and if
$A\subset B$
holds then since $G$ is an $H $-principal wood of $V$, $A$
is
a subset of a $\left( V\cup H\right) $-firm over $V$
component of $B$, hence either $\left( A\,\vdash \uparrow
\left( f_{ A } ,\left\{ T\right\} \right) \right) \subseteq
\left( B\,\vdash \uparrow \left( f_{ B } ,\left\{ R\right\}
\right) \right) $ holds or $\left( B\,\vdash \uparrow
\left( f_{ B } ,\left\{ R\right\} \right) \right) \cap
\left(
A\,\vdash \uparrow \left( f_{ A } ,\left\{ T\right\} \right)
\right) =\emptyset $ holds, and similarly if $B\subset A$
holds
then either $\left( B\,\vdash \uparrow \left( f_{ B }
,\left\{ R\right\} \right) \right) \subseteq \left(
A\,\vdash
\uparrow \left( f_{ A } ,\left\{ T\right\} \right) \right) $
holds or $\left( B\,\vdash \uparrow \left( f_{ B } ,\left\{
R\right\} \right) \right) \cap \left( A\,\vdash \uparrow
\left( f_{ A } ,\left\{ T\right\} \right) \right)
=\emptyset $
holds.   And finally suppose $A$ is any member of
$\mathbb{B} \left( G\right) $ and $T$ and $R$ are any two
\emph{distinct} $\left( V\cup H\right) $-keys of $A$ such
that neither $T$ nor $R$ is a member of $V$.   Then by
analogy with Lemma \ref{Lemma 28},
with the $i$ of Lemma \ref{Lemma 28} taken
as $f_{ A } $, we define a
\label{Start of original page 167}
 relation $\to $ among the $\left( V\cup H\right) $-keys of
$A$ as follows: if $B$ and $C$ are $\left( V\cup H\right)
$-keys of $A$, then $B\to C$ holds ifif there \emph{exists}
a partition of $A$ into two nonempty parts such that $C$
intersects both parts, $C$ is the \emph{only} member of
$\left( V\cup H\right) $ to intersect both parts, and $B$
does \emph{not} intersect the part that has $f_{ A } $ as a
member.   Then by Lemma \ref{Lemma 28} (i), at most one
of $T\to R$ and $R\to T$ can hold.   Suppose first that
$T\to R$ holds and that $\left\{ B,C\right\} $ is a
partition
of $A$ into two nonempty parts such that $f_{ A } \in B$
holds, $R$ intersects both $B$ and $C$, $R$ is the
\emph{only} member of $\left( V\cup H\right) $ to intersect
both $B$ and $C$, and $T$ does \emph{not} intersect $B$.
Then it directly follows from the definition of $\uparrow
\left( f_{ A } ,\left\{ R\right\} \right) $ that $\uparrow
\left( f_{ A } ,\left\{ R\right\} \right) \subseteq B$
holds,
hence that $C\subseteq \left( A\,\vdash \uparrow \left( f_{
A } ,\left\{ R\right\} \right) \right) $ holds.   Now let
$\left\{ D,E\right\} $ be any partition of $A$ into two
nonempty parts such that $f_{ A } \in D$ holds, $T$
intersects both $D$ and $E$, and $T$ is the \emph{only}
member of $\left( V\cup H\right) $ to intersect both $D$
and $E$.   Then since $R\neq T$ holds by assumption, $R$
does \emph{not} intersect both $D$ and $E$.   Now if $R\cap
D$ was empty, then since $T\cap B$ is empty, $D\cap B$
would be empty by Lemma \ref{Lemma 26}.   But $D\cap B$ has
the
member $f_{ A } $ hence is nonempty, hence $R\cap E$ is
empty, hence by Lemma \ref{Lemma 26}, $B\cap E$ is empty,
hence
$B\subseteq D$ holds.   Hence for every partition $\left\{
D,E\right\} $ of $A$ into two nonempty parts such that $f_{
A } \in D$ holds, $T$ intersects both $D$ and $E$, and $T$
is the \emph{only} member of $\left( V\cup H\right) $ to
intersect both $D$ and $E$, $B\subseteq D$ holds.   Hence
$B\subseteq \uparrow \left( f_{ A } ,\left\{ T\right\}
\right) $ holds, hence $\left( A\,\vdash \uparrow \left(
f_{ A } ,\left\{ T\right\} \right) \right) \subseteq C$
holds, hence $\left( A\,\vdash \uparrow \left( f_{ A }
,\left\{ T\right\} \right) \right) \subseteq \left(
A\,\vdash
\uparrow \left( f_{ A } ,\left\{ R\right\} \right) \right) $
holds, and furthermore, since $\uparrow \left( f_{ A }
,\left\{ R\right\} \right) \subseteq B$ holds and $T\cap B$
is empty, $T\cap \uparrow \left( f_{ A } ,\left\{ R\right\}
\right) $ is empty, hence since, by
Lemma \ref{Lemma 28} (viii), $T$ \emph{does} intersect
$\uparrow \left(
f_{ A } ,\left\{ T\right\} \right) $, $\uparrow \left( f_{ A
} ,\left\{ R\right\} \right) $ is not equal to $\uparrow
\left( f_{ A } ,\left\{ T\right\} \right) $, hence $\left(
A\,\vdash \uparrow \left( f_{ A } ,\left\{ T\right\} \right)
\right) \subset \left( A\,\vdash \uparrow \left( f_{ A }
,\left\{ R\right\} \right) \right) $ holds.   And
analogously, if $R\to T$ holds, then $\left( A\,\vdash
\uparrow \left( f_{ A } ,\left\{ R\right\} \right) \right)
\subset \left( A\,\vdash \uparrow \left( f_{ A } ,\left\{
T\right\} \right) \right) $ holds.   Now suppose, finally,
that \emph{neither} $T\to R$ \emph{nor} $R\to T$ holds.
Then for every partition $\left\{ B,C\right\} $ of $A$ into
two nonempty parts such that $R$ intersects both parts and
$R$ is the \emph{only} member of $\left( V\cup H\right) $
to intersect both parts, $T$ \emph{does} intersect the part
that has $f_{ A } $ as a member, hence since, by
Lemma \ref{Lemma 28} (viii), $\left\{ \uparrow \left( f_{ A
}
,\left\{ R\right\} \right) ,\left( A\,\vdash \uparrow \left(
f_{ A } ,\left\{ R\right\} \right) \right) \right\} $ is
such
a partition of $A$ and $f_{ A } \in \uparrow \left( f_{ A }
,\left\{ R\right\} \right) $ holds, $T\cap \uparrow \left(
f_{ A } ,\left\{ R\right\} \right) $ is nonempty, hence,
since by assumption $T\neq R$ holds, $T\cap \left(
A\,\vdash \uparrow \left( f_{ A } ,\left\{ R\right\} \right)
\right) $ \emph{is} empty.   And analogously, $R\cap \left(
A\,\vdash \uparrow \left( f_{ A } ,\left\{ T\right\} \right)
\right) $ is empty.   Hence by Lemma \ref{Lemma 26}, $\left(
A\,\vdash \uparrow \left( f_{ A } ,\left\{ R\right\} \right)
\right) \cap \left( A\,\vdash \uparrow \left( f_{ A }
,\left\{ T\right\} \right) \right) =\emptyset $ holds.

We next note that if $A$ is any member of $\mathbb{B}
\left( G\right) $, $T$ is any $\left( V\cup H\right) $-key
of $A$ such that $T$ is \emph{not} a member of $V$, and we
define $Z$
\label{Start of original page 168}
 to be the set whose members are all the $\left( V\cup
H\right) $-keys $R$ of $A$ such that $R\notin V$ and $R\to
T$ both hold, where the relation $\to $ among the $\left(
V\cup H\right) $-keys of $A$ is defined as in the preceding
paragraph, then the set $\left( A\,\vdash \uparrow \left(
f_{ A } ,\left\{ T\right\} \right) \right) \,\vdash \left(
 \bigcup_{R\in
Z } \left( A\,\vdash \uparrow \left( f_{ A }
,\left\{ R\right\} \right) \right) \right) $, or in other
words, the
set $\left( A\,\vdash \uparrow \left( f_{ A } ,\left\{
T\right\} \right) \right) \cap \left( \bigcap_{ R\in Z }
\uparrow \left( f_{
A } ,\left\{ R\right\} \right) \right) $, is a $\left( V\cup
H\right) $-firm over $V$ component of $A$, (and hence, in
particular, is nonempty).   For by
Lemma \ref{Lemma 28} (viii), $T\cap \left( A\,\vdash
\uparrow \left( f_{
A } ,\left\{ T\right\} \right) \right) $ is nonempty.   Let
$i$ be any member of $T\cap \left( A\,\vdash \uparrow
\left( f_{ A } ,\left\{ T\right\} \right) \right) $, and let
$X$ be the set whose members are \emph{all} the $\left(
V\cup H\right) $-keys of $A$ that are not members of $V$.
We first note that if $R$ is any member of $X$, then each
$\left( V\cup H\right) $-firm over $V$ component of $A$ is
either a subset of $\left( A\,\vdash \uparrow \left( f_{ A
} ,\left\{ R\right\} \right) \right) $ or else does not
intersect $\left( A\,\vdash \uparrow \left( f_{ A } ,\left\{
R\right\} \right) \right) $, hence each $\left( V\cup
H\right) $-firm over $V$ component of $A$ is either a
subset of $\left( A\,\vdash \uparrow \left( f_{ A } ,\left\{
T\right\} \right) \right) \,\vdash \left( \bigcup_{R\in Z }
 \left(
A\,\vdash \uparrow \left( f_{ A } ,\left\{ R\right\} \right)
\right) \right) $ or else does not intersect $\left(
A\,\vdash
\uparrow \left( f_{ A } ,\left\{ T\right\} \right) \right)
\,\vdash \left( \bigcup_{R\in Z } \left( A\,\vdash \uparrow
 \left(
f_{ A } ,\left\{ R\right\} \right) \right) \right) $.
Now let $j$
be any member of $\left( A\,\vdash \uparrow \left( f_{ A }
,\left\{ T\right\} \right) \right) \,\vdash \left(
 \bigcup_{ R\in Z }
\left( A\,\vdash \uparrow \left( f_{ A } ,\left\{ R\right\}
\right) \right) \right) $, let $R$ be any member of $X$,
and let
$\left\{ B,C\right\} $ be any partition of $A$ into two
nonempty parts such that $R$ intersects both parts and $R$
is the \emph{only} member of $\left( V\cup H\right) $ to
intersect both parts, and suppose furthermore that $f_{ A }
\in B$ holds.   Suppose first that $R$ is \emph{not} a
member of $Z$, hence that $R\to T$ does \emph{not} hold.
Then since by Lemma \ref{Lemma 28} (viii), $\left\{
\uparrow \left( f_{ A } ,\left\{ T\right\} \right) ,\left(
A\,\vdash \uparrow \left( f_{ A } ,\left\{ T\right\} \right)
\right) \right\} $ is a partition of $A$ into two nonempty
parts such that $T$ intersects both parts and $T$ is the
\emph{only} member of $\left( V\cup H\right) $ to intersect
both parts, and moreover $f_{ A } $ is a member of
$\uparrow \left( f_{ A } ,\left\{ T\right\} \right) $, $R$
\emph{does} intersect $\uparrow \left( f_{ A } ,\left\{
T\right\} \right) $, hence either $R=T$ holds or else $R$
does \emph{not} intersect $\left( A\,\vdash \uparrow \left(
f_{ A } ,\left\{ T\right\} \right) \right) $.   Suppose
first
that $R=T$ holds.   Then since it follows directly from our
assumptions on $V$ and $H$ that $T$ is a member of $W$,
hence that $\#\left( T\right) =2$ holds, it follows directly
from Lemma \ref{Lemma 36} that there is \emph{exactly one}
partition
of $A$ into two nonempty parts such that $T$ intersects
both parts and $T$ is the \emph{only} member of $\left(
V\cup H\right) $ to intersect both parts.   Hence $B$ is
equal to $\uparrow \left( f_{ A } ,\left\{ T\right\} \right)
$ and $C$ is equal to $\left( A\,\vdash \uparrow \left( f_{
A } ,\left\{ T\right\} \right) \right) $, hence, since
\label{Start of original page 169}
 $\left( A\,\vdash \uparrow \left( f_{ A } ,\left\{
T\right\}
\right) \right) \,\vdash \left( \bigcup_{R\in Z } \left(
A\,\vdash
\uparrow \left( f_{ A } ,\left\{ R\right\} \right) \right)
\right) $
is a subset of $\left( A\,\vdash \uparrow \left( f_{ A }
,\left\{ T\right\} \right) \right) $ hence both $i$ and $j$
are members of $\left( A\,\vdash \uparrow \left( f_{ A }
,\left\{ T\right\} \right) \right) $, $i$ and $j$ are
members
of the \emph{same} member of $\left\{ B,C\right\} $.   Now
suppose that $R$ is \emph{not} equal to $T$, hence that $R$
does \emph{not} intersect $\left( A\,\vdash \uparrow \left(
f_{ A } ,\left\{ T\right\} \right) \right) $, and
furthermore
that $T$ does \emph{not} intersect both $B$ and $C$.   Then
it directly follows from Lemma \ref{Lemma 26} that one of
$B\cap
\left( A\,\vdash \uparrow \left( f_{ A } ,\left\{ T\right\}
\right) \right) $ and $C\cap \left( A\,\vdash \uparrow
\left( f_{ A } ,\left\{ T\right\} \right) \right) $ is
empty,
hence that $\left( A\,\vdash \uparrow \left( f_{ A }
,\left\{ T\right\} \right) \right) $ is a subset of one
member of $\left\{ B,C\right\} $, hence again that $i$ and
$j$ are members of the \emph{same} member of $\left\{
B,C\right\} $.   Now suppose that $R$ \emph{is} a member of
$Z$, hence that $R\to T$ holds, (hence that $R$ is
\emph{not} equal to T).   We first note that by
Lemma \ref{Lemma 28} (viii), $\left\{ \uparrow \left( f_{ A
}
,\left\{ R\right\} \right) ,\left( A\,\vdash \uparrow \left(
f_{ A } ,\left\{ R\right\} \right) \right) \right\} $ is a
partition of $A$ into two nonempty parts such that $R$
intersects both parts and $R$ is the \emph{only} member of
$\left( V\cup H\right) $ to intersect both parts, hence $T$
does \emph{not} intersect both $\uparrow \left( f_{ A }
,\left\{ R\right\} \right) $ and $\left( A\,\vdash \uparrow
\left( f_{ A } ,\left\{ R\right\} \right) \right) $, and
furthermore $R$ does not intersect $\uparrow \left( f_{ A }
,\left\{ T\right\} \right) $, for $R\to T$ implies that
there
exists a partition $\left\{ D,E\right\} $ of $A$ into two
nonempty parts such that $f_{ A } \in D$ holds, $T$
intersects both parts, $T$ is the \emph{only} member of
$\left( V\cup H\right) $ to intersect both parts, and $R$
does \emph{not} intersect $D$, and if $\left\{ D,E\right\} $
is such a partition of $A$, then by the definition of
$\uparrow \left( f_{ A } ,\left\{ T\right\} \right) $,
$\uparrow \left( f_{ A } ,\left\{ T\right\} \right)
\subseteq
D$ holds, hence $R$ does not intersect $\uparrow \left( f_{
A } ,\left\{ T\right\} \right) $.
Hence by Lemma \ref{Lemma 26}, if
$T\cap \uparrow \left( f_{ A } ,\left\{ R\right\} \right) $
was empty, then $\uparrow \left( f_{ A } ,\left\{ R\right\}
\right) \cap \uparrow \left( f_{ A } ,\left\{ T\right\}
\right) $ would be empty, but $\uparrow \left( f_{ A }
,\left\{ R\right\} \right) \cap \uparrow \left( f_{ A }
,\left\{ T\right\} \right) $ has the member $f_{ A } $ hence
is nonempty, hence $T\cap \left( A\,\vdash \uparrow \left(
f_{ A } ,\left\{ R\right\} \right) \right) $ is empty.
Hence, since $i$ is a member of $T\cap \left( A\,\vdash
\uparrow \left( f_{ A } ,\left\{ T\right\} \right) \right) $
hence is a member of $T$, $i$ is \emph{not} a member of
$\left( A\,\vdash \uparrow \left( f_{ A } ,\left\{ R\right\}
\right) \right) $, hence, since this is true for \emph{all}
members $R$ of $Z$, $i$ \emph{is} a member of $\left(
A\,\vdash \uparrow \left( f_{ A } ,\left\{ T\right\} \right)
\right) \,\vdash \left( \bigcup_{R\in Z } \left( A\,\vdash
\uparrow \left( f_{ A } ,\left\{ R\right\} \right) \right)
\right) $,
hence $\left( A\,\vdash \uparrow \left( f_{ A } ,\left\{
T\right\} \right) \right) \,\vdash \left( \bigcup_{R\in Z }
\left(
A\,\vdash \uparrow \left( f_{ A } ,\left\{ R\right\} \right)
\right) \right) $ is nonempty.   And furthermore, with $R$
 still
being any member of $Z$, and $\left\{ B,C\right\} $ being
any
partition of $A$ into two nonempty parts such that $f_{ A }
\in B$ holds, $R$ intersects both parts, and $R$ is the
\emph{only} member of $\left( V\cup H\right) $ to intersect
both parts, it follows directly from the definition of
$\uparrow \left( f_{ A } ,\left\{ R\right\} \right) $ that
$\uparrow \left( f_{ A } ,\left\{ R\right\} \right)
\subseteq
B$ holds, hence since, as just shown, $i$ is a member of
$\left( A\,\vdash \uparrow \left( f_{ A } ,\left\{ T\right\}
\right) \right) \,\vdash \left( \bigcup_{R\in Z }
\left( A\,\vdash
\uparrow \left( f_{ A } ,\left\{ R\right\} \right) \right)
\right)
=\left( A\,\vdash \uparrow \left( f_{ A } ,\left\{ T\right\}
\right) \right) \cap \left( \bigcap_{ R\in Z }\uparrow
 \left( f_{ A }
,\left\{ R\right\} \right) \right) $,
\label{Start of original page 170}
 and $j$ is by definition a member of this set, both $i$
and $j$ are members of $\uparrow \left( f_{ A } ,\left\{
R\right\} \right) $, hence both $i$ and $j$ are members of
$B$, hence both $i$ and $j$ are members of the same member
of $\left\{ B,C\right\} $.   Hence for \emph{every} member
$R$ of $X$, or in other words, for every $\left( V\cup
H\right) $-key $R$ of $A$ such that $R$ is \emph{not} a
member of $V$, and for every partition $\left\{ B,C\right\}
$
of $A$ into two nonempty parts such that $R$ intersects
both parts and $R$ is the \emph{only} member of $\left(
V\cup H\right) $ to intersect both parts, $j$ is a member
of the \emph{same} member of $\left\{ B,C\right\} $ as $i$
is, hence, since this is true for \emph{every} member $j$
of $\left( A\,\vdash \uparrow \left( f_{ A } ,\left\{
T\right\} \right) \right) \,\vdash \left( \bigcup_{R\in Z }
 \left(
A\,\vdash \uparrow \left( f_{ A } ,\left\{ R\right\} \right)
\right) \right) $, and furthermore, as shown above,
$i$ \emph{is} a
member of $\left( A\,\vdash \uparrow \left( f_{ A } ,\left\{
T\right\} \right) \right) \,\vdash \left( \bigcup_{R\in Z }
\left(
A\,\vdash \uparrow \left( f_{ A } ,\left\{ R\right\} \right)
\right) \right) $, it follows directly from
Lemmas \ref{Lemma 29}
and \ref{Lemma 31}
that $\left( A\,\vdash \uparrow \left( f_{ A } ,\left\{
T\right\} \right) \right) \,\vdash \left( \bigcup_{R\in Z }
 \left(
A\,\vdash \uparrow \left( f_{ A } ,\left\{ R\right\} \right)
\right) \right) $ is a subset of the unique
$\left( V\cup H\right)
$-firm over $V$ component of $A$ that has $i$ as a member.
 Hence since, as noted above, each $\left( V\cup H\right)
$-firm over $V$ component of $A$ is either a subset of
$\left( A\,\vdash \uparrow \left( f_{ A } ,\left\{ T\right\}
\right) \right) \,\vdash \left( \bigcup_{R\in Z }
\left( A\,\vdash
\uparrow \left( f_{ A } ,\left\{ R\right\} \right) \right)
\right) $
or else does not intersect this set,
and by Lemma \ref{Lemma 31},
distinct $\left( V\cup H\right) $-firm over $V$ components
of $A$ do not intersect one another, $\left( A\,\vdash
\uparrow \left( f_{ A } ,\left\{ T\right\} \right) \right)
\,\vdash \left( \bigcup_{R\in Z } \left( A\,\vdash \uparrow
 \left(
f_{ A } ,\left\{ R\right\} \right) \right) \right) $ is
 \emph{equal}
to the unique $\left( V\cup H\right) $-firm over $V$
component of $A$ that has $i$ as a member, which, in the
notation of Lemma \ref{Lemma 28}, is $\uparrow \left(
i,X\right) $.

From now on until the end of the proof of
Theorem \ref{Theorem 2}, and
for each member $A$ of $\mathbb{B} \left( G\right) $, we
define the relation $\to $ among the $\left( V\cup H\right)
$-keys of $A$ as in Lemma \ref{Lemma 28},
with the $i$ of Lemma \ref{Lemma 28}
taken as the member $f_{ A } $ of $A$, or in other words,
if $T$ and $R$ are any $\left( V\cup H\right) $-keys of
$A$, then we define $R\to T$ to hold ifif there
\emph{exists} a partition $\left\{ B,C\right\} $ of $A$ into
two nonempty parts such that $T$ intersects both $B$ and
$C$, $T$ is the \emph{only} member of $\left( V\cup
H\right) $ to intersect both $B$ and $C$, and $R$ does
\emph{not} intersect the member of $\left\{ B,C\right\} $
that has $f_{ A } $ as a member.

And from now until the end of the proof of
Theorem \ref{Theorem 2} we
also define, as in Lemma \ref{Lemma 28}, with the set $U$
of Lemma \ref{Lemma 28} taken as the set
\label{Start of original page 171}
 $\left( V\cup H\right) $, and for each member $A$ of
$\mathbb{B} \left( G\right) $, and for each member $i$ of
$A$, and for each subset $X$ of the set of all the $\left(
V\cup H\right) $-keys of $A$, the set $\uparrow \left(
i,X\right) $ to be the subset of $A$ whose members are all
the members $j$ of $A$ such that if $T$ is any member of
$X$, and $\left\{ B,C\right\} $ is any partition of $A$ into
two nonempty parts such that $T$ intersects both $B$ and
$C$ and $T$ is the \emph{only} member of $\left( V\cup
H\right) $ to intersect both $B$ and $C$, then $j$ is a
member of the \emph{same} member of $\left\{ B,C\right\} $
as
$i$ is.   We note that the definition of $\uparrow \left(
i,X\right) $ depends implicitly on $A$, and we will always
make clear which member $A$ of $\mathbb{B} \left( G\right)
$ $\uparrow \left( i,X\right) $ is being defined with
respect to.

We next note that it follows immediately from the foregoing
that if $A$ is any member of $\mathbb{B} \left( G\right) $
and $T$ is any $\left( V\cup H\right) $-key of $A$ such
that $T$ is \emph{not} a member of $V$, and if furthermore
$\left( A\,\vdash \uparrow \left( f_{ A } ,\left\{ T\right\}
\right) \right) $ is \emph{not} a member of $P$, or in
other words, if $\left( A\,\vdash \uparrow \left( f_{ A }
,\left\{ T\right\} \right) \right) $ is a member of $\left(
J\,\vdash P\right) $, then there is at least one member $B$
of $\mathcal{P} \left( P,A\right) $ such that $B$ is a
member of $\mathcal{P} \left( J,\left( A\,\vdash \uparrow
\left( f_{ A } ,\left\{ T\right\} \right) \right) \right) $
and $B$ is a subset of the $\left( V\cup H\right) $-firm
over $V$ component $\left( A\,\vdash \uparrow \left( f_{ A
} ,\left\{ T\right\} \right) \right) \,\vdash \left(
 \bigcup_{R\in Z
} \left( A\,\vdash \uparrow \left( f_{ A } ,\left\{
R\right\}
\right) \right) \right) $ of $A$, where $Z$ is defined as
above to
be the set of all the $\left( V\cup H\right) $-keys $R$ of
$A$ such that $R\notin V$ and $R\to T$ both hold.   For if
$Z$ is nonempty then $C\equiv \left( A\,\vdash \uparrow
\left( f_{ A } ,\left\{ T\right\} \right) \right) \,\vdash
\left( \bigcup_{R\in Z } \left( A\,\vdash \uparrow
\left( f_{ A }
,\left\{ R\right\} \right) \right) \right) $ is a
\emph{strict}
subset of $\left( A\,\vdash \uparrow \left( f_{ A } ,\left\{
T\right\} \right) \right) $ and furthermore there is no
member $D$ of $\left( J\,\vdash P\right) $ such that
$C\subset D\subset \left( A\,\vdash \uparrow \left( f_{ A }
,\left\{ T\right\} \right) \right) $ holds, hence each of
the
one or more members of $\mathcal{P} \left( P,A\right) $
that is a subset of the $\left( V\cup H\right) $-firm over
$V$ component $C$ of $A$ satisfies the stated conditions on
$B$, and if $Z$ is empty then $\left( A\,\vdash \uparrow
\left( f_{ A } ,\left\{ T\right\} \right) \right) $ is
itself
a $\left( V\cup H\right) $-firm over $V$ component of $A$,
and is furthermore by assumption not a member of $P$, hence
each of the two or more members of $\mathcal{P} \left(
P,A\right) $ that are subsets of $\left( A\,\vdash \uparrow
\left( f_{ A } ,\left\{ T\right\} \right) \right) $
satisfies
the stated conditions on $B$.

We next extend the map $S$ to the domain $\mathbb{B} \left(
\bar{ J } \right) $ by choosing, for each
ordered pair $\left(A,T\right) $
of a member $A$ of $\mathbb{B} \left( G\right) $ and a
$\left( V\cup H\right) $-key $T$ of $A$ such that $T$ is
\emph{not} a member of $V$ and $\left( A\,\vdash \uparrow
\left( f_{ A } ,\left\{ T\right\} \right) \right) $ is
\emph{not} a member of $P$, $S_{ \left( A\,\vdash \uparrow
\left( f_{ A } ,\left\{ T\right\} \right) \right) } $ to be
a member $B$ of $\mathcal{P} \left( J,\left( A\,\vdash
\uparrow \left( f_{ A } ,\left\{ T\right\} \right) \right)
\right) $ such that $B$ is a member of $\mathcal{P} \left(
P,A\right) $ that is a subset of the $\left( V\cup H\right)
$-firm over $V$ component
\label{Start of original page 172}
 $\left( A\,\vdash \uparrow \left( f_{ A } ,\left\{
T\right\}
\right) \right) \,\vdash \left( \bigcup_{R\in Z }
\left( A\,\vdash
\uparrow \left( f_{ A } ,\left\{ R\right\} \right) \right)
\right) $
of $A$, and we note that by the previous paragraph, such a
$B$ always exists.

And we note that the extended map $S$ is such that $S_{ A }
$ is a member of $\mathcal{P} \left( J,A\right) $ for
\emph{every} member $A$ of $\mathbb{B} \left( \bar{ J }
\right) $, for this is so by construction if $A$ is a
member of $\left( \bar{ J } \,\vdash \bar{ P } \right) $,
while
if $A$ is a member of $\left( \bar{ P } \,\vdash G\right) $
then $S_{ A } $ is a member of $\mathcal{P} \left(
P,A\right) $ and no member of $\left( J\,\vdash P\right) $
is a member of $\Xi \left( \mathcal{P} \left( P,A\right)
\right) $, hence $S_{ A } $ is a member of $\mathcal{P}
\left( J,A\right) $, and if $A$ is a member of $\mathbb{B}
\left( G\right) $ then, as noted on
page \pageref{Start of original page 165}, $S_{ A } $
is a member of $\mathcal{P} \left( P,A\right) $ that is a
subset of the $\left( V\cup H\right) $-firm over $V$
component of $A$ that has $f_{ A } $ as a member, hence
$S_{ A } $ is not a subset of $\left( A\,\vdash \uparrow
\left( f_{ A } ,\left\{ T\right\} \right) \right) $ for
\emph{any} $\left( V\cup H\right) $-key $T$ of $A$ that is
not a member of $V$.

We next note that if $A$ is any member of $\mathbb{B}
\left( G\right) $, and $B$ is any $\left( V\cup H\right)
$-firm over $V$ component of $A$ such that $f_{ A } $ is
\emph{not} a member of $B$, then it follows directly from
Lemma \ref{Lemma 32} that there exists a \emph{unique}
$\left( V\cup
H\right) $-key $T$ of $A$ such $T$ is not a member of $V$,
$T$ intersects $B$, and there exists a partition $\left\{
D,E\right\} $ of $A$ into two nonempty parts such that $f_{
A } \in D$ holds, $B\subseteq E$ holds, $T$ intersects both
$D$ and $E$, and $T$ is the \emph{only} member of $\left(
V\cup H\right) $ to intersect both $D$ and $E$, and we note
that it also follows directly from Lemma \ref{Lemma 32}
that if $T$
is this unique $\left( V\cup H\right) $-key of $A$, then
$R\to T$ holds for every $\left( V\cup H\right) $-key $R$
of $A$, different from $T$, such that $R\notin V$ and
$R\cap B\neq \emptyset $ both hold, and $T\to R$ does not
hold
for \emph{any} $\left( V\cup H\right) $-key $R$ of $A$ such
that $R\notin V$ and $R\cap B\neq \emptyset $ both hold, and
furthermore if $R$ and $\tilde{ R }$ are \emph{any}
$\left( V\cup
H\right) $-keys of $A$ such that neither $R$ nor
$\tilde{ R }$ is
a member of $V$, both $R$ and $\tilde{ R }$ intersect $B$,
and
$R\to \tilde{ R }$ holds, then $\tilde{ R } =T$ holds.

We note furthermore that since it follows directly from our
assumptions on $V$ and $H$ that if $A$ is any member of
$\mathbb{B} \left( G\right) $, and $T$ is any $\left( V\cup
H\right) $-key of $A$ such that $T$ is \emph{not} a member
of $V$, then $T$ is a member of $W$ hence has exactly two
members, that in particular, if $B$ is any $\left( V\cup
H\right) $-firm over $V$ component of $A$ such that $f_{ A
} $ is \emph{not} a member of $B$, and $T$ is the unique
$\left( V\cup H\right) $-key of $A$ that satisfies the
conditions of the previous paragraph with reference to $B$,
then $T$ has exactly two members and $T\cap B$ has
\emph{exactly one member.}

If $A$ is any member of $\mathbb{B} \left( G\right) $ and
$B$ is any $\left( V\cup H\right) $-firm over $V$ component
of $A$ such that $f_{ A } $ is \emph{not} a member of $B$,
then we define the \emph{key member of }$B $ to be the
unique
member of $B$ that is a member of $T$, where $T$ is the
unique $\left( V\cup H\right) $-key of $A$ such that $T$
satisfies
\label{Start of original page 173}
 the conditions, with reference to $B$, stated in the
paragraph before the previous one.

We note that if $A$ is any member of $\mathbb{B} \left(
G\right) $ and $X$ is the set whose members are all the
$\left( V\cup H\right) $-keys of $A$ that are \emph{not}
members of $V$, then $X$ is a subset of $W$ hence is a
partition, and no two distinct members of $X$ intersect one
another.

We note furthermore that if $A$ is any member of
$\mathbb{B} \left( G\right) $, $T$ is any $\left( V\cup
H\right) $-key of $A$ such that $T$ is \emph{not} a member
of $V$, and $Z$ is the set of all the $\left( V\cup
H\right) $-keys $R$ of $A$ such that $R\notin V$ and $R\to
T$ both hold, then the unique $\left( V\cup H\right)
$-key of $A $ which satisfies, with reference to the
$\left( V\cup
H\right) $-firm over $V $ component $\left(
A\,\vdash \uparrow
\left( f_{ A } ,\left\{ T\right\} \right) \right) \,\vdash
\left( \bigcup_{R\in Z } \left( A\,\vdash \uparrow
\left( f_{ A }
,\left\{ R\right\} \right) \right) \right) $ of $A $, the
conditions stated in the paragraph before the previous
three,
is $T $ itself.

And we note furthermore that if $A$ is any member of
$\mathbb{B} \left( G\right) $, $T$ is any $\left( V\cup
H\right) $-key of $A$ such that $T$ is \emph{not} a member
of $V$, and $i$ is a member of $T$ such that $i$ is the key
member of some $\left( V\cup H\right) $-firm over $V$
component of $A$, then it follows directly from the
definition of the key member of a $\left( V\cup H\right)
$-firm over $V$ component of $A$, that if $\left\{
B,C\right\} $ is the unique, (by Lemma \ref{Lemma 36}),
partition of $A$ into two nonempty parts such that $T$
intersects both $B$ and $C$, and $T$ is the \emph{only}
member of $\left( V\cup H\right) $ to intersect both $B$
and $C$, then $i$ is a member of the member of $\left\{
B,C\right\} $ that does \emph{not} have $f_{ A } $ as a
member.   Hence only one of the two members of $T$ can be
the key member of a $\left( V\cup H\right) $-firm over $V$
component of $A$.

Hence if $A$ is any member of $\mathbb{B} \left( G\right)
$, then for each $\left( V\cup H\right) $-key $T$ of $A$
such that $T$ is \emph{not} a member of $V$, there is
\emph{exactly one} $\left( V\cup H\right) $-firm over $V$
component $B$ of $A$ such that $f_{ A } \notin B$ holds and
the key member of $B$ is a member of $T$.   Hence the
number of $\left( V\cup H\right) $-firm over $V$ components
of $A$ is equal to one plus the number of $\left( V\cup
H\right) $-keys of $A$ that are \emph{not} members of $V$.

We recall from
page \pageref{Start of original page 7} that for any
ordered pair $\left(F,E\right) $ of
a wood $F$, and a set $E$ such that every member of $E$ is
a set, we define $\mathcal{O} \left( F,E\right) $ to be the
set whose members are all the members $i$ of $\mathcal{U}
\left( F\right) $ such that there \emph{exists} a member
$A$ of $F$ such that $i\in A$ holds and there is \emph{no}
member $B$ of $E$ such that $i\in B$ and $B\subseteq A$
both hold, and we note that it immediately follows from
page \pageref{Start of original page 73}
that in the present case, $\mathcal{O} \left(
J,H\right) =\mathcal{O} \left( P,H\right) =\mathcal{O}
\left( Q,H\right) =\mathcal{O} \left( G,H\right)
=\mathcal{O} \left( V,H\right) $ holds.

And we recall from
page \pageref{Start of original page 78}
that $\mathbb{G} \left(
Q,H\right) $ is by definition the map
\label{Start of original page 174}
 whose domain is the set of all ordered
 pairs $\left(i,B\right) $ of a
member $i$ of $\mathcal{O} \left( Q,H\right) $ and a member
$B$ of $Q$, and such that for each member $\left(i,B\right)
$ of
$\mathcal{D} \left( \mathbb{G} \left( Q,H\right) \right) $,
$\mathbb{G}_{ iB } \left( Q,H\right) \equiv \left(
\mathbb{G}
\left( Q,H\right) \right) _{ \left( i,B\right) } $ is the
set whose members are all the members $C$ of $Q$ such that
$B\subseteq C$ and $C\subseteq \mathcal{Z} \left(
Q,H,i\right) $ both hold.   (Thus $\mathbb{G}_{ iB }
\left(
Q,H\right) $ is the empty set $\emptyset $ if $B\subseteq
\mathcal{Z} \left( Q,H,i\right) $ does not hold.)

Thus, in particular, if $i$ is any member of $\mathcal{O}
\left( V,H\right) =\mathcal{O} \left( Q,H\right) $, then
$\mathbb{G} _{ i\mathcal{Z} \left( P,H,i\right) }
\hspace{-1.63pt} \left(
Q,H\right) $ is the set whose members are all the
members $C$ of $Q$ such that $\mathcal{Z} \left(
P,H,i\right) \subseteq C\subseteq \mathcal{Z} \left(
Q,H,i\right) $ holds, and is equal to $\left( \left\{
\mathcal{Z} \left( P,H,i\right) \right\} \cup \mathbb{Y}
\left( Q,\mathcal{Z} \left( P,H,i\right) ,\mathcal{Z}
\left( Q,H,i\right) \right) \right) $.

And we recall from
pages \pageref{Start of original page 80} and
\pageref{Start of original page 81}
that if $x$ is any
member of $\mathbb{U} _{ d } \left( V,\omega \right) $, $i$
is any member of $\mathcal{O} \left( V,H\right) $, and
$\rho $ is any member of $\mathbb{D} $, or in other words,
$\rho $ is any member of $\mathbb{R}^{ \left( Q\,\vdash
P\right) } $ such that $0\leq \rho _{ A } \leq 1$ holds for
every member $A$ of $\left( Q\,\vdash P\right) $, and if we
define the member $u$ of $\mathbb{R}^{ \left( \mathbb{G}_{
i\mathcal{Z} \left( P,H,i\right) } \left( Q,H\right)
\right) }
=\mathbb{R}^{ \left( \left\{ \mathcal{Z} \left( P,H,i\right)
\right\} \cup \mathbb{Y} \left( Q,\mathcal{Z} \left(
P,H,i\right) ,\mathcal{Z} \left( Q,H,i\right) \right)
\right) } $ by
\[
u_{ \mathcal{Z} \left( P,H,i\right) } \equiv \left(
 \prod_{B\in
\mathbb{Y} \left( Q,\mathcal{Z} \left( P,H,i\right)
,\mathcal{Z} \left( Q,H,i\right) \right) } \rho _{
B } \right)
\]
and by
\[
u_{ C } \equiv \left( \prod_{B\in \mathbb{Y}
\left( Q,C,\mathcal{Z}
\left( Q,H,i\right) \right) } \rho _{ B } \right) \left(
1-\rho _{ C } \right)
\]
for all $C\in \mathbb{Y} \left( Q,\mathcal{Z} \left(
P,H,i\right) ,\mathcal{Z} \left( Q,H,i\right) \right) $,
then
\[
\mu _{ i } \left( P,Q,H,x,\mathcal{X} \left( P,Q,H,\rho
\right) \right) = \sum_{C\in \left(
\mathbb{G}_{ i\mathcal{Z} \left(
P,H,i\right) } \left( Q,H\right) \right) } u_{ C }
x_{ C }
\]
holds and
\[
 \sum_{C\in \left(
 \mathbb{G}_{ i\mathcal{Z} \left( P,H,i\right) }
\left(
Q,H\right) \right) } u_{ C } =1
\]
holds, and furthermore, $u_{ C } \geq 0$ holds for all
members $C$ of $\mathbb{G} _{ i\mathcal{Z} \left(
P,H,i\right) } \left( Q,H\right) $.

Now if $A$ is any member of $\mathbb{B} \left( G\right) $,
$B$ is any $\left( V\cup H\right) $-firm over $V$ component
of $A$ such that $f_{ A } $ is \emph{not} a member of $B$,
and $i$ is the key member of $B$, then $i$ is certainly a
member of $\mathcal{O} \left( V,H\right) $, for since $H$
is a partition, the \emph{only} member of $H$ that has $i$
as a member is the unique $\left( V\cup H\right) $-key $T$
of $A$ that has $i$ as a member and is \emph{not} a member
of $V$, and this $\left( V\cup H\right) $-key $T$ of $A$ is
certainly not a subset of any member of $V$.   And
furthermore, if $T$ is the unique
\label{Start of original page 175}
 $\left( V\cup H\right) $-key of $A$ that has $i$ as a
member and is \emph{not} a member of $V$, then $\mathcal{Z}
\left( P,H,i\right) $ is the largest member of $P$ that has
$i$ as a member but does not have $T$ as a subset, and
$\mathcal{Z} \left( Q,H,i\right) $ is the largest member of
$Q$ that has $i$ as a member but does not have $T$ as a
subset, hence, since $A$ is a member of both $P$ and $Q$
that \emph{does} have $T$ as a subset, both $\mathcal{Z}
\left( P,H,i\right) $ and $\mathcal{Z} \left( Q,H,i\right)
$ are strict subsets of $A$, hence, since both $P$ and $Q$
are members of $\mathbb{O} \left( G,H\right) $, both
$\mathcal{Z} \left( P,H,i\right) $ and $\mathcal{Z} \left(
Q,H,i\right) $ are subsets of the $\left( V\cup H\right)
$-firm over $V$ component $B$ of $A$.   And furthermore,
since $\mathcal{K} \left( P,A,i\right) $, which by
definition is the unique member of $\mathcal{P} \left(
P,A\right) $ that has $i$ as a member, is a member of $P$
that is a strict subset of $A$ and that intersects $B$,
$\mathcal{K} \left( P,A,i\right) $ is a subset of $B$,
hence $T$ is \emph{not} a subset of $\mathcal{K} \left(
P,A,i\right) $, hence since, by the definition of
$\mathcal{P} \left( P,A\right) $, there is no member $C$ of
$P$ such that $\mathcal{K} \left( P,A,i\right) \subset
C\subset A$ holds, $\mathcal{K} \left( P,A,i\right) $ is
the \emph{largest} member of $P$ that has $i$ as a member
but does not have $T$ as a subset, hence $\mathcal{K}
\left( P,A,i\right) =\mathcal{Z} \left( P,H,i\right) $
holds, and furthermore, since $\mathcal{Z} \left(
P,H,i\right) \subseteq \mathcal{Z} \left( Q,H,i\right) $
holds, $\mathcal{K} \left( P,A,i\right) \subseteq
\mathcal{Z} \left( Q,H,i\right) $ holds.

For each member $C$ of $\mathbb{G} _{ i\mathcal{Z} \left(
P,H,i\right) } \left( Q,H\right) $, or in other words, for
each member $C$ of $Q$ such that $\mathcal{Z} \left(
P,H,i\right) \subseteq C\subseteq \mathcal{Z} \left(
Q,H,i\right) $ holds, let $F_{ C } $ be the set whose
members are all the members $D$ of $P$ such that
$D\subseteq C$ holds and there is \emph{no} member $E$ of
$P$ such that $D\subset E\subseteq C$ holds.   Thus $F_{ C
} $ is equal to $\left\{ C\right\} $ if $C$ is a member of
$P$, or in other words, if $C$ is equal to $\mathcal{Z}
\left( P,H,i\right) $, and $F_{ C } $ is equal to
$\mathcal{P} \left( P,C\right) $ if $C$ is \emph{not} a
member of $P$, or in other words, if $C$ is a member of
$\mathbb{Y} \left( Q,\mathcal{Z} \left( P,H,i\right)
,\mathcal{Z} \left( Q,H,i\right) \right) $.   And we also
define $F_{ B } $ to be the set whose members are all the
members $D$ of $P$ such that $D\subseteq B$ holds and there
is \emph{no} member $E$ of $P$ such that $D\subset
E\subseteq B$ holds, so that $F_{ B } $ is equal to $\left\{
B\right\} $ if $B$ is a member of $P$, (in which case
$\mathcal{Z} \left( P,H,i\right) =\mathcal{Z} \left(
Q,H,i\right) =B$ holds), and $F_{ B } $ is equal to
$\mathcal{P} \left( P,B\right) $ if $B$ is \emph{not} a
member of $P$.

Then by Lemma \ref{Lemma 4}, and for each member $C$ of
$\mathbb{G}
_{ i\mathcal{Z} \left( P,H,i\right) } \left( Q,H\right) $,
\[
x_{ C } = \sum_{D\in F_{ C } } \omega _{ CD } x_{ D }
\]
holds.   Hence
\[
\mu _{ i } \left( P,Q,H,x,\mathcal{X} \left( P,Q,H,\rho
\right) \right) = \sum_{C\in \left(
\mathbb{G}_{ i\mathcal{Z} \left(
P,H,i\right) } \left( Q,H\right) \right) } u_{ C } x_{ C }
\hspace{2.0cm}
\]
\label{Start of original page 176}
\[
\hspace{5.0cm}
= \sum_{C\in \left(
\mathbb{G}_{ i\mathcal{Z} \left( P,H,i\right) }
\left( Q,H\right) \right) } \left(
 u_{ C } \sum_{ D\in F_{ C } }
\omega _{ CD } x_{ D } \right)
\]
\[
\hspace{6.0cm}
= \sum_{D\in F_{ B } } \left(
 \sum_{C\in \left( \mathbb{G} _{ iD } \left(
Q,H\right) \cap \mathbb{G}_{ i\mathcal{Z} \left(
P,H,i\right)
 } \left( Q,H\right) \right) } u_{ C } \omega _{ CD }
 \right) x_{ D }
\]
holds, where in obtaining the last line we noted that, for
each member $C$ of \\
$\mathbb{G} _{ i\mathcal{Z} \left(
P,H,i\right) } \left( Q,H\right) $, $F_{ C } \subseteq F_{
B } $ holds, and furthermore, that for each member $D$ of
$F_{ B } $ such that $D\subseteq \mathcal{Z} \left(
Q,H,i\right) $ does \emph{not} hold, $\mathbb{G} _{ iD }
\left( Q,H\right) $ is the empty set $\emptyset $.

For each member $D$ of $F_{ B } $, we define
\[
r_{ D } \equiv \left( \sum_{C\in \left( \mathbb{G} _{ iD }
 \left( Q,H\right)
\cap \mathbb{G}_{ i\mathcal{Z} \left( P,H,i\right) } \left(
Q,H\right) \right) } u_{ C } \omega _{ CD } \right).
\]

Then it immediately follows from the foregoing that
\[
\mu _{ i } \left( P,Q,H,x,\mathcal{X} \left( P,Q,H,\rho
\right) \right) = \sum_{D\in F_{ B } } r_{ D } x_{ D }
\]
holds, and furthermore that
\[
 \sum_{D\in F_{ B } } r_{ D } =1
\]
holds, and furthermore that $r_{ D } \geq 0$ holds for
every member $D$ of $F_{ B } $, and furthermore that $r_{ D
} =0$ holds for every member $D$ of $F_{ B } $ such that
$D\subseteq \mathcal{Z} \left( Q,H,i\right) $ does
\emph{not} hold.

We next define, for each member $A$ of $\mathbb{B} \left(
G\right) $, and for each ordered
pair $\left(K,D\right) $ of a member $K$
of $\left( J\,\vdash P\right) $ such that $K$ is a member
of $\Xi \left( \mathcal{P} \left( P,A\right) \right) $, or
in other words, a member $K$ of $\left( J\,\vdash P\right)
$ such that $K$ is equal to the set $\left( A\,\vdash
\uparrow \left( f_{ A } ,\left\{ T\right\} \right) \right) $
for some $\left( V\cup H\right) $-key $T$ of $A$ such that
$T$ is \emph{not} a member of $V$, and a member $D$ of
$\mathcal{P} \left( J,K\right) $, the real number $\nu _{
KD } $ as follows.   Let $T$ be the unique $\left( V\cup
H\right) $-key of $A$ such that $K $ is equal to
$\left( A\,\vdash
\uparrow \left( f_{ A } ,\left\{ T\right\} \right)
\right)$.  (We note that by Lemma \ref{Lemma 28}
(viii),
$T$ may be identified as the unique $\left( V\cup H\right)
$-key of $ A $ that intersects both $K$ and $\left(
A\,\vdash
K\right) $.)   Let $B$ be equal to the $\left( V\cup
H\right) $-firm over $V$ component $\left( A\,\vdash
\uparrow \left( f_{ A } ,\left\{ T\right\} \right) \right)
\,\vdash \left( \bigcup_{R\in Z } \left( A\,\vdash
\uparrow \left(
f_{ A } ,\left\{ R\right\} \right) \right) \right) $ of
$A$, (where
$Z$ is the set of all the $\left( V\cup H\right) $-keys $R$
of $A$ such that $R\notin V$ and $R\to T$ both hold), or in
other words, let $B$ be the unique $\left( V\cup H\right)
$-firm over $V$ component of $A$
whose key member is a member of $T $.  Let $i $ be the key
member of $B $,
\label{Start of original page 177}
 and let the set $F_{ B } $ be defined as above to be the
set whose members are all the members $D$ of $P$ such that
$D\subseteq B$ holds and there is \emph{no} member $E$ of
$P$ such that $D\subset E\subseteq B$ holds, so that $F_{ B
} $ is equal to $\left\{ B\right\} $ if $B$ is a member of
$P$, and $F_{ B } $ is equal to $\mathcal{P} \left(
P,B\right) $ if $B$ is \emph{not} a member of $P$.   We
note that $\mathcal{P} \left( J,K\right) $ is equal to the
set whose members are the members of $F_{ B } $, together
with, for each $\left( V\cup H\right) $-key $R$ of $A$ such
that the three conditions $R\notin V$, $R\to T$, and $R\cap
B\neq \emptyset $ all hold, (if there \emph{are} any such
$\left( V\cup H\right) $-keys $R$ of A), the set $\left(
A\,\vdash \uparrow \left( f_{ A } ,\left\{ R\right\} \right)
\right) $.   Then for $D\in \mathcal{P} \left( J,K\right)
$, we define $\nu _{ KD } $ to be equal to $0$ if $D$ is
\emph{not} a member of $F_{ B } $, while if $D$ \emph{is} a
member of $F_{ B } $, then we define $\nu _{ KD } $ to be
equal to $r_{ D } $, where $r_{ D } $ was defined, with
reference to the key member $i$ of $B$, on
page \pageref{Start of original page 176}.
(Thus $\nu _{ KD } $ depends on $\rho $ through the
coefficients $u_{ C } $, $C\in \left( \mathbb{G} _{ iD }
\left(
Q,H\right) \cap \mathbb{G}_{ i\mathcal{Z}
\left( P,H,i\right)
 } \left( Q,H\right) \right) $.)   We note that it follows
immediately from this definition that $  \sum_{D\in
\mathcal{P}
\left( J,K\right) } \nu _{ KD } =1$ holds, and
that $\nu _{ KD } $ is equal to $0$ unless $D$ is a member
of $\mathcal{P} \left( P,A\right) $, (hence in particular,
$\nu _{ KD } $ is equal to $0$ if $D$ is a member of
$\left( J\,\vdash P\right) $), and furthermore, that
\[
\mu _{ i } \left( P,Q,H,x,\mathcal{X} \left( P,Q,H,\rho
\right) \right) =
\sum_{D\in \mathcal{P} \left( J,K\right) } \nu _{ KD }
x_{ D } = \sum_{D\in F_{ B } } \nu _{
KD } x_{ D }
\]
holds, where $i$ is the key member of $B$.

We next extend the definition of $z$ over $\left( J\,\vdash
\left\{ \mathcal{U} \left( V\right) \right\} \right) $ by
defining, for each member $K$ of $\left( J\,\vdash P\right)
$:
\[
z_{ K } \equiv
\sum_{D\in \mathcal{P} \left( J,K\right) } \nu _{ KD }
z_{ D } .
\]

We note that this is actually a direct, rather than
inductive, definition, due to $\nu _{ KD } $ being equal to
$0$ for $D\in \left( J\,\vdash P\right) $, and we also note
that it directly follows from the definition of $J$ on
pages \pageref{Start of original page 165} and
\pageref{Start of original page 166}
that $\mathcal{U} \left( V\right) $
is \emph{not} a member of $\left( J\,\vdash P\right) $.

Finally we complete the second step, started on
page \pageref{Start of original page 165},
in the definition of our new integration variables as
follows.

For each member $B$ of $\left( J\,\vdash \left\{
\mathcal{U}
\left( V\right) \right\} \right) $ such that $B$ is
\emph{not} a member of \\
$J\cap \left( \Xi \left(
\mathcal{P} \left( P,A\right) \right) \vdash
\hspace{-0.14pt} \left\{
A\right\} \right) $ for
some member $A$ of the subset $\mathbb{B} \left( G\right) $
of $P$, we define $a_{ B } \equiv z_{ B } $.

And for each member $A$ of $\mathbb{B} \left( G\right) $,
and for each member $B$ of $J\cap \left( \Xi \left(
\mathcal{P} \left( P,A\right) \right) \vdash
\hspace{-0.14pt} \left\{
A\right\} \right) $, we define $a_{ B } \equiv \left( z_{ B
} -z_{ S_{ C } } \right) $, where $C$ is the
\emph{smallest} member of $J$ to contain $B$ as a
\emph{strict} subset.   (We recall
\label{Start of original page 178}
 that the map $S$ was extended to the domain $\mathbb{B}
\left( \bar{ J } \right) $ on
pages \pageref{Start of original page 171} and
\pageref{Start of original page 172}.)
We note that if the smallest member of $J$ to contain $B$
as a strict subset is in fact $A$, (which occurs when $B$
is equal to $\left( A\,\vdash \uparrow \left( f_{ A }
,\left\{ T\right\} \right) \right) $, where $T$ is a $\left(
V\cup H\right) $-key of $A$ such that $T\notin V$ holds and
$T$ \emph{does} intersect the unique $\left( V\cup H\right)
$-firm over $V$ component of $A$ that has $f_{ A } $ as a
member and contains $S_{ A } $ as a subset), then since
$z_{ S_{ A } }=0 $ holds, $a_{ B } =z_{ B } $ holds.

We choose as our new set of integration variables the $a_{
B } $, \\
$B\in \left( J\,\vdash \left( \mathcal{R} \left(
S\right) \cup \left\{ \mathcal{U} \left( V\right) \right\}
\right) \right) $, and we note that, exactly as on
pages \pageref{Start of original page 118}
and \pageref{Start of original page 119},
the transformation to this new set of
integration variables may be expressed as a sequence of
transformations of the type considered
in Lemma \ref{Lemma 24}, hence
the linear transformation to this new set of integration
variables has determinant equal to $1$.

Now for each member $l$ of $\mathcal{U} \left( V\right) $
we have, exactly as on
page \pageref{Start of original page 118}, that
\[
x_{ \mathcal{C} \left( V,l\right) } =x_{ \mathcal{U} \left(
V\right) } + \sum_{A\in \mathbb{Y} \left( \bar{ P }
,\mathcal{C}
\left(
V,l\right) ,\mathcal{U} \left( V\right) \right) } \left(
z_{ \mathcal{K} \left( P,A,l\right) } - \sum_{C\in
\mathcal{P}
\left(
P,A\right) } \omega _{ AC } z_{ C } \right)
\]
holds, and furthermore, if $L$ is any member of $P$, and
$l$ is any member of $L$, that
\[
x_{ L } =x_{ \mathcal{U} \left( V\right) } +\sum_{A\in
\mathbb{Y}
\left( \bar{ P } ,L,\mathcal{U} \left( V\right) \right) }
\left( z_{ \mathcal{K} \left( P,A,l\right) } - \sum_{C\in
 \mathcal{P}
\left( P,A\right) } \omega _{ AC } z_{ C } \right)
\]
holds.

And furthermore in these equations, exactly as on
pages \pageref{Start of original page 118}
and \pageref{Start of original page 119},
$x_{ \mathcal{U} \left( V\right) } $ may be
expressed in terms of the $z$ variables by use of the
particular equation of the above form for $x_{ \mathcal{C}
\left( V,h\right) } $, where $h$ is the particular member of
$\mathcal{U} \left( V\right) $ such that $x_{ \mathcal{C}
\left( V,h\right) } $ has the fixed value $b$ in the
definition of the integration domain $\mathbb{W} $.

Now let $A$ be any member of $\mathbb{B} \left( G\right) $,
and $K$ be any member of \\
$J\cap \left( \Xi \left(
\mathcal{P} \left( P,A\right) \right) \,\vdash \mathcal{P}
\left( P,A\right) \right) $.   Then
\[
   \sum_{D\in \mathcal{P} \left( J,K\right) } \nu _{
KD } a_{ D } =
\sum_{D\in \mathcal{P} \left( J,K\right) } \nu _{ KD }
\left( z_{ D } -z_{ S_{ K } } \right)
=\left( z_{ K } -z_{ S_{ K } } \right)
\]
holds, hence $z_{ S_{ K } } = z_{ K } -
\sum_{D\in \mathcal{P} \left(
J,K\right) } \nu _{ KD } a_{ D } $ holds,
hence for all members $D$ of $\mathcal{P} \left( J,K\right)
$ we have
\[
z_{ D } =a_{ D } +z_{ S_{ K } } = z_{ K } + \left(
a_{ D } -
\sum_{C\in
\mathcal{P} \left( J,K\right) } \nu _{ KC } a_{ C
} \right).
\]
\label{Start of original page 179}

For each member $A$ of $\mathbb{B} \left( G\right) $ we
now define, for each member $D$ of \\
$J\cap \left( \Xi \left(
\mathcal{P} \left( P,A\right) \right) \,\vdash \left\{
A\right\} \right) $, $M_{ D } $ to be the \emph{largest}
member of $J$ that contains $D$ as a subset and is a
\emph{strict} subset of $A$.   Thus $M_{ D } $ is equal to
$\left( A\,\vdash \uparrow \left( f_{ A } ,\left\{ R\right\}
\right) \right) $, where $R$ is the unique $\left( V\cup
H\right) $-key of $A$ that satisfies the three conditions
that $R$ is \emph{not} a member of $V$, that $R$
\emph{does} intersect the unique $\left( V\cup H\right)
$-firm over $V$ component of $A$ that has $f_{ A } $ as a
member, and that $D\subseteq \left( A\,\vdash \uparrow
\left( f_{ A } ,\left\{ R\right\} \right) \right) $ holds.
(We recall that by Lemma \ref{Lemma 28} (x) and Lemma
\ref{Lemma 28}
(xi), $A$ is equal to the disjoint union
of the unique $\left( V\cup H\right) $-firm over $V$
component of $A$ that has $f_{ A } $ as a member, together
with the sets $\left( A\,\vdash \uparrow \left( f_{ A }
,\left\{ T\right\} \right) \right) $ for all the $\left(
V\cup H\right) $-keys $T$ of $A$ such that $T$ intersects
the unique $\left( V\cup H\right) $-firm over $V$ component
of $A$ that has $f_{ A } $ as a member and $T$ is
\emph{not} a member of $V $.)

Then if $A$ is any member of $\mathbb{B} \left( G\right) $,
$D$ is any member of $J\cap \left( \Xi \left( \mathcal{P}
\left( P,A\right) \right) \,\vdash \left\{ A\right\} \right)
$, and $l$ is any member of $D$, it immediately follows
from the foregoing, together with the fact that $z_{ S_{
A } } =0 $ holds, hence that $a_{ M_{ D } }
=z_{ M_{ D } } $
holds, that
\[
z_{ D } =a_{ M_{ D } }+ \sum_{K\in \mathbb{Y}
\left( J,D,M_{ D }
\right) } \left( a_{ \mathcal{K} \left( J,K,l\right) }
- \sum_{C\in
\mathcal{P} \left( J,K\right) } \nu _{ KC } a_{ C
} \right)
\]
holds.   And this is true in particular for any member $D$
of $\mathcal{P} \left( P,A\right) $.   Hence the
expressions for the $x_{ L } $, $L\in P$, in terms of the
variables $a_{ B } $, $B\in \left( J\,\vdash \left(
\mathcal{R} \left( S\right) \cup \left\{ \mathcal{U} \left(
V\right) \right\} \right) \right) $, may be obtained by
substituting into the expressions on
page \pageref{Start of original page 178} the
equations above for all members $D$ of $P$ such that $D$ is
a member of $\mathcal{P} \left( P,A\right) $ for some
member $A$ of $\mathbb{B} \left( G\right) $, together with
the equations $z_{ D } =a_{ D } $ for the members $D$ of
$\left( P\,\vdash \left\{ \mathcal{U} \left( V\right)
\right\} \right) $ such that $D$ is \emph{not} a member of
$\mathcal{P} \left( P,A\right) $ for any member $A$ of
$\mathbb{B} \left( G\right) $.

\vspace{2.5ex}

We now make the following observations:

\begin{bphzobservation} \label{Observation 30}
\end{bphzobservation}
\vspace{-6.143ex}

\noindent \hspace{2.6ex}{\bf ) }Let $A$ be any member of
$\bar{ P } $, $L$ be any member of
$\mathcal{P} \left( P,A\right) $, and $l$ be
any member of $L $.  Then $x_{ L } $ is equal to
$z_{ L } $ plus a term $x_{ S_{
A } } $ that is the same for \emph{all} members $D$ of
$\mathcal{P} \left( P,A\right) $.   For by
page \pageref{Start of original page 178} we
have that
\[
x_{ L } =x_{\mathcal{U} \left( V\right) }
+ \left( \sum_{N\in \mathbb{Y}
\left( \bar{ P } ,A,\mathcal{U}
\left( V \right) \right) } \hspace{-0.1cm}
 \left( z_{\mathcal{K} \left(P,N,l
\right) }- \hspace{-0.1cm}
\sum_{C\in \mathcal{P} \left( P,N \right)
 } \hspace{-0.2cm} \omega_{ NC } z_{ C } \right)
 \hspace{-0.1cm} \right)
  + \left( z_{ \mathcal{K} \left(P,A,l
\right) } - \hspace{-0.1cm} \sum_{C\in
\mathcal{P} \left( P,A \right) } \hspace{-0.2cm}
\omega_{ AC } z_{ C } \hspace{-0.32pt}
\right)
\]
 holds, and the
\emph{only} term that differs for different members $L$ of
$\mathcal{P} \left( P,A\right) $ is the term $z_{
\mathcal{K}
\left( P,A,l\right) } $, which is equal to $z_{ L } .$
\label{Start of original page 180}

\begin{bphzobservation} \label{Observation 31}
\end{bphzobservation}
\vspace{-6.143ex}

\noindent \hspace{2.6ex}{\bf ) }Let $A$ be any member of
$\mathbb{B} \left(
G\right) $, $T$ be any $\left( V\cup H\right) $-key of $A$
such that $T$ is \emph{not} a member\hspace{\stretch{1}}
of\hspace{\stretch{1}} $V$,\hspace{\stretch{1}}
$B$\hspace{\stretch{1}} be\hspace{\stretch{1}}
equal\hspace{\stretch{1}} to\hspace{\stretch{1}}
$\left( A\,\vdash \uparrow \left( f_{ A } ,\left\{
T\right\} \right) \right) $,\hspace{\stretch{1}}
and\hspace{\stretch{1}} $L$\hspace{\stretch{1}}
be\hspace{\stretch{1}} any\hspace{\stretch{1}}
member\hspace{\stretch{1}} of\hspace{\stretch{1}}
$\mathcal{P} \left( P,A\right) $.
\newpage
\noindent Then the coefficient of
$a_{ B } $ in $z_{ L } $ is equal to $1$ if $L\subseteq B$
holds, and equal to $0$ otherwise.   For if $l$ is any
member of $L$, then by
page \pageref{Start of original page 179} we have that
\[
z_{ L } =a_{ M_{ L } } + \sum_{K\in \mathbb{Y}
\left( J,L,M_{ L }
\right) } \left( a_{ \mathcal{K} \left( J,K,l\right) }
 -  \sum_{C\in
\mathcal{P} \left( J,K\right) } \nu _{ KC } a_{ C
} \right)
\]
holds.   And furthermore it follows directly from the
definition, on
pages \pageref{Start of original page 176} and
\pageref{Start of original page 177},
of the real numbers
$\nu _{ KC } $, for each ordered pair
$\left(K,C\right) $ of a member $K$
of $\left( J\,\vdash P\right) \cap \Xi \left( \mathcal{P}
\left( P,A\right) \right) $ and a member $C$ of
$\mathcal{P} \left( J,K\right) $, that $\nu _{ KC } =0$
holds if $C$ is equal to $\left( A\,\vdash \uparrow \left(
f_{ A } ,\left\{ R\right\} \right) \right) $ for any $\left(
V\cup H\right) $-key $R$ of $A$ such that $R$ is not a
member of $V$, hence $a_{ B } $ does not occur in the term
$ \sum_{C\in \mathcal{P} \left( J,K\right) } \nu _{
KC } a_{ C } $ for any member $K$ of $\mathbb{Y} \left(
J,L,M_{ L } \right) $.   And the remaining terms in $z_{ L
} $ comprise the sum $\sum_{D\in X } a_{ D } $, where $X$
is the set whose members are all the members $D$ of $J$
such that $L\subseteq D\subseteq M_{ L } $ holds, and $a_{
B } $ occurs in this sum with coefficient $1$ if
$L\subseteq B$ holds, and with coefficient $0$ otherwise.

\begin{bphzobservation} \label{Observation 32}
\end{bphzobservation}
\vspace{-6.143ex}

\noindent \hspace{2.6ex}{\bf ) }Let $A$ be any member of
$\mathbb{B} \left( G\right) $,
$T$ be any $\left( V\cup H\right) $-key of $A$ such that
$T$ is \emph{not} a member of $V$, $B$ be equal to $\left(
A\,\vdash \uparrow \left( f_{ A } ,\left\{ T\right\} \right)
\right) $, and $L$ be any member of $\left( P\,\vdash
\mathcal{P} \left( P,A\right) \right) $.   Then the
coefficient of $a_{ B } $ in $z_{ L } $ is equal to $0$.
This follows directly from the above formula for $z_{ L } $.

\vspace{2.5ex}

We shall now prove that if
$\left(B,i,n,j,s,E,v\right) $ is any ordered
septuple as on
page \pageref{Start of original page 163},
such that the condition on
page \pageref{Start of original page 164}
 holds, and such that $\left( \#\left( \mathcal{D}
\left( j\right) \right) +\#\left( \mathcal{D} \left(
E\right) \right) \right) $ is not greater than the integer
$N$ defined on
page \pageref{Start of original page 158},
then there exists a finite
collection of ordered septuples $\left( \tilde{ B }
,\tilde{ i }
,\tilde{ n },\tilde{ j } ,\tilde{ s },\tilde{ E } ,\tilde{
v }\right) $ as on
page \pageref{Start of original page 163}, each
satisfying the condition on
page \pageref{Start of original page 164},
and each satisfying
the requirement that $\left( \#\left(
\mathcal{D} \left( \tilde{ j }
 \right) \right) +\#\left( \mathcal{D} \left( \tilde{ E }
\right) \right) \right) $ is not greater than the integer
$N$ defined on
page \pageref{Start of original page 158},
and each satisfying the further
requirement that if $A$ is any member of $\mathbb{B} \left(
G\right) $, and $T\equiv \left\{ l,m\right\} $ is any
$\left(
V\cup H\right) $-key of $A$ such that $T$ is \emph{not} a
member of $V$, then there is \emph{no} member $\beta $ of
$\mathcal{D} \left( \tilde{ j } \right) $ such that
 $\tilde{ j } _{
\beta } =l$ holds and there is \emph{no} member $\beta $
of $\mathcal{D} \left( \tilde{ j } \right) $ such that
 $\tilde{ j }
_{ \beta } = m $ holds, such that for each member $k $
of $\mathbb{N} $ ,
$I \rule{0pt}{2.55ex}^{ \hspace{-0.75ex} \circ }_{ k }
\left( B,i,n,j,s,E,v,\rho \right) $ is
equal to the sum, over this finite collection of
septuples $\left(
\tilde{B } ,\tilde{ i } ,\tilde{ n },\tilde{ j } ,
\tilde{ s },\tilde{ E } ,\tilde{ v }\right) $, of
$I \rule{0pt}{2.55ex}^{ \hspace{-0.75ex} \circ }_{ k }
\left( \tilde{ B } ,\tilde{ i } ,
\tilde{ n },\tilde{ j } ,\tilde{ s },\tilde{ E }
,\tilde{ v },\rho \right) $ times a finite real-number
 coefficient
that is \emph{independent of }$ k $, and either independent
of
$\rho $ or else equal to the
\label{Start of original page 181}
sum of a finite number of terms, each of which is the
product of finite
 powers, all $\geq 0$ and independent of $k$ and $\rho $,
of the $\rho _{ C } $, $C\in \left( Q\,\vdash P\right) $.
\enlargethispage{1.4ex}

We first note that if $A$ is any member of $\mathbb{B}
\left( G\right) $, and $T\equiv \left\{ l,m\right\} $ is any
$\left( V\cup H\right) $-key of $A$ such that $T$ is
\emph{not} a member of $V$, then $T$ is a member of $W$,
and furthermore, by
pages \pageref{Start of original page 172} and
\pageref{Start of original page 173},
\emph{exactly one} of the two members $l$ and $m$ of $T$ is
the key member of some $\left( V\cup H\right) $-firm over
$V$ component of $A$ that does not have $f_{ A } $ as a
member.

If $\left(B,i,n,j,s,E,v\right) $ is
an ordered septuple as on
page \pageref{Start of original page 163},
such that the condition on
page \pageref{Start of original page 164}
 holds, and such that
$\left( \#\left( \mathcal{D} \left( j\right) \right)
+\#\left( \mathcal{D} \left( E\right) \right) \right) $ is
not greater than the integer $N$ defined on
page \pageref{Start of original page 158}, we
begin by expressing, for each member $l$ of $\mathcal{U}
\left( W\right) $ such that for some member $A$ of
$\mathbb{B} \left( G\right) $, $\mathcal{C} \left(
W,l\right) $, (or in other words, the unique member of $W$
that has $l$ as a member), is a $\left( V\cup H\right)
$-key of $A$ such that $\mathcal{C} \left( W,l\right) $ is
\emph{not} a member of $V$, and $l$ is the member of
$\mathcal{C} \left( W,l\right) $ that is \emph{not} the key
member of any $\left( V\cup H\right) $-firm over $V$
component of $A$ that does not have $f_{ A } $ as a member,
$\hat{ y }_{ l } $ as $\hat{ y }_{ l } =\left( \hat{ y }_{
l } +\hat{ y }_{ m } \right)
-\hat{ y }_{ m } =t_{ \mathcal{C} \left( W,l\right) }
 -\hat{ y
}_{ m } $,
where $m$ is the other member of $\mathcal{C} \left(
W,l\right) $, and thus \emph{is} the key member of the
$\left( V\cup H\right) $-firm over $V$ component $\left(
A\,\vdash \uparrow \left( f_{ A } ,\left\{ \mathcal{C}
\left(
W,l\right) \right\} \right) \right) \,\vdash \left(
\bigcup_{ R\in Z
} \left( A\,\vdash \uparrow \left( f_{ A } ,\left\{
R\right\}
\right) \right) \right) $ of $A$, where $Z$ is the set whose
members are all the $\left( V\cup H\right) $-keys $R$ of
$A$ such that $R\notin V$ and $R\to \mathcal{C} \left(
W,l\right) $ both hold, and the differential operator $t_{
D } $ was defined on
pages \pageref{Start of original page 158} and
\pageref{Start of original page 163}, for each member
$D$ of $W$, to be equal to $\left( \hat{ y }_{ n } +\hat{ y
}_{ o }
\right) $, where $n$ and $o$ are the two members of $D$, or
in other words, where $D=\left\{ n,o\right\} $ holds.   We
do
this for every member $\beta $ of $\mathcal{D} \left(
j\right) $ such that $j_{ \beta } $ is equal to such a
member $l$ of $\mathcal{U} \left( W\right) $, thus showing,
for each member $k$ of $\mathbb{N} $ and for each member
$\rho $ of $\mathbb{D} $, that $I \rule{0pt}{2.55ex}^{
\hspace{-0.75ex} \circ } _{ k } \left(
B,i,n,j,s,E,v,\rho \right) $ is equal to the sum, over a
finite collection of septuples $\left( \tilde{ B } ,\tilde{
i }
,\tilde{ n },\tilde{ j } ,\tilde{ s },\tilde{ E } ,\tilde{
v }\right) $ as on
page \pageref{Start of original page 163}, each
satisfying the condition on
page \pageref{Start of original page 164},
and each satisfying
the requirement that $\left( \#\left(
 \mathcal{D} \left( \tilde{ j }
 \right) \right) +\#\left( \mathcal{D} \left( \tilde{ E }
\right) \right) \right) $ is not greater than the integer
$N$ defined on
page \pageref{Start of original page 158},
and each satisfying the further
requirement that if $A$ is any member of $\mathbb{B} \left(
G\right) $, and $T\equiv \left\{ l,m\right\} $ is any
$\left(
V\cup H\right) $-key of $A$ such that $T$ is \emph{not} a
member of $V$, and $l$ is the member of $T$ that is
\emph{not} the key member of some $\left( V\cup H\right)
$-firm over $V$ component of $A$ that does not have $f_{ A
} $ as a member, then there is \emph{no} member $\beta $ of
$\mathcal{D} \left( \tilde{ j } \right) $ such that
$\tilde{ j } _{
\beta } =l$ holds, of $I \rule{0pt}{2.55ex}^{
\hspace{-0.75ex} \circ } _{ k } \left(
 \tilde{ B } ,\tilde{ i
} ,\tilde{ n },\tilde{ j } ,\tilde{ s },\tilde{ E }
,\tilde{ v },\rho \right) $ times a
finite integer coefficient, (positive or negative), that is
\label{Start of original page 182}
 independent of $k$ and independent of $\rho $.

Now let $\left(B,i,n,j,s,E,v\right) $ be any
ordered septuple as on
page \pageref{Start of original page 163},
such that the condition on
page \pageref{Start of original page 164} holds, and
such that $\left( \#\left( \mathcal{D} \left( j\right)
\right) +\#\left( \mathcal{D} \left( E\right) \right)
\right) $ is not greater than the integer $N$ defined on
page \pageref{Start of original page 158},
and such that if $A$ is any member of
$\mathbb{B} \left( G\right) $, $T$ is any $\left( V\cup
H\right) $-key of $A$ such that $T$ is \emph{not} a member
of $V$, and $l$ is the member of $T$ that is \emph{not} the
key member of any $\left( V\cup H\right) $-firm over $V$
component of $A$ that does not have $f_{ A } $ as a member,
then there is \emph{no} member $\beta $ of $\mathcal{D}
\left( j\right) $ such that $j_{ \beta } =l$ holds.

Let $A$ be any member of $\mathbb{B} \left( G\right) $, let
$T$ be any $\left( V\cup H\right) $-key of $A$ such that
$T$ is \emph{not} a member of $V$, let $m$ be the member of
$T$ that \emph{is} the key member of a $\left( V\cup
H\right) $-firm over $V$ component of $A$ that does not
have $f_{ A } $ as a member, (so that, specifically, $m$ is
the key member of the $\left( V\cup H\right) $-firm over
$V$ component $\left( A\,\vdash \uparrow \left( f_{ A }
,\left\{ T\right\} \right) \right) \,\vdash \left(
\bigcup_{ R\in Z }
\left( A\,\vdash \uparrow \left( f_{ A } ,\left\{ R\right\}
\right) \right) \right) $ of $A$, where $Z$ is the set whose
members are all the $\left( V\cup H\right) $-keys $R$ of
$A$ such that $R\notin V$ and $R\to T$ both hold), and let
$\beta $ be any member of $\mathcal{D} \left( j\right) $
such that $j_{ \beta } =m$ holds.

Let $\tilde{ j } $ be the map whose domain is equal
 to $\left(
\mathcal{D} \left( j\right) \,\vdash \left\{ \beta \right\}
\right) $, and such that for each member $\gamma $ of
$\mathcal{D} \left( \tilde{ j } \right) $,
 $\tilde{ j }_{ \gamma }
=j_{
\gamma } $ holds, (so that $\tilde{ j } $ is equal
to $\left(
j\,\vdash \left\{ \left( \beta ,j_{ \beta } \right)
\right\}
\right) =\left( j\,\vdash \left\{ \left( \beta ,m\right)
\right\} \right) $).

Let $C$ be the member $\left( A\,\vdash \uparrow \left( f_{
A } ,\left\{ T\right\} \right) \right) $ of $J$.   We note
that it directly follows from the fact that, by
page \pageref{Start of original page 165},
$S_{ A } $ is a subset of the $\left( V\cup H\right) $-firm
over $V$ component of $A$ that has $f_{ A } $ as a member,
and from the fact that, by
pages \pageref{Start of original page 171} and
\pageref{Start of original page 172}, $C$ is
\emph{not} equal to $S_{ D } $ for any member $D$ of
$\left( J\,\vdash P\right) $, that $C$ is a member of
$\left( J\,\vdash \left( \mathcal{R} \left( S\right) \cup
\left\{ \mathcal{U} \left( V\right) \right\} \right) \right)
$, hence that by
page \pageref{Start of original page 178},
$a_{ C } $ is one of our new
independent integration variables.

Then it directly follows from our assumptions on
$\tilde{ \mathcal{J} } $, that for any member $k$ of
$\mathbb{N} $, and for any member $\rho $ of $\mathbb{D} $,
the integral over $\mathbb{W} $, of $\hat{ a }_{ C } $
acting on
the integrand of $I \rule{0pt}{2.55ex}^{
\hspace{-0.75ex} \circ } _{ k } \left( B,i,n,\tilde{ j }
,s,E,v,\rho \right) $, is equal to $0$, or in other words,
that
\[
0=\int_{ \mathbb{W} } \left( \prod_{D\in \left( V\,\vdash
\left\{ O\right\}
\right) } d^{ d } x_{ D } \right) \times \hspace{10.0cm}
\]
\[
\times \hspace{-3.0pt} \left( \hspace{-3.0pt}
 \hat{ a }_{ C } \hspace{-3.0pt} \left(
 \hspace{-3.0pt} \left( \prod_{B\in
\left(
Y\,\vdash G\right) } \hspace{-10.0pt} \mathcal{E}
\hspace{-2.0pt} \left( \left(
P\cap \Xi \hspace{-2.0pt}
 \left( \mathcal{P} \hspace{-2.0pt}
  \left( G,B\right) \right)
\right) ,\left( Q\cap \Xi \hspace{-2.0pt}
 \left( \mathcal{P} \left(
G,B\right) \right) \right) ,H,\sigma ,R,\downarrow
\hspace{-4.0pt} \left(
x,\Xi \hspace{-2.0pt} \left( \mathcal{P} \hspace{-2.0pt}
\left( G,B\right) \right) \right)
\right) \hspace{-3.0pt} \right) \hspace{-3.6pt}
\times \rule{0pt}{5.5ex} \right. \right.
\]
\label{Start of original page 183}
\[
 \times \left( \left( \prod_{\alpha \in \mathcal{D}
\left( B\right) } \left( \left( x_{ \mathcal{K}
\left( Q,B_{ \alpha } ,i_{ \alpha } \right) } -x_{ B_{
\alpha } } \right) .n_{ \alpha } \right) \right) \left(
  \prod_{\beta \in
\mathcal{D} \left( \tilde{ j } \right) } \left( s_{ \beta
}.\hat{ y }_{ \tilde{ j } _{ \beta } } \right) \right)
\times \right.
\]
\[
\hspace{6.0cm} \hspace{-11.78pt}
\left. \left. \left. \rule{0pt}{5.5ex} \times
\left( \prod_{\gamma \in
\mathcal{D}
\left( E\right) } \left( v_{ \gamma } .t_{ E_{ \gamma } }
\right) \right) \tilde{ \mathcal{J} } _{ kQ }
\left( y\right) \right) { \rule[-3.75ex]{0pt}{3.75ex} }_{
\hspace{-5.4pt}
y=\mu \left( P,Q,H,x,\mathcal{X} \left( P,Q,H,\rho \right)
\right) } \right) \right)
\]
holds.

\vspace{2.5ex}

We now make the following observations:

\begin{bphzobservation} \label{Observation 33}
\end{bphzobservation}
\vspace{-6.143ex}

\noindent \hspace{2.6ex}{\bf ) }When we substitute the
expressions on
pages \pageref{Start of original page 179} and
\pageref{Start of original page 180}
for the $z_{ L } $ variables, $L\in \left( P\,\vdash
\left( \mathcal{R} \left( S\right) \cup \left\{ \mathcal{U}
\left( V\right) \right\} \right)
\right) $,\hspace{0.1ex}
in\hspace{0.1ex} terms\hspace{0.1ex} of\hspace{0.1ex}
the $a_{ D } $
variables, $D\in \left( J\,\vdash \left(
\mathcal{R} \left( S\right) \cup \left\{ \mathcal{U} \left(
V\right) \right\} \right) \right) $,
\newpage
\noindent into the expressions on
page \pageref{Start of original page 178}
 for the $x_{ L } $ variables in terms of the
$z_{ D } $ variables, the only terms in the sums over $A\in
\mathbb{Y} \left( \bar{ P } ,\mathcal{C} \left( V,l\right)
,\mathcal{U} \left( V\right) \right) $ and $A\in \mathbb{Y}
\left( \bar{ P } ,L,\mathcal{U} \left( V\right) \right) $,
in
the expressions on
page \pageref{Start of original page 178}
 for $x_{ \mathcal{C} \left(
V,l\right) } $ and $x_{ L } $, that get altered more than
simply by re-writing each letter ``$ z $'' as the letter
 ``$ a $'',
 are
the terms where $A$ is a member of $\mathbb{B} \left(
G\right) $.   This follows directly from the formula on
page \pageref{Start of original page 179}
 for the $z_{ D } $ variables in terms of the
$a_{ L } $ variables when $D$ is a member of $J\cap \left(
\Xi \left( \mathcal{P} \left( P,A\right) \right) \,\vdash
\left\{ A\right\} \right) $ for some member $A$ of
$\mathbb{B} \left( G\right) $, together with the fact that,
as defined on
page \pageref{Start of original page 177},
if $D $ is any member of $\left( J\,\vdash
\left\{ \mathcal{U} \left( V\right) \right\} \right) $
such that $D $ is \emph{not} a member of
$J\cap \left( \Xi \left(
\mathcal{P} \left( P,A\right) \right) \,\vdash \left\{
A\right\} \right) $ for some member $A $ of
$\mathbb{B} \left( G\right) $, then $
a_{ D } \equiv z_{ D } $ holds.
\enlargethispage{2.75ex}

\begin{bphzobservation} \label{Observation 34}
\end{bphzobservation}
\vspace{-6.143ex}

\noindent \hspace{2.6ex}{\bf ) }Let $D$ be any member of
$\left( Y\,\vdash G\right) $.  Then
\[
\mathcal{E} \left( \left( P\cap \Xi \left( \mathcal{P}
\left( G,D\right) \right) \right) ,\left( Q\cap \Xi \left(
\mathcal{P} \left( G,D\right) \right) \right) ,H,\sigma
,R,\downarrow \left( x,\Xi \left( \mathcal{P} \left(
G,D\right) \right) \right) \right)
\]
is completely independent of $a_{ C } $, (where $C$ is as
defined on
page \pageref{Start of original page 182}), hence
\[
\hat{ a }_{ C } \mathcal{E} \left( \left( P\cap \Xi \left(
\mathcal{P} \left( G,D\right) \right) \right) ,\left( Q\cap
\Xi \left( \mathcal{P} \left( G,D\right) \right) \right)
,H,\sigma ,R,\downarrow \left( x,\Xi \left( \mathcal{P}
\left( G,D\right) \right) \right) \right)
\]
is identically equal to $0$.   For the expression
\[
\mathcal{E} \left( \left( P\cap \Xi \left(\mathcal{P}
\left(
G,D\right) \right) \right) ,\left( Q\cap \Xi
\left( \mathcal{P} \left( G,D\right) \right) \right),
H,\sigma ,R,\downarrow \! \left(x,\Xi
\left( \mathcal{P} \left( G,D\right) \right) \right)
\right)
\]
depends on $x$ only through differences $\left( x_{ K }
-x_{ L } \right) $, where both $K$ and $L$ are members of
$\Xi \left( \mathcal{P} \left( G,D\right) \right) $, and by
observations \ref{Observation 30})
to \ref{Observation 33}) above, $a_{ C } $
occurs with the \emph{same} coefficient in $x_{ K } $ for
every member $K$ of $\Xi \left( \mathcal{P} \left(
G,D\right) \right) $, hence every such difference is
completely independent of $a_{ C } $.

\begin{bphzobservation} \label{Observation 35}
\end{bphzobservation}
\vspace{-6.143ex}

\noindent \hspace{2.6ex}{\bf ) }Let $D$ be any member of
$\mathbb{B} \left( Q\right) $
such that $D$ is \emph{not} a member of $\mathbb{B} \left(
G\right) $, and let $l$ be any member of $D$.   Then the
expression $\left( x_{ \mathcal{K} \left( Q,D,l\right) }
-x_{ D
} \right) $ is completely independent of $a_{ C } $, where
$C$ is defined as on
page \pageref{Start of original page 182},
hence $\hat{ a }_{ C } \left(
x_{ \mathcal{K} \left( Q,D,l\right) } -x_{ D } \right) $ is
equal to $0$.   For if $D$ is a subset of the member $A$ of
$\mathbb{B} \left( G\right) $ with respect to which $C$ was
defined on
page \pageref{Start of original page 182},
then $D$ is a strict subset of $A$, hence since $Q$
\label{Start of original page 184}
 is a member of $\mathbb{O} \left( G,H\right) $, $D$ is a
subset of some $\left( V\cup H\right) $-firm over $V$
component of $A$.   Hence the independence of $\left(
x_{ \mathcal{K} \left( Q,D,l\right) } -x_{ D } \right) $
from
$a_{ C } $ follows directly from
observations \ref{Observation 30})
to \ref{Observation 33}) above,
together with the fact that both
$x_{ \mathcal{K} \left( Q,D,l\right) } $ and $x_{ D } $ are
linear combinations, with coefficients summing to $1$, of
the $x_{ K } $, $K\in \mathcal{P} \left( P,D\right) $.

\begin{bphzobservation} \label{Observation 36}
\end{bphzobservation}
\vspace{-6.143ex}

\noindent \hspace{2.6ex}{\bf ) }Let $D$ be any member of
$\mathbb{B} \left( G\right) $
and $l$ be any member of $D$.   Then $\left( x_{ \mathcal{K}
\left( Q,D,l\right) } -x_{ D } \right) $
\newpage
\noindent is independent of
$a_{ C } $, where $C$ is defined as on
page \pageref{Start of original page 182}, unless
$D$ is equal to $A$, where $A$ is the member of $\mathbb{B}
\left( G\right) $ with reference to which $C$ was defined
on
page \pageref{Start of original page 182},
and if $D$ \emph{is} equal to $A$, then
$\hat{ a }_{ C } \left( x_{ \mathcal{K} \left( Q,D,l\right)
} -x_{ D }
\right) $ is equal to the sum of a finite number of terms,
each of which has the form of a finite real-number
coefficient times the product of finite powers, all $\geq
0$, of the $\rho _{ K } $, $K\in \left( Q\,\vdash P\right)
$.   For it directly follows from
observations \ref{Observation 30})
to \ref{Observation 33}) above,
together with the fact that if $D$
is not equal to $A$, then either $D\cap A=\emptyset $
holds, or
$D\subset A$ holds, in which case $D$ is a subset of a
$\left( V\cup H\right) $-firm over $V$ component of $A$,
(since $G$ is an $H $-principal wood of V), or $A\subset D$
holds, in which case $A$ is a subset of a $\left( V\cup
H\right) $-firm over $V$ component of $D$, that $\left(
x_{ \mathcal{K} \left( Q,D,l\right) } -x_{ D } \right) $ is
independent of $a_{ C } $ unless $D$ is equal to $A$.   And
if $D$ \emph{is} equal to $A$, then since both
$x_{ \mathcal{K} \left( Q,D,l\right) } $ and $x_{ D } $ are
linear combinations, with $\rho $-independent coefficients,
of the $x_{ K } $, $K\in \mathcal{P} \left( P,A\right) $,
and these $x_{ K } $ are expressed in terms of the $a_{ L }
$ variables by the formulae on
pages \pageref{Start of original page 178} to
\pageref{Start of original page 180}, the
only $\rho $-dependence of the expressions for
$x_{ \mathcal{K} \left( Q,D,l\right) } $ and $x_{ D } $ in
terms
of the $a_{ L } $ variables is through the $\nu _{ KL } $
coefficients, whose only $\rho $-dependence, by
pages \pageref{Start of original page 176}
and \pageref{Start of original page 177},
is through the $r_{ L } $ coefficients, whose
only $\rho $-dependence, by
page \pageref{Start of original page 176},
is through the $u_{
N } $ coefficients, whose only $\rho $-dependence, by
page \pageref{Start of original page 174},
is through the sum of a finite number, (in fact, at
the most, two), of terms, each of which has the form of a
product of finite powers, all $\geq 0$, of the $\rho _{ R }
$, $R\in \left( Q\,\vdash P\right) $.

\begin{bphzobservation} \label{Observation 37}
\end{bphzobservation}
\vspace{-6.143ex}

\noindent \hspace{2.6ex}{\bf ) }From observations
\ref{Observation 35})
and \ref{Observation 36}) above it
immediately follows that the action of $\hat{ a }_{ C } $,
where
$C$ is defined as on
page \pageref{Start of original page 182}, on the factor
$\left( \prod_{\alpha \in
\mathcal{D} \left( B\right) } \left( \left( x_{
\mathcal{K} \left( Q,B_{ \alpha } ,i_{ \alpha } \right)
} -x_{ B_{ \alpha } } \right) .n_{ \alpha } \right) \right)
$ in the integrand of $I \rule{0pt}{2.55ex}^{
\hspace{-0.75ex} \circ } _{ k } \left( B,i,n,\tilde{ j
}
,s,E,v,\rho \right) $, is equal to the sum of a finite
number of terms, each of which has the form of a finite
real number, independent of $k$ and $\rho $, times a
product of finite powers, all $\geq 0$ and independent of
$k$ and $\rho $, of the
\label{Start of original page 185}
 $\rho _{ R } $, $R\in \left( Q\,\vdash P\right) $, times
$I \rule{0pt}{2.55ex}^{ \hspace{-0.75ex} \circ } _{ k }
\left( \tilde{ B } ,\tilde{ i } ,
\tilde{ n },\tilde{ j }
,s,E,v,\rho \right) $, for some ordered septuple $\left(
\tilde{B } ,\tilde{ i } ,\tilde{ n },\tilde{ j } ,
s,E,v\right) $ as on
page \pageref{Start of original page 163},
such that the condition on
page \pageref{Start of original page 164} holds, and
such that $\left( \tilde{ B } ,\tilde{ i } ,\tilde{ n },
\tilde{ j }
,s,E,v\right) $ differs from $\left( B,i,n,\tilde{ j }
,s,E,v\right) $ at most in the first three components.

\begin{bphzobservation} \label{Observation 38}
\end{bphzobservation}
\vspace{-6.143ex}

\noindent \hspace{2.6ex}{\bf ) }With the definitions of
page \pageref{Start of original page 182}, let $l$ be the
member of $T$ that is \emph{not} the key member of any
$\left( V\cup H\right) $-firm over $V$ component of $A$
that does not have $f_{ A } $ as a member, (so that $T$ is
equal to $\left\{ l,m\right\} $), and let $\left\{
r,s\right\} $ be any member of $W$ such that $\left\{
r,s\right\} $ is \emph{not} equal to $T$.   Then
\[
\hat{ a }_{ C } \mu _{ r } \left( P,Q,H,x,\mathcal{X} \left(
P,Q,H,\rho \right) \right) =\hat{ a }_{ C } \mu _{ s }
\left(
P,Q,H,x,\mathcal{X} \left( P,Q,H,\rho \right) \right)
\]
holds, and moreover $\hat{ a }_{ C } \mu _{ r } \left(
P,Q,H,x,\mathcal{X} \left( P,Q,H,\rho \right) \right) $ is
equal to a Kronecker delta in the undisplayed $\mathbb{E}
_{ d } $ indices, times the sum of a finite number of
terms, each of which has the form of a finite real number,
independent of $\rho $, times a product of finite powers,
all $\geq 0$ and independent of $\rho $, of the $\rho _{ K
} $, $K\in \left( Q\,\vdash P\right) $.

And furthermore
\[
\hat{ a }_{ C } \mu _{ m } \left( P,Q,H,x,\mathcal{X} \left(
P,Q,H,\rho \right) \right) =1+\hat{ a }_{ C } \mu _{ l }
\left(
P,Q,H,x,\mathcal{X} \left( P,Q,H,\rho \right) \right)
\]
holds, where the $1$ displayed in the right-hand side is to
be understood to be multiplied by a Kronecker delta in the
undisplayed $\mathbb{E} _{ d } $ indices, and moreover \\
$\hat{ a }_{ C } \mu _{ l } \left( P,Q,H,x,\mathcal{X}
\left(
P,Q,H,\rho \right) \right) $ is equal to a Kronecker delta
in the undislayed $\mathbb{E} _{ d } $ indices, times the
sum of a finite number of terms, each of which has the form
of a finite real number, independent of $\rho $, times a
product of finite powers, all $\geq 0$ and independent of
$\rho $, of the $\rho _{ K } ,K\in \left( Q\,\vdash
P\right) $.

For if $v$ is any member of $\mathcal{U} \left( W\right) $
then $v$ is certainly a member of $\mathcal{O} \left(
V,H\right) $, hence by
pages \pageref{Start of original page 80},
\pageref{Start of original page 81}, and
\pageref{Start of original page 174}, $\mu _{
v } \left( P,Q,H,x,\mathcal{X} \left( P,Q,H,\rho \right)
\right) $ is a linear combination, with coefficients
summing to $1$, of the $x_{ D } $, $D\in \mathbb{G} _{
v\mathcal{Z} \left( P,H,v\right) } \left( Q,H\right) $, or
in other words, of the $x_{ D } $ for the members $D$ of
$Q$ such that $\mathcal{Z} \left( P,H,v\right) \subseteq
D\subseteq \mathcal{Z} \left( Q,H,v\right) $ holds, and
each coefficient in this sum has the form of the sum of a
finite number, (in fact, at the most, two) of terms, each
of which has the form of a finite real number, (in fact,
$+1$ or $-1 $), times a product of finite powers,
independent of $\rho $, of the $\rho _{ K } $, $K\in \left(
Q\,\vdash P\right) $.

We recall from
page \pageref{Start of original page 7}
 that $\mathcal{T} \left( A,H\right)
$ is by definition the subset of $A$ whose members are all
the members $e$ of $A$ such that there is \emph{no} member
$R$ of $H$ such that $e\in R$ and $R\subseteq A$ both hold.
  And we recall from
page \pageref{Start of original page 157}
 that $H$ is a partition
such that if $R$ is any member of $H$ such that $R$
intersects \emph{more} than one member of $V$, then $R$ has
\label{Start of original page 186}
 \emph{exactly} two members, and furthermore that $W$ is
the subset of $H$ whose members are all the members $R$ of
$H$ such that $R$ intersects exactly two members of $V$, so
that $W$ is a partition such that every member of $W$ has
exactly two members.

Now if $v$ is any member of $\mathcal{U} \left( W\right) $,
then exactly one of the four possibilities $v\in \left(
C\,\vdash \mathcal{T} \left( A,H\right) \right) $, $v\in
\left( \left( A\,\vdash C\right) \,\vdash \mathcal{T}
\left( A,H\right) \right) $, $v\in \mathcal{T} \left(
A,H\right) $, and $v\in \left( \mathcal{U} \left( V\right)
\,\vdash A\right) $ holds.

Suppose first that $v\in \left( C\,\vdash \mathcal{T}
\left( A,H\right) \right) $ holds.   (We note that this
case includes, in particular, the case $v=m $.)
Then $v$ is a member of $C$ such that there exists a member
$R$ of $H$ such that $v\in R$ and $R\subseteq A$ both hold.
  Hence $\mathcal{Z} \left( Q,H,v\right) \subset A$ holds,
hence $\mathcal{Z} \left( Q,H,v\right) \subseteq C$ holds,
since $Q\in \mathbb{O} \left( G,H\right) $ and $\mathcal{Z}
\left( Q,H,v\right) \subset A$ imply together that
$\mathcal{Z} \left( Q,H,v\right) $ is a subset of some
$\left( V\cup H\right) $-firm over $V$ component of $A$,
and each $\left( V\cup H\right) $-firm over $V$ component
of $A$ is either a subset of $C$ or else does not intersect
$C$, and $\mathcal{Z} \left( Q,H,v\right) \cap C$ has the
member $v$ hence is nonempty.   And the fact that
$\mathcal{Z} \left( Q,H,v\right) \subseteq C$ holds implies
immediately that \emph{every} member $D$ of $\mathbb{G} _{
v\mathcal{Z} \left( P,H,v\right) } \left( Q,H\right) $, or
in other words, every member $D$ of $Q$ such that
$\mathcal{Z} \left( P,H,v\right) \subseteq D\subseteq
\mathcal{Z} \left( Q,H,v\right) $ holds, is a subset of
$C$.   Hence it immediately follows from observations
\ref{Observation 30}) to
\ref{Observation 33}) above, together with the fact
that if $\mathcal{Z} \left( P,H,v\right) \subset
\mathcal{Z} \left( Q,H,v\right) $ holds, then for every
member $D$ of $Q$ such that $\mathcal{Z} \left(
P,H,v\right) \subseteq D\subseteq \mathcal{Z} \left(
Q,H,v\right) $ holds, $x_{ D } $ is by
Lemma \ref{Lemma 4} a linear
combination, with coefficients summing to $1$, of the $x_{
L } $, $L\in \mathcal{P} \left( P,\mathcal{Z} \left(
Q,H,v\right) \right) $, that if $D$ is any member of $Q$
such that $\mathcal{Z} \left( P,H,v\right) \subseteq
D\subseteq \mathcal{Z} \left( Q,H,v\right) $ holds, then
$x_{ D } $ is equal to $\left( a_{ C } +x_{ S_{ A }
} \right) $ plus a term that is independent of
$a_{ C } $.   Hence
$\mu _{ v } \left( P,Q,H,x,\mathcal{X} \left( P,Q,H,\rho
\right) \right) $ is itself equal to $\left( a_{ C } +x_{
S_{ A } } \right) $ plus a term that is independent of $a_{
C } $.

Suppose now that $v\in \left( \left( A\,\vdash C\right)
\,\vdash \mathcal{T} \left( A,H\right) \right) $ holds.
(We note that this case includes, in particular, the case
$v=l $.)   Then $v$ is a member of $\left( A\,\vdash
C\right) $ such that there exists a member $R$ of $H$ such
that $v\in R$ and $R\subseteq A$ both hold.   Hence
$\mathcal{Z} \left( Q,H,v\right) \subset A$ holds, hence
$\mathcal{Z} \left( Q,H,v\right) \subseteq \left( A\,\vdash
C\right) $ holds, since $Q\in \mathbb{O} \left( G,H\right)
$ and $\mathcal{Z} \left( Q,H,v\right) \subset A$ imply
together that $\mathcal{Z} \left( Q,H,v\right) $ is a
subset of some $\left( V\cup H\right) $-firm over $V$
component of $A$, and each $\left( V\cup H\right) $-firm
over $V$ component of $A$ is either a subset of $\left(
A\,\vdash C\right) $ or else does not intersect $\left(
A\,\vdash C\right) $, and $\mathcal{Z} \left( Q,H,v\right)
\cap \left( A\,\vdash C\right) $ has the member $v$ hence
is nonempty.   And the fact that $\mathcal{Z} \left(
Q,H,v\right) \subseteq \left( A\,\vdash C\right) $ holds
implies immediately that \emph{every} member $D$ of
$\mathbb{G} _{ v\mathcal{Z} \left( P,H,v\right) } \left(
Q,H\right) $, or in other words, every member $D$ of $Q$
\label{Start of original page 187}
 such that $\mathcal{Z} \left( P,H,v\right) \subseteq
D\subseteq \mathcal{Z} \left( Q,H,v\right) $ holds, is a
subset of $\left( A\,\vdash C\right) $.   Hence it
immediately follows from
observations \ref{Observation 30}) to
\ref{Observation 33}) above,
together with the fact that if
$\mathcal{Z} \left( P,H,v\right) \subset \mathcal{Z} \left(
Q,H,v\right) $ holds, then for every member $D$ of $Q$ such
that $\mathcal{Z} \left( P,H,v\right) \subseteq D\subseteq
\mathcal{Z} \left( Q,H,v\right) $ holds, $x_{ D } $ is by
Lemma \ref{Lemma 4} a linear combination, with coefficients
summing
to $1$, of the $x_{ L } $, $L\in \mathcal{P} \left(
P,\mathcal{Z} \left( Q,H,v\right) \right) $, that if $D$ is
any member of $Q$ such that $\mathcal{Z} \left(
P,H,v\right) \subseteq D\subseteq \mathcal{Z} \left(
Q,H,v\right) $ holds, then $x_{ D } $ is equal to $x_{ S_{
A } } $ plus a term that is independent of $a_{ C } $.
Hence $\mu _{ v } \left( P,Q,H,x,\mathcal{X} \left(
P,Q,H,\rho \right) \right) $ is itself equal to $x_{ S_{ A
} } $ plus a term that is independent of $a_{ C } $.

Now suppose that $v\in \mathcal{T} \left( A,H\right) $
holds.   Then $v$ is a member of $A$ such that there is
\emph{no} member $R$ of $H$ such that $v\in R$ and
$R\subseteq A$ both hold.   Hence $A\subseteq \mathcal{Z}
\left( P,H,v\right) $ holds, hence \emph{no} member $D$ of
$\mathbb{G} _{ v\mathcal{Z} \left( P,H,v\right) } \left(
Q,H\right) $ is a strict subset of $A$, hence it
immediately follows from
observations \ref{Observation 30}) to
\ref{Observation 33}) above
that for every member $D$ of $\mathbb{G}
_{ v\mathcal{Z} \left( P,H,v\right) } \left( Q,H\right) $,
$x_{ D } $ is equal to $x_{ \mathcal{U} \left( V\right) } $
plus
a term that is independent of $a_{ C } $, hence $\mu _{ v }
\left( P,Q,H,x,\mathcal{X} \left( P,Q,H,\rho \right)
\right) $ is itself equal to $x_{ \mathcal{U} \left(
V\right) }
$ plus a term that is independent of $a_{ C } $.

And finally suppose that $v\in \left( \mathcal{U} \left(
V\right) \,\vdash A\right) $ holds.   Then $\mathcal{Z}
\left( P,H,v\right) \subset A$ certainly does \emph{not}
hold, hence again \emph{no} member $D$ of $\mathbb{G} _{
v\mathcal{Z} \left( P,H,v\right) } \left( Q,H\right) $ is
a strict subset of $A$, hence again, for every member $D$
of $\mathbb{G} _{ v\mathcal{Z} \left( P,H,v\right) } \left(
Q,H\right) $, $x_{ D } $ is equal to $x_{ \mathcal{U}
\left( V\right) } $ plus a term that is independent of $a_{
C
} $, hence $\mu _{ v } \left( P,Q,H,x,\mathcal{X} \left(
P,Q,H,\rho \right) \right) $ is itself equal to
$x_{ \mathcal{U} \left( V\right) } $ plus a term that is
independent of $a_{ C } $.

Now let $\left\{ r,s\right\} $ be any member of $W$.

Suppose first that $r\in \left( C\,\vdash \mathcal{T}
\left( A,H\right) \right) $ holds.   Then either $r=m$
holds, (in which case $s=l$ holds and $\left\{ r,s\right\} $
is equal to T), or else $s$ is \emph{also} a member of
$\left( C\,\vdash \mathcal{T} \left( A,H\right) \right) $.
 Hence if $r\in \left( C\,\vdash \mathcal{T} \left(
A,H\right) \right) $ holds and $\left\{ r,s\right\} $ is
\emph{not} equal to $T$, then it immediately follows from
the foregoing that
\[
\hat{ a }_{ C } \mu _{ r } \left( P,Q,H,x,\mathcal{X} \left(
P,Q,H,\rho \right) \right) =\hat{ a }_{ C } \mu _{ s }
\left(
P,Q,H,x,\mathcal{X} \left( P,Q,H,\rho \right) \right)
\]
holds, and moreover that $\hat{ a }_{ C } \mu _{ r } \left(
P,Q,H,x,\mathcal{X} \left( P,Q,H,\rho \right) \right) $ is
equal to a Kronecker delta in the undisplayed $\mathbb{E}
_{ d } $ indices, times the sum of a finite number of
terms, each of which has the form of a finite real number,
independent of $\rho $, times a product of finite powers,
all $\geq 0$ and independent of $\rho $, of the $\rho _{ K
} $, $K\in \left( Q\,\vdash P\right) $.

Now suppose that $r\in \left( \left( A\,\vdash C\right)
\,\vdash \mathcal{T} \left( A,H\right) \right) $ holds.
Then either $r=l$
\label{Start of original page 188}
 holds, (in which case $s=m$ holds and $\left\{ r,s\right\}
$
is equal to T), or else $s$ is \emph{also} a member of
$\left( \left( A\,\vdash C\right) \,\vdash \mathcal{T}
\left( A,H\right) \right) $.   Hence if $r\in \left( \left(
A\,\vdash C\right) \,\vdash \mathcal{T} \left( A,H\right)
\right) $ holds and $\left\{ r,s\right\} $ is \emph{not}
equal to $T$, then again it immediately follows from the
foregoing that
\[
\hat{ a }_{ C } \mu _{ r } \left( P,Q,H,x,\mathcal{X} \left(
P,Q,H,\rho \right) \right) =\hat{ a }_{ C } \mu _{ s }
\left(
P,Q,H,x,\mathcal{X} \left( P,Q,H,\rho \right) \right)
\]
holds, and moreover that $\hat{ a }_{ C } \mu _{ r } \left(
P,Q,H,x,\mathcal{X} \left( P,Q,H,\rho \right) \right) $ is
equal to a Kronecker delta in the undisplayed $\mathbb{E}
_{ d } $ indices, times the sum of a finite number of
terms, each of which has the form of a finite real number,
independent of $\rho $, times a product of finite powers,
all $\geq 0$ and independent of $\rho $, of the $\rho _{ K
} $, $K\in \left( Q\,\vdash P\right) $.

Now suppose that $r$ is either a member of $\mathcal{T} \!
\left( A, \! H\right) $ or else is a member of $\left(
\mathcal{U} \left( V\right) \vdash \! A\right) $.   Then if
$r$ is a member of $\mathcal{T} \left( A,H\right) $, $s$ is
a member of $\left( \mathcal{U} \left( V\right) \,\vdash
A\right) $, while if $r$ is a member of $\left( \mathcal{U}
\left( V\right) \,\vdash A\right) $, then $s$ is either a
member of $\mathcal{T} \left( A,H\right) $ or else is a
member of $\left( \mathcal{U} \left( V\right) \,\vdash
A\right) $.   Hence if $r$ is either a member of
$\mathcal{T} \left( A,H\right) $ or else is a member of
$\left( \mathcal{U} \left( V\right) \,\vdash A\right) $,
then it again immediately follows from the foregoing that
\[
\hat{ a }_{ C } \mu _{ r } \left( P,Q,H,x,\mathcal{X} \left(
P,Q,H,\rho \right) \right) =\hat{ a }_{ C } \mu _{ s }
\left(
P,Q,H,x,\mathcal{X} \left( P,Q,H,\rho \right) \right)
\]
holds, and moreover that $\hat{ a }_{ C } \mu _{ r } \left(
P,Q,H,x,\mathcal{X} \left( P,Q,H,\rho \right) \right) $ is
equal to a Kronecker delta in the undisplayed $\mathbb{E}
_{ d } $ indices, times the sum of a finite number of
terms, each of which has the form of a finite real number,
independent of $\rho $, times a product of finite powers,
all $\geq 0$ and independent of $\rho $, of the $\rho _{ K
} $, $K\in \left( Q\,\vdash P\right) $.
\enlargethispage{2.0ex}

And finally we note that $m$ is a member of $\left(
C\,\vdash \mathcal{T} \left( A,H\right) \right) $ and $l$
is a member of $\left( \left( A\,\vdash C\right) \,\vdash
\mathcal{T} \left( A,H\right) \right) $, hence it
immediately follows from the foregoing that
\[
\hat{ a }_{ C } \mu _{ m } \left( P,Q,H,x,\mathcal{X} \left(
P,Q,H,\rho \right) \right) =1+\hat{ a }_{ C } \mu _{ l }
\left(
P,Q,H,x,\mathcal{X} \left( P,Q,H,\rho \right) \right)
\]
holds, where the $1$ displayed in the right-hand side is to
be understood to be multiplied by a Kronecker delta in the
undisplayed $\mathbb{E} _{ d } $ indices,
and moreover that \\
$\hat{ a }_{ C } \mu _{ l } \left( P,Q,H,x,\mathcal{X}
\left(
P,Q,H,\rho \right) \right) $ is equal to a Kronecker delta
in the undisplayed $\mathbb{E} _{ d } $ indices, times the
sum of a finite number of terms, each of which has the form
of a finite real number, independent of $\rho $, times a
product of finite powers, all $\geq 0$ and independent of
$\rho $, of the $\rho _{ K } $, $K\in \left( Q\,\vdash
P\right) $.

\begin{bphzobservation} \label{Observation 39}
\end{bphzobservation}
\vspace{-6.143ex}

\noindent \hspace{2.6ex}{\bf ) }With the definitions of
page \pageref{Start of original page 182},
\[
\hat{ a }_{ C } \tilde{ \mathcal{J} }_{ kQ }
\left( \mu \left(
P,Q,H,x,\mathcal{X} \left( P,Q,H,\rho \right) \right)
\right) =\hat{ a }_{ C } \left( \left(
 \tilde{ \mathcal{J} }_{ kQ }\left(
y\right) \right)_{
 y=\mu
\left( P,Q,H,x,\mathcal{X} \left( P,Q,H,\rho \right)
\right) } \right)
\]
is equal to
\label{Start of original page 189}
\[
\left( \left(
\hat{ y }_{ m } \tilde{ \mathcal{J} }_{ kQ }
\left( y\right) \right)_{
y=\mu
\left(
P,Q,H,x,\mathcal{X} \left( P,Q,H,\rho \right) \right) }
\right)
\]
plus the sum, over the members $R$ of $W$, of a finite
coefficient times
\[
\left( \left(
t_{ R } \tilde{ \mathcal{J} }_{ kQ }
\left( y\right) \right)_{ y=\mu \left(
P,Q,H,x,\mathcal{X} \left( P,Q,H,\rho \right) \right) }
\right) ,
\]
where for each member $R$ of $W$, the finite coefficient is
equal to the sum of a finite number of terms, each of which
has the form of a finite real number, independent of $x$
and $\rho $, times a product of finite powers, all $\geq 0$
and independent of $x$ and $\rho $, of the $\rho _{ K } $,
$K\in \left( Q\,\vdash P\right) $, and, as defined on
pages \pageref{Start of original page 158}
and \pageref{Start of original page 163},
$t_{ R } $ is the differential operator
$t_{ R } \equiv \left( \hat{ y }_{ r } +\hat{ y }_{ s }
\right) $, where
$r$ and $s$ are the two members of $R$, or in other words,
$R=\left\{ r,s\right\} $ holds,
plus the sum, over the members $v$ of $\left( \mathcal{U}
\left( V\right) \,\vdash \mathcal{U} \left( W\right)
\right) $, of a finite coefficent times
\[
\left( \left(
\hat{ y }_{ v } \tilde{ \mathcal{J} }_{ kQ }
\left( y\right) \right)_{
y=\mu
\left(
P,Q,H,x,\mathcal{X} \left( P,Q,H,\rho \right) \right) }
\right) ,
\]
where for each member $v$ of $\left( \mathcal{U} \left(
V\right) \,\vdash \mathcal{U} \left( W\right) \right) $,
the finite coefficient is equal to the sum of a finite
number of terms, each of which has the form of a finite
real number, independent of $x$ and $\rho $, times a
product of finite powers, all $\geq 0$ and independent of
$x$ and $\rho $, of the $\rho _{ K } $, $K\in \left(
Q\,\vdash P\right) $.

This follows immediately from
observation \ref{Observation 38})
above, together with the fact that, by the chain rule for
differentiating a function of a function,
\[
\hat{ a }_{ C } \tilde{ \mathcal{J} }_{ kQ }
\left( \mu \left(
P,Q,H,x,\mathcal{X} \left( P,Q,H,\rho \right) \right)
\right) =\hat{ a }_{ C } \left( \left(
 \tilde{ \mathcal{J} }_{ kQ }\left(
y\right) \right)_{
 y=\mu
\left( P,Q,H,x,\mathcal{X} \left( P,Q,H,\rho \right)
\right) }\right)
\]
is equal to
\[
  \sum_{v\in \mathcal{U} \left( V\right) } \left( \left(
   \hat{ a }_{ C }
\mu _{ v } \left( P,Q,H,x,\mathcal{X} \left( P,Q,H,\rho
\right) \right) \right) . \left( \left(
\hat{ y }_{ v } \tilde{ \mathcal{J} }_{
kQ } \left(
y\right) \right)_{ y=\mu \left( P,Q,H,x,\mathcal{X}
\left( P,Q,H,\rho
\right) \right) } \right) \right).
\]

\vspace{2.5ex}

We now observe that, with the definitions of
page \pageref{Start of original page 182}, it
follows directly from the equation displayed on
page \pageref{Start of original page 183}, together with
observations \ref{Observation 34}), \ref{Observation 37}),
and \ref{Observation 39})
above, that $I \rule{0pt}{2.55ex}^{ \hspace{-0.75ex} \circ
} _{ k } \left( B,i,n,j,s,E,v,\rho
\right) $, where $\left(B,i,n,j,s,E,v\right) $ is
 any ordered septuple as
on
page \pageref{Start of original page 182},
and $k$ is any member of $\mathbb{N} $, is
equal to the sum, over a finite number of ordered septuples
$\left( \tilde{ B } ,\tilde{ i } ,\tilde{ n },
\tilde{\tilde{ j } } ,\tilde{ s },\tilde{ E }
,\tilde{ v }\right) $ as on
page \pageref{Start of original page 163},
each satisfying the condition on
page \pageref{Start of original page 164}, and each
\label{Start of original page 190}
 satisfying the condition that $\left( \#\left( \mathcal{D}
\left( \tilde{ \tilde{ j } } \right) \right) +
\#\left( \mathcal{D} \left(
\tilde{ E } \right) \right) \right) $ is not greater than
the integer $N$ defined on
page \pageref{Start of original page 158},
and each satisfying the
requirement that if $A$ is any member of $\mathbb{B} \left(
G\right) $, $T$ is any $\left( V\cup H\right) $-key of $A$
such that $T$ is \emph{not} a member of $V$, and $l$ is the
member of $T$ that is \emph{not} the key member of any
$\left( V\cup H\right) $-firm over $V$ component of $A$
that does not have $f_{ A } $ as a member, then there is
\emph{no} member $\beta $ of $\mathcal{D}
\left( \tilde{ \tilde{ j } }
\right) $ such that $\tilde{ \tilde{ j } } _{ \beta } =l$
 holds, and
each satisfying the requirement that if $v$ is any member
of $\mathcal{U} \left( W\right) $, then the number of
members $\beta $ of $\mathcal{D} \left(
\tilde{ \tilde{ j } } \right) $
such that $\tilde{ \tilde{ j } } _{ \beta } =v$ holds
is $\leq $ the
number of members $\beta $ of $\mathcal{D} \left( j\right)
$ such that $j_{ \beta } =v$ holds, and each satisfying
the requirement that if $A$ is the particular member of
$\mathbb{B} \left( G\right) $ chosen arbitrarily on
page \pageref{Start of original page 182},
and $T$ is the particular $\left( V\cup H\right)
$-key of $A$ that is not a member of $V$, chosen
arbitrarily on
page \pageref{Start of original page 182}, and, as on
page \pageref{Start of original page 182}, $m$ is
the member of $T$ that \emph{is} the key member of a
$\left( V\cup H\right) $-firm over $V$ component of $A$
that does not have $f_{ A } $ as a member, then the number
of members $\beta $ of $\mathcal{D} \left( \tilde{
 \tilde{ j } } \right)
$ such that $\tilde{ \tilde{ j } } _{ \beta } =m$ holds,
 is one
\emph{less} than the number of members $\beta $ of
$\mathcal{D} \left( j\right) $ such that $j_{ \beta } =m$
holds, of $I \rule{0pt}{2.55ex}^{ \hspace{-0.75ex} \circ }
_{ k } \left( \tilde{ B } ,
\tilde{ i } ,\tilde{ n },\tilde{ \tilde{ j } }
 ,\tilde{ s },\tilde{ E } ,\tilde{ v },\rho
\right) $ times a finite
coefficient which is equal to the sum of a finite number of
terms, each of which has the form of a finite real number,
independent of $k$ and $\rho $, times a product of finite
powers, all $\geq 0$ and independent of $k$ and $\rho $, of
the $\rho _{ K } $, $K\in \left( Q\,\vdash P\right) $.

(We note that when we use
observations \ref{Observation 34}),
\ref{Observation 37}),
and \ref{Observation 39}) above to write out the
expanded form of the equation displayed on
page \pageref{Start of original page 183},
and take the dot product of $\hat{ a }_{
C } $ with the unit $d $-vector
$s_{ \beta } $, where $\beta
$ is the particular member of $\mathcal{D} \left( j\right)
$, such that $j_{ \beta } =m$ holds, that was arbitrarily
selected on
page \pageref{Start of original page 182},
the term \\
$\left( \left(
s_{ \beta } .\hat{ y }_{ m }
\tilde{ \mathcal{J} }_{ kQ }\left( y\right) \right)_{
 y=\mu \left(
P,Q,H,x,\mathcal{X} \left( P,Q,H,\rho \right) \right) }
\right) $,
which occurs in \\
$s_{ \beta } .\hat{ a }_{ C } \left(
\left( \tilde{
\mathcal{J}
}_{
kQ }\left( y\right) \right)_{
 y=\mu \left( P,Q,H,x,\mathcal{X} \left(
P,Q,H,\rho \right) \right) } \right) $ by
observation \ref{Observation 39})
above, exactly reproduces the given
integral $I \rule{0pt}{2.55ex}^{ \hspace{-0.75ex} \circ }_{
k }\left( B,i,n,j,s,E,v,\rho \right) $.)

And by using this result a sufficiently large, but finite,
number of times, after first having used the result on
page \pageref{Start of original page 181},
we finally obtain the result stated on
page \pageref{Start of original page 180},
namely that if $\left(B,i,n,j,s,E,v\right) $ is any
ordered septuple as on
page \pageref{Start of original page 163},
such that the condition on
page \pageref{Start of original page 164} holds,
and such that $\left( \#\left( \mathcal{D} \left( j\right)
\right) +\#\left( \mathcal{D} \left( E\right) \right)
\right) $ is not greater
\label{Start of original page 191}
 than the integer $N$ defined on
page \pageref{Start of original page 158}, then there
exists a finite collection of ordered
 septuples $\left( \tilde{ B }
 ,\tilde{ i } ,\tilde{ n },\tilde{ j } ,
\tilde{ s },\tilde{ E } ,\tilde{ v }\right) $ as on
page \pageref{Start of original page 163},
each satisfying the condition on
page \pageref{Start of original page 164},
and each satisfying the requirement that $\left( \#\left(
\mathcal{D} \left( \tilde{ j } \right) \right) +\#\left(
\mathcal{D} \left( \tilde{ E } \right) \right)
\right) $ is not
greater than the integer $N$ defined on
page \pageref{Start of original page 158}, and
each satisfying the further requirement that if $A$ is any
member of $\mathbb{B} \left( G\right) $, and $T\equiv
\left\{ r,s\right\} $ is any $\left( V\cup H\right) $-key of
$A$ such that $T$ is \emph{not} a member of $V$, then there
is \emph{no} member $\beta $ of $\mathcal{D}
\left( \tilde{ j }
\right) $ such that $\tilde{ j } _{ \beta } =r$ holds and
there is \emph{no} member $\beta $ of $\mathcal{D} \left(
\tilde{j } \right) $ such that
 $\tilde{ j } _{ \beta } =s$ holds,
such that for each member $k$ of $\mathbb{N} $,
$I \rule{0pt}{2.55ex}^{ \hspace{-0.75ex} \circ }
_{ k }\left( B,i,n,j,s,E,v,\rho \right) $ is equal to the
sum, over this finite collection of
septuples $\left( \tilde{ B }
 ,\tilde{ i } ,\tilde{ n },\tilde{ j } ,
\tilde{ s },\tilde{ E } ,\tilde{ v }\right) $, of
$I \rule{0pt}{2.55ex}^{ \hspace{-0.75ex} \circ }
_{ k }\left( \tilde{ B } ,\tilde{ i } ,
\tilde{ n },\tilde{ j } ,\tilde{ s },\tilde{ E }
,\tilde{ v },\rho \right) $ times a finite real-number
 coefficient
that is \emph{independent of} $k $, and equal to the sum of
a
finite number of terms, each of which has the form of a
finite real number, independent of $k$ and $\rho $, times a
product of finite powers, all $\geq 0$ and independent of
$k$ and $\rho $, of the $\rho _{ K } $, $K\in \left(
Q\,\vdash P\right) $.

Finally we shall prove that
if $\left(B,i,n,j,s,E,v\right) $ is any
ordered septuple as on
page \pageref{Start of original page 163},
such that the condition on
page \pageref{Start of original page 164}
 holds, and such that $\left( \#\left(
\mathcal{D} \left( j\right) \right) +\#\left( \mathcal{D}
\left( E\right) \right) \right) $ is not greater than the
integer $N$ defined on
page \pageref{Start of original page 158},
and such that if $A$ is
any member of $\mathbb{B} \left( G\right) $, and $T$ is any
$\left( V\cup H\right) $-key of $A$ such that $T$ is
\emph{not} a member of $V$, then there is \emph{no} member
$\beta $ of $\mathcal{D} \left( j\right) $ such that $j_{
\beta } \in T$ holds, then the integrals with respect to
$\rho $, over $\mathbb{D} $, of any k-independent product
of finite powers, all $\geq 0$ and independent of $k$ and
$\rho $, of the $\rho _{ K } $, $K\in \left( Q\,\vdash
P\right) $, times the
$I \rule{0pt}{2.55ex}^{ \hspace{-0.75ex} \circ } _{ k }
\left(
B,i,n,j,s,E,v,\rho \right) $, $k\in \mathbb{N} $, form a
Cauchy sequence.

We shall do this in two steps.   We first prove that if
$\left(B,i,n,j,s,E,v\right) $ is any
ordered septuple satisfying the
conditions just specified, and if we define
$I \rule{0pt}{2.55ex}^{ \hspace{-0.75ex} * }
\left(
B,i,n,j,s,E,v,\rho \right) $ by
\[
I \rule{0pt}{2.55ex}^{ \hspace{-0.75ex} * }
\left( B,i,n,j,s,E,v,\rho \right) \equiv \int_{
\mathbb{W} } \left(
 \prod_{A\in \left( V\,\vdash \left\{ O\right\} \right) }
 d^{
d } x_{ A } \right) \times \hspace{-5.0pt} \hspace{7.0cm}
\]
\[
\times \hspace{-2.0pt} \left( \hspace{-2.0pt} \left(
 \prod_{C\in \left( Y\,\vdash G\right) } \hspace{-6.0pt}
\mathcal{E} \left( \left( P\cap \Xi \left( \mathcal{P}
\left( G,C\right) \right) \right) ,\left( Q\cap \Xi \left(
\mathcal{P} \left( G,C\right) \right) \right) ,H,\sigma
,R,\downarrow \left( x,\Xi \left( \mathcal{P} \left(
G,C\right) \right) \right) \right) \hspace{-2.0pt}
 \right) \hspace{-2.0pt} \rule{0pt}{5.5ex} \times \right.
\]
\[
\times \left( \left(
 \prod_{\alpha \in \mathcal{D} \left( B\right) }
\left( \left( x_{ \mathcal{K} \left( Q,B_{ \alpha } ,i_{
\alpha } \right) } -x_{ B_{ \alpha } } \right) .n_{ \alpha }
\right) \right) \left(
 \prod_{ \beta \in \mathcal{D} \left( j\right) }
\left( s_{ \beta }.\hat{ y }_{ j_{ \beta } }\right) \right)
\times \right.
\]
\[
\hspace{7.0cm} \hspace{-31.0pt} \left. \left. \times
\left( \prod_{\gamma
\in
\mathcal{D} \left( E\right) } \left( v_{ \gamma }.t_{
E_{ \gamma } } \right) \right) \mathcal{J} _{ Q }
\left( y\right) \right)
_{ y=\mu \left( P,Q,H,x,\mathcal{X} \left( P,Q,H,\rho
\right) \right) } \right) ,
\]
\label{Start of original page 192}
 then $I \rule{0pt}{2.55ex}^{ \hspace{-0.75ex} * }\left(
B,i,n,j,s,E,v,\rho \right) $ is finite and
\emph{absolutely} convergent, and is moreover bounded above
in magnitude, for all $\rho \in \mathbb{D} $, by a finite,
$\rho $-independent constant.

Let $\left(B,i,n,j,s,E,v\right) $ be any
ordered septuple satisfying the
conditions just specified, and let $I \rule{0pt}{2.55ex}^{
\hspace{-0.75ex} * }\left(
B,i,n,j,s,E,v,\rho \right) $ be defined, for any member
$\rho $ of $\mathbb{D} $, by the equation just given.
Then in order to prove the finiteness, absolute
convergence, and $\rho $-independent bound on the magnitude
of $I \rule{0pt}{2.55ex}^{ \hspace{-0.75ex} * }\left(
B,i,n,j,s,E,v,\rho \right) $ it is sufficient,
by Lemma \ref{Lemma 23}, to prove the corresponding results
for the
integral over $\mathbb{W}$ of some function $ F(x) $ such
that
for all $x\in \mathbb{W} $, $ F(x) $ is greater than or
equal
to the magnitude of the integrand of $I
\rule{0pt}{2.55ex}^{ \hspace{-0.75ex} * }\left(
B,i,n,j,s,E,v,\rho \right) $.

For any ordered triple $\left( x,r,\rho \right) $ of a
member $x$ of $\mathbb{U} _{ d } \left( V,\omega \right) $,
a member $r$ of $\mathcal{U} \left( V\right) $, and a
member $\rho $ of $\mathbb{D} $, we define $w\left(
x,r,\rho \right) $ to be equal to
\[
\left| \mu _{ r } \left( P,Q,H,x,\mathcal{X} \left(
P,Q,H,\rho
\right) \right) -\mu _{ s } \left( P,Q,H,x,\mathcal{X}
\left( P,Q,H,\rho \right) \right) \right|
\]
if $r$ is a member of the member $\left\{ r,s\right\} $ of
the partition $W$, and to be equal to $1$ if $r$ is not a
member of any member of $W$.

Then it immediately follows from the assumed properties of
$\mathcal{J} $, that for all $x\in \mathbb{W} $, the
integrand of the following integral
$\bar{I} \hspace{-0.45ex} \bar{ \rule{0pt}{1.9ex} }
 \left( B,i,j,\rho
\right) $ is greater than or equal to the magnitude of the
integrand of $I \rule{0pt}{2.55ex}^{ \hspace{-0.75ex} * }
\left( B,i,n,j,s,E,v,\rho \right) $:
\[
\bar{I} \hspace{-0.25ex} \bar{ \rule{0pt}{1.8ex} }
\left( B,i,j,\rho \right) \equiv \int_{
\mathbb{W} } \left(
 \prod_{A\in \left( V\,\vdash \left\{ O\right\} \right) }
d^{
d } x_{ A } \right) \times \hspace{-10.7pt} \hspace{9.0cm}
\]
\[
\times \hspace{-2.0pt} \left( \hspace{-2.0pt} \left(
 \prod_{C\in \left( Y\,\vdash G\right) } \hspace{-6.0pt}
\mathcal{E} \left( \left( P\cap \Xi \left( \mathcal{P}
\left( G,C\right) \right) \right) ,\left( Q\cap \Xi \left(
\mathcal{P} \left( G,C\right) \right) \right) ,H,\sigma
,R,\downarrow \left( x,\Xi \left( \mathcal{P} \left(
G,C\right) \right) \right) \right) \hspace{-2.0pt}
 \right) \hspace{-2.0pt} \times \right.
\]
\[
\times M \left(
 \prod_{\alpha \in \mathcal{D} \left( B\right) }
\left| x_{ \mathcal{K} \left( Q,B_{ \alpha } ,i_{ \alpha }
\right) } -x_{ B_{ \alpha } } \right| \right) \left(
 \prod_{\beta \in
\mathcal{D}
\left( j\right) } \left( \frac{ 1 }{ w\left( x,
j_{ \beta } ,\rho
\right) } \right) \right) \times
\]
\[
\times \left(
 \prod_{\Delta \equiv \left\{ r,s\right\} \in W }
\left| \mu _{ r } \left( P,Q,H,x,\mathcal{X} \left(
P,Q,H,\rho
\right) \right) -\mu _{ s } \left( P,Q,H,x,\mathcal{X}
\left( P,Q,H,\rho \right) \right) \right|^{ -\theta _{
\Delta } } \right)
\times
\]
\[
\left. \times \hspace{-2.0pt} \left(
\prod_{\Delta \equiv \left\{ r,s\right\} \in W }
\hspace{-6.0pt}
\mathbb{S} \left( T-\left| \mu _{ r } \left(
P,Q,H,x,\mathcal{X}
\left( P,Q,H,\rho \right) \right) -\mu _{ s } \left(
P,Q,H,x,\mathcal{X} \left( P,Q,H,\rho \right) \right)
\right| \right) \hspace{-0.7pt} \right)
\hspace{-2.0pt} \right)
\]

Hence to prove the finiteness and absolute convergence of
$I \rule{0pt}{2.55ex}^{ \hspace{-0.75ex} * }
\left( B,i,n,j,s,E,v,\rho \right) $, and a $\rho
$-independent bound, for $\rho \in \mathbb{D} $, on the
magnitude of $I \rule{0pt}{2.55ex}^{ \hspace{-0.75ex} * }
\left( B,i,n,j,s,E,v,\rho \right) $,
it is sufficient to prove the convergence of
$\bar{I} \hspace{-0.25ex} \bar{ \rule{0pt}{1.9ex} }
\left(
B,i,j,\rho \right) $, and a $\rho $-independent upper
bound, for
\label{Start of original page 193}
 $\rho \in \mathbb{D} $, on
 $\bar{I} \hspace{-0.25ex} \bar{ \rule{0pt}{1.9ex} }
 \left( B,i,j,\rho \right)
$.

We now change to the new integration variables $a_{ C } $,
$C\in \left( J\,\vdash \left( \mathcal{R} \left( S\right)
\cup \left\{ \mathcal{U} \left( V\right) \right\} \right)
\right) $, as defined on
pages \pageref{Start of original page 178} and
\pageref{Start of original page 179},
and we note that, as observed on
page \pageref{Start of original page 179},
the Jacobian determinant
for the transformation to this new set of integration
variables is equal to $1$.

We recall that $G$ is an $H $-principal wood of $V$, and
that, as defined on
page \pageref{Start of original page 160},
$Y$ is the set whose members are
all the $\left( V\cup H\right) $-firm over $V$ components
of members of $\mathbb{B} \left( G\right) $, together with
the set $\mathcal{U} \left( V\right) $, and we note that,
by Lemma \ref{Lemma 35} (iii), the set $G\cup Y$ is a
wood of $V$.

\vspace{2.5ex}

We now make the following observations:

\begin{bphzobservation} \label{Observation 40}
\end{bphzobservation}
\vspace{-6.143ex}

\noindent \hspace{2.6ex}{\bf ) }Let $T$ be any member of
$W$.   Then if $\mathcal{Y}
\left( \left( G\cup Y\right) ,T\right) \in G$ holds,
$\mathcal{Y} \left( \left( G\cup Y\right) ,T\right) $ is a
member of $\mathbb{B} \left( G\right) $ and $T$ is a
$\left( V\cup H\right) $-key of $\mathcal{Y} \left( \left(
G\cup Y\right) ,T\right) $ such that $T$ is \emph{not} a
member of $V$, while if $\mathcal{Y} \left( \left( G\cup
Y\right) ,T\right) \in \left( Y\,\vdash G\right) $ holds,
then $T$ is not a $\left( V\cup H\right) $-key of
\emph{any} member of $\mathbb{B} \left( G\right) $.   For
$\mathcal{U} \left( V\right) $ is a member of $Y$ hence a
member of $\left( G\cup Y\right) $, hence $\left( G\cup
Y\right) $ certainly has at least one member, namely
$\mathcal{U} \left( V\right) $, that contains $T$ as a
subset, hence by definition $\mathcal{Y} \left( \left(
G\cup Y\right) ,T\right) $ is the \emph{smallest} member of
$\left( G\cup Y\right) $ that contains $T$ as a subset.
Suppose first that $\mathcal{Y} \left( \left( G\cup
Y\right) ,T\right) \in G$ holds.   Now $T$ is a member of
$W$ hence $T$ intersects two members of $V$, hence $T$ is
\emph{not} a subset of any member of $V$, (since $V$ is a
partition), hence $\mathcal{Y} \left( \left( G\cup Y\right)
,T\right) $ is not a member of $V$, hence $\mathcal{Y}
\left( \left( G\cup Y\right) ,T\right) $ is a member of
$\mathbb{B} \left( G\right) $.   Hence every $\left( V\cup
H\right) $-firm over $V$ component of $\mathcal{Y} \left(
\left( G\cup Y\right) ,T\right) $ is a member of $Y$, hence
$T$ is \emph{not} a subset of any $\left( V\cup H\right)
$-firm over $V$ component of $\mathcal{Y} \left( \left(
G\cup Y\right) ,T\right) $, hence, since by
Lemma \ref{Lemma 31} the
set whose members are all the $\left( V\cup H\right) $-firm
over $V$ components of $\mathcal{Y} \left( \left( G\cup
Y\right) ,T\right) $, is a partition of $\mathcal{Y} \left(
\left( G\cup Y\right) ,T\right) $, and furthermore $T$ is a
member of $W$, hence $T$ has exactly two members, $T$
intersects exactly two distinct $\left( V\cup H\right)
$-firm over $V$ components of $\mathcal{Y} \left( \left(
G\cup Y\right) ,T\right) $, say $C$ and $D$.   Let $r$ be
the member of $T$ that is a member of $C$, let $s$ be the
member of $T$ that is a member of $D$, and let $X$ be the
set whose members are all the $\left( V\cup H\right) $-keys
of $\mathcal{Y} \left( \left( G\cup Y\right) ,T\right) $
that are \emph{not} members of $V$.   Then by
Lemma \ref{Lemma 29},
$\uparrow \left( r,X\right) $, where $\uparrow \left(
r,X\right) $ is defined as on
pages \pageref{Start of original page 170} and
\pageref{Start of original page 171} with
reference to the member $\mathcal{Y} \left( \left( G\cup
Y\right) ,T\right) $ of $\mathbb{B} \left( G\right) $, is a
$\left( V\cup H\right) $-firm over $V$ component of
$\mathcal{Y} \left( \left( G\cup Y\right) ,T\right) $,
hence by Lemma \ref{Lemma 31} $\uparrow \left( r,X\right) $
is the
unique $\left( V\cup H\right) $-firm over $V$ component of
$\mathcal{Y} \left( \left( G\cup Y\right) ,T\right) $
\label{Start of original page 194}
 that has $r$ as a member, hence $C$ is equal to $\uparrow
\left( r,X\right) $.   Let $Z$ be the set whose members are
all the members of $X$ that intersect $C=\uparrow \left(
r,X\right) $.   Then by Lemma \ref{Lemma 28} (xi),
$\mathcal{Y} \left( \left( G\cup Y\right) ,T\right) $ is
equal to the disjoint union of $C=\uparrow \left(
r,X\right) $ and the sets $\left( \mathcal{Y} \left( \left(
G\cup Y\right) ,T\right) \,\vdash \uparrow \left( r,\left\{
R\right\} \right) \right) $ for all the members $R$ of $Z$,
hence, since $s$ is \emph{not} a member of $C$, $s$ is a
member of the set $\left( \mathcal{Y} \left( \left( G\cup
Y\right) ,T\right) \,\vdash \uparrow \left( r,\left\{
R\right\} \right) \right) $ for some member $R$ of $Z$.
Let $R$ be the member of $Z$ such that $s\in \left(
\mathcal{Y} \left( \left( G\cup Y\right) ,T\right) \,\vdash
\uparrow \left( r,\left\{ R\right\} \right) \right) $ holds.
 Then $T$ is a member of $\left( \left( V\cup H\right)
\,\vdash V\right) $ such that $T$ intersects both
$C=\uparrow \left( r,X\right) $ and $\left( \mathcal{Y}
\left( \left( G\cup Y\right) ,T\right) \,\vdash \uparrow
\left( r,\left\{ R\right\} \right) \right) $, hence, since
$C=\uparrow \left( r,X\right) $ is a subset of $\uparrow
\left( r,\left\{ R\right\} \right) $, and by
Lemma \ref{Lemma 28} (viii),
 $R$ is the \emph{only} member of $\left(
V\cup H\right) $ to intersect both $\uparrow \left(
r,\left\{ R\right\} \right) $ and $\left( \mathcal{Y} \left(
\left( G\cup Y\right) ,T\right) \,\vdash \uparrow \left(
r,\left\{ R\right\} \right) \right) $, $R$ is equal to $T$,
hence $T$ is a $\left( V\cup H\right) $-key of $\mathcal{Y}
\left( \left( G\cup Y\right) ,T\right) $ such that $T$ is
\emph{not} a member of $V$.   Now suppose that $\mathcal{Y}
\left( \left( G\cup Y\right) ,T\right) $ is a member of
$\left( Y\,\vdash G\right) $.   Then $\mathcal{Y} \left(
\left( G\cup Y\right) ,T\right) $ is either equal to
$\mathcal{U} \left( V\right) $, or else $\mathcal{Y} \left(
\left( G\cup Y\right) ,T\right) $ is equal to a $\left(
V\cup H\right) $-firm over $V$ component of some member $A$
of $\mathbb{B} \left( G\right) $.   Suppose first that
$\mathcal{Y} \left( \left( G\cup Y\right) ,T\right) $ is
equal to $\mathcal{U} \left( V\right) $.   Then
$\mathcal{U} \left( V\right) $ is \emph{not} a member of
$G$, and $T$ is not a subset of any member of $\mathbb{B}
\left( G\right) $, hence since $T$ has exactly two members,
at most one of the two members of $T$ can be a member of
any particular member of $\mathbb{B} \left( G\right) $,
hence $T$ is certainly not a $\left( V\cup H\right) $-key
of any member of $\mathbb{B} \left( G\right) $.   Now
suppose that $\mathcal{Y} \left( \left( G\cup Y\right)
,T\right) $ is \emph{not} equal to $\mathcal{U} \left(
V\right) $.   Then $\mathcal{Y} \left( \left( G\cup
Y\right) ,T\right) $ is a member of $\left( Y\,\vdash
\left( V\cup \left\{ \mathcal{U} \left( V\right) \right\}
\right) \right) $, hence by
Lemma \ref{Lemma 35} (i),
$\mathcal{Y} \left( G,\mathcal{Y} \left( \left( G\cup
Y\right) ,T\right) \right) $ is a member of $\mathbb{B}
\left( G\right) $, and $\mathcal{Y} \left( \left( G\cup
Y\right) ,T\right) $ is a $\left( V\cup H\right) $-firm
over $V$ component of $\mathcal{Y} \left( G,\mathcal{Y}
\left( \left( G\cup Y\right) ,T\right) \right) $.   Hence
$T$ is a subset of a $\left( V\cup H\right) $-firm over $V$
component of $\mathcal{Y} \left( G,\mathcal{Y} \left(
\left( G\cup Y\right) ,T\right) \right) $, hence since $T$
is a member of $\left( \left( V\cup H\right) \,\vdash
V\right) $, Lemma \ref{Lemma 28} (xvi) and
Lemma \ref{Lemma 29}
imply that $T$ is \emph{not} a $\left( V\cup H\right) $-key
of $\mathcal{Y} \left( G,\mathcal{Y} \left( \left( G\cup
Y\right) ,T\right) \right) $.   And if $A$ is any member of
$\mathbb{B} \left( G\right) $ such that $\mathcal{Y} \left(
G,\mathcal{Y} \left( \left( G\cup Y\right) ,T\right)
\right) \subset A$ holds, then since $G$ is an $H
$-principal
wood of $V$, $\mathcal{Y} \left( G,\mathcal{Y} \left(
\left( G\cup Y\right) ,T\right) \right) $ is a subset of a
$\left( V\cup H\right) $-firm over $V$ component of $A$,
hence $T$ is a subset of a $\left( V\cup H\right) $-firm
over $V$ component of $A$, hence again
Lemma \ref{Lemma 28} (xvi) and
Lemma \ref{Lemma 29} imply that $T$ is \emph{not} a
$\left( V\cup H\right) $-key of $A$.   Hence since
$\mathcal{Y} \left( \left( G\cup Y\right) ,T\right) \in
\left( Y\,\vdash G\right) $ implies that $\mathcal{Y}
\left( G,\mathcal{Y} \left( \left( G\cup Y\right) ,T\right)
\right) $ is equal to $\mathcal{Y} \left( G,T\right) $, $T$
is not a $\left( V\cup H\right) $-key of any member $A$ of
$\mathbb{B} \left( G\right) $ such that $T\subseteq A$
holds.   And finally, if $A$ is any member of $\mathbb{B}
\left( G\right) $ such that $T\subseteq A$ does \emph{not}
hold, then at most one of the two members of $T$ can
\label{Start of original page 195}
 be a member of $A$, hence $T$ is certainly not a $\left(
V\cup H\right) $-key of $A$.

\begin{bphzobservation} \label{Observation 41}
\end{bphzobservation}
\vspace{-6.143ex}

\noindent \hspace{2.6ex}{\bf ) }Let $\left\{ r,s\right\} $
be any member of $W$ such that
$\mathcal{Y} \left( \left( G\cup Y\right) ,\left\{
r,s\right\} \right) $ is a member of $\left( Y\,\vdash
G\right) $, and let $C\equiv \mathcal{Y} \left( \left(
G\cup Y\right) ,\left\{ r,s \right\} \right) $.   Then the
dependence of
\[
\left( \mu _{ r } \left( P,Q,H,x,\mathcal{X} \left(
P,Q,H,\rho \right) \right) -\mu _{ s } \left(
P,Q,H,x,\mathcal{X} \left( P,Q,H,\rho \right) \right)
\right)
\]
on the $a_{ K } $ variables is limited to dependence on the
$a_{ K } $ such that $K$ is a member of \\
$P\cap \left( \Xi
\left( \mathcal{P} \left( G,C\right) \right) \,\vdash
\left\{ C\right\} \right) $.   For if $v$ is any member of
$\mathcal{U} \left( W\right) $, then \\
$\mu _{ v } \left(
P,Q,H,x,\mathcal{X} \left( P,Q,H,\rho \right) \right) $ is
a linear combination, with coefficients summing to $1$, of
the $x_{ K } $ such that $K$ is a member of $Q$ such that
$\mathcal{Z} \left( P,H,v\right) \subseteq K\subseteq
\mathcal{Z} \left( Q,H,v\right) $ holds.   Let $D\equiv
\mathcal{Y} \left( P,\left\{ r,s\right\} \right) $.   Then
since $P$ is a member of $\mathbb{O} \left( G,H\right) $,
either $D$ is equal to the member $\mathcal{Y} \left(
G,C\right) =\mathcal{Y} \left( G,\left\{ r,s\right\} \right)
$ of $\mathbb{B} \left( G\right) $, or else $D$ is a subset
of $C$ such that $D$ is \emph{not} a subset of any member
of $\mathcal{P} \left( G,C\right) $.   Suppose first that
$D$ is equal to the member $\mathcal{Y} \left( G,C\right)
=\mathcal{Y} \left( G,\left\{ r,s\right\} \right) $ of
$\mathbb{B} \left( G\right) $.   Then since $\mathcal{Z}
\left( Q,H,r\right) \subset D$ and $\mathcal{Z} \left(
Q,H,s\right) \subset D$ both hold, and $Q$ is a member of
$\mathbb{O} \left( G,H\right) $, both $\mathcal{Z} \left(
Q,H,r\right) $ and $\mathcal{Z} \left( Q,H,s\right) $ are
subsets of $C$, hence both $\mathcal{Z} \left( Q,H,r\right)
$ and $\mathcal{Z} \left( Q,H,s\right) $ are strict subsets
of $C$, since $\mathcal{Z} \left( Q,H,r\right) \cap
\mathcal{Z} \left( Q,H,s\right) =\emptyset $ holds and
neither
$\mathcal{Z} \left( Q,H,r\right) $ nor $\mathcal{Z} \left(
Q,H,s\right) $ is empty.   And furthermore, both
$\mathcal{Z} \left( P,H,r\right) $ and $\mathcal{Z} \left(
P,H,s\right) $ are members of $\mathcal{P} \left(
P,D\right) $ that are strict subsets of $C$.   (In fact
$\mathcal{Z} \left( P,H,r\right) $ is equal to $\mathcal{K}
\left( P,D,r\right) $ and $\mathcal{Z} \left( P,H,s\right)
$ is equal to $\mathcal{K} \left( P,D,s\right) $.)
Hence by Lemma \ref{Lemma 4}, if $K$ is any member of $Q$
such that
$\mathcal{Z} \left( P,H,r\right) \subseteq K\subseteq
\mathcal{Z} \left( Q,H,r\right) $ holds or $\mathcal{Z}
\left( P,H,s\right) \subseteq K\subseteq \mathcal{Z} \left(
Q,H,s\right) $ holds, then $x_{ K } $ is equal to a linear
combination, with coefficients summing to $1$, of the $x_{
L } $ such that $L$ is a member of $\mathcal{P} \left(
P,D\right) $ such that $L\subseteq C$ holds, hence by
page \pageref{Start of original page 165},
if $K$ is any such member of $Q$, then $x_{ K } $ is
equal to $x_{ S_{ D } } $ plus a linear combination, with
coefficients summing to $1$, of the $z_{ L } $ such that
$L$ is a member of $\mathcal{P} \left( P,D\right) $ such
that $L\subseteq C$ holds.   Now $D$ is a member of
$\mathbb{B} \left( G\right) $ in the present case, hence by
page \pageref{Start of original page 177},
if $L$ is any member of $\mathcal{P} \left(
P,D\right) $, then $z_{ L } $ is equal to $a_{ L } $ plus
$z_{ S_{ E } } $, where $E$ is the \emph{smallest} member
of $J$ to contain $L$ as a \emph{strict} subset, where $J$
is the wood defined on
page \pageref{Start of original page 165}.
Now since $C\subset D$
holds in the present case, $C$ is \emph{not} a member of
$P$ in the present case, hence every member $L$ of
$\mathcal{P} \left( P,D\right) $ such that $L\subseteq C$
holds, is a \emph{strict} subset of $C$, hence if $L$ is
any member of $\mathcal{P} \left( P,D\right) $ such that
$L\subseteq C$ holds, then the smallest member $E$ of $J$
to contain $L$ as a strict subset, is equal to the smallest
member of $J$ that contains $C$ as a subset, and is the
\label{Start of original page 196}
 \emph{same} for \emph{every} member $L$ of $\mathcal{P}
\left( P,D\right) $ such that $L\subseteq C$ holds.   Let
$E$ be the smallest member of $J$ to contain $C$ as a
subset.   Then by the foregoing, if $K$ is any member of
$Q$ such that $\mathcal{Z} \left( P,H,r\right) \subseteq
K\subseteq \mathcal{Z} \left( Q,H,r\right) $ holds or
$\mathcal{Z} \left( P,H,s\right) \subseteq K\subseteq
\mathcal{Z} \left( Q,H,s\right) $ holds, then $x_{ K } $ is
equal to $\left( x_{ S_{ D } } +z_{ S_{ E } } \right) $ plus
a linear combination, with coefficients summing to $1$, of
the $a_{ L } $ such that $L$ is a member of $\mathcal{P}
\left( P,D\right) $ such that $L\subset C$ holds.   Hence
$\mu _{ r } \left( P,Q,H,x,\mathcal{X} \left( P,Q,H,\rho
\right) \right) $ is equal to $\left( x_{ S_{ D } }
+z_{ S_{
E } } \right) $ plus a linear combination, with
coefficents summing to $1$, of the $a_{ L } $ such that $L$
is a member of $\mathcal{P} \left( P,D\right) $ such that
$L\subset C$ holds, and $\mu _{ s } \left(
P,Q,H,x,\mathcal{X} \left( P,Q,H,\rho \right) \right) $ is
also equal to $\left( x_{ S_{ D } } +z_{ S_{ E } } \right) $
plus a linear combination, with coefficients summing to
$1$, of the $a_{ L } $ such that $L$ is a member of
$\mathcal{P} \left( P,D\right) $ such that $L\subset C$
holds, hence
\[
\left( \mu _{ r } \left( P,Q,H,x,\mathcal{X} \left(
P,Q,H,\rho \right) \right) -\mu _{ s } \left(
P,Q,H,x,\mathcal{X} \left( P,Q,H,\rho \right) \right)
\right)
\]
is equal to a linear combination, with coefficients summing
to $0$, of the $a_{ L } $ such that $L$ is a member of
$\mathcal{P} \left( P,D\right) $ such that $L\subset C$
holds, and every such member $L$ of $\mathcal{P} \left(
P,D\right) $ is a member of $P\cap \left( \Xi \left(
\mathcal{P} \left( G,C\right) \right) \,\vdash \left\{
C\right\} \right) $   (We note in passing that since every
member $L$ of $\mathcal{P} \left( P,D\right) $ that can
contribute to $\mu _{ r } \left( P,Q,H,x,\mathcal{X} \left(
P,Q,H,\rho \right) \right) $, is a subset of $\mathcal{Z}
\left( Q,H,r\right) $, and every member $L$ of $\mathcal{P}
\left( P,D\right) $ that can contribute to $\mu _{ s }
\left( P,Q,H,x,\mathcal{X} \left( P,Q,H,\rho \right)
\right) $ is a subset of $\mathcal{Z} \left( Q,H,s\right)
$, and $\mathcal{Z} \left( Q,H,r\right) \cap \mathcal{Z}
\left( Q,H,s\right) $ is equal to the empty set $\emptyset
$,
the coefficients in $\left( \mu _{ r } \left(
P,Q,H,x,\mathcal{X} \left( P,Q,H,\rho \right) \right) -\mu
_{ s } \left( P,Q,H,x,\mathcal{X} \left( P,Q,H,\rho \right)
\right) \right) $ do \emph{not} vanish identically.)
   Now suppose that $D\equiv \mathcal{Y} \left( P,\left\{
r,s\right\} \right) $ is a subset of $C$ such that $D$ is
\emph{not} a subset of any member of $\mathcal{P} \left(
G,C\right) $.   Then no member of $\mathcal{P} \left(
P,D\right) $ is a strict subset of any member of
$\mathcal{P} \left( G,C\right) $, and $\mathcal{Z} \left(
P,H,r\right) $ is equal to $\mathcal{K} \left( P,D,r\right)
$ and $\mathcal{Z} \left( P,H,s\right) $ is equal to
$\mathcal{K} \left( P,D,s\right) $, and $\mathcal{Z} \left(
Q,H,r\right) $ and $\mathcal{Z} \left( Q,H,s\right) $ are
strict subsets of $D$ such that $\mathcal{Z} \left(
Q,H,r\right) \cap \mathcal{Z} \left( Q,H,s\right)
=\emptyset $
holds, hence if $K$ is any member of $Q$ such that
$\mathcal{Z} \left( P,H,r\right) \subseteq K\subseteq
\mathcal{Z} \left( Q,H,r\right) $ holds or $\mathcal{Z}
\left( P,H,s\right) \subseteq K\subseteq \mathcal{Z} \left(
Q,H,s\right) $ holds, then $x_{ K } $ is equal to a linear
combination, with coefficients summing to $1$, of the $x_{
L } $, $L\in \mathcal{P} \left( P,D\right) $, hence by
page \pageref{Start of original page 165},
if $K$ is any such member of $Q$, then $x_{ K } $ is
equal to $x_{ S_{ D } } $ plus a linear combination, with
coefficients summing to $1$, of the $z_{ L } $, $L\in
\mathcal{P} \left( P,D\right) $.   And furthermore no
member $L$ of $\mathcal{P} \left( P,D\right) $ is a member
of $\mathcal{P} \left( P,A\right) $ for any member $A$ of
$\mathbb{B} \left( G\right) $, hence by
page \pageref{Start of original page 177}, if $L$
is any member of $\mathcal{P} \left( P,D\right) $, then
$z_{ L } =a_{ L } $ holds, hence if $K$ is any member of
$Q$ such that $\mathcal{Z} \left( P,H,r\right) \subseteq
K\subseteq \mathcal{Z} \left( Q,H,r\right) $ holds or
$\mathcal{Z} \left( P,H,s\right) \subseteq K\subseteq
\mathcal{Z} \left( Q,H,s\right) $ holds, then
\label{Start of original page 197}
 $x_{ K } $ is equal to $x_{ S_{ D } } $ plus a linear
combination, with coefficients summing to $1$, of the $a_{
L } $, $L\in \mathcal{P} \left( P,D\right) $, hence
\[
\left( \mu _{ r } \left( P,Q,H,x,\mathcal{X} \left(
P,Q,H,\rho \right) \right) -\mu _{ s } \left(
P,Q,H,x,\mathcal{X} \left( P,Q,H,\rho \right) \right)
\right)
\]
is equal to a linear combination, with coefficients summing
to $0$, of the $a_{ L } $, $L\in \mathcal{P} \left(
P,D\right) $, and every member $L$ of $\mathcal{P} \left(
P,D\right) $ is a member of $\left( \Xi \left( \mathcal{P}
\left( G,C\right) \right) \,\vdash \left\{ C\right\} \right)
$.   (And, as in the previous case, the fact that
$\mathcal{Z} \left( Q,H,r\right) \cap \mathcal{Z} \left(
Q,H,s\right) =\emptyset $ holds implies that the
coefficients do
\emph{not} vanish identically.)

\begin{bphzobservation} \label{Observation 42}
\end{bphzobservation}
\vspace{-6.143ex}

\noindent \hspace{2.6ex}{\bf ) }Let $T\equiv
\left\{ l,m\right\} $ be any member of $W$
such that $\mathcal{Y} \left( \left( G\cup Y\right)
,T\right) \in G$ holds, (so that, by
observation \ref{Observation 40})
 above, $\mathcal{Y} \left( \left( G\cup Y\right)
,T\right) $ is a member of $\mathbb{B} \left( G\right) $
and $T$ is a $\left( V\cup H\right) $-key of $\mathcal{Y}
\left( \left( G\cup Y\right) ,T\right) $ such that $T$ is
\emph{not} a member of V), let $C\equiv \mathcal{Y} \left(
\left( G\cup Y\right) ,T\right) $, let $D\equiv \left(
C\,\vdash \uparrow \left( f_{ C } ,\left\{ T\right\} \right)
\right) $, where for any member $r$ of $C$ and any subset
$X$ of the set of all the $\left( V\cup H\right) $-keys of
$C$ that are \emph{not} members of $V$, $\uparrow \left(
r,X\right) $ is defined as on
pages \pageref{Start of original page 170} and
\pageref{Start of original page 171} with
reference to the member $C$ of $\mathbb{B} \left( G\right)
$, (so that, by the definition on
page \pageref{Start of original page 165} of the wood
$J$, $D$ is a member of J), and let $m$ be the member of
$T$ that is the key member of the $\left( V\cup H\right)
$-firm over $V$ component $\left( C\,\vdash \uparrow \left(
f_{ C } ,\left\{ T\right\} \right) \right) \,\vdash \left(
 \bigcup_{R\in
Z } \left( C\,\vdash \uparrow \left( f_{ C }
,\left\{ R\right\} \right) \right) \right) $ of $C$, where
$Z$ is the
set whose members are all the $\left( V\cup H\right) $-keys
$R$ of $C$ such that $R\notin V$ and $R\to T$ both hold,
where the relation $\to $ among the $\left( V\cup H\right)
$-keys of $C$ is defined with reference to the member $f_{
C } $ of $C$, as on
page \pageref{Start of original page 170}.   Then
\[
\left( \mu _{ m } \left( P,Q,H,x,\mathcal{X} \left(
P,Q,H,\rho \right) \right) -\mu _{ l } \left(
P,Q,H,x,\mathcal{X} \left( P,Q,H,\rho \right) \right)
\right)
\]
is equal to $a_{ D } $ minus a linear combination, with
coefficients summing to $1$, of the $a_{ L } $ such that
$L$ is a member of $\mathcal{P} \left( P,C\right) $ such
that $L$ is a subset of the unique $\left( V\cup H\right)
$-firm over $V$ component $N$ of $C$ that has $l$ as a
member.   (We note that $N\cap D=\emptyset $ holds, so that
$L\cap D=\emptyset $ holds for every such member $L$ of
$\mathcal{P} \left( P,C\right) $.)

For let $N$ be the unique $\left( V\cup H\right) $-firm
over $V$ component of $C$ that has $l$ as a member.   Then
since $P$ is a member of $\mathbb{O} \left( G,H\right) $,
and $\left\{ l,m\right\} $ is a subset of $C$ but is
\emph{not} a subset of $N$, $\mathcal{Z} \left(
P,H,l\right) $ is equal to $\mathcal{K} \left( P,C,l\right)
$, hence $\mathcal{Z} \left( P,H,l\right) $ is a member of
$\mathcal{P} \left( P,C\right) $ that is a subset of $N$,
and furthermore since $Q$ is a member of $\mathbb{O} \left(
G,H\right) $, $\mathcal{Z} \left( Q,H,l\right) $ is also a
subset of $N$, hence if $K$ is any member of $Q$ such that
$\mathcal{Z} \left( P,H,l\right) \subseteq K\subseteq
\mathcal{Z} \left( Q,H,l\right) $ holds, then by
Lemma \ref{Lemma 4},
$x_{ K } $ is equal to a linear combination, with
coefficients summing to $1$, of the $x_{ L } $ such
\label{Start of original page 198}
 that $L$ is a member of $\mathcal{P} \left( P,C\right) $
such that $L\subseteq N$ holds, hence by
page \pageref{Start of original page 165}, $x_{ K
} $ is equal to $x_{ S_{ C } } $ plus a linear combination,
with coefficients summing to $1$, of the $z_{ L } $ such
that $L$ is a member of $\mathcal{P} \left( P,C\right) $
such that $L\subseteq N$ holds.

We next note that $N$ is \emph{not} equal to $\left(
C\,\vdash \uparrow \left( f_{ C } ,\left\{ R\right\} \right)
\right) $ for \emph{any} $\left( V\cup H\right) $-key $R$
of $C$ such that $R$ is not a member of $V$, from which it
follows immediately, from the definition of the wood $J$ on
page \pageref{Start of original page 165},
that if $N$ is a member of $J$ then $N$ is a
member of $P$.   For if $R$ was a $\left( V\cup
H\right) $-key of $C$ such that $ R\notin V $ held
and $N $ was equal to $\left( C\,\vdash
\uparrow \left( f_{ C } ,\left\{ R\right\} \right) \right)
$,
then by Lemma \ref{Lemma 28} (viii), $R$ would be the
\emph{only} member of $\left( V\cup H\right) $ to
intersect both $\left( C\,\vdash N\right)
$and $N $.  But $l $ is a member of $N $ and $m $
is \emph{not} a
member of $N $, hence $T $ intersects both
$\left( C\,\vdash N\right)
$ and $N $, and $R $ is certainly
not equal to $T $, for
$D\equiv \left(C\,\vdash \uparrow \left( f_{ C } ,\left\{
T\right\} \right) \right) $ is \emph{disjoint} from
 $N $, and moreover $m$ is a member
of $D$ and $l$ is \emph{not} a member of $D$.   (We refer
to the definition, on
pages \pageref{Start of original page 172} and
\pageref{Start of original page 173},
of the key member of any $\left(
V\cup H\right) $-firm over $V$ component of $C$ that does
not have $f_{ C } $ as a member, and we note in passing
that if $R$ is any $\left( V\cup H\right) $-key of $C$ such
that $R\notin V$ holds,then the set $\left( C\,\vdash
\uparrow \left( f_{ C } ,\left\{ R\right\} \right) \right) $
is $\left( V\cup H\right) $-firm over $V$ ifif there are
\emph{no} $\left( V\cup H\right) $-keys $\tilde{ R } $
 of $C$ such
that $\tilde{ R } \notin V$ and $\tilde{ R } \to R$ both
hold.)

Now let $U$ be the \emph{smallest} member of $J$ to contain
$N$ as a \emph{strict} subset.   Then it immediately
follows from the foregoing that if $L$ is any member of
$\mathcal{P} \left( P,C\right) $ such that $L\subseteq N$
holds, then $U$ is the smallest member of $J$ to contain
$L$ as a strict subset.   For if $N$ is a member of $J$,
then as just shown, $N$ is a member of $P$, hence there is
exactly one member $L$ of $\mathcal{P} \left( P,C\right) $
such that $L\subseteq N$ holds, namely the set $L=N$, and
$U$ is by definition the smallest member of $J$ to contain
$N$ as a strict subset.   And if $N$ is \emph{not} a member
of $J$, then $N$ is not a member of $P$, (since $P$ is a
subset of J), hence if $L$ is any member of $\mathcal{P}
\left( P,C\right) $ such that $L\subseteq N$ holds, then
$L\subset N$ holds, and moreover $L$ is not a strict subset
of any member of $\mathcal{P} \left( G,C\right) $, (since
$G$ is a subset of P), hence, since it follows directly
from the definition of $J$ that every member of $\left(
J\,\vdash P\right) $ that is a strict subset of $N$, is a
subset of some member of $\mathcal{P} \left( G,C\right) $
that is a subset of $N$, $U$ is again the smallest member
of $J$ to contain $L$ as a strict subset.

Hence by
page \pageref{Start of original page 177},
if $L$ is any member of $\mathcal{P}
\left( P,C\right) $ such that $L\subseteq N$ holds, then
$z_{ L } $ is equal to $\left( z_{ S_{ U } } +a_{ L }
\right)
$, hence by the foregoing, if $K$ is any member of $Q$
such that $\mathcal{Z} \left( P,H,l\right) \subseteq
K\subseteq \mathcal{Z} \left( Q,H,l\right) $ holds, then
$x_{ K } $ is
\label{Start of original page 199}
 equal to $\left( x_{ S_{ C } }+z_{ S_{ U } }\right)  $
plus a linear combination, with coefficients summing to
$1$, of the $a_{ L } $ such that $L$ is a member of
$\mathcal{P} \left( P,C\right) $ such that $L\subseteq N$
holds, hence $\mu _{ l } \left( P,Q,H,x,\mathcal{X} \left(
P,Q,H,\rho \right) \right) $ is equal to $\left( x_{ S_{
C } }+z_{ S_{ U } }\right)  $ plus a linear combination,
with coefficients summing to $1$, of the $a_{ L } $ such
that $L$ is a member of $\mathcal{P} \left( P,C\right) $
such that $L\subseteq N$ holds.

We next note that $D\subset U$ holds, and that $U$ is the
\emph{smallest} member of $J$ to contain $D$ as a
\emph{strict} subset.   For $U$ is either equal to $C$, or
else is equal to $\left( C\,\vdash \uparrow \left( f_{ C }
,\left\{ R\right\} \right) \right) $ for some $\left( V\cup
H\right) $-key $R$ of $C$ such that $R$ is \emph{not} a
member of $V$.   And furthermore, by
pages \pageref{Start of original page 165} to
\pageref{Start of original page 170},
if $U$ is equal to $C$, then $N$ is equal to the unique
$\left( V\cup H\right) $-firm over $V$ component of $C$
that has $f_{ C } $ as a member, hence $T$ intersects the
unique $\left( V\cup H\right) $-firm over $V$ component of
$C$ that has $f_{ C } $ as a member, hence there is
\emph{no} $\left( V\cup H\right) $-key $R$ of $C$ such that
$R\notin V$ and $T\to R$ both hold, hence there is
\emph{no} member $K$ of $\left( J\,\vdash P\right) $ such
that $D\subset K$ and $K\subset D$ both hold, and
furthermore $D\subset C$ certainly holds hence $D\subset U$
holds, while if $U$ is equal to $\left( C\,\vdash \uparrow
\left( f_{ C } ,\left\{ R\right\} \right) \right) $ for some
$\left( V\cup H\right) $-key $R$ of $C$ such that $R$ is
\emph{not} a member of $V$, then $f_{ C } $ is \emph{not} a
member of $N$, (for the unique $\left( V\cup H\right)
$-firm over $V$ component of $C$ that has $f_{ C } $ as a
member is not a strict subset of \emph{any} member of $J$
that is a strict subset of C), and $R$, (which by
Lemma \ref{Lemma 28} (viii) is identified uniquely as the
only
member of $\left( V\cup H\right) $ to intersect both
$\left( C\,\vdash U\right) $ and U), is equal, by
page \pageref{Start of original page 172},
to the unique $\left( V\cup H\right) $-key of $C$
that is \emph{not} a member of $V$ and which has as a
member the key member of $N$, hence again by
page \pageref{Start of original page 172},
$T\to R$ holds, hence by
pages \pageref{Start of original page 165} to
\pageref{Start of original page 170}, $D\subset U$
holds, and furthermore, since $T$ intersects $N$, there is
\emph{no} $\left( V\cup H\right) $-key $ \tilde{ R } $
 of $C$ such
that $\tilde{ R } \notin V$ and $T\to \tilde{ R } $ and
$\tilde{ R } \to
R$ all hold, hence there is \emph{no} member $K$ of $J$
such that $D\subset K\subset U$ holds.

Now let $Z$ be the set whose members are all the $\left(
V\cup H\right) $-keys $R$ of $C$ such that $R\notin V$ and
$R\to T$ both hold, and let $M\equiv \left( C\,\vdash
\uparrow \left( f_{ C } ,\left\{ T\right\} \right) \right)
\,\vdash \left( \bigcup_{R\in Z } \left( C\,\vdash
\uparrow \left(
f_{ C } ,\left\{ R\right\} \right) \right) \right) $ be
the unique
$\left( V\cup H\right) $-firm over $V$ component of $C$
that has $m$ as its key member.   Then since $P$ is a
member of $\mathbb{O} \left( G,H\right) $, and $\left\{
l,m\right\} $ is a subset of $C$ but is \emph{not} a subset
of $M$, $\mathcal{Z} \left( P,H,m\right) $ is equal to
$\mathcal{K} \left( P,C,m\right) $, hence $\mathcal{Z}
\left( P,H,m\right) $ is a member of $\mathcal{P} \left(
P,C\right) $ that is a subset of $M$, and furthermore since
$Q$ is a member of $\mathbb{O} \left( G,H\right) $,
$\mathcal{Z} \left( Q,H,m\right) $ is also a subset of $M$,
hence if $K$ is any member of $Q$ such that $\mathcal{Z}
\left( P,H,m\right) \subseteq K\subseteq \mathcal{Z} \left(
Q,H,m\right) $ holds, then by
\label{Start of original page 200}
 Lemma \ref{Lemma 4}, $x_{ K } $ is equal to a linear
combination,
with coefficients summing to $1$, of the $x_{ L } $ such
that $L$ is a member of $\mathcal{P} \left( P,C\right) $
such that $L\subseteq M$ holds, hence by
page \pageref{Start of original page 165}, $x_{ K
} $ is equal to $x_{ S_{ C } } $ plus a linear combination,
with coefficients summing to $1$, of the $z_{ L } $ such
that $L$ is a member of $\mathcal{P} \left( P,C\right) $
such that $L\subseteq M$ holds.

Now, in accordance with the notation of
pages \pageref{Start of original page 174} to
\pageref{Start of original page 176},
but with the set $A$ of those pages taken as the
present $C$, and the set $B$ of those
pages taken as the present $M$, so that the $i$ of those
pages is the present
$m$, let $F_{ M } $ be the set whose members are all the
members $L$ of $\mathcal{P} \left( P,C\right) $ such that
$L\subseteq M$ holds, so that $F_{ M } $ is equal to
$\left\{ M\right\} $ if $M$ \emph{is} a member of $P$, and
$F_{ M } $ is equal to $\mathcal{P} \left( P,M\right) $ if
$M$ is \emph{not} a member of $P$.

Then by
page \pageref{Start of original page 177},
\[
\mu _{ m } \left( P,Q,H,x,\mathcal{X} \left( P,Q,H,\rho
\right) \right) = \sum_{L\in F_{ M } } \nu _{ DL } x_{ L }
=x_{ S_{ C } } + \sum_{L\in F_{ M } } \nu _{ DL } z_{ L }
=x_{ S_{ C } }+z_{ D }
\]
holds if $D=\left( C\,\vdash \uparrow \left( f_{ C }
,\left\{ T\right\} \right) \right) $ is a member of $\left(
J\,\vdash P\right) $, (we note that our present $D$
corresponds to the $K$ of
pages \pageref{Start of original page 176} and
\pageref{Start of original page 177}, and that
the sets $D$ of
pages \pageref{Start of original page 176} and
\pageref{Start of original page 177} correspond to our
present sets $ L $), while if $D$ is a member of $P$,
then the
set $Z$ defined above is empty, $D$ is equal to $M$, and
moreover both $\mathcal{Z} \left( P,H,m\right) $ and
$\mathcal{Z} \left( Q,H,m\right) $ are also equal to $M$
hence equal to $D$, hence $\mu _{ m } \left(
P,Q,H,x,\mathcal{X} \left( P,Q,H,\rho \right) \right) $ is
equal to $x_{ D } $, hence again
\[
\mu _{ m } \left( P,Q,H,x,\mathcal{X} \left( P,Q,H,\rho
\right) \right) =x_{ S_{ C } } +z_{ D }
\]
holds.   Hence by the foregoing and
page \pageref{Start of original page 177},
\[
\mu _{ m } \left( P,Q,H,x,\mathcal{X} \left( P,Q,H,\rho
\right) \right) =x_{ S_{ C } } +z_{ S_{ U } } +a_{ D }
\]
holds, hence, as stated,
\[
\left( \mu _{ m } \left( P,Q,H,x,\mathcal{X} \left(
P,Q,H,\rho \right) \right) -\mu _{ l } \left(
P,Q,H,x,\mathcal{X} \left( P,Q,H,\rho \right) \right)
\right)
\]
is equal to $a_{ D } $ minus a linear combination, with
coefficients summing to $1$, of the $a_{ L } $ such that
$L$ is a member of $\mathcal{P} \left( P,C\right) $ such
that $L$ is a subset of the unique $\left( V\cup H\right)
$-firm over $V$ component $N$ of $C$ that has $l$ as a
member.

\begin{bphzobservation} \label{Observation 43}
\end{bphzobservation}
\vspace{-6.143ex}

\noindent \hspace{2.6ex}{\bf ) }Let $C$ be any member of
$\left( Y\,\vdash G\right) $.  Then the dependence of
\[
\mathcal{E} \left( \left( P\cap \Xi \left( \mathcal{P}
\left( G,C\right) \right) \right) ,\left( Q\cap \Xi \left(
\mathcal{P} \left( G,C\right) \right) \right) ,H,\sigma
,R,\downarrow \left( x,\Xi \left( \mathcal{P} \left(
G,C\right) \right) \right) \right)
\]
on the $a_{ D } $ variables is limited to dependence on the
$a_{ D } $ such that $D$ is a member of $P\cap \left( \Xi
\left( \mathcal{P} \left( G,C\right) \right) \,\vdash
\left\{ C\right\} \right) $.   (We note that this
observation
is related to observation \ref{Observation 34}) above on
page \pageref{Start of original page 183}.)
For the expression \\
$\mathcal{E} \left(
\left( P\cap \Xi \left( \mathcal{P} \left( G,C\right)
\right) \right) ,\left( Q\cap \Xi \left( \mathcal{P} \left(
G,C\right) \right) \right) ,H,\sigma ,R,\downarrow \left(
x,\Xi \left( \mathcal{P} \left( G,C\right) \right) \right)
\right) $\hspace{\stretch{1}}
\label{Start of original page 201}
depends\hspace{\stretch{1}} on\hspace{\stretch{1}} $x$ \\
only through differences $\left( x_{ K }
-x_{ L } \right) $, where both $K$ and $L$ are members of
$\Xi \left( \mathcal{P} \left( G,C\right) \right) $, and by
observations \ref{Observation 30})
to \ref{Observation 33}) above, if $D$ is
any member of $\left( J\,\vdash \left( \mathcal{R} \left(
S\right) \cup \left\{ \mathcal{U} \left( V\right) \right\}
\right) \right) $ such that $D$ is \emph{not} a member of
$P\cap \left( \Xi \left( \mathcal{P} \left( G,C\right)
\right) \,\vdash \left\{ C\right\} \right) $, then either
$x_{ K } $ is independent of $a_{ D } $ for every member
$K$ of $\Xi \left( \mathcal{P} \left( G,C\right) \right) $,
or else $a_{ D } $ occurs in $x_{ K } $ with the
\emph{same} coefficient for every member $K$ of $\Xi \left(
\mathcal{P} \left( G,C\right) \right) $, so that $a_{ D } $
cancels out of every difference $\left( x_{ K } -x_{ L }
\right) $, where both $K$ and $L$ are members of $\Xi
\left( \mathcal{P} \left( G,C\right) \right) $.

\begin{bphzobservation} \label{Observation 44}
\end{bphzobservation}
\vspace{-6.143ex}

\noindent \hspace{2.6ex}{\bf ) }Let $D$ be any member of
$\mathbb{B} \left( Q\right) $
such that $D$ is \emph{not} a member of $\mathbb{B} \left(
G\right) $, let $C\equiv \mathcal{Y} \left( \left( G\cup
Y\right) ,D\right) $, and let $l$ be any member of $D$.
Then $C=\mathcal{Y} \left( \left( G\cup Y\right) ,D\right)
$ is a member of $\left( Y\,\vdash G\right) $, and the
dependence of $\left( x_{ \mathcal{K}
\left( Q,D,l\right) } -x_{
D } \right) $ on the $a_{ K } $ variables is limited to
dependence on the $a_{ K } $ such that $K$ is a member of
$P\cap \left( \Xi \left( \mathcal{P} \left( G,C\right)
\right) \,\vdash \left\{ C\right\} \right) $.   (We note
that
this observation is related to
observation \ref{Observation 35}) above on
page \pageref{Start of original page 183}.)
For $D$ is a member of
$\left( Q\,\vdash G\right) $ and $Q$ is a member of
$\mathbb{O} \left( G,H\right) $, so if $A$ is any member of
$G$ such that $D\subseteq A$ holds, then $A$ is a member of
$\mathbb{B} \left( G\right) $, (since $D$ is not a member
of $G$ hence is not a member of $ V $), hence $D\subset A$
holds, hence $D$ is a subset of a $\left( V\cup H\right)
$-firm over $V$ component of $A$, and by the definition of
$Y$ on
page \pageref{Start of original page 160},
that $\left( V\cup H\right) $-firm over
$V$ component of $A$ is a member of $Y$.   Hence if there
\emph{exists} any member $A$ of $G$ such that $D\subseteq
A$ holds, then $C=\mathcal{Y} \left( \left( G\cup Y\right)
,D\right) $ is a $\left( V\cup H\right) $-firm over $V$
component of the member $\mathcal{Y} \left( G,D\right) $ of
$\mathbb{B} \left( G\right) $, hence $C$ is a member of
$Y$, and moreover $C$ is not a member of $V$ since $D$ is
not a member of $V$, hence $C$ is not a member of $G$,
(since \emph{no} $\left( V\cup H\right) $-firm over $V$
component of any member of $\mathbb{B} \left( G\right) $ is
a member of $G$, unless it is a member of $ V $),
hence $C$ is
a member of $\left( Y\,\vdash G\right) $.   And if there is
\emph{no} member $A$ of $G$ such that $D\subseteq A$ holds,
then $\mathcal{U} \left( V\right) $ is \emph{not} a member
of $G$, and $C=\mathcal{Y} \left( \left( G\cup Y\right)
,D\right) $ is equal to the member $\mathcal{U} \left(
V\right) $ of $\left( Y\,\vdash G\right) $.   Now both
$\mathcal{K} \left( Q,D,l\right) $ and $D$ are members of
$\Xi \left( \mathcal{P} \left( G,C\right) \right) $, for
both $\mathcal{K} \left( Q,D,l\right) $ and $D$ are subsets
of $C$, and both $\mathcal{K} \left( Q,D,l\right) $ and $D$
are members of $Q$ hence overlap no member of $G$, hence in
particular overlap no member of $\mathcal{P} \left(
G,C\right) $, and by the definition of $C\equiv \mathcal{Y}
\left( \left( G\cup Y\right) ,D\right) $, $D$ is \emph{not}
a subset of any member of $\mathcal{P} \left( G,C\right) $,
hence if $F$ is the set whose members are all the members
$L$ of $\mathcal{P} \left( G,C\right) $ such that
$L\subseteq D$ holds, then $F$ is a partition of $D$ and
every member of $F$ is a \emph{strict} subset of $D$, (and
in fact $F$ is equal to $\mathcal{P} \left( G,D\right)
 $), hence $D$ is certainly a member of $\Xi \left(
\mathcal{P} \left( G,C\right) \right) $, and furthermore
$\mathcal{K} \left( G,D,l\right) \subseteq \mathcal{K}
\left( Q,D,l\right) $ holds, hence $\mathcal{K} \left(
Q,D,l\right) $ is \emph{not} a
\label{Start of original page 202}
 strict subset of any member of $\mathcal{P} \left(
G,C\right) $, hence $\mathcal{K} \left( Q,D,l\right) $ is
also a member of $\Xi \left( \mathcal{P} \left( G,C\right)
\right) $.   Hence exactly as in
observation \ref{Observation 43})
above, $\left( x_{ \mathcal{K} \left( Q,D,l\right) } -x_{ D
}
\right) $ does \emph{not} depend on any $a_{ K } $ such
that $K$ is \emph{not} a member of $P\cap \left( \Xi \left(
\mathcal{P} \left( G,C\right) \right) \,\vdash \left\{
C\right\} \right) $.

\begin{bphzobservation} \label{Observation 45}
\end{bphzobservation}
\vspace{-6.143ex}

\noindent \hspace{2.6ex}{\bf ) }Let $D$ be any member of
$\mathbb{B} \left( G\right) $
and $l$ be any member of $D$.   Then whenever the product
of factors\rule[-3.5ex]{0pt}{3.5ex}
\newpage
\[
\left(
 \prod_{C\in \left( Y\,\vdash G\right) } \mathcal{E} \left(
\left( P\cap \Xi \left( \mathcal{P} \left( G,C\right)
\right) \right) ,\left( Q\cap \Xi \left( \mathcal{P} \left(
G,C\right) \right) \right) ,H,\sigma ,R,\downarrow \left(
x,\Xi \left( \mathcal{P} \left( G,C\right) \right) \right)
\right) \right) \times \hspace{-18.7pt} \hspace{1.0cm}
\]
\[
\times \left(
 \prod_{\Delta \equiv \left\{ r,s\right\} \in W }
\mathbb{S} \left( T-\left| \mu _{ r } \left(
P,Q,H,x,\mathcal{X}
\left( P,Q,H,\rho \right) \right) -\mu _{ s } \left(
P,Q,H,x,\mathcal{X} \left( P,Q,H,\rho \right) \right)
\right| \right) \right)
\]
is not equal to $0$, $\left| x_{ \mathcal{K} \left(
Q,D,l\right) }
-x_{ D } \right| \leq \frac{ \left(
\#\left( V\right) -1\right)
T }{
\left( 1-2\lambda \right) } $ holds.

For we note first that both $\mathcal{K} \left(
Q,D,l\right) $ and $D$ are members of $\Xi \left(
\mathcal{P} \left( P,D\right) \right) $, hence since
$\mathbb{W}$ is a subset of $\mathbb{U} _{ d } \left(
V,\omega \right) $ which, by
page \pageref{Start of original page 15}, is a subset of
$\mathbb{F} _{ d } \left( V\right) $, both $x_{ \mathcal{K}
\left( Q,D,l\right) } $ and $x_{ D } $ are members of
$\mathbb{V} \left( \downarrow \left( x,\mathcal{P} \left(
P,D\right) \right) \right) $, or in other words, both
$x_{ \mathcal{K} \left( Q,D,l\right) } $ and $x_{ D } $ are
members of the convex hull of the $x_{ L } $, $L\in
\mathcal{P} \left( P,D\right) $.   Hence by
Lemma \ref{Lemma 3},
$\left| x_{ \mathcal{K} \left( Q,D,l\right) } -x_{ D }
\right|
\leq \hspace{1.5ex} \max_{ \hspace{-7.0ex}
\begin{array}{c} \\[-3.1ex]
\scriptstyle{ K\in
\mathcal{P} \left( P,D\right) } \\[-1.2ex]
\scriptstyle{ L\in \mathcal{P} \left(
P,D\right) }
\end{array} } \hspace{-1.2ex}
\left| x_{ K } -x_{ L } \right| =\mathbb{L}
\left(
P,D,x\right) $ holds.

We next note that since $P$ is a member of $\mathbb{O}
\left( G,H\right) $, every member of $\mathcal{P} \left(
P,D\right) $ is a subset of some $\left( V\cup H\right)
$-firm over $V$ component of $D$.

Let $C$ be any $\left( V\cup H\right) $-firm over $V$
component of $D$, (which implies, by the definition of the
set $Y$ on
page \pageref{Start of original page 160},
that $C$ is a member of $ Y $), and
suppose that $C$ is \emph{not} a member of $P$.   Then $C$
is \emph{not} a member of $G$, hence $C$ is a member of
$\left( Y\,\vdash G\right) $, hence whenever the above
product of factors is not equal to $0$, and by Lemmas
\ref{Lemma 14} and \ref{Lemma 15} and
page \pageref{Start of original page 80},
and by analogy with
pages \pageref{Start of original page 96}
and \pageref{Start of original page 116},
$\mathbb{L} \left( P,C,x\right) \leq
\frac{ \left(
\#\left( \mathcal{P} \left( P,C\right) \right)
-1\right) T }{
\left( 1-2\lambda \right) } $ holds.

We now define, for each $\left( V\cup H\right) $-firm over
$V$ component $C$ of $D$, $F_{ C } $ to be the set whose
members are all the members $L$ of $\mathcal{P} \left(
P,D\right) $ such that $L\subseteq C$ holds, so that $F_{ C
} $ is equal to $\mathcal{P} \left( P,C\right) $ if $C$ is
\emph{not} a member of $P$, and $F_{ C } $ is equal to
$\left\{ C\right\} $ if $C$ \emph{is} a member of $P$.
Then
it immediately follows from the previous paragraph,
together with the fact that if $C$ is any $\left( V\cup
H\right) $-firm over $V$ component of $D$ such that $C$
\emph{is} a member of $P$, and $K$ and $L$ are any members
of $F_{ C } $, then $K=L=C$ holds so that $\left| x_{ K }
-x_{ L
} \right| =0$ holds, that if $C$ is \emph{any} $\left( V\cup
H\right) $-firm over $V$
\label{Start of original page 203}
 component of $D$, and $K$ and $L$ are any members of $F_{
C } $, then $\left| x_{ K } -x_{ L } \right| \leq
\frac{ \left( \#\left( F_{
C } \right) -1\right) T }{ \left(
1-2\lambda \right) } $ holds.

Now the set $\left( G\cup Y\right) $ is a wood of $V$ by
Lemma \ref{Lemma 35} (iii), and the members of
$\mathcal{P} \left( \left( G\cup Y\right) ,D\right) $ are
the $\left( V\cup H\right) $-firm over $V$ components of
$D$, hence by use of Lemma \ref{Lemma 15} with the
set $V$ of Lemma \ref{Lemma 15} taken as the set
$\mathcal{P} \left( \left( G\cup
Y\right) ,D\right) $, (and noting that the fact that $D$ is
$\left( V\cup H\right) $-connected implies that $D$ is
$\left( \mathcal{P} \left( \left( G\cup Y\right) ,D\right)
\cup H\right) $-connected), and noting, moreover,
that if $\left\{ r,s\right\} $ is any $\left( V\cup H\right)
$-key of $D$ such that $\left\{ r,s\right\} $ is \emph{not}
a
member of $V$, then $\mu _{ r } \left( P,Q,H,x,\mathcal{X}
\left( P,Q,H,\rho \right) \right) $ is a member of the
convex hull of the $x_{ L } $ such that $L$ is a member of
$\mathcal{P} \left( P,D\right) $ such that $L$ is a subset
of the unique $\left( V\cup H\right) $-firm over $V$
component of $D$ that has $r$ as a member, and $\mu _{ s }
\left( P,Q,H,x,\mathcal{X} \left( P,Q,H,\rho \right)
\right) $ is a member of the convex hull of the $x_{ L } $
such that $L$ is a member of $\mathcal{P} \left( P,D\right)
$ such that $L$ is a subset of the unique $\left( V\cup
H\right) $-firm over $V$ component of $D$ that has $s$ as a
member, we find that if $K$ and $L$ are \emph{any} members
of $\mathcal{P} \left( P,D\right) $, then $\left| x_{ K }
-x_{ L
} \right| $ is less than or equal to $\left( \#\left(
\mathcal{P}
\left( \left( G\cup Y\right) ,D\right) \right) -1\right) T$
plus the sum, over the members $n$ of the domain of the map
$M$ constructed as in Lemma \ref{Lemma 15}, (so that
$\#\left(
\mathcal{D} \left( M\right) \right) \leq \#\left(
\mathcal{P} \left( \left( G\cup Y\right) ,D\right) \right)
$ holds and $\mathcal{R} \left( M\right) $ is a subset of
$\mathcal{P} \left( \left( G\cup Y\right) ,D\right)
$), of $ \frac{ \left( \#\left( F_{ M_{ n } } \right)
 -1 \right) T }{ \left( 1-2\lambda \right) } $,
hence, since it follows directly from the construction of
Lemma \ref{Lemma 15} that we may assume that no member $C$
of
$\mathcal{P} \left( \left( G\cup Y\right) ,D\right) $, (or
in other words, no $\left( V\cup H\right) $-firm over $V$
component $C$ of $ D $), is equal to $M_{ n } $
for \emph{more}
than one member $n$ of $\mathcal{D} \left( M\right) $,
$\left| x_{ K } -x_{ L } \right| $ is less than or equal to
$ \left( -T\right) $ plus
the sum, over the members $C$ of $\mathcal{P} \left( \left(
G\cup Y\right) ,D\right) $, of $ \frac{ \# \left(
F_{ C } \right) T }{ \left(
1-2\lambda \right) } $, hence $\left| x_{ K } -x_{ L }
\right| $ is
less than or equal to $ \left( -T\right) $ plus
$ \frac{ \#\left( \mathcal{P}
\left( P,D\right) \right) T }{ \left( 1-2\lambda \right)
} $, hence $\left| x_{ K } -x_{ L } \right| $ is less than
or equal to $
\frac{ \left( \#\left( \mathcal{P} \left( P,D\right) \right)
-1\right) T }{ \left( 1-2\lambda \right) } $, hence
$\left| x_{ K } -x_{ L } \right| $ is less than or equal to
$ \frac{ \left(
\#\left( V\right) -1\right) T }{ \left( 1-2\lambda \right)
 } $.

\vspace{2.5ex}

Now let $\left(B,i,n,j,s,E,v\right) $ be any
 ordered septuple
$\left(B,i,n,j,s,E,v\right) $ as on
pages \pageref{Start of original page 191} and
\pageref{Start of original page 192},
and satisfying
the conditions specified on
page \pageref{Start of original page 191}.   We define
 $\tilde{ B } $
to be the map obtained from $B$ by \emph{deleting} from $B$
all members $\left( \alpha ,B_{ \alpha } \right) $ of $B$
such that $B_{ \alpha } $ is a member of $\mathbb{B}
\left( G\right) $, and \emph{retaining,} as members
of $\tilde{ B }
 $, all members $\left( \alpha ,B_{ \alpha } \right) $
of $B$ such that $B_{ \alpha } $ is
\label{Start of original page 204}
 a member of $\left( Q\,\vdash G\right) $.   It then
follows immediately from Lemma \ref{Lemma 23} and
observation \ref{Observation 45}) above that for all
$\rho \in \mathbb{D} $,
$\bar{I} \hspace{-0.34ex} \bar{ \rule{0pt}{1.8ex} }
\left( B,i,j,\rho \right) $ is less than or equal to
$ \left( \frac{ \left( \#\left( V\right) -1\right) T }{
\left( 1-2\lambda
\right) } \right)^{ \left( \#\left( \mathcal{D}
\left( B\right)
\right) -\#\left( \mathcal{D} \left(
B \hspace{-0.5ex} \tilde{ \rule{0pt}{1.2ex} }
\hspace{0.5ex}
 \right) \right)
\right) } $ times $\bar{I} \hspace{-0.34ex} \bar{
\rule{0pt}{1.9ex} } \left( \tilde{ B } ,i,j,\rho \right) $,
hence to prove the finiteness of $\bar{I} \hspace{-0.34ex}
\bar{ \rule{0pt}{1.9ex} } \left( B,i,j,\rho
\right) $ and a $\rho $-independent bound on $\bar{I}
\hspace{-0.34ex} \bar{ \rule{0pt}{1.9ex} } \left(
B,i,j,\rho \right) $, it is sufficient to prove the
finiteness of $\bar{I} \hspace{-0.34ex} \bar{
\rule{0pt}{1.9ex} } \left( \tilde{ B } ,i,j,\rho
 \right) $ and a
$\rho $-independent bound on $\bar{I} \hspace{-0.34ex}
\bar{ \rule{0pt}{1.9ex} } \left(
\tilde{ B } ,i,j,\rho
\right) $.

We note furthermore that the condition on
page \pageref{Start of original page 164} is in
fact a condition on the two members $B$ and $j$ of the
septuple $\left(B,i,n,j,s,E,v\right) $, and
that it immediately follows
from the fact that $B$ and $j$ satisfy the condition on
page \pageref{Start of original page 164},
that $\tilde{ B }$ and $j$ satisfy the
condition on
page \pageref{Start of original page 164}
 for all members $A$ of $\mathbb{B} \left(
Q\right) $ \emph{except} for the members $A$ of $\mathbb{B}
\left( G\right) $, or in other words, that $\tilde{ B }$
and $j$ satisfy the condition on
page \pageref{Start of original page 164}
 when that condition is
modified to apply just to the members $A$ of $\left(
Q\,\vdash G\right) $ rather than to all the members $A$ of
$\mathbb{B} \left( Q\right) $.   For if $A$ is any member
of $\left( Q\,\vdash G\right) =\left( \mathbb{B} \left(
Q\right) \,\vdash \mathbb{B} \left( G\right) \right) $,
then it follows directly from the fact that $G\subseteq
P\subseteq Q$ holds, that if $C$ is any member of $G$, then
either $A\subset C$ holds or $A\cap C=\emptyset $ holds or
$C$
is a subset of some member of $\mathcal{P} \left(
P,A\right) $, (for if $C\subset A$ holds and $D$ is a
member of $\mathcal{P} \left( P,A\right) $ such that $D\cap
C$ is nonempty, then $D\subset C$ cannot hold hence
$C\subseteq D$ holds), hence $C$ is \emph{not} a member of
$\left( \Xi \left( \mathcal{P} \left( P,A\right) \right)
\,\vdash \mathcal{P} \left( P,A\right) \right) $, hence the
number of members $\alpha $ of $\mathcal{D} \left( \tilde{
B }
\right) $ such that $\tilde{ B } _{ \alpha } \in \left( \Xi
\left( \mathcal{P} \left( P,A\right) \right) \,\vdash
\mathcal{P} \left( P,A\right) \right) $ holds, is equal to
the number of members $\alpha $ of $\mathcal{D} \left(
B\right) $ such that $B_{ \alpha } \in \left( \Xi \left(
\mathcal{P} \left( P,A\right) \right) \,\vdash \mathcal{P}
\left( P,A\right) \right) $ holds.

We now express $\bar{I} \hspace{-0.34ex} \bar{
\rule{0pt}{1.9ex} } \left( \tilde{ B } ,i,j,\rho \right) $
in
terms of the $a_{ K } $ variables, \\
$K\in \left( J\,\vdash
\left( \mathcal{R} \left( S\right) \cup \left\{ \mathcal{U}
\left( V\right) \right\} \right) \right) $, and we note that
if $U\equiv \left\{ l,m\right\} $ is any member of $W$ such
that $\mathcal{Y} \left( \left( G\cup Y\right) ,U\right)
\in G$ holds, so that by
observation \ref{Observation 40}) above,
$\mathcal{Y} \left( \left( G\cup Y\right) ,U\right) $ is a
member of $\mathbb{B} \left( G\right) $ and $U$ is a
$\left( V\cup H\right) $-key of $\mathcal{Y} \left( \left(
G\cup Y\right) ,U\right) $ such that $U$ is \emph{not} a
member of $V$, then it follows directly from observations
\ref{Observation 41}) to
\ref{Observation 44}) above that if we define
$C\equiv \mathcal{Y} \left( \left( G\cup Y\right) ,U\right)
$ and $D\equiv \left( C\,\vdash \uparrow \left( f_{ C }
,\left\{ U\right\} \right) \right) $, (so that, by
page \pageref{Start of original page 165},
$D$ is a member of $J$ and, by
pages \pageref{Start of original page 165},
\pageref{Start of original page 171},
and \pageref{Start of original page 172},
$D$ is \emph{not} a member of $\mathcal{R}
\left( S\right) $), then the \emph{only} dependence
of $\bar{I} \hspace{-0.34ex} \bar{ \rule{0pt}{1.9ex} }
\left( \tilde{ B } ,i,j,\rho \right) $ on
$a_{ D } $ is
through the two factors
\[
\left| \mu _{ m } \left( P,Q,H,x,\mathcal{X} \left(
P,Q,H,\rho
\right) \right) -\mu _{ l } \left( P,Q,H,x,\mathcal{X}
\left( P,Q,H,\rho \right) \right)
\right|^{ -\theta _{ U } }
\times \hspace{-7.0pt} \hspace{3.5cm}
\]
\[
\hspace{2.5cm} \hspace{-5.6pt}
\times \mathbb{S} \left( T-\left| \mu _{ m } \left(
P,Q,H,x,\mathcal{X} \left( P,Q,H,\rho \right) \right) -\mu
_{ l } \left( P,Q,H,x,\mathcal{X} \left( P,Q,H,\rho \right)
\right) \right| \right),
\]
and furthermore, if $m$ is the member of $U$ that is a
member of $D$, then
\label{Second line of original page 205}
by observation \ref{Observation 42}) above, if
we define $N$ to be the unique $\left( V\cup H\right)
$-firm over $V$ component of $C$ that has $l$ as a member,
and if we moreover define $F_{ N } $ to be the set whose
members are all the members $L$ of $\mathcal{P} \left(
P,C\right) $ such that $L\subseteq N$ holds, and if we
moreover define $e_{ L } $, $L\in F_{ N } $, to be the
coefficients, summing to $1$, of the $a_{ L } $, $L\in F_{
N } $, referred to in
observation \ref{Observation 42}) above, then
the above two factors are equal to
\[
\left| a_{ D } - \left(
\sum_{L\in F_{ N } } e_{ L } a_{ L }
\right) \right|^{ -\theta
_{ U } }
\mathbb{S} \left( T-\left| a_{ D } - \left(
 \sum_{L\in F_{ N }
} e_{
L } a_{ L } \right) \right| \right)
\]
hence, since by assumption $\theta _{ R } <d$ holds for
every member $R$ of $W$, hence in particular, $\theta _{ U
} <d$ holds, the integral over $a_{ D } $ is an integral
with $ d $-dimensional spherical symmetry, and has the
finite, $\rho $-independent value:
\[
\left\{ \begin{array}{cc}
\displaystyle{
\frac{ 2\left( 2\pi \right)^{ \left( \frac{ d-1 }{ 2 }
\right) } }{ \left( d-2\right) !! }
\frac{ T^{ \left( d-\theta _{
U } \right) } }{ \left( d-\theta _{ U } \right) } } &
 \left( d\textrm{ odd and }\geq 1\right) \\
\displaystyle{ \frac{ 2\pi^{
\left( \frac{ d }{ 2 } \right) } }{ \left( \frac{
 d-2 }{ 2 }
\right) ! } \frac{ T^{ \left( d-\theta _{ U } \right) } }
{ \left( d-\theta _{ U } \right) } } & \left(
d\textrm{ even and }\geq 2\right)
\end{array} \right\} ,
\]
where $\left( -1\right) !!\equiv 1$, and for $d$ odd,
$d\geq 3$, $\left( d-2\right) !!\equiv \left( d-2\right)
\left( d-4\right) !!$.

We use this result for \emph{every} member $U$ of $W$ such
that $\mathcal{Y} \left( \left( G\cup Y\right) ,U\right)
\in G$ holds, or in other words, by
observation \ref{Observation 40})
 above, for \emph{every} member $U$ of $W$ such that $U$
is a $\left( V\cup H\right) $-key, such that $U$ is
\emph{not} a member of $V$, of some member of $\mathbb{B}
\left( G\right) $, or in other words, since \emph{no}
member of $W$ is a member of $V$, (since every member of
$W$ intersects exactly \emph{two} members of $V$, and $V$
is a partition), for \emph{every} member $U$ of $W$ such
that $U$ is a $\left( V\cup H\right) $-key of some member
of $\mathbb{B} \left( G\right) $, after which we find, by
observations \ref{Observation 41})
to \ref{Observation 44}) above, that $\bar{I}
\hspace{-0.34ex} \bar{ \rule{0pt}{1.9ex} }
\left( \tilde{ B } ,i,j,\rho \right) $ completely factorizes
into a product of finite, $\rho $-independent real numbers,
times the product, over the members $C$ of $\left(
Y\,\vdash G\right) $, of an integral
$ \hspace{0.2ex}
{ \rule{0pt}{2.8ex} }^{ \circ } \hspace{-1.4ex} \bar{I} _{
C }
\hspace{-0.3ex} \left(
\hspace{0.8ex} \tilde{ \rule{0pt}{2.0ex} } \hspace{-1.2ex}
\tilde{B}
 _{ C } ,\tilde{ i } _{ C } ,
\tilde{ j } _{ C } ,\rho
\right) $, where $\hspace{0.8ex} \tilde{ \rule{0pt}{2.0ex}
} \hspace{-1.2ex} \tilde{B} _{ C } $ is defined
to be the
map that is the subset of the map $\tilde{ B }$ which
includes
precisely all the members $\left( \alpha ,
\tilde{ B } _{ \alpha
 } \right) $ of $\tilde{ B }$ such that
 $\tilde{ B } _{ \alpha } \in
\left( \Xi \left( \mathcal{P} \left( G,C\right) \right)
\,\vdash \mathcal{P} \left( G,C\right) \right) $ holds, (or
in other words, by
observation \ref{Observation 44}) above and the
fact that
\label{Start of original page 206}
 $\mathcal{R} \left( \tilde{ B } \right) $ is a subset of
$\left( Q\,\vdash G\right) =\left( \mathbb{B} \left(
Q\right) \,\vdash \mathbb{B} \left( G\right) \right) $,
such that $\mathcal{Y} \left( \left( G\cup Y\right) ,
\tilde{ B }
_{ \alpha } \right) =C$ holds),
$\tilde{ i } _{ C } $ is
defined to be the map that is the subset of the map $i$
which includes precisely all the members $\left( \alpha
,i_{ \alpha } \right) $ of $i$ such that $\alpha \in
\mathcal{D} \left( \hspace{0.8ex} \tilde{ \rule{0pt}{2.0ex}
} \hspace{-1.2ex} \tilde{B} _{ C } \right) $
holds, and $
\tilde{j } _{ C } $ is defined to be the map
that is the subset
of the map $j$ which includes precisely all the members
$\left( \beta ,j_{ \beta } \right) $ of $j$ such that $j_{
\beta } $ is a member of $\mathcal{U} \left( W\right) $,
and the unique member $\mathcal{C} \left( W,j_{ \beta }
\right) $ of $W$ that has $j_{ \beta } $ as a member,
satisfies the requirement that $\mathcal{Y} \left( \left(
G\cup Y\right) ,\mathcal{C} \left( W,j_{ \beta } \right)
\right) =C$ holds, and for any member $C$ of $\left(
Y\,\vdash G\right) $, and for any ordered quadruple $\left(
B \rule{0pt}{2.55ex}^{ \hspace{-1.1ex} * },i
\rule{0pt}{2.55ex}^{ \hspace{-0.75ex} * },j
\rule{0pt}{2.55ex}^{ \hspace{-0.75ex} * },\rho
\right) $ of a map $B \rule{0pt}{2.55ex}^{ \hspace{-1.1ex}
* }$ such that
$\mathcal{D} \left( B \rule{0pt}{2.55ex}^{ \hspace{-1.1ex}
* }\right) $ is finite and $\mathcal{R}
\left( B \rule{0pt}{2.55ex}^{ \hspace{-1.1ex} * }\right)
\subseteq \left( Q\cap \left( \Xi \left(
\mathcal{P} \left( G,C\right) \right) \,\vdash \mathcal{P}
\left( G,C\right) \right) \right) $ holds, a map $i
\rule{0pt}{2.55ex}^{ \hspace{-0.75ex} * }$ such
that $\mathcal{D} \left( B \rule{0pt}{2.55ex}^{
\hspace{-1.1ex} * }\right) \subseteq \mathcal{D}
\left( i \rule{0pt}{2.55ex}^{ \hspace{-0.75ex} * }\right) $
holds and for each member $\alpha $ of
$\mathcal{D} \left( B \rule{0pt}{2.55ex}^{ \hspace{-1.1ex}
* }\right) $, $i \rule{0pt}{2.55ex}^{ \hspace{-0.75ex} *
}_{ \alpha } $ is a
member of $B \rule{0pt}{2.55ex}^{ \hspace{-1.1ex} * }_{
\alpha } $, a map $j \rule{0pt}{2.55ex}^{ \hspace{-0.75ex}
* }$ such that
$\mathcal{D} \left( j \rule{0pt}{2.55ex}^{ \hspace{-0.75ex}
* }\right) $ is finite and for each
member $\beta $ of $\mathcal{D} \left( j
\rule{0pt}{2.55ex}^{ \hspace{-0.75ex} * }\right) $, $j
\rule{0pt}{2.55ex}^{ \hspace{-0.75ex} * }_{
\beta } $ is either a member of $\mathcal{U} \left(
W\right) $ such that $\mathcal{Y} \left( \left( G\cup
Y\right) ,\mathcal{C} \left( W,j \rule{0pt}{2.55ex}^{
\hspace{-0.75ex} * }_{ \beta } \right)
\right) =C$ holds, or else is a member of $\left( C\,\vdash
\mathcal{U} \left( W\right) \right) $, and a member $\rho $
of $\mathbb{D} $, the integral ${ \rule{0pt}{2.8ex} }^{
\circ } \hspace{-1.4ex} \bar{I} _{ C } \left(
B \rule{0pt}{2.55ex}^{ \hspace{-1.1ex} * },i
\rule{0pt}{2.55ex}^{ \hspace{-0.75ex} * },j
\rule{0pt}{2.55ex}^{ \hspace{-0.75ex} * },\rho
\right) $ is defined by
\[
{ \rule{0pt}{2.8ex} }^{ \circ } \hspace{-1.4ex}
\bar{I}_{ C }\left( B \rule{0pt}{2.55ex}^{ \hspace{-1.1ex}
* },i \rule{0pt}{2.55ex}^{ \hspace{-0.75ex} * },j
\rule{0pt}{2.55ex}^{ \hspace{-0.75ex} * },\rho \right)
\equiv
\int_{ \tilde{ \mathbb{W}
}_{ C } } \left( \prod_{A\in N_{ C } } d^{ d } a_{ A }
\right) \times \hspace{-3.7pt} \hspace{9.0cm}
\]
\[
\times \left(
\mathcal{E} \left( \left( P\cap \Xi \left( \mathcal{P}
\left( G,C\right) \right) \right) ,\left( Q\cap \Xi \left(
\mathcal{P} \left( G,C\right) \right) \right) ,H,\sigma
,R,\downarrow \left( x,\Xi \left( \mathcal{P} \left(
G,C\right) \right) \right) \right) \rule{0pt}{5.5ex}
\times \right.
\]
\[
\times \left(
 \prod_{\alpha \in \mathcal{D} \left( B
\rule{0pt}{1.8ex}^{ \hspace{-0.9ex} * }\right) }
\left| x_{ \mathcal{K} \left( Q,B \rule{0pt}{1.8ex}^{
\hspace{-0.9ex} * }_{ \alpha } ,i \rule{0pt}{1.8ex}^{
\hspace{-0.55ex} * }_{ \alpha }
\right) } -x_{ B \rule{0pt}{1.8ex}^{ \hspace{-0.9ex} * }_{
\alpha } } \right| \right)
\left( \prod_{\beta \in
\mathcal{D}
\left( j \rule{0pt}{1.8ex}^{ \hspace{-0.6ex} * }\right) }
\left( \frac{ 1 }{ w\left(
x,j \rule{0pt}{2.55ex}^{ \hspace{-0.75ex} * }_{ \beta }
,\rho \right) } \right) \right) \times
\]
\[
\times \left(
\prod_{\Delta \equiv \left\{ r,s\right\}
\in \tilde{ W } _{ C }
} \left| \mu _{ r } \left( P,Q,H,x,\mathcal{X} \left(
P,Q,H,\rho \right) \right) -\mu _{ s } \left(
P,Q,H,x,\mathcal{X} \left( P,Q,H,\rho \right) \right)
\right|^{ -\theta _{ \Delta } } \right) \times
\]
\[
\left.
\times \rule{0pt}{5.5ex} \hspace{-3.0pt} \left(
 \prod_{\Delta \equiv \left\{ r,s\right\} \in
 \tilde{ W }_{ C }
 } \hspace{-11.0pt}
  \mathbb{S} \left( T-\left| \mu _{ r } \left(
P,Q,H,x,\mathcal{X} \left( P,Q,H,\rho \right) \right) -\mu
_{ s } \left( P,Q,H,x,\mathcal{X} \left( P,Q,H,\rho \right)
\right) \right| \right) \hspace{-2.0pt} \right)
\hspace{-3.0pt} \right) \hspace{-2.0pt} ,
\]
where $N_{ C } $ is the set $\left( \left( P\cap \left( \Xi
\left( \mathcal{P} \left( G,C\right) \right) \,\vdash
\left\{ C\right\} \right) \right) \,\vdash
\left( \mathcal{R}
\left( S\right) \cup \left\{ \mathcal{U} \left( V\right)
\right\} \right) \right) $,
$\tilde{ \mathbb{W} }_{ C } $ is the set
$\mathbb{E} _{ d }^{ \left( N_{ C } \right) } $, and
$\tilde{ W } _{ C } $ is
the set whose members are all the members $\left\{
r,s\right\} $ of $W$ such that $\mathcal{Y} \left( \left(
G\cup Y\right) ,\left\{ r,s\right\} \right) =C$ holds.

We note that in consequence of the manner in which the set
$S$ was extended to the domain $\mathbb{B}
\left( \bar{ J }
\right) $ on
page \pageref{Start of original page 171},
$N_{ C } $ includes, for each
member $D$ of $\mathbb{B} \left( P\cap \left( \Xi \left(
\mathcal{P} \left( G,C\right) \right) \right) \right)
=\left( \left( P\cap \Xi \left( \mathcal{P} \left(
G,C\right) \right) \right) \,\vdash \mathcal{P} \left(
G,C\right) \right) $, exactly $\left( \#\left( \mathcal{P}
\left( P,D\right) \right) -1\right) $ of the $\#\left(
\mathcal{P} \left( P,D\right) \right) $ members of
$\mathcal{P} \left( P,D\right) $, hence the $a_{ K } $
\label{Start of original page 207}
 variables, $K\in N_{ C } $, are in the same relation to
the wood $P\cap \Xi \left( \mathcal{P} \left( G,C\right)
\right) $, as the $z_{ K } $ variables, defined on
pages \pageref{Start of original page 117}
and \pageref{Start of original page 118}
for Theorem \ref{Theorem 1}, are to the wood $P$ of
Theorem \ref{Theorem 1}, and furthermore, by
page \pageref{Start of original page 6} and
by analogy with
pages \pageref{Start of original page 117} and
\pageref{Start of original page 118},
$\#\left( N_{ C } \right) $ is
equal to $\left( \#\left( \mathcal{P} \left( G,C\right)
\right) -1\right) $.

And we note that every variable $x$ in the above formula
for ${ \rule{0pt}{2.8ex} }^{ \circ } \hspace{-1.4ex}
\bar{I} _{ C } \left( B \rule{0pt}{2.55ex}^{
\hspace{-1.1ex} * },i \rule{0pt}{2.55ex}^{ \hspace{-0.75ex}
* },j \rule{0pt}{2.55ex}^{ \hspace{-0.75ex} * },\rho
\right) $ is to be
expressed in terms of the $a_{ K } $ variables by the
formulae on
pages \pageref{Start of original page 178} and
\pageref{Start of original page 179},
and that by observations \ref{Observation 41}),
\ref{Observation 43}), and
\ref{Observation 44}) above, and in
consequence of the particular combinations in which the $x$
variables occur in the formula for ${ \rule{0pt}{2.8ex} }^{
\circ } \hspace{-1.4ex} \bar{I} _{ C } \left(
B \rule{0pt}{2.55ex}^{ \hspace{-1.1ex} * },i
\rule{0pt}{2.55ex}^{ \hspace{-0.75ex} * },j
\rule{0pt}{2.55ex}^{ \hspace{-0.75ex} * },\rho
\right) $, when we \emph{do} express the $x$
variables in the integrand of ${ \rule{0pt}{2.8ex} }^{
\circ } \hspace{-1.4ex} \bar{I} _{ C } \left(
B \rule{0pt}{2.55ex}^{ \hspace{-1.1ex} * },i
\rule{0pt}{2.55ex}^{ \hspace{-0.75ex} * },j
\rule{0pt}{2.55ex}^{ \hspace{-0.75ex} * },\rho
\right) $ in terms of the $a_{ K } $
variables, the dependence of the integrand of ${
\rule{0pt}{2.8ex} }^{ \circ } \hspace{-1.4ex} \bar{I} _{ C
} \left( B \rule{0pt}{2.55ex}^{ \hspace{-1.1ex} *
},i \rule{0pt}{2.55ex}^{ \hspace{-0.75ex} * },j
\rule{0pt}{2.55ex}^{ \hspace{-0.75ex} * },\rho \right) $ on
the $a_{ K } $
variables is limited to dependence on the $a_{ K } $ such
that $K$ is a member of $N_{ C } $.

We next note that if $B$ and $j$ are any maps such that
$\mathcal{D} \left( B\right) $ is finite and $\mathcal{R}
\left( B\right) \subseteq \mathbb{B} \left( Q\right) $
holds, $\mathcal{D} \left( j\right) $ is finite and
$\mathcal{R} \left( j\right) \subseteq \mathcal{U} \left(
V\right) $ holds, the condition on
page \pageref{Start of original page 164} holds for $B$
and $j$, and if $A$ is any member of $\mathbb{B} \left(
G\right) $, and $T$ is any $\left( V\cup H\right) $-key of
$A$ such that $T$ is \emph{not} a member of $V$, then there
is \emph{no} member $\beta $ of $\mathcal{D} \left(
j\right) $ such that $j_{ \beta } \in T$ holds,
and if $\tilde{ B
} $ is defined in terms on $B$, as on
page \pageref{Start of original page 203}, by
\emph{deleting} from $B$ all members $\left( \alpha ,B_{
\alpha } \right) $ of $B$ such that $B_{ \alpha } $ is a
member of $\mathbb{B} \left( G\right) $, and
\emph{retaining,} as members of $\tilde{ B } $, all members
$\left( \alpha ,B_{ \alpha } \right) $ of $B$ such that
$B_{ \alpha } $ is a member of $\left( Q\,\vdash G\right)
=\left( \mathbb{B} \left( Q\right) \,\vdash \mathbb{B}
\left( G\right) \right) $, (so that, as noted on
page \pageref{Start of original page 204},
$\tilde{ B }$ and $j$ satisfy the condition on
page \pageref{Start of original page 164}
when that condition is modified to apply just to the
members $A$ of $\left( Q\,\vdash G\right) =\left(
\mathbb{B} \left( Q\right) \,\vdash \mathbb{B} \left(
G\right) \right) $ rather than to all members $A$ of
$\mathbb{B} \left( Q\right) $), and if $C$ is any
member of $\left( Y\,\vdash G\right) $, and if
$\hspace{0.8ex} \tilde{ \rule{0pt}{2.0ex} }
\hspace{-1.2ex} \tilde{B} _{ C } $
is defined in terms of $\tilde{ B }$ as on
page \pageref{Start of original page 206} to be
the map that is the subset of the map
$\tilde{ B }$ that includes
precisely all the members $\left( \alpha ,
\tilde{ B } _{ \alpha
 } \right) $ of $\tilde{ B }$ such that
 $\tilde{ B }_{ \alpha } \in
\left( \Xi \left( \mathcal{P} \left( G,C\right) \right)
\,\vdash \mathcal{P} \left( G,C\right) \right) $
holds, (or
in other words, such that $\mathcal{Y}
\left( \left( G\cup
Y\right) ,\tilde{ B }_{ \alpha } \right) =C$ holds),
and if $\tilde{ j }
_{ C } $ is defined in terms of $j$ as on
page \pageref{Start of original page 206} to
be the map that is the subset of the map $j$ that includes
precisely all the members $\left( \beta ,j_{ \beta }
\right) $ of $j$ such that $j_{ \beta } $ is a member of
$\mathcal{U} \left( W\right) $, and the unique member
$\mathcal{C} \left( W,j_{ \beta } \right) $ of $W$ that
has $j_{ \beta } $ as a member, satisfies the requirement
that $\mathcal{Y} \left( \left( G\cup Y\right) ,\mathcal{C}
\left( W,j_{ \beta } \right) \right) =C$ holds, (or in
other words, such that $j_{ \beta } $ is a member of
$\mathcal{U} \left( \tilde{ W }_{ C } \right) $),
 and if $A$
is any member of $\mathbb{B} \left( Q\right) \cap \left(
\Xi \left( \mathcal{P} \left( G,C\right) \right) \,\vdash
\mathcal{P} \left( G,C\right) \right) $, then the
\label{Start of original page 208}
 condition on
page \pageref{Start of original page 164}
 for $A$ depends on $B$ and $j$
\emph{only} through the maps $\hspace{0.8ex} \tilde{
\rule{0pt}{2.0ex} } \hspace{-1.2ex} \tilde{B} _{ C }
$ and $\tilde{ j }
_{ C } $, (which are subsets of the maps $B$ and $j$
respectively), and furthermore the condition on
page \pageref{Start of original page 164}
for $A$ is equivalent to the modification of the condition
on
page \pageref{Start of original page 164}
 for $A$, obtained by replacing, in the condition on
page \pageref{Start of original page 164}
 for $A$, all occurrences of the map
$B$ by the map $\hspace{0.8ex} \tilde{ \rule{0pt}{2.0ex} }
\hspace{-1.2ex} \tilde{B} _{ C } $, and all
occurrences of
the map $j$ by the map $\tilde{ j } _{ C } $.   For the
condition on
page \pageref{Start of original page 164}
 for $A$ only depends on the map $B$
through the number of members $\alpha $ of $\mathcal{D}
\left( B\right) $ such that $B_{ \alpha } \in \left( \Xi
\left( \mathcal{P} \left( P,A\right) \right) \,\vdash
\mathcal{P} \left( P,A\right) \right) $ holds, and $A\in
\left( \Xi \left( \mathcal{P} \left( G,C\right) \right)
\,\vdash \mathcal{P} \left( G,C\right) \right) $ and
$G\subseteq P$ together imply that $\left( \Xi \left(
\mathcal{P} \left( P,A\right) \right) \,\vdash \mathcal{P}
\left( P,A\right) \right) \subseteq \left( \Xi \left(
\mathcal{P} \left( G,C\right) \right) \,\vdash \mathcal{P}
\left( G,C\right) \right) $ holds, (for if $D$ is any
member of $\left( \Xi \left( \mathcal{P} \left( P,A\right)
\right) \,\vdash \mathcal{P} \left( P,A\right) \right) $,
then $D$ is a subset of $A$ hence $D$ is a subset of $C$,
and each member of $\mathcal{P} \left( P,A\right) $ either
does not intersect $D$ or else is a \emph{strict} subset of
$D$, and if $E$ is any member of $\mathcal{P} \left(
G,C\right) $, then either $E\cap A=\emptyset $ holds or else
$E\subset A$ holds, and if $E\cap A=\emptyset $ holds then
$E\cap D=\emptyset $ holds, while if $E\subset A$ holds,
then since $E$ is a member of $G$ hence a member of $P$,
page \pageref{Start of original page 5}
implies that $E$ is a subset of a unique member of
$\mathcal{P} \left( P,A\right) $, and if that unique member
of $\mathcal{P} \left( P,A\right) $ does not intersect $D$,
then $E\cap D=\emptyset $ holds, while if that unique
member of
$\mathcal{P} \left( P,A\right) $ is a strict subset of $D$,
then $E\subset D$ holds), hence every such member $\alpha $
of $\mathcal{D} \left( B\right) $ is a member of
$\mathcal{D} \left( \hspace{0.8ex} \tilde{
\rule{0pt}{2.0ex} } \hspace{-1.2ex} \tilde{B} _{ C }
 \right) $, and the
condition on
page \pageref{Start of original page 164}
 for $A$ only depends on the map $j$
through the number of members $\beta $ of $\mathcal{D}
\left( j\right) $ such that $j_{ \beta } \in \mathcal{U}
\left( W\right) $ and $\left\{ \mathcal{Z} \left( P,H,j_{
\beta } \right) ,\mathcal{Z} \left( P,H,l\right) \right\}
\in \mathcal{Q} \left( \mathcal{P} \left( P,A\right)
\right) $ both hold, where $l$ is the other member of the
unique member $\mathcal{C} \left( W,j_{ \beta } \right) $
of $W$ that has $j_{ \beta } $ as a member, and the facts
that $G\subseteq P$ holds and $A$ is a member of $\left(
\Xi \left( \mathcal{P} \left( G,C\right) \right) \,\vdash
\mathcal{P} \left( G,C\right) \right) $ together imply that
every member $\left\{ r,s\right\} $ of $W$ such
 that $\left\{
\mathcal{Z} \left( P,H,r\right) ,\mathcal{Z} \left(
P,H,s\right) \right\} \in \mathcal{Q} \left( \mathcal{P}
\left( P,A\right) \right) $ holds, is such that
$\mathcal{Y} \left( \left( G\cup Y\right) ,\left\{
r,s\right\} \right) =C$ holds, hence is such that $\left\{
r,s\right\} $ is a member of $\tilde{ W } _{ C } $,
 (for $\left\{
\mathcal{Z} \left( P,H,r\right) ,\mathcal{Z} \left(
P,H,s\right) \right\} \in \mathcal{Q} \left( \mathcal{P}
\left( P,A\right) \right) $ implies that $\mathcal{Z}
\left( P,H,r\right) $ is equal to $\mathcal{K} \left(
P,A,r\right) $ and that $\mathcal{Z} \left( P,H,s\right) $
is equal to $\mathcal{K} \left( P,A,s\right) $, hence that
$\left\{ r,s\right\} $ is \emph{not} a subset of any member
of $\mathcal{P} \left( P,A\right) $, and that $\left\{
r,s\right\} $ \emph{is} a subset of $A$ hence that $\left\{
r,s\right\} $ is a subset of $C$, and $A\in \left( \Xi
\left( \mathcal{P} \left( G,C\right) \right) \,\vdash
\mathcal{P} \left( G,C\right) \right) $ furthermore implies
that if $E$ is any member of $\mathcal{P} \left( G,C\right)
$, then either $E\cap A=\emptyset $ holds or $E\subset A$
holds, hence by
page \pageref{Start of original page 5}
 and the fact that $G\subseteq P$ holds,
either $E\cap A=\emptyset $ holds or $E$ is a subset of a
unique
member of $\mathcal{P} \left( P,A\right) $, hence $\left\{
r,s\right\} $ is \emph{not} a subset of any member of
$\mathcal{P} \left( G,C\right) $, hence, since it follows
directly from the
\label{Start of original page 209}
 definition of $Y$ on
page \pageref{Start of original page 160},
that there is \emph{no}
member $D$ of $\left( Y\,\vdash G\right) $ such that
$D\subset C$ holds and $D$ is \emph{not} a subset of any
member of $\mathcal{P} \left( G,C\right) $, $\mathcal{Y}
\left( \left( G\cup Y\right) ,\left\{ r,s\right\} \right) $
is equal to $ C $), hence every such member $\beta $ of
$\mathcal{D} \left( j\right) $, is a member of $\mathcal{D}
\left( \tilde{ j } _{ C } \right) $.

We shall next show that if $C$ is any member of $\left(
Y\,\vdash G\right) $, and if
$\left(B,i,j\right) $ is any ordered triple
of a map $B$ such that $\mathcal{D} \left( B\right) $ is
finite and $\mathcal{R} \left( B\right) \subseteq \left(
Q\cap \left( \Xi \left( \mathcal{P} \left( G,C\right)
\right) \, \hspace{-0.8pt}
\vdash \mathcal{P} \left( G,C\right) \right)
\right) $ holds, a map $i$ such that $\mathcal{D} \left(
B\right) \subseteq \mathcal{D} \left( i\right) $ holds and
for each member $\alpha $ of $\mathcal{D} \left( B\right)
$, $i_{ \alpha } $ is a member of $B_{ \alpha } $, and a
map $j$ such that $\mathcal{D} \left( j\right) $ is finite
and \\
$\mathcal{R} \left( j\right) \subseteq \left(
\mathcal{U} \left( \tilde{ W } _{ C } \right) \cup \left(
C\,\vdash \mathcal{U} \left( W\right) \right) \right) $
holds, such that for each member $A$ of $\mathbb{B} \left(
Q\right) \cap \left( \Xi \left( \mathcal{P} \left(
G,C\right) \right) \,\vdash \mathcal{P} \left( G,C\right)
\right) $, $B$ and $j$ satisfy the condition on
page \pageref{Start of original page 164}
for $A$, and if $\rho $ is any member of $\mathbb{D} $,
then the integral ${ \rule{0pt}{2.8ex} }^{ \circ }
\hspace{-1.4ex} \bar{I} _{ C } \left( B,i,j,\rho \right)
$ is finite and is bounded above by a finite real number,
\emph{independent of} $ \rho $, and we note that, by the
foregoing, it will immediately follow from this that if
$\left(B,i,n,j,s,E,v\right) $ is any ordered
septuple as on
pages \pageref{Start of original page 191}
and \pageref{Start of original page 192},
and satisfying the conditions specified on
page \pageref{Start of original page 191},
then the integral $I \rule{0pt}{2.55ex}^{
\hspace{-0.75ex} * }\left( B,i,n,j,s,E,v,\rho
\right) $, as defined on
page \pageref{Start of original page 191}, is finite and
\emph{absolutely} convergent, and is moreover bounded above
in magnitude, for all $\rho \in \mathbb{D} $, by a finite,
$\rho $-independent constant.

Let $C$ be any member of $\left( Y\,\vdash G\right) $, let
$\left(B,i,j\right) $ be any ordered
triple of a map $B$ such that
$\mathcal{D} \left( B\right) $ is finite and $\mathcal{R}
\left( B\right) \subseteq \left( Q\cap \left( \Xi \left(
\mathcal{P} \left( G,C\right) \right) \,\vdash \mathcal{P}
\left( G,C\right) \right) \right) $ holds, a map $i$ such
that $\mathcal{D} \left( B\right) \subseteq \mathcal{D}
\left( i\right) $ holds and for each member $\alpha $ of
$\mathcal{D} \left( B\right) $, $i_{ \alpha } $ is a
member of $B_{ \alpha } $, and a map $j$ such that
$\mathcal{D} \left( j\right) $ is finite and $\mathcal{R}
\left( j\right) \subseteq \left( \mathcal{U}
 \left( \tilde{ W }
_{ C } \right) \cup \left( C\,\vdash \mathcal{U} \left(
W\right) \right) \right) $ holds, such that for each member
$A$ of $\mathbb{B} \left( Q\right) \cap \left( \Xi \left(
\mathcal{P} \left( G,C\right) \right) \,\vdash \mathcal{P}
\left( G,C\right) \right) $, $B$ and $j$ satisfy the
condition on
page \pageref{Start of original page 164}
for $A$, and let $\rho $ be any
member of $\mathbb{D} $.

We first note that it follows directly from the facts that
$\mathcal{U} \left( V\right) $ is $\left( V\cup H\right)
$-connected and that if $E$ is any member of $H$ such that
$E$ intersects \emph{more} than one member of $V$, then $E$
has \emph{exactly} two members, that $C=\mathcal{U} \left(
\mathcal{P} \left( G,C\right) \right) $ is $\left(
\mathcal{P} \left( G,C\right) \cup H\right) $-connected and
that if $E$ is any member of $H$ such that $E$ intersects
\emph{more} than one member of $\mathcal{P} \left(
G,C\right) $, then $E$ has \emph{exactly} two members, so
that the partitions $\mathcal{P} \left( G,C\right) $ and
$H$ satisfy the relationships assumed to hold, in
Theorem \ref{Theorem 1}, between the partition $V$
of Theorem \ref{Theorem 1}, and $H$.

And we note furthermore that the partition
$\tilde{ W } _{ C } $, which was defined on
page \pageref{Start of original page 206}
 to be the subset of $W$ whose members are all
\label{Start of original page 210}
 the members $\left\{ r,s\right\} $ of $W$ such that
$\mathcal{Y} \left( \left( G\cup Y\right) ,\left\{
r,s\right\} \right) =C$ holds, is equal to the subset
 of $H$
whose members are all the members $E$ of $H$ such that $E$
intersects exactly two members of $\mathcal{P} \left(
G,C\right) $, for if $\left\{ r,s\right\} $ is
any member of
$\tilde{ W } _{ C } $ then $\left\{ r,s\right\} $
is a subset of
$C$ but is \emph{not} a subset of any member of
$\mathcal{P} \left( G,C\right) $, hence
$\left\{ r,s\right\}
$ intersects exactly two members of $\mathcal{P} \left(
G,C\right) $, and if $E$ is any member of $H$ such that $E$
intersects exactly two members of $\mathcal{P} \left(
G,C\right) $, then $E$ intersects more than one member of
$V$ hence $E$ has exactly two members, hence $E$ intersects
exactly two members of $V$ hence $E$ is a member of $W$,
and furthermore $E$ is a subset of $C$ and $\mathcal{Y}
\left( \left( G\cup Y\right) ,E\right) =C$ holds, hence $E$
is a member of $\tilde{ W }_{ C } $.   Hence the
relation between
the partition $\tilde{ W }_{ C }$ and the partitions
$\mathcal{P}
\left( G,C\right) $ and $H$, is the same as the relation
between the partition $W$ of Theorem \ref{Theorem 1}, and
the
partitions $V$ and $H$ of Theorem \ref{Theorem 1}.

We define $\tilde{ P }_{ C } \equiv \left( P\cap \Xi \left(
\mathcal{P} \left( G,C\right) \right) \right) $ and
$\tilde{ Q }_{ C
} \equiv \left( Q\cap \Xi \left( \mathcal{P} \left(
G,C\right) \right) \right) $, so that $\tilde{ P }_{ C } $
and
$\tilde{ Q }_{ C } $
are woods of $\mathcal{P} \left( G,C\right) $, and we now
proceed by exact analogy with steps in
\mbox{Theorem \ref{Theorem 1}.}

We first note that, if $\left\{ r,s\right\} $ is any member
of $\tilde{ W }_{ C } $, then by
page \pageref{Start of original page 80} and
Lemma \ref{Lemma 14}, and by
analogy with
page \pageref{Start of original page 96},
for all members $ a $
of $\tilde{ \mathbb{W} }_{
C }$ such that \\
$\mathcal{E} \left( \tilde{ P }_{ C }
,\tilde{ Q }_{ C }
,H,\sigma ,R,\downarrow \left( x,\Xi \left( \mathcal{P}
\left( G,C\right) \right) \right) \right) $ is not equal to
$0$,
\[
\left| \mu _{ r } \left( P,Q,H,x,\mathcal{X} \left(
P,Q,H,\rho
\right) \right) -\mu _{ s } \left( P,Q,H,x,\mathcal{X}
\left( P,Q,H,\rho \right) \right) \right|^{ -\alpha }
 \leq \hspace{-44.2pt} \hspace{5.0cm}
\]
\[
\hspace{9.0cm} \hspace{-13.1pt} \leq \left( \frac{ 1 }{
1-2\lambda } \right)^{ \left| \alpha \right| } \left|
x_{ \mathcal{Z} \left(
P,H,r\right) } -x_{ \mathcal{Z} \left( P,H,s\right) }
\right|^{
-\alpha }
\]
holds for all $\rho \in \mathbb{D} $ and for all finite
real numbers $\alpha $, and we also note that $\mathcal{Z}
\left( P,H,r\right) $ is equal to $\mathcal{Z}
\left( \tilde{ P }
_{ C } ,H,r\right) $ and that $\mathcal{Z} \left(
P,H,s\right) $ is equal to $\mathcal{Z} \left( \tilde{ P
}_{ C }
,H,s\right) $, and we also note that, since both
$\mathcal{Z} \left( P,H,r\right) $ and $\mathcal{Z} \left(
P,H,s\right) $ are members of $\mathcal{P} \left(
P,A\right) =\mathcal{P} \left( \tilde{ P }_{ C } ,A\right)
$ for
some member $A$ of $\mathbb{B} \left( \overline{
 \tilde{ P }_{ C } }
\right) $,
$\left( x_{ \mathcal{Z} \left( P,H,r\right) } -x_{
\mathcal{Z}
\left( P,H,s\right) } \right) =\left( a_{ \mathcal{Z} \left(
P,H,r\right) } -a_{ \mathcal{Z} \left( P,H,s\right) }
 \right) $
holds.   (This last observation is true even when $C$ is a
$\left( V\cup H\right) $-firm over $V$ component of a
member $D$ of $\mathbb{B} \left( G\right) $, and $\left\{
r,s\right\} $ is a member of $\tilde{ W }_{ C }$ such that
$\mathcal{Y}
\left( P,\left\{ r,s\right\} \right) =D$ holds, since in
that
case, by the definition of the wood $J$ on
page \pageref{Start of original page 165}, the
smallest member of $J$ to contain $\mathcal{Z} \left(
P,H,r\right) $ as a strict subset is equal to the smallest
member of $J$ to contain $\mathcal{Z} \left( P,H,s\right) $
as a strict subset.)

We next use the above inequality, by analogy with
pages \pageref{Start of original page 96}
and \pageref{Start of original page 97},
to bound the integrand of $ \hspace{0.5ex} {
\rule{0pt}{2.8ex} }^{ \circ } \hspace{-1.4ex}
\bar{I}_{ C } \left(
B,i,j,\rho \right) $ by a product of finite real numbers,
independent of $\rho $ and the $a_{ K } $, times a new
integrand in which all occurrences of \\
$\left| \mu_{ r } \left(
P,Q,H,x,\mathcal{X} \left( P,Q,H,\rho \right) \right)
-\mu_{
s } \left( P,Q,H,x,\mathcal{X} \left( P,Q,H,\rho \right)
\right) \right| $
\label{Start of original page 211}
 for any member $\left\{ r,s\right\} $ of $\tilde{ W }_{ C
}$ have been
replaced by the corresponding $\left| a_{ \mathcal{Z} \left(
P,H,r\right) } -a_{ \mathcal{Z}
\left( P,H,s\right) } \right| $,
(together with a factor $\left( 1-2\lambda \right) $ inside
the step function $\mathbb{S} $, as on
page \pageref{Start of original page 97}).

We then proceed, by analogy with
pages \pageref{Start of original page 97} and
\pageref{Start of original page 98}, to
bound our new integrand by the sum of a finite number of
terms, in each of which, for every member $\alpha $ of
$\mathcal{D} \left( B\right) $, the factor $\left|
x_{ \mathcal{K}
\left( Q,B_{ \alpha } ,i_{ \alpha } \right) } -x_{ B_{
\alpha } } \right| $ associated with the member $\alpha $ of
$\mathcal{D} \left( B\right) $, has been replaced by some
choice of a factor $\left| x_{ K } -x_{ L } \right| =\left|
a_{ K } -a_{ L }
\right| $, where both $K$ and $L$ are members of
$\mathcal{P}
\left( P,B_{ \alpha } \right) =\mathcal{P} \left( \tilde{ P
}_{ C }
,B_{ \alpha } \right) $.

We next consider an arbitrarily chosen such term, and
define, by analogy with
page \pageref{Start of original page 99},
$\alpha $ to be the map
whose domain is $ \displaystyle \bigcup_{A\in \mathbb{B}
\left( \overline{ \tilde{ P }_{
C } } \right)
  } \mathcal{Q} \left( \mathcal{P} \left(
\tilde{ P }_{ C
} ,A\right) \right) = \displaystyle \bigcup_{A\in
\mathbb{B} \left(
\overline{ \tilde{ P
}_{ C } }
\right)  } \mathcal{Q} \left( \mathcal{P}
\left( P,A\right) \right) $, and such that for each member
$\left\{ K,L\right\} $ of $\mathcal{D} \left( \alpha \right)
$, $\alpha_{ \left\{ K,L\right\} } $
is equal to the negative of
the total power of $\left| x_{ K } -x_{ L } \right| =\left|
a_{ K } -a_{ L }
\right| $ in the integrand for this term, excluding the
$\mathcal{E} $ function and the step-function factors.
Then for each member $A$ of $\mathbb{B} \left(
\overline{ \tilde{ P }_{ C } }
\right) $, $\alpha $ is a set of powers for $A$, and, as on
page \pageref{Start of original page 100},
the integrand for this term is equal to a
finite real number, independent of $\rho $ and the $a_{ K }
$, times the $\mathcal{E} $ function, times the
step-function factors, times the product, over the members
$A$ of $\mathbb{B} \left( \overline{
 \tilde{ P }_{ C } } \right) $, of
the factor
$\Psi \left( \downarrow \left( x,\mathcal{P} \left( \tilde{
P }_{ C
} ,A\right) \right) ,\alpha \right) =\Psi \left( \downarrow
\left( x,\mathcal{P} \left( P,A\right) \right) ,\alpha
\right) $, where we note that in this factor, every
interval $\left| x_{ K } -x_{ L } \right| $, $\left\{
K,L\right\} \in
\mathcal{P} \left( P,A\right) $, is equal to the
corresponding interval $\left| a_{ K } -a_{ L } \right| $.
 (We may
write the factor $\Psi \left( \downarrow \left(
x,\mathcal{P} \left( P,A\right) \right) ,\alpha \right) $
as $\Psi \left( \downarrow \left( a,\mathcal{P} \left(
P,A\right) \right) ,\alpha \right) $, if we note that for
the particular member $S_{ A } $ of $\mathcal{P} \left(
P,A\right) $, $a_{ S_{ A } } $ is identically equal to
$0 $.)

We next note, by the condition on
page \pageref{Start of original page 164} for the
members $A$ of \\
$\mathbb{B} \left( Q\right) \cap \left( \Xi
\left( \mathcal{P} \left( G,C\right) \right) \,\vdash
\mathcal{P} \left( G,C\right) \right) $, and by analogy
with pages \pageref{Start of original page 100} and
\pageref{Start of original page 101},
and taking into account that
for each member $A$ of $\left( \Xi \left( V\right) \,\vdash
V\right) $, the present definition of $D_{ A } $, given on
page \pageref{Start of original page 158},
differs from the definition of $D_{ A } $ in
Theorem \ref{Theorem 1} by the inclusion, in the
contribution
associated with the member $\left\{ r,s\right\} $ of $\left(
W\cap \mathcal{Q} \left( A\right) \right) $, of the term
$\left( \nu _{ gr } +\nu _{ gs } \right) $ that takes into
account the differential operator $\left( \prod_{\alpha \in
\mathcal{D}
\left( g\right) } c_{ \alpha } .\hat{ y }_{ g_{
\alpha } } \right) $
in the integrand of the integral on
page \pageref{Start of original page 159}, where $g$
is the map introduced on
page \pageref{Start of original page 157},
that if $A$ is any
member of $\left( \tilde{ Q }_{ C } \,\vdash \tilde{ P }_{
C } \right) $,
\label{Start of original page 212}
 then in exact correspondence with
 observation \ref{Observation 19})
on pages \pageref{Start of original page 109} and
\pageref{Start of original page 110},
the following inequality holds
for $\Gamma \left( \alpha ,\mathcal{P} \left( \tilde{ P }_{
C }
,A\right) \right) =\Gamma \left( \alpha ,\mathcal{P} \left(
P,A\right) \right) $:
\[
\Gamma \left( \alpha ,\mathcal{P} \left( P,A\right) \right)
\leq d\left( \#\left( \mathcal{P} \left( P,A\right) \right)
-1\right) -1- \sum_{C\in \left( \mathcal{P} \left(
P,A\right)
\,\vdash V\right) } \left( D_{ C }
- \sum_{\left( i,B\right)
\in \mathbb{I} _{ C } \left( Q,H\right) }
u_{ iB } \right) ,
\]
and if $A$ is any member of $\mathbb{B} \left( \tilde{ P
}_{ C }
\right) $, then in exact correspondence with
observation \ref{Observation 20}) on
page \pageref{Start of original page 110}, the
following inequality holds
for $\Gamma \left( \alpha ,\mathcal{P} \left( \tilde{ P }_{
C }
,A\right) \right) =\Gamma \left( \alpha ,\mathcal{P} \left(
P,A\right) \right) $:
\[
\Gamma \left( \alpha ,\mathcal{P} \left( P,A\right) \right)
\leq \left(
 d\left( \#\left( \mathcal{P} \left( P,A\right) \right)
-1\right) +D_{ A } - \left(
 \sum_{C\in \left( \mathcal{P} \left(
P,A\right) \,\vdash V\right) } D_{ C } \right) + \right.
\hspace{-60.6pt} \hspace{5.0cm}
\]
\[
\hspace{4.0cm} \hspace{-86.9pt} \left.
+ \left(
 \sum_{C\in \left( \mathcal{P} \left( P,A\right) \,\vdash
V\right) } \hspace{0.2cm}
 \sum_{\left( i,B\right) \in \left(
\mathbb{I} _{ C } \left( Q,H\right) \,\vdash \mathbb{I} _{
A } \left( Q,H\right) \right) } u_{ iB } \right)
 - \left( \sum_{i\in
\mathcal{T} \left( A,H\right) } \hspace{0.2cm}
 \sum_{B\in \mathbb{Y}
\left( Q,\mathcal{K} \left( P,A,i\right) ,A\right) }
u_{ iB } \right) \right) .
\]

We next note, by Lemma \ref{Lemma 10}, and by analogy with
page \pageref{Start of original page 111},
that if $a $ is any member of
$\tilde{ \mathbb{W} }_{ C }$ such
that $\mathcal{E} \left( \tilde{ P }_{ C } ,\tilde{ Q }_{ C
} ,H,\sigma
,R,\downarrow \left( x,\Xi \left( \mathcal{P} \left(
G,C\right) \right) \right) \right) $ is not equal to $0$,
and $A$ is any member of $\mathbb{B} \left( \tilde{ P }_{ C
}
\right) $, then $\mathbb{M} \left( \left( \tilde{ P }_{ C }
\,\vdash
\left\{ A\right\} \right) ,\tilde{ P }_{ C } ,
H,A,\sigma ,R,x\right) $
does \emph{not} hold, which implies, by analogy with
pages \pageref{Start of original page 111}
to \pageref{Start of original page 113},
and noting that $\mathbb{L} \left( \tilde{
P }_{ C }
,A,x\right) $ is equal to $\mathbb{L} \left( P,A,x\right)
$, that either $\mathbb{L} \left( P,A,x\right) \geq R$
holds, or else there exists a member $D$ of $\mathcal{P}
\left( P,E\right) $, where $E$ is the smallest member $E$
of $ \overline{ \tilde{ P }_{ C } } $ to contain
$A$ as a \emph{strict}
subset,
such that $D\neq A$ holds, $\mathbb{L} \left( P,A,x\right)
\geq \sigma \left| x_{ A } -x_{ D } \right| $ holds, and
there exists a
member $T$ of $\tilde{ W }_{ C }$ such that $T$ intersects
both $A$
and $D$, (and we note that if $A$ is equal to $C$, then
\emph{no} member of $ \overline{ \tilde{ P }_{ C } } $
contains $A$ as a
\emph{strict} subset, hence $\mathbb{L} \left( P,A,x\right)
\geq R$ holds if $A$ is equal to $ C $).

We now define $r$, \emph{exactly} as on
page \pageref{Start of original page 112}, to be
the member of $\mathbb{R}^{ \mathbb{B} \left( P\right) } $
such that for each member $A$ of $\mathbb{B} \left(
P\right) $,
\[
r_{ A } \equiv \left(
 D_{ A } - \sum_{\left( i,B\right) \in
\mathbb{I} _{
A } \left( Q,H\right) } u_{ iB } \right)
 + \frac{\#\left(
\mathbb{B} \left( P\right) \cap \Xi \left( \mathcal{P}
\left( V,A\right) \right) \right)}{\#\left( \mathbb{B}
\left( \bar{ P } \right) \right)}   ,
\]
where we note that $\mathbb{B} \left( P\right) \cap \Xi
\left( \mathcal{P} \left( V,A\right) \right) $ is the set
of all the members $B$ of $\mathbb{B} \left( P\right) $
such that $B\subseteq A$ holds, and we note that, as
observed on
page \pageref{Start of original page 112},
$r_{ A } \geq 0$ holds for every
member $A$ of $\mathbb{B} \left( P\right) $.

We now bound our present integrand, by analogy with
pages \pageref{Start of original page 111}
to \pageref{Start of original page 114},
by a sum of a finite number of terms, each
of which has the form of our present integrand, times the
product, over the members
\label{Start of original page 213}
 $A$ of $\mathbb{B} \left( \tilde{ P }_{ C } \right) $, of
a factor
$\left( \frac{ n_{ A } }{ d_{ A } } \right)^{ r_{ A } } $,
 where $n_{ A } $
is equal to $\left| x_{ K } -x_{ L } \right| =\left| a_{ K
} -a_{ L } \right| $ for
some choice of a member $\left\{ K,L\right\} $ of
$\mathcal{P} \left( \tilde{ P }_{ C } ,A\right)
=\mathcal{P} \left(
P,A\right) $, and $d_{ A } $ is either equal to $R$ or else
is equal to $\sigma \left| x_{ A } -x_{ D } \right| =\sigma
\left| a_{ A }
-a_{ D } \right| $ for some choice of a member $D$ of
$\mathcal{P}
\left( \tilde{ P }_{ C } ,E\right) =\mathcal{P} \left(
P,E\right) $
such that $D\neq A$ holds and there exists a member $T$ of
$\tilde{ W }_{ C }$ such that $T$ intersects both $A$ and
$D$, where
$E$ is the smallest member of $ \overline{
\tilde{ P }_{ C } } $ to
contain $A$
as a \emph{strict} subset, (and if $A=C$ holds then $d_{ A
} $ is equal to $ R $).

We next consider an arbitrarily chosen one of these terms,
and define, by analogy with
page \pageref{Start of original page 113},
$\beta $ to be the
map whose domain is equal to $\mathcal{D} \left( \alpha
\right) = \hspace{-6.0pt}
 \displaystyle \bigcup_{A\in \mathbb{B} \left(
\overline{ \tilde{ P }_{ C } } \right)
 } \hspace{-6.0pt}
  \mathcal{Q} \left( \mathcal{P} \left( \tilde{ P
}_{ C }
,A\right) \right) $, and such that for each member $\left\{
K,L\right\} $ of $\mathcal{D} \left( \beta \right)
=\mathcal{D} \left( \alpha \right) $, $\beta_{ \left\{ K,L
\right\} } $ is equal to
$\alpha_{ \left\{ K,L\right\} } $ plus
the negative of the total power of $\left| x_{ K } -x_{ L }
\right| =\left| a_{ K } -a_{ L } \right| $ that occurs in
the \emph{new} factors in our chosen term.  Then just as on
pages \pageref{Start of original page 113}
and \pageref{Start of original page 114},
the integrand of our chosen term is equal to a
finite real number, independent of $\rho $ and the $a_{ K }
$, times the $\mathcal{E} $ function, times the
step-function factors, times the product, over the members
$A$ of $\mathbb{B} \left( \overline{
 \tilde{ P }_{ C } } \right) $, of
the factor
$\Psi \left( \downarrow \left( a,\mathcal{P} \left(
P,A\right) \right) ,\beta \right) $, where we note, as on
page \pageref{Start of original page 211},
that for the particular member $S_{ A } $ of
$\mathcal{P} \left( P,A\right) $, $a_{ S_{ A } } $ is
identically equal to $0$.

We next note that if $A$ is any member of $\left( \Xi
\left( \mathcal{P} \left( G,C\right) \right) \,\vdash
\mathcal{P} \left( G,C\right) \right) $, hence in
particular if $A$ is any member of $\mathbb{B}
\left( \tilde{ Q }
_{ C } \right) =\left( \tilde{ Q }_{ C } \,\vdash
\mathcal{P} \left(
G,C\right) \right) $, then $\mathcal{P} \left( P,A\right)
\cap V$ is a subset of $\mathcal{P} \left( P,A\right) \cap
\mathcal{P} \left( G,C\right) $, for if $D$ is any member
of $\mathcal{P} \left( P,A\right) \cap V$, then $D$ is a
strict subset of $A$ hence $D$ is a strict subset of $C$,
and $D$ is a member of $V$ hence $D$ is a member of $G$,
hence by
page \pageref{Start of original page 5},
$D$ is a subset of a unique member of
$\mathcal{P} \left( G,C\right) $.   Let $E$ be the unique
member of $\mathcal{P} \left( G,C\right) $ such that
$D\subseteq E$ holds.   Then since $A$ is a member of
$\left( \Xi \left( \mathcal{P} \left( G,C\right) \right)
\,\vdash \mathcal{P} \left( G,C\right) \right) $, either
$E\cap A=\emptyset $ holds or else $E\subset A$ holds, and
$E\cap A$ has the nonempty subset $D$, hence $E\subset A$
holds.   Hence, since $D\subset E$ and $E\subset A$ cannot
\emph{both} hold, (since $E$ is a member of $G$ hence $E$
is a member of $P$, and $D$ is a member of $\mathcal{P}
\left( P,A\right) $), $D=E$ holds, hence $D$ is a
member of $\mathcal{P} \left( G,C\right) $, hence $D$ is a
member of $\mathcal{P} \left( P,A\right) \cap \mathcal{P}
\left( G,C\right) $.

And we note furthermore that it follows immediately from
the
\label{Start of original page 214}
 preceding paragraph, that if $A$ is any member of $\left(
\Xi \left( \mathcal{P} \left( G,C\right) \right) \,\vdash
\mathcal{P} \left( G,C\right) \right) $, hence in
particular, if $A$ is any member of $\mathbb{B}
\left( \tilde{ Q }
_{ C } \right) $, then $\left( \mathcal{P} \left( P,A\right)
\,\vdash \mathcal{P} \left( G,C\right) \right) $ is a
subset of $\left( \mathcal{P} \left( P,A\right) \,\vdash
V\right) $.

We next note that if $A$ is any member of $\mathbb{B}
\left( \tilde{ P }_{ C } \right) $, then by analogy with
observations \ref{Observation 22})
to \ref{Observation 24}) on
page \pageref{Start of original page 114},
the contribution to $\Gamma \left( \beta ,\mathcal{P}
\left( \tilde{ P }_{ C } ,A\right) \right) $ from the
\emph{new}
factors in our chosen term is $\leq \left(
\displaystyle \sum_{D\in \left(
\mathcal{P} \left( P,A\right) \,\vdash \mathcal{P} \left(
G,C\right) \right) } r_{ D } \right) -r_{ A } $, hence
since, as noted on
page \pageref{Start of original page 212},
$r_{ D } \geq 0$ holds for
\emph{every} member $D$ of $\mathbb{B} \left( P\right) $,
and, as just shown, $\left( \mathcal{P} \left( P,A\right)
\,\vdash \mathcal{P} \left( G,C\right) \right) $ is a
subset of $\left( \mathcal{P} \left( P,A\right) \,\vdash
V\right) $, the contribution to $\Gamma \left( \beta
,\mathcal{P} \left( \tilde{ P }_{ C } ,A\right) \right) $
from the
\emph{new} factors in our chosen term is $\leq
\left( \displaystyle \sum_{D\in
\left( \mathcal{P} \left( P,A\right) \,\vdash V\right) }
r_{ D } \right) -r_{ A } $, exactly as in
observation \ref{Observation 24}) on
page \pageref{Start of original page 114}.  Hence
by the foregoing, and by
observations \ref{Observation 24}) and
\ref{Observation 25}) on
pages \pageref{Start of original page 114}
and \pageref{Start of original page 115},
if $A$ is any member of $\mathbb{B} \left( \tilde{ P }
_{ C } \right) $, then since $\mathcal{P} \left( \tilde{ P
}_{ C }
,A\right) =\mathcal{P} \left( P,A\right) $ holds, the
following inequality holds:
\[
\Gamma \left( \beta ,\mathcal{P} \left( \tilde{ P }_{ C }
,A\right)
\right) \leq d\left( \#\left( \mathcal{P} \left( \tilde{ P
}_{ C }
,A\right) \right) -1\right) - \frac{ 1 }{ \#\left(
\mathbb{B} \left( \bar{ P }
\right) \right) } .
\]

And we note furthermore that if $A$ is any member of
$\left( \tilde{ Q }_{ C } \,\vdash \tilde{ P }_{ C }
\right) $, then by
analogy with observation \ref{Observation 23}) on
page \pageref{Start of original page 114} and
observation \ref{Observation 26}) on
page \pageref{Start of original page 115}, the
contribution to
$\Gamma \left( \beta ,\mathcal{P} \left( \tilde{ P }_{ C }
,A\right)
\right) $ from the \emph{new} factors in our chosen term is
$\leq \left( \displaystyle \sum_{D\in \left(
\mathcal{P} \left( P,A\right)
\,\vdash
\mathcal{P} \left( G,C\right) \right) } r_{ D } \right)
$, hence again since $r_{ D } \geq 0$ holds for \emph{all}
members $D$ of $\mathbb{B} \left( P\right) $, and $\left(
\mathcal{P} \left( P,A\right) \,\vdash \mathcal{P} \left(
G,C\right) \right) $ is a subset of $\left( \mathcal{P}
\left( P,A\right) \,\vdash V\right) $, the contribution to
$\Gamma \left( \beta ,\mathcal{P} \left( \tilde{ P }_{ C }
,A\right)
\right) $ from the \emph{new} factors in our chosen term is
$\leq \left( \displaystyle \sum_{D\in \left(
\mathcal{P} \left( P,A\right)
\,\vdash
V\right) } r_{ D } \right) $, exactly as in
observation \ref{Observation 26}) on
page \pageref{Start of original page 115}.   Hence
by the foregoing, and
by observations \ref{Observation 26})
and \ref{Observation 27}) on
pages \pageref{Start of original page 115}
and \pageref{Start of original page 116},
if $A$ is any member of $\left( \tilde{ Q }_{ C }
\,\vdash \tilde{ P }_{ C } \right) $, then since
$\mathcal{P} \left(
\tilde{ P }_{ C } ,A\right) =\mathcal{P} \left( P,A\right)
$ holds,
the following inequality holds for $\Gamma \left( \beta
,\mathcal{P} \left( \tilde{ P }_{ C } ,A\right) \right) $:
\[
\Gamma \left( \beta ,\mathcal{P} \left( \tilde{ P }_{ C }
,A\right)
\right) \leq d\left( \#\left( \mathcal{P} \left( \tilde{ P
}_{ C }
,A\right) \right) -1\right) - \frac{ 1 }{ \#\left(
\mathbb{B} \left( \bar{ P }
\right) \right) } .
\]

Hence if $A$ is \emph{any} member of $\mathbb{B}
\left( \tilde{ Q }
_{ C } \right) $, then exactly as in
observation \ref{Observation 28}) on
page \pageref{Start of original page 116},
the following inequality holds for $\Gamma
\left( \beta ,\mathcal{P} \left( \tilde{ P }_{ C }
,A\right) \right) $:
\label{Start of original page 215}
\[
\Gamma \left( \beta ,\mathcal{P} \left( \tilde{ P }_{ C }
,A\right) \right) \leq d\left( \#\left( \mathcal{P}
\left( \tilde{ P
}_{ C } ,A\right) \right) -1\right) - \frac{ 1 }{ \#\left(
\mathbb{B}
\left( \bar{ P } \right) \right) } .
\]

We may now complete the proof of the absolute convergence
of ${ \rule{0pt}{2.8ex} }^{ \circ } \hspace{-1.4ex} \bar{I}
_{ C } \left( B,i,j,\rho \right) $, and a $\rho
$-independent bound on the magnitude of ${
\rule{0pt}{2.8ex} }^{ \circ } \hspace{-1.4ex} \bar{I} _{ C }
\left( B,i,j,\rho \right) $, by exact analogy with the
remainder of the proof of Theorem \ref{Theorem 1}, taking
$V$, $W$,
$P$, and $Q$ of Theorem \ref{Theorem 1} respectively as our
present
$\mathcal{P} \left( G,C\right) $, $ \tilde { W }_{ C } $,
 $\tilde{ P }_{ C } $,
and $\tilde{ Q }_{ C }$ respectively, with the only
changes being
firstly the fact that we have $- \displaystyle \frac{
1 }{ \#\left(
\mathbb{B} \left( \bar{ P
} \right) \right) } $ rather than $- \displaystyle \frac{
1 }{
\#\left( \mathbb{B}
\left( \overline{ \tilde{ P }_{ C } } \right) \right) } $
in the bound just
obtained on
$\Gamma \left( \beta ,\mathcal{P} \left(
 \tilde{ P }_{ C } ,A\right)
\right) $, and this requires \emph{no} changes to the
procedure of proof, and secondly that since, by
page \pageref{Start of original page 206},
we are already using the $a_{ K } $ variables, $K\in \left(
\left( P\cap \left( \Xi \left( \mathcal{P} \left(
G,C\right) \right) \,\vdash \left\{ C\right\} \right)
\right)
\,\vdash \left( \mathcal{R} \left( S\right) \cup \left\{
\mathcal{U} \left( V\right) \right\} \right) \right) $, or
in other words, $K\in \left( \tilde{ P }_{ C } \,\vdash
\left(
\mathcal{R} \left( S\right) \cup \left\{ C\right\} \right)
\right) $, there is no need to make any change of variables
corresponding to the change to the $z_{ K } $ variables on
pages \pageref{Start of original page 117} to
\pageref{Start of original page 119}.

We next note that if $C$ is any member of $\left( Y\,\vdash
G\right) $, then since, by
pages \pageref{Start of original page 176} to
\pageref{Start of original page 179}, when the
$x_{ L } $ variables are expressed in terms of the $a_{ D }
$ variables, their only $\rho $-dependence is through the
coefficients $\nu _{ KD } $, and since the factors \\
$\mathcal{E} \left( \tilde{ P }_{ C } ,\tilde{ Q }_{ C }
,H,\sigma
,R,\downarrow \left( x,\Xi \left( \mathcal{P} \left(
G,C\right) \right) \right) \right) $ and $\left(
x_{ \mathcal{K} \left( Q,N,l\right) }
 -x_{ N } \right) $, where
$N$ is any member of $\left( Q\cap \left( \Xi \left(
\mathcal{P} \left( G,C\right) \right) \,\vdash \mathcal{P}
\left( G,C\right) \right) \right) $ and $l$ is any member
of $N$, depend on the $x_{ L } $ variables only through
linear combinations, each with the sum of all its
coefficients being equal to $0$, of the $x_{ L } $, $L\in
\mathcal{P} \left( G,C\right) $, and since, again by
pages \pageref{Start of original page 176}
to \pageref{Start of original page 179},
when any $x_{ L } $ variable, $L\in
\mathcal{P} \left( G,C\right) $, is expressed in terms of
the $a_{ D } $ variables, the only terms which occur that
depend on any of the $\nu _{ KD } $ coefficients, are terms
that occur identically in \emph{every} $x_{ L } $, $L\in
\mathcal{P} \left( G,C\right) $, these factors,
$\mathcal{E} \left( \tilde{ P }_{ C } ,\tilde{ Q }_{ C }
,H,\sigma
,R,\downarrow \left( x,\Xi \left( \mathcal{P} \left(
G,C\right) \right) \right) \right) $ and $\left(
x_{ \mathcal{K} \left( Q,N,l\right) }
 -x_{ N } \right) $, where
$N$ is any member of $\left( Q\cap \left( \Xi \left(
\mathcal{P} \left( G,C\right) \right) \,\vdash \mathcal{P}
\left( G,C\right) \right) \right) $ and $l$ is any member
of $N$, are \emph{completely independent of }$ \rho $
when they
are expressed in terms of the $a_{ D } $ variables.   Hence
if $\left(B,i,j\right) $ is any
ordered \emph{triple} as on
page \pageref{Start of original page 209},
and satisfying the conditions, with reference to $C$,
specified for the ordered \emph{triple}
$\left(B,i,j\right) $ on
page \pageref{Start of original page 209},
then the $\rho $-independence of our bound on
$ \hspace{0.5ex}
{ \rule{0pt}{2.8ex} }^{ \circ } \hspace{-1.4ex} \bar{I} _{
C } \left( B,i,j,\rho \right) $ was in fact
attained immediately we had carried out the step, on
pages \pageref{Start of original page 210}
and \pageref{Start of original page 211},
of replacing all occurrences of
\label{Start of original page 216}
 $\left| \mu _{ r } \left( P,Q,H,x,\mathcal{X} \left(
P,Q,H,\rho
\right) \right) -\mu _{ s } \left( P,Q,H,x,\mathcal{X}
\left( P,Q,H,\rho \right) \right) \right| $ by the
corresponding
$\left| a_{ \mathcal{Z} \left( P,H,r\right) }
 -a_{ \mathcal{Z}
\left(
P,H,s\right) } \right| $ for each member $\left\{
r,s\right\}
$
of $\tilde{ W
}_{ C } $.

This completes the first of the two steps mentioned on
page \pageref{Start of original page 191},
namely the proof that if
$\left(B,i,n,j,s,E,v\right) $ is any
ordered septuple as on
page \pageref{Start of original page 191},
and satisfying the conditions specified on
page \pageref{Start of original page 191},
then the integral
$I \rule{0pt}{2.55ex}^{ \hspace{-0.75ex} * }\left(
B,i,n,j,s,E,v,\rho \right) $ defined on
page \pageref{Start of original page 191},
is finite and \emph{absolutely} convergent, and is
moreover bounded in magnitude, for all $\rho \in \mathbb{D}
$, by a finite, $\rho $-independent constant.

Finally we shall carry out the second of the two steps
mentioned on
page \pageref{Start of original page 191},
by showing that the result just
obtained, together with our assumptions on the map
$\tilde{ \mathcal{J} } $ and its relation to the map
$\mathcal{J}
$, imply that if $\left(B,i,n,j,s,E,v\right) $ is
any ordered septuple as on
page \pageref{Start of original page 191},
and satisfying the conditions specified on
page \pageref{Start of original page 191},
then the integrals with respect to $\rho $, over
$\mathbb{D} $, of any $ k $-independent product of finite
powers, all $\geq 0$ and independent of $k$ and $\rho $, of
the $\rho _{ K } $, $K\in \left( Q\,\vdash P\right) $,
times the $I \rule{0pt}{2.55ex}^{ \hspace{-0.75ex} \circ }
_{ k } \left( B,i,n,j,s,E,v,\rho
\right) $, $k\in \mathbb{N} $, form a Cauchy sequence.

Let $\left(B,i,n,j,s,E,v\right) $ be any
ordered septuple as on
page \pageref{Start of original page 191},
and satisfying the conditions specified on
page \pageref{Start of original page 191},
and let $\rho $ be any member of $\mathbb{D} $.

Now for each member $k$ of $\mathbb{N} $, $I
\rule{0pt}{2.55ex}^{ \hspace{-0.75ex} \circ }_{ k }
\left( B,i,n,j,s,E,v,\rho \right) $, as defined on
page \pageref{Start of original page 163},
is obtained from $I \rule{0pt}{2.55ex}^{
\hspace{-0.75ex} * }\left( B,i,n,j,s,E,v,\rho
\right) $, as defined on
page \pageref{Start of original page 191}, by replacing the
$\mathcal{J} _{ Q } \left( y\right) $ that occurs in the
integrand of $I \rule{0pt}{2.55ex}^{ \hspace{-0.75ex} *
}\left( B,i,n,j,s,E,v,\rho \right) $, by
$\tilde{ \mathcal{J} } _{ kQ } \left( y\right) $.   Hence
since, by assumption, and for every member $k$ of
$\mathbb{N} $, $\tilde{ \mathcal{J} } _{ kQ }
 \left( y\right) $
satisfies all the assumptions made for $\mathcal{J} _{ Q }
\left( y\right) $, it immediately follows from our proof of
the finiteness and \emph{absolute} convergence of $I
\rule{0pt}{2.55ex}^{ \hspace{-0.75ex} * }\left(
B,i,n,j,s,E,v,\rho \right) $, that for every member $k$ of
$\mathbb{N} $, $I \rule{0pt}{2.55ex}^{ \hspace{-0.75ex}
\circ }_{ k } \left( B,i,n,j,s,E,v,\rho
\right) $ is finite and \emph{absolutely} convergent, and
moreover satisfies the \emph{same} $\rho $-independent
bound on its magnitude that we obtained for $I
\rule{0pt}{2.55ex}^{ \hspace{-0.75ex} * }\left(
B,i,n,j,s,E,v,\rho \right) $.

We shall next show that if $\varepsilon $ is any given real
number $>0$, then there exists a member $k$ of $\mathbb{N}
$, \emph{independent of }$\rho $, such that for all members
$l$
of $\mathbb{N} $ such that $l\geq k$ holds, and for all
members $\rho $ of $\mathbb{D} $, $\left| I
\rule{0pt}{2.55ex}^{ \hspace{-0.75ex} \circ }_{ l }
\left(
B,i,n,j,s,E,v,\rho \right) -I \rule{0pt}{2.55ex}^{
\hspace{-0.75ex} * }\left( B,i,n,j,s,E,v,\rho
\right) \right| \leq \varepsilon $ holds.

We first note that, by assumption, for any given real
number $\delta >0$ and for any given real number $r>0$,
there exists a member $k$ of $\mathbb{N} $ such that for
all members $l$ of $\mathbb{N} $ such that $l\geq k$ holds,
and
\label{Start of original page 217}
 for all members $y$ of $\mathbb{E}_{ d }^{ \mathcal{U}
 \left(
V\right) } $ such that $\left| y_{ e } -y_{ f } \right|
\geq r$ holds
for all members $\left\{ e,f\right\} $ of $W$, and for all
maps $t$ and $u$ such that $\mathcal{D} \left( t\right) $
is finite, $\#\left( \mathcal{D} \left( t\right) \right)
\leq N$ holds, (where $N$ is the integer defined on
page \pageref{Start of original page 158}),
$\mathcal{R} \left( t\right) \subseteq
\mathcal{U} \left( V\right) $ holds, $\mathcal{D} \left(
t\right) \subseteq \mathcal{D} \left( u\right) $ holds, and
for each member $\alpha $ of $\mathcal{D} \left( t\right)
$, $u_{ \alpha } $ is a unit $d $-vector, $\left| \left(
\prod_{\alpha \in \mathcal{D} \left( t\right) }
u_{ \alpha } .\hat{ y }_{ t_{ \alpha } } \right)
 \left( \tilde{ \mathcal{J} } _{ lQ } \left( y\right)
-\mathcal{J} _{ Q } \left( y\right) \right) \right| \leq
\delta $ holds.

We next note, by analogy with
pages \pageref{Start of original page 116} and
\pageref{Start of original page 117}, and
pages \pageref{Start of original page 202} to
\pageref{Second line of original page 205},
that the total $d\left( \#\left(
V\right) -1\right) $-volume of the subset of $\mathbb{W}
$ where our integrands are not \emph{forced} to be equal to
$0$ by our assumption that $\mathcal{J} _{ Q } \left(
y\right) $, (and consequently also all the
$\tilde{ \mathcal{J} }
_{ kQ } \left( y\right) $), vanishes whenever
$\left| y_{ e } -y_{ f } \right| \geq T$ holds for any
member $\left\{
e,f\right\} $ of $W$, is \emph{finite,} and is moreover
bounded above by a finite real number, independent of $\rho
$, times $T^{ d\left( \#\left( V\right) -1\right) } $.

And we observe furthermore that if $\rho $ is any member of
$\mathbb{D} $, then by the preceding paragraph and by
analogy with
page \pageref{Start of original page 90},
and by the facts that for any
member $\left\{ e,f\right\} $ of $W$, $\mathcal{Z} \left(
Q,H,e\right) \cap \mathcal{Z} \left( Q,H,f\right)
=\emptyset $
holds, and $\mu _{ e } \left( P,Q,H,x,\mathcal{X} \left(
P,Q,H,\rho \right) \right) $ is equal to a linear
combination, with coefficients summing to $1$, of $x_{ K }
$ for the members $K$ of $P$ such that $K\subseteq
\mathcal{Z} \left( Q,H,e\right) $ holds and there is
\emph{no} member $M$ of $P$ such that $K\subset M\subseteq
\mathcal{Z} \left( Q,H,e\right) $ holds, and $\mu _{ f }
\left( P,Q,H,x,\mathcal{X} \left( P,Q,H,\rho \right)
\right) $ is equal to a linear combination, with
coefficients summing to $1$, of $x_{ L } $ for the members
$L$ of $P$ such that $L\subseteq \mathcal{Z} \left(
Q,H,f\right) $ holds and there is \emph{no} member $M$ of
$P$ such that $L\subset M\subseteq \mathcal{Z} \left(
Q,H,f\right) $ holds, (so that there is \emph{no} member
$U$ of $P$ such that $x_{ U } $ occurs with nonzero
coefficient in \emph{both} $\mu _{ e } \left(
P,Q,H,x,\mathcal{X} \left( P,Q,H,\rho \right) \right) $
\emph{and} $\mu _{ f } \left( P,Q,H,x,\mathcal{X} \left(
P,Q,H,\rho \right) \right) $), the subset of
$\mathbb{W}$ whose members are all the members $x$ of
$\mathbb{W}$ such that
\[
\left| \mu _{ e } \left(
P,Q,H,x,\mathcal{X} \left( P,Q,H,\rho \right) \right) -\mu
_{ f } \left( P,Q,H,x,\mathcal{X} \left( P,Q,H,\rho \right)
\right) \right| \leq T
\]
holds for \emph{all} members $\left\{ e,f\right\} $ of $W$
\emph{and}
\[
\left| \mu _{ e } \left( P,Q,H,x,\mathcal{X} \left(
P,Q,H,\rho
\right) \right) -\mu _{ f } \left( P,Q,H,x,\mathcal{X}
\left( P,Q,H,\rho \right) \right) \right| =0
\]
holds for \emph{at least one} member $\left\{ e,f\right\} $
of
$W$, may be enclosed in an open subset of $\mathbb{W}
$ whose $d\left( \#\left( V\right) -1\right) $-volume is not
greater than any \emph{given} real number that is $>0$, and
furthermore if $\delta $ is any given real number $>0$,
then there exists a real number $r>0$ such that the subset
of $\mathbb{W}$ whose members are all the members $x$ of
$\mathbb{W}$ such that
\[
\left| \mu _{ e } \left( P,Q,H,x,\mathcal{X} \left(
P,Q,H,\rho
\right) \right) -\mu _{ f } \left( P,Q,H,x,\mathcal{X}
\left( P,Q,H,\rho \right) \right) \right| \leq T
\]
\label{Start of original page 218}
 holds for \emph{all} members $\left\{ e,f\right\} $ of $W$
\emph{and}
\[
\left| \mu _{ e } \left( P,Q,H,x,\mathcal{X} \left(
P,Q,H,\rho
\right) \right) -\mu _{ f } \left( P,Q,H,x,\mathcal{X}
\left( P,Q,H,\rho \right) \right) \right| <r
\]
holds for \emph{at least one} member $\left\{ e,f\right\} $
of
$W$, has $d\left( \#\left( V\right) -1\right) $-volume not
greater than $\delta $.

Hence by analogy with Lemma \ref{Lemma 23}, it directly
follows from
the assumed properties of the maps $\mathcal{J} $ and
$\tilde{ \mathcal{J} } $, together with our proof of the
\emph{absolute} convergence of $I \rule{0pt}{2.55ex}^{
\hspace{-0.75ex} * }\left( B,i,n,j,s,E,v,\rho
\right) $ and our $\rho $-independent bound on the
magnitude of \\
$I \rule{0pt}{2.55ex}^{ \hspace{-0.75ex} * }\left(
B,i,n,j,s,E,v,\rho \right) $, that
if $\delta $ is any given real number $>0$, then there
exists a real number $r>0$, \emph{independent of }$\rho $,
such that for all $\rho \in \mathbb{D} $, the contribution
to $I \rule{0pt}{2.55ex}^{ \hspace{-0.75ex} * }\left(
B,i,n,j,s,E,v,\rho \right) $ from the subset
of $\mathbb{W}$ whose members are all the members $x$ of
$\mathbb{W}$ such that
\[
\left| \mu _{ e } \left( P,Q,H,x,\mathcal{X} \left(
P,Q,H,\rho
\right) \right) -\mu _{ f } \left( P,Q,H,x,\mathcal{X}
\left( P,Q,H,\rho \right) \right) \right| <r
\]
holds for \emph{at least one} member $\left\{ e,f\right\} $
of
$W$, is not greater than $\delta $, \emph{and} such that
for all $\rho \in \mathbb{D} $ and for all $k\in \mathbb{N}
$, the contribution to $I \rule{0pt}{2.55ex}^{
\hspace{-0.75ex} \circ }_{ k } \left(
B,i,n,j,s,E,v,\rho \right) $ from that same subset of
$\mathbb{W} $, is also not greater than $\delta $.

So if $\varepsilon $ is any given real number $>0$, we
choose, by the preceding paragraph, a real number $r>0$,
independent of $\rho $, such that for all $\rho \in
\mathbb{D} $, the contribution to $I \rule{0pt}{2.55ex}^{
\hspace{-0.75ex} * }\left(
B,i,n,j,s,E,v,\rho \right) $ from the subset of $\mathbb{W}
$ whose members are all the members $x$ of $\mathbb{W}$ such
that
\[
\left| \mu _{ e } \left( P,Q,H,x,\mathcal{X} \left(
P,Q,H,\rho
\right) \right) -\mu _{ f } \left( P,Q,H,x,\mathcal{X}
\left( P,Q,H,\rho \right) \right) \right| <r
\]
holds for \emph{at least one} member $\left\{ e,f\right\} $
of
$W$, is not greater than $ \frac{ \varepsilon }{ 3 } $,
$and_{ } $ such that for all $\rho \in \mathbb{D} $ and
for all $k\in \mathbb{N} $, the contribution to $I
\rule{0pt}{2.55ex}^{ \hspace{-0.75ex} \circ }
_{ k } \left( B,i,n,j,s,E,v,\rho \right) $ from that same
subset of $\mathbb{W} $, is also not greater than
$ \frac{
\varepsilon }{ 3 } $.

And we then use the assumption mentioned in the third
paragraph on
page \pageref{Start of original page 217},
taking the $r$ of that assumption to be equal to the $r$ we
have just chosen, and the $\delta $ of that assumption to
be equal to $\varepsilon $ divided by ($3 $ times an
upper bound, independent of $\rho $, on the $d\left(
\#\left( V\right) -1\right) $-volume of the subset of
$\mathbb{W}$ whose members are all the members $x$ of
$\mathbb{W}$ such that
\[
\left| \mu _{ e } \left( P,Q,H,x,\mathcal{X} \left(
P,Q,H,\rho
\right) \right) -\mu _{ f } \left( P,Q,H,x,\mathcal{X}
\left( P,Q,H,\rho \right) \right) \right| \leq T
\]
holds for every member $\left\{ e,f\right\} $ of $W$,
(obtained by Lemma \ref{Lemma 14} and by analogy with
pages \pageref{Start of original page 116} and
\pageref{Start of original page 117}, and
pages \pageref{Start of original page 202} to
\pageref{Second line of original page 205}, as mentioned on
page \pageref{Start of original page 217}),
times the product, over the members $\alpha
$ of $\mathcal{D} \left( B\right) $, of an
\label{Start of original page 219}
 upper bound, independent of $\rho $, on $\left|
x_{ \mathcal{K}
\left( Q,B_{ \alpha } ,i_{ \alpha } \right) } -x_{ B_{
\alpha } } \right| $, for all members $x$ of $\mathbb{W}$
such
that
\[
\left| \mu _{ e } \left( P,Q,H,x,\mathcal{X} \left(
P,Q,H,\rho
\right) \right) -\mu _{ f } \left( P,Q,H,x,\mathcal{X}
\left( P,Q,H,\rho \right) \right) \right| \leq T
\]
holds for every member $\left\{ e,f\right\} $ of $W$, (again
obtained by Lemma \ref{Lemma 14} and by analogy with
pages \pageref{Start of original page 116} and
\pageref{Start of original page 117}, and
pages \pageref{Start of original page 202} to
\pageref{Second line of original page 205})).

And by the foregoing and that assumption, with the $r$ and
$\delta $ of that assumption taken as just specified, we
find by that assumption a member $k$ of $\mathbb{N} $,
\emph{independent of }$\rho $, such that for all members $l$
of
$\mathbb{N} $ such that $l\geq k$ holds, and for all
members $\rho $ of $\mathbb{D} $,
\[
\left| I \rule{0pt}{2.55ex}^{ \hspace{-0.75ex} \circ }_{ l
} \left( B,i,n,j,s,E,v,\rho \right)
-I \rule{0pt}{2.55ex}^{ \hspace{-0.75ex} * }\left(
B,i,n,j,s,E,v,\rho \right) \right| \leq
\frac{ \varepsilon }{ 3 } +
\frac{ \varepsilon }{ 3 } + \frac{ \varepsilon }{ 3 }
=\varepsilon
\]
holds.

And it follows immediately from this, together with our
$\rho $-independent bound on the magnitudes of $I
\rule{0pt}{2.55ex}^{ \hspace{-0.75ex} * }\left(
B,i,n,j,s,E,v,\rho \right) $ and all the $I
\rule{0pt}{2.55ex}^{ \hspace{-0.75ex} \circ }_{ k }
\left( B,i,n,j,s,E,v,\rho \right) $, $k\in \mathbb{N} $,
that the integrals with respect to $\rho $, over
$\mathbb{D} $, of any $ k $-independent product of finite
powers, all $\geq 0$ and independent of $k$ and $\rho $, of
the $\rho _{ K } $, $K\in \left( Q\,\vdash P\right) $,
times the $I \rule{0pt}{2.55ex}^{ \hspace{-0.75ex} \circ
}_{ k } \left( B,i,n,j,s,E,v,\rho \right)
$, $k\in \mathbb{N} $, form a Cauchy sequence, and this
completes the proof of Theorem \ref{Theorem 2}.

\section{Applications.}
\label{Section 8}

When the position-space $R $-operation is applied to a
single
Feynman subdiagram whose vertex positions are represented
by the member $x$ of $\mathbb{E} _{ d }^{ V } $, where $V$
is a finite set that has one member for each vertex of the
diagram, the unrenormalized integrand usually has the form
of a sum of a finite number of terms, each of which has the
form of a product of an internal function $I\left( x\right)
$ and an
external function $E\left( x\right) $.   The internal
function $I\left( x\right) $ is
assumed to depend on no position variables other than the
$x_{ A } $, $A\in V$, whereas the external function
$E\left( x\right) $
may depend also on other position variables that are not
involved in the renormalization of the subdiagram
represented by the finite set $V$.   The internal
propagators of the subdiagram are included in $I\left(
x\right) $, but
whether any other factors that occur are included in
$I\left( x\right) $
or in $E\left( x\right) $ may depend on the application.
For example,
any propagators with one end in the subdiagram represented
by $V$ and their other end \emph{not} in the subdiagram
represented by $V$, will be included as factors in $E\left(
x\right) $,
(and if the positions of their outer ends have not been
integrated over, they will be among the
\label{Start of original page 220}
 undisplayed position variables on which $E\left( x\right)
$ may depend),
while position-dependent coupling constants might be
included in $I\left( x\right) $, and position-dependent
background
classical fields might be included in $E\left( x\right) $.
 There may
also be, for example, functions forming part of a smooth
partition of unity, introduced for technical reasons in the
course of calculations.

The position-space $R $-operation for a single such term
consists of either leaving $I\left( x\right) $ unaltered,
or else
replacing it by a function $\tilde{ I }
 \left( x\right) $ that
is identical to $I\left( x\right) $ when all $\left| x_{ A
} -x_{ B }
\right| $,
$\left\{ A,B\right\} \in \mathcal{Q} \left( V\right) $, are
small enough, (for example, in our Theorems, smaller than a
fixed real number $S>0 $), but which may differ from
$I\left( x\right) $ by the inclusion of sufficiently smooth
\emph{long-distance} cutoffs, and replacing $E\left(
x\right) $ by its
Taylor expansion, up to a certain finite degree, about a
member $\tilde{ x }$ of $\mathbb{E} _{ d }^{ V } $,
where $\tilde{ x }_{ A }
\equiv \sum_{B\in V } \lambda _{ B } x_{ B } $ holds for
\emph{all} members $A$ of $V$, where the $\lambda _{ B } $,
$B\in V$, are real numbers, depending on $V$ and on $B$ but
on \emph{nothing else,} such that $\sum_{B\in V } \lambda _{
B } \equiv 1$ holds.   (For our convergence Theorems we
also assume that $\lambda _{ B } \geq 0$ holds for every
member $B$ of $V$, so that $\sum_{B\in V } \lambda _{ B }
x_{ B } $ is a member of the convex hull of the $x_{ A } $,
$A\in V $.)

The $R $-operation algebra is most conveniently carried out
by
introducing independent variables $y_{ i } \in \mathbb{E}
_{ d } $ for all the separate factors making up the
unrenormalized Feynman integrand, (for example, an
independent variable $y_{ i } \in \mathbb{E} _{ d } $ for
each propagator end, an independent variable $y_{ i } \in
\mathbb{E} _{ d } $ for each position-dependent classical
field or background field, and an independent variable $y_{
i } \in \mathbb{E} _{ d } $ for each position-dependent
coupling constant), and when we do this, it is natural to
take the set $V$ representing the vertices to be a
\emph{partition,} such that the members of the member of
$V$ corresponding to a given vertex of the Feynman diagram
correspond to all the separate propagator ends that meet at
that vertex in the unrenormalized Feynman diagram, together
with any other variables, (such as position-dependent
coupling constants), associated with that vertex.   (This
in fact arises naturally from the perturbation expansion
that generates the sum of Feynman diagrams.)
Then, as noted on
page \pageref{Start of original page 3},
each nonempty subset of the
partition $V$ corresponds to a member of $\Xi \left(
V\right) $.

In our Theorems we include all the factors making up the
\label{Start of original page 221}
 integrand of the unrenormalized Feynman diagram, (or all
the factors making up one of the finite number of terms in
that integrand), into a single function $\mathcal{J}_{
V } \left( y\right) $, $y\in \mathbb{E}_{ d }^{
 \mathcal{U} \left( V\right) } $,
where $V$ is a partition that has one member for each
vertex of the Feynman diagram, and the members of the
member of $V$ that corresponds to any given vertex of the
Feynman diagram, may be taken as corresponding to all the
independent position variables or factors associated with
that vertex, as just described, and we assume that the
information that determines which of the members $i$ of a
given member of $V$ are internal to, and which are external
to, a given subdiagram that includes the vertex
corresponding to that member of $V$, is coded into the
partition $H$, and interpreted by the function $\mathcal{T}
\left( A,H\right) $, as defined on
page \pageref{Start of original page 7}, where $A$ is
the member of $\Xi \left( V\right) $ corresponding to that
subdiagram, (so $A$ is equal to the union of the members of
$V$ associated with the vertices of that subdiagram).
Specifically, a member $i$ of $A$ is treated as
corresponding to an argument of the \emph{internal}
function of the subdiagram associated with $A$ if $i$ is a
member of $\left( A\,\vdash \mathcal{T} \left( A,H\right)
\right) $, (or in other words, if there \emph{does} exist a
member $B$ of $H$ such that $i\in B$ and $B\subseteq A$
both hold), and a member $i$ of $A$ is treated as
corresponding to an argument of the \emph{external}
function of the subdiagram associated with $A$ if $i$ is a
member of $\mathcal{T} \left( A,H\right) $, (or in other
words, if there is \emph{no} member $B$ of $H$ such that
$i\in B$ and $B\subseteq A$ both hold).   Propagators
correspond to two-member members of $H $;
position-dependent coupling constants, if they are required
to be included in the \emph{internal} function of
\emph{every} subdiagram that includes the vertex they are
associated with, may be coded as one-member members of
$H $; and any members $i$ of $\mathcal{U} \left( V\right) $
such that their corresponding variable $y_{ i } $ is
required to be treated as an argument of the
\emph{external} function of every subdiagram whose
associated member of $\Xi \left( V\right) $ has $i$ as a
member, may be identified as members of $\mathcal{U} \left(
V\right) $ that are either not members of \emph{any} member
of $H$, or else are not members of any member of $H$ that
is a subset of $\mathcal{U} \left( V\right) $.

In QCD it is sometimes necessary to impose
\emph{long-distance} cutoffs on the internal propagators of
a counterterm in order to avoid \emph{long-distance}
divergences in that counterterm.   These
\emph{long-distance} cutoffs on the propagators inside
counterterms determine the strong-interaction mass scale,
and they are the reason for allowing the replacement of the
internal function $I\left( x\right) $ of
\label{Start of original page 222}
 a single renormalized subdiagram by a modified function $
\tilde{I } \left( x\right) $ as on
page \pageref{Start of original page 220} above.   (Ward
identities may be restored at each order of perturbation
theory by the addition of finite counterterms of
non-BRS-invariant structure.)

It may be sufficient, for QCD, to have just two versions of
each propagator, (one with, and one without, the
long-distance cutoff), but in our $R $-operation we allow,
each time we contract a new subdiagram, the replacement of
each internal propagator of that subdiagram that is
\emph{not} an internal propagator of any previously
contracted subdiagram, by a modified version of that
propagator that agrees with the unmodified version whenever
the positions $y_{ i } \in \mathbb{E} _{ d } $ and $y_{ j }
\in \mathbb{E} _{ d } $ of the ends of that propagator are
such that $\left| y_{ i } -y_{ j } \right| \leq S$ holds,
where $S$ is a
fixed real number $>0$, and we allow the details of the
modification to depend on the identity of the subdiagram we
are contracting when the modification is made.   (We note
that this is certainly adequate for QCD.)

Thus if $V$ and $H$ satisfy the assumptions of Theorems
\ref{Theorem 1}
and \ref{Theorem 2}, and if $W$ is defined, as
on pages \pageref{Start of original page 93} and
\pageref{Start of original page 157},
to be the subset of $H$ whose members are all the members
$E$ of $H$ such that $E$ intersects exactly two members of
$V$, then for each member $\left\{ i,j\right\} $ of $W$, we
have the propagator associated with
$\left\{ i,j\right\} $ in
the unrenormalized Feynman diagram, which we denote by $G_{
\emptyset } $ and which is a member of $\left(
\mathbb{R} \cup
\left\{
+\infty ,-\infty \right\} \right)^{ \left( \mathbb{E} _{
d }^{ \left\{ i,j\right\} } \right) }
 $, and we also have, for each member $A$ of $\Xi
\left( V\right) $ such that $A$ is $\left( V\cup H\right)
$-connected and $\left\{ i,j\right\} \subseteq A$ holds, a
propagator $G_{ A } $ which is also a member of $\left(
\mathbb{R}
\cup \left\{ +\infty ,-\infty \right\} \right)^{ \left(
 \mathbb{E} _{
d }^{ \left\{ i,j\right\} } \right) } $.   For each
member $\left\{
\left( i,y_{ i } \right) ,\left( j,y_{ j } \right) \right\}
$ of $\mathbb{E} _{ d }^{ \left\{ i,j\right\} } $, and for
each member $A$ of $\left\{ \emptyset \right\} \cup \Xi
\left(
V\right) $ such that either $A=\emptyset $ holds or else
$A$ is
$\left( V\cup H\right) $-connected and $\left\{ i,j\right\}
\subseteq A$ holds, we define $G_{ Aij } \left( y_{ i }
,y_{ j } \right) \equiv G_{ Aji } \left( y_{ j } ,y_{ i }
\right) \equiv \left( G_{ A } \right)_{ \left\{ \left(
i,y_{ i } \right)
,\left( j,y_{ j } \right) \right\} } $.   (We note that
$\left\{ i,j\right\} $ may be identified from $G_{ A } $ for
any such $A$ by the fact that if $y$ is any member of
$\mathcal{D} \left( G_{ A } \right) $, then $\left\{
i,j\right\} =\mathcal{D} \left( y\right) $.)   We
then find, for each wood $F$ of $V$ such that every member
of $F$ is $\left( V\cup H\right) $-connected, (or in other
words, for each member $F$ of $\mathcal{G} \left(
V,H\right) $), that corresponding
\label{Start of original page 223}
 to the function $\mathcal{J}_{ V } \left( y\right) $ on
page \pageref{Start of original page 221},
we should take the function
\[
\mathcal{J}_{ F } \left( y\right) \equiv \left(
 \prod_{\Delta \equiv
\left\{ i,j\right\} \in W } G_{ \mathcal{Y} \left(
F,\left\{ i,j\right\} \right) ij } \left( y_{ i } ,
y_{ j } \right) \right) E\left(
\downarrow \left( y,\left( \mathcal{U} \left( V\right)
\,\vdash \mathcal{U} \left( W\right) \right) \right)
\right) ,
\]
where we note that, by the definition on
page \pageref{Start of original page 6} of the
function $\mathcal{Y} $, and for each member $\left\{
i,j\right\} $ of $W$, $\mathcal{Y} \left( F,\left\{
i,j\right\} \right) $ is equal to the \emph{smallest} member
$A$ of $F$ such that $\left\{ i,j\right\} \subseteq A$
holds,
if any members $A$ of $F$ exist such that $\left\{
i,j\right\} \subseteq A$ holds, and $\mathcal{Y} \left(
F,\left\{ i,j\right\} \right) $ is equal to $\emptyset $ if
there
are \emph{no} members $A$ of $F$ such that $\left\{
i,j\right\} \subseteq A$ holds, and where the factor
$E\left( \downarrow \left( y,\left( \mathcal{U} \left(
V\right) \,\vdash \mathcal{U} \left( W\right) \right)
\right) \right) $ contains all factors making up the
particular term under consideration of the integrand of the
Feynman diagram under consideration, \emph{other} than the
internal propagators of the diagram.

We note that the index $F$ on $\mathcal{J}_{ F }
\left( y\right)
$ thus arises \emph{entirely} because we have allowed
propagators to be modified inside a subdiagram when we
contract that subdiagram by the $R $-operation.

(We may avoid, if required, including $+\infty $ and
$-\infty $ in the range of the propagators $G_{ A } $, by
defining $G_{ Aij } \left( y_{ i } ,y_{ j } \right) $ to be
any arbitrary member of $\mathbb{R} $ on the measure-zero
subset of $\mathbb{E} _{ d }^{ \left\{ i,j\right\} } $ whose
members are all the members $\left\{ \left( i,y_{ i }
\right) ,\left( j,y_{ j } \right) \right\} $ of $\mathbb{E}
_{ d }^{ \left\{ i,j\right\} } $ such that
$\left| y_{ i } -y_{ j } \right| =0$ holds.)

Now with $\mathcal{J}_{ F } \left( y\right) $ defined as
above
for each member $F$ of $\mathcal{G} \left( V,H\right) $,
let $F$ and $G$ be any members of $\mathcal{G} \left(
V,H\right) $, and $y$ be any member of $\mathbb{E} _{
d }^{ \mathcal{U} \left( V\right) } $, such that for every
member $\left\{ i,j\right\} $ of $W$, either
 $\left| y_{ i } -y_{ j
} \right| \leq S$ holds or $\mathcal{Y} \left( F,
\left\{ i,j\right\}
\right) =\mathcal{Y} \left( G,\left\{ i,j\right\} \right) $
holds.   Then since we assume, for every member $\left\{
i,j\right\} $ of $W$, that if $A$ and $B$ are any members of
$\left\{ \emptyset \right\} \cup \Xi \left( V\right) $ such
that
$A$ is either equal to $\emptyset $ or else is $\left( V\cup
H\right) $-connected and such that $\left\{ i,j\right\}
\subseteq A$ holds, and $B$ is either equal to $\emptyset $
or
else is $\left( V\cup H\right) $-connected and such that
$\left\{ i,j\right\} \subseteq B$ holds, then $G_{ Aij }
\left( y_{ i } ,y_{ j } \right) =G_{ Bij } \left( y_{ i }
,y_{ j } \right) $ holds for all $y_{ i } \in \mathbb{E} _{
d } $ and all $y_{ j } \in \mathbb{E} _{ d } $ such that
$\left| y_{ i } -y_{ j } \right| \leq S$ holds, we find
immediately that
$\mathcal{J}_{ F } \left( y\right) =\mathcal{J}_{ G }
\left( y\right)
$ holds.   Hence the $\mathcal{J}_{ F } \left( y\right) $,
constructed as above, satisfy the relation that is required
to hold among the $\mathcal{J}_{ F } \left( y\right) $, for
distinct members $F$ of $\mathcal{G} \left( V,H\right) $,
in Lemma \ref{Lemma 22}, Theorem \ref{Theorem 1}, and
\mbox{Theorem \ref{Theorem 2}.}

If oversubtractions are required, they may be obtained in
Theorem \ref{Theorem 1} simply by choosing the integers
$\theta_{ \left\{
i,j\right\} } $ to be larger
\label{Start of original page 224}
 than they are required to be by the actual behaviour of
the propagators, while to obtain oversubtractions in
Theorem \ref{Theorem 2}, we must, due to the assumption
made in Theorem \ref{Theorem 2} that
$\theta_{ \left\{ i,j\right\} } <d$ holds for every
member $\left\{ i,j\right\} $ of $W$, introduce a second set
of integers $\tilde{ \theta }_{ \left\{ i,j\right\} } $
 such that
$\theta_{ \left\{ i,j\right\} } \leq \tilde{ \theta }_{
 \left\{
i,j\right\} } $ holds for every member
$\left\{ i,j\right\} $ of
$W$, and use the $\tilde{ \theta }_{
\left\{ i,j\right\} } $ in
place of the $\theta_{ \left\{ i,j\right\} } $
in the definition
of the integers $D_{ A } $ on
page \pageref{Start of original page 158}.   We may check
that this does not affect the proof of
Theorem \ref{Theorem 2}, since
the \emph{only} use made of the assumption that $\theta_{
\left\{ i,j\right\} } <d$ holds for every member $\left\{
i,j\right\} $ of $W$, is to obtain the convergence, on
pages \pageref{Start of original page 204}
to \pageref{Start of original page 206},
of the integrals over the key ends of the
key propagators of the members of $\mathbb{B} \left(
G\right) $.

We note that the condition $\theta_{
 \left\{ i,j\right\} } <d$
imposed on the integers
$\theta_{ \left\{ i,j\right\} } $ in
Theorem \ref{Theorem 2}, is compensated by the inclusion of
the
differential operator $ \left(
 \prod_{\alpha \in \mathcal{D}
\left(
g\right) } c_{ \alpha } .\hat{ y }_{ g_{
\alpha } } \right) $,
involving the arbitrary maps $g$ and $c$ as on
page \pageref{Start of original page 157},
in the integrand of the integral on
page \pageref{Start of original page 159}.   This is
consistent with the fact that the power-counting behaviour
of Feynman diagrams, including the Feynman diagrams for
QCD, arises from core propagators, each of which
contributes strictly less than $d$ to the degree of
divergence of the diagram, together with derivatives acting
on those core propagators, where the derivatives may either
act out of the vertices of the unrenormalized diagram, or
else may be part of the definition of the propagator.
(For example, for $d=4$ the Proca propagator has terms that
contribute $+4$ to the degree of divergence of a diagram,
but those terms have the form of two derivatives acting on
a core propagator that only contributes $+2$ to the degree
of divergence of a diagram.)

We do not require translation invariance for either of our
Theorems, but for Theorem \ref{Theorem 2} we require our
propagators to
have the property that if $v$ is any map such that
$\mathcal{D} \left( v\right) $ is finite, $\#\left(
\mathcal{D} \left( v\right) \right) \leq N$ holds, where
$N$ is the integer defined on
page \pageref{Start of original page 158}, and for each
member $\beta $ of $\mathcal{D} \left( v\right) $, $v_{
\beta } $ is a unit $d $-vector, then the power-counting
behaviour of $ \left( \prod_{\beta
\in \mathcal{D} \left( v\right)
} \left( v_{ \beta } .\left(
\hat{ y }_{ i } +\hat{ y }_{ j }
\right) \right) \right)
G_{ Aij } \left( y_{ i } ,y_{ j } \right) $, as
$\left| y_{ i } -y_{ j } \right| $ tends to $0$, where
$\left\{
i,j\right\} $ is any member of $W$, is \emph{not worse} than
the power-counting behaviour of $G_{ Aij } \left( y_{ i }
,y_{ j } \right) $ as $\left| y_{ i } -y_{ j } \right| $
tends to
\label{Start of original page 225}
$0$.   This immediately gives the inequality in the
third paragraph
of page \pageref{Start of original page 159}
 for the $\mathcal{J}_{ F } \left( y\right) $
as defined at the bottom of
page \pageref{Start of original page 223},
provided that the
function $E\left( \downarrow \left( y,\left( \mathcal{U}
\left( V\right) \,\vdash \mathcal{U} \left( W\right)
\right) \right) \right) $ is sufficiently well-behaved.

As an example of this behaviour, which we call ``translation
smooth'', we note that if $f$ is any \emph{completely
smooth}
function of $x\in \mathbb{E} _{ d } $ and $y\in \mathbb{E}
_{ d } $, $\alpha $ is any real number, and $u$ is any unit
$d $-vector, then $\left( u.\hat{ x }+u.\hat{ y }
\right) \left( f\left|
x-y\right|^{
-\alpha } \right) $ is equal to $\left( \left(
u.\hat{ x }+u.\hat{ y } \right) f\right)
\left| x-y\right|^{ -\alpha } $,
and that
since $\left( \left( u.\hat{ x }+u.\hat{ y }
\right) f\right) $ is also a
completely smooth function of $x$ and $y$, we may iterate
this result any finite number of times with any finite
sequence of unit $d $-vectors.

We recall from
page \pageref{Start of original page 7} that for any
ordered pair $\left(A,H\right) $ of
a set $A$, and a set $H$ such that every member of $H$ is a
set, we define $\mathcal{T} \left( A,H\right) $ to be the
subset of $A$ whose members are all the members $i$ of $A$
such that there is \emph{no} member $B$ of $H$ such that
$i\in B$ and $B\subseteq A$ both hold, and we also recall
from page \pageref{Start of original page 7}
that for any ordered
pair $\left(F,H\right) $ of a wood
$F$, and a set $H$ such that every member of $H$ is a set,
we define $\mathcal{O} \left( F,H\right) $ to be the set
whose members are all the members $i$ of $\mathcal{U}
\left( F\right) $ such that there \emph{exists} a member
$A$ of $F$ such that $i\in A$ holds and there is \emph{no}
member $B$ of $H$ such that $i\in B$ and $B\subseteq A$
both hold, and we recall from
page \pageref{Start of original page 73}
 that if $F$ and $G$
are any woods such that $\mathcal{M} \left( F\right)
=\mathcal{M} \left( G\right) $ holds, and $H$ is a set such
that every member of $H$ is a set, then $\mathcal{O} \left(
F,H\right) =\mathcal{O} \left( G,H\right) $ holds.

And we recall from
page \pageref{Start of original page 3} that for any
ordered pair $\left(F,i\right) $
of a set $F$ such that every member of $F$ is a set, and a
member $i$ of $\mathcal{U} \left( F\right) $, we define
$\mathcal{C} \left( F,i\right) $ to be the intersection of
all the members $A$ of $F$ such that $i\in A$ holds, and we
recall from
page \pageref{Start of original page 4}
 that if $F$ is a \emph{wood,} then
$\mathcal{C} \left( F,i\right) $ is equal to the unique
member of $\mathcal{M} \left( F\right) $ that has $i$ as a
member.

And we recall from
page \pageref{Start of original page 8}
 that for any ordered triple
$\left(F,A,B\right) $ of a wood $F$, a
\emph{nonempty} set $A$, and a set
$B$, we define $\mathbb{Y} \left( F,A,B\right) $ to be the
set whose members are all the members $C$ of $F$ such that
$A\subset C$ and $C\subseteq B$ both hold, and we note that
$\mathbb{Y} \left( F,A,B\right) $ is equal to the empty set
$\emptyset $ if $A\subset B$ does \emph{not} hold.

And we recall from
page \pageref{Start of original page 74}
 that for any ordered pair
$\left(F,H\right) $ of a wood $F$ and a
set $H$ such that every member of
$H$ is a set, we define $\mathbb{I} \left( F,H\right) $ to
be the map whose domain is $\mathbb{B} \left( F\right)
=\left( F\,\vdash \mathcal{M} \left( F\right) \right) $,
and such that for each member $A$ of $\mathbb{B} \left(
F\right) $, $\mathbb{I} _{ A } \left( F,H\right) \equiv
\left( \mathbb{I} \left( F,H\right) \right) _{ A } $ is the
set whose members are all the ordered
pairs $\left(i,B\right) $ of a
member $i$ of
\label{Start of original page 226}
 $\mathcal{T} \left( A,H\right) $ and a member $B$ of
$\mathbb{Y} \left( F,\mathcal{C} \left( F,i\right)
,A\right) $.

And we recall from
page \pageref{Start of original page 7}
 that for any ordered triple
$\left(F,H,i\right) $ of a wood $F$, a set $H$ such that
every member of
$H$ is a set, and a member $i$ of $\mathcal{O} \left(
F,H\right) $, we define $\mathcal{Z} \left( F,H,i\right) $
to be the \emph{largest} member $A$ of $F$ such that $i\in
A$ holds and there is \emph{no} member $B$ of $H$ such that
$i\in B$ and $B\subseteq A$ both hold.

And we note that it follows directly from
Lemma \ref{Lemma 20}, with
the wood $Q$ of Lemma \ref{Lemma 20} set equal to the wood
$F$ of
Lemma \ref{Lemma 20}, and from
page \pageref{Start of original page 75}, that if $F$ is
any wood,
and $H$ is any set such that every member of $H$ is a set,
then $\mathcal{U} \left( \mathcal{R} \left( \mathbb{I}
\left( F,H\right) \right) \right) $ is the set whose
members are all the ordered
pairs $\left(i,B\right) $ of a member $i$ of
$\mathcal{O} \left( F,H\right) $ and a member $B$ of
$\mathbb{Y} \left( F,\mathcal{C} \left( F,i\right)
,\mathcal{Z} \left( F,H,i\right) \right) $.

And we also recall from
page \pageref{Start of original page 7}
that for any ordered
triple $\left(F,A,i\right) $ of a wood $F$, a member
$A$ of $\left( \Xi
\left( \mathcal{M} \left( F\right) \right) \,\vdash
\mathcal{M} \left( F\right) \right) $, and a member $i$ of
$A$, we define $\mathcal{K} \left( F,A,i\right) $ to be the
unique member $B$ of $\mathcal{P} \left( F,A\right) $ such
that $i\in B$ holds.

Now in the position-space $R $-operation for a subdiagram
$A$,
such that $D_{ A } =0$ holds, the inner ends of the
\emph{legs} of $A$ are detached from the vertices of $A$
and shifted to the contraction point $x_{ A } $ of $A$.
And for general $D_{ A } $ the $R $-operation for $A$
Taylor-expands, up to a specified finite degree $\leq D_{ A
} $, the product of the leg functions of $A$, (or more
generally, the \emph{external} function of A), with respect
to the positions of their inner ends, \emph{about} the
position where all the inner ends are at $x_{ A } $.   The
\emph{internal} lines of $A$ are \emph{unaffected} by the
position space $R $-operation for $A$, apart from possible
replacements of the line functions $G_{ \emptyset ij }
\left(
x,y\right) $, $x\in \mathbb{E} _{ d } $, $y\in \mathbb{E}
_{ d } $, by $G_{ Aij } \left( x,y\right) $, as described
above.   The \emph{position arguments} $x\in \mathbb{E} _{ d
} $ and $y\in \mathbb{E} _{ d } $ of the ends of the
internal lines of $A$ are \emph{completely unaffected} by
the position-space $R $-operation for $A$.

Thus the $ i $-end of line $\left\{ i,j\right\} $,
where $\left\{
i,j\right\} $ is a member of $W$, gets shifted by the
$R $-operation for every member $A$ of $F$ such that $i\in
A$
and $j\notin A$ both hold.   Thus if $i\in A$ and $j\notin
A$ both hold, then the $R $-operation for $A$ shifts
the $ i $-end
of line $\left\{ i,j\right\} $ from $x_{ \mathcal{K} \left(
F,A,i\right) } $ to $x_{ A } $.

And the final position of the $ i $-end of line $\left\{
i,j\right\} $ is the contraction point of the largest member
of $F$ that has $i$ as a member but does \emph{not} have
$j$ as a member, which, if $V$, $H$ and $W$ are as in
Theorems $1$ and $2$, so that, in particular, $H$ is a
partition, so $\left\{ i,j\right\} $
\label{Start of original page 227}
 is the \emph{only} member of $H$ that has $i$ as a member,
and $\left\{ i,j\right\} $ is \emph{not} a subset of any
member of $V$, will be $x_{ \mathcal{Z} \left( F,H,i\right)
 } $.

And furthermore, if $i$ is any member of $\left(
\mathcal{U} \left( V\right) \,\vdash \mathcal{O} \left(
V,H\right) \right) =\left( \mathcal{U} \left( F\right)
\,\vdash \mathcal{O} \left( F,H\right) \right) $, then
there is \emph{no} member $A$ of $F$ such that $i\in
\mathcal{T} \left( A,H\right) $ holds, and $i$ is an
argument of the \emph{internal} function for \emph{every}
member $A$ of $F$ such that $i\in A$ holds, hence the
position argument associated with $i$ does \emph{not} get
shifted by the $R $-operation for \emph{any} member $A$ of
$F$, hence the final position of the position argument
associated with $i$, is $x_{ \mathcal{C}
\left( F,i\right) }
=x_{ \mathcal{C} \left( V,i\right) } $.

Hence for any member $i$ of $ \hspace{0.5ex}
\mathcal{U} \left( V\right) $,
the \emph{final} value of the position argument associated
with $i$, after doing the $R $-operations for \emph{all} the
members of $\mathbb{B} \left( F\right) $, where $F$ is any
wood of $V$ such that every member of $F$ is $\left( V\cup
H\right) $-connected, is $\eta _{ i } \left( F,H,x\right)
$, where for any ordered triple $\left(F,H,x\right) $ of
a wood $F$, a
set $H$ such that every member of $H$ is a set, and a
member $x$ of $\mathbb{F} _{ d } \left( \mathcal{M} \left(
F\right) \right) $, where $d$ is an integer $\geq 1$, we
defined the member $\eta \left( F,H,x\right) $ of
$\mathbb{E} _{ d }^{ \mathcal{U} \left( F\right) } $ on
page \pageref{Start of original page 78} by:
\[
\eta _{ i } \left( F,H,x\right) \equiv \left( \eta \left(
F,H,x\right) \right) _{ i } \equiv \left\{
\begin{array}{cl} x_{ \mathcal{Z} \left(
F,H,i\right) } & \hspace{2.0ex} \textrm{if }i\in
\mathcal{O} \left( F,H\right) \\
 x_{ \mathcal{C} \left( F,i\right) } & \hspace{2.0ex}
\textrm{if }i\in \left( \mathcal{U}
\left( F\right) \,\vdash \mathcal{O} \left( F,H\right)
\right) \end{array} \right.
\]

Now to be of practical use for QCD our $R $-operation must
not
generate any terms corresponding to counterterms in the
action whose total number of derivatives $+$ gluon fields
$+ \hspace{0.5ex} \frac{ 3 }{ 2 } $ times
quark fields is greater than
$4$.   Therefore, before we assign a degree of divergence
to a subdiagram, we must break up the Feynman integrand for
the subdiagram into a sum of terms such that in each term,
every derivative acting out of a vertex acts on a
\emph{specific} line.

Furthermore, when we apply the $R $-operation to a nest of
subdiagrams, we must, after each successive $R $-operation,
break up the sum of Taylor terms generated for the smaller
subdiagrams, and recalculate the new degree of divergence
for the next largest subdiagram separately for each term.
(If we do not do this, then for a nest of $n$ tripeds, all
of which share a common leg, we can generate a counterterm
with three gauge fields and $\left( n+1\right) $
derivatives.)

Now let $V$ and $H$ be partitions as in Theorems $1$ and
$2$, and let $F$ be any member of $\mathcal{G} \left(
V,H\right) $, or in other words, let $F$ be any wood of $V$
\label{Start of original page 228}
 such that every member of $F$ is $\left( V\cup H\right)
$-connected.

And for each member $\left(i,A\right) $ of $\mathcal{U}
\left(
\mathcal{R} \left( \mathbb{I} \left( F,H\right) \right)
\right) $, or in other words, for each
ordered pair $\left(i,A\right) $
of a member $i$ of $\mathcal{O} \left( F,H\right)
=\mathcal{O} \left( V,H\right) $, and a member $A$ of \\
$\mathbb{Y} \left( F,\mathcal{C} \left( F,i\right)
,\mathcal{Z} \left( F,H,i\right) \right) $, let $n_{ iA } $
be the number of derivatives, in a given BPHZ term, that
are produced by the $R $-operation for $A$ and act on the
position argument associated with the member $i$ of
$\mathcal{U} \left( V\right) $.

Then the most direct application of our rule just stated
would would restrict the $n_{ iA } $ for a given member $A$
of $\mathbb{B} \left( F\right) $, when we do the
$R $-operation for $A$, by the following \emph{Rule 1}:

\begin{bphzrule} \label{Rule 1}
\end{bphzrule}

\[
   \sum_{i\in \mathcal{T} \left( A,H\right) } n_{ iA
} \leq D_{ A } -  \sum_{B\in \left( \mathcal{P} \left(
F,A\right) \,\vdash V\right) } \left(
 D_{ B } - \sum_{\left(
i,C\right) \in \left( \mathbb{I} _{ B } \left( F,H\right)
\,\vdash \mathbb{I} _{ A } \left( F,H\right) \right)
} n_{ iC } \right) ,
\]
since the right-hand side of this inequality is the new
degree of divergence of $A$, for the particular BPHZ Taylor
terms for the members $C$ of $\mathbb{B} \left( F\right) $
such that $C\subset A$ holds, that are defined by the
integers $n_{ iC } \geq 0$ that occur in the right-hand
side of this inequality, calculated after having carried
out the $R $-operations (and the corresponding subdiagram
contractions) for all the members $C$ of $\mathbb{B} \left(
F\right) $ such that $C\subset A$ holds.

Now if we add to both sides of the above Rule \ref{Rule 1}
inequality the expression
\[
\left(
 \sum_{i\in \mathcal{T} \left( A,H\right) } \hspace{0.2cm}
   \sum_{C\in
\mathbb{Y} \left( F,\mathcal{C} \left( V,i\right)
,\mathcal{K} \left( F,A,i\right) \right) } n_{ iC }
\right)
= \sum_{B\in \left( \mathcal{P} \left( F,A\right) \,\vdash
V\right) } \left(
 \sum_{i\in \left( \mathcal{T} \left(
A,H\right) \cap B\right) } \hspace{0.2cm}
 \sum_{C\in \mathbb{Y} \left(
F,\mathcal{C} \left( V,i\right) ,B\right) } n_{ iC }
\right) =
\]
\[
\hspace{5.0cm} = \left(
 \sum_{B\in \left( \mathcal{P} \left( F,A\right) \,\vdash
V\right) } \hspace{0.2cm}
 \sum_{\left( i,C\right) \in \left( \mathbb{I}
_{ A } \left( F,H\right) \cap \mathbb{I} _{ B } \left(
F,H\right) \right) } n_{ iC } \right) ,
\]
(c.f.
page \pageref{Start of original page 109}),
we find that the above Rule \ref{Rule 1}
inequality is equivalent to
\[
\sum_{\left( i,C\right) \in \mathbb{I} _{ A }
\left( F,H\right)
 } n_{ iC } \leq D_{ A } -  \sum_{B\in \left( \mathcal{P}
\left( F,A\right) \,\vdash V\right) } \left( D_{ B }
-\sum_{\left( i,C\right) \in \mathbb{I} _{ B }
\left( F,H\right)
 } n_{ iC } \right) .
\]

However it is \emph{not} necessary to use a rule for the
Taylor degrees \emph{quite} as restrictive
as Rule \ref{Rule 1} in
order to avoid unwanted counterterms, and we choose
\emph{not} to use Rule \ref{Rule 1}, but rather to use the
following
\emph{Rule 2}:
\label{Start of original page 229}

\begin{bphzrule} \label{Rule 2}
\end{bphzrule}

\[
  \sum_{i\in \mathcal{T} \left( A,H\right) } n_{ iA }
\leq D_{ A } - \left(
 \sum_{i\in \mathcal{T} \left( A,H\right)
} \hspace{0.2cm} \sum_{C\in \mathbb{Y} \left( F,\mathcal{C}
\left( V,i\right)
,\mathcal{K} \left( F,A,i\right) \right) } n_{ iC }
\right) .
\]

Now if we add to both sides of this Rule \ref{Rule 2}
inequality the
same expression as before, we find
that this Rule \ref{Rule 2}
inequality is equivalent to
\[
\sum_{\left( i,C\right) \in \mathbb{I} _{ A }
\left( F,H\right)
 } n_{ iC } \leq D_{ A } .
\]

Now $ \displaystyle \sum_{\left( i,C\right)
\in \mathbb{I} _{ A } \left(
F,H\right) } n_{ iC } $ is equal to the total number
of derivatives, acting on the fields in the counterterm for
$A$, that are produced by the $R $-operations for all the
members $B$ of $\mathbb{B} \left( F\right) $ such that
$B\subseteq A$ holds.   Hence Rule \ref{Rule 2} is precisely
equivalent to requiring that the \emph{total} number of
derivatives acting on the fields in the counterterm for
$A$, or in other words, the total number of derivatives
acting on the fields in the counterterm for $A$ that are
produced by the $R $-operations for all the members $B$ of
$\mathbb{B} \left( F\right) $ such that $B\subseteq A$
holds, \emph{plus} the total number of derivatives that act
out of vertices of $A$ onto the legs of $A$, (or more
generally, onto the external function of $ A $), in the
\emph{unrenormalized} Feynman diagram, be $\leq $ $D_{ A }
$ plus the total number of derivatives that act out of
vertices of $A$ onto the legs of $A$, (or, more generally,
onto the external function of $ A $), in the unrenormalized
Feynman diagram.

Now for QCD with $d=4$, and bearing in mind our rule that
before determining the degree of divergence of any
subdiagram of the unrenormalized Feynman diagram, we break
up the integrand for the diagram into the sum of a finite
number of terms, in each of which, every derivative acting
out of a vertex acts on a \emph{specific} line or other
factor, (such as an external field), and we consider each
such term \emph{separately,} the degree of divergence $D_{
A } $ of any connected subdiagram $A$ of the unrenormalized
Feynman diagram, such that $A$ contains at least two
vertices of the unrenormalized Feynman diagram, (or in
other words, such that $A$ is a member of $\left( \Xi
\left( V\right) \,\vdash V\right) =\mathbb{B}
\left( \Xi \left( V\right) \right) $), is equal to
$4$ minus (the number of gluon legs of $A$ plus
$ \frac{ 3 }{
2 } $ times the number of quark legs of $A$ plus the number
of derivatives that act out of vertices of $A$ onto legs of
$A$, or more
\label{Start of original page 230}
 generally onto the external function of $A $).   Hence,
bearing in mind that our rules also ensure that there is
\emph{no} counterterm for any subdiagram $A$ such that $D_{
A } <0$ holds, (for there are then \emph{no} acceptable
integers $n_{ iA } \geq 0 $), we find, for QCD with
$d=4$, that Rule \ref{Rule 2} is precisely equivalent to
allowing
all BPHZ Taylor terms consistent with the requirement that
in the counterterm for $A$, the total number of gluon
fields plus $ \frac{ 3 }{ 2 } $ times quark fields plus
derivatives be $\leq 4$.

We note that, as is well known, the above formula for $D_{
A } $ for QCD with $d=4$ arises since each internal gluon
line contributes $+2$ to $D_{ A } $, each internal quark
line contributes $+3$ to $D_{ A } $, and each vertex of $A$
contributes, in the form of derivatives, $4$ minus (the
number of gluon legs of that vertex plus
$ \frac{ 3 }{ 2 } $
times the number of quark legs of that vertex plus the
number of derivatives that act out of that vertex onto the
legs of $A$, or more generally, onto the external function
of $ A $).   And furthermore, twice the number of internal
gluon lines of $A$ is equal to the total number of gluon
legs of the vertices of $A$, minus the number of external
gluon legs of $A$, and twice the number of internal quark
lines of $A$ is equal to the total number of quark legs of
the vertices of $A$, minus the number of external quark
legs of $A$.   Hence the total of the contributions to $D_{
A } $ listed so far, which corresponds to the total
contribution from the map $\theta $ in the notation of
Theorem \ref{Theorem 1}, and to the total contribution from
the maps
$\theta $ and $g$ in the notation of
Theorem \ref{Theorem 2}, is equal
to ($ 4$ times the number of vertices of $ A $) minus (the
number of gluon legs of $A$ plus $ \frac{ 3 }{ 2 } $ times
the
number of quark legs of $A$ plus the number of derivatives
acting from vertices of $A$ onto the legs of $A$, or more
generally, onto the external function of $ A $).   Hence
when
we add the remaining contribution, namely $-4$ times (one
less than the number of vertices of $ A $), we obtain the
stated result.

We note that if $A$ is any member of $\mathbb{B} \left(
F\right) $ such that $D_{ A } \geq 0$ holds and such that
there exists a member $B$ of $\mathcal{P} \left( F,A\right)
$ such that $B$ is a member of $\mathbb{B} \left( F\right)
$ and $D_{ B } \geq 1$ holds, then Rule \ref{Rule 2} allows
terms
with $n_{ iC } =0$ for all members $ \left( i,C\right) $
of $\mathcal{U}
\left( \mathcal{R} \left( \mathbb{I} \left( F,H\right)
\right) \right) $ such that $C\subset A$ holds, and $
\displaystyle \sum_{i\in \mathcal{T}
\left( A,H\right) } n_{ iA }
=D_{ A } $, which are forbidden by Rule \ref{Rule 1}.

We now recall from
page \pageref{Start of original page 59}
 that for any ordered pair
$\left( M,\theta \right) $ of a map $M$ such that
$\mathcal{D} \left( M\right) $ is finite, and every member
of $\mathcal{R} \left( M\right) $ is a
\label{Start of original page 231}
 finite set, and a map $\theta $ such that $\mathcal{D}
\left( M\right) \subseteq \mathcal{D} \left( \theta \right)
$ holds, and $\mathcal{R} \left( \theta \right) $ is a
subset of the set $\mathbb{Z} $ of all the integers, we
define $\mathbb{X} \left( M,\theta \right) $ to be the set
whose members are all the maps $p$ such that $\mathcal{D}
\left( p\right) =\mathcal{U} \left( \mathcal{R} \left(
M\right) \right) $ holds, $\mathcal{R} \left( p\right)
\subseteq \mathbb{N} $ holds, and for every member $A$ of
$\mathcal{D} \left( M\right) $, $ \sum_{\alpha \in M_{ A }
} p_{ \alpha } \leq \theta _{ A } $ holds.

And we note that this definition has the immediate
consequence that if, for any member $A$ of $\mathcal{D}
\left( M\right) $, $\theta _{ A } <0$ holds, then
$\mathbb{X} \left( M,\theta \right) $ is the empty set,
since there is no map $p$ with the required properties.

Thus $\mathbb{X} \left( \mathbb{I} \left( F,H\right)
,D\right) $, where $D$ is the map whose domain is $\left(
\Xi \left( V\right) \,\vdash V\right) $, and such that for
each member $A$ of $\left( \Xi \left( V\right) \,\vdash
V\right) $, $D_{ A } $ is the degree of divergence of $A$
as defined above, is the set whose members are all the maps
$n$ such that $\mathcal{D} \left( n\right) =\mathcal{U}
\left( \mathcal{R} \left( \mathbb{I} \left( F,H\right)
\right) \right) $ holds, (or in other words, such that
$\mathcal{D} \left( n\right) $ is the set of all the
ordered pairs $\left(i,B\right) $ of a
member $i$ of $\mathcal{O} \left(
F,H\right) =\mathcal{O} \left( V,H\right) $, and a member
$B$ of $\mathbb{Y} \left( F,\mathcal{C} \left( F,i\right)
,\mathcal{Z} \left( F,H,i\right) \right) $),
$\mathcal{R} \left( n\right) \subseteq \mathbb{N} $ holds,
and for every member $A$ of $\mathcal{D} \left( \mathbb{I}
\left( F,H\right) \right) $, or in other words, for every
member $A$ of $\mathbb{B} \left( F\right) =\left( F\,\vdash
V\right) $, $ \displaystyle \sum_{\left( i,B\right)
\in \mathbb{I} _{ A }
\left(
F,H\right) } n_{ iB } \leq D_{ A } $ holds.

We complete the definition of our $R $-operation by
specifying
that, when we do the $R $-operation for a member $A$ of
$\mathbb{B} \left( F\right) $ such that there exist one or
more members $B$ of $\mathbb{B} \left( F\right) $ such that
$B\subset A$ holds, (so that, by the rule for sequencing
$R $-operations, the $R $-operations for those members $B$
of
$\mathbb{B} \left( F\right) $ will have been done
\emph{before} we do the $R $-operation for $ A $), any
Taylor
numerator factors, such as $\left( x_{ \mathcal{K} \left(
F,B,i\right) } -x_{ B } \right) $, produced by the
$R $-operations for members $B$ of $\mathbb{B} \left(
F\right)
$ such that $B\subset A$ holds, are to be taken as part of
the \emph{internal} function for $A$ when we do the
$R $-operation for $A$, so they do \emph{not} get affected
or
operated on when we do the $R $-operation for $A$.

Finally we note that the $R $-operation for each member $A$
of
$\mathbb{B} \left( F\right) $ introduces a factor of $-1$,
resulting in an overall factor of $\left( -1\right)^{
\#\left( \mathbb{B} \left( F\right) \right) } $.

We may now confirm by induction that, in the notation of
Theorem \ref{Theorem 1}, the result of all the $R
$-operations for the
member $F$ of $\mathcal{G} \left( V,H\right) $, is
\[
\left(
-1\right)^{ \#\left( \mathbb{B} \left( F\right) \right) }
\hspace{-7.0pt}
 \sum_{n\in \mathbb{X} \left( \mathbb{I} \left( F,H\right)
,D\right) } \hspace{-3.0pt} \left( \hspace{-3.0pt} \left(
 \prod_{\left( i,B\right) \in \mathcal{U}
\left( \mathcal{R} \left( \mathbb{I} \left( F,H\right)
\right) \right) } \hspace{-2.0pt}
 \left( \frac{ \left( \left( x_{ \mathcal{K}
\left( F,B,i\right) } -x_{ B } \right) .\hat{ y }_{ i }
\right)^{
n_{ iB } } }{ n_{ iB } ! } \right) \hspace{-2.6pt} \right)
\hspace{-2.0pt}
\mathcal{J} _{ F } \left( y\right) \hspace{-2.0pt}
 \right)_{ \hspace{-3.0pt}
y=\eta \left( F,H,x\right) }.
\]
\label{Start of original page 232}

 We first note that the rule for sequencing the
$R $-operations for the members of $\mathbb{B} \left(
F\right)
$ is that, if $A$ and $B$ are any members of $\mathbb{B}
\left( F\right) $ such that $B\subset A$ holds, then we
must do the $R $-operation for $B$ \emph{before} we do the
$R $-operation for $A$.

Now our result is certainly true when $F$ is equal to $V$,
(or in other words, when $\mathbb{B} \left( F\right) $ is
empty), hence it will be sufficient to prove that if $F$ is
any member of $\mathcal{G} \left( V,H\right) $, and $A$ is
any $\left( V\cup H\right) $-connected member of $\left(
\Xi \left( V\right) \,\vdash V\right) $ such that if $B$ is
any member of $F$, then $B$ is either a \emph{strict}
subset of $A$, or else $B\cap A$ is empty, then our result
is true for $F\cup \left\{ A\right\} $ if it is true
 for $F$.

Let $F$ be any member of $\mathcal{G} \left( V,H\right) $,
and let $A$ be any $\left( V\cup H\right) $-connected
member of $\left( \Xi \left( V\right) \,\vdash V\right) $
such that if $B$ is any member of $F$, then $B$ is either a
\emph{strict} subset of $A$, or else $B\cap A$ is empty,
and assume that our result is true for $F$.

We first note that when we replace any internal propagators
of $A$, that are \emph{not} internal propagators of any
member of $F$, by modified propagators as described on
pages \pageref{Start of original page 221} to
\pageref{Start of original page 223},
(making the replacements
\emph{inside} the $ y $-derivatives), the result is the
replacement of $\mathcal{J}_{ F } \left( y\right) $ by
$\mathcal{J}_{ \left( F\cup \left\{ A\right\}
\right) } \left(
y\right) $, defined in accordance with
page \pageref{Start of original page 223}.

We next note that if $B$ is any member of $\mathbb{B}
\left( F\right) $, and $i$ is any member of $B$, then
$\mathcal{K} \left( F,B,i\right) $ is equal $\mathcal{K}
\left( \left( F\cup \left\{ A\right\} \right) ,B,i\right) $.

Now let $i$ be any member of $\mathcal{T} \left( A,H\right)
$.   Then $\mathcal{Z} \left( F,H,i\right) $, which by
definition is the largest member of $F$ to have $i$ as a
member but not contain as a subset any member of $H$ that
has $i$ as a member, is equal to $\mathcal{K} \left( \left(
F\cup \left\{ A\right\} \right) ,A,i\right) $, and
$\mathcal{Z} \left( \left( F\cup \left\{ A\right\} \right)
,H,i\right) $ is equal to $A$.   Hence $\left( x_{
\mathcal{Z} \left( F,H,i\right) }
 -x_{ \mathcal{Z} \left( \left(
F\cup \left\{ A\right\} \right) ,H,i\right) } \right) $ is
equal to $\left( x_{ \mathcal{K} \left( \left( F\cup \left\{
A\right\} \right) ,A,i\right) } -x_{ A } \right) $.

Furthermore, if $i$ is any member of $\mathcal{O} \left(
F,H\right) =\mathcal{O} \left( V,H\right) $ such that $i$
is \emph{not} a member of $\mathcal{T} \left( A,H\right) $,
then $\mathcal{Z} \left( \left( F\cup \left\{ A\right\}
\right) ,H,i\right) $ is equal to $\mathcal{Z} \left(
F,H,i\right) $, and furthermore, for \emph{every} member
$i$ of $\mathcal{U} \left( V\right) $, $\mathcal{C} \left(
\left( F\cup \left\{ A\right\} \right) ,i\right)
 =\mathcal{C}
\left( F,i\right) =\mathcal{C} \left( V,i\right) $ holds.

Hence, by the definition of $\eta \left( F,H,x\right) $ on
page \pageref{Start of original page 78} and on
page \pageref{Start of original page 227}, \\
$\eta _{ i } \left( \left(
F\cup \left\{ A\right\} \right) ,H,x\right) =\eta _{ i }
\left( F,H,x\right) $ holds for every member $i$ of
$\mathcal{U} \left( V\right) $ such that $i$ is \emph{not}
a member of $\mathcal{T} \left( A,H\right) $, while if $i$
\emph{is} a member of $\mathcal{T} \left( A,H\right) $,
then $\eta _{ i } \left( \left( F\cup \left\{ A\right\}
\right) ,H,x\right) =x_{ A } $ holds.

We next note that by our rule stated on
page \pageref{Start of original page 231}, all
Taylor numerators \\
$\left( x_{ \mathcal{K}
\left( F,B,i\right) }
-x_{ B } \right) $ coming from the $R $-operations for
\label{Start of original page 233}
 members $B$ of $F$ such that $B\subset A$ holds, are to be
included in the \emph{internal} function for $A$ when we do
the $R $-operation for $A$, hence are \emph{unaffected} by
the
$R $-operation for $A$, (other than re-writing $\mathcal{K}
\left( F,B,i\right) $ as $\mathcal{K} \left( \left( F\cup
\left\{ A\right\} \right) ,B,i\right) $ by the
identity noted
above).

This means that for each individual Taylor term produced by
the $R $-operation for $A$, (where each derivative
associated
with $A$ in that term, acts on the position variable
associated with a \emph{specific} member $i$ of
$\mathcal{T} \left( A,H\right) $), we may realize
the differentiations by \emph{restoring} the independent
$y_{ i } $ variables for all the members $i$ of
$\mathcal{U} \left( V\right) $, (which we may do
unambiguously since it simply corresponds to allowing all
the $\#\left( \mathcal{U} \left( V\right) \right) $
position-arguments of $\mathcal{J}_{ \left( F\cup \left\{
A\right\} \right) } \left( y\right) $ to be independent
variables again), then acting with the differential
operator
\[
\left(
 \prod_{i\in \mathcal{T} \left( A,H\right) } \left(
 \frac{ \left(
\left( x_{ \mathcal{Z} \left( F,H,i\right) }
 -x_{ \mathcal{Z}
\left( \left( F\cup \left\{ A\right\} \right) ,H,i\right) }
\right) .\hat{ y }_{ i } \right)^{ n_{ iA } } }{
n_{ iA } ! } \right) \right) = \hspace{-0.15pt}
 \hspace{6.0cm}
\]
\[
\hspace{6.0cm}
= \left(
 \prod_{i\in \mathcal{T} \left( A,H\right) } \left(
 \frac{ \left(
\left( x_{ \mathcal{K} \left( \left( F\cup \left\{ A\right\}
\right) ,A,i\right) } -x_{ A } \right) .\hat{ y }_{ i }
\right)^{ n_{
iA } } }{ n_{ iA } ! } \right) \right),
\]
where the integers $n_{ iA } \geq 0$, $i\in \mathcal{T}
\left( A,H\right) $, define the specific Taylor term for
$A$ which we are considering, then \emph{evaluating} the
$y_{ i } $ variables, $i\in \mathcal{U} \left( V\right) $,
by $y_{ i } =\eta _{ i } \left( \left( F\cup \left\{
A\right\} \right) ,H,x\right) $ for all members $i$ of
$\mathcal{U} \left( V\right) $, or in other words, $y=\eta
\left( \left( F\cup \left\{ A\right\}
\right) ,H,x\right) $.

We next note that $\mathbb{I} \left( \left( F\cup \left\{
A\right\} \right) ,H\right) $ is the map whose domain is
$\mathbb{B} \left( F\cup \left\{ A\right\} \right)
=\mathbb{B} \left( F\right) \cup \left\{ A\right\} $, and
such that for each member $C$ of $\mathbb{B} \left(
F\right) \cup \left\{ A\right\} $, $\mathbb{I} _{ C } \left(
\left( F\cup \left\{ A\right\} \right) ,H\right) $
is the set
whose members are all the ordered
pairs $\left(i,B\right) $ of a member
$i$ of $\mathcal{T} \left( C,H\right) $ and a member $B$ of
$\mathbb{Y} \left( \left( F\cup \left\{ A\right\} \right)
,\mathcal{C} \left( V,i\right) ,C\right) $.   Hence for
every member $C$ of $\mathbb{B} \left( F\right) $,
$\mathbb{I} _{ C } \left( \left( F\cup \left\{ A\right\}
\right) ,H\right) $ is equal to $\mathbb{I} _{ C } \left(
F,H\right) $.

And furthermore, $\mathcal{U} \left( \mathcal{R} \left(
\mathbb{I} \left( \left( F\cup \left\{ A\right\} \right)
,H\right) \right) \right) $ is the set whose members are
all the ordered pairs $\left(i,B\right) $ of a member
$ i $ of $\mathcal{O}\left( V,H \right) $ and  member $ B $
of \\
$ \mathbb{Y} \left( \left( F \cup \left\{ A \right\}
\right), \mathcal{C} \left( V,i \right), \mathcal{Z} \left(
\left( F \cup \left\{ A \right\} \right), H, i \right)
\right) $, hence $\mathcal{U} \left( \mathcal{R} \left(
\mathbb{I} \left( \left( F \cup \left\{ A \right\} \right)
, H \right) \right) \right) $ is equal to the disjoint
union of $\mathcal{U} \left( \mathcal{R} \left( \mathbb{I}
\left( F, H \right) \right) \right) $ and the set whose
members are all the ordered pairs $ \left( i, A\right) $,
$i \in \mathcal{T} \left( A,H \right) $.

Now by our induction assumption, the integers $ n_{ iC }
\geq 0 $, $\left( i,C \right) \in \mathcal{U} \left(
\mathcal{R} \left( \mathbb{I} \left( F,H \right) \right)
\right) $, are to be summed over the set $\mathbb{X} \left(
\mathbb{I} \left( F,H \right) , D \right) $,
\label{Start of original page 234}
which is the set whose members are all the maps $ n $ such
that $ \mathcal{D} \left( n \right) $ is equal to
$\mathcal{U} \left( \mathcal{R} \left( \mathbb{I} \left(
F,H \right) \right) \right) $, $\mathcal{R} \left( n
\right) \subseteq \mathbb{N} $ holds, and for every member
$ C $ of $\mathbb{B}\left( F \right) $, $\displaystyle
\sum_{ \left( i,B \right) \in \mathbb{I}_{ C } \left( F, H
\right) } n_{ iB } \leq D_{ C } $ holds, and by our
fundamental rule, stated on
page \pageref{Start of original page 227},
when we do the $ R
$-operation for $ A $, we are to break up the sum of Taylor
terms resulting from previous $ R $-operations into a sum
of terms in each of which every derivative resulting from
previous $ R $-operations acts on the position variable
associated with a \emph{specific} member $ i $ of
$\mathcal{U} \left( V \right) $, and determine the
summation range of the integers $ n_{ iA } \geq 0 $, $ i
\in \mathcal{T} \left( A,H \right) $, \emph{separately} for
each such term, or in other words, \emph{separately} for
each member $ n $ of $ \mathbb{X} \left( \mathbb{I} \left(
F,H \right) , D \right) $, and furthermore, as stated on
page \pageref{Start of original page 228},
we choose to determine the summation range of the
$ n_{ iA } \geq 0 $, $ i \in \mathcal{T} \left( A,H \right)
$, by \emph{Rule \ref{Rule 2}} as stated on
page \pageref{Start of original page 229},
(where we must replace the $ F $ in the Rule \ref{Rule 2}
inequality by $\left( F \cup \left\{ A \right\} \right) $
in the present case).

This means that after performing the $ R $-operation for $
A $ we have a double summation, where the \emph{inner}
summation, to be performed \emph{first}, is over the $ n_{
iA } \geq 0 $, $i \in \mathcal{T} \left( A,H \right) $,
subject to the Rule \ref{Rule 2} inequality as stated on
page \pageref{Start of original page 229},
with the $ F $ of that inequality replaced by
$\left( F \cup \left\{ A \right\} \right) $, and the
\emph{outer} summation, to be performed \emph{second}, is
over the members $ n $ of $\mathbb{X} \left( \mathbb{I}
\left( F,H \right) , D \right) $.

Thus we may now complete the proof of the induction step by
the use of Lemma \ref{Lemma 17}, with the map $ V $ of
Lemma \ref{Lemma 17} taken as our present map $\mathbb{I}
\left( \left( F \cup \left\{ A \right\} \right) , H \right)
$, the set $ J $ of Lemma \ref{Lemma 17} taken as the
subset $\mathbb{B} \left( F \right) $ of $\mathcal{D}
\left( \mathbb{I} \left( \left( F \cup \left\{ A \right\}
\right) , H \right) \right) $, (so that the set $
\downarrow \hspace{-1.0ex} \left( V,J \right) $
of Lemma \ref{Lemma 17} is
our present map $\mathbb{I} \left( F,H \right) $), the set
$ K $ of Lemma \ref{Lemma 17} taken as the one-member set $
\left\{ A \right\} $, the map $ W $ of Lemma \ref{Lemma 17}
taken as the map whose domain is the one-member set $
\left\{ A \right\} $, and such that $ W_{ A } $ is the set
whose members are all the ordered pairs $\left( i,B \right)
$ of a member $ i $ of $\mathcal{T} \left( A,H \right) $,
and a member $ B $ of $\left( \mathbb{Y} \left( \left( F
\cup \left\{ A \right\} \right) , \mathcal{C} \left( V,i
\right) , A \right) \, \vdash \mathbb{Y} \left( F,
\mathcal{C} \left( V,i \right) , \mathcal{Z} \left( F,H,i
\right) \right) \right) = \left\{ A \right\} $, (hence $
W_{ A } $ is the set whose members are all the ordered
pairs $ \left( i,A \right) $, $i \in \mathcal{T} \left( A,H
\right) $), the map $ \theta $ of Lemma \ref{Lemma 17}
taken as our present map $ D $, and for each member $ u $
of $\mathbb{N}^{ \mathcal{U} \left( \mathcal{R} \left(
\mathbb{I} \left( F,H \right) \right) \right) } $, the map
$ \zeta \left( u \right) $ of Lemma \ref{Lemma 17} taken as
the map whose domain is the one-member set $ \left\{ A
\right\} $, and such that
\[
\zeta_{ A } \left( u \right) \equiv \sum_{ \left( i,B
\right) \in \left( \mathbb{I}_{ A } \left( \left( F \cup
\left\{ A \right\} \right) , H \right) \cap \mathcal{U}
\left( \mathcal{R} \left( \mathbb{I} \left( F,H \right)
\right) \right) \right) } = \hspace{-21.3pt}
 \hspace{9.0cm}
\]
\label{Start of original page 235}
\[
= \left( \sum_{ i \in \mathcal{T} \left( A,H \right) }
\hspace{0.2cm} \sum_{ B \in \mathbb{Y} \left( F,
\mathcal{C} \left( V,i \right) , \mathcal{Z} \left( F,H,i
\right) \right) } u_{ iB } \right) =
\]
\[
\hspace{7.0cm} \hspace{-2.8pt}
= \left( \sum_{i \in \mathcal{T} \left( A,H \right) }
\hspace{0.2cm} \sum_{ B \in \mathbb{Y} \left( \left( F \cup
\left\{ A \right\} \right) , \mathcal{C} \left( V,i \right)
, \mathcal{K} \left( \left( F \cup \left\{ A \right\}
\right) , A,i \right) \right) } u_{ iB } \right)
\]
holds, and the map $ Y $ of Lemma \ref{Lemma 17} taken as
the map whose domain is equal to $\mathcal{U} \left(
\mathcal{R} \left( \mathbb{I} \left( \left( F \cup \left\{
A \right\} \right) , H \right) \right) \right) $, or in
other words, whose domain is equal to the disjoint union of
$\mathcal{U} \left( \mathcal{R} \left( \mathbb{I} \left(
F,H \right) \right) \right) $ and the set whose members are
all the ordered pairs $ \left( i,A \right) $, $ i \in
\mathcal{T} \left( A,H \right) $, and such that for each
member $ \left( i,B \right) $ of $\mathcal{U} \left(
\mathcal{R} \left( \mathbb{I} \left( \left( F \cup \left\{
A \right\} \right) , H \right) \right) \right) $, $ Y_{ iB
} $ is equal to the differential operator $\left( \left(
x_{ \mathcal{K} \left( \left( F \cup \left\{ A \right\}
\right) , B,i \right) } - x_{ B } \right) . \hat{ y }_{ i }
\right) $.

We note that when we make these choices for the sets that
occur in Lemma \ref{Lemma 17}, the set $\mathcal{U} \left(
\mathcal{R} \left( W \right) \right) $ of Lemma
\ref{Lemma 17} is the set whose members are all the ordered
pairs $\left( i,A \right) $, $ i \in \mathcal{T} \left( A,H
\right) $, and the set $ \mathbb{X} \left( W, \left( \theta
- \zeta \left( u \right) \right) \right) $ of Lemma
\ref{Lemma 17}, where $ u $ is any member of $\mathbb{N}^{
\mathcal{U} \left( \mathcal{R} \left( \mathbb{I} \left( F,H
\right) \right) \right) } $, is the set whose members are
all the maps $ v $ such that $\mathcal{D} \left( v \right)
$ is the set $\mathcal{U} \left( \mathcal{R} \left( W
\right) \right) $ of all the ordered pairs $ \left( i,A
\right) $, $ i \in \mathcal{T} \left( A,H \right) $,
$\mathcal{R} \left( v \right) \subseteq \mathbb{N} $ holds,
and the \mbox{Rule \ref{Rule 2}} inequality holds for
$ A $ in the form
\[
\sum_{ i \in \mathcal{T} \left( A,H \right) } v_{ iA } \leq
D_{ A } - \left( \sum_{ i \in \mathcal{T} \left( A,H
\right) } \hspace{0.2cm} \sum_{ B \in \mathbb{Y} \left(
\left( F \cup \left\{ A \right\} \right) , \mathcal{C}
\left( V,i \right) , \mathcal{K} \left( \left( F \cup
\left\{ A \right\} \right) , A,i \right) \right) } u_{ iB }
\right) .
\]

\vspace{0.6cm}

We note that, as is well known, if $ a $ and $ b $ are any
members of $\mathbb{R} $ such that $a < b $ holds, then
thre exist members $ f $ of $\mathbb{R}^{ \mathbb{R} } $
such that $ f\left( r \right) $ is equal to $ 1 $ for all
members $ r $ of $\mathbb{R} $ such that $ r \leq a $
holds, $f \left( r \right) $ is equal to $ 0 $ for all
members $ r $ of $\mathbb{R} $ such that $r \geq b $ holds,
and $ f $ is infinitely differentiable with respect to $ r
$ for \emph{all} members $ r $ of $\mathbb{R} $.  For
example, a possible form of $ f\left( r \right) $ for $ a
\leq r \leq b $, is $ f\left( r \right)
= \displaystyle \frac{ 1 }{ 1 +
e^{ \frac{ -1 }{ r - a } } e^{ \frac{ 1 }{ b - r } } } $.
(We note that if we choose $ f\left( r \right) $ for $a
\leq r \leq b $ as in this example, then the identity $
f\left( r \right) + f\left( a + b - r \right) = 1 $ holds
for \emph{all} $r \in \mathbb{R} $.)

Then for $\mathcal{J}_{ F } \left( y \right) $, $ F \in
\mathcal{G} \left( V,H \right) $, defined as on
page \pageref{Start of original page 223}, we may
\label{Start of original page 236}
construct sequences $\mathcal{J}_{ kF } \left( y \right) $,
$ k \in \mathbb{N} $, satisfying the requirements of
Theorem \ref{Theorem 2}, by replacing the propagators $ G_{
Aij } \left( y_{ i }, y_{ j } \right) $ by sequences $ G_{
kAij } \left( y_{ i }, y_{ j } \right) $, $ k \in
\mathbb{N} $, of regularized propagators, which may be
obtained from the propagators $ G_{ Aij } \left( y_{ i },
y_{ j } \right) $ either by convoluting these propagators
with a sequence of normalized, spherically symmetric
smearing functions, differentiable at least $ N $ times,
where $ N $ is the integer defined on
page \pageref{Start of original page 158}, and which
vanish identically outside a sequence of radii $ b_{ k } $
which tend to $ 0 $ as $ k $ tends to $ \infty $, or else
simply by \emph{multiplying} these propagators by a
sequence $ f_{ k } \left( \left| y_{ i } - y_{ j } \right|
\right) $ of functions, each differentiable at least $ N $
times, where $ N $ is the integer defined on
page \pageref{Start of original page 158}, such
that for all $ k \in \mathbb{N} $, real numbers $ a_{ k } $
and $ b_{ k } $ exist such that $ 0 < a_{ k } < b_{ k } $
holds, $ f_{ k } \left( r \right) = 0 $ holds for all $ 0
\leq r \leq a_{ k } $, and $ f_{ k } \left( r \right) = 1 $
holds for all $
r \geq b_{ k } $, and the sequences $ a_{ k } $ and $ b_{ k
} $, $ k \in \mathbb{N} $, tend to $ 0 $ as $ k $ tends to
$ \infty $.

We observe that we made just three uses of the regularized
integrands in \mbox{Theorem \ref{Theorem 2}.}

Firstly we needed the regularization to justify the use, on
page \pageref{Start of original page 162},
of the Taylor remainder result of Lemma
\ref{Lemma 22}.  For unlike the case of Theorem
\ref{Theorem 1}, where the good set of woods conditions
and Lemma \ref{Lemma 14} ensured that no propagator
affected by the Taylor remainder had its arguments
coinciding for any $\rho \in \mathbb{D} $, in Theorem
\ref{Theorem 2} it \emph{is} possible for the arguments of
a \emph{key} propagator of a member of $\mathbb{B} \left( G
\right) $ to coincide for some $\rho \in \mathbb{D} $, so
we need the regularization to justify the use of Taylor
remainder in this instance.  (We separated the analysis of
the convergence of the integrals associated with the key
propagators of the members of $\mathbb{B} \left( G \right)
$, from the analysis of the convergence of the integrals
associated with the firm components of the members of
$\mathbb{B} \left( G \right) $, by transforming to the $
a_{ K } $ integration variables as defined on
pages \pageref{Start of original page 178} and
\pageref{Start of original page 179}.)

Secondly we used the regularizations on
page \pageref{Start of original page 163} to justify
swapping the order of the $\rho $-integrations and the $ x
$-integrations, to enable us to do the $ x $-integrals
before doing the $\rho $-integrals.

And\hspace{\stretch{1}} thirdly\hspace{\stretch{1}}
we\hspace{\stretch{1}} used\hspace{\stretch{1}}
the\hspace{\stretch{1}} regularizations\hspace{\stretch{1}}
on\hspace{\stretch{1}} pages\hspace{\stretch{1}}
\pageref{Start of original page 182}\hspace{\stretch{1}}
and\hspace{\stretch{1}}
\pageref{Start of original page 183}\hspace{\stretch{1}}
to\hspace{\stretch{1}} obtain\hspace{\stretch{1}} the\\
integration-by-parts result stated on those pages.

We did not at any stage need the actual evaluation of any
regularized integral, and furthermore, as demonstrated on
pages \pageref{Start of original page 191} to
\pageref{Start of original page 216},
our integrals, after carrying out all the
integrations by parts, as done on
pages \pageref{Start of original page 180} to
\pageref{Start of original page 191}, to
remove all the derivatives
\label{Start of original page 237}
from all the \emph{key} propagators of the members of
$\mathbb{B} \left( G \right) $, are \emph{absolutely}
convergent.

Thus in practical calculations it may be possible to
proceed without introducing any explicit regularization,
allowing the use of Taylor remainder, swapping the order of
the $\rho $-integrations and the $ x $-integrations, and
integrating by parts as above, \emph{as if} our integrand
was regularized, provided that we strictly restrict our use
of such manipulations to the manner in which they are used
in our proof of Theorem \ref{Theorem 2}, where they are
justified by the introduction of regularizations.  This
would be analogous to the possibility of carrying out a
certain strictly limited class of manipulations with the
Dirac $\delta $-function, without the necessity of defining
it formally, via sequences of functions, as a generalized
function.

Finally we note that the crucial properties of the set
$\Omega \left( H, \sigma, R, x \right) $, defined on
page \pageref{Start of original page 20},
that enable us to prove Theorem \ref{Theorem 1},
are Lemma \ref{Lemma 7}, which with Lemma \ref{Lemma 8}
makes possible the definition of good sets of woods, Lemma
\ref{Lemma 10}, which gives lower bounds, used on
page \pageref{Start of original page 112},
on $\mathbb{L} \left( P,A,x \right) $ for the members $ A $
of $\mathbb{B} \left( P \right) $ if $\left( P,Q \right) $
generates a good set of woods for $ \left( H, \sigma, R, x
\right) $, and Lemma \ref{Lemma 11}, used on
page \pageref{Start of original page 121},
which ensures that if $ A $ is any member of $\mathbb{B}
\left( \bar{ P } \right) $ and $ X $ is any $\sigma
$-cluster of $\downarrow \hspace{-0.5ex}
\left( x, \mathcal{P} \left( P,A
\right) \right) $, then $\left( V \cup H \right)
$-connected components of $\mathcal{U} \left( X \right) $
will under appropriate conditions be members of $\left( Q
\, \vdash P \right) $.

There may be alternative definitions of $\Omega \left( H,
\sigma, R, x \right) $, \emph{not} equivalent to that given
on page \pageref{Start of original page 20},
which nevertheless allow analogues of
these three properties to be derived.  One example of an
alternative definition, which may possibly be adequate,
would say that a member $\left( P,Q \right) $ of
$\mathcal{N} \left( V,H \right) $ is a member of $\Omega
\left( H, \sigma, R, x \right) $ ifif $\mathbb{L} \left(
Q,A,x \right) < R $ holds for all $ A \in \left( Q \,
\vdash P \right) $, and for all $ \left\{ i,j \right\} \in
W $ such that $\mathcal{Z} \left( Q,H,i \right) \in \left(
Q \, \vdash P \right) $ holds, there exists a member $ B $
of $ P $ such that $ B \subseteq \mathcal{Z} \left( Q,H,j
\right) $ holds, there is \emph{no} member $ C $ of $ P $
such that $ B \subset C \subseteq \mathcal{Z} \left( Q,H,j
\right) $ holds, and $\mathbb{L} \left( Q, \mathcal{Z}
\left( Q,H,i \right) , x \right) < \sigma \left| x_{
\mathcal{Z} \left( Q,H,i \right) } - x_{ B } \right| $
holds.

\end{document}